\documentclass{pasj01}

\RequirePackage{mathpazo}
\RequirePackage[T1]{fontenc}

\usepackage[OT2,T1]{fontenc}

\def\commenta{$^*$}
\def\commentb{$^\dagger$}
\def\commentc{$^\ddagger$}
\def\commentd{$^\S$}
\def\commente{$^\|$}
\def\commentf{$^\#$}

\newcounter{author}
\def\authorcount#1#2{\refstepcounter{author}\label{#1}
                     \altaffiltext{\ref{#1}}{#2}}

\def\Isogaiprep{K. Isogai et al. in preparation}

\def\Wakamatsuprep{Y. Wakamatsu et al. in preparation}

\begin{document}
\SetRunningHead{T. Kato et al.}{Period Variations in SU UMa-Type Dwarf Novae IX}

\Received{201X/XX/XX}
\Accepted{201X/XX/XX}

\title{Survey of Period Variations of Superhumps in SU UMa-Type Dwarf Novae.
    IX: The Ninth Year (2016--2017)}

\author{Taichi~\textsc{Kato},\altaffilmark{\ref{affil:Kyoto}*}
        Keisuke~\textsc{Isogai},\altaffilmark{\ref{affil:Kyoto}}
        Franz-Josef~\textsc{Hambsch},\altaffilmark{\ref{affil:GEOS}}$^,$\altaffilmark{\ref{affil:BAV}}$^,$\altaffilmark{\ref{affil:Hambsch}}
        Tonny~\textsc{Vanmunster},\altaffilmark{\ref{affil:Vanmunster}}
        Hiroshi~\textsc{Itoh},\altaffilmark{\ref{affil:Ioh}}
        Berto~\textsc{Monard},\altaffilmark{\ref{affil:Monard}}$^,$\altaffilmark
{\ref{affil:Monard2}}
        Tam\'as~\textsc{Tordai},\altaffilmark{\ref{affil:Polaris}}
        Mariko~\textsc{Kimura},\altaffilmark{\ref{affil:Kyoto}}
        Yasuyuki~\textsc{Wakamatsu},\altaffilmark{\ref{affil:Kyoto}}
        Seiichiro~\textsc{Kiyota},\altaffilmark{\ref{affil:Kis}}
        Ian~\textsc{Miller},\altaffilmark{\ref{affil:Miller}}
        Peter~\textsc{Starr},\altaffilmark{\ref{affil:Starr}}
        Kiyoshi~\textsc{Kasai},\altaffilmark{\ref{affil:Kai}}
        Sergey~Yu.~\textsc{Shugarov},\altaffilmark{\ref{affil:Sternberg}}$^,$\altaffilmark{\ref{affil:Slovak}}
        Drahomir~\textsc{Chochol},\altaffilmark{\ref{affil:Slovak}}
        Natalia~\textsc{Katysheva},\altaffilmark{\ref{affil:Sternberg}}
        Anna~M.~\textsc{Zaostrojnykh},\altaffilmark{\ref{affil:Kazan}}
        Matej~\textsc{Seker\'a\v{s}},\altaffilmark{\ref{affil:Slovak}}
        Yuliana~G.~\textsc{Kuznyetsova},\altaffilmark{\ref{affil:MainUkraine}}
        Eugenia~S.~\textsc{Kalinicheva},\altaffilmark{\ref{affil:Moscow}}
        Polina~\textsc{Golysheva},\altaffilmark{\ref{affil:Sternberg}}
        Viktoriia~\textsc{Krushevska},\altaffilmark{\ref{affil:MainUkraine}}
        Yutaka~\textsc{Maeda},\altaffilmark{\ref{affil:Mdy}}
        Pavol~A.~\textsc{Dubovsky},\altaffilmark{\ref{affil:Dubovsky}}
        Igor~\textsc{Kudzej},\altaffilmark{\ref{affil:Dubovsky}}
        Elena~P.~\textsc{Pavlenko},\altaffilmark{\ref{affil:CrAO}}$^,$\altaffilmark{\ref{affil:CrimeanFU}}
        Kirill~A.~\textsc{Antonyuk},\altaffilmark{\ref{affil:CrAO}}
        Nikolaj~V.~\textsc{Pit},\altaffilmark{\ref{affil:CrAO}}
        Aleksei~A.~\textsc{Sosnovskij},\altaffilmark{\ref{affil:CrAO}}
        Oksana~I.~\textsc{Antonyuk},\altaffilmark{\ref{affil:CrAO}}
        Aleksei~V.~\textsc{Baklanov},\altaffilmark{\ref{affil:CrAO}}
        Roger~D.~\textsc{Pickard},\altaffilmark{\ref{affil:BAAVSS}}$^,$\altaffilmark{\ref{affil:Pickard}}
        Naoto~\textsc{Kojiguchi},\altaffilmark{\ref{affil:OKU}}
        Yuki~\textsc{Sugiura},\altaffilmark{\ref{affil:OKU}}
        Shihei~\textsc{Tei},\altaffilmark{\ref{affil:OKU}}
        Kenta~\textsc{Yamamura},\altaffilmark{\ref{affil:OKU}}
        Katsura~\textsc{Matsumoto},\altaffilmark{\ref{affil:OKU}}
        Javier~\textsc{Ruiz},\altaffilmark{\ref{affil:Ruiz1}}$^,$\altaffilmark{\ref{affil:Ruiz2}}$^,$\altaffilmark{\ref{affil:Ruiz3}}
        Geoff~\textsc{Stone},\altaffilmark{\ref{affil:AAVSO}}
        Lewis~M.~\textsc{Cook},\altaffilmark{\ref{affil:LewCook}}
        Enrique~de~\textsc{Miguel},\altaffilmark{\ref{affil:Miguel}}$^,$\altaffilmark{\ref{affil:Miguel2}}
        Hidehiko~\textsc{Akazawa},\altaffilmark{\ref{affil:OUS}}
        William~N.~\textsc{Goff},\altaffilmark{\ref{affil:Goff}}
        Etienne~\textsc{Morelle},\altaffilmark{\ref{affil:Morelle}}
        Stella~\textsc{Kafka},\altaffilmark{\ref{affil:AAVSO}}
        Colin~\textsc{Littlefield},\altaffilmark{\ref{affil:LCO}}
        Greg~\textsc{Bolt},\altaffilmark{\ref{affil:Bolt}}
        Franky~\textsc{Dubois},\altaffilmark{\ref{affil:Dubois}}
        Stephen~M.~\textsc{Brincat},\altaffilmark{\ref{affil:Brincat}}
        Hiroyuki~\textsc{Maehara},\altaffilmark{\ref{affil:OAO}}
        Takeshi~\textsc{Sakanoi},\altaffilmark{\ref{affil:TohokuPlanetary}}
        Masato~\textsc{Kagitani},\altaffilmark{\ref{affil:TohokuPlanetary}}
        Akira~\textsc{Imada},\altaffilmark{\ref{affil:Hamburg}}$^,$\altaffilmark{\ref{affil:HidaKwasan}}
        Irina~B.~\textsc{Voloshina},\altaffilmark{\ref{affil:Sternberg}}
        Maksim~V.~\textsc{Andreev},\altaffilmark{\ref{affil:Terskol}}$^,$\altaffilmark{\ref{affil:ICUkraine}}
        Richard~\textsc{Sabo},\altaffilmark{\ref{affil:Sabo}}
        Michael~\textsc{Richmond},\altaffilmark{\ref{affil:RIT}}
        Tony~\textsc{Rodda},\altaffilmark{\ref{affil:Rodda}}
        Peter~\textsc{Nelson},\altaffilmark{\ref{affil:Nelson}}
        Sergey~\textsc{Nazarov},\altaffilmark{\ref{affil:CrAO}}
        Nikolay~\textsc{Mishevskiy},\altaffilmark{\ref{affil:AAVSO}}
        Gordon~\textsc{Myers},\altaffilmark{\ref{affil:Myers}}
        Denis~\textsc{Denisenko},\altaffilmark{\ref{affil:Sternberg}}
        Krzysztof~Z.~\textsc{Stanek},\altaffilmark{\ref{affil:Ohio}}
        Joseph~V.~\textsc{Shields},\altaffilmark{\ref{affil:Ohio}}
        Christopher~S.~\textsc{Kochanek},\altaffilmark{\ref{affil:Ohio}}
        Thomas~W.-S.~\textsc{Holoien},\altaffilmark{\ref{affil:Ohio}}
        Benjamin~\textsc{Shappee},\altaffilmark{\ref{affil:Carnegie}}
        Jos\'e~L.~\textsc{Prieto},\altaffilmark{\ref{affil:DiegoPortales}}$^,$\altaffilmark{\ref{affil:Princeton}}
        Koh-ichi~\textsc{Itagaki},\altaffilmark{\ref{affil:Itagaki}}
        Koichi~\textsc{Nishiyama},\altaffilmark{\ref{affil:MiyakiObs}}
        Fujio~\textsc{Kabashima},\altaffilmark{\ref{affil:MiyakiObs}}
        Rod~\textsc{Stubbings},\altaffilmark{\ref{affil:Stubbings}}
        Patrick~\textsc{Schmeer},\altaffilmark{\ref{affil:Schmeer}}
        Eddy~\textsc{Muyllaert},\altaffilmark{\ref{affil:VVSBelgium}}
        Tsuneo~\textsc{Horie},\altaffilmark{\ref{affil:Heo}}
        Jeremy~\textsc{Shears},\altaffilmark{\ref{affil:Shears}}$,$\altaffilmark{\ref{affil:BAAVSS}}
        Gary~\textsc{Poyner},\altaffilmark{\ref{affil:Poyner}}
        Masayuki~\textsc{Moriyama},\altaffilmark{\ref{affil:Myy}}
}

\authorcount{affil:Kyoto}{
     Department of Astronomy, Kyoto University, Kyoto 606-8502, Japan}
\email{$^*$tkato@kusastro.kyoto-u.ac.jp}

\authorcount{affil:GEOS}{
     Groupe Europ\'een d'Observations Stellaires (GEOS),
     23 Parc de Levesville, 28300 Bailleau l'Ev\^eque, France}

\authorcount{affil:BAV}{
     Bundesdeutsche Arbeitsgemeinschaft f\"ur Ver\"anderliche Sterne
     (BAV), Munsterdamm 90, 12169 Berlin, Germany}

\authorcount{affil:Hambsch}{
     Vereniging Voor Sterrenkunde (VVS), Oude Bleken 12, 2400 Mol, Belgium}

\authorcount{affil:Vanmunster}{
     Center for Backyard Astrophysics Belgium, Walhostraat 1A,
     B-3401 Landen, Belgium}

\authorcount{affil:Ioh}{
     Variable Star Observers League in Japan (VSOLJ),
     1001-105 Nishiterakata, Hachioji, Tokyo 192-0153, Japan}

\authorcount{affil:Monard}{
     Bronberg Observatory, Center for Backyard Astrophysics Pretoria,
     PO Box 11426, Tiegerpoort 0056, South Africa}

\authorcount{affil:Monard2}{
     Kleinkaroo Observatory, Center for Backyard Astrophysics Kleinkaroo,
     Sint Helena 1B, PO Box 281, Calitzdorp 6660, South Africa}

\authorcount{affil:Polaris}{
     Polaris Observatory, Hungarian Astronomical Association,
     Laborc utca 2/c, 1037 Budapest, Hungary}

\authorcount{affil:Kis}{
     VSOLJ, 7-1 Kitahatsutomi, Kamagaya, Chiba 273-0126, Japan}

\authorcount{affil:Miller}{
     Furzehill House, Ilston, Swansea, SA2 7LE, UK}

\authorcount{affil:Starr}{
     Warrumbungle Observatory, Tenby, 841 Timor Rd,
     Coonabarabran NSW 2357, Australia}

\authorcount{affil:Kai}{
     Baselstrasse 133D, CH-4132 Muttenz, Switzerland}

\authorcount{affil:Sternberg}{
     Sternberg Astronomical Institute, Lomonosov Moscow State University, 
     Universitetsky Ave., 13, Moscow 119992, Russia}

\authorcount{affil:Slovak}{
     Astronomical Institute of the Slovak Academy of Sciences,
     05960 Tatranska Lomnica, Slovakia}

\authorcount{affil:Kazan}{
     Institute of Physics, Kazan Federal University,
     Ulitsa Kremlevskaya 16a, Kazan 420008, Russia}

\authorcount{affil:MainUkraine}{
     Main Astronomical Observatory of National Academy of
     Sciences of Ukraine, 27 Akademika Zabolotnoho St.,
     Kyiv 03143, Ukraine}

\authorcount{affil:Moscow}{
     Faculty of Physics, Lomonosov Moscow State University, 
     Leninskie gory, Moscow 119991, Russia}

\authorcount{affil:Mdy}{
     Kaminishiyamamachi 12-14, Nagasaki, Nagasaki 850-0006, Japan}

\authorcount{affil:Dubovsky}{
     Vihorlat Observatory, Mierova 4, 06601 Humenne, Slovakia}

\authorcount{affil:CrAO}{
     Federal State Budget Scientific Institution ``Crimean Astrophysical
     Observatory of RAS'', Nauchny, 298409, Republic of Crimea}

\authorcount{affil:CrimeanFU}{
     V. I. Vernadsky Crimean Federal University, 4 Vernadskogo Prospekt,
     Simferopol, 295007, Republic of Crimea}

\authorcount{affil:BAAVSS}{
     The British Astronomical Association, Variable Star Section (BAA VSS),
     Burlington House, Piccadilly, London, W1J 0DU, UK}

\authorcount{affil:Pickard}{
     3 The Birches, Shobdon, Leominster, Herefordshire, HR6 9NG, UK}

\authorcount{affil:OKU}{
     Osaka Kyoiku University, 4-698-1 Asahigaoka, Osaka 582-8582, Japan}

\authorcount{affil:Ruiz1}{
     Observatorio de C\'antabria, Ctra. de Rocamundo s/n, Valderredible, 
     39220 Cantabria, Spain}

\authorcount{affil:Ruiz2}{
     Instituto de F\'{\i}sica de Cantabria (CSIC-UC), Avenida Los Castros s/n, 
     E-39005 Santander, Cantabria, Spain}

\authorcount{affil:Ruiz3}{
     Agrupaci\'on Astron\'omica C\'antabria, Apartado 573,
     39080, Santander, Spain}

\authorcount{affil:AAVSO}{
     American Association of Variable Star Observers, 49 Bay State Rd.,
     Cambridge, MA 02138, USA}

\authorcount{affil:LewCook}{
     Center for Backyard Astrophysics Concord, 1730 Helix Ct. Concord,
     California 94518, USA}

\authorcount{affil:Miguel}{
     Departamento de Ciencias Integradas, Facultad de Ciencias
     Experimentales, Universidad de Huelva,
     21071 Huelva, Spain}

\authorcount{affil:Miguel2}{
     Center for Backyard Astrophysics, Observatorio del CIECEM,
     Parque Dunar, Matalasca\~nas, 21760 Almonte, Huelva, Spain}

\authorcount{affil:OUS}{
     Department of Biosphere-Geosphere System Science, Faculty of Informatics,
     Okayama University of Science, 1-1 Ridai-cho, Okayama,
     Okayama 700-0005, Japan}

\authorcount{affil:Goff}{
     13508 Monitor Ln., Sutter Creek, California 95685, USA}

\authorcount{affil:Morelle}{
     9 rue Vasco de GAMA, 59553 Lauwin Planque, France}

\authorcount{affil:LCO}{
     Department of Physics, University of Notre Dame, 
     225 Nieuwland Science Hall, Notre Dame, Indiana 46556, USA}

\authorcount{affil:Bolt}{
     Camberwarra Drive, Craigie, Western Australia 6025, Australia}

\authorcount{affil:Dubois}{
     Public observatory Astrolab Iris, Verbrandemolenstraat 5,
     B 8901 Zillebeke, Belgium}

\authorcount{affil:Brincat}{
     Flarestar Observatory, San Gwann SGN 3160, Malta}

\authorcount{affil:OAO}{
     Okayama Astrophysical Observatory, National Astronomical Observatory 
     of Japan, Asakuchi, Okayama 719-0232, Japan}

\authorcount{affil:TohokuPlanetary}{
     Planetary Plasma and Atmospheric Research Center, Graduate School of
     Science, Tohoku University, Sendai 980-8578, Japan}

\authorcount{affil:Hamburg}{
     Hamburger Sternwarte, Universit\"at Hamburg, Gojenbergsweg 112,
     D-21029 Hamburg, Germany}

\authorcount{affil:HidaKwasan}{
     Kwasan and Hida Observatories, Kyoto University, Yamashina,
     Kyoto 607-8471, Japan}

\authorcount{affil:Terskol}{
     Terskol Branch of Institute of Astronomy, Russian Academy of Sciences,
     361605, Peak Terskol, Kabardino-Balkaria Republic, Russia}

\authorcount{affil:ICUkraine}{
     International Center for Astronomical, Medical and Ecological Research
     of NASU, Ukraine 27 Akademika Zabolotnoho Str. 03680 Kyiv,
     Ukraine}

\authorcount{affil:Sabo}{
     2336 Trailcrest Dr., Bozeman, Montana 59718, USA}

\authorcount{affil:RIT}{
     Physics Department, Rochester Institute of Technology, Rochester,
     New York 14623, USA}

\authorcount{affil:Rodda}{
     1, Rivermede, Ponteland, Newcastle upon Tyne, NE20 9XA, UK}

\authorcount{affil:Nelson}{
     1105 Hazeldean Rd, Ellinbank 3820, Australia}

\authorcount{affil:Myers}{
     Center for Backyard Astrophysics San Mateo, 5 inverness Way,
     Hillsborough, CA 94010, USA}

\authorcount{affil:Ohio}{
     Department of Astronomy, the Ohio State University, Columbia,
     OH 43210, USA}

\authorcount{affil:Carnegie}{
     Carnegie Observatories, 813 Santa Barbara Street, Pasadena,
     CA 91101, USA}

\authorcount{affil:DiegoPortales}{
     N\'ucleo de Astronom\'ia de la Facultad de Ingenier\'ia, Universidad
     Diego Portales, Av. Ej\'ercito 441, Santiago, Chile}

\authorcount{affil:Princeton}{
     Department of Astrophysical Sciences, Princeton University,
     NJ 08544, USA}

\authorcount{affil:Itagaki}{
     Itagaki Astronomical Observatory, Teppo-cho, Yamagata 990-2492}

\authorcount{affil:MiyakiObs}{
     Miyaki-Argenteus Observatory, Miyaki, Saga 840-1102, Japan}

\authorcount{affil:Stubbings}{
     Tetoora Observatory, 2643 Warragul-Korumburra Road, Tetoora Road,
     Victoria 3821, Australia}

\authorcount{affil:Schmeer}{
     Bischmisheim, Am Probstbaum 10, 66132 Saarbr\"{u}cken, Germany}

\authorcount{affil:VVSBelgium}{
     Vereniging Voor Sterrenkunde (VVS), Moffelstraat 13 3370
     Boutersem, Belgium}

\authorcount{affil:Heo}{
     759-10 Tokawa, Hadano-shi, Kanagawa 259-1306, Japan}

\authorcount{affil:Shears}{
     ``Pemberton'', School Lane, Bunbury, Tarporley, Cheshire, CW6 9NR, UK}

\authorcount{affil:Poyner}{
     BAA Variable Star Section, 67 Ellerton Road, Kingstanding,
     Birmingham B44 0QE, UK}

\authorcount{affil:Myy}{
     290-383, Ogata-cho, Sasebo, Nagasaki 858-0926, Japan}



\maketitle

\begin{abstract}
Continuing the project described by Kato et al. (2009,
PASJ, 61, S395), we collected times of superhump maxima for
127 SU UMa-type dwarf novae observed mainly during
the 2016--2017 season and characterized these objects.
We provide updated statistics of relation between
the orbital period and the variation of superhumps, 
the relation between period variations and the rebrightening 
type in WZ Sge-type objects.  We obtained the period minimum 
of 0.05290(2)~d and confirmed the presence of the period gap
above the orbital period $\sim$ 0.09~d.
We note that four objects
(NY Her, 1RXS J161659.5$+$620014, CRTS J033349.8$-$282244
and SDSS J153015.04$+$094946.3) have supercycles shorter
than 100~d but show infrequent normal outbursts.
We consider that these objects are similar to V503 Cyg,
whose normal outbursts are likely suppressed by a disk tilt.
These four objects are excellent candidates to search for
negative superhumps.  DDE 48 appears to be a member of
ER UMa-type dwarf novae.  We identified a new eclipsing
SU UMa-type object MASTER OT J220559.40$-$341434.9.
We observed 21 WZ Sge-type dwarf novae during this
interval and reported 18 out of them in this paper.
Among them, ASASSN-16js is a good candidate for
a period bouncer.  ASASSN-16ia showed a precursor
outburst for the first time in a WZ Sge-type superoutburst.
ASASSN-16kg, CRTS J000130.5$+$050624 and
SDSS J113551.09$+$532246.2 are located in the period gap.
We have newly obtained 15 orbital periods, 
including periods from early superhumps.
\end{abstract}

\section{Introduction}

   This is a continuation of series of papers \citet{Pdot},
\citet{Pdot2}, \citet{Pdot3}, \citet{Pdot4}, \citet{Pdot5},
\citet{Pdot6}, \citet{Pdot7} and \citet{Pdot8} reporting
new observations of superhumps in SU UMa-type dwarf novae.
SU UMa-type dwarf novae are a class of cataclysmic
variables (CVs) which are close binary systems
transferring matter from a low-mass dwarf secondary to
a white dwarf, forming an accretion disk
[see e.g. \citet{war95book} for CVs in general].

   In SU UMa-type dwarf novae, there are two types of
outbursts (normal outbursts and superoutbursts).
Outbursts and superoutbursts in SU UMa-type dwarf novae
are considered to be a result of the combination of
thermal and tidal instabilities [thermal-tidal instability (TTI)
model by \citet{osa89suuma}; \citet{osa96review}].

   During superoutbursts, semi-periodic variations called
superhumps are observed whose period (superhump period,
$P_{\rm SH}$) is a few percent longer than the orbital period
($P_{\rm orb}$).  Superhumps are considered to
originate from a precessing eccentric (or flexing)
disk in the gravity field of the rotating binary,
and the eccentricity in the disk is believed to be
a consequence of the 3:1 resonance in the accretion disk
[see e.g. \citet{whi88tidal}; \citet{hir90SHexcess};
\citet{lub91SHa}; \citet{woo11v344lyr}].

\begin{figure}
  \begin{center}
    \FigureFile(80mm,110mm){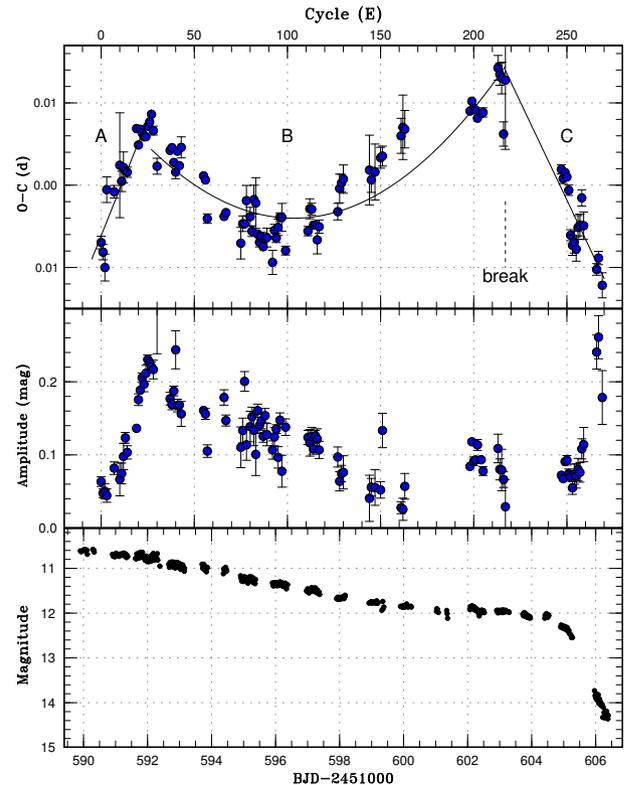}
  \end{center}
  \caption{Representative $O-C$ diagram showing three stages (A--C)
  of $O-C$ variation.  The data were taken from the 2000 superoutburst
  of SW UMa.  (Upper:) $O-C$ diagram.  Three distinct stages
  (A -- evolutionary stage with a longer superhump period, 
  B -- middle stage, and C -- stage after
  transition to a shorter period) and the location of the period break
  between stages B and C are shown. (Middle): Amplitude of superhumps.
  During stage A, the amplitude of the superhumps grew.
  (Lower:) Light curve.
  (Reproduction of figure 1 in \cite{kat13qfromstageA})}
  \label{fig:stagerev}
\end{figure}

   It has become evident since \citet{Pdot} that
the superhump periods systematically vary in a way
common to many objects.  \citet{Pdot} introduced
superhump stages (stages A, B and C):
initial growing stage with a long period (stage A) and
fully developed stage with a systematically
varying period (stage B) and later stage C with a shorter,
almost constant period (see figure \ref{fig:stagerev}).

   It has recently been proposed by \citet{osa13v344lyrv1504cyg}
that stage A superhumps reflect the dynamical precession rate
at the 3:1 resonance radius and that the rapid decrease
of the period (stage B) reflects the pressure effect
which has an effect of retrograde precession
(\cite{lub92SH}; \cite{hir93SHperiod};
\cite{mur98SH}; \cite{mon01SH}; \cite{pea06SH}).
As proposed by \citet{kat13qfromstageA}
stage A superhumps can be then used to ``dynamically''
determine the mass ratio ($q$), which had been difficult
to measure except for eclipsing systems and systems with
bright secondaries to detect radial-velocity variations.
It has been confirmed that this stage A method gives
$q$ values as precise as in eclipsing systems.
There have been more than 50 objects whose
$q$ values are determined by this method and it has
been proven to be an especially valuable tool in depicting
the terminal stage of CV evolution (cf. \cite{Pdot7};
\cite{kat15wzsge}).

   In this paper, we present new observations of
SU UMa-type dwarf novae mainly obtained in 2016--2017.
We present basic observational materials
and discussions in relation to individual objects.
Starting from \citet{Pdot6}, we have been intending
these series of papers to be also a source of compiled
information, including historical, of individual dwarf novae.

   The material and methods of analysis are given in
section \ref{sec:obs}, observations and analysis of
individual objects are given in section \ref{sec:individual},
including discussions particular to the objects.
General discussions are given in section
\ref{sec:discuss} and the summary is given in section
\ref{sec:summary}.  Some tables and figures are available
online only.

\section{Observation and Analysis}\label{sec:obs}

\subsection{Data Source}

   The data were obtained under campaigns led by 
the VSNET Collaboration \citep{VSNET}.
We also used the public data from
the AAVSO International Database\footnote{
   $<$http://www.aavso.org/data-download$>$.
}.
Outburst detections of many new and known objects
relied on the ASAS-SN CV patrol \citep{dav15ASASSNCVAAS}\footnote{
   $<$http://cv.asassn.astronomy.ohio-state.edu/$>$.
}, the MASTER network \citep{MASTER}, and
Catalina Real-time Transient Survey
(CRTS; \cite{CRTS})\footnote{
   $<$http://nesssi.cacr.caltech.edu/catalina/$>$.
   For the information of the individual Catalina CVs, see
   $<$http://nesssi.cacr.caltech.edu/catalina/AllCV.html$>$.
} in addition to outburst detections reported to
VSNET, AAVSO\footnote{
  $<$https://www.aavso.org/$>$.
}, BAAVSS alert\footnote{
  $<$https://groups.yahoo.com/neo/groups/baavss-alert/$>$.
} and cvnet-outburst.\footnote{
  $<$https://groups.yahoo.com/neo/groups/cvnet-outburst/$>$.
}

   For objects detected in CRTS, we preferably used the names 
provided in \citet{dra14CRTSCVs} and \citet{cop16DNCRTS}.
If these names are not yet available,
we used the International Astronomical
Union (IAU)-format names provided by the CRTS team 
in the public data release\footnote{
  $<$http://nesssi.cacr.caltech.edu/DataRelease/$>$.
}
Since \citet{Pdot}, we have used coordinate-based 
optical transient (OT) designations for some objects, such as
apparent dwarf nova candidates reported in
the Transient Objects Confirmation Page of
the Central Bureau for Astronomical Telegrams\footnote{
   $<$http://www.cbat.eps.harvard.edu/unconf/tocp.html$>$.
} and CRTS objects without registered designations
in \citet{dra14CRTSCVs} or in the CRTS public data release
and listed the original identifiers in table \ref{tab:outobs}.

   We provided coordinates from astrometric catalogs
for ASAS-SN \citep{ASASSN} CVs and two objects
without precise coordinate-based names other than listed
in the General Catalog of Variable Stars \citep{GCVS}
in table \ref{tab:coord}.
We mainly used Gaia DR1 \citep{GaiaDR1}, Sloan Digital Sky Survey
(SDSS, \cite{SDSS9}),
the Initial Gaia Source List (IGSL, \cite{IGSL}) and 
Guide Star Catalog 2.3.2 (GSC 2.3.2, \cite{GSC232}).
Some objects were detected as transients by
Gaia\footnote{
  $<$http://gsaweb.ast.cam.ac.uk/alerts/alertsindex$>$ and
  Gaia identifications supplied by the AAVSO VSX.
}
and CRTS and we used their coordinates.
The coordinates used in this paper are J2000.0.
We also supplied SDSS $g$, Gaia $G$ and GALEX
NUV magnitudes when counterparts are present.

\subsection{Observations and Basic Reduction}

   The majority of the data were acquired
by time-resolved CCD photometry by using 20--60cm telescopes
located world-wide.
The list of outbursts and observers is summarized in 
table \ref{tab:outobs}.
The data analysis was performed in the same way described
in \citet{Pdot} and \citet{Pdot6} and we mainly used
R software\footnote{
   The R Foundation for Statistical Computing:\\
   $<$http://cran.r-project.org/$>$.
} for data analysis.

   In de-trending the data, we mainly used locally-weighted
polynomial regression (LOWESS: \cite{LOWESS})
and sometimes lower (1--3rd order) polynomial fitting
when the observation baseline was short.
The times of superhumps maxima were determined by
the template fitting method as described in \citet{Pdot}.
The times of all observations are expressed in 
barycentric Julian days (BJD).

   In figures, the points are accompanied by 1$\sigma$
error bars whenever available, which are omitted
when the error is smaller than the plot mark or
the errors were not available (as in some raw light
curves of superhumps).

\subsection{Abbreviations and Terminology}

   The abbreviations used in this paper are the same
as in \citet{Pdot6}: we used
$\epsilon \equiv P_{\rm SH}/P_{\rm orb}-1$ for 
the fractional superhump excess.
We have used since \citet{osa13v1504cygKepler}
the alternative fractional superhump excess in the frequency unit
$\epsilon^* \equiv 1-P_{\rm orb}/P_{\rm SH} = \epsilon/(1+\epsilon)$
because this fractional superhump excess
is a direct measure of the precession rate.  We therefore
used $\epsilon^*$ in discussing the precession rate.

   The $P_{\rm SH}$, $P_{\rm dot}$ and other parameters
are listed in table \ref{tab:perlist} in same format as in
\citet{Pdot}.  The definitions of parameters $P_1, P_2, E_1, E_2$
and $P_{\rm dot}$ are the same as in \citet{Pdot}:
$P_1$ and $P_2$ represent periods in stage B and C, respectively
($P_1$ is averaged during the entire course of the observed
segment of stage B),
and $E_1$ and $E_2$ represent intervals (in cycle numbers)
to determine $P_1$ and $P_2$, respectively.\footnote{
   The intervals ($E_1$ and $E_2$) for the stages B and C given in the table
   sometimes overlap because there is sometimes observational
   ambiguity (usually due to the lack of observations
   and errors in determining the times of maxima)
   in determining the stages.
}
Some superoutbursts are not listed in table \ref{tab:perlist}
due to the lack of observations (e.g. single-night
observations with less than two superhump maxima
or poor observations for the object with already
well measured $P_{\rm SH}$).

   We used the same terminology of superhumps summarized in
\citet{Pdot3}.  We especially call attention to
the term ``late superhumps''.  We only used
the concept of ``traditional'' late superhumps when
there is an $\sim$0.5 phase shift
[\citet{vog83lateSH}; see also table 1 in \citet{Pdot3} 
for various types of superhumps], 
since we suspect that many of the past
claims of detections of ``late superhumps'' were likely
stage C superhumps before it became evident that
there are complex structures in the $O-C$ diagrams
of superhumps (see discussion in \cite{Pdot}).

\subsection{Period Analysis}

   We used phase dispersion minimization (PDM; \cite{PDM})
for period analysis and 1$\sigma$ errors for the PDM analysis
was estimated by the methods of \citet{fer89error} and \citet{Pdot2}.
We have used a variety of bootstrapping in
estimating the robustness of the result of the PDM analysis
since \citet{Pdot3}.
We analyzed 100 samples which randomly contain 50\% of
observations, and performed PDM analysis for these samples.
The bootstrap result is shown as a form of 90\% confidence intervals
in the resultant PDM $\theta$ statistics.
If this paper provides the first solid presentation of
a new SU UMa-type classification, we provide the result
of PDM period analysis and averaged superhump profile.

\subsection{$O-C$ Diagrams}

   Comparisons of $O-C$ diagrams between different
superoutbursts are also presented whenever available.
This figure not only provides information about
the difference of $O-C$ diagrams between different
superoutbursts but also helps identifying superhump
stages especially when observations were insufficient
or the start of the outburst was missed.
In drawing combined $O-C$ diagrams, we usually used
$E=$0 for the start of the superoutburst, which usually
refers to the first positive detection of the outburst.
This epoch usually has an accuracy of $\sim$1~d for
well-observed objects, and if the outburst was not sufficiently
observed, we mentioned in the figure caption how to estimate
$E$ in such an outburst.
In some cases, this $E=$0 is defined as the appearance
of superhumps.  This treatment is necessary since
some objects have a long waiting time before
appearance of superhumps.
We also note that there is sometimes an ambiguity in
selecting the true period among aliases.  In some
cases, this can be resolved by the help of
the $O-C$ analysis.  The procedure and example
are shown in subsection 2.2 in \citet{Pdot7}.

\begin{table*}
\caption{List of Superoutbursts.}\label{tab:outobs}
\begin{center}
\begin{tabular}{ccccl}
\hline
Subsection & Object & Year & Observers or references\commenta & \multicolumn{1}{c}{ID\commentb} \\
\hline
\ref{obj:v1047aql}   & V1047 Aql   & 2016 & Trt & \\
\ref{obj:bbari}      & BB Ari      & 2016 & Kis, AAVSO, SRI, RPc, Ioh, & \\
                     &             &      & Shu, RAE & \\
\ref{obj:v391cam}    & V391 Cam    & 2017 & Trt, DPV & \\
\ref{obj:oycar}      & OY Car      & 2016 & SPE, HaC, MGW & \\
--                   & HT Cas      & 2016 & \Wakamatsuprep & \\
\ref{obj:gscet}      & GS Cet      & 2016 & Kis, OKU, HaC, Shu, CRI, & \\
                     &             &      & KU, Ioh, Trt & \\
\ref{obj:gzcet}      & GZ Cet      & 2016 & OKU & \\
\ref{obj:akcnc}      & AK Cnc      & 2016 & Aka & \\
\ref{obj:gzcnc}      & GZ Cnc      & 2017 & KU, Mdy, HaC & \\
\ref{obj:gpcvn}      & GP CVn      & 2016 & RPc, Kai, Trt, IMi, Kis, & \\
                     &             &      & CRI, deM, AAVSO & \\
\ref{obj:v337cyg}    & V337 Cyg    & 2016 & Kai & \\
\ref{obj:v1113cyg}   & V1113 Cyg   & 2016 & OKU, Ioh, Kis & \\
\ref{obj:ixdra}      & IX Dra      & 2016 & Kis, SGE, COO & \\
\hline
  \multicolumn{5}{l}{\parbox{500pt}{\commenta Key to observers:
Aka (H. Akazawa, OUS),
BSM\commentc (S. Brincat),
COO (L. Cook),
CRI (Crimean Astrophys. Obs.),
DDe (D. Denisenko),
deM (E. de Miguel),
DPV (P. Dubovsky),
Dub (F. Dubois team),
GBo (G. Bolt),
GFB\commentc (W. Goff),
HaC (F.-J. Hambsch, remote obs. in Chile),
IMi\commentc (I. Miller),
Ioh (H. Itoh),
KU (Kyoto U., campus obs.),
Kai (K. Kasai),
Kis (S. Kiyota),
LCO (C. Littlefield),
MEV\commentc (E. Morelle),
NGW (G. Myers),
MLF (B. Monard),
MNI (N. Mishevskiy),
Mdy (Y. Maeda),
Mhh (H. Maehara),
NKa (N. Katysheva and S. Shugarov),
Naz (S. Nazarov),
Nel (P. Nelson),
OKU (Osaya Kyoiku U.),
RAE (T. Rodda),
RIT (M. Richmond),
RPc\commentc (R. Pickard),
Rui (J. Ruiz),
SGE\commentc (G. Stone),
SPE\commentc (P. Starr),
SRI\commentc (R. Sabo),
Shu (S. Shugarov team),
T60 (Haleakala Obs. T60 telescope),
Trt (T. Tordai),
Van (T. Vanmunster),
Vol (I. Voloshina),
AAVSO (AAVSO database)
}} \\
  \multicolumn{5}{l}{\commentb Original identifications, discoverers or data source.} \\
  \multicolumn{5}{l}{\commentc Inclusive of observations from the AAVSO database.} \\
\end{tabular}
\end{center}
\end{table*}

\addtocounter{table}{-1}
\begin{table*}
\caption{List of Superoutbursts (continued).}
\begin{center}
\begin{tabular}{ccccl}
\hline
Subsection & Object & Year & Observers or references\commenta & \multicolumn{1}{c}{ID\commentb} \\
\hline
\ref{obj:irgem}      & IR Gem      & 2016 & Kai, Aka, CRI, BSM, AAVSO, & \\
                     &             &      & Trt & \\
                     &             & 2017 & Kai, Trt & \\
\ref{obj:nyher}      & NY Her      & 2016 & GFB, Ioh, DPV, Trt, COO, & \\
                     &             &      & IMi, SGE & \\
\ref{obj:mnlac}      & MN Lac      & 2016 & Van & \\
\ref{obj:v699oph}    & V699 Oph    & 2016 & Kis, Ioh & \\
\ref{obj:v344pav}    & V344 Pav    & 2016 & HaC & \\
\ref{obj:v368peg}    & V368 Peg    & 2016 & Trt & \\
\ref{obj:v893sco}    & V893 Sco    & 2016 & GBo, HaC, Kis, Aka & \\
\ref{obj:v493ser}    & V493 Ser    & 2016 & Shu & \\
\ref{obj:awsge}      & AW Sge      & 2016 & DPV & \\
\ref{obj:v1389tau}   & V1389 Tau   & 2016 & HaC, KU, Ioh & \\
\ref{obj:suuma}      & SU UMa      & 2017 & Trt & \\
\ref{obj:hvvir}      & HV Vir      & 2016 & HaC, DPV, AAVSO, deM, Mdy, & \\
                     &             &      & KU, RPc, GBo, Aka, IMi, & \\
                     &             &      & BSM, Kis & \\
\ref{obj:nsv2026}    & NSV 2026    & 2016b & Trt, Dub & \\
\ref{obj:nsv14681}   & NSV 14681   & 2016 & Van & \\
\ref{obj:j1616}      & 1RXS J161659 & 2016 & deM, MEV, IMi, Van, Trt & 1RXS J161659.5$+$620014 \\
                     &             & 2016b & MEV, DPV, IMi & \\
\ref{obj:asassn13ak} & ASASSN-13ak & 2016 & Trt, Kis & \\
\ref{obj:asassn13al} & ASASSN-13al & 2016 & Van & \\
\ref{obj:asassn13bc} & ASASSN-13bc & 2015 & LCO, Rui, Trt & \\
                     &             & 2016 & SGE, Shu, NKa, Ioh, Rui & \\
\ref{obj:asassn13bj} & ASASSN-13bj & 2016 & Kai, OKU, Trt, SGE, DPV, & \\
                     &             &      & IMi, KU \\
\ref{obj:asassn13bo} & ASASSN-13bo & 2016 & IMi, Shu & \\
\ref{obj:asassn13cs} & ASASSN-13cs & 2016 & SGE, KU, COO & \\
\ref{obj:asassn13cz} & ASASSN-13cz & 2016 & Kai, Trt, Rui, DPV & \\
\ref{obj:asassn14gg} & ASASSN-14gg & 2016 & Van, GFB & \\
\hline
\end{tabular}
\end{center}
\end{table*}

\addtocounter{table}{-1}
\begin{table*}
\caption{List of Superoutbursts (continued).}
\begin{center}
\begin{tabular}{ccccl}
\hline
Subsection & Object & Year & Observers or references\commenta & \multicolumn{1}{c}{ID\commentb} \\
\hline
\ref{obj:asassn15cr} & ASASSN-15cr & 2017 & DPV, Ioh, Shu, CRI & \\
\ref{obj:asassn16da} & ASASSN-16da & 2016 & deM, Van, GFB, SGE, Kai & \\
\ref{obj:asassn16dk} & ASASSN-16dk & 2016 & HaC & \\
\ref{obj:asassn16ds} & ASASSN-16ds & 2016 & MLF, HaC, SPE & \\
--                   & ASASSN-16dt & 2016 & \citet{kim17asassn16dt16hg} & \\
\ref{obj:asassn16dz} & ASASSN-16dz & 2016 & Van & \\
--                   & ASASSN-16eg & 2016 & \citet{wak17asassn16eg} & \\
\ref{obj:asassn16ez} & ASASSN-16ez & 2016 & DPV, Ioh, Kis, MEV, IMi, & \\
                     &             &      & Van, KU \\
\ref{obj:asassn16fr} & ASASSN-16fr & 2016 & KU, Ioh, HaC & \\
\ref{obj:asassn16fu} & ASASSN-16fu & 2016 & HaC, MLF & \\
--                   & ASASSN-16fy & 2016 & \Isogaiprep & \\
\ref{obj:asassn16gh} & ASASSN-16gh & 2016 & MLF & \\
\ref{obj:asassn16gj} & ASASSN-16gj & 2016 & MLF, HaC & \\
\ref{obj:asassn16gl} & ASASSN-16gl & 2016 & MLF, HaC, DDe & \\
--                   & ASASSN-16hg & 2016 & \citet{kim17asassn16dt16hg} & \\
\ref{obj:asassn16hi} & ASASSN-16hi & 2016 & HaC & \\
\ref{obj:asassn16hj} & ASASSN-16hj & 2016 & HaC, KU & \\
\ref{obj:asassn16ia} & ASASSN-16ia & 2016 & GFB, Ioh, Ter, Van, SGE, & \\
                     &             &      & CRI, COO, Trt \\
\ref{obj:asassn16ib} & ASASSN-16ib & 2016 & MLF, HaC & \\
\ref{obj:asassn16ik} & ASASSN-16ik & 2016 & MLF, HaC & \\
\ref{obj:asassn16is} & ASASSN-16is & 2016 & Shu, IMi, Van, Ioh, Rui & \\
\ref{obj:asassn16iu} & ASASSN-16iu & 2016 & HaC, MLF & \\
\ref{obj:asassn16iw} & ASASSN-16iw & 2016 & HaC, SPE, NKa, Kis, Van, & \\
                     &             &      & Ioh \\
\ref{obj:asassn16jb} & ASASSN-16jb & 2016 & MLF, HaC, SPE & \\
\ref{obj:asassn16jd} & ASASSN-16jd & 2016 & HaC, Ioh & \\
\ref{obj:asassn16jk} & ASASSN-16jk & 2016 & CRI, Van & \\
\ref{obj:asassn16js} & ASASSN-16js & 2016 & HaC, MLF, SPE & \\
\ref{obj:asassn16jz} & ASASSN-16jz & 2016 & Van & \\
\hline
\end{tabular}
\end{center}
\end{table*}

\addtocounter{table}{-1}
\begin{table*}
\caption{List of Superoutbursts (continued).}
\begin{center}
\begin{tabular}{ccccl}
\hline
Subsection & Object & Year & Observers or references\commenta & \multicolumn{1}{c}{ID\commentb} \\
\hline
\ref{obj:asassn16kg} & ASASSN-16kg & 2016 & MLF, HaC & \\
\ref{obj:asassn16kx} & ASASSN-16kx & 2016 & HaC, MLF & \\
\ref{obj:asassn16le} & ASASSN-16le & 2016 & KU, Ioh & \\
\ref{obj:asassn16lj} & ASASSN-16lj & 2016 & Van & \\
\ref{obj:asassn16lo} & ASASSN-16lo & 2016 & KU, IMi, OKU, Ioh & \\
\ref{obj:asassn16mo} & ASASSN-16mo & 2016 & OKU, KU, Trt, Dub, Van & \\
\ref{obj:asassn16my} & ASASSN-16my & 2016 & HaC, Ioh & \\
\ref{obj:asassn16ni} & ASASSN-16ni & 2016 & KU, Ioh, Trt & \\
\ref{obj:asassn16nq} & ASASSN-16nq & 2016 & Kis, Ioh, RPc, Van, Trt & \\
\ref{obj:asassn16nr} & ASASSN-16nr & 2016 & MLF, HaC & \\
\ref{obj:asassn16nw} & ASASSN-16nw & 2016 & Kai & \\
\ref{obj:asassn16ob} & ASASSN-16ob & 2016 & MLF, HaC, SPE & \\
\ref{obj:asassn16oi} & ASASSN-16oi & 2016 & MLF, HaC, SPE & \\
\ref{obj:asassn16os} & ASASSN-16os & 2016 & MLF, HaC, SPE & \\
\ref{obj:asassn16ow} & ASASSN-16ow & 2016 & Ioh, Van, NKa, Mdy, MEV, & \\
                     &             &      & Kis, Kai & \\
\ref{obj:asassn17aa} & ASASSN-17aa & 2017 & MLF, SPE, HaC & \\
\ref{obj:asassn17ab} & ASASSN-17ab & 2017 & HaC & \\
\ref{obj:asassn17az} & ASASSN-17az & 2017 & MLF & \\
\ref{obj:asassn17bl} & ASASSN-17bl & 2017 & HaC, SPE & \\
\ref{obj:asassn17bm} & ASASSN-17bm & 2017 & MLF, HaC & \\
\ref{obj:asassn17bv} & ASASSN-17bv & 2017 & MLF, SPE, HaC & \\
\ref{obj:asassn17ce} & ASASSN-17ce & 2017 & SPE. MLF, HaC & \\
\ref{obj:asassn17ck} & ASASSN-17ck & 2017 & HaC & \\
\ref{obj:asassn17cn} & ASASSN-17cn & 2017 & MLF, SPE, HaC, Ioh & \\
\ref{obj:asassn17cx} & ASASSN-17cx & 2017 & Mdy & \\
\ref{obj:asassn17dg} & ASASSN-17dg & 2017 & HaC, MLF, SPE & \\
\ref{obj:asassn17dq} & ASASSN-17dq & 2017 & HaC, MLF & \\
\ref{obj:j0001}      & CRTS J000130 & 2016 & Van, Shu & CRTS J000130.5$+$050624  \\
\ref{obj:j0153}      & CRTS J015321 & 2016 & Kai & CRTS J015321.5$+$340857 \\
\ref{obj:j0333}      & CRTS J033349 & 2016 & MLF, HaC, KU & CRTS J033349.8$-$282244 \\
\hline
\end{tabular}
\end{center}
\end{table*}

\addtocounter{table}{-1}
\begin{table*}
\caption{List of Superoutbursts (continued).}
\begin{center}
\begin{tabular}{ccccl}
\hline
Subsection & Object & Year & Observers or references\commenta & \multicolumn{1}{c}{ID\commentb} \\
\hline
\ref{obj:j0236}      & CRTS J023638 & 2016 & CRI, Trt, Shu, Rui & CRTS J023638.0$+$111157 \\
\ref{obj:j0446}      & CRTS J044637 & 2017 & Ioh, KU & CRTS J044636.9$+$083033 \\
\ref{obj:j0826}      & CRTS J082603 & 2017 & Van & CRTS J082603.7$+$113821 \\
\ref{obj:j0851}      & CRTS J085113 & 2008 & Mhh & CRTS J085113.4$+$344449 \\
                     &              & 2016 & KU, Trt & \\
\ref{obj:j0856}      & CRTS J085603 & 2016 & Van, Ioh & CRTS J085603.8$+$322109 \\
\ref{obj:j1649}      & CRTS J164950 & 2015 & RIT, Van & CRTS J164950.4$+$035835 \\
                     &             & 2016 & CRI, Rui & \\
\ref{obj:j0624}      & CSS J062450 & 2017 & Trt, Van & CSS131223:062450$+$503111 \\
\ref{obj:dde26}      & DDE 26      & 2016 & Ioh, IMi, Shu, RPc & \\
\ref{obj:dde48}      & DDE 48      & 2016 & MNI, IMi & \\
\ref{obj:j0213}      & MASTER J021315 & 2016 & Van & MASTER OT J021315.37$+$533822.7 \\
\ref{obj:j0302}      & MASTER J030205 & 2016 & OKU, deM, Van, COO, Ioh, & MASTER OT J030205.67$+$254834.3 \\
                     &                &      & Mdy, T60, NKa, RPc, Trt, & \\
                     &                &      & Naz & \\
\ref{obj:j0426}      & MASTER J042609 & 2016 & Kis, Ioh, Kai, Shu, Trt & MASTER OT J042609.34$+$354144.8 \\
\ref{obj:j0432}      & MASTER J043220 & 2017 & Van & MASTER OT J043220.15$+$784913.8 \\
\ref{obj:j0439}      & MASTER J043915 & 2016 & Ioh, CRI & MASTER OT J043915.60$+$424232.3 \\
\ref{obj:j0547}      & MASTER J054746 & 2016 & Van & MASTER OT J054746.81$+$762018.9 \\
\ref{obj:j0553}      & MASTER J055348 & 2017 & Van, Mdy & MASTER OT J055348.98$+$482209.0 \\
\ref{obj:j0558}      & MASTER J055845 & 2016 & Shu & MASTER OT J055845.55$+$391533.4 \\
\ref{obj:j0647}      & MASTER J064725 & 2016 & Ioh, RPc, CRI & MASTER OT J064725.70$+$491543.9 \\
\ref{obj:j0653}      & MASTER J065330 & 2017 & Van, Ioh & MASTER OT J065330.46$+$251150.9 \\
\ref{obj:j0754}      & MASTER J075450 & 2017 & Van & MASTER OT J075450.18$+$091020.2 \\
\ref{obj:j1505}      & MASTER J150518 & 2017 & HaC & MASTER OT J150518.03$-$143933.6 \\
\ref{obj:j1511}      & MASTER J151126 & 2016 & HaC, MLF & MASTER OT J151126.74$-$400751.9 \\
\ref{obj:j1623}      & MASTER J162323 & 2015 & Van & MASTER OT J162323.48$+$782603.3 \\
                     &                & 2016 & COO, Trt, IMi & \\
\ref{obj:j1651}      & MASTER J165153 & 2017 & Van & MASTER OT J165153.86$+$702525.7 \\
\ref{obj:j1748}      & MASTER J174816 & 2016 & Van, Mdy & MASTER OT J174816.22$+$501723.3 \\
--                   & MASTER J191841 & 2016 & \Isogaiprep & MASTER OT J191841.98$+$444914.5 \\
\hline
\end{tabular}
\end{center}
\end{table*}

\addtocounter{table}{-1}
\begin{table*}
\caption{List of Superoutbursts (continued).}
\begin{center}
\begin{tabular}{ccccl}
\hline
Subsection & Object & Year & Observers or references\commenta & \multicolumn{1}{c}{ID\commentb} \\
\hline
\ref{obj:j2113}      & MASTER J211322 & 2016 & Van & MASTER OT J211322.92$+$260647.4 \\
\ref{obj:j2205}      & MASTER J220559 & 2016 & MLF, HaC & MASTER OT J220559.40$-$341434.9 \\
--                   & OT J002656  & 2016 & \citet{kat17j0026} & CSS101212:002657$+$284933 \\
\ref{obj:sbs1108}    & SBS 1108    & 2016 & Ioh, COO, Vol, Kai, KU & SBS 1108$+$574 \\
\ref{obj:j0320}      & SDSS J032015 & 2016 & Van, IMi & SDSS J032015.29$+$441059.3 \\
\ref{obj:j0910}      & SDSS J032015 & 2016 & Van & SDSS J091001.63$+$164820.0 \\
\ref{obj:j1135}      & SDSS J113551 & 2017 & Van, Mdy & SDSS J113551.09$+$532246.2 \\
\ref{obj:j1152}      & SDSS J115207 & 2009 & \citet{Pdot2} & SDSS J115207.00$+$404947.8 \\
                     &              & 2017 & Mdy, KU, LCO, Ioh, DPV, & \\
                     &              &      & Kis & \\
\ref{obj:j1314}      & SDSS J131432 & 2017 & Mdy, Van & SDSS J131432.10$+$444138.7 \\
\ref{obj:j1530}      & SDSS J153015 & 2017 & Van & SDSS J153015.04$+$094946.3 \\
\ref{obj:j1557}      & SDSS J155720 & 2016 & HaC, Kis & SDSS J155720.75$+$180720.2 \\
--                   & SDSS J173047 & 2016 & \Isogaiprep & SDSS J173047.59$+$554518.5 \\
\ref{obj:j1348}      & SSS J134850 & 2016 & MLF, HaC & SSS J134850.1$-$310835 \\
\ref{obj:j0137}      & TCP J013758 & 2016 & Kis, IMi, Ioh, RPc, Shu, & TCP J01375892$+$4951055 \\
                     &             &      & CRI, Rui, Trt & \\
\ref{obj:j1800}      & TCP J180018 & 2016 & HaC, Nel, SPE & TCP J18001854$-$3533149 \\
\hline
\end{tabular}
\end{center}
\end{table*}

\begin{table*}
\caption{Coordinates of objects without coordinate-based names.}\label{tab:coord}
\begin{center}
\begin{tabular}{ccccccc}
\hline
Object & Right Ascention & Declination & Source\commenta & SDSS $g$ & Gaia G & GALEX NUV \\
\hline
ASASSN-13ak & \timeform{17h 48m 27.87s} & \timeform{+50D 50' 39.8''} & Gaia & 19.89(2) & 19.06 & -- \\
ASASSN-13al & \timeform{19h 32m 06.39s} & \timeform{+67D 27' 40.4''} & GSC2.3.2 & -- & -- & 21.5(4) \\
ASASSN-13bc & \timeform{18h 02m 22.44s} & \timeform{+45D 52' 44.6''} & Gaia & 19.53(2) & 18.40 & 19.4(1) \\
ASASSN-13bj & \timeform{16h 00m 20.52s} & \timeform{+70D 50' 07.2''} & Gaia & -- & 18.43 & -- \\
ASASSN-13bo & \timeform{01h 43m 54.23s} & \timeform{+29D 01' 03.8''} & SDSS & 20.94(4) & -- & 21.8(2) \\
ASASSN-13cs & \timeform{17h 11m 38.40s} & \timeform{+05D 39' 51.0''} & Gaia & -- & 19.80 & 20.9(2) \\
ASASSN-13cz & \timeform{15h 27m 55.11s} & \timeform{+63D 27' 54.2''} & Gaia & 18.94(1) & 18.74 & -- \\
ASASSN-14gg & \timeform{18h 21m 38.61s} & \timeform{+61D 59' 04.0''} & Gaia & -- & 19.74 & 19.4(1) \\
ASASSN-15cr & \timeform{07h 34m 42.71s} & \timeform{+50D 42' 29.0''} & Gaia & -- & 19.33 & 20.2(1) \\
ASASSN-16da & \timeform{12h 56m 09.83s} & \timeform{+62D 37' 04.4''} & SDSS & 21.55(5) & -- & 21.6(4) \\
ASASSN-16dk & \timeform{10h 20m 53.48s} & \timeform{-86D 17' 29.77''} & Gaia & -- & 20.41 & 19.31(7) \\
ASASSN-16ds & \timeform{18h 25m 09.96s} & \timeform{-46D 20' 17.9''} & ASAS-SN & -- & -- & -- \\
ASASSN-16dz & \timeform{06h 42m 25.58s} & \timeform{+08D 25' 46.6''} & Gaia & -- & 19.10 & -- \\
ASASSN-16ez & \timeform{15h 31m 29.87s} & \timeform{+21D 38' 30.2''} & SDSS & 21.28(4) & -- & -- \\
ASASSN-16fr & \timeform{16h 42m 51.80s} & \timeform{-08D 52' 41.0''} & SDSS & 20.97(4) & -- & -- \\
ASASSN-16fu & \timeform{22h 14m 05.03s} & \timeform{-09D 04' 19.4''} & SDSS & 21.64(7) & -- & -- \\
ASASSN-16gh & \timeform{18h 15m 57.62s} & \timeform{-72D 40' 38.1''} & ASAS-SN & -- & -- & -- \\
ASASSN-16gj & \timeform{09h 59m 58.97s} & \timeform{-19D 01' 00.0''} & GSC2.3.2 & -- & -- & 21.3(3) \\
ASASSN-16gl & \timeform{18h 27m 16.25s} & \timeform{-52D 47' 44.1''} & ASAS-SN & -- & -- & -- \\
ASASSN-16hi & \timeform{21h 38m 58.01s} & \timeform{-73D 19' 17.5''} & Gaia & -- & 18.86 & 20.9(2) \\
ASASSN-16ia & \timeform{20h 51m 59.24s} & \timeform{+34D 49' 46.1''} & Gaia & -- & -- & -- \\
ASASSN-16ib & \timeform{14h 32m 03.74s} & \timeform{-33D 08' 13.9''} & IGSL & -- & -- & 21.5(4) \\
ASASSN-16ik & \timeform{19h 27m 45.88s} & \timeform{-67D 15' 16.7''} & IGSL & -- & -- & 21.8(5) \\
ASASSN-16is & \timeform{18h 31m 03.63s} & \timeform{+11D 32' 02.9''} & Gaia & -- & 20.36 & -- \\
ASASSN-16iu & \timeform{01h 43m 47.87s} & \timeform{-70D 17' 01.1''} & Gaia & -- & 19.99 & 20.39(9) \\
ASASSN-16iw & \timeform{00h 58m 11.10s} & \timeform{-01D 07' 50.9''} & SDSS & 21.9(1) & -- & -- \\
\hline
  \multicolumn{7}{l}{\parbox{440pt}{\commenta source of the coordinates:
2MASS (2MASS All-Sky Catalog of Point Sources; \cite{2MASS}),
ASAS-SN (ASAS-SN measurements),
CRTS (CRTS measurements),
Gaia (Gaia DR1, \cite{GaiaDR1} and outburst detections),
GSC2.3.2 (The Guide Star Catalog, Version 2.3.2, \cite{GSC232}),
IGSL (The Initial Gaia Source List 3, \cite{IGSL}),
IPHAS DR2 (INT/WFC Photometric H$\alpha$ Survey, \cite{wit08IPHAS}),
SDSS (The SDSS Photometric Catalog, Release 9, \cite{SDSS9}).
}} \\
\end{tabular}
\end{center}
\end{table*}

\addtocounter{table}{-1}
\begin{table*}
\caption{Coordinates of objects without coordinate-based names (continued).}
\begin{center}
\begin{tabular}{ccccccc}
\hline
Object & Right Ascention & Declination & Source\commenta & SDSS $g$ & Gaia G & GALEX NUV \\
\hline
ASASSN-16jb & \timeform{17h 50m 44.99s} & \timeform{-25D 58' 37.1''} & ASAS-SN & -- & -- & -- \\
ASASSN-16jd & \timeform{18h 50m 33.33s} & \timeform{-26D 50' 40.8''} & ASAS-SN & -- & -- & -- \\
ASASSN-16jk & \timeform{15h 40m 24.84s} & \timeform{+23D 07' 50.8''} & Gaia & 20.73(3) & 20.68 & 21.7(3) \\
ASASSN-16js & \timeform{00h 51m 19.17s} & \timeform{-65D 57' 17.0''} & Gaia & -- & 20.08 & 22.1(2) \\
ASASSN-16jz & \timeform{19h 18m 53.39s} & \timeform{+79D 32' 16.0''} & IGSL & -- & -- & -- \\
ASASSN-16kg & \timeform{21h 36m 29.86s} & \timeform{-25D 13' 48.3''} & CRTS & -- & -- & -- \\
ASASSN-16kx & \timeform{06h 17m 18.72s} & \timeform{-49D 38' 57.3''} & ASAS-SN & -- & -- & -- \\
ASASSN-16le & \timeform{23h 34m 35.56s} & \timeform{+54D 33' 25.5''} & Gaia & -- & 18.83 & -- \\
ASASSN-16lj & \timeform{20h 15m 46.04s} & \timeform{+75D 47' 41.7''} & Gaia & 20.99(5) & 20.17 & 21.5(2) \\
ASASSN-16lo & \timeform{18h 08m 41.02s} & \timeform{+46D 19' 34.9''} & IGSL & -- & -- & -- \\
ASASSN-16mo & \timeform{02h 56m 56.67s} & \timeform{+49D 27' 47.1''} & Gaia & -- & 20.19 & -- \\
ASASSN-16my & \timeform{07h 41m 08.46s} & \timeform{-30D 03' 17.9''} & Gaia & -- & 18.52 & -- \\
ASASSN-16ni & \timeform{05h 05m 00.32s} & \timeform{+60D 45' 53.7''} & ASAS-SN & -- & -- & -- \\
ASASSN-16nq & \timeform{23h 22m 09.25s} & \timeform{+39D 50' 07.8''} & Gaia & -- & 19.10 & 21.1(3) \\
ASASSN-16nr & \timeform{07h 09m 49.33s} & \timeform{-49D 09' 03.6''} & GSC2.3.2 & -- & -- & -- \\
ASASSN-16nw & \timeform{01h 53m 49.09s} & \timeform{+52D 52' 05.1''} & IGSL & -- & -- & -- \\
ASASSN-16ob & \timeform{06h 47m 18.89s} & \timeform{-64D 37' 07.3''} & Gaia & -- & -- & -- \\
ASASSN-16oi & \timeform{06h 21m 32.38s} & \timeform{-62D 58' 15.6''} & GSC2.3.2 & -- & -- & 22.0(5) \\
ASASSN-16os & \timeform{08h 43m 05.59s} & \timeform{-84D 53' 45.6''} & GSC2.3.2 & -- & -- & -- \\
ASASSN-16ow & \timeform{06h 30m 47.05s} & \timeform{+02D 39' 31.4''} & IPHAS & -- & -- & -- \\
ASASSN-17aa & \timeform{04h 23m 56.40s} & \timeform{-74D 05' 27.5''} & ASAS-SN & -- & -- & -- \\
ASASSN-17ab & \timeform{10h 40m 51.25s} & \timeform{-37D 03' 30.2''} & Gaia & -- & -- & -- \\
ASASSN-17az & \timeform{00h 15m 09.31s} & \timeform{-69D 45' 49.2''} & ASAS-SN & -- & -- & -- \\
ASASSN-17bl & \timeform{12h 31m 50.86s} & \timeform{-50D 25' 07.4''} & ASAS-SN & -- & -- & -- \\
ASASSN-17bm & \timeform{10h 55m 27.84s} & \timeform{-48D 04' 27.4''} & GSC2.3.2 & -- & -- & -- \\
ASASSN-17bv & \timeform{09h 08m 45.65s} & \timeform{-62D 37' 11.0''} & IGSL & -- & -- & -- \\
ASASSN-17ce & \timeform{13h 24m 24.46s} & \timeform{-54D 09' 21.7''} & Gaia & -- & 18.52 & -- \\
ASASSN-17ck & \timeform{08h 30m 46.29s} & \timeform{-28D 58' 13.5''} & GSC2.3.2 & -- & -- & -- \\
ASASSN-17cn & \timeform{09h 31m 22.60s} & \timeform{-35D 20' 54.3''} & Gaia & -- & -- & -- \\
ASASSN-17cx & \timeform{10h 59m 57.97s} & \timeform{-11D 57' 56.8''} & GSC2.3.2 & -- & -- & 20.8(2) \\
ASASSN-17dg & \timeform{16h 02m 33.49s} & \timeform{-60D 32' 50.3''} & 2MASS & -- & -- & -- \\
ASASSN-17dq & \timeform{09h 01m 25.26s} & \timeform{-59D 31' 40.1''} & ASAS-SN & -- & -- & -- \\
DDE 26 & \timeform{22h 03m 28.21s} & \timeform{+30D 56' 36.5''} & Gaia & 19.61(1) & 19.32 & -- \\
SBS 1108$+$574 & \timeform{11h 11m 26.83s} & \timeform{+57D 12' 38.6''} & Gaia & 19.22(1) & 19.26 & 19.5(1) \\
\hline
\end{tabular}
\end{center}
\end{table*}

\begin{table*}
\caption{Superhump Periods and Period Derivatives}\label{tab:perlist}
\begin{center}
\begin{tabular}{c@{\hspace{7pt}}c@{\hspace{7pt}}c@{\hspace{7pt}}c@{\hspace{7pt}}c@{\hspace{7pt}}c@{\hspace{7pt}}c@{\hspace{7pt}}c@{\hspace{7pt}}c@{\hspace{7pt}}c@{\hspace{7pt}}c@{\hspace{7pt}}c@{\hspace{7pt}}c@{\hspace{7pt}}c}
\hline
Object & Year & $P_1$ (d) & err & \multicolumn{2}{c}{$E_1$\commenta} & $P_{\rm dot}$\commentb & err\commentb & $P_2$ (d) & err & \multicolumn{2}{c}{$E_2$\commenta} & $P_{\rm orb}$ (d)\commentc & Q\commentd \\
\hline
V1047 Aql & 2016 & 0.073666 & 0.000054 & 0 & 14 & -- & -- & -- & -- & -- & -- & -- & C \\
BB Ari & 2016 & 0.072491 & 0.000026 & 27 & 70 & 19.7 & 4.2 & 0.072179 & 0.000019 & 70 & 115 & -- & A \\
OY Car & 2016 & 0.064653 & 0.000028 & 0 & 104 & 9.9 & 1.7 & 0.064440 & 0.000049 & 103 & 159 & 0.063121 & B \\
HT Cas & 2016 & 0.076333 & 0.000005 & 19 & 62 & -- & -- & 0.075886 & 0.000005 & 72 & 145 & 0.073647 & A \\
GS Cet & 2016 & 0.056645 & 0.000014 & 14 & 156 & 6.3 & 0.6 & -- & -- & -- & -- & 0.05597 & AE \\
GZ Cet & 2016 & 0.056702 & 0.000028 & 0 & 54 & 11.4 & 2.8 & 0.056409 & 0.000006 & 141 & 425 & 0.055343 & B \\
AK Cnc & 2016 & 0.067454 & 0.000030 & 0 & 76 & -- & -- & -- & -- & -- & -- & 0.0651 & C \\
GZ Cnc & 2017 & 0.092881 & 0.000022 & 32 & 91 & $-$0.9 & 4.9 & 0.092216 & 0.000291 & 91 & 113 & 0.08825 & C \\
GP CVn & 2016 & 0.064796 & 0.000027 & 17 & 96 & 9.5 & 2.5 & -- & -- & -- & -- & 0.062950 & B \\
V1113 Cyg & 2016 & 0.078848 & 0.000028 & 52 & 141 & $-$2.4 & 2.9 & -- & -- & -- & -- & -- & B \\
IX Dra & 2016 & 0.066895 & 0.000045 & 0 & 92 & 4.7 & 4.6 & -- & -- & -- & -- & -- & C \\
IR Gem & 2016 & 0.071090 & 0.000047 & 0 & 33 & -- & -- & 0.070633 & 0.000047 & 56 & 104 & 0.0684 & C \\
IR Gem & 2017 & 0.071098 & 0.000020 & 25 & 56 & -- & -- & -- & -- & -- & -- & 0.0684 & C \\
NY Her & 2016 & 0.075832 & 0.000043 & 0 & 42 & -- & -- & 0.075525 & 0.000051 & 49 & 114 & -- & B \\
V699 Oph & 2016 & 0.070212 & 0.000096 & 0 & 28 & -- & -- & -- & -- & -- & -- & -- & C \\
V344 Pav & 2016 & 0.079878 & 0.000031 & 0 & 76 & $-$8.8 & 2.7 & -- & -- & -- & -- & -- & CG \\
V893 Sco & 2016 & 0.074666 & 0.000326 & 0 & 26 & -- & -- & -- & -- & -- & -- & 0.075961 & C2 \\
V493 Ser & 2016 & -- & -- & -- & -- & -- & -- & 0.082730 & 0.000129 & 0 & 13 & 0.08001 & C \\
V1389 Tau & 2016 & 0.080456 & 0.000081 & 0 & 35 & -- & -- & 0.079992 & 0.000025 & 34 & 121 & -- & C \\
\hline
  \multicolumn{14}{l}{\commenta Interval used for calculating the period (corresponding to $E$ in section \ref{sec:individual}).} \\
  \multicolumn{14}{l}{\commentb Unit $10^{-5}$.} \\
  \multicolumn{14}{l}{\parbox{440pt}{\commentc References: \\
GZ Cet \citep{pre04j0137},
AK Cnc \citep{are98akcnc},
GZ Cnc \citep{tap03gzcnc},
IR Gem \citep{fei88irgem},
V493 Ser \citep{tho15SDSSCVs},
HV Vir \citep{pat03suumas},
SBS 1108 \citep{Pdot4},
OY Car, GS Cet, GP CVn, V893 Sco,
ASASSN-16da, ASASSN-16fu, ASASSN-16ia,
ASASSN-16is, ASASSN-16jb, ASASSN-16js,
ASASSN-16lo, ASASSN-16oi, ASASSN-16os,
ASASSN-17bl, ASASSN-17cn, MASTER J042609,
MASTER J220559, SDSS J115207 (this work)
  }} \\
  \multicolumn{14}{l}{\parbox{440pt}{\commentd Data quality and comments. A: excellent, B: partial coverage or slightly low quality, C: insufficient coverage or observations with large scatter, G: $P_{\rm dot}$ denotes global $P_{\rm dot}$, M: observational gap in middle stage, U: uncertainty in alias selection, 2: late-stage coverage, the listed period may refer to $P_2$, a: early-stage coverage, the listed period may be contaminated by stage A superhumps, E: $P_{\rm orb}$ refers to the period of early superhumps, P: $P_{\rm orb}$ refers to a shorter stable periodicity recorded in outburst.}} \\
\end{tabular}
\end{center}
\end{table*}

\addtocounter{table}{-1}
\begin{table*}
\caption{Superhump Periods and Period Derivatives (continued)}
\begin{center}
\begin{tabular}{c@{\hspace{7pt}}c@{\hspace{7pt}}c@{\hspace{7pt}}c@{\hspace{7pt}}c@{\hspace{7pt}}c@{\hspace{7pt}}c@{\hspace{7pt}}c@{\hspace{7pt}}c@{\hspace{7pt}}c@{\hspace{7pt}}c@{\hspace{7pt}}c@{\hspace{7pt}}c@{\hspace{7pt}}c}
\hline
Object & Year & $P_1$ & err & \multicolumn{2}{c}{$E_1$} & $P_{\rm dot}$ & err & $P_2$ & err & \multicolumn{2}{c}{$E_2$} & $P_{\rm orb}$ & Q \\
\hline
HV Vir & 2016 & 0.058244 & 0.000009 & 31 & 227 & 3.1 & 0.4 & -- & -- & -- & -- & 0.057069 & A \\
NSV 2026 & 2016b & 0.069906 & 0.000022 & 0 & 13 & -- & -- & -- & -- & -- & -- & -- & C \\
NSV 14681 & 2016 & 0.090063 & 0.000008 & 0 & 77 & $-$0.5 & 0.8 & -- & -- & -- & -- & -- & C \\
1RXS J161659 & 2016 & 0.071370 & 0.000063 & 0 & 43 & -- & -- & 0.071063 & 0.000054 & 56 & 74 & -- & C \\
1RXS J161659 & 2016b & 0.071229 & 0.000056 & 0 & 58 & -- & -- & -- & -- & -- & -- & -- & C \\
ASASSN-13al & 2016 & 0.0783 & 0.0002 & 0 & 3 & -- & -- & -- & -- & -- & -- & -- & C \\
ASASSN-13bc & 2015 & 0.070393 & 0.000118 & 0 & 16 & -- & -- & -- & -- & -- & -- & -- & C \\
ASASSN-13bc & 2016 & 0.070624 & 0.000100 & 0 & 39 & -- & -- & 0.070101 & 0.000046 & 39 & 85 & -- & C \\
ASASSN-13bj & 2016 & 0.072553 & 0.000047 & 0 & 21 & -- & -- & 0.071918 & 0.000053 & 23 & 44 & -- & C \\
ASASSN-13bo & 2016 & 0.071860 & 0.000025 & 0 & 41 & -- & -- & -- & -- & -- & -- & -- & CU \\
ASASSN-13cs & 2016 & 0.077105 & 0.000098 & 0 & 20 & -- & -- & -- & -- & -- & -- & -- & C \\
ASASSN-13cz & 2016 & 0.080135 & 0.000044 & 0 & 13 & -- & -- & 0.079496 & 0.000368 & 62 & 76 & -- & C \\
ASASSN-14gg & 2016 & 0.059311 & 0.000035 & 0 & 89 & 13.1 & 2.9 & -- & -- & -- & -- & -- & B \\
ASASSN-15cr & 2017 & 0.061554 & 0.000021 & 16 & 149 & 7.8 & 1.5 & 0.061260 & 0.000005 & 146 & 217 & -- & B \\
ASASSN-16da & 2016 & 0.057344 & 0.000024 & 10 & 175 & 7.5 & 0.9 & 0.056994 & 0.000062 & 203 & 239 & 0.05610 & BE \\
ASASSN-16dk & 2016 & -- & -- & -- & -- & -- & -- & 0.075923 & 0.000047 & 0 & 67 & -- & C \\
ASASSN-16ds & 2016 & 0.067791 & 0.000027 & 33 & 195 & 7.1 & 0.6 & 0.067228 & 0.000051 & -- & -- & -- & B \\
ASASSN-16dt & 2016 & 0.064507 & 0.000005 & 62 & 214 & $-$1.6 & 0.5 & -- & -- & -- & -- & 0.064197 & AE \\
ASASSN-16dz & 2016 & 0.066260 & 0.000170 & 0 & 16 & -- & -- & -- & -- & -- & -- & -- & CU \\
ASASSN-16eg & 2016 & 0.077880 & 0.000003 & 15 & 106 & 10.4 & 0.8 & 0.077589 & 0.000007 & 120 & 181 & 0.075478 & AE \\
ASASSN-16ez & 2016 & 0.057621 & 0.000017 & 0 & 77 & 2.1 & 2.9 & -- & -- & -- & -- & -- & C \\
ASASSN-16fr & 2016 & 0.071394 & 0.000144 & 0 & 35 & -- & -- & -- & -- & -- & -- & -- & C \\
ASASSN-16fu & 2016 & 0.056936 & 0.000013 & 35 & 195 & 4.6 & 0.6 & -- & -- & -- & -- & 0.05623 & BE \\
ASASSN-16gh & 2016 & 0.061844 & 0.000017 & 16 & 100 & 6.7 & 2.7 & -- & -- & -- & -- & -- & B \\
ASASSN-16gj & 2016 & 0.057997 & 0.000022 & 74 & 208 & 7.0 & 1.0 & -- & -- & -- & -- & -- & B \\
ASASSN-16gl & 2016 & 0.055834 & 0.000010 & 0 & 118 & 1.6 & 1.2 & -- & -- & -- & -- & -- & B \\
ASASSN-16hg & 2016 & 0.062371 & 0.000014 & 15 & 115 & 0.6 & 1.7 & -- & -- & -- & -- & -- & B \\
ASASSN-16hi & 2016 & 0.059040 & 0.000024 & 0 & 121 & 8.6 & 1.5 & 0.058674 & 0.000023 & 118 & 188 & -- & B \\
ASASSN-16hj & 2016 & 0.055644 & 0.000041 & 20 & 145 & 11.3 & 1.3 & 0.055465 & 0.000036 & 144 & 324 & 0.05499 & BE \\
\hline
\end{tabular}
\end{center}
\end{table*}

\addtocounter{table}{-1}
\begin{table*}
\caption{Superhump Periods and Period Derivatives (continued)}
\begin{center}
\begin{tabular}{c@{\hspace{7pt}}c@{\hspace{7pt}}c@{\hspace{7pt}}c@{\hspace{7pt}}c@{\hspace{7pt}}c@{\hspace{7pt}}c@{\hspace{7pt}}c@{\hspace{7pt}}c@{\hspace{7pt}}c@{\hspace{7pt}}c@{\hspace{7pt}}c@{\hspace{7pt}}c@{\hspace{7pt}}c}
\hline
Object & Year & $P_1$ & err & \multicolumn{2}{c}{$E_1$} & $P_{\rm dot}$ & err & $P_2$ & err & \multicolumn{2}{c}{$E_2$} & $P_{\rm orb}$ & Q \\
\hline
ASASSN-16ib & 2016 & 0.058855 & 0.000015 & 47 & 144 & 2.2 & 2.0 & -- & -- & -- & -- & -- & C \\
ASASSN-16ik & 2016 & 0.064150 & 0.000018 & 33 & 126 & 1.0 & 2.1 & -- & -- & -- & -- & -- & B \\
ASASSN-16is & 2016 & 0.058484 & 0.000015 & 0 & 105 & 4.2 & 1.7 & -- & -- & -- & -- & 0.05762 & CE \\
ASASSN-16iu & 2016 & 0.058720 & 0.000062 & 0 & 104 & 26.7 & 3.3 & 0.058661 & 0.000300 & 34 & 53 & -- & C \\
ASASSN-16iw & 2016 & 0.065462 & 0.000039 & 42 & 153 & 10.0 & 3.2 & -- & -- & -- & -- & 0.06495 & BE \\
ASASSN-16jb & 2016 & 0.064397 & 0.000021 & 30 & 193 & 5.9 & 0.7 & 0.064170 & 0.000075 & 193 & 232 & 0.06305 & AE \\
ASASSN-16jd & 2016 & 0.058163 & 0.000039 & 34 & 223 & 7.9 & 0.6 & 0.057743 & 0.000159 & 223 & 258 & -- & B \\
ASASSN-16jk & 2016 & 0.061391 & 0.000028 & 16 & 146 & 8.6 & 1.3 & -- & -- & -- & -- & -- & C \\
ASASSN-16js & 2016 & 0.060934 & 0.000015 & 48 & 173 & 4.9 & 1.0 & -- & -- & -- & -- & 0.06034 & AE \\
ASASSN-16jz & 2016 & 0.060936 & 0.000014 & 0 & 51 & -- & -- & -- & -- & -- & -- & -- & C \\
ASASSN-16kg & 2016 & 0.100324 & 0.000189 & 0 & 30 & -- & -- & -- & -- & -- & -- & -- & CU \\
ASASSN-16kx & 2016 & 0.080760 & 0.000036 & 0 & 54 & $-$6.4 & 6.5 & 0.080536 & 0.000041 & 79 & 153 & -- & C \\
ASASSN-16le & 2016 & 0.0808 & 0.0013 & 0 & 2 & -- & -- & -- & -- & -- & -- & -- & C \\
ASASSN-16lj & 2016 & 0.0857 & 0.0004 & 0 & 2 & -- & -- & -- & -- & -- & -- & -- & C \\
ASASSN-16lo & 2016 & 0.054608 & 0.000036 & 38 & 86 & -- & -- & -- & -- & -- & -- & 0.05416 & CE \\
ASASSN-16mo & 2016 & 0.066477 & 0.000016 & 0 & 84 & 3.9 & 2.3 & -- & -- & -- & -- & -- & C \\
ASASSN-16my & 2016 & 0.087683 & 0.000049 & 23 & 92 & 3.0 & 5.7 & -- & -- & -- & -- & -- & C \\
ASASSN-16ni & 2016 & 0.115242 & 0.000442 & 0 & 11 & -- & -- & -- & -- & -- & -- & -- & CU \\
ASASSN-16nq & 2016 & 0.079557 & 0.000045 & 0 & 39 & 0.0 & 9.3 & 0.079069 & 0.000035 & 59 & 161 & -- & B \\
ASASSN-16nr & 2016 & 0.082709 & 0.000080 & 0 & 59 & $-$19.8 & 10.1 & -- & -- & -- & -- & -- & CG \\
ASASSN-16nw & 2016 & 0.072813 & 0.000045 & 0 & 43 & -- & -- & -- & -- & -- & -- & -- & C \\
ASASSN-16ob & 2016 & 0.057087 & 0.000014 & 52 & 249 & 1.8 & 0.5 & -- & -- & -- & -- & -- & B \\
ASASSN-16oi & 2016 & 0.056241 & 0.000017 & 12 & 122 & 5.0 & 1.7 & -- & -- & -- & -- & 0.05548 & BE \\
ASASSN-16os & 2016 & 0.054992 & 0.000013 & 39 & 168 & 0.3 & 1.4 & -- & -- & -- & -- & 0.05494 & BE \\
ASASSN-16ow & 2016 & 0.089311 & 0.000052 & 0 & 40 & -- & -- & 0.088866 & 0.000022 & 55 & 102 & -- & B \\
ASASSN-17aa & 2017 & 0.054591 & 0.000013 & 0 & 182 & 2.8 & 0.3 & -- & -- & -- & -- & 0.05393 & BE \\
ASASSN-17ab & 2017 & 0.070393 & 0.000016 & 15 & 88 & 3.6 & 2.5 & -- & -- & -- & -- & -- & C \\
ASASSN-17az & 2017 & 0.056492 & 0.000038 & 0 & 36 & -- & -- & -- & -- & -- & -- & -- & CU \\
ASASSN-17bl & 2017 & 0.055367 & 0.000010 & 53 & 237 & 3.6 & 0.6 & -- & -- & -- & -- & 0.05467 & CE \\
\hline
\end{tabular}
\end{center}
\end{table*}

\addtocounter{table}{-1}
\begin{table*}
\caption{Superhump Periods and Period Derivatives (continued)}
\begin{center}
\begin{tabular}{c@{\hspace{7pt}}c@{\hspace{7pt}}c@{\hspace{7pt}}c@{\hspace{7pt}}c@{\hspace{7pt}}c@{\hspace{7pt}}c@{\hspace{7pt}}c@{\hspace{7pt}}c@{\hspace{7pt}}c@{\hspace{7pt}}c@{\hspace{7pt}}c@{\hspace{7pt}}c@{\hspace{7pt}}c}
\hline
Object & Year & $P_1$ & err & \multicolumn{2}{c}{$E_1$} & $P_{\rm dot}$ & err & $P_2$ & err & \multicolumn{2}{c}{$E_2$} & $P_{\rm orb}$ & Q \\
\hline
ASASSN-17bm & 2017 & 0.082943 & 0.000056 & 0 & 53 & -- & -- & -- & -- & -- & -- & -- & C \\
ASASSN-17bv & 2017 & 0.082690 & 0.000021 & 12 & 52 & $-$6.3 & 3.9 & 0.082489 & 0.000048 & 58 & 103 & -- & B \\
ASASSN-17ce & 2017 & 0.081293 & 0.000111 & 0 & 22 & -- & -- & 0.080796 & 0.000042 & 21 & 139 & -- & C \\
ASASSN-17ck & 2017 & 0.083 & 0.001 & 0 & 1 & -- & -- & -- & -- & -- & -- & -- & C \\
ASASSN-17cn & 2017 & 0.053991 & 0.000014 & 0 & 137 & 5.6 & 0.8 & -- & -- & -- & -- & 0.05303 & BE \\
ASASSN-17cx & 2017 & 0.0761 & 0.0007 & 0 & 2 & -- & -- & -- & -- & -- & -- & -- & C \\
ASASSN-17dg & 2017 & -- & -- & -- & -- & -- & -- & 0.066482 & 0.000046 & 0 & 36 & -- & C \\
ASASSN-17dq & 2017 & 0.058052 & 0.000034 & 0 & 93 & 9.3 & 3.5 & 0.057660 & 0.000076 & 90 & 142 & -- & C \\
CRTS J000130 & 2016 & 0.094749 & 0.000066 & 0 & 63 & -- & -- & -- & -- & -- & -- & -- & C \\
CRTS J023638 & 2016 & 0.073703 & 0.000057 & 0 & 42 & -- & -- & 0.073504 & 0.000053 & 40 & 80 & -- & C \\
CRTS J033349 & 2016 & -- & -- & -- & -- & -- & -- & 0.076159 & 0.000049 & 0 & 60 & -- & C \\
CRTS J082603 & 2017 & 0.0719 & 0.0004 & 0 & 1 & -- & -- & -- & -- & -- & -- & -- & C \\
CRTS J085113 & 2016 & 0.08750 & 0.00009 & 0 & 1 & -- & -- & -- & -- & -- & -- & -- & C \\
CRTS J085603 & 2016 & 0.060043 & 0.000193 & 0 & 18 & -- & -- & -- & -- & -- & -- & -- & C \\
CRTS J164950 & 2016 & 0.064905 & 0.000091 & 0 & 61 & -- & -- & -- & -- & -- & -- & -- & C \\
CSS J044637 & 2017 & 0.093 & 0.001 & 0 & 1 & -- & -- & -- & -- & -- & -- & -- & C \\
CSS J062450 & 2017 & 0.077577 & 0.000094 & 0 & 14 & -- & -- & -- & -- & -- & -- & -- & C \\
DDE 26 & 2016 & 0.088804 & 0.000067 & 0 & 44 & -- & -- & -- & -- & -- & -- & -- & C \\
MASTER J021315 & 2016 & 0.105124 & 0.000252 & 10 & 21 & -- & -- & -- & -- & -- & -- & -- & C \\
MASTER J030205 & 2016 & 0.061553 & 0.000022 & 1 & 96 & 8.4 & 2.5 & -- & -- & -- & -- & -- & B \\
MASTER J042609 & 2016 & 0.067624 & 0.000016 & 0 & 64 & 6.4 & 2.7 & 0.067221 & 0.000051 & 64 & 122 & 0.065502 & B \\
MASTER J043220 & 2017 & 0.0640 & 0.0006 & 0 & 1 & -- & -- & -- & -- & -- & -- & -- & C \\
MASTER J043915 & 2016 & 0.062428 & 0.000045 & 0 & 112 & -- & -- & -- & -- & -- & -- & -- & C \\
MASTER J054746 & 2016 & 0.0555 & 0.0004 & 0 & 3 & -- & -- & -- & -- & -- & -- & -- & C \\
MASTER J055348 & 2017 & 0.0750 & 0.0001 & 0 & 24 & -- & -- & -- & -- & -- & -- & -- & CU \\
MASTER J064725 & 2016 & 0.067584 & 0.000020 & 0 & 108 & 1.2 & 3.5 & -- & -- & -- & -- & -- & CG \\
MASTER J065330 & 2017 & 0.064012 & 0.000167 & 0 & 13 & -- & -- & -- & -- & -- & -- & -- & C \\
MASTER J075450 & 2017 & 0.0664 & 0.0050 & 0 & 1 & -- & -- & -- & -- & -- & -- & -- & C \\
MASTER J150518 & 2017 & 0.071145 & 0.000125 & 0 & 56 & $-$29.5 & 1.0 & -- & -- & -- & -- & -- & CGU \\
\hline
\end{tabular}
\end{center}
\end{table*}

\addtocounter{table}{-1}
\begin{table*}
\caption{Superhump Periods and Period Derivatives (continued)}
\begin{center}
\begin{tabular}{c@{\hspace{7pt}}c@{\hspace{7pt}}c@{\hspace{7pt}}c@{\hspace{7pt}}c@{\hspace{7pt}}c@{\hspace{7pt}}c@{\hspace{7pt}}c@{\hspace{7pt}}c@{\hspace{7pt}}c@{\hspace{7pt}}c@{\hspace{7pt}}c@{\hspace{7pt}}c@{\hspace{7pt}}c}
\hline
Object & Year & $P_1$ & err & \multicolumn{2}{c}{$E_1$} & $P_{\rm dot}$ & err & $P_2$ & err & \multicolumn{2}{c}{$E_2$} & $P_{\rm orb}$ & Q \\
\hline
MASTER J151126 & 2016 & 0.058182 & 0.000016 & 16 & 171 & 4.5 & 0.6 & -- & -- & -- & -- & -- & C \\
MASTER J055845 & 2016 & 0.058070 & 0.000081 & 0 & 19 & -- & -- & -- & -- & -- & -- & -- & C2 \\
MASTER J162323 & 2016 & 0.09013 & 0.00007 & 0 & 4 & -- & -- & -- & -- & -- & -- & -- & Ca \\
MASTER J165153 & 2017 & 0.071951 & 0.000079 & 0 & 31 & -- & -- & -- & -- & -- & -- & -- & C \\
MASTER J174816 & 2016 & 0.083328 & 0.000120 & 0 & 21 & -- & -- & -- & -- & -- & -- & -- & CU \\
MASTER J191841 & 2016 & 0.022076 & 0.000007 & 0 & 51 & -- & -- & -- & -- & -- & -- & -- & B \\
MASTER J220559 & 2016 & 0.061999 & 0.000067 & 0 & 83 & 28.4 & 6.5 & 0.061434 & 0.000078 & 81 & 116 & 0.061286 & C \\
OT J002656 & 2016 & 0.132240 & 0.000054 & 30 & 112 & 16.4 & 1.6 & -- & -- & -- & -- & -- & B \\
SBS 1108 & 2016 & 0.039051 & 0.000008 & 0 & 72 & -- & -- & -- & -- & -- & -- & 0.038449 & CP \\
SDSS J032015 & 2016 & 0.073757 & 0.000028 & 0 & 137 & 2.5 & 4.2 & -- & -- & -- & -- & -- & CG \\
SDSS J091001 & 2017 & 0.0734 & 0.0002 & 0 & 2 & -- & -- & -- & -- & -- & -- & -- & C \\
SDSS J113551 & 2017 & 0.0966 & 0.0001 & 0 & 18 & -- & -- & -- & -- & -- & -- & -- & CU \\
SDSS J115207 & 2009 & 0.070028 & 0.000088 & 0 & 68 & -- & -- & -- & -- & -- & -- & 0.067750 & CG \\
SDSS J115207 & 2017 & 0.070362 & 0.000044 & 0 & 52 & -- & -- & 0.069914 & 0.000019 & 52 & 131 & 0.067750 & B \\
SDSS J131432 & 2017 & 0.065620 & 0.000034 & 0 & 55 & 18.3 & 8.6 & -- & -- & -- & -- & -- & C \\
SDSS J153015 & 2017 & 0.075241 & 0.000039 & 0 & 41 & -- & -- & -- & -- & -- & -- & -- & C \\
SDSS J155720 & 2016 & 0.085565 & 0.000131 & 0 & 29 & -- & -- & -- & -- & -- & -- & -- & C \\
SDSS J173047 & 2016 & 0.024597 & 0.000007 & 0 & 329 & 0.8 & 0.3 & -- & -- & -- & -- & -- & B \\
SSS J134850 & 2016 & 0.084534 & 0.000017 & 0 & 80 & $-$3.0 & 1.6 & -- & -- & -- & -- & -- & CG \\
TCP J013758 & 2016 & 0.061692 & 0.000024 & 31 & 142 & 12.6 & 0.8 & 0.061408 & 0.000032 & 140 & 208 & -- & B \\
TCP J180018 & 2016 & 0.058449 & 0.000024 & 26 & 233 & 5.7 & 0.7 & -- & -- & -- & -- & -- & B \\
\hline
\end{tabular}
\end{center}
\end{table*}

\section{Individual Objects}\label{sec:individual}

\subsection{V1047 Aquilae}\label{obj:v1047aql}

   V1047 Aql was discovered as a dwarf nova (S 8191)
by \citet{hof64an28849}.  \citet{hof64an28849} reported
a blue color in contrast to the nearby stars.
\citet{mas03faintCV} obtained a spectrum typical for
a quiescent dwarf nova.
According to R. Stubbings, the observation by Greg Bolt
during the 2005 August outburst detected superhumps,
and the superhump period was about 0.074~d
(see \cite{kat12DNSDSS}).
The object shows rather frequent outbursts (approximately
once in 50~d), and a number of outbursts have been 
detected mainly by R. Stubbings visually since 2004.

\begin{figure}
  \begin{center}
    \FigureFile(85mm,110mm){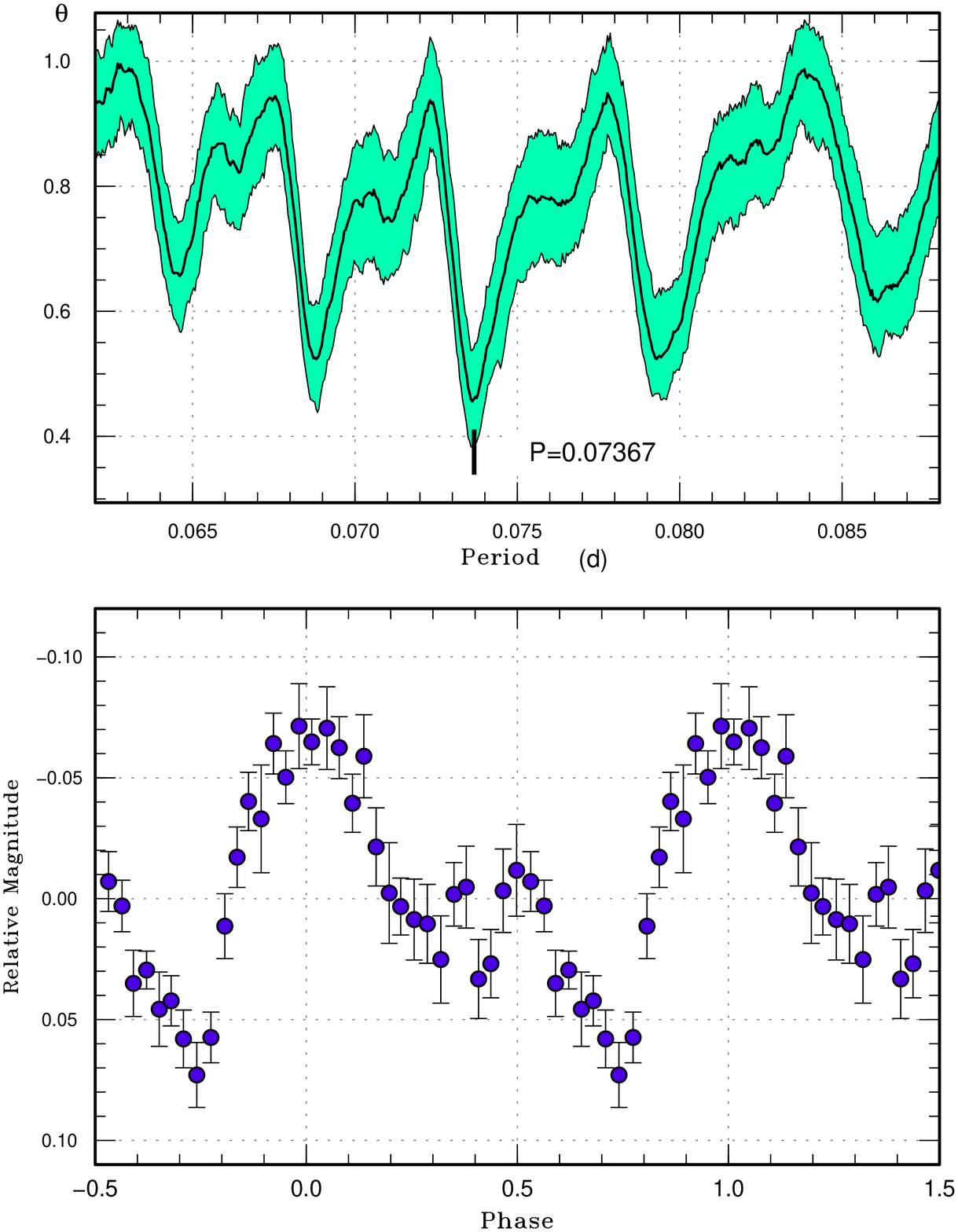}
  \end{center}
  \caption{Superhumps in V1047 Aql (2016).
     (Upper): PDM analysis.
     (Lower): Phase-averaged profile.}
  \label{fig:v1047aqlshpdm}
\end{figure}

   The 2016 superoutburst was detected by R. Stubbings
at a visual magnitude of 15.0 on July 8.  Subsequent
observations detected superhumps (vsnet-alert 19974;
figure \ref{fig:v1047aqlshpdm}).
Using the 2005 period, we could identify two maxima
on two nights: $E$=0, BJD 2457581.3853(7) ($N$=74)
and $E$=14, BJD 2457582.4190(11) ($N$=72).
The period given in table \ref{tab:perlist} is
determined by the PDM method.

   Although observations are not sufficient,
visual observations by R. Stubbings suggest a supercycle
of $\sim$90~d, which would make V1047 Aql one of
ordinary SU UMa-type dwarf novae with shortest
supercycles.


\subsection{BB Arietis}\label{obj:bbari}

   This object was discovered as a variable star
(Ross 182, NSV 907) on a plate on 1926 November 26
\citep{ros27VS5}.  The dwarf nova-type nature was 
suspected by the association with an ROSAT source
(Kato, vsnet-chat 3317).  The SU UMa-type nature
was confirmed during the 2004 superoutburst.
For more information, see \citet{Pdot6}.

\begin{figure}
  \begin{center}
    \FigureFile(88mm,70mm){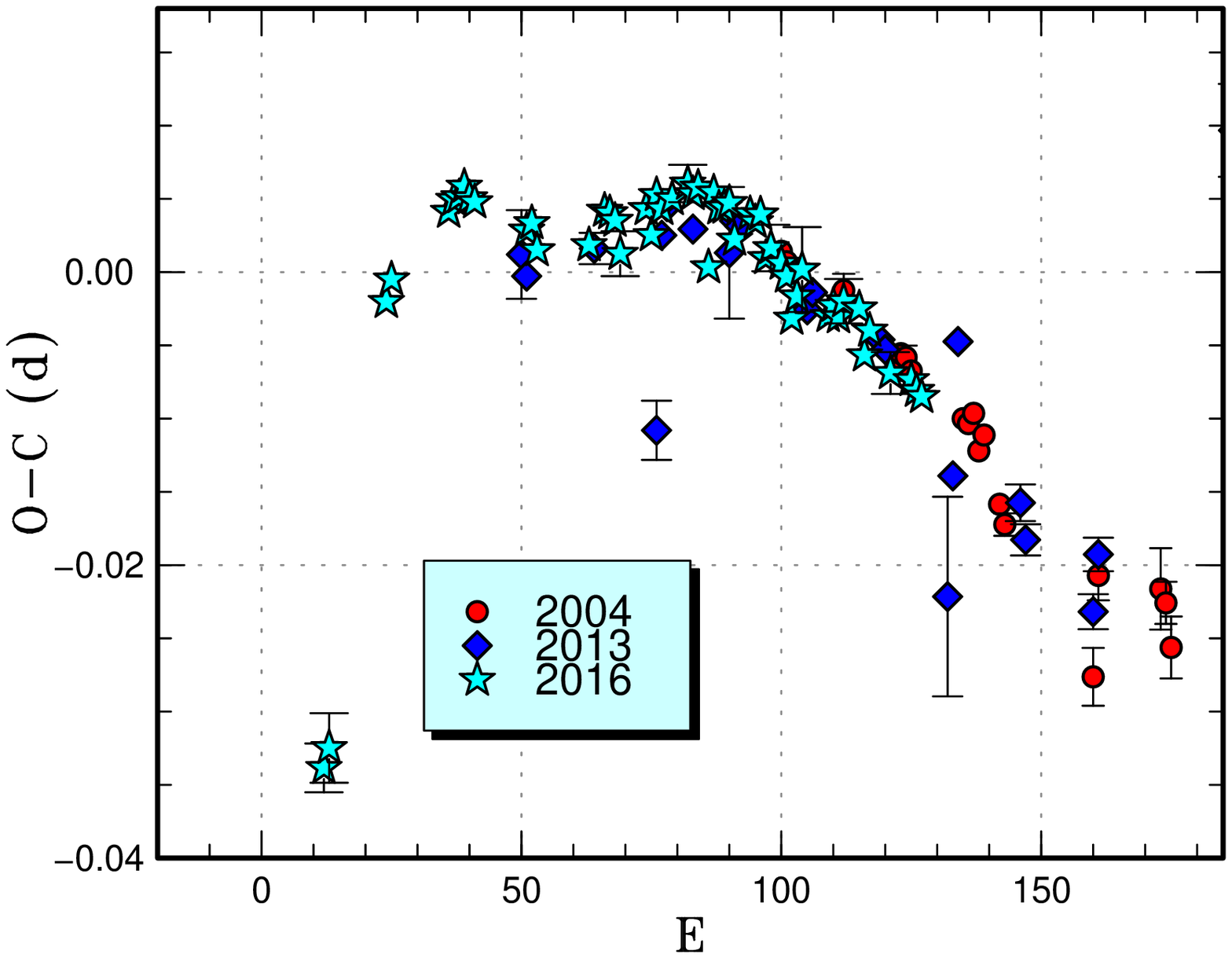}
  \end{center}
  \caption{Comparison of $O-C$ diagrams of BB Ari between different
  superoutbursts.  A period of 0.07249~d was used to draw this figure.
  Approximate cycle counts ($E$) after the starts of outbursts
  were used.  The definition is different from
  the corresponding figure in \citet{Pdot6}.
  The 2013 superoutburst had a separate precursor outburst
  and the cycle count is different by 44 from
  that used in \citet{Pdot6}.  The value suggests that
  superhumps during the 2013 superoutburst evolved
  3~d after the precursor outburst.
  Since the start of the 2004 superoutburst
  was not well constrained, we shifted the $O-C$ diagram
  to best fit the 2016 one.
  }
  \label{fig:bbaricomp2}
\end{figure}

   The 2016 superoutburst was detected by P. Schmeer
at a visual magnitude of 13.2 on October 30
(vsnet-alert 20273).  Thanks to the early detection
(this visual detection was 1~d earlier than
the ASAS-SN detection), stage A growing superhumps
were detected (vsnet-alert 20292).
At the time of the initial observation,
the object was fading from a precursor outburst.
Further observations recorded development of
superhumps clearly (vsnet-alert 20312, 20321).
The times of superhump maxima are listed in
table \ref{tab:bbari2016}.
There were clear stages A--C (figure \ref{fig:bbaricomp2}).
The 2013 superoutburst had a separate precursor outburst
and a comparison of the $O-C$ diagrams suggests
a difference of 44 cycle count from
that used in \citet{Pdot6}.  The value suggests that
superhumps during the 2013 superoutburst evolved
3~d after the precursor outburst.

\begin{table}
\caption{Superhump maxima of BB Ari (2016)}\label{tab:bbari2016}
\begin{center}
\begin{tabular}{rp{55pt}p{40pt}r@{.}lr}
\hline
\multicolumn{1}{c}{$E$} & \multicolumn{1}{c}{max\commenta} & \multicolumn{1}{c}{error} & \multicolumn{2}{c}{$O-C$\commentb} & \multicolumn{1}{c}{$N$\commentc} \\
\hline
0 & 57692.5399 & 0.0017 & $-$0&0323 & 194 \\
1 & 57692.6138 & 0.0024 & $-$0&0310 & 72 \\
12 & 57693.4416 & 0.0003 & $-$0&0008 & 129 \\
13 & 57693.5156 & 0.0003 & 0&0008 & 125 \\
24 & 57694.3177 & 0.0003 & 0&0051 & 188 \\
25 & 57694.3910 & 0.0002 & 0&0059 & 162 \\
26 & 57694.4636 & 0.0002 & 0&0061 & 182 \\
27 & 57694.5368 & 0.0002 & 0&0068 & 194 \\
28 & 57694.6085 & 0.0002 & 0&0059 & 80 \\
29 & 57694.6808 & 0.0002 & 0&0057 & 153 \\
39 & 57695.4037 & 0.0004 & 0&0035 & 80 \\
40 & 57695.4767 & 0.0004 & 0&0040 & 81 \\
41 & 57695.5474 & 0.0005 & 0&0021 & 69 \\
51 & 57696.2726 & 0.0003 & 0&0022 & 82 \\
54 & 57696.4924 & 0.0006 & 0&0045 & 42 \\
55 & 57696.5648 & 0.0005 & 0&0043 & 115 \\
56 & 57696.6367 & 0.0007 & 0&0038 & 128 \\
57 & 57696.7070 & 0.0015 & 0&0015 & 100 \\
62 & 57697.0725 & 0.0006 & 0&0044 & 69 \\
63 & 57697.1432 & 0.0005 & 0&0026 & 48 \\
64 & 57697.2184 & 0.0011 & 0&0054 & 20 \\
65 & 57697.2899 & 0.0012 & 0&0044 & 41 \\
67 & 57697.4356 & 0.0006 & 0&0050 & 47 \\
70 & 57697.6541 & 0.0013 & 0&0060 & 75 \\
71 & 57697.7259 & 0.0012 & 0&0053 & 93 \\
72 & 57697.7988 & 0.0011 & 0&0057 & 113 \\
74 & 57697.9384 & 0.0005 & 0&0002 & 134 \\
75 & 57698.0160 & 0.0004 & 0&0053 & 153 \\
76 & 57698.0874 & 0.0009 & 0&0042 & 216 \\
77 & 57698.1599 & 0.0004 & 0&0042 & 242 \\
78 & 57698.2327 & 0.0003 & 0&0045 & 241 \\
79 & 57698.3027 & 0.0006 & 0&0020 & 133 \\
82 & 57698.5219 & 0.0008 & 0&0037 & 23 \\
83 & 57698.5939 & 0.0004 & 0&0031 & 86 \\
84 & 57698.6669 & 0.0003 & 0&0036 & 77 \\
85 & 57698.7364 & 0.0011 & 0&0006 & 124 \\
86 & 57698.8096 & 0.0016 & 0&0012 & 71 \\
88 & 57698.9535 & 0.0007 & 0&0001 & 135 \\
89 & 57699.0251 & 0.0005 & $-$0&0008 & 135 \\
90 & 57699.0947 & 0.0005 & $-$0&0037 & 127 \\
91 & 57699.1687 & 0.0005 & $-$0&0022 & 135 \\
92 & 57699.2430 & 0.0029 & $-$0&0004 & 49 \\
97 & 57699.6023 & 0.0007 & $-$0&0036 & 33 \\
98 & 57699.6755 & 0.0003 & $-$0&0030 & 74 \\
99 & 57699.7472 & 0.0011 & $-$0&0038 & 111 \\
100 & 57699.8208 & 0.0015 & $-$0&0027 & 83 \\
103 & 57700.0378 & 0.0004 & $-$0&0033 & 129 \\
104 & 57700.1071 & 0.0006 & $-$0&0065 & 127 \\
105 & 57700.1812 & 0.0005 & $-$0&0048 & 134 \\
109 & 57700.4683 & 0.0014 & $-$0&0078 & 40 \\
113 & 57700.7578 & 0.0011 & $-$0&0084 & 144 \\
114 & 57700.8295 & 0.0011 & $-$0&0091 & 90 \\
115 & 57700.9016 & 0.0004 & $-$0&0096 & 67 \\
\hline
  \multicolumn{6}{l}{\commenta BJD$-$2400000.} \\
  \multicolumn{6}{l}{\commentb Against max $= 2457692.5722 + 0.072513 E$.} \\
  \multicolumn{6}{l}{\commentc Number of points used to determine the maximum.} \\
\end{tabular}
\end{center}
\end{table}

\subsection{V391 Camelopardalis}\label{obj:v391cam}

   This object (=1RXS J053234.9$+$624755) was discovered
as a dwarf nova by \citet{ber05j0532}. 
\citet{kap06j0532} provided a radial-velocity
study and yielded an orbital period of 0.05620(4) d.
The SU UMa-type nature was established during
the 2005 superoutburst \citep{ima09j0532}.
See \citet{Pdot} for more history.  The 2009 superoutburst
was also studied in \citet{Pdot2}.

   The 2017 superoutburst was detected by P. Schmeer
at a visual magnitude of 11.4 and also by the ASAS-SN
team at $V$=11.82 on March 15.  Single superhump
was recorded at BJD 2457829.3171(2) ($N$=236).
Although there were observations on three nights
immediately after the superoutburst, we could neither
detect superhump nor orbital periods.

\subsection{OY Carinae}\label{obj:oycar}

\begin{figure}
  \begin{center}
    \FigureFile(85mm,110mm){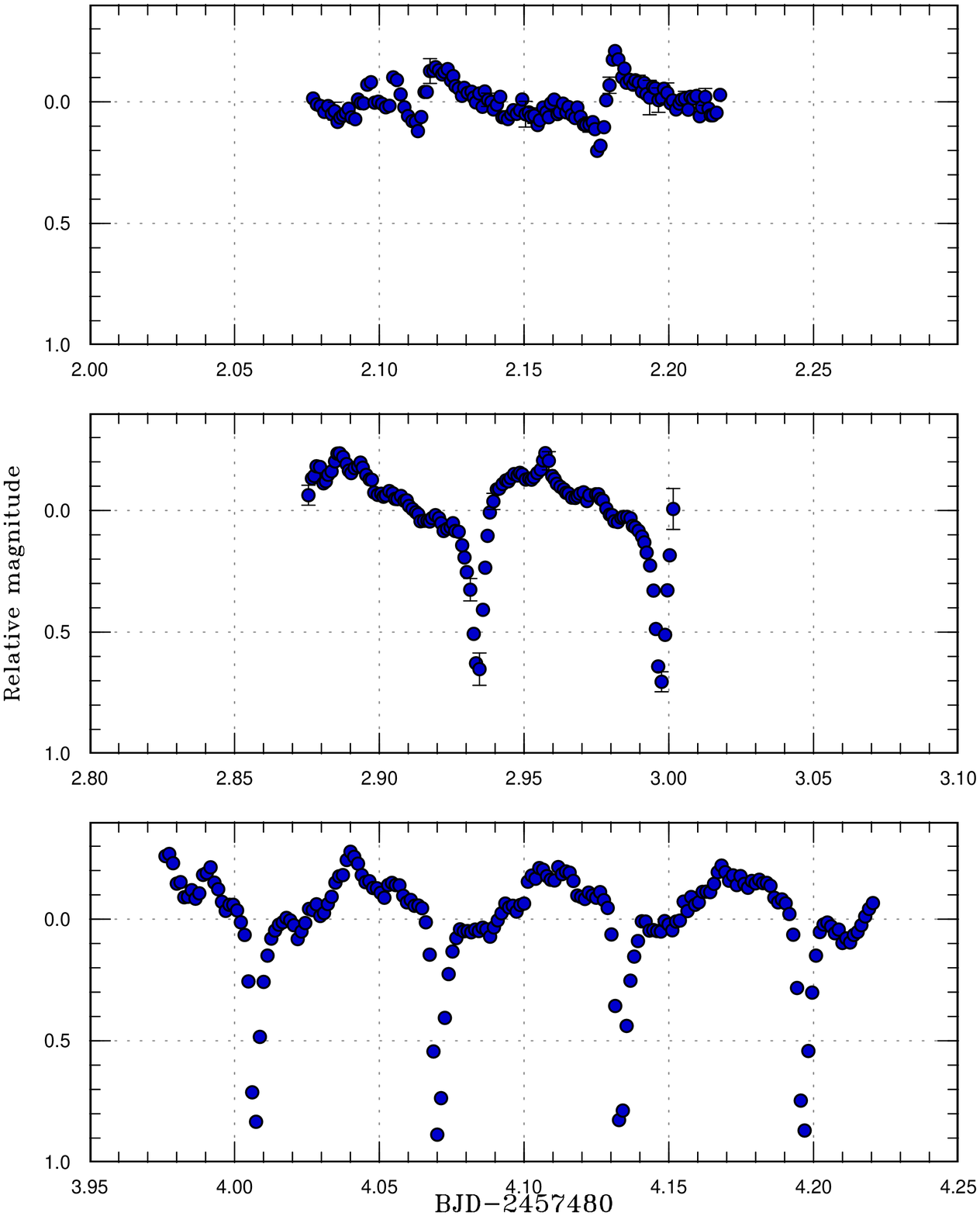}
  \end{center}
  \caption{Eclipses and superhumps in OY Car
  in the earliest phase (2016).
  The data were binned to 0.001~d.
  During the first run (upper panel), eclipses were very
  shallow since they overlapped with superhumps.
  }
  \label{fig:oycarshlc}
\end{figure}

\begin{figure}
  \begin{center}
    \FigureFile(85mm,70mm){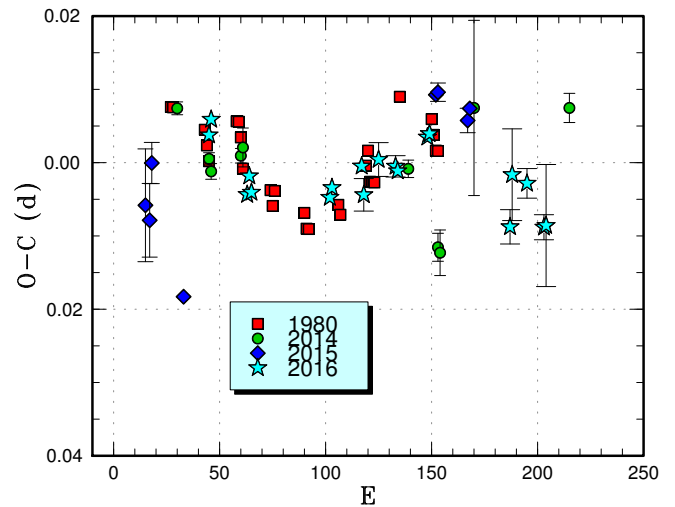}
  \end{center}
  \caption{Comparison of $O-C$ diagrams of OY Car between different
  superoutbursts.  A period of 0.06465~d was used to draw this figure.
  Approximate cycle counts ($E$) after the starts of outbursts
  were used.  The 2015 superoutburst with a separate precursor outburst
  was shifted by 15 cycles to best match the others.
  Since the start of the 2016 superoutburst was not well
  constrained, the values were shifted by 45 cycles
  to best match the others.  This shift suggests that
  the actual start of the 2016 superoutburst occurred
  2~d before the initial detection.
  }
  \label{fig:oycarcomp2}
\end{figure}

   See \citet{Pdot7} for the history of this well-known
eclipsing SU UMa-type dwarf nova.
The 2016 superoutburst was detected by R. Stubbings
at a visual magnitude of 11.6 on April 2 (vsnet-alert 19676).
Due to an accidental delay in the start of observations,
the earliest time-resolved CCD observations were obtained
on April 3 (vsnet-alert 19706).  On that night, superhumps
(likely in the growing phase) unfortunately overlapped
with eclipses (figure \ref{fig:oycarshlc}, upper panel).
Distinct superhumps were recorded on April 4
(vsnet-alert 19692; figure \ref{fig:oycarshlc}, middle panel).
A further analysis suggested that stage A superhumps
escaped detection before April 4 (due to the lack of
observations and overlapping eclipses).
At the time of April 4, the superhumps were already
likely stage B (table \ref{tab:oycaroc2016},
maxima outside eclipses).
We could, however, confirmed a positive $P_{\rm dot}$
for stage B superhumps (cf. figure \ref{fig:oycarcomp2}),
whose confirmation had been still awaited (cf. \cite{Pdot7}).

   The combined data used in \citet{Pdot7} and new observations,
we have obtained the eclipse ephemeris for the use of
defining the orbital phases in this paper
using the MCMC analysis \citep{Pdot4}:
\begin{equation}
{\rm Min(BJD)} = 2457120.49413(2) + 0.0631209131(5) E .
\label{equ:oycarecl}
\end{equation}
The epoch corresponds to the center of the entire observation.
The mean period, however, did not show a secular decrease
(e.g. \cite{han15oycar}; \cite{Pdot7}).  It may be that
period changes in this system are sporadic and do not reflect
the secular CV evolution.

\begin{table}
\caption{Superhump maxima of OY Car (2016)}\label{tab:oycaroc2016}
\begin{center}
\begin{tabular}{rp{50pt}p{30pt}r@{.}lcr}
\hline
$E$ & max\commenta & error & \multicolumn{2}{c}{$O-C$\commentb} & phase\commentc & $N$\commentd \\
\hline
0 & 57482.8885 & 0.0006 & 0&0027 & 0.27 & 47 \\
1 & 57482.9553 & 0.0006 & 0&0049 & 0.33 & 49 \\
18 & 57484.0441 & 0.0007 & $-$0&0047 & 0.58 & 36 \\
19 & 57484.1113 & 0.0006 & $-$0&0022 & 0.65 & 33 \\
20 & 57484.1737 & 0.0006 & $-$0&0044 & 0.63 & 33 \\
57 & 57486.5651 & 0.0009 & $-$0&0037 & 0.52 & 25 \\
58 & 57486.6311 & 0.0012 & $-$0&0023 & 0.56 & 24 \\
72 & 57487.5391 & 0.0009 & 0&0012 & 0.95 & 15 \\
73 & 57487.5998 & 0.0022 & $-$0&0027 & 0.91 & 19 \\
80 & 57488.0572 & 0.0023 & 0&0024 & 0.16 & 29 \\
88 & 57488.5735 & 0.0015 & 0&0018 & 0.34 & 21 \\
89 & 57488.6375 & 0.0009 & 0&0012 & 0.35 & 15 \\
103 & 57489.5472 & 0.0010 & 0&0063 & 0.76 & 21 \\
104 & 57489.6123 & 0.0010 & 0&0068 & 0.80 & 20 \\
142 & 57492.0563 & 0.0023 & $-$0&0044 & 0.51 & 39 \\
143 & 57492.1281 & 0.0063 & 0&0027 & 0.65 & 33 \\
150 & 57492.5795 & 0.0020 & 0&0018 & 0.80 & 17 \\
158 & 57493.0907 & 0.0017 & $-$0&0039 & 0.90 & 28 \\
159 & 57493.1556 & 0.0083 & $-$0&0036 & 0.93 & 14 \\
\hline
  \multicolumn{7}{l}{\commenta BJD$-$2400000.} \\
  \multicolumn{7}{l}{\commentb Against max $= 2457482.8858 + 0.064612 E$.} \\
  \multicolumn{7}{l}{\commentc Orbital phase.} \\
  \multicolumn{7}{l}{\commentd Number of points used to determine the maximum.} \\
\end{tabular}
\end{center}
\end{table}

\subsection{GS Ceti}\label{obj:gscet}

   This object (SDSS J005050.88$+$000912.6) was selected
as a CV during the course of the SDSS \citep{szk05SDSSCV4}.
The spectrum was that of a quiescent dwarf nova.
\citet{sou07SDSSCV2} obtained 8~hr of photometry
giving a suspected orbital period of $\sim$76~min. 

\begin{figure}
  \begin{center}
    \FigureFile(85mm,110mm){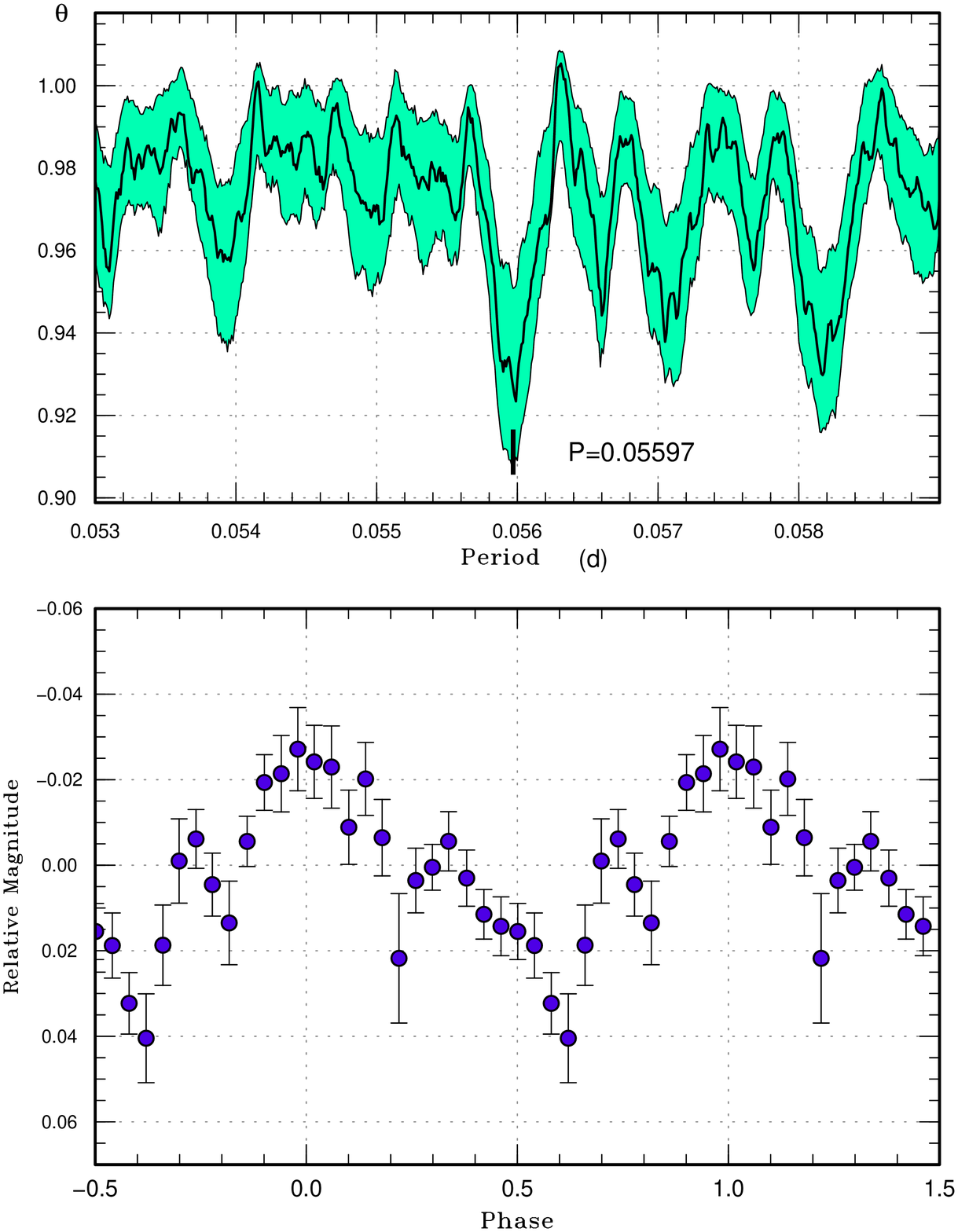}
  \end{center}
  \caption{Early superhumps in GS Cet (2016).
     (Upper): PDM analysis.
     (Lower): Phase-averaged profile.}
  \label{fig:gsceteshpdm}
\end{figure}

\begin{figure}
  \begin{center}
    \FigureFile(85mm,110mm){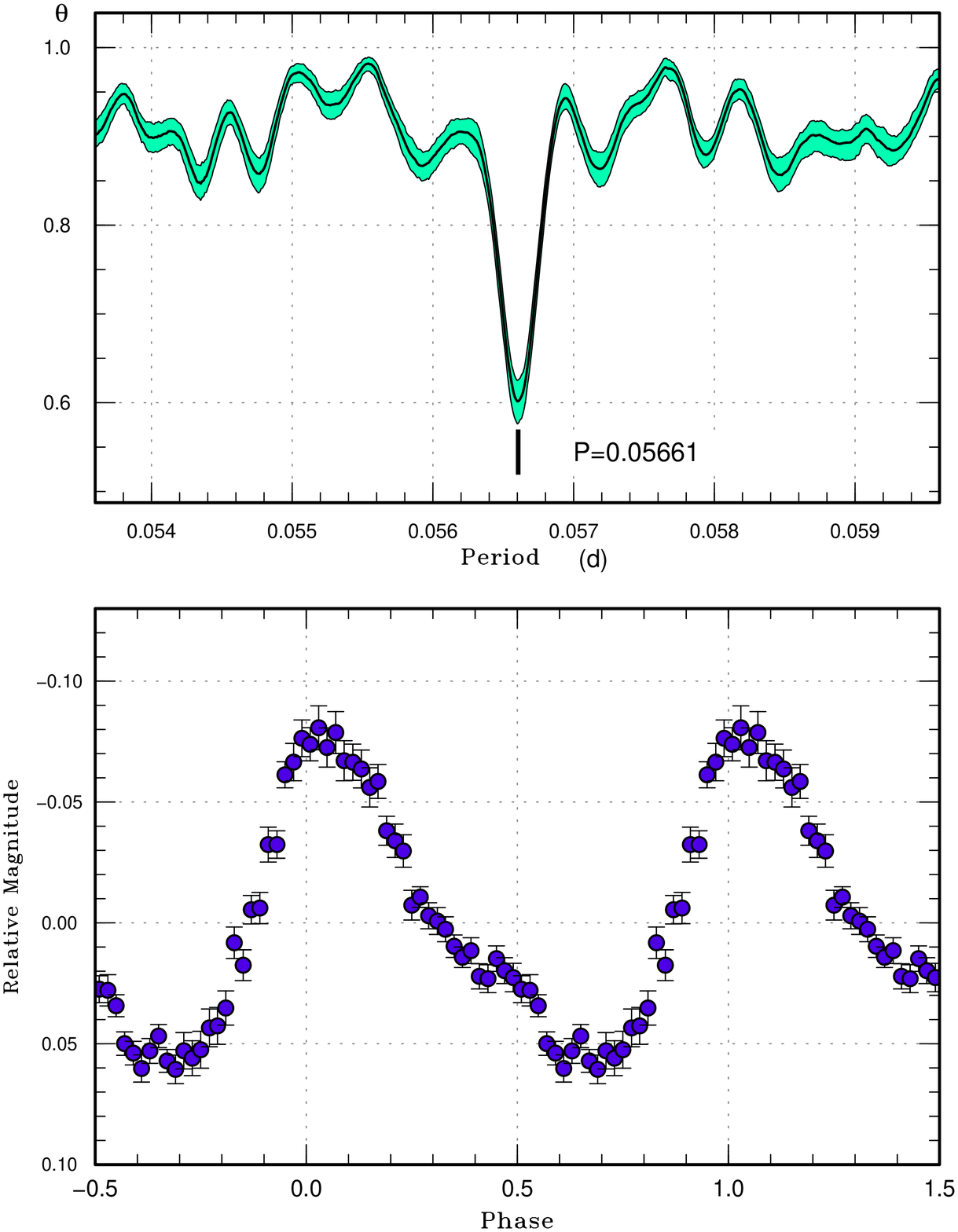}
  \end{center}
  \caption{Ordinary superhumps in GS Cet (2016).
     (Upper): PDM analysis.
     (Lower): Phase-averaged profile.}
  \label{fig:gscetshpdm}
\end{figure}

   Although there were no secure outburst record
in the past, the object was detected in bright
outburst on 2016 November 9 at $V$=13.0 by the ASAS-SN team
(vsnet-alert 20328).
Subsequent observations detected early superhumps
(vsnet-alert 20334, 20342).
Although the profile was not doubly peaked as in
many WZ Sge-type dwarf novae (cf. \cite{kat15wzsge}),
we consider the signal to be that of early superhumps
since it was seen before the appearance of ordinary
superhumps and the period was close to the suggested
orbital period by quiescent photometry
(figure \ref{fig:gsceteshpdm}).
The object started to show ordinary superhumps
on November 17 (vsnet-alert 20368, 20381,
20395, 20404; figure \ref{fig:gscetshpdm}).
The times of superhump maxima are listed in
table \ref{tab:gscetoc2016}.  There were clear
stages A and B.

   The best period of early superhumps by the PDM
method was 0.05597(3)~d.  Combined with the period
of stage A superhumps, the $\epsilon^*$ of
0.0288(8) corresponds to $q$=0.078(2).
Although the object is a WZ Sge-type dwarf nova,
it is not a very extreme one as judged from
the relatively large $P_{\rm dot}$ of stage B
superhumps and the lack of the feature of
an underlying white dwarf in the optical spectra
in quiescence (\cite{szk05SDSSCV4}; \cite{sou07SDSSCV2}).
Although there were some post-superoutburst
observations, the quality of the data
was not sufficient to detect superhumps.


\begin{table}
\caption{Superhump maxima of GS Cet (2016)}\label{tab:gscetoc2016}
\begin{center}
\begin{tabular}{rp{55pt}p{40pt}r@{.}lr}
\hline
\multicolumn{1}{c}{$E$} & \multicolumn{1}{c}{max\commenta} & \multicolumn{1}{c}{error} & \multicolumn{2}{c}{$O-C$\commentb} & \multicolumn{1}{c}{$N$\commentc} \\
\hline
0 & 57709.1297 & 0.0005 & $-$0&0088 & 62 \\
7 & 57709.5380 & 0.0008 & 0&0029 & 12 \\
8 & 57709.5897 & 0.0008 & $-$0&0021 & 19 \\
9 & 57709.6481 & 0.0020 & $-$0&0004 & 15 \\
14 & 57709.9371 & 0.0005 & 0&0054 & 138 \\
15 & 57709.9942 & 0.0004 & 0&0059 & 191 \\
16 & 57710.0499 & 0.0002 & 0&0049 & 158 \\
17 & 57710.1048 & 0.0002 & 0&0032 & 222 \\
18 & 57710.1611 & 0.0003 & 0&0027 & 143 \\
25 & 57710.5575 & 0.0006 & 0&0026 & 21 \\
26 & 57710.6147 & 0.0007 & 0&0032 & 21 \\
27 & 57710.6706 & 0.0009 & 0&0024 & 23 \\
39 & 57711.3486 & 0.0010 & 0&0005 & 100 \\
40 & 57711.4036 & 0.0002 & $-$0&0011 & 322 \\
43 & 57711.5759 & 0.0007 & 0&0012 & 22 \\
44 & 57711.6342 & 0.0014 & 0&0029 & 14 \\
52 & 57712.0830 & 0.0010 & $-$0&0016 & 59 \\
73 & 57713.2710 & 0.0006 & $-$0&0033 & 28 \\
74 & 57713.3268 & 0.0008 & $-$0&0041 & 24 \\
78 & 57713.5523 & 0.0015 & $-$0&0052 & 21 \\
79 & 57713.6095 & 0.0013 & $-$0&0047 & 20 \\
80 & 57713.6660 & 0.0005 & $-$0&0049 & 23 \\
89 & 57714.1783 & 0.0031 & $-$0&0025 & 18 \\
90 & 57714.2345 & 0.0007 & $-$0&0030 & 93 \\
91 & 57714.2918 & 0.0006 & $-$0&0023 & 70 \\
92 & 57714.3491 & 0.0005 & $-$0&0016 & 31 \\
102 & 57714.9150 & 0.0010 & $-$0&0023 & 150 \\
103 & 57714.9737 & 0.0004 & $-$0&0002 & 179 \\
107 & 57715.1981 & 0.0005 & $-$0&0024 & 32 \\
108 & 57715.2538 & 0.0004 & $-$0&0034 & 33 \\
109 & 57715.3118 & 0.0005 & $-$0&0020 & 57 \\
110 & 57715.3695 & 0.0006 & $-$0&0010 & 60 \\
113 & 57715.5381 & 0.0039 & $-$0&0023 & 13 \\
114 & 57715.5945 & 0.0011 & $-$0&0026 & 21 \\
115 & 57715.6503 & 0.0018 & $-$0&0035 & 20 \\
131 & 57716.5634 & 0.0011 & 0&0032 & 22 \\
132 & 57716.6198 & 0.0031 & 0&0029 & 13 \\
133 & 57716.6754 & 0.0038 & 0&0019 & 14 \\
140 & 57717.0650 & 0.0019 & $-$0&0052 & 97 \\
149 & 57717.5867 & 0.0053 & 0&0066 & 22 \\
150 & 57717.6428 & 0.0063 & 0&0061 & 12 \\
155 & 57717.9271 & 0.0017 & 0&0072 & 50 \\
156 & 57717.9815 & 0.0021 & 0&0049 & 81 \\
\hline
  \multicolumn{6}{l}{\commenta BJD$-$2400000.} \\
  \multicolumn{6}{l}{\commentb Against max $= 2457709.1385 + 0.056654 E$.} \\
  \multicolumn{6}{l}{\commentc Number of points used to determine the maximum.} \\
\end{tabular}
\end{center}
\end{table}

\subsection{GZ Ceti}\label{obj:gzcet}

   This object was originally selected as a CV
(SDSS J013701.06$-$091234.9) during the course of
the SDSS \citep{szk03SDSSCV2}.  \citet{szk03SDSSCV2}
obtained spectra showing broad absorption surrounding
the emission lines of H$\beta$ and higher members of
the Balmer series.  The object showed the TiO bandheads
of an M dwarf secondary.  A radial-velocity study by
\citet{szk03SDSSCV2} suggested an orbital period of
80--86 min.  There was a superoutburst in 2003
December and \citet{pre04j0137} reported the orbital
and superhump periods of 79.71(1) min and 81.702(7) min,
respectively.  \citet{pre04j0137} reported the period
variation of superhumps, which can be now interpreted
as stages B and C.  \citet{pre04j0137} suggested that
this object has a low mass-transfer rate.
The same superoutburst was studied by \citet{ima06j0137},
who reported the superhump period of 0.056686(12)~d.
\citet{ima06j0137} noticed the unusual presence of
the TiO bands for this short-$P_{\rm SH}$ object
and discussed that the secondary should be luminous.
\citet{ish07CVIR} obtained an infrared spectrum
dominated by the secondary component.  \citet{ish07CVIR}
suggested that the evolutionary path of GZ Cet is
different from that of ordinary CVs, and that it is
a candidate of a member of EI Psc-like systems.
EI Psc-like systems are CVs below the period minimum
showing hydrogen (likely somewhat reduced in abundance)
in their spectra (cf. \cite{tho02j2329};
\cite{uem02j2329letter}; \cite{lit13sbs1108}) and are
consider to be evolving towards AM CVn-type objects.
Superhump observations during the superoutbursts in
2009 and 2011 were also reported in \citet{Pdot} and
\citet{Pdot4}, respectively.

\begin{figure}
  \begin{center}
    \FigureFile(88mm,70mm){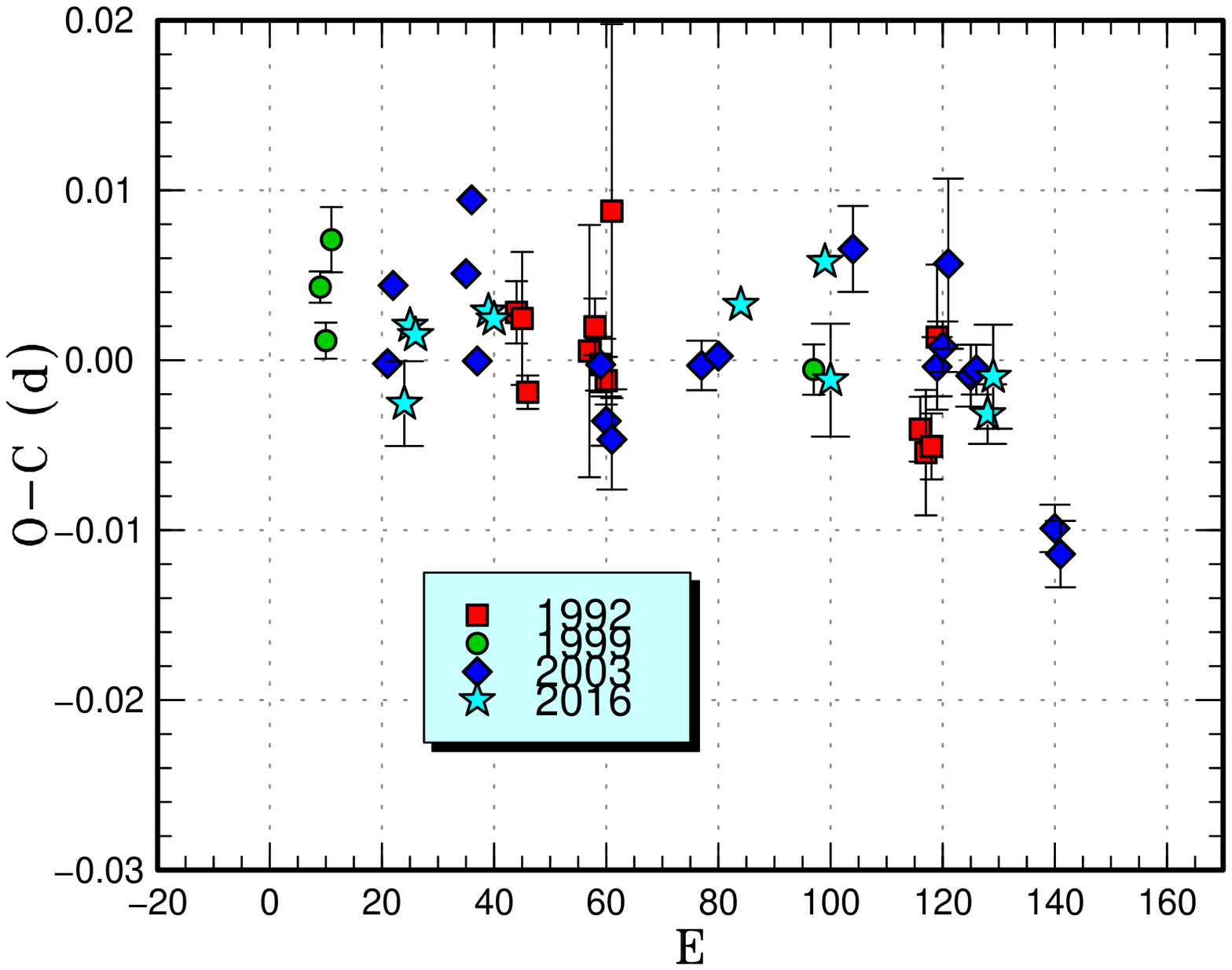}
  \end{center}
  \caption{Comparison of $O-C$ diagrams of GZ Cet between different
  superoutbursts.  A period of 0.05672~d was used to draw this figure.
  Approximate cycle counts ($E$) after the start of the superoutburst
  were used.
  }
  \label{fig:gzcetcomp2}
\end{figure}

   The 2016 superoutburst was detected by R. Stubbings
at a visual magnitude of 12.6 on December 18
(vsnet-alert 20493).  The ASAS-SN team also recorded
the outburst at $V$=12.66 on December 17.
This superoutburst was observed in its relatively late
phase to the post-superoutburst phase
(vsnet-alert 20594).  There was also a post-superoutburst
rebrightening on 2017 January 15 (vsnet-alert 20569).
The times of superhump maxima are listed in
table \ref{tab:gzcetoc2016}.  The times after $E$=266
represent post-superoutburst superhumps.
The maxima for $E \le$54 were stage B superhumps
and ``textbook'' stage C superhumps continued even
during the post-superoutburst phase without
a phase jump as in traditional late superhumps
(figure \ref{fig:gzcetcomp2}).

\begin{table}
\caption{Superhump maxima of GZ Cet (2016)}\label{tab:gzcetoc2016}
\begin{center}
\begin{tabular}{rp{55pt}p{40pt}r@{.}lr}
\hline
\multicolumn{1}{c}{$E$} & \multicolumn{1}{c}{max\commenta} & \multicolumn{1}{c}{error} & \multicolumn{2}{c}{$O-C$\commentb} & \multicolumn{1}{c}{$N$\commentc} \\
\hline
0 & 57743.0047 & 0.0001 & $-$0&0090 & 108 \\
1 & 57743.0605 & 0.0001 & $-$0&0097 & 121 \\
2 & 57743.1185 & 0.0013 & $-$0&0082 & 27 \\
17 & 57743.9670 & 0.0001 & $-$0&0070 & 121 \\
18 & 57744.0226 & 0.0002 & $-$0&0079 & 117 \\
19 & 57744.0801 & 0.0003 & $-$0&0069 & 80 \\
54 & 57746.0667 & 0.0005 & 0&0027 & 86 \\
141 & 57750.9873 & 0.0003 & 0&0088 & 72 \\
159 & 57752.0029 & 0.0002 & 0&0077 & 121 \\
160 & 57752.0582 & 0.0005 & 0&0064 & 67 \\
177 & 57753.0200 & 0.0003 & 0&0079 & 120 \\
193 & 57753.9237 & 0.0002 & 0&0079 & 98 \\
194 & 57753.9799 & 0.0002 & 0&0076 & 120 \\
195 & 57754.0336 & 0.0002 & 0&0047 & 116 \\
212 & 57754.9954 & 0.0004 & 0&0062 & 78 \\
213 & 57755.0499 & 0.0006 & 0&0043 & 103 \\
229 & 57755.9562 & 0.0003 & 0&0068 & 120 \\
230 & 57756.0114 & 0.0004 & 0&0055 & 118 \\
247 & 57756.9729 & 0.0003 & 0&0067 & 121 \\
248 & 57757.0278 & 0.0004 & 0&0051 & 120 \\
266 & 57758.0403 & 0.0004 & 0&0008 & 49 \\
299 & 57759.9016 & 0.0018 & $-$0&0020 & 21 \\
300 & 57759.9609 & 0.0008 & 0&0009 & 43 \\
301 & 57760.0189 & 0.0006 & 0&0023 & 42 \\
371 & 57763.9655 & 0.0008 & $-$0&0052 & 27 \\
372 & 57764.0213 & 0.0010 & $-$0&0058 & 38 \\
424 & 57766.9494 & 0.0014 & $-$0&0152 & 42 \\
425 & 57767.0057 & 0.0015 & $-$0&0153 & 21 \\
\hline
  \multicolumn{6}{l}{\commenta BJD$-$2400000.} \\
  \multicolumn{6}{l}{\commentb Against max $= 2457743.0137 + 0.056488 E$.} \\
  \multicolumn{6}{l}{\commentc Number of points used to determine the maximum.} \\
\end{tabular}
\end{center}
\end{table}

\subsection{AK Cancri}\label{obj:akcnc}

   AK Cnc was discovered as a short-period variable star
(AN 77.1933) with a photographic range of
14 to fainter than 15.5 \citep{mor33newVS}.
\citet{mor33newVS} detected two maxima on 48 plates
between JD 2425323 and 2426763.
\citet{GCVS2sup2} classified this object to be
a U Gem-type variable without a particular remark.
\citet{wil83CVspec1} reported a G-type spectrum
unlike for a CV.  The identification was later
found to be incorrect (\cite{how90faintCV3};
\cite{wen93akcnc}).
The identification chart by \citet{vog82atlas}
was correct.  Amateur observers, particularly
AAVSO and VSOLJ observers, made regular monitoring
since 1986 and detected several outbursts.
Time-resolved CCD observation by \citet{how90faintCV3}
recorded a declining part of an outburst.
\citet{szk92CVspec} obtained a spectrum in quiescence,
which was characteristic to a dwarf nova.
\citet{wen93akcnc} and \citet{wen93akcnccycle} reported
observations using photographic archival materials
and discussed outburst properties.
\citet{wen93akcnc} also gave a summary of confusing
history of the identification of this object.

   \citet{kat94akcnc} was the first to identify
this object to be an SU UMa-type dwarf nova by
observing the 1992 superoutburst.  \citet{men96akcnc}
reported another superoutburst in 1995.
The orbital period was spectroscopically measured
to be 0.0651(2)~d \citep{are98akcnc}.
\citet{Pdot} provided analyses of the 1999 and 2003
superoutbursts.  \citet{Pdot4} further reported
observations of the 2012 superoutburst.

   The 2016 superoutburst was detected at a visual
magnitude of 13.5 by G. Poyner on April 5.
The times of superhump maxima are listed in
table \ref{tab:akcncoc2016}.  Due to the rather
poor coverage, we could not determine $P_{\rm dot}$
for stage B although the distinction between 
stages B and C was clear.  Although positive $P_{\rm dot}$
for stage B is expected for this $P_{\rm orb}$, it
still awaits better observations (figure \ref{fig:akcnccomp2}).

\begin{figure}
  \begin{center}
    \FigureFile(88mm,70mm){akcnccomp2.eps}
  \end{center}
  \caption{Comparison of $O-C$ diagrams of AK Cnc between different
  superoutbursts.  A period of 0.06743~d was used to draw this figure.
  Approximate cycle counts ($E$) after the start of the superoutburst
  were used.
  }
  \label{fig:akcnccomp2}
\end{figure}


\begin{table}
\caption{Superhump maxima of AK Cnc (2016)}\label{tab:akcncoc2016}
\begin{center}
\begin{tabular}{rp{55pt}p{40pt}r@{.}lr}
\hline
\multicolumn{1}{c}{$E$} & \multicolumn{1}{c}{max\commenta} & \multicolumn{1}{c}{error} & \multicolumn{2}{c}{$O-C$\commentb} & \multicolumn{1}{c}{$N$\commentc} \\
\hline
0 & 57485.9732 & 0.0025 & $-$0&0043 & 17 \\
1 & 57486.0453 & 0.0003 & 0&0004 & 38 \\
2 & 57486.1121 & 0.0010 & $-$0&0002 & 24 \\
15 & 57486.9901 & 0.0005 & 0&0014 & 28 \\
16 & 57487.0571 & 0.0006 & 0&0010 & 38 \\
60 & 57490.0249 & 0.0007 & 0&0025 & 38 \\
75 & 57491.0388 & 0.0009 & 0&0053 & 21 \\
76 & 57491.0993 & 0.0033 & $-$0&0017 & 22 \\
104 & 57492.9853 & 0.0018 & $-$0&0033 & 26 \\
105 & 57493.0550 & 0.0031 & $-$0&0011 & 26 \\
\hline
  \multicolumn{6}{l}{\commenta BJD$-$2400000.} \\
  \multicolumn{6}{l}{\commentb Against max $= 2457485.9775 + 0.067415 E$.} \\
  \multicolumn{6}{l}{\commentc Number of points used to determine the maximum.} \\
\end{tabular}
\end{center}
\end{table}

\subsection{GZ Cancri}\label{obj:gzcnc}

   GZ Cnc was discovered by K. Takamizawa as a variable star
(=TmzV34).  The object was confirmed as a dwarf nova
(\cite{kat01gzcnc}; \cite{kat02gzcncnsv10934}).  \citet{tap03gzcnc}
obtained the orbital period of 0.08825(28)~d by
radial-velocity observations.  The SU UMa-type nature
was established during the 2010 \citep{Pdot2}.
See \citet{Pdot6} for more information.

\begin{figure}
  \begin{center}
    \FigureFile(88mm,70mm){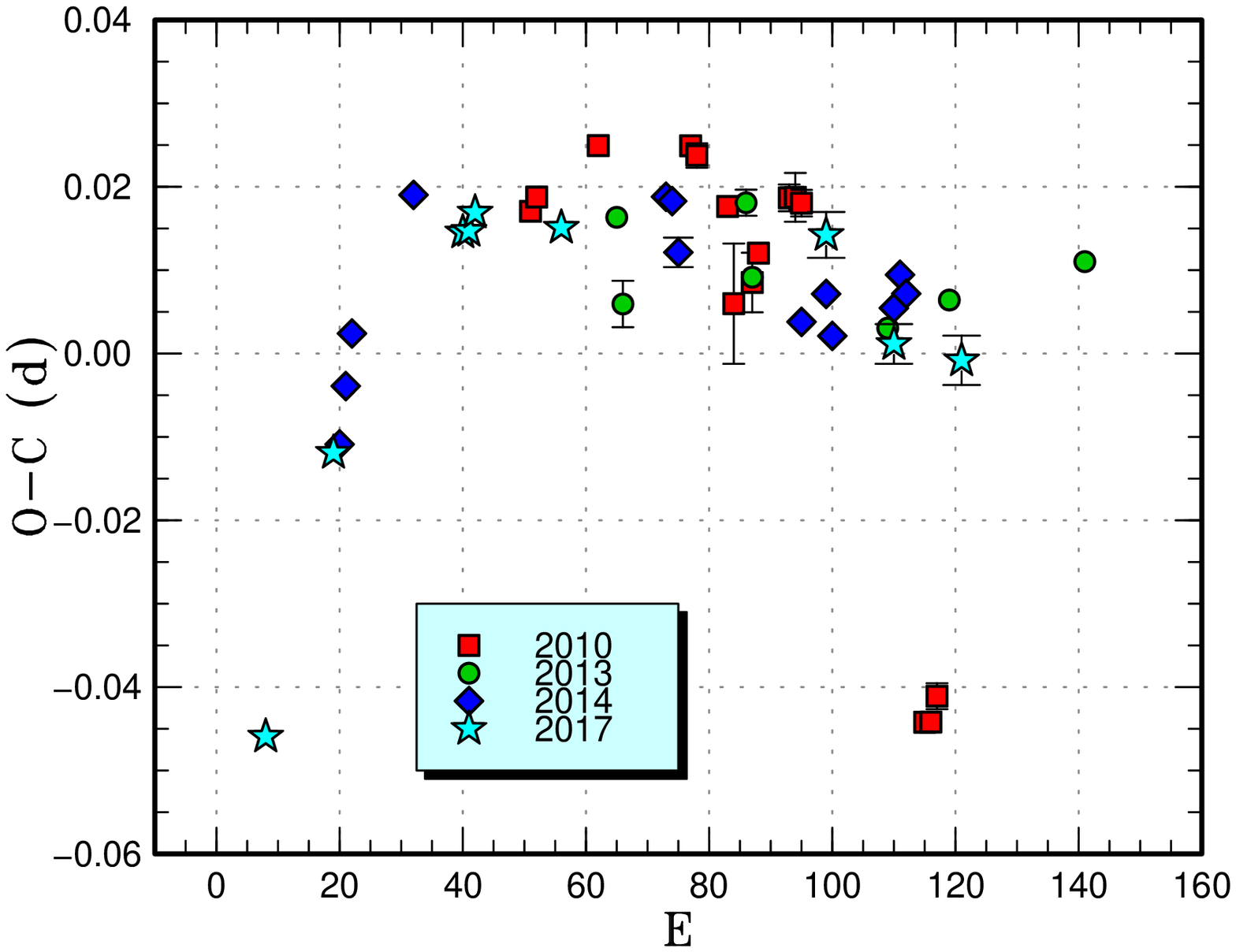}
  \end{center}
  \caption{Comparison of $O-C$ diagrams of GZ Cnc between different
  superoutbursts.  A period of 0.09290~d was used to draw this figure.
  Approximate cycle counts ($E$) after the start of the superoutburst
  were used.
  }
  \label{fig:gzcnccomp3}
\end{figure}

   The 2017 superoutburst was detected by R. Stubbings
at a visual magnitude of 13.0 on February 2
and on the same night at 12.5 mag by T. Horie.
Subsequent observations detected growing superhumps
on February 3 and 4.  Superhumps grew further on
February 6 (vsnet-alert 20642).
The times of superhump maxima are listed in
table \ref{tab:gzcncoc2017}.  Thanks to the early
detection of the outburst, stage A superhumps
were clearly detected (figure \ref{fig:gzcnccomp3}).
The $\epsilon^*$ for stage A superhumps [0.081(3)]
corresponds to $q$=0.27(2).

\begin{table}
\caption{Superhump maxima of GZ Cnc (2017)}\label{tab:gzcncoc2017}
\begin{center}
\begin{tabular}{rp{55pt}p{40pt}r@{.}lr}
\hline
\multicolumn{1}{c}{$E$} & \multicolumn{1}{c}{max\commenta} & \multicolumn{1}{c}{error} & \multicolumn{2}{c}{$O-C$\commentb} & \multicolumn{1}{c}{$N$\commentc} \\
\hline
0 & 57788.0546 & 0.0015 & $-$0&0381 & 208 \\
11 & 57789.1105 & 0.0007 & $-$0&0062 & 120 \\
32 & 57791.0878 & 0.0002 & 0&0162 & 156 \\
33 & 57791.1809 & 0.0003 & 0&0162 & 185 \\
34 & 57791.2760 & 0.0007 & 0&0182 & 88 \\
48 & 57792.5748 & 0.0008 & 0&0138 & 19 \\
91 & 57796.5686 & 0.0028 & 0&0047 & 33 \\
102 & 57797.5774 & 0.0024 & $-$0&0104 & 23 \\
113 & 57798.5974 & 0.0030 & $-$0&0145 & 34 \\
\hline
  \multicolumn{6}{l}{\commenta BJD$-$2400000.} \\
  \multicolumn{6}{l}{\commentb Against max $= 2457788.0927 + 0.093090 E$.} \\
  \multicolumn{6}{l}{\commentc Number of points used to determine the maximum.} \\
\end{tabular}
\end{center}
\end{table}

\subsection{GP Canum Venaticorum}\label{obj:gpcvn}

   This object was originally selected as a CV
(SDSS J122740.83$+$513925.0) during the course of
the SDSS \citep{szk06SDSSCV5}.  \citet{szk06SDSSCV5}
obtained a spectrum showing an underlying white dwarf.
\citet{lit08eclCV} clarified that this object is
an eclipsing dwarf nova with a short orbital period.
The object underwent the first-recorded superoutburst
in 2007 June.  This 2007 superoutburst was analyzed
by \citet{she08j1227} and \citet{Pdot}.
\citet{Pdot3} reported on the 2011 superoutburst
and provided a corrected eclipse ephemeris.
\citet{sav11CVeclmass} reported the orbital parameters
(including $q$) by modeling the eclipse profile.
Although \citet{zen10v849eclCVs} suspected cyclic $O-C$
variation of eclipses, their result was doubtful
due to the very low time-resolution of observations
and very few points on the $O-C$ diagram.

   The 2016 superoutburst was detected by the ASAS-SN
team at $V$=15.29 on April 25.  Both superhumps and
eclipses were recorded (vsnet-alert 19778).
Using the combined data of 2007, 2011 and 2016 observations,
we have refined the eclipse ephemeris
by the MCMC modeling \citep{Pdot4}:
\begin{equation}
{\rm Min(BJD)} = 2455395.37115(4) + 0.0629503676(9) E .
\label{equ:gpcvnecl}
\end{equation}
The epoch in \citet{lit08eclCV} corresponds to an $O-C$
value of 0.00168~d against this ephemeris.
The ephemeris in \citet{lit08eclCV} predicts eclipses
to occur 0.0096~d later than our actual observations
in 2016.

\begin{figure}
  \begin{center}
    \FigureFile(85mm,70mm){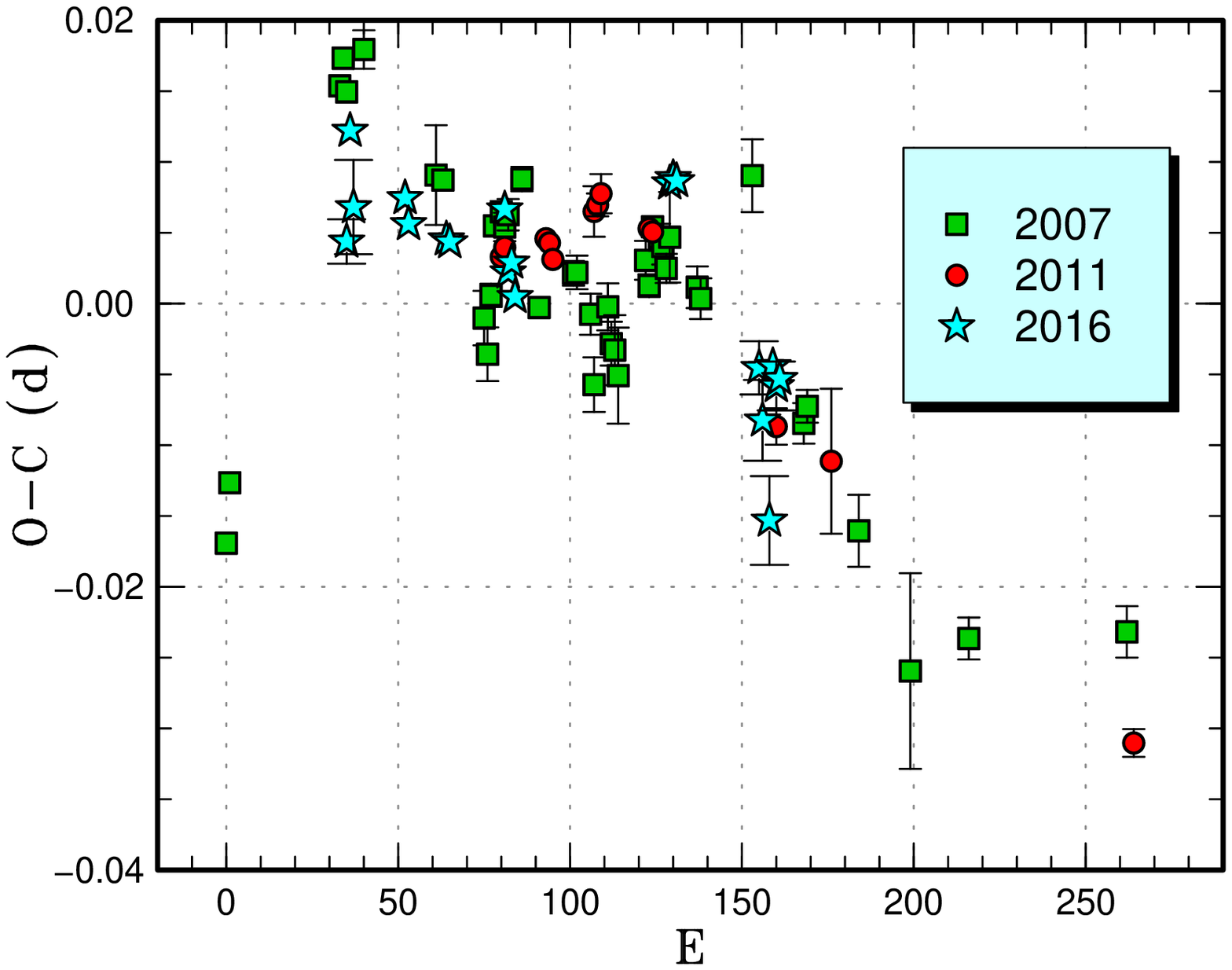}
  \end{center}
  \caption{Comparison of $O-C$ diagrams of GP CVn between different
  superoutbursts.  A period of 0.05828~d was used to draw this figure.
  Approximate cycle counts ($E$) after the appearance of
  superhumps were used.  Note that this treatment is
  different from the corresponding figure in \citet{Pdot3}.
  Since the 2007 observation apparently caught the early
  part of stage A, we set the initial superhump of 2007
  to be $E$=0 in this figure.  Other superoutbursts have been
  shifted to best match the 2007 one.  The shift value suggests
  that the ASAS-SN detection of the 2016 superoutburst
  occurred $\sim$13 cycles after the appearance of
  superhumps.
  }
  \label{fig:gpcvncomp2}
\end{figure}

   The times of superhump maxima during the 2016
superoutburst are listed in table \ref{tab:gpcvnoc2016}.
Stage B with a positive $P_{\rm dot}$ and a transition
to stage C superhumps were recorded
(see also figure \ref{fig:gpcvncomp2}).

\begin{table}
\caption{Superhump maxima of GP CVn (2016)}\label{tab:gpcvnoc2016}
\begin{center}
\begin{tabular}{rp{50pt}p{30pt}r@{.}lcr}
\hline
$E$ & max\commenta & error & \multicolumn{2}{c}{$O-C$\commentb} & phase\commentc & $N$\commentd \\
\hline
0 & 57505.4014 & 0.0016 & $-$0&0043 & 0.95 & 20 \\
1 & 57505.4740 & 0.0004 & 0&0036 & 0.10 & 41 \\
2 & 57505.5333 & 0.0033 & $-$0&0016 & 0.05 & 21 \\
17 & 57506.5052 & 0.0002 & 0&0005 & 0.49 & 140 \\
18 & 57506.5681 & 0.0002 & $-$0&0012 & 0.48 & 143 \\
29 & 57507.2793 & 0.0003 & $-$0&0012 & 0.78 & 43 \\
30 & 57507.3439 & 0.0005 & $-$0&0013 & 0.81 & 47 \\
46 & 57508.3822 & 0.0007 & 0&0027 & 0.30 & 63 \\
47 & 57508.4425 & 0.0004 & $-$0&0017 & 0.26 & 135 \\
48 & 57508.5079 & 0.0003 & $-$0&0009 & 0.30 & 117 \\
49 & 57508.5702 & 0.0003 & $-$0&0032 & 0.29 & 103 \\
94 & 57511.4920 & 0.0006 & 0&0094 & 0.70 & 59 \\
95 & 57511.5571 & 0.0004 & 0&0099 & 0.74 & 57 \\
96 & 57511.6217 & 0.0010 & 0&0098 & 0.76 & 39 \\
120 & 57513.1625 & 0.0019 & $-$0&0010 & 0.24 & 54 \\
121 & 57513.2235 & 0.0029 & $-$0&0046 & 0.21 & 41 \\
123 & 57513.3460 & 0.0031 & $-$0&0115 & 0.15 & 24 \\
124 & 57513.4216 & 0.0005 & $-$0&0005 & 0.36 & 54 \\
125 & 57513.4850 & 0.0017 & $-$0&0018 & 0.36 & 61 \\
126 & 57513.5502 & 0.0008 & $-$0&0012 & 0.40 & 63 \\
\hline
  \multicolumn{7}{l}{\commenta BJD$-$2400000.} \\
  \multicolumn{7}{l}{\commentb Against max $= 2457505.4057 + 0.064648 E$.} \\
  \multicolumn{7}{l}{\commentc Orbital phase.} \\
  \multicolumn{7}{l}{\commentd Number of points used to determine the maximum.} \\
\end{tabular}
\end{center}
\end{table}

\subsection{V337 Cygni}\label{obj:v337cyg}

   V337 Cyg was discovered as a long-period variable
(AN 101.1928).  The dwarf nova-type nature was
confirmed in 1996.  The SU UMa-type nature was
established during the 2006 superoutburst
(cf. \cite{boy07v337cyg}).  See \citet{Pdot7} for
more history.

   The 2016 superoutburst was detected by M. Moriyama
at an unfiltered CCD magnitude of 15.5 on November 17.
Observations on a single night yielded three
superhumps (table \ref{tab:v337cygoc2016}).
The maximum $E$=2 suffered from large atmospheric
extinction and the quality of this measurement
was poor.  The $P_{\rm SH}$ is omitted from
table \ref{tab:perlist} since there were observations
with much more accurate values in the past.


\begin{table}
\caption{Superhump maxima of V337 Cyg (2016)}\label{tab:v337cygoc2016}
\begin{center}
\begin{tabular}{rp{55pt}p{40pt}r@{.}lr}
\hline
\multicolumn{1}{c}{$E$} & \multicolumn{1}{c}{max\commenta} & \multicolumn{1}{c}{error} & \multicolumn{2}{c}{$O-C$\commentb} & \multicolumn{1}{c}{$N$\commentc} \\
\hline
0 & 57722.2200 & 0.0011 & 0&0015 & 68 \\
1 & 57722.2925 & 0.0021 & $-$0&0030 & 76 \\
2 & 57722.3739 & 0.0022 & 0&0015 & 65 \\
\hline
  \multicolumn{6}{l}{\commenta BJD$-$2400000.} \\
  \multicolumn{6}{l}{\commentb Against max $= 2457722.2185 + 0.076935 E$.} \\
  \multicolumn{6}{l}{\commentc Number of points used to determine the maximum.} \\
\end{tabular}
\end{center}
\end{table}

\subsection{V1113 Cygni}\label{obj:v1113cyg}

   V1113 Cyg was discovered as a dwarf nova by \citet{hof66an289139}.
The SU UMa-type nature was identified by \citet{kat96v1113cyg}.
See \citet{Pdot8} for more history.

\begin{figure}
  \begin{center}
    \FigureFile(85mm,70mm){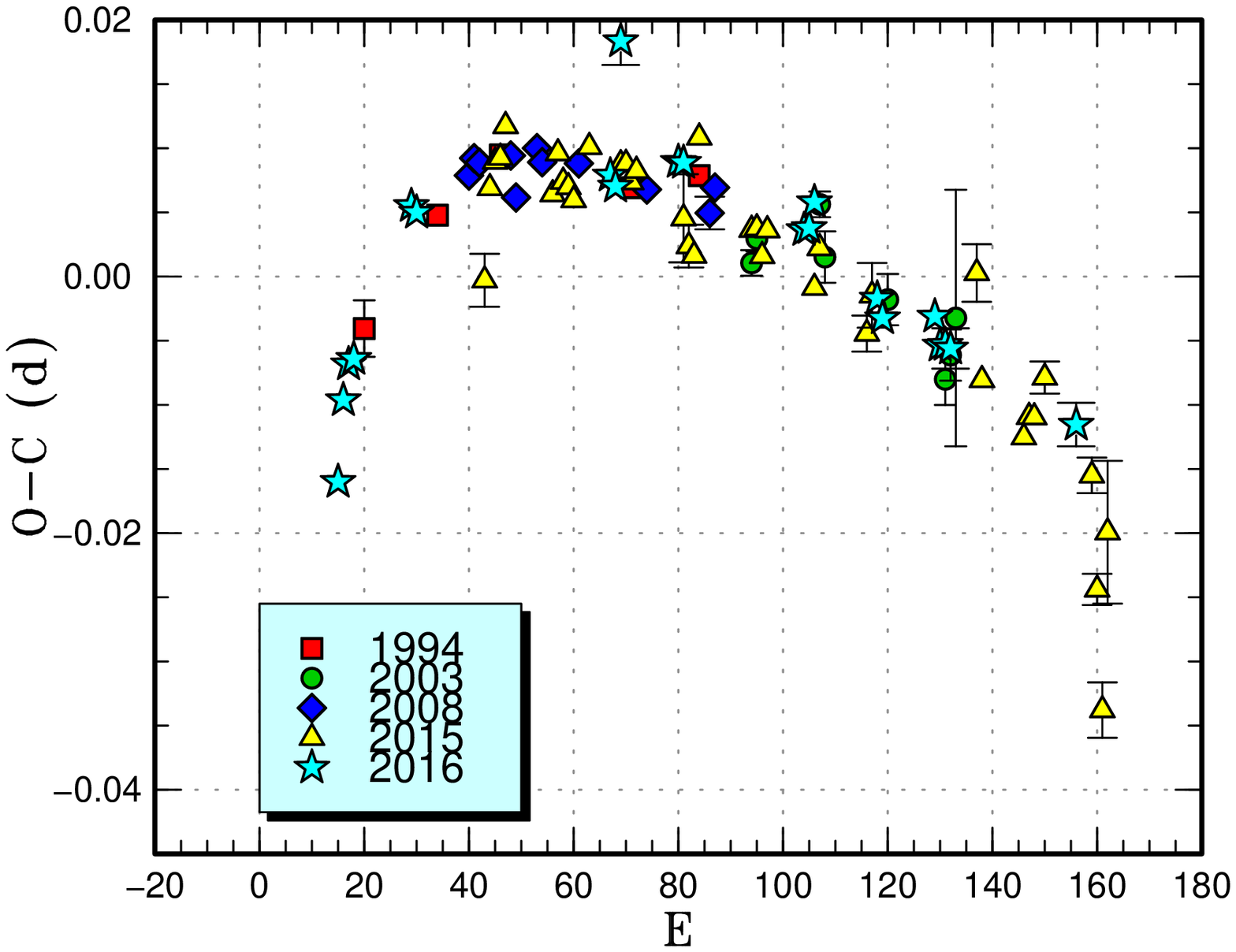}
  \end{center}
  \caption{Comparison of $O-C$ diagrams of V1113 Cyg between different
  superoutbursts.  A period of 0.07911~d was used to draw this figure.
  Approximate cycle counts ($E$) after the peak of the superoutburst
  were used.  Since the start of the 2016 superoutburst was
  very well defined, we used the peak of the superoutburst
  and redefined the cycle counts.  The other outbursts were
  shifted to best match the 2016 one.
  }
  \label{fig:v1113cygcomp3}
\end{figure}

   The 2016 superoutburst was detected by H. Maehara
at a visual magnitude of 14.3 on July 27 (vsnet-alert 20003).
A visual observation by P. Dubovsky on the same night and
ASAS-SN detection on the next night indicated
further brightening (vsnet-alert 20011, 20015).
Thanks to the early detection and notification,
growing superhumps were detected (vsnet-alert 20022).
The times of superhump maxima are listed in
table \ref{tab:v1113cygoc2016}, which clearly indicate
the presence of stage A superhumps
(figure \ref{fig:v1113cygcomp3}).  It may be noteworthy
that stage A lasted nearly 40 cycles
(figure \ref{fig:v1113cygcomp3}), which may be analogous
to long-$P_{\rm orb}$ SU UMa-type dwarf novae with
slowly evolving superhumps (such as V1006 Cyg: \cite{kat16v1006cyg};
V452 Cas: \cite{Pdot8}).
Since stage A superhumps were observed, a spectroscopic
radial-velocity study is desired to determine $q$
using the stage A superhump method.

\begin{table}
\caption{Superhump maxima of V1113 Cyg (2016)}\label{tab:v1113cygoc2016}
\begin{center}
\begin{tabular}{rp{55pt}p{40pt}r@{.}lr}
\hline
\multicolumn{1}{c}{$E$} & \multicolumn{1}{c}{max\commenta} & \multicolumn{1}{c}{error} & \multicolumn{2}{c}{$O-C$\commentb} & \multicolumn{1}{c}{$N$\commentc} \\
\hline
0 & 57599.0116 & 0.0013 & $-$0&0162 & 63 \\
1 & 57599.0971 & 0.0005 & $-$0&0099 & 64 \\
2 & 57599.1790 & 0.0003 & $-$0&0071 & 201 \\
3 & 57599.2586 & 0.0004 & $-$0&0066 & 148 \\
14 & 57600.1407 & 0.0002 & 0&0053 & 238 \\
15 & 57600.2193 & 0.0003 & 0&0048 & 158 \\
52 & 57603.1492 & 0.0003 & 0&0078 & 144 \\
53 & 57603.2275 & 0.0006 & 0&0070 & 157 \\
54 & 57603.3179 & 0.0019 & 0&0183 & 55 \\
65 & 57604.1787 & 0.0006 & 0&0089 & 158 \\
66 & 57604.2578 & 0.0006 & 0&0089 & 157 \\
89 & 57606.0721 & 0.0006 & 0&0037 & 87 \\
90 & 57606.1513 & 0.0005 & 0&0039 & 96 \\
91 & 57606.2324 & 0.0011 & 0&0059 & 92 \\
103 & 57607.1743 & 0.0008 & $-$0&0016 & 452 \\
104 & 57607.2518 & 0.0013 & $-$0&0031 & 104 \\
114 & 57608.0431 & 0.0006 & $-$0&0030 & 96 \\
115 & 57608.1199 & 0.0007 & $-$0&0052 & 92 \\
116 & 57608.1991 & 0.0005 & $-$0&0052 & 97 \\
117 & 57608.2779 & 0.0016 & $-$0&0054 & 60 \\
141 & 57610.1706 & 0.0017 & $-$0&0113 & 98 \\
\hline
  \multicolumn{6}{l}{\commenta BJD$-$2400000.} \\
  \multicolumn{6}{l}{\commentb Against max $= 2457599.0279 + 0.079107 E$.} \\
  \multicolumn{6}{l}{\commentc Number of points used to determine the maximum.} \\
\end{tabular}
\end{center}
\end{table}

\subsection{IX Draconis}\label{obj:ixdra}

   IX Dra is one of ER UMa-type dwarf novae
\citep{ish01ixdra}.  See \citet{Pdot6} and \citet{ole04ixdra}
for the history.

\begin{figure}
  \begin{center}
    \FigureFile(85mm,70mm){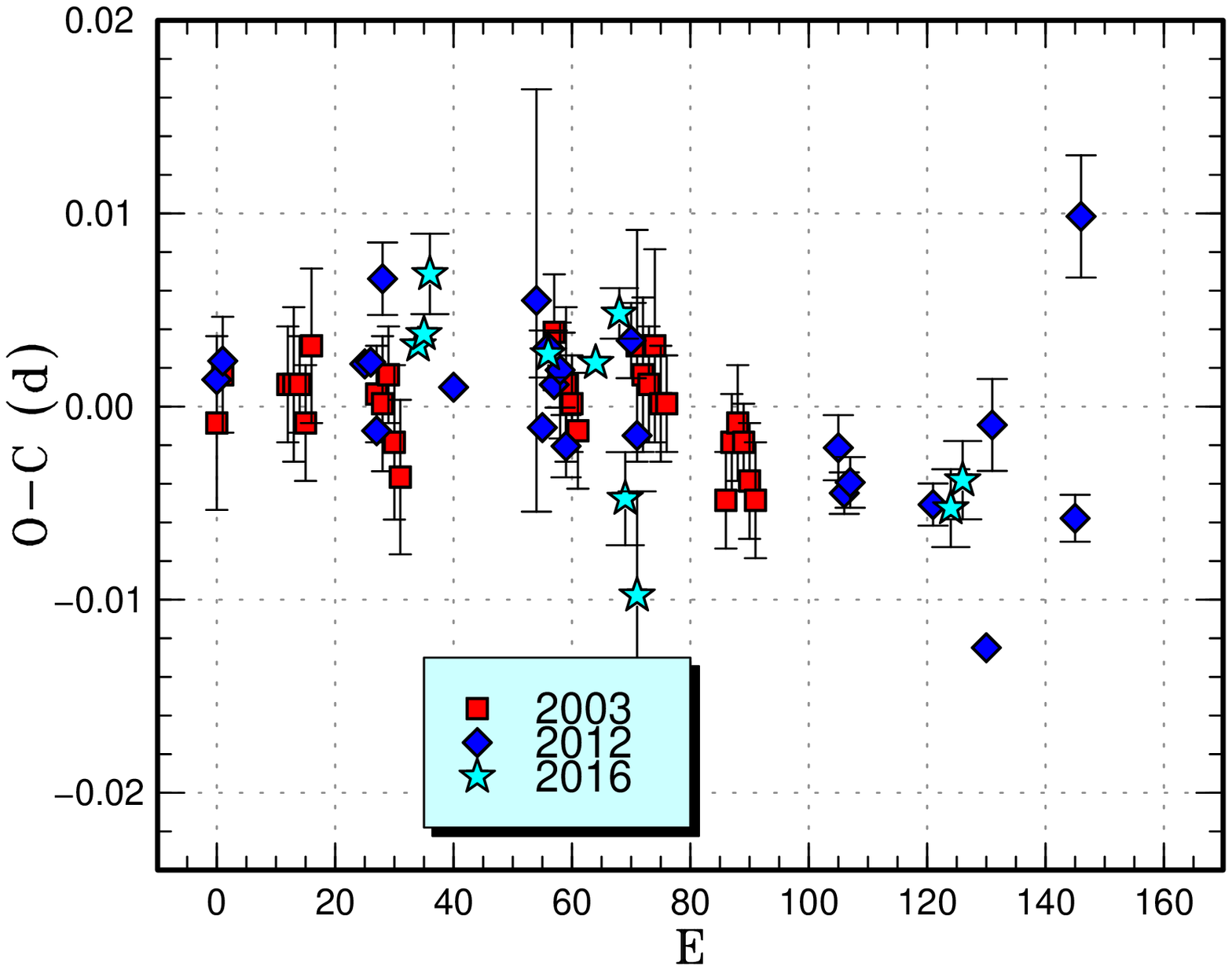}
  \end{center}
  \caption{Comparison of $O-C$ diagrams of IX Dra between different
  superoutbursts.  A period of 0.06700~d was used to draw this figure.
  Approximate cycle counts ($E$) after the start of the superoutburst
  were used.
  }
  \label{fig:ixdracomp2}
\end{figure}

   The 2016 May superoutburst was detected by P. Dubovsky
at a visual magnitude of 15.2 on May 29.  Subsequent observations
detected superhumps (vsnet-alert 19868).
The times of superhump maxima are listed in
table \ref{tab:ixdraoc2016}.
A combined $O-C$ diagram (figure \ref{fig:ixdracomp2})
did not show a strong sign of a stage transition.

   In order to determine the change in the supercycle
(cf. \cite{otu13ixdra}), we have extracted nine
maxima of superoutbursts since 2015 April, when
the ASAS-SN team started a good coverage of this field.
The mean supercycle between JD 2457142 and 2457305
(2015 April to October) was 54.4(3)~d, while it
increased to 58.9(3)~d between JD 2457420 and 2457657
(2016 February to September).
These values are much shorter than what is predicted
(should be longer than 62~d by 2015)
by a claimed secular trend in \citet{otu13ixdra}.
The rapid variation suggests that snapshot values
as in \citet{otu13ixdra} probably did not reflect
the long-term trend well.

\begin{table}
\caption{Superhump maxima of IX Dra (2016)}\label{tab:ixdraoc2016}
\begin{center}
\begin{tabular}{rp{55pt}p{40pt}r@{.}lr}
\hline
\multicolumn{1}{c}{$E$} & \multicolumn{1}{c}{max\commenta} & \multicolumn{1}{c}{error} & \multicolumn{2}{c}{$O-C$\commentb} & \multicolumn{1}{c}{$N$\commentc} \\
\hline
0 & 57540.7847 & 0.0005 & $-$0&0004 & 56 \\
1 & 57540.8523 & 0.0008 & 0&0003 & 62 \\
2 & 57540.9224 & 0.0021 & 0&0035 & 26 \\
22 & 57542.2582 & 0.0012 & 0&0014 & 95 \\
30 & 57542.7938 & 0.0006 & 0&0018 & 71 \\
34 & 57543.0643 & 0.0013 & 0&0048 & 123 \\
35 & 57543.1217 & 0.0024 & $-$0&0047 & 123 \\
37 & 57543.2507 & 0.0054 & $-$0&0095 & 117 \\
90 & 57546.8062 & 0.0020 & 0&0006 & 64 \\
92 & 57546.9417 & 0.0020 & 0&0022 & 58 \\
\hline
  \multicolumn{6}{l}{\commenta BJD$-$2400000.} \\
  \multicolumn{6}{l}{\commentb Against max $= 2457540.7852 + 0.066895 E$.} \\
  \multicolumn{6}{l}{\commentc Number of points used to determine the maximum.} \\
\end{tabular}
\end{center}
\end{table}

\subsection{IR Geminorum}\label{obj:irgem}

   IR Gem was discovered as a U Gem-type variable star
(AN S5423) by \citet{pop61kraur}.  Although little
was known other than outbursts with an interval
of $\sim$75~d and amplitudes of $\sim$2.5 mag
(\cite{pop60irgem}; \cite{mei76irgem}),\footnote{
   There is a close companion star and old literature
   often referred to combined magnitudes.
}
this object has been well monitored by AAVSO observers
since its discovery.
Several outbursts were already recorded
in the 1960s \citep{may68UG}.
\citet{bon78bluevar2} obtained a spectrum typical
for an outbursting dwarf nova.  \citet{bur79DNspec}
reported a dwarf nova-type spectrum in quiescence.
\citet{sha84irgem} identified this object to be
an SU UMa-type dwarf nova by detecting superhumps.
\citet{sha84irgem} suggested a small mass ratio
(either a massive white dwarf or an undermassive
secondary) based on a radial-velocity study.
Although \citet{fei88irgem}, \citet{laz90irgem}
and \citet{laz91irgem}
reported more detailed spectroscopic studies,
the orbital period was not well measured.
Observations of superhumps during the 1991 superoutburst
were reported in \citet{kat01irgem}.
\citet{Pdot} reanalyzed this superoutburst
and reported another one in 2009.
Another superoutburst in 2010 was reported in \citet{Pdot2}.

   The 2016 superoutburst was detected by the ASAS-SN
team at $V$=12.95 on March 22 and $V$=12.00 on March 24.
Subsequent observations detected superhumps
(vsnet-alert 19645).
The times of superhump maxima are listed in
table \ref{tab:irgemoc2016}.  The observation started
two days later than the announcement and stage A
superhumps were not recorded.

\begin{figure}
  \begin{center}
    \FigureFile(85mm,70mm){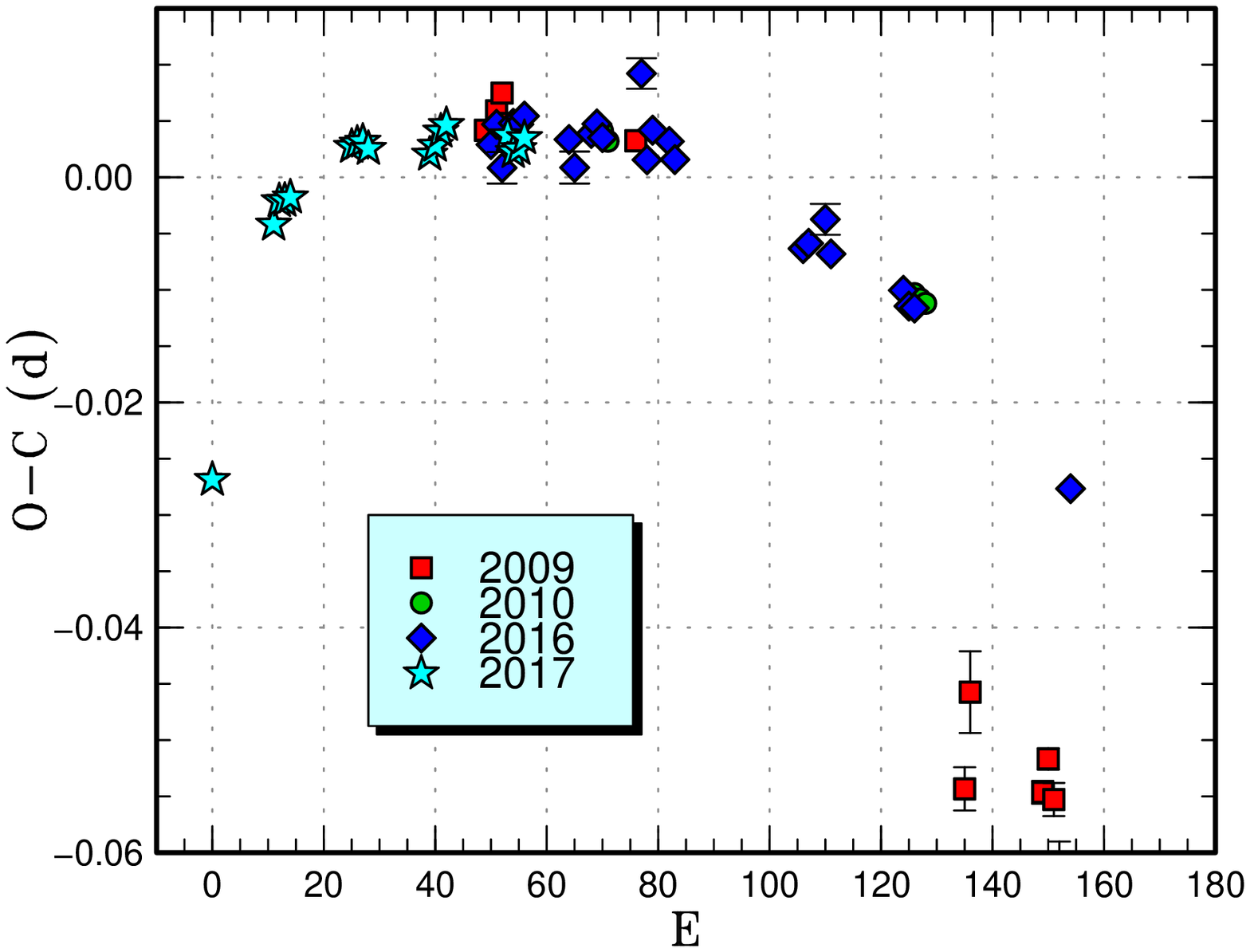}
  \end{center}
  \caption{Comparison of $O-C$ diagrams of IR Gem between different
  superoutbursts.  A period of 0.07109~d was used to draw this figure.
  Approximate cycle counts ($E$) after the start of the superoutburst
  were used.  The 2010 superoutburst was preceded by a separate
  precursor.  We shifted the $O-C$ values to best fit the 2016 ones.
  The result suggests that superhumps started to evolve
  20 cycles after the peak of the precursor outburst.
  The final points in the 2009 superoutbursts probably
  correspond to traditional late superhumps.  
  }
  \label{fig:irgemcomp}
\end{figure}

   The 2017 superoutburst was detected by K. Kasai
on March 12 (vsnet-alert 20763) while observing KaiV36, 
an ellipsoidal variable star in the field of IR Gem.
The outburst was detected early enough and stage A superhumps
were observed (figure \ref{fig:irgemcomp}).
The object was still in quiescence
on March 10.  The times of superhump maxima are
listed in table \ref{tab:irgemoc2017}.
The observations were not long enough and $P_{\rm dot}$
was not determined.  The $\epsilon^*$ for stage A
superhumps is 0.068(11), whose errors mainly comes
from the uncertainty in the orbital period
[0.0684(6)~d] \citep{fei88irgem}.
This $\epsilon^*$ corresponds to $q$=0.22(4).
Accurate determination of the orbital period
is desired since the object is bright enough and
its behavior during superoutbursts has been well
documented.

\begin{table}
\caption{Superhump maxima of IR Gem (2016)}\label{tab:irgemoc2016}
\begin{center}
\begin{tabular}{rp{55pt}p{40pt}r@{.}lr}
\hline
\multicolumn{1}{c}{$E$} & \multicolumn{1}{c}{max\commenta} & \multicolumn{1}{c}{error} & \multicolumn{2}{c}{$O-C$\commentb} & \multicolumn{1}{c}{$N$\commentc} \\
\hline
0 & 57474.0108 & 0.0006 & $-$0&0046 & 79 \\
1 & 57474.0837 & 0.0007 & $-$0&0026 & 78 \\
2 & 57474.1509 & 0.0014 & $-$0&0062 & 57 \\
4 & 57474.2971 & 0.0003 & $-$0&0017 & 176 \\
5 & 57474.3681 & 0.0003 & $-$0&0015 & 192 \\
6 & 57474.4399 & 0.0003 & $-$0&0006 & 164 \\
14 & 57475.0065 & 0.0012 & $-$0&0007 & 50 \\
15 & 57475.0751 & 0.0014 & $-$0&0029 & 33 \\
18 & 57475.2914 & 0.0003 & 0&0009 & 142 \\
19 & 57475.3633 & 0.0003 & 0&0020 & 176 \\
20 & 57475.4332 & 0.0004 & 0&0010 & 145 \\
27 & 57475.9366 & 0.0014 & 0&0085 & 35 \\
28 & 57476.0000 & 0.0007 & 0&0011 & 53 \\
29 & 57476.0737 & 0.0011 & 0&0039 & 52 \\
32 & 57476.2860 & 0.0005 & 0&0037 & 60 \\
33 & 57476.3554 & 0.0005 & 0&0023 & 42 \\
56 & 57477.9826 & 0.0009 & 0&0002 & 101 \\
57 & 57478.0542 & 0.0010 & 0&0009 & 89 \\
60 & 57478.2696 & 0.0014 & 0&0038 & 34 \\
61 & 57478.3376 & 0.0004 & 0&0010 & 63 \\
74 & 57479.2585 & 0.0004 & 0&0010 & 103 \\
75 & 57479.3282 & 0.0004 & $-$0&0002 & 127 \\
76 & 57479.3991 & 0.0004 & $-$0&0001 & 152 \\
104 & 57481.3736 & 0.0007 & $-$0&0091 & 67 \\
\hline
  \multicolumn{6}{l}{\commenta BJD$-$2400000.} \\
  \multicolumn{6}{l}{\commentb Against max $= 2457474.0154 + 0.070840 E$.} \\
  \multicolumn{6}{l}{\commentc Number of points used to determine the maximum.} \\
\end{tabular}
\end{center}
\end{table}


\begin{table}
\caption{Superhump maxima of IR Gem (2017)}\label{tab:irgemoc2017}
\begin{center}
\begin{tabular}{rp{55pt}p{40pt}r@{.}lr}
\hline
\multicolumn{1}{c}{$E$} & \multicolumn{1}{c}{max\commenta} & \multicolumn{1}{c}{error} & \multicolumn{2}{c}{$O-C$\commentb} & \multicolumn{1}{c}{$N$\commentc} \\
\hline
0 & 57825.4884 & 0.0014 & $-$0&0181 & 72 \\
11 & 57826.2930 & 0.0001 & 0&0015 & 314 \\
12 & 57826.3662 & 0.0002 & 0&0033 & 78 \\
13 & 57826.4373 & 0.0003 & 0&0031 & 78 \\
14 & 57826.5087 & 0.0005 & 0&0031 & 60 \\
25 & 57827.2952 & 0.0001 & 0&0045 & 232 \\
26 & 57827.3666 & 0.0002 & 0&0045 & 218 \\
27 & 57827.4378 & 0.0002 & 0&0044 & 116 \\
28 & 57827.5083 & 0.0004 & 0&0035 & 68 \\
39 & 57828.2898 & 0.0005 & $-$0&0000 & 78 \\
40 & 57828.3617 & 0.0009 & 0&0005 & 56 \\
41 & 57828.4340 & 0.0004 & 0&0014 & 78 \\
42 & 57828.5057 & 0.0005 & 0&0017 & 67 \\
53 & 57829.2867 & 0.0003 & $-$0&0022 & 65 \\
54 & 57829.3563 & 0.0004 & $-$0&0040 & 78 \\
55 & 57829.4277 & 0.0005 & $-$0&0040 & 79 \\
56 & 57829.4998 & 0.0005 & $-$0&0033 & 72 \\
\hline
  \multicolumn{6}{l}{\commenta BJD$-$2400000.} \\
  \multicolumn{6}{l}{\commentb Against max $= 2457825.5065 + 0.071368 E$.} \\
  \multicolumn{6}{l}{\commentc Number of points used to determine the maximum.} \\
\end{tabular}
\end{center}
\end{table}

\subsection{NY Herculis}\label{obj:nyher}

   NY Her was originally discovered by \citet{hof49newvar} as
a Mira-type variable.  Based on photographic observations
by \citet{pas88nyher} and the CRTS detection on 2011 June 10,
the object was identified as an SU UMa-type dwarf nova
with a short supercycle \citep{Pdot4}.
For more history, see \citet{Pdot4}.

\begin{figure}
  \begin{center}
    \FigureFile(85mm,110mm){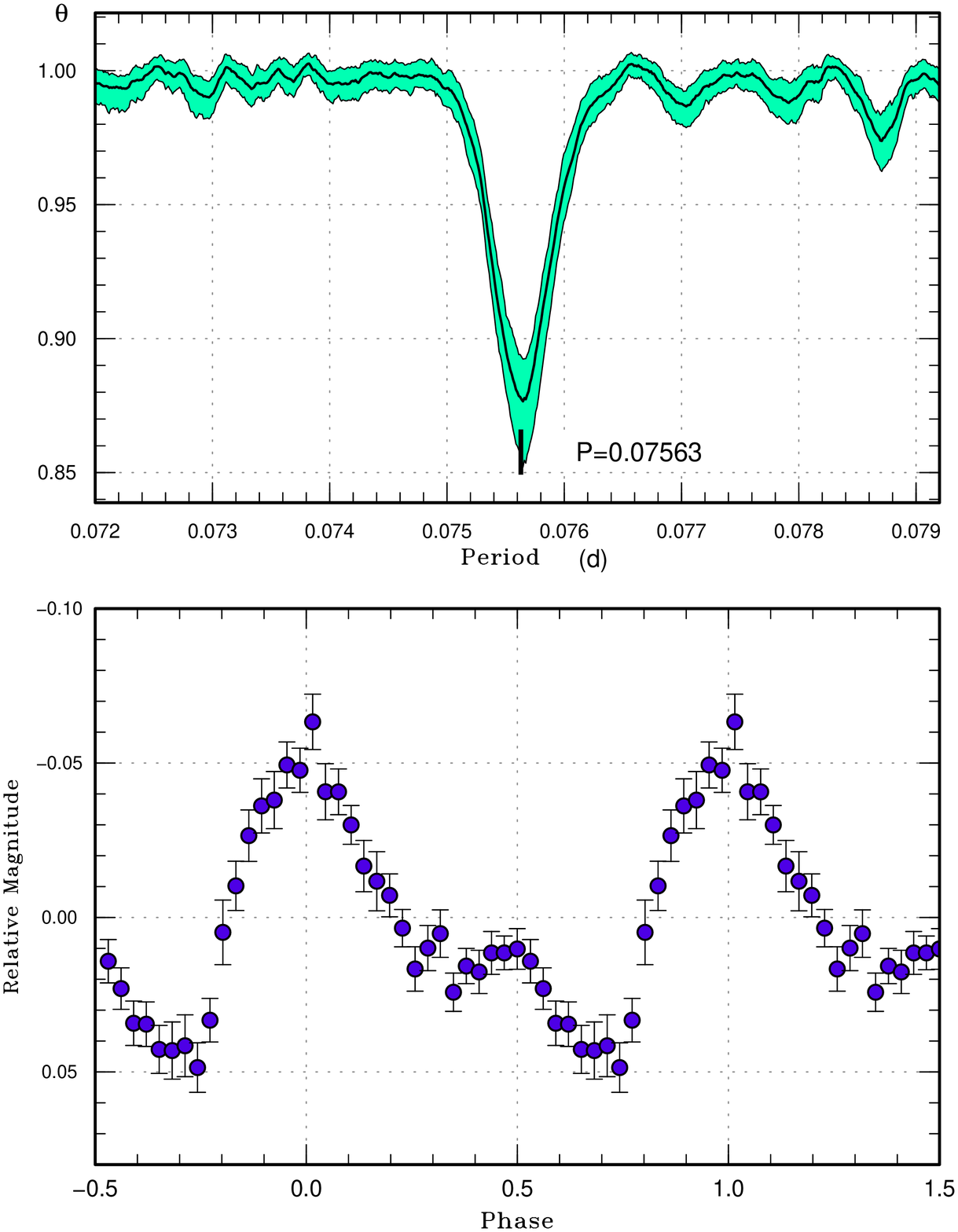}
  \end{center}
  \caption{Superhumps in NY Her during the superoutburst
     plateau (2016).
     (Upper): PDM analysis.
     (Lower): Phase-averaged profile.}
  \label{fig:nyher2016shpdm}
\end{figure}

\begin{figure}
  \begin{center}
    \FigureFile(88mm,70mm){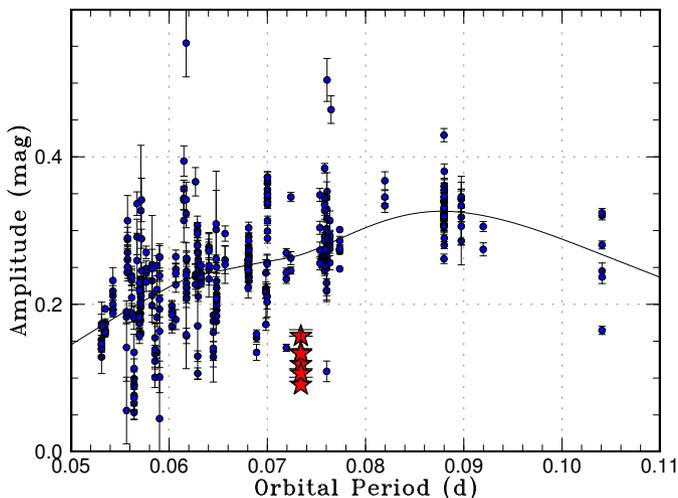}
  \end{center}
  \caption{Dependence of superhump amplitudes on orbital period.
    The superoutburst samples are described in subsection 4.7.1
    in \citet{Pdot3}.
    We selected the range of $-5 < E < 10$ respect to
    the peak superhump amplitude to illustrate
    the maximum superhump amplitudes.  The curve indicates
    a spline-smoothed interpolation of the sample in \citet{Pdot3}.
    The location of NY Her (reflecting the first night of
    observation; we consider that these observations were
    early enough to make a comparison in this figure)
    is shown by stars.
    The single point right to NY Her is a superhump
    of QY Per in 1999.  The other superhumps of the same
    superoutburst had amplitudes larger than 0.2 mag and
    this measurement does not reflect the characteristic
    amplitude of superhumps in QY Per.}
  \label{fig:humpampporb2}
\end{figure}

   The 2016 June superoutburst was detected by the ASAS-SN
team at $V$=16.19 on June 28.  Subsequent observations
detected superhumps (vsnet-alert 19938, 19939, 19948).
The times of superhump maxima are listed in
table \ref{tab:nyheroc2016}.
There was a rather smooth transition from stage B to C.
Since the 2016 observations was much better than the 2011 one,
we provide an updated superhump profile
in figure \ref{fig:nyher2016shpdm}.
It is noteworthy that the mean superhump amplitude
(0.10 mag) is much smaller than most of SU UMa-type
dwarf novae with similar $P_{\rm SH}$ (or $P_{\rm orb}$)
(see figure \ref{fig:humpampporb2}).  Such an unusual
low superhump amplitude is commonly seen in
period bouncers and it may be a signature that NY Her
is in a different evolutionary location from
the standard one with this $P_{\rm orb}$.

\begin{figure}
  \begin{center}
    \FigureFile(88mm,70mm){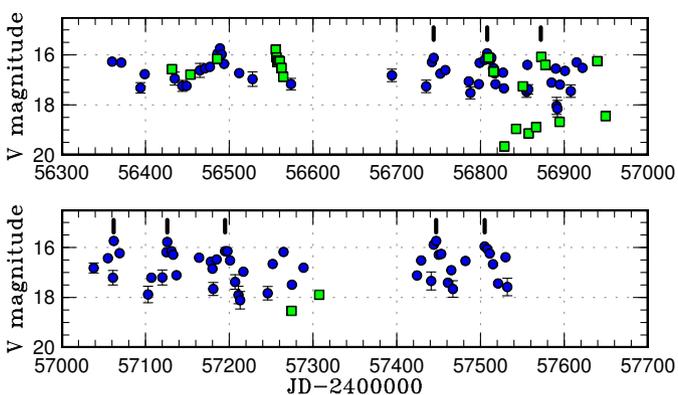}
  \end{center}
  \caption{ASAS-SN and unfiltered CCD light curve of NY Her.
    Filled circles and squares represent ASAS-SN and Poyner's
    measurements, respectively.
    Although details of each outburst are not very clear,
    bright outbursts (likely superoutbursts) tend
    to occur in every $\sim$60--70~d.
    The maxima of bright outbursts listed in table
    \ref{tab:nyherout} and covered by observations in
    this figure are shown by ticks.
    }
  \label{fig:nyherasas}
\end{figure}

   ASAS-SN light curve suggest that bright outbursts
(likely superoutbursts) tend to occur in every $\sim$60--70~d
(figure \ref{fig:nyherasas}).  We selected long outbursts
(presumable superoutbursts) from the ASAS-SN and Poyner's
observations and listed in table \ref{tab:nyherout}.
Note that we selected the brightest points of outbursts
and they do not necessarily reflect the starts of
the outbursts.  These maxima can be well expressed by
a period of 63.5(2)~d with residuals less than 5~d.
We consider that this period is the supercycle
of this system.  The entire durations of superoutbursts
were less than 10~d, which are much shorter than
those in ER UMa-type dwarf nova (cf. \cite{kat95eruma};
\cite{rob95eruma}) but are similar to that of V503 Cyg
with a supercycle of 89~d \citep{har95v503cyg}.
Although the supercycle is between ER UMa-type dwarf novae
and ordinary SU UMa-type dwarf novae, it is not clear
whether NY Her fills a gap between them since NY Her
does not have intermediate properties between them.
NY Her may be classified as an unique object with
a short supercycle and a small superhump amplitude
despite the relatively long $P_{\rm SH}$.


\begin{table}
\caption{Superhump maxima of NY Her (2016)}\label{tab:nyheroc2016}
\begin{center}
\begin{tabular}{rp{55pt}p{40pt}r@{.}lr}
\hline
\multicolumn{1}{c}{$E$} & \multicolumn{1}{c}{max\commenta} & \multicolumn{1}{c}{error} & \multicolumn{2}{c}{$O-C$\commentb} & \multicolumn{1}{c}{$N$\commentc} \\
\hline
0 & 57568.7208 & 0.0015 & $-$0&0081 & 85 \\
1 & 57568.8020 & 0.0010 & $-$0&0026 & 141 \\
2 & 57568.8771 & 0.0009 & $-$0&0032 & 106 \\
3 & 57568.9548 & 0.0039 & $-$0&0011 & 40 \\
9 & 57569.4083 & 0.0010 & $-$0&0016 & 39 \\
10 & 57569.4829 & 0.0009 & $-$0&0026 & 33 \\
13 & 57569.7107 & 0.0050 & $-$0&0018 & 38 \\
14 & 57569.7885 & 0.0017 & 0&0004 & 74 \\
15 & 57569.8671 & 0.0028 & 0&0033 & 75 \\
16 & 57569.9368 & 0.0015 & $-$0&0027 & 67 \\
27 & 57570.7743 & 0.0011 & 0&0025 & 74 \\
29 & 57570.9240 & 0.0023 & 0&0010 & 75 \\
36 & 57571.4560 & 0.0013 & 0&0033 & 54 \\
37 & 57571.5267 & 0.0015 & $-$0&0017 & 52 \\
40 & 57571.7561 & 0.0013 & 0&0007 & 73 \\
41 & 57571.8337 & 0.0014 & 0&0027 & 66 \\
42 & 57571.9149 & 0.0016 & 0&0082 & 74 \\
49 & 57572.4354 & 0.0010 & $-$0&0009 & 38 \\
50 & 57572.5121 & 0.0012 & 0&0001 & 36 \\
53 & 57572.7474 & 0.0027 & 0&0084 & 72 \\
54 & 57572.8150 & 0.0018 & 0&0004 & 66 \\
55 & 57572.8892 & 0.0017 & $-$0&0011 & 74 \\
56 & 57572.9733 & 0.0048 & 0&0074 & 29 \\
58 & 57573.1254 & 0.0084 & 0&0081 & 73 \\
59 & 57573.1964 & 0.0032 & 0&0035 & 74 \\
63 & 57573.4910 & 0.0012 & $-$0&0046 & 103 \\
64 & 57573.5729 & 0.0019 & 0&0016 & 48 \\
66 & 57573.7205 & 0.0019 & $-$0&0020 & 56 \\
67 & 57573.8032 & 0.0024 & 0&0050 & 66 \\
68 & 57573.8722 & 0.0015 & $-$0&0017 & 72 \\
69 & 57573.9475 & 0.0030 & $-$0&0020 & 50 \\
75 & 57574.3993 & 0.0028 & $-$0&0042 & 19 \\
76 & 57574.4757 & 0.0012 & $-$0&0035 & 38 \\
80 & 57574.7808 & 0.0012 & $-$0&0010 & 68 \\
81 & 57574.8619 & 0.0014 & 0&0044 & 74 \\
82 & 57574.9277 & 0.0016 & $-$0&0054 & 52 \\
106 & 57576.7464 & 0.0051 & $-$0&0026 & 24 \\
114 & 57577.3474 & 0.0036 & $-$0&0069 & 30 \\
\hline
  \multicolumn{6}{l}{\commenta BJD$-$2400000.} \\
  \multicolumn{6}{l}{\commentb Against max $= 2457568.7289 + 0.075661 E$.} \\
  \multicolumn{6}{l}{\commentc Number of points used to determine the maximum.} \\
\end{tabular}
\end{center}
\end{table}

\begin{table}
\caption{List of recent superoutbursts of NY Her}\label{tab:nyherout}
\begin{center}
\begin{tabular}{ccc}
\hline
Cycle & JD$-$2400000 & magnitude \\
\hline
0 & 56744 & 16.12 \\
1 & 56808 & 15.94 \\
2 & 56872 & 16.08 \\
5 & 57062 & 15.74 \\
6 & 57126 & 15.78 \\
7 & 57195 & 16.15 \\
8 & 57258 & 15.94 \\
11 & 57447 & 15.74 \\
12 & 57505 & 15.96 \\
13 & 57568 & 16.19 \\
\hline
\end{tabular}
\end{center}
\end{table}

\subsection{MN Lacertae}\label{obj:mnlac}

   This object (=VV 381) was discovered by \citet{mil71cygvar}.
Relatively frequent outbursts were recorded
in \citet{mil71cygvar} and the object was originally
considered to be a Z Cam-type dwarf nova.
T. Kato, however, noted a very faint quiescence
during a systematic survey of $I$-band photometry
of dwarf novae (1990, unpublished) and he suggested
that the outburst amplitude should be comparable to
those of SU UMa-type dwarf novae.

   Since this object was initially cataloged as
a Z Cam-type dwarf nova, \citet{sim11zcamcamp1}
included it as a target for ``Z CamPaign'' project.
As a result, the outburst behavior was relatively
well recorded in the AAVSO database, particularly
in 2010--2012.  The possibility of an SU UMa-type
dwarf nova was particularly noted after a long
outburst in 2011 June (vsnet-alert 13420, 13424).
During this outburst, accurate astrometry was
obtained confirming that the true quiescent
magnitude is indeed faint (22nd mag or even fainter).
There was another outburst in 2012 October,
during which a call for observations of superhumps
was issued (vsnet-alert 15063).  Following this
outburst, the object was withdrawn from the
Z CamPaign project and it has not been observed
as frequently as before.

\begin{figure}
  \begin{center}
    \FigureFile(85mm,110mm){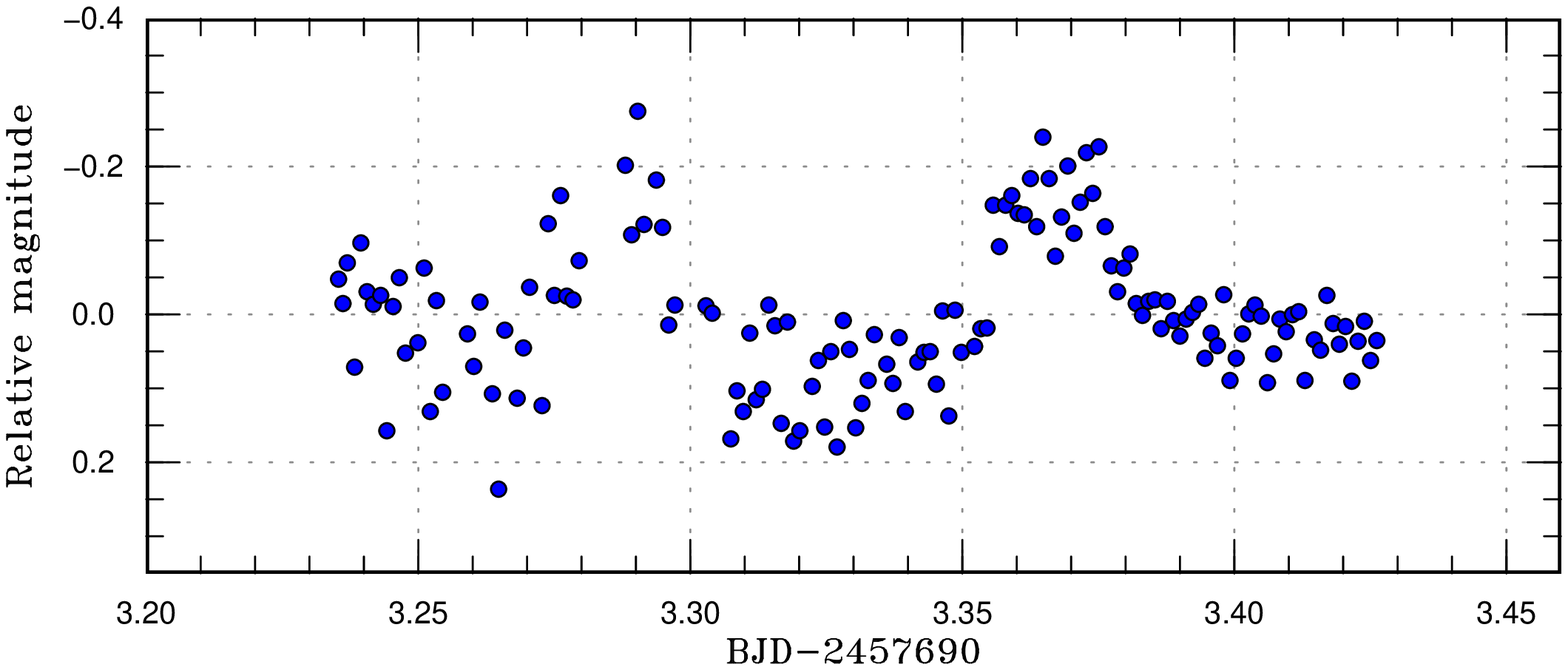}
  \end{center}
  \caption{Superhumps in MN Lac (2016).
  }
  \label{fig:mnlacshlc}
\end{figure}

   The 2016 bright outburst was detected by the ASAS-SN
team at $V$=15.32 on October 30.  Single-night observations
on October 31 indeed detected superhumps
(vsnet-alert 20283; figure \ref{fig:mnlacshlc}).
The times of superhump maxima were BJD 2457693.2873(15)
($N$=37) and 2457693.3684(8) ($N$=53).  The best
superhump period by the PDM method is 0.080(1)~d.
Although the SU UMa-type nature was confirmed,
more observations are needed to establish
a more accurate superhump period.

   Thanks to the excellent coverage in 2010--2012,
we could determine the supercycle.  The maxima
of superoutbursts (table \ref{tab:mnlacout}) can be
expressed by a supercycle of 180(8)~d with
the maximum $|O-C|$ of 14~d.  The result is consistent
with the high outburst frequency reported in
\citet{mil71cygvar}.

\begin{table}
\caption{List of likely superoutbursts of 
         MN Lac in 2010--2012}\label{tab:mnlacout}
\begin{center}
\begin{tabular}{ccccc}
\hline
Year & Month & Day & max\commenta & $V$-mag \\
\hline
2010 & 11 &  6 & 55506 & 15.93 \\
2011 &  5 & 31 & 55713 & 15.74 \\
2011 & 11 & 24 & 55890 & 16.12 \\
2012 &  4 & 30 & 56048 & 15.94 \\
\hline
  \multicolumn{5}{l}{\commenta JD$-$2400000.} \\
\end{tabular}
\end{center}
\end{table}

\subsection{V699 Ophiuchi}\label{obj:v699oph}

   This object was discovered as a dwarf nova
(HV 10577) with a photographic range of 13.8 to
fainter than 16.0 \citep{boy42v699oph}.
\citet{boy42v699oph} recorded five outbursts
between 1937 June 5 and 1940 July 5.
The intervals of the first four outbursts were
in the range of 320--390~d.
Although \citet{wal58CVchart} presented a finding chart,
later spectroscopic studies have shown that
the marked object is a normal star
(\cite{zwi96CVspec}; \cite{liu99CVspec2}).

   On 1999 April 16, A. Pearce detected an outburst
(vsnet-alert 2877).  Accurate astrometry and photometry of
the outbursting object indicated that the true V699 Oph is
an unresolved companion to a 16-th magnitude star
(vsnet-alert 2878, vsnet-chat 1810, 1868).
The first confirmed superoutburst was recorded in 2003.
This outburst was preceded by a separate precursor
and followed by a rebrightening, forming a ``triple outburst''.
\citep{Pdot}.  The 2008 and 2010 superoutbursts were
also reported in \citet{Pdot} and \citet{Pdot2}, respectively.

\begin{figure}
  \begin{center}
    \FigureFile(88mm,70mm){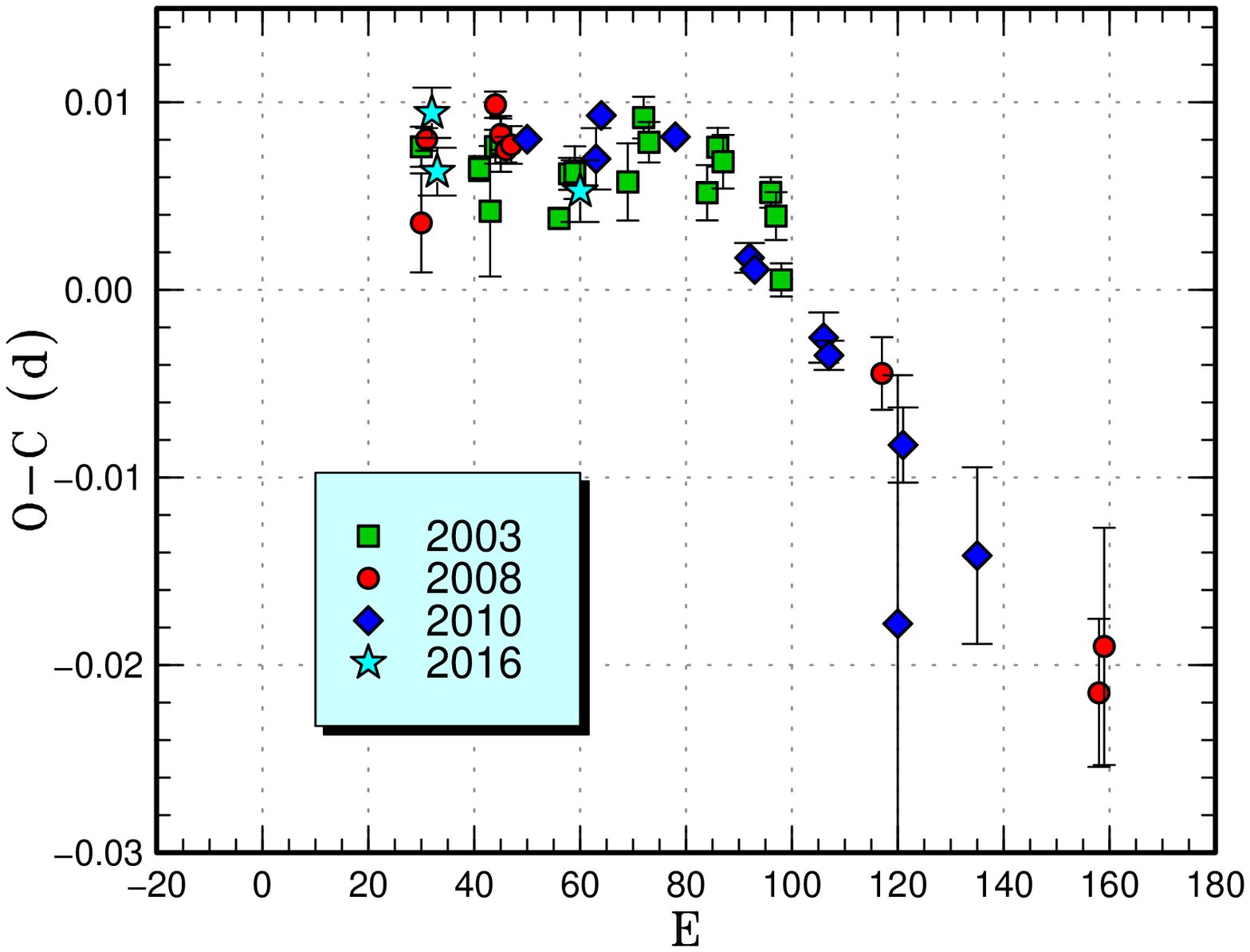}
  \end{center}
  \caption{Comparison of $O-C$ diagrams of V699 Oph
  between different superoutbursts.
  A period of 0.07031~d was used to draw this figure.
  Approximate cycle counts ($E$) after the start of the superoutburst
  were used.
  }
  \label{fig:v699ophcomp2}
\end{figure}

   The 2016 superoutburst was detected by the ASAS-SN team
at $V$=14.56 on May 15 and by R. Stubbings
at a visual magnitude of 14.4 on the same night.
Time-resolved photometric observations were obtained
on two nights and the times of superhump maxima
are listed in table \ref{tab:v699ophoc2016}.
The 2016 observation probably recorded the early part of
stage B superhumps (figure \ref{fig:v699ophcomp2}).

\begin{table}
\caption{Superhump maxima of V699 Oph (2016)}\label{tab:v699ophoc2016}
\begin{center}
\begin{tabular}{rp{55pt}p{40pt}r@{.}lr}
\hline
\multicolumn{1}{c}{$E$} & \multicolumn{1}{c}{max\commenta} & \multicolumn{1}{c}{error} & \multicolumn{2}{c}{$O-C$\commentb} & \multicolumn{1}{c}{$N$\commentc} \\
\hline
0 & 57527.1699 & 0.0013 & 0&0015 & 122 \\
1 & 57527.2371 & 0.0013 & $-$0&0015 & 128 \\
28 & 57529.1344 & 0.0016 & 0&0001 & 77 \\
\hline
  \multicolumn{6}{l}{\commenta BJD$-$2400000.} \\
  \multicolumn{6}{l}{\commentb Against max $= 2457527.1684 + 0.070212 E$.} \\
  \multicolumn{6}{l}{\commentc Number of points used to determine the maximum.} \\
\end{tabular}
\end{center}
\end{table}

\subsection{V344 Pavonis}\label{obj:v344pav}

   This dwarf nova was discovered in outburst on 1990
July 21.  The object was spectroscopically confirmed
as a dwarf nova.  There were two outbursts recorded in
archival plates between 1979 May and 1984 September
\citep{maz90v344paviauc}.
\citet{mas03faintCV} obtained a typical spectrum
of a dwarf nova in quiescence.
\citet{uem04v344pav} studied the 2004 outburst and
identified the SU UMa-type nature.
The analysis was refined in \citet{Pdot}.

\begin{figure}
  \begin{center}
    \FigureFile(88mm,70mm){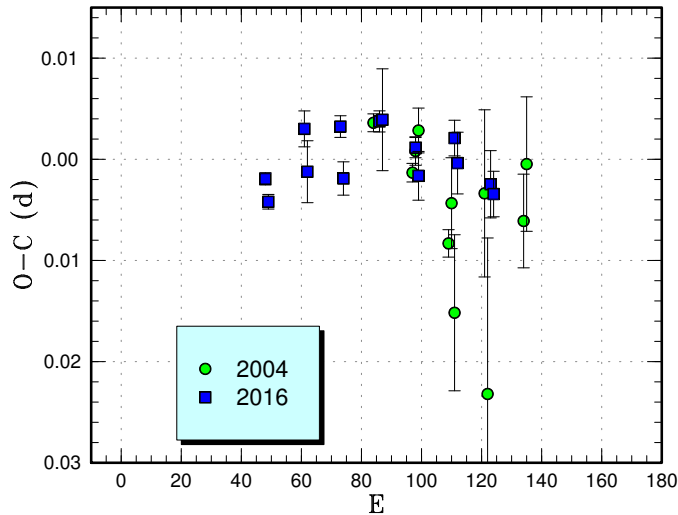}
  \end{center}
  \caption{Comparison of $O-C$ diagrams of V344 Pav
  between different superoutbursts.
  A period of 0.07988~d was used to draw this figure.
  Approximate cycle counts ($E$) after the start of the superoutburst
  were used.  Since the start of the 2004 outburst
  was not well constrained, we shifted the $O-C$ diagram
  so that the rapid fading of the two superoutbursts
  match each other.
  }
  \label{fig:v344pavcomp}
\end{figure}

   The 2016 superoutburst was detected by R. Stubbings
at a visual magnitude of 14.4 on April 25.
Subsequent observations detected superhumps
(vsnet-alert 19796).  The times of superhump
maxima are listed in table \ref{tab:v344pavoc2016}.
Time-resolved photometry was obtained only
in the later phase of the superoutbursts 
both in 2004 and 2016.  The superhump stage has been
therefore unclear (figure \ref{fig:v344pavcomp}).
We listed a global $P_{\rm dot}$ in table \ref{tab:perlist}.
Observations in the earlier phase of the superoutburst
are needed to characterize superhumps of this object
better.

\begin{table}
\caption{Superhump maxima of V344 Pav (2016)}\label{tab:v344pavoc2016}
\begin{center}
\begin{tabular}{rp{55pt}p{40pt}r@{.}lr}
\hline
\multicolumn{1}{c}{$E$} & \multicolumn{1}{c}{max\commenta} & \multicolumn{1}{c}{error} & \multicolumn{2}{c}{$O-C$\commentb} & \multicolumn{1}{c}{$N$\commentc} \\
\hline
0 & 57507.7957 & 0.0006 & $-$0&0020 & 20 \\
1 & 57507.8734 & 0.0007 & $-$0&0043 & 22 \\
13 & 57508.8391 & 0.0018 & 0&0030 & 22 \\
14 & 57508.9148 & 0.0031 & $-$0&0013 & 6 \\
25 & 57509.7979 & 0.0011 & 0&0032 & 23 \\
26 & 57509.8726 & 0.0016 & $-$0&0019 & 21 \\
38 & 57510.8369 & 0.0010 & 0&0037 & 22 \\
39 & 57510.9169 & 0.0050 & 0&0039 & 8 \\
50 & 57511.7928 & 0.0010 & 0&0012 & 22 \\
51 & 57511.8699 & 0.0024 & $-$0&0016 & 21 \\
63 & 57512.8322 & 0.0018 & 0&0022 & 23 \\
64 & 57512.9096 & 0.0031 & $-$0&0003 & 10 \\
75 & 57513.7862 & 0.0033 & $-$0&0024 & 20 \\
76 & 57513.8651 & 0.0022 & $-$0&0034 & 21 \\
\hline
  \multicolumn{6}{l}{\commenta BJD$-$2400000.} \\
  \multicolumn{6}{l}{\commentb Against max $= 2457507.7977 + 0.079878 E$.} \\
  \multicolumn{6}{l}{\commentc Number of points used to determine the maximum.} \\
\end{tabular}
\end{center}
\end{table}

\subsection{V368 Pegasi}\label{obj:v368peg}

   V368 Peg is a dwarf nova (Antipin Var 63) discovered by
\citet{ant99v368pegftcamv367pegv2209cyg}.
See \citet{Pdot8} for the summary of the history.
The 2016 superoutburst was detected by P. Schmeer
at a visual magnitude of 13.0 on September 28.
Time-resolved photometry was performed only on
a single night.  The resultant superhump maxima
were BJD 2457661.4175(5) ($N$=66) and 2457661.4883(4)
($N$=76).

\subsection{V893 Sco}\label{obj:v893sco}

   V893 Sco was discovered as a variable star by
\citet{sat72v893sco}.  The variable had been lost
for a long time, and was rediscovered by K. Haseda
\citep{kat98v893sco}.  For more historical information,
see \citet{Pdot6}.  This object is an eclipsing SU UMa-type
dwarf nova (cf. \cite{bru00v893sco}; \cite{mat00v893sco}.

\begin{figure}
  \begin{center}
    \FigureFile(85mm,110mm){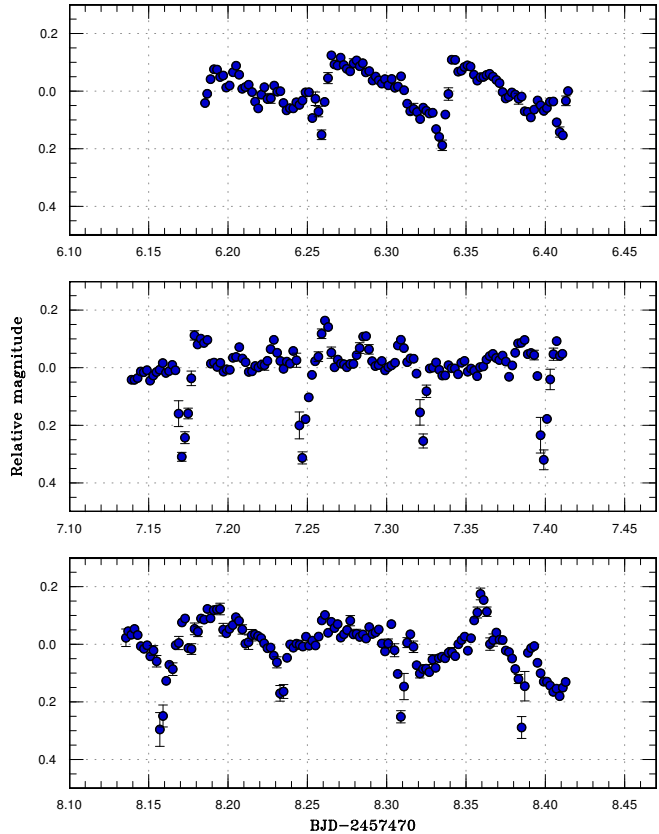}
  \end{center}
  \caption{Eclipses and superhumps in V893 Sco (2016).
  The data were binned to 0.002~d.
  }
  \label{fig:v893scoshlc}
\end{figure}

   The 2016 superoutburst was detected by R. Stubbings
at a visual magnitude of 12.8 on March 21.  It once
faded to $V$=13.64 on the same night and brightened
to $V$=12.37 on March 25 (vsnet-alert 19652).
The outburst on March 21 should have been a precursor.
Our time-resolved photometry started on March 28
and detected superhumps (vsnet-alert 19661;
figure \ref{fig:v893scoshlc}).
Since our observation started relatively late,
we could record only the final part of the superoutburst.
Later observations were dominated by the orbital humps
and we could only extract a small number of
superhump maxima outside the eclipses
(table \ref{tab:v893scooc2016}).
We obtained the eclipse ephemeris for the use of
defining the orbital phases in this paper
\begin{equation}
{\rm Min(BJD)} = 2454173.3030(3) + 0.0759614600(16) E
\label{equ:v893scoecl}
\end{equation}
using the MCMC modeling \citep{Pdot4} using the data
up to \citet{Pdot6} and current set of observation.

\begin{table}
\caption{Superhump maxima of V893 Sco (2016)}\label{tab:v893scooc2016}
\begin{center}
\begin{tabular}{rp{50pt}p{30pt}r@{.}lcr}
\hline
$E$ & max\commenta & error & \multicolumn{2}{c}{$O-C$\commentb} & phase\commentc & $N$\commentd \\
\hline
0 & 57476.1964 & 0.0017 & $-$0&0062 & 0.10 & 92 \\
1 & 57476.2774 & 0.0007 & 0&0002 & 0.05 & 118 \\
2 & 57476.3528 & 0.0007 & 0&0009 & 0.06 & 117 \\
13 & 57477.1799 & 0.0011 & 0&0067 & 0.09 & 115 \\
14 & 57477.2601 & 0.0049 & 0&0122 & 0.11 & 111 \\
15 & 57477.3120 & 0.0015 & $-$0&0105 & 0.07 & 112 \\
16 & 57477.3987 & 0.0016 & 0&0015 & 0.11 & 69 \\
26 & 57478.1390 & 0.0013 & $-$0&0049 & 0.17 & 42 \\
\hline
  \multicolumn{7}{l}{\commenta BJD$-$2400000.} \\
  \multicolumn{7}{l}{\commentb Against max $= 2457476.2025 + 0.074666 E$.} \\
  \multicolumn{7}{l}{\commentc Orbital phase.} \\
  \multicolumn{7}{l}{\commentd Number of points used to determine the maximum.} \\
\end{tabular}
\end{center}
\end{table}

\subsection{V493 Serpentis}\label{obj:v493ser}

   This object (=SDSS J155644.24$-$000950.2) was selected as
a dwarf nova by SDSS \citep{szk02SDSSCVs}.  The SU UMa-type
nature was identified by observations of
the 2006 and 2007 superoutbursts \citep{Pdot}.
See \citet{Pdot5} for more history.

\begin{figure}
  \begin{center}
    \FigureFile(88mm,70mm){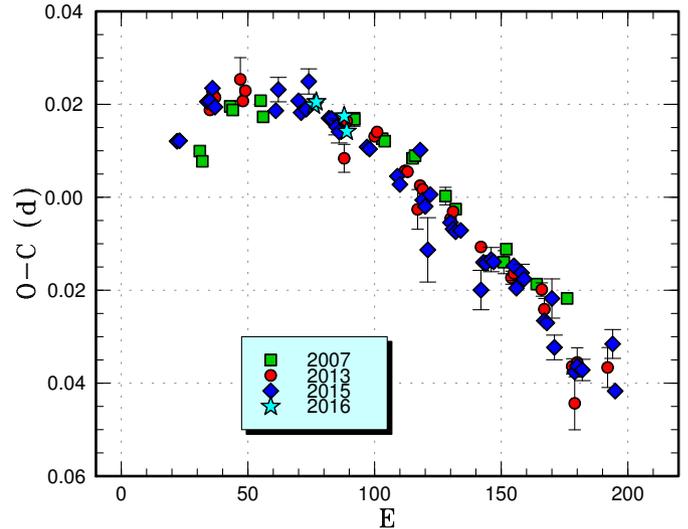}
  \end{center}
  \caption{Comparison of $O-C$ diagrams of V493 Ser
  between different superoutbursts.
  A period of 0.08310~d was used to draw this figure.
  Approximate cycle counts ($E$) after the start of the superoutburst
  were used.
  }
  \label{fig:v493sercomp3}
\end{figure}

   The 2016 superoutburst was detected by T. Horie
at a visual magnitude of 12.5 on June 5.
It was pointed out by H. Maehara the outburst already started
on June 1 (vsnet-alert 19872).
Time-resolved photometry was carried out on
two nights, yielding superhump maxima in
table \ref{tab:v493seroc2016}.
A comparison of $O-C$ diagrams (figure \ref{fig:v493sercomp3})
suggest that these observations recorded
the early phase of stage C.

\begin{table}
\caption{Superhump maxima of V493 Ser (2016)}\label{tab:v493seroc2016}
\begin{center}
\begin{tabular}{rp{55pt}p{40pt}r@{.}lr}
\hline
\multicolumn{1}{c}{$E$} & \multicolumn{1}{c}{max\commenta} & \multicolumn{1}{c}{error} & \multicolumn{2}{c}{$O-C$\commentb} & \multicolumn{1}{c}{$N$\commentc} \\
\hline
0 & 57547.3780 & 0.0013 & $-$0&0006 & 26 \\
1 & 57547.4618 & 0.0008 & 0&0005 & 25 \\
12 & 57548.3728 & 0.0012 & 0&0015 & 22 \\
13 & 57548.4526 & 0.0008 & $-$0&0014 & 25 \\
\hline
  \multicolumn{6}{l}{\commenta BJD$-$2400000.} \\
  \multicolumn{6}{l}{\commentb Against max $= 2457547.3786 + 0.082730 E$.} \\
  \multicolumn{6}{l}{\commentc Number of points used to determine the maximum.} \\
\end{tabular}
\end{center}
\end{table}

\subsection{AW Sagittae}\label{obj:awsge}

   AW Sge was discovered as a dwarf nova by \citet{wol06awsge}.
The object was identified as an SU UMa-type dwarf nova
during the 2000 outburst \citep{Pdot}.  See \citet{Pdot6}
for more history.

   The 2016 superoutburst was detected by R. Stubbings
at a visual magnitude of 14.6 on June 14.
Time-resolved photometric observations were carried out
on a single night and yielded the superhumps maxima:
BJD 2457558.3859(5) ($N$=75) and 2457558.4606(9) ($N$=50).

\subsection{V1389 Tauri}\label{obj:v1389tau}

   This object was discovered by K. Itagaki at an unfiltered
CCD magnitude of 14.1 on 2008 August 7 \citep{yam08j0406cbet1463}.
There was an X-ray counterpart (1RXS J040700.2$+$005247)
and the dwarf nova-type classification was readily
suggested.  The object was recorded already in outburst
at $V$=13.5 on August 4 in the ASAS-3 \citep{ASAS3} data 
(vsnet-alert 10419).
There were two past outbursts (2004 October 20 
and 2006 March 16) recorded in the ASAS-3 data
(vsnet-alert 10419).
Subsequent observations detected superhumps
(vsnet-alert 10422, 10423).  This outburst was studied
in \citet{Pdot}.  Another superoutburst in 2010 was
studied in \citet{Pdot2}.

\begin{figure}
  \begin{center}
    \FigureFile(88mm,70mm){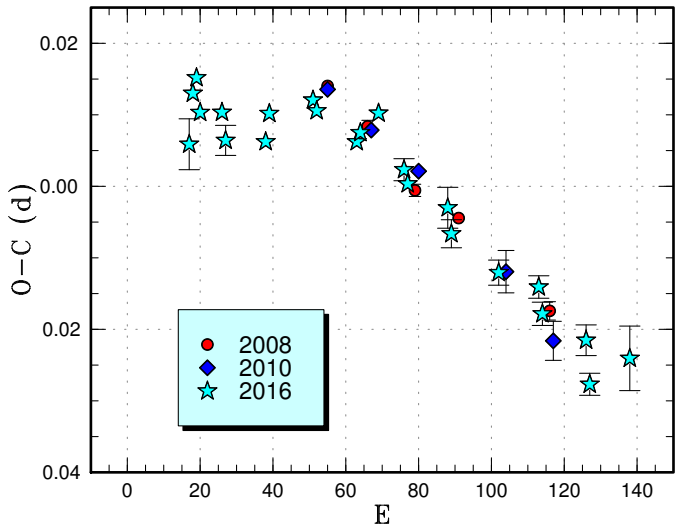}
  \end{center}
  \caption{Comparison of $O-C$ diagrams of V1389 Tau
  between different superoutbursts.
  A period of 0.08046~d was used to draw this figure.
  Approximate cycle counts ($E$) after the start of the superoutburst
  were used.  Since the start of the 2010 superoutburst
  was not known, we have shifted the $O-C$ diagram to
  best fit the others.
  }
  \label{fig:v1389taucomp}
\end{figure}

   The 2016 superoutburst was detected by the ASAS-SN
team at $V$=13.52 on October 23.  Subsequent observations
detected superhumps (vsnet-alert 20267).
The times of superhump maxima are listed in
table \ref{tab:v1389tauoc2016}.
As in other typical long-$P_{\rm SH}$ systems
(cf. figure 4 in \cite{Pdot}), stage B was relatively short.
A comparison of the $O-C$ diagrams has confirmed
that the superhumps recorded in 2008 were indeed
stage C ones (figure \ref{fig:v1389taucomp}).
Although individual superhump maxima were not
measured, a PDM analysis of the post-superoutburst
data (4.5~d segment after BJD 2457697) detected a period
of 0.08000(11)~d.  This value suggests that stage C
superhump lasted even after the termination
of the superoutburst.

\begin{table}
\caption{Superhump maxima of V1389 Tau (2016)}\label{tab:v1389tauoc2016}
\begin{center}
\begin{tabular}{rp{55pt}p{40pt}r@{.}lr}
\hline
\multicolumn{1}{c}{$E$} & \multicolumn{1}{c}{max\commenta} & \multicolumn{1}{c}{error} & \multicolumn{2}{c}{$O-C$\commentb} & \multicolumn{1}{c}{$N$\commentc} \\
\hline
0 & 57686.0276 & 0.0036 & $-$0&0098 & 97 \\
1 & 57686.1152 & 0.0006 & $-$0&0023 & 176 \\
2 & 57686.1979 & 0.0004 & 0&0002 & 178 \\
3 & 57686.2735 & 0.0008 & $-$0&0044 & 104 \\
9 & 57686.7562 & 0.0009 & $-$0&0025 & 25 \\
10 & 57686.8328 & 0.0021 & $-$0&0061 & 21 \\
21 & 57687.7177 & 0.0013 & $-$0&0029 & 17 \\
22 & 57687.8021 & 0.0010 & 0&0014 & 21 \\
34 & 57688.7695 & 0.0009 & 0&0069 & 23 \\
35 & 57688.8484 & 0.0006 & 0&0058 & 17 \\
46 & 57689.7291 & 0.0010 & 0&0048 & 15 \\
47 & 57689.8109 & 0.0008 & 0&0064 & 16 \\
52 & 57690.2159 & 0.0009 & 0&0107 & 98 \\
59 & 57690.7712 & 0.0015 & 0&0050 & 20 \\
60 & 57690.8497 & 0.0007 & 0&0033 & 16 \\
71 & 57691.7314 & 0.0029 & 0&0033 & 22 \\
72 & 57691.8082 & 0.0020 & 0&0000 & 20 \\
85 & 57692.8488 & 0.0018 & $-$0&0014 & 16 \\
96 & 57693.7318 & 0.0016 & 0&0000 & 22 \\
97 & 57693.8085 & 0.0016 & $-$0&0034 & 20 \\
109 & 57694.7704 & 0.0021 & $-$0&0034 & 20 \\
110 & 57694.8447 & 0.0015 & $-$0&0093 & 16 \\
121 & 57695.7334 & 0.0045 & $-$0&0022 & 22 \\
\hline
  \multicolumn{6}{l}{\commenta BJD$-$2400000.} \\
  \multicolumn{6}{l}{\commentb Against max $= 2457686.0374 + 0.080150 E$.} \\
  \multicolumn{6}{l}{\commentc Number of points used to determine the maximum.} \\
\end{tabular}
\end{center}
\end{table}

\subsection{SU Ursae Majoris}\label{obj:suuma}

   This object is the prototype of SU UMa-type
dwarf novae.  See \citet{Pdot7} for the history.
The 2017 superoutburst was detected by E. Muyllaert
at a visual magnitude of 11.3 on February 23.
Only single superhump at BJD 2457810.5647(3) ($N$=89)
was observed.

\subsection{HV Virginis}\label{obj:hvvir}

   HV Vir was originally discovered by \citet{sch31hvvir}
in outburst on 1929 February 11.  The object was also
given a designation of NSV 6201 as a suspected variable.
\citet{due84hvvir} and \citet{due87novaatlas} classified
it as a classical nova and provided a light curve 
of the 1929 outburst based on his examination of
archival plates.  Amateur observers, particularly by
the Variable Star Observers' League in Japan (VSOLJ),
suspected it to be a dwarf nova and started monitoring
since 1987 [i.e. following the publication
of \citet{due87novaatlas}].  The object was caught in
outburst by P. Schmeer on 1992 April 20 at a visual
magnitude of 12.0 \citep{sch92hvviriauc}.
The 1992 outburst was extensively studied
(\cite{bar92hvvir}; \cite{lei94hvvir}; \cite{kat01hvvir}).
It might be worth noting that \citet{bar92hvvir}
recorded low-amplitude variations with a period
corresponding to the orbital period, their interpretation
(originating from the hot spot as in quiescence)
was strongly affected by \citet{pat81wzsge}.
Although \citet{szk92hvviriauc} reported the detection of
superhumps, the detailed result has not been published.
\citet{lei94hvvir} reported the detection of historical
outbursts in 1939, 1970 and 1981 in archival plates.
Although \citet{lei94hvvir} noted chaotic
``early superhump variability'', its period was not
precisely determined.  \citet{lei94hvvir} recorded
superhumps and reported a negative $P_{\rm dot}$,
which was incorrect due to an error in cycle counts
probably misguided by the received wisdom at that time
that SU UMa-type dwarf novae universally show
negative $P_{\rm dot}$ (cf. \cite{war85suuma};
\cite{pat93vyaqr}).  Using additional observations
and all available data, \citet{kat01hvvir} clarified
that this object showed two types of superhumps
(early superhumps and ordinary superhumps) and
the $P_{\rm dot}$ for ordinary superhumps was positive.
\citet{kat01hvvir} proposed the close similarity to
AL Com (cf. \cite{kat96alcom}) and WZ Sge,
giving a basis of the modern concept of WZ Sge-type
dwarf novae \citep{kat15wzsge}.

   The object underwent another superoutburst in 2002.
This outburst was also extensively studied
by \citet{ish03hvvir}, who established the positive
$P_{\rm dot}$ using a much more complete set of observations
than in 1992.  \citet{pat03suumas} also reported
the superhump period of the same outburst and
the orbital period of 0.057069(6)~d from quiescent photometry.
There was another superoutburst in
2008, which was reported in \citet{Pdot}.

\begin{figure}
  \begin{center}
    \FigureFile(85mm,70mm){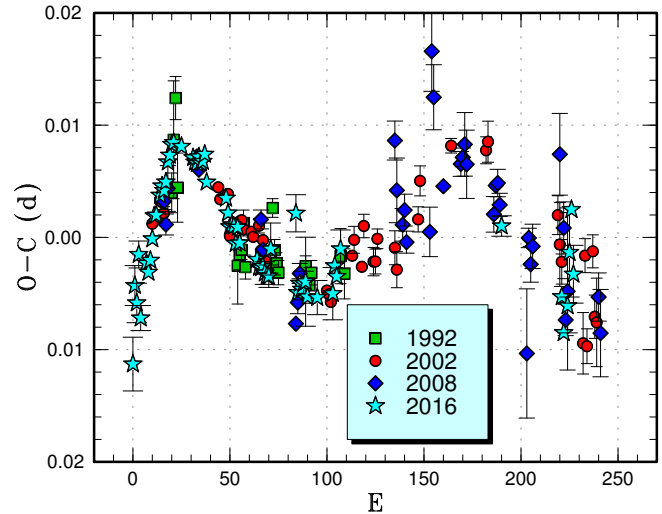}
  \end{center}
  \caption{Comparison of $O-C$ diagrams of HV Vir between different
  superoutbursts.  A period of 0.05828~d was used to draw this figure.
  Approximate cycle counts ($E$) after the emergence of ordinary
  superhumps were used.  After the high-quality observations
  in 2016, it became apparent that the emergence of ordinary
  superhumps was not well recorded in the past superoutbursts.
  The cycle counts were shifted by 20, 10 and 15 for the 1992,
  2002 and 2008 superoutbursts, respectively, to match
  the 2016 observations.
  }
  \label{fig:hvvircomp2}
\end{figure}

\begin{figure}
  \begin{center}
    \FigureFile(85mm,110mm){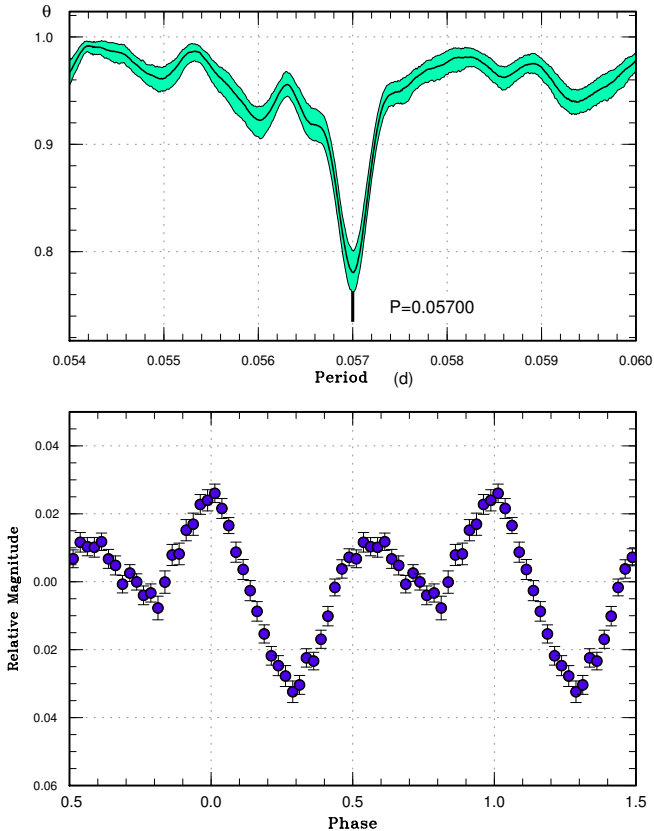}
  \end{center}
  \caption{Early superhumps in HV Vir (2016).
     (Upper): PDM analysis.
     (Lower): Phase-averaged profile.}
  \label{fig:hvvir2016eshpdm}
\end{figure}

   The 2016 superoutburst was detected by the ASAS-SN
team at $V$=12.0 on March 10 (cf. vsnet-alert 19571).
Initial observations already detected early superhumps
(vsnet-alert 19573, 19576, 19589;
figure \ref{fig:hvvir2016eshpdm}).
The object then developed ordinary superhumps
(vsnet-alert 19581, 19599, 19633).
The times of superhump maxima are listed in
table \ref{tab:hvviroc2016}.  The data very clearly
demonstrate the presence of stages A and B,
although there was an observational gap in the middle
of stage B.
The superhump period of stage A was very ideally
determined to be 0.05907(6)~d
(cf. figure \ref{fig:hvvircomp2}).  This period gives
the fractional superhump excess of $\epsilon^*$=0.034(1),
which corresponds to $q$=0.093(3).  
This value supersedes the earlier determination
by the same method to be $q$=0.072(1) using
the less extensive 2002 data.  The period was determined
for the 2002 data from single-night observations
assuming that stage A continued up to the first
observation of stage B while the present observation
obtained an almost complete coverage of stage A
(see figure \ref{fig:hvvircomp2}).  It was likely
that the error was underestimated in the 2002 superoutburst.
The outburst started rapid fading on March 29--30
and the entire duration of the superoutburst
was at least 20~d.  Despite dense observations,
no post-outburst rebrightening was recorded.

\begin{figure}
  \begin{center}
    \FigureFile(85mm,110mm){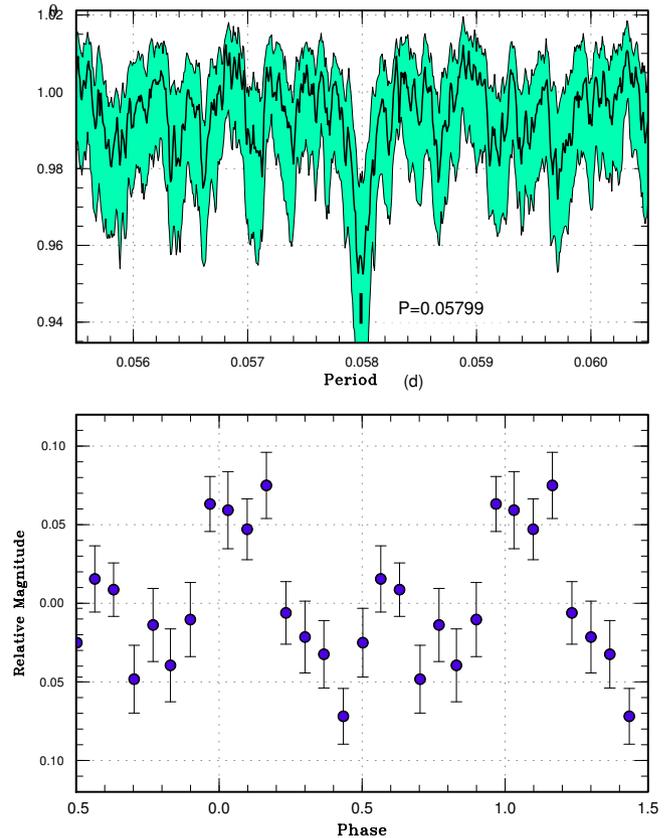}
  \end{center}
  \caption{Post-superoutburst superhumps in HV Vir (2016).
     (Upper): PDM analysis.
     The data for BJD 2457478--2457494 were used.
     (Lower): Phase-averaged profile.}
  \label{fig:hvvir2016postpdm}
\end{figure}

   A PDM analysis of the post-superoutburst observations
yielded a period of 0.05799(2)~d
(figure \ref{fig:hvvir2016postpdm}).  This period
corresponds to a disk radius of 0.33$a$
assuming that the precession rate is not affected
by the pressure effect.  The value is in the range of
0.30--0.38$a$ determined for well-observed
WZ Sge-type dwarf novae \citep{kat13qfromstageA}.

   The period of early superhumps [0.057000(8)~d]
is in agreement with 0.056996(9)~d determined
from the 2008 observation (from the observations reported
in \cite{Pdot}).  The quality of past observations
were lower: 0.057085~d (without error estimate)
for the 1992 outburst \citep{kat01hvvir}, which was
based only on published times of maxima,
and 0.0569(1)~d for the 2002
outburst \citep{ish03hvvir}.  The current observations,
combined with the 2008 data, established the period of
early superhumps of this object to a precision
directly comparable to the orbital period
for the first time.  The 2016 and 2008 periods
were 0.13(2)\% and 0.13(3)\% shorter than
the orbital period, respectively.

\begin{table*}
\caption{Superhump maxima of HV Vir (2016)}\label{tab:hvviroc2016}
\begin{center}
\begin{tabular}{rp{55pt}p{40pt}r@{.}lrrp{55pt}p{40pt}r@{.}lr}
\hline
\multicolumn{1}{c}{$E$} & \multicolumn{1}{c}{max\commenta} & \multicolumn{1}{c}{error} & \multicolumn{2}{c}{$O-C$\commentb} & \multicolumn{1}{c}{$N$\commentc} & \multicolumn{1}{c}{$E$} & \multicolumn{1}{c}{max\commenta} & \multicolumn{1}{c}{error} & \multicolumn{2}{c}{$O-C$\commentb} & \multicolumn{1}{c}{$N$\commentc} \\
\hline
0 & 57463.6347 & 0.0024 & $-$0&0131 & 24 & 52 & 57466.6775 & 0.0004 & 0&0006 & 28 \\
1 & 57463.6999 & 0.0018 & $-$0&0061 & 29 & 53 & 57466.7342 & 0.0005 & $-$0&0009 & 23 \\
2 & 57463.7567 & 0.0018 & $-$0&0076 & 24 & 54 & 57466.7940 & 0.0005 & 0&0006 & 19 \\
3 & 57463.8193 & 0.0012 & $-$0&0033 & 20 & 55 & 57466.8508 & 0.0004 & $-$0&0008 & 21 \\
4 & 57463.8719 & 0.0011 & $-$0&0089 & 23 & 64 & 57467.3739 & 0.0002 & $-$0&0020 & 108 \\
7 & 57464.0516 & 0.0006 & $-$0&0039 & 52 & 65 & 57467.4315 & 0.0003 & $-$0&0027 & 118 \\
8 & 57464.1091 & 0.0005 & $-$0&0047 & 48 & 66 & 57467.4897 & 0.0003 & $-$0&0027 & 152 \\
9 & 57464.1684 & 0.0008 & $-$0&0036 & 33 & 67 & 57467.5481 & 0.0004 & $-$0&0025 & 59 \\
10 & 57464.2286 & 0.0005 & $-$0&0017 & 58 & 68 & 57467.6058 & 0.0010 & $-$0&0031 & 48 \\
11 & 57464.2889 & 0.0005 & 0&0004 & 60 & 69 & 57467.6638 & 0.0005 & $-$0&0033 & 73 \\
12 & 57464.3473 & 0.0003 & 0&0005 & 47 & 70 & 57467.7221 & 0.0010 & $-$0&0033 & 23 \\
14 & 57464.4656 & 0.0003 & 0&0023 & 137 & 71 & 57467.7828 & 0.0023 & $-$0&0009 & 14 \\
15 & 57464.5247 & 0.0002 & 0&0032 & 172 & 84 & 57468.5436 & 0.0017 & 0&0027 & 41 \\
16 & 57464.5826 & 0.0002 & 0&0028 & 172 & 85 & 57468.5948 & 0.0006 & $-$0&0044 & 65 \\
17 & 57464.6416 & 0.0002 & 0&0036 & 150 & 86 & 57468.6539 & 0.0008 & $-$0&0035 & 27 \\
18 & 57464.7017 & 0.0003 & 0&0054 & 40 & 87 & 57468.7117 & 0.0009 & $-$0&0040 & 24 \\
19 & 57464.7606 & 0.0004 & 0&0060 & 32 & 88 & 57468.7705 & 0.0011 & $-$0&0035 & 20 \\
20 & 57464.8198 & 0.0005 & 0&0070 & 19 & 89 & 57468.8289 & 0.0040 & $-$0&0033 & 20 \\
21 & 57464.8783 & 0.0005 & 0&0072 & 25 & 90 & 57468.8858 & 0.0005 & $-$0&0046 & 26 \\
25 & 57465.1110 & 0.0006 & 0&0070 & 50 & 95 & 57469.1772 & 0.0015 & $-$0&0045 & 30 \\
31 & 57465.4597 & 0.0002 & 0&0061 & 166 & 103 & 57469.6438 & 0.0023 & $-$0&0040 & 28 \\
32 & 57465.5176 & 0.0003 & 0&0058 & 81 & 104 & 57469.7045 & 0.0008 & $-$0&0015 & 25 \\
33 & 57465.5758 & 0.0003 & 0&0057 & 119 & 105 & 57469.7619 & 0.0008 & $-$0&0024 & 21 \\
34 & 57465.6346 & 0.0002 & 0&0063 & 134 & 107 & 57469.8809 & 0.0017 & 0&0001 & 26 \\
35 & 57465.6924 & 0.0004 & 0&0058 & 46 & 190 & 57474.7202 & 0.0009 & 0&0045 & 25 \\
36 & 57465.7508 & 0.0003 & 0&0060 & 38 & 221 & 57476.5205 & 0.0007 & $-$0&0010 & 52 \\
37 & 57465.8097 & 0.0007 & 0&0066 & 15 & 222 & 57476.5756 & 0.0009 & $-$0&0041 & 50 \\
38 & 57465.8655 & 0.0007 & 0&0042 & 23 & 224 & 57476.6946 & 0.0024 & $-$0&0017 & 14 \\
48 & 57466.4469 & 0.0005 & 0&0030 & 56 & 225 & 57476.7576 & 0.0040 & 0&0031 & 15 \\
49 & 57466.5038 & 0.0004 & 0&0017 & 57 & 226 & 57476.8197 & 0.0008 & 0&0069 & 33 \\
50 & 57466.5609 & 0.0004 & 0&0005 & 58 & 227 & 57476.8722 & 0.0026 & 0&0012 & 39 \\
51 & 57466.6193 & 0.0003 & 0&0007 & 79 & \multicolumn{1}{c}{--} & \multicolumn{1}{c}{--} & \multicolumn{1}{c}{--} & \multicolumn{2}{c}{--} & \multicolumn{1}{c}{--}\\
\hline
  \multicolumn{12}{l}{\commenta BJD$-$2400000.} \\
  \multicolumn{12}{l}{\commentb Against max $= 2457463.6478 + 0.058252 E$.} \\
  \multicolumn{12}{l}{\commentc Number of points used to determine the maximum.} \\
\end{tabular}
\end{center}
\end{table*}

\subsection{NSV 2026}\label{obj:nsv2026}

   This object was discovered as a variable star
(=HV 6907) by \citet{hof35newvar}.  The SU UMa-type
nature was confirmed during the 2015 superoutburst.
For more history, see \citet{Pdot8}.

\begin{figure}
  \begin{center}
    \FigureFile(85mm,70mm){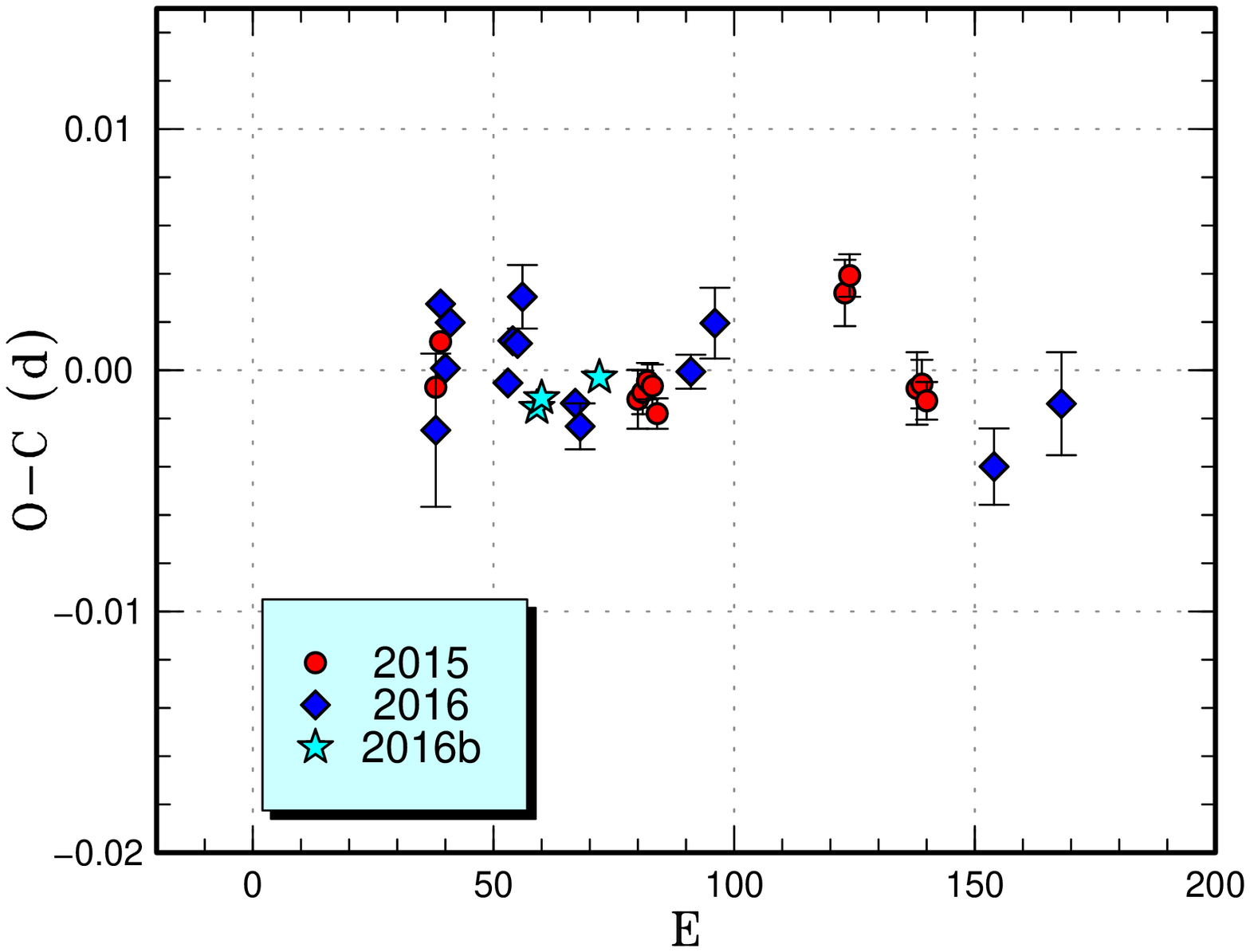}
  \end{center}
  \caption{Comparison of $O-C$ diagrams of NSV 2026 between different
  superoutbursts.  A period of 0.06982~d was used to draw this figure.
  Approximate cycle counts ($E$) after the starts of the outbursts
  were used.  The start of the 2016 outburst refers to
  the precursor outburst.  Since the start of the 2015 outburst
  was not well constrained, the $O-C$ curve was shifted
  as in the 2016 one.
  }
  \label{fig:nsv2026comp2}
\end{figure}

   There was a superoutburst in 2016 February \citep{Pdot8}.
Another superoutburst occurred in 2016 November,
which was detected by J. Shears at an unfiltered
CCD magnitude of 14.19 and by E. Muyllaert at
a visual magnitude of 14.0 on November 25.
The object was further observed
to brighten to a visual magnitude of 13.2 on November 26.
The times of superhump maxima are listed in
table \ref{tab:nsv2026oc2016b}.
These superhumps were likely stage B ones
(figure \ref{fig:nsv2026comp2}).
As judged from the interval of two superoutbursts
in 2016 and the supercycle of $\sim$95~d \citep{Pdot8},
two superoutbursts were likely missed between
the two superoutbursts in 2016.

\begin{table}
\caption{Superhump maxima of NSV 2026 (2016b)}\label{tab:nsv2026oc2016b}
\begin{center}
\begin{tabular}{rp{55pt}p{40pt}r@{.}lr}
\hline
\multicolumn{1}{c}{$E$} & \multicolumn{1}{c}{max\commenta} & \multicolumn{1}{c}{error} & \multicolumn{2}{c}{$O-C$\commentb} & \multicolumn{1}{c}{$N$\commentc} \\
\hline
0 & 57722.4877 & 0.0005 & $-$0&0002 & 78 \\
1 & 57722.5579 & 0.0006 & 0&0002 & 64 \\
13 & 57723.3966 & 0.0005 & $-$0&0000 & 59 \\
\hline
  \multicolumn{6}{l}{\commenta BJD$-$2400000.} \\
  \multicolumn{6}{l}{\commentb Against max $= 2457722.4878 + 0.069906 E$.} \\
  \multicolumn{6}{l}{\commentc Number of points used to determine the maximum.} \\
\end{tabular}
\end{center}
\end{table}

\subsection{NSV 14681}\label{obj:nsv14681}

   NSV 14681 was discovered as a variable star (SVS 749)
of unknown type with a photographic range of 14 to
fainter than 14.5 \citep{bel36nsv14681}.
The CRTS team detected an outburst
at an unfiltered CCD magnitude of 15.6 on 2007 June 13
and it was readily identified with NSV 14681
\citep{dra14CRTSCVs}.  The CV is a fainter component
of a close pair \citep{kat12DNSDSS}.
The CRTS team detected another outburst at 16.4 mag
on 2009 September 14.

\begin{figure}
  \begin{center}
    \FigureFile(85mm,110mm){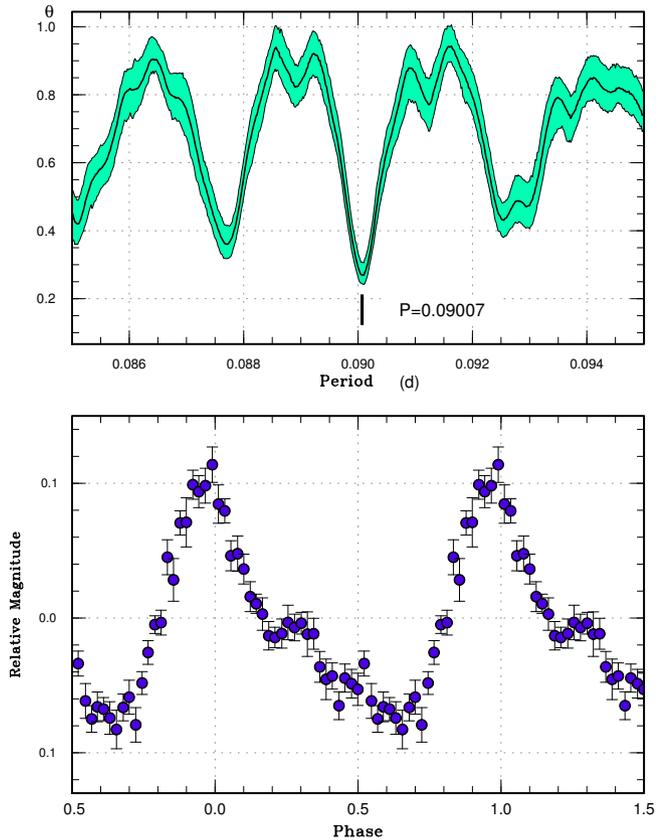}
  \end{center}
  \caption{Superhumps in NSV 14681 (2016).
     (Upper): PDM analysis.
     (Lower): Phase-averaged profile.}
  \label{fig:nsv14681shpdm}
\end{figure}

   The 2016 outburst was detected by the ASAS-SN team
at $V$=14.35 on October 19.  Subsequent observations
detected superhumps (vsnet-alert 20245, 20256;
figure \ref{fig:nsv14681shpdm}).
The times of superhump maxima are listed in
table \ref{tab:nsv14681oc2016}.
The superhump stage is unknown.  The object is on
the lower edge of the period gap.


\begin{table}
\caption{Superhump maxima of NSV 14681 (2016)}\label{tab:nsv14681oc2016}
\begin{center}
\begin{tabular}{rp{55pt}p{40pt}r@{.}lr}
\hline
\multicolumn{1}{c}{$E$} & \multicolumn{1}{c}{max\commenta} & \multicolumn{1}{c}{error} & \multicolumn{2}{c}{$O-C$\commentb} & \multicolumn{1}{c}{$N$\commentc} \\
\hline
0 & 57684.5223 & 0.0004 & $-$0&0002 & 90 \\
34 & 57687.5852 & 0.0005 & 0&0005 & 90 \\
35 & 57687.6745 & 0.0013 & $-$0&0002 & 59 \\
77 & 57691.4572 & 0.0010 & $-$0&0001 & 76 \\
\hline
  \multicolumn{6}{l}{\commenta BJD$-$2400000.} \\
  \multicolumn{6}{l}{\commentb Against max $= 2457684.5225 + 0.090063 E$.} \\
  \multicolumn{6}{l}{\commentc Number of points used to determine the maximum.} \\
\end{tabular}
\end{center}
\end{table}

\subsection{1RXS J161659.5$+$620014}\label{obj:j1616}

   This object (hereafter 1RXS J161659) was initially
identified as an X-ray selected variable
(also given a name as MASTER OT J161700.81$+$620024.9),
which was first detected in bright state on
2012 September 11 at an unfiltered CCD magnitude
of 14.4 \citep{bal13j1616}.  The dwarf nova-type
variability was confirmed by analysis of the CRTS data
(\cite{bal13j1616}; see also vsnet-alert 16079, 16720).

\begin{figure}
  \begin{center}
    \FigureFile(85mm,110mm){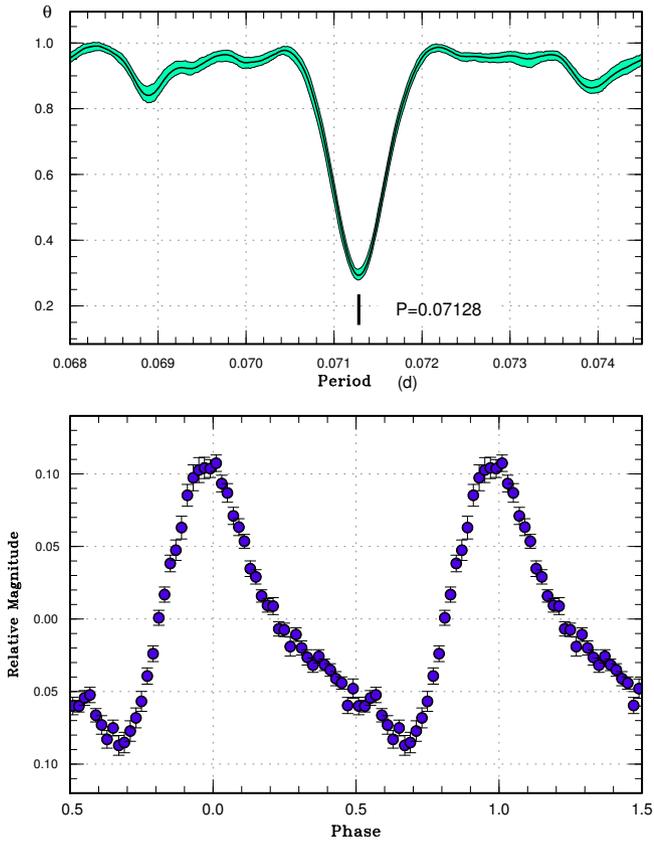}
  \end{center}
  \caption{Superhumps in 1RXS J161659 during the superoutburst
     plateau (2016).
     (Upper): PDM analysis.
     (Lower): Phase-averaged profile.}
  \label{fig:j1616shpdm}
\end{figure}

   The 2016 April outburst was detected by the ASAS-SN
team at $V$=14.74 on April 22.  Subsequent observations
detected superhumps (vsnet-alert 19763, 19765, 19772;
figure \ref{fig:j1616shpdm}).
The times of superhump maxima are listed in
table \ref{tab:j1616oc2016}.
The nature of the humps for $E \ge$155 (post-superoutburst)
is unclear due to the gap in the observation.
These humps may be either traditional late superhumps
or the extension of stage C superhumps (if it is
the case, the cycle count should be increased by one).
We consider the latter possibility less likely,
since this interpretation requires the period of
stage C superhumps to be 0.07065(2)~d, which appears
to be too short (by $\sim$1\%) shorter than
that of stage B superhumps.
We do not use these maxima in obtaining the periods
in table \ref{tab:perlist}.

\begin{figure}
  \begin{center}
    \FigureFile(88mm,70mm){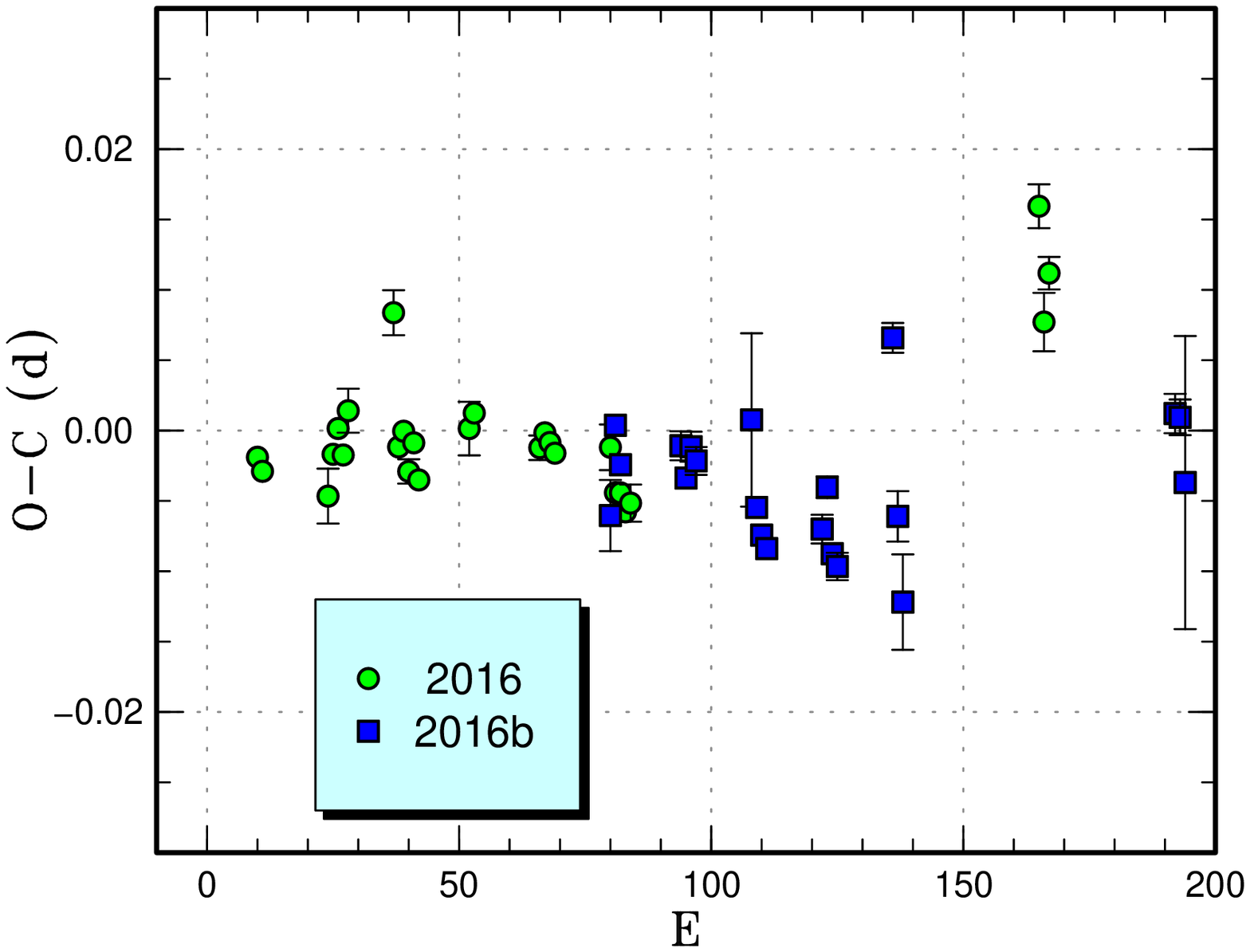}
  \end{center}
  \caption{Comparison of $O-C$ diagrams of 1RXS J161659
  between different superoutbursts.
  A period of 0.07130~d was used to draw this figure.
  Approximate cycle counts ($E$) after the start of the superoutburst
  were used.
  }
  \label{fig:j1616comp}
\end{figure}

   The 2016 July outburst was detected by the CRTS
team at an unfiltered CCD magnitude of 14.63 on
July 10 (cf. vsnet-alert 19970).  Although it was
considered to be too early for a next superoutburst, subsequent
observations detected superhumps (vsnet-alert 19996).
The times of superhump maxima are listed in
table \ref{tab:j1616oc2016b}.  As in the superoutburst
in 2016 April, the nature of maxima for $E \ge$112
(post-superoutburst) was unclear.
A comparison of $O-C$ diagrams between two superoutbursts
is given in figure \ref{fig:j1616comp}.

   These observations indicate that the supercycle
is only $\sim$80~d.  We studied past ASAS-SN observations
and detected outbursts (table \ref{tab:j1616out}).
The outburst pattern became more regular since
the 2015 July (it may have been due to the change
in the variability in this system or the improvement
of observations in ASAS-SN) and we obtained
a mean supercycle of 89(1)~d from five most recent
superoutbursts (with $|O-C|$ values less than 8~d). 
Despite the shortness of the supercycle, normal
outbursts are not as frequent as in ER UMa-type
dwarf novae (\cite{kat95eruma}; \cite{rob95eruma})
or active SU UMa-type dwarf novae, such as SS UMi
(\cite{kat00ssumi}; \cite{ole06ssumi}) and BF Ara
\citep{kat01bfara}.  The object resembles V503 Cyg
with a supercycle of 89~d with a few normal outbursts
between superoutbursts \citep{har95v503cyg}.
V503 Cyg is known to show different states
\citep{kat02v503cyg}, which is now considered to
be a result of the disk tilt suppressing normal
outbursts (\cite{ohs12eruma}; \cite{osa13v1504cygKepler};
\cite{osa13v344lyrv1504cyg}).  A search for
negative superhumps in 1RXS J161659 would be
fruitful.


\begin{table}
\caption{Superhump maxima of 1RXS J161659 (2016)}\label{tab:j1616oc2016}
\begin{center}
\begin{tabular}{rp{55pt}p{40pt}r@{.}lr}
\hline
\multicolumn{1}{c}{$E$} & \multicolumn{1}{c}{max\commenta} & \multicolumn{1}{c}{error} & \multicolumn{2}{c}{$O-C$\commentb} & \multicolumn{1}{c}{$N$\commentc} \\
\hline
0 & 57502.3747 & 0.0003 & 0&0019 & 72 \\
1 & 57502.4450 & 0.0004 & 0&0008 & 72 \\
14 & 57503.3701 & 0.0020 & $-$0&0019 & 26 \\
15 & 57503.4444 & 0.0004 & 0&0010 & 64 \\
16 & 57503.5175 & 0.0007 & 0&0028 & 65 \\
17 & 57503.5870 & 0.0007 & 0&0008 & 70 \\
18 & 57503.6614 & 0.0016 & 0&0039 & 41 \\
27 & 57504.3101 & 0.0016 & 0&0102 & 47 \\
28 & 57504.3718 & 0.0005 & 0&0006 & 127 \\
29 & 57504.4442 & 0.0005 & 0&0016 & 149 \\
30 & 57504.5127 & 0.0009 & $-$0&0013 & 68 \\
31 & 57504.5860 & 0.0007 & 0&0007 & 70 \\
32 & 57504.6547 & 0.0006 & $-$0&0021 & 50 \\
42 & 57505.3713 & 0.0019 & 0&0009 & 37 \\
43 & 57505.4437 & 0.0006 & 0&0019 & 76 \\
56 & 57506.3682 & 0.0009 & $-$0&0015 & 74 \\
57 & 57506.4405 & 0.0004 & $-$0&0005 & 149 \\
58 & 57506.5111 & 0.0004 & $-$0&0013 & 92 \\
59 & 57506.5817 & 0.0005 & $-$0&0021 & 68 \\
70 & 57507.3664 & 0.0016 & $-$0&0025 & 50 \\
71 & 57507.4345 & 0.0005 & $-$0&0058 & 90 \\
72 & 57507.5058 & 0.0005 & $-$0&0059 & 83 \\
73 & 57507.5757 & 0.0005 & $-$0&0073 & 79 \\
74 & 57507.6476 & 0.0013 & $-$0&0068 & 47 \\
155 & 57513.4440 & 0.0016 & 0&0084 & 27 \\
156 & 57513.5071 & 0.0021 & 0&0001 & 32 \\
157 & 57513.5819 & 0.0012 & 0&0035 & 36 \\
\hline
  \multicolumn{6}{l}{\commenta BJD$-$2400000.} \\
  \multicolumn{6}{l}{\commentb Against max $= 2457502.3728 + 0.071373 E$.} \\
  \multicolumn{6}{l}{\commentc Number of points used to determine the maximum.} \\
\end{tabular}
\end{center}
\end{table}


\begin{table}
\caption{Superhump maxima of 1RXS J161659 (2016b)}\label{tab:j1616oc2016b}
\begin{center}
\begin{tabular}{rp{55pt}p{40pt}r@{.}lr}
\hline
\multicolumn{1}{c}{$E$} & \multicolumn{1}{c}{max\commenta} & \multicolumn{1}{c}{error} & \multicolumn{2}{c}{$O-C$\commentb} & \multicolumn{1}{c}{$N$\commentc} \\
\hline
0 & 57585.3665 & 0.0025 & 0&0033 & 31 \\
1 & 57585.4442 & 0.0005 & 0&0095 & 69 \\
2 & 57585.5127 & 0.0007 & 0&0065 & 67 \\
14 & 57586.3697 & 0.0010 & 0&0059 & 47 \\
15 & 57586.4387 & 0.0006 & 0&0035 & 68 \\
16 & 57586.5122 & 0.0011 & 0&0055 & 67 \\
17 & 57586.5825 & 0.0010 & 0&0044 & 46 \\
28 & 57587.3697 & 0.0062 & 0&0055 & 25 \\
29 & 57587.4348 & 0.0005 & $-$0&0009 & 66 \\
30 & 57587.5041 & 0.0006 & $-$0&0031 & 71 \\
31 & 57587.5745 & 0.0007 & $-$0&0042 & 46 \\
42 & 57588.3602 & 0.0010 & $-$0&0046 & 48 \\
43 & 57588.4344 & 0.0008 & $-$0&0018 & 130 \\
44 & 57588.5010 & 0.0007 & $-$0&0067 & 138 \\
45 & 57588.5714 & 0.0010 & $-$0&0077 & 79 \\
56 & 57589.3720 & 0.0011 & 0&0067 & 30 \\
57 & 57589.4306 & 0.0018 & $-$0&0062 & 35 \\
58 & 57589.4958 & 0.0034 & $-$0&0124 & 34 \\
112 & 57593.3594 & 0.0014 & $-$0&0079 & 34 \\
113 & 57593.4304 & 0.0013 & $-$0&0083 & 38 \\
114 & 57593.4971 & 0.0104 & $-$0&0131 & 37 \\
126 & 57594.3679 & 0.0008 & 0&0001 & 38 \\
127 & 57594.4345 & 0.0012 & $-$0&0048 & 38 \\
128 & 57594.5075 & 0.0023 & $-$0&0031 & 19 \\
168 & 57597.4031 & 0.0040 & 0&0339 & 39 \\
\hline
  \multicolumn{6}{l}{\commenta BJD$-$2400000.} \\
  \multicolumn{6}{l}{\commentb Against max $= 2457585.3633 + 0.071464 E$.} \\
  \multicolumn{6}{l}{\commentc Number of points used to determine the maximum.} \\
\end{tabular}
\end{center}
\end{table}


\begin{table*}
\caption{List of outbursts of 1RXS J161659}\label{tab:j1616out}
\begin{center}
\begin{tabular}{cccccc}
\hline
Year & Month & Day & max\commenta & $V$-mag & type \\
\hline
2013 &  4 & 25 & 56408 & 15.02 & super \\
2013 &  7 & 12 & 56486 & 15.26 & ? \\
2013 &  8 & 21 & 56526 & 15.20 & super \\
2014 &  6 & 17 & 56826 & 15.44 & normal? \\
2014 &  8 &  4 & 56874 & 15.01 & super \\
2014 &  9 & 16 & 56917 & 15.62 & ? \\
2015 &  3 & 18 & 57100 & 15.41 & normal \\
2015 &  4 &  7 & 57120 & 15.14 & super \\
2015 &  5 &  6 & 57149 & 15.89 & normal? \\
2015 &  6 & 27 & 57201 & 15.56 & normal \\
2015 &  7 & 18 & 57222 & 15.27 & normal \\
2015 &  7 & 31 & 57235 & 14.64 & super \\
2016 &  1 &  7 & 57395 & 15.58 & normal \\
2016 &  1 & 24 & 57412 & 14.61 & super \\
2016 &  2 &  9 & 57428 & 15.84 & normal \\
2016 &  2 & 23 & 57442 & 15.58 & normal \\
2016 &  3 &  2 & 57450 & 15.66 & normal \\
2016 &  4 & 21 & 57500 & 14.66 & super \\
2016 &  6 & 26 & 57566 & 16.16 & normal \\
2016 &  7 & 11 & 57581 & 14.86 & super \\
2016 &  7 & 26 & 57596 & 15.75 & normal \\
2017 &  1 & 17 & 57771 & 14.84 & super \\
\hline
  \multicolumn{6}{l}{\commenta JD$-$2400000.} \\
\end{tabular}
\end{center}
\end{table*}

\subsection{ASASSN-13ak}\label{obj:asassn13ak}

   This object was detected as a transient at $V$=15.4
on 2013 May 23 by the ASAS-SN team \citep{sta13asassn13akatel5082}.
There was an independent detection by the MASTER network
\citep{shu13asassn13akatel5083}.  The SU UMa-type
nature was identified during the 2015 superoutburst
\citep{Pdot8}.

   The 2016 superoutburst was detected by the ASAS-SN
team at $V$=14.43 on August 2.  We obtained time-resolved
observations on two nights, yielding superhump maxima
in table \ref{tab:asassn13akoc2016}.
The resultant period is longer than that in 2015 \citep{Pdot8} and
these superhumps may have been stage A ones, despite
that the amplitudes were already large since
the 2016 observations were obtained in the earlier phase
than in 2015.


\begin{table}
\caption{Superhump maxima of ASASSN-13ak (2016)}\label{tab:asassn13akoc2016}
\begin{center}
\begin{tabular}{rp{55pt}p{40pt}r@{.}lr}
\hline
\multicolumn{1}{c}{$E$} & \multicolumn{1}{c}{max\commenta} & \multicolumn{1}{c}{error} & \multicolumn{2}{c}{$O-C$\commentb} & \multicolumn{1}{c}{$N$\commentc} \\
\hline
0 & 57604.4248 & 0.0004 & $-$0&0019 & 97 \\
1 & 57604.5177 & 0.0004 & 0&0021 & 35 \\
8 & 57605.1371 & 0.0010 & $-$0&0003 & 88 \\
\hline
  \multicolumn{6}{l}{\commenta BJD$-$2400000.} \\
  \multicolumn{6}{l}{\commentb Against max $= 2457604.4267 + 0.088838 E$.} \\
  \multicolumn{6}{l}{\commentc Number of points used to determine the maximum.} \\
\end{tabular}
\end{center}
\end{table}

\subsection{ASASSN-13al}\label{obj:asassn13al}

   This object was detected as a transient
at $V$=15.2 on 2013 June 1 by the ASAS-SN team
\citep{pri13asassn13clatel5102} (The ASAS-SN 
Transients page gave a magnitude of 16.03 with
a ``BADCAL'' flag).  The CV-type nature was confirmed
by spectroscopy.\footnote{
  $<$http://www.astronomy.ohio-state.edu/$\sim$assassin/followup/spec\_asassn13al.png.$>$
}

\begin{figure}
  \begin{center}
    \FigureFile(85mm,110mm){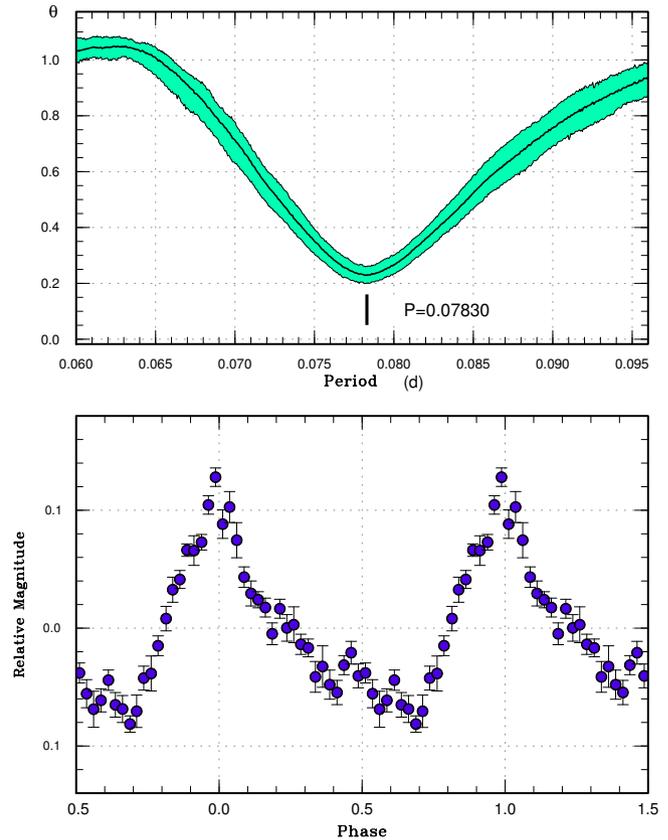}
  \end{center}
  \caption{Superhumps in ASASSN-13al (2016).
     (Upper): PDM analysis.
     (Lower): Phase-averaged profile.}
  \label{fig:asassn13alshpdm}
\end{figure}

   The 2016 outburst was detected by the ASAS-SN team
at $V$=14.77 on October 9.  There was a previous
detection in the ASAS-SN data at $V$=14.75 on
2012 June 7.
Subsequent observations detected superhumps
(vsnet-alert 20220; figure \ref{fig:asassn13alshpdm}).
The time of superhump maxima are listed in
table \ref{tab:asassn13cloc2016}.  The period was
not very well determined since the observations
were undertaken only on a single night.
The best superhump period by the PDM method is
0.0783(2)~d.


\begin{table}
\caption{Superhump maxima of ASASSN-13al (2016)}\label{tab:asassn13cloc2016}
\begin{center}
\begin{tabular}{rp{55pt}p{40pt}r@{.}lr}
\hline
\multicolumn{1}{c}{$E$} & \multicolumn{1}{c}{max\commenta} & \multicolumn{1}{c}{error} & \multicolumn{2}{c}{$O-C$\commentb} & \multicolumn{1}{c}{$N$\commentc} \\
\hline
0 & 57672.4399 & 0.0007 & 0&0004 & 79 \\
1 & 57672.5178 & 0.0006 & $-$0&0001 & 80 \\
2 & 57672.5954 & 0.0007 & $-$0&0009 & 78 \\
3 & 57672.6755 & 0.0008 & 0&0007 & 56 \\
\hline
  \multicolumn{6}{l}{\commenta BJD$-$2400000.} \\
  \multicolumn{6}{l}{\commentb Against max $= 2457672.4395 + 0.078452 E$.} \\
  \multicolumn{6}{l}{\commentc Number of points used to determine the maximum.} \\
\end{tabular}
\end{center}
\end{table}

\subsection{ASASSN-13bc}\label{obj:asassn13bc}

   This object was detected as a transient
at $V$=16.9 on 2013 July 4 by the ASAS-SN team.
A number of past outbursts were recorded in
the CRTS data.

   The 2015 outburst was detected by the ASAS-SN team
at $V$=14.83 on July 30.  Subsequent observations
detected superhumps (vsnet-alert 18921, 18930).
The times of superhump maxima are listed in
table \ref{tab:asasn13bcoc2015}.

\begin{figure}
  \begin{center}
    \FigureFile(85mm,110mm){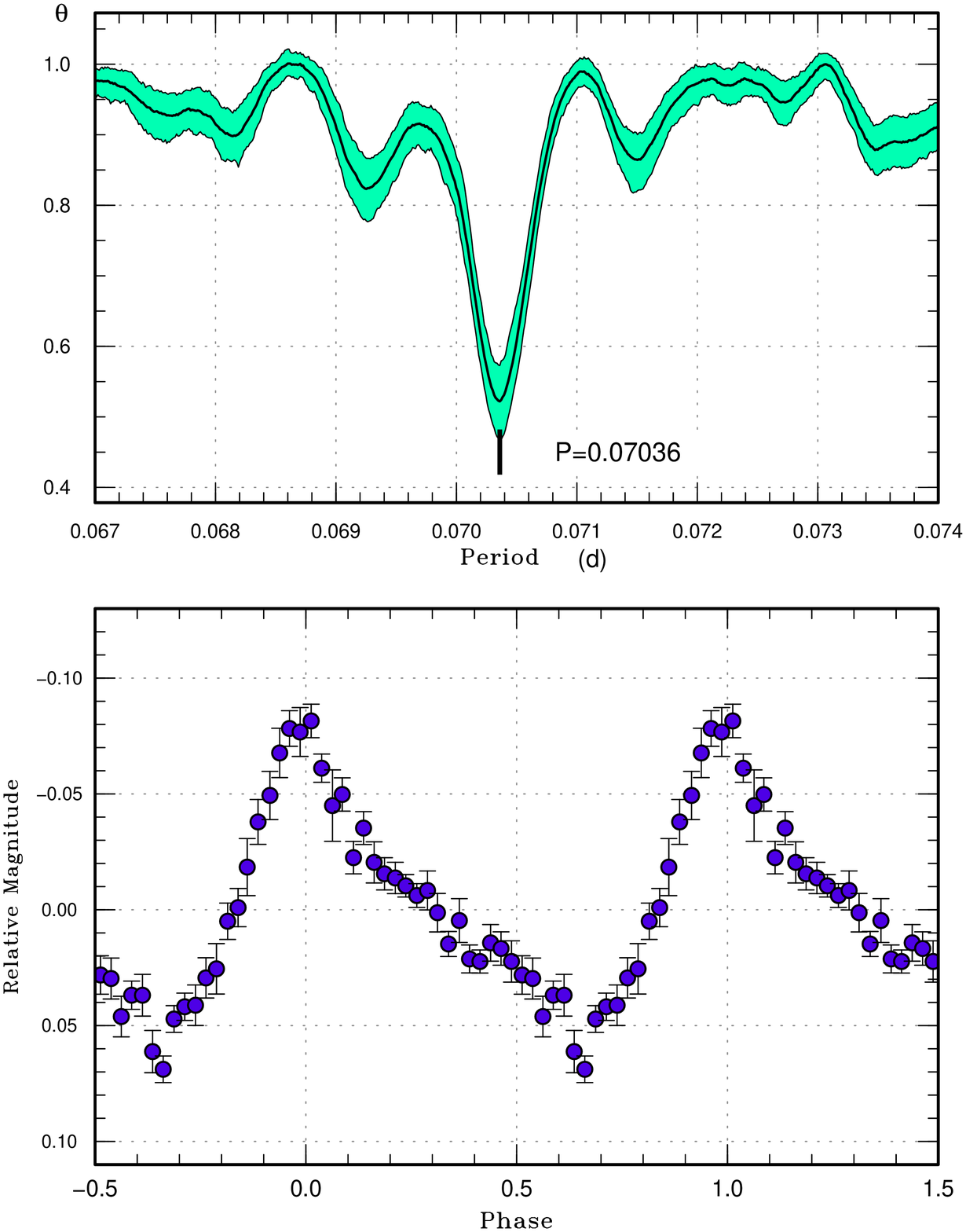}
  \end{center}
  \caption{Superhumps in ASASSN-13bc during the superoutburst
     plateau (2016).
     (Upper): PDM analysis.
     (Lower): Phase-averaged profile.}
  \label{fig:asassn13bcshpdm}
\end{figure}

\begin{figure}
  \begin{center}
    \FigureFile(88mm,70mm){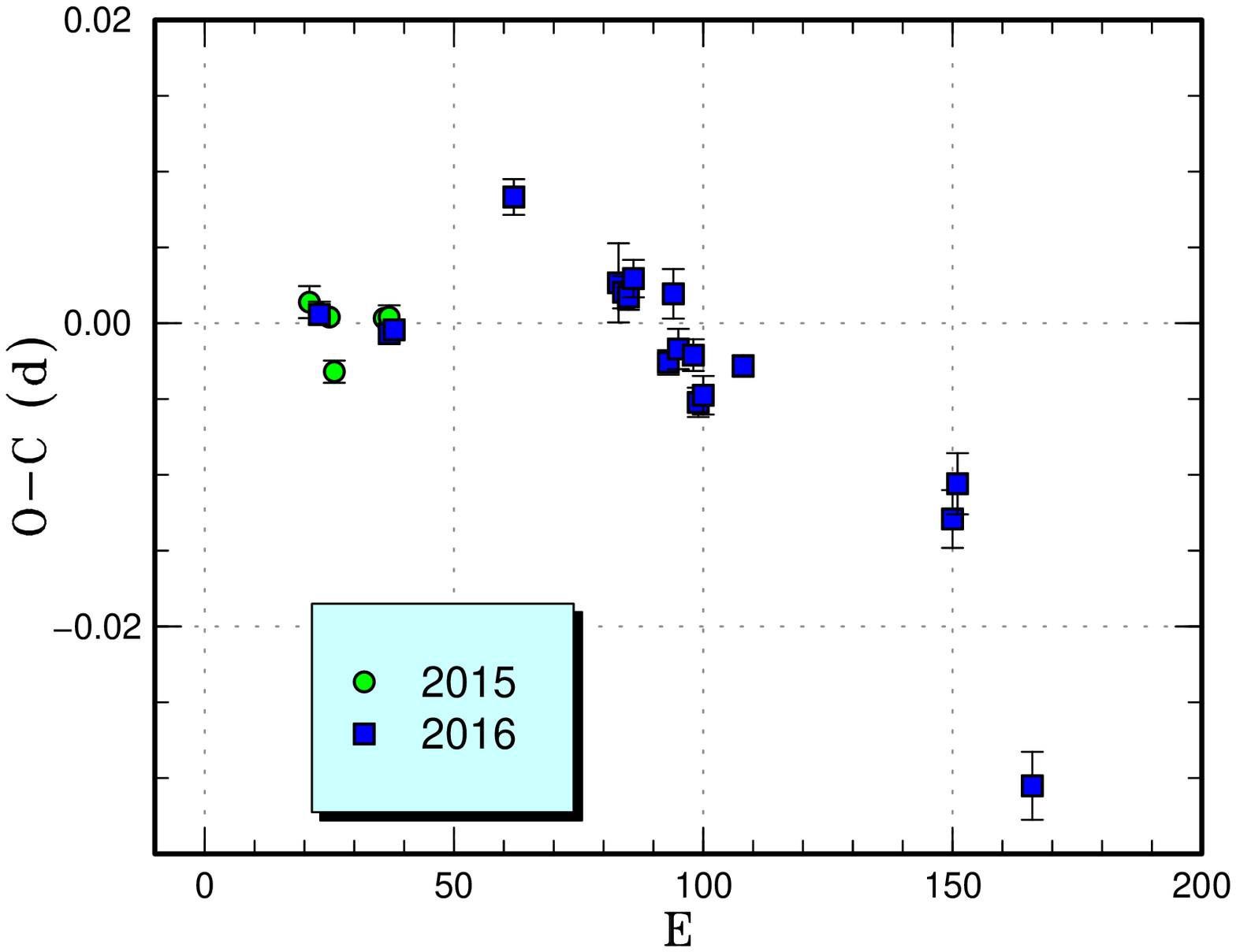}
  \end{center}
  \caption{Comparison of $O-C$ diagrams of ASASSN-13bc
  between different superoutbursts.
  A period of 0.07040~d was used to draw this figure.
  Approximate cycle counts ($E$) after the start of the superoutburst
  were used.
  }
  \label{fig:asassn13bccomp}
\end{figure}

   The 2016 outburst was detected by the ASAS-SN team
at $V$=15.19 on May 24.  Superhumps were also observed
(vsnet-alert 19843, 19867).
The object underwent a post-superoutburst rebrightening
at $V$=16.13 on June 9 (cf. vsnet-alert 19883).
The times of superhump maxima are listed in
table \ref{tab:asasn13bcoc2016}.
The superhump profile is given for the better observed
2016 one (figure \ref{fig:asassn13bcshpdm}).
A combined $O-C$ diagram (figure \ref{fig:asassn13bccomp})
suggests that the 2015 observations covered
the early phase of stage B and the 2016 ones recorded
both stages B and C, although the later part of
stage B was not well recorded due to the lack
of observations.


\begin{table}
\caption{Superhump maxima of ASASSN-13bc (2015)}\label{tab:asasn13bcoc2015}
\begin{center}
\begin{tabular}{rp{55pt}p{40pt}r@{.}lr}
\hline
\multicolumn{1}{c}{$E$} & \multicolumn{1}{c}{max\commenta} & \multicolumn{1}{c}{error} & \multicolumn{2}{c}{$O-C$\commentb} & \multicolumn{1}{c}{$N$\commentc} \\
\hline
0 & 57235.4745 & 0.0011 & 0&0013 & 79 \\
2 & 57235.6146 & 0.0007 & 0&0006 & 47 \\
4 & 57235.7551 & 0.0006 & 0&0004 & 407 \\
5 & 57235.8219 & 0.0007 & $-$0&0032 & 226 \\
15 & 57236.5294 & 0.0004 & 0&0004 & 138 \\
16 & 57236.5999 & 0.0008 & 0&0005 & 94 \\
\hline
  \multicolumn{6}{l}{\commenta BJD$-$2400000.} \\
  \multicolumn{6}{l}{\commentb Against max $= 2457235.4732 + 0.070393 E$.} \\
  \multicolumn{6}{l}{\commentc Number of points used to determine the maximum.} \\
\end{tabular}
\end{center}
\end{table}


\begin{table}
\caption{Superhump maxima of ASASSN-13bc (2016)}\label{tab:asasn13bcoc2016}
\begin{center}
\begin{tabular}{rp{55pt}p{40pt}r@{.}lr}
\hline
\multicolumn{1}{c}{$E$} & \multicolumn{1}{c}{max\commenta} & \multicolumn{1}{c}{error} & \multicolumn{2}{c}{$O-C$\commentb} & \multicolumn{1}{c}{$N$\commentc} \\
\hline
0 & 57534.4878 & 0.0006 & $-$0&0109 & 21 \\
14 & 57535.4722 & 0.0004 & $-$0&0090 & 80 \\
15 & 57535.5428 & 0.0005 & $-$0&0085 & 90 \\
39 & 57537.2412 & 0.0012 & 0&0057 & 77 \\
60 & 57538.7139 & 0.0026 & 0&0048 & 59 \\
61 & 57538.7837 & 0.0011 & 0&0044 & 54 \\
62 & 57538.8538 & 0.0008 & 0&0043 & 54 \\
63 & 57538.9254 & 0.0012 & 0&0058 & 40 \\
70 & 57539.4127 & 0.0008 & 0&0018 & 35 \\
71 & 57539.4876 & 0.0016 & 0&0066 & 28 \\
72 & 57539.5544 & 0.0013 & 0&0032 & 17 \\
75 & 57539.7652 & 0.0010 & 0&0034 & 56 \\
76 & 57539.8324 & 0.0010 & 0&0005 & 53 \\
77 & 57539.9033 & 0.0013 & 0&0012 & 39 \\
85 & 57540.4684 & 0.0007 & 0&0050 & 63 \\
127 & 57543.4152 & 0.0019 & 0&0044 & 41 \\
128 & 57543.4879 & 0.0020 & 0&0069 & 17 \\
142 & 57544.4432 & 0.0011 & $-$0&0202 & 28 \\
143 & 57544.5240 & 0.0022 & $-$0&0096 & 21 \\
\hline
  \multicolumn{6}{l}{\commenta BJD$-$2400000.} \\
  \multicolumn{6}{l}{\commentb Against max $= 2457534.4987 + 0.070174 E$.} \\
  \multicolumn{6}{l}{\commentc Number of points used to determine the maximum.} \\
\end{tabular}
\end{center}
\end{table}

\subsection{ASASSN-13bj}\label{obj:asassn13bj}

   This object was detected as a transient
at $V$=16.2 on 2013 July 10 by the ASAS-SN team.
Two superhump maxima were obtained during the 2013
superoutburst \citep{Pdot5}.

\begin{figure}
  \begin{center}
    \FigureFile(85mm,110mm){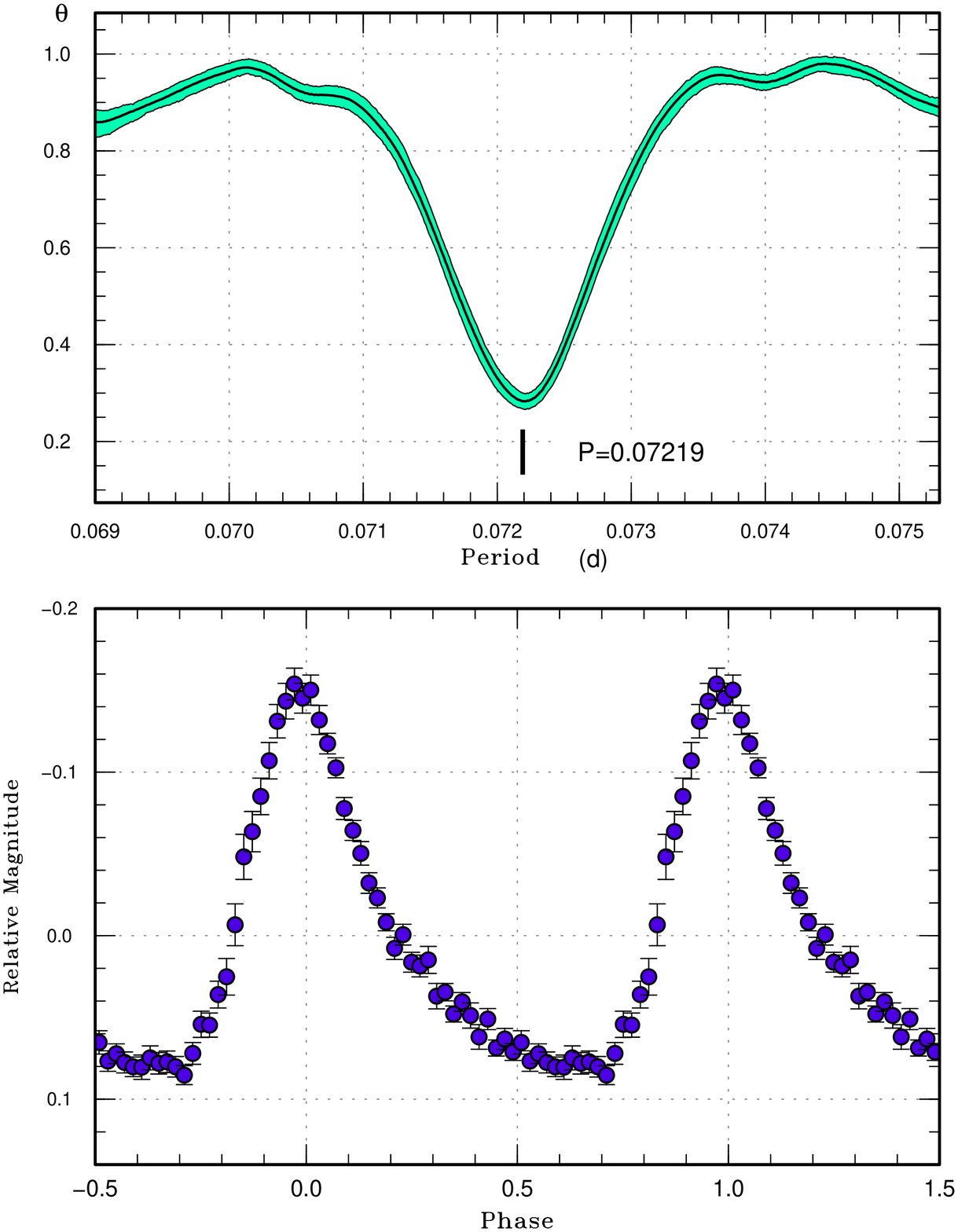}
  \end{center}
  \caption{Superhumps in ASASSN-13bj (2016).
     (Upper): PDM analysis.
     (Lower): Phase-averaged profile.}
  \label{fig:asassn13bjshpdm}
\end{figure}

   The 2016 superoutburst was detected by the ASAS-SN
team at $V$=14.98 on July 3.  Subsequent observations
detected superhumps (vsnet-alert 19957, 19965, 19975;
figure \ref{fig:asassn13bjshpdm}).
The times of superhump maxima are listed in
table \ref{tab:asassn13bjoc2016}.
There was a marked decrease in the superhump period
and we tentatively identified a stage B-C transition
around $E=$22.  The accuracy of the resultant periods
was not sufficiently high since they were determined
by only short baselines.


\begin{table}
\caption{Superhump maxima of ASASSN-13bj (2016)}\label{tab:asassn13bjoc2016}
\begin{center}
\begin{tabular}{rp{55pt}p{40pt}r@{.}lr}
\hline
\multicolumn{1}{c}{$E$} & \multicolumn{1}{c}{max\commenta} & \multicolumn{1}{c}{error} & \multicolumn{2}{c}{$O-C$\commentb} & \multicolumn{1}{c}{$N$\commentc} \\
\hline
0 & 57574.3923 & 0.0003 & $-$0&0048 & 41 \\
1 & 57574.4669 & 0.0004 & $-$0&0024 & 50 \\
2 & 57574.5373 & 0.0007 & $-$0&0043 & 49 \\
15 & 57575.4823 & 0.0003 & 0&0017 & 71 \\
16 & 57575.5555 & 0.0004 & 0&0026 & 68 \\
19 & 57575.7716 & 0.0005 & 0&0021 & 44 \\
20 & 57575.8436 & 0.0006 & 0&0018 & 45 \\
21 & 57575.9157 & 0.0006 & 0&0016 & 46 \\
23 & 57576.0605 & 0.0005 & 0&0020 & 77 \\
24 & 57576.1313 & 0.0006 & 0&0006 & 77 \\
25 & 57576.2077 & 0.0017 & 0&0047 & 35 \\
27 & 57576.3482 & 0.0003 & 0&0007 & 117 \\
28 & 57576.4226 & 0.0003 & 0&0030 & 170 \\
29 & 57576.4943 & 0.0006 & 0&0023 & 112 \\
30 & 57576.5656 & 0.0008 & 0&0015 & 72 \\
41 & 57577.3560 & 0.0007 & $-$0&0028 & 45 \\
42 & 57577.4288 & 0.0005 & $-$0&0022 & 117 \\
43 & 57577.4992 & 0.0004 & $-$0&0040 & 119 \\
44 & 57577.5715 & 0.0007 & $-$0&0039 & 71 \\
\hline
  \multicolumn{6}{l}{\commenta BJD$-$2400000.} \\
  \multicolumn{6}{l}{\commentb Against max $= 2457574.3971 + 0.072236 E$.} \\
  \multicolumn{6}{l}{\commentc Number of points used to determine the maximum.} \\
\end{tabular}
\end{center}
\end{table}

\subsection{ASASSN-13bo}\label{obj:asassn13bo}

\begin{figure}
  \begin{center}
    \FigureFile(85mm,110mm){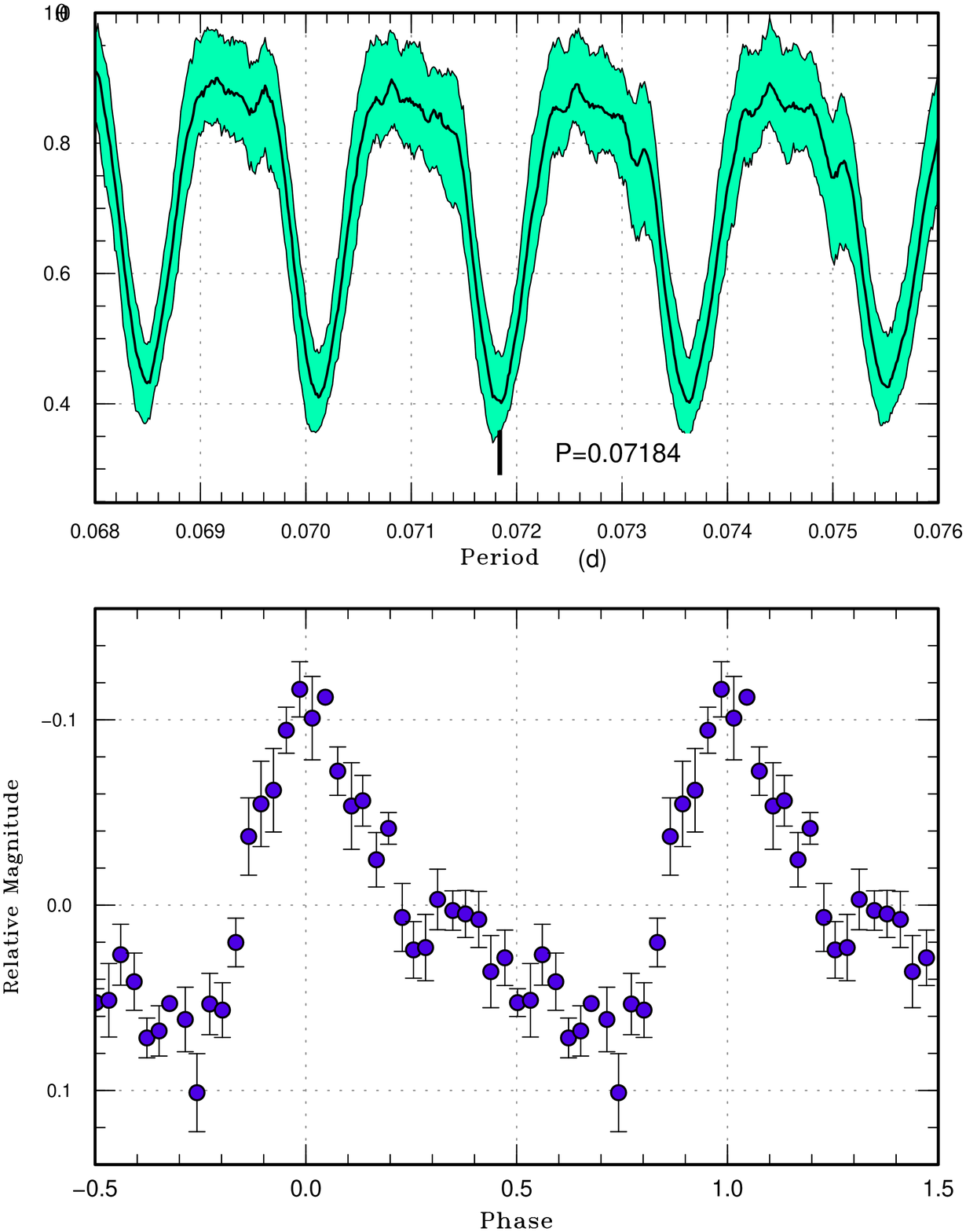}
  \end{center}
  \caption{Superhumps in ASASSN-13bo (2016).
     (Upper): PDM analysis.
     (Lower): Phase-averaged profile.}
  \label{fig:asassn13boshpdm}
\end{figure}

   This object was detected as a transient at $V$=15.96
on 2013 July 13 by the ASAS-SN team.
The 2016 outburst was detected by the ASAS-SN team
at $V$=15.19 on August 1.  The 2016 outburst was
the brightest recorded one and a superoutburst was
suspected.  Subsequent observations detected superhumps
(vsnet-alert 20053, 20068).
There was a 3-d gap in the observations and alias
periods are possible (figure \ref{fig:asassn13boshpdm}).
The period in table \ref{tab:perlist} refers to
the one giving smallest residuals and it was determined
by the PDM method.  The object faded to $\sim$20 mag
on August 13.


\begin{table}
\caption{Superhump maxima of ASASSN-13bo (2016)}\label{tab:asassn13booc2016}
\begin{center}
\begin{tabular}{rp{55pt}p{40pt}r@{.}lr}
\hline
\multicolumn{1}{c}{$E$} & \multicolumn{1}{c}{max\commenta} & \multicolumn{1}{c}{error} & \multicolumn{2}{c}{$O-C$\commentb} & \multicolumn{1}{c}{$N$\commentc} \\
\hline
0 & 57606.6548 & 0.0017 & 0&0000 & 28 \\
40 & 57609.5212 & 0.0009 & $-$0&0006 & 68 \\
41 & 57609.5941 & 0.0009 & 0&0006 & 70 \\
\hline
  \multicolumn{6}{l}{\commenta BJD$-$2400000.} \\
  \multicolumn{6}{l}{\commentb Against max $= 2457606.6548 + 0.071675 E$.} \\
  \multicolumn{6}{l}{\commentc Number of points used to determine the maximum.} \\
\end{tabular}
\end{center}
\end{table}

\subsection{ASASSN-13cs}\label{obj:asassn13cs}

   This object was detected as a transient
at $V$=14.9 on 2013 September 2 by the ASAS-SN team.
The object was spectroscopically identified
as a dwarf nova in outburst.\footnote{
  $<$http://www.astronomy.ohio-state.edu/$\sim$assassin/followup/spec\_asassn13cs.png$>$.
}

\begin{figure}
  \begin{center}
    \FigureFile(85mm,110mm){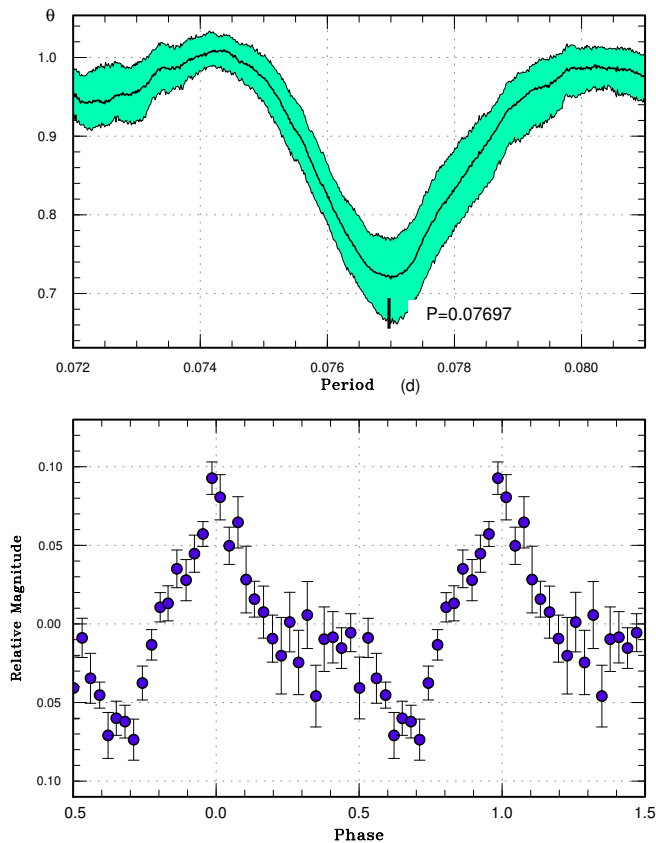}
  \end{center}
  \caption{Superhumps in ASASSN-13cs (2016).
     (Upper): PDM analysis.
     (Lower): Phase-averaged profile.}
  \label{fig:asassn13csshpdm}
\end{figure}

   The 2016 outburst was detected by the ASAS-SN
team at $V$=15.03 on June 21.  The light curves
of past outbursts suggested an SU UMa-type dwarf nova.
Subsequent observations detected superhumps
(vsnet-alert 19920, 19923; figure \ref{fig:asassn13csshpdm}).
The times of superhump maxima are listed in
table \ref{tab:asassn13csoc2016}.


\begin{table}
\caption{Superhump maxima of ASASSN-13cs (2016)}\label{tab:asassn13csoc2016}
\begin{center}
\begin{tabular}{rp{55pt}p{40pt}r@{.}lr}
\hline
\multicolumn{1}{c}{$E$} & \multicolumn{1}{c}{max\commenta} & \multicolumn{1}{c}{error} & \multicolumn{2}{c}{$O-C$\commentb} & \multicolumn{1}{c}{$N$\commentc} \\
\hline
0 & 57562.4140 & 0.0027 & $-$0&0035 & 17 \\
1 & 57562.4950 & 0.0015 & 0&0003 & 36 \\
2 & 57562.5705 & 0.0017 & $-$0&0012 & 26 \\
4 & 57562.7279 & 0.0015 & 0&0020 & 35 \\
5 & 57562.8043 & 0.0039 & 0&0013 & 23 \\
8 & 57563.0344 & 0.0015 & 0&0000 & 167 \\
13 & 57563.4238 & 0.0017 & 0&0039 & 17 \\
14 & 57563.4957 & 0.0030 & $-$0&0013 & 22 \\
15 & 57563.5781 & 0.0042 & 0&0040 & 10 \\
17 & 57563.7267 & 0.0011 & $-$0&0016 & 45 \\
18 & 57563.8052 & 0.0015 & $-$0&0002 & 39 \\
19 & 57563.8825 & 0.0016 & 0&0000 & 45 \\
20 & 57563.9559 & 0.0031 & $-$0&0037 & 32 \\
\hline
  \multicolumn{6}{l}{\commenta BJD$-$2400000.} \\
  \multicolumn{6}{l}{\commentb Against max $= 2457562.4175 + 0.077105 E$.} \\
  \multicolumn{6}{l}{\commentc Number of points used to determine the maximum.} \\
\end{tabular}
\end{center}
\end{table}

\subsection{ASASSN-13cz}\label{obj:asassn13cz}

   This object was detected as a transient
at $V$=14.9 on 2013 September 14 by the ASAS-SN team.
The 2013 outburst turned out to be a superoutburst
by the detection of superhumps \citep{Pdot6}.

\begin{figure}
  \begin{center}
    \FigureFile(88mm,70mm){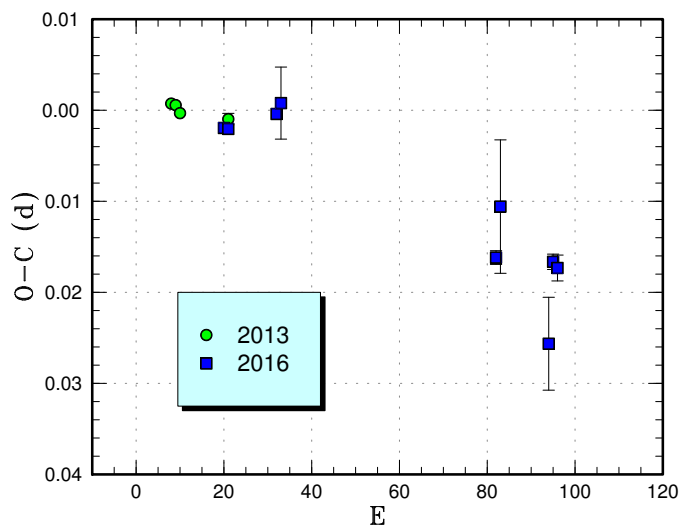}
  \end{center}
  \caption{Comparison of $O-C$ diagrams of ASASSN-13cz
  between different superoutbursts.
  A period of 0.07995~d was used to draw this figure.
  Approximate cycle counts ($E$) after the start of the superoutburst
  were used.
  }
  \label{fig:asassn13czcomp}
\end{figure}

\begin{figure}
  \begin{center}
    \FigureFile(85mm,110mm){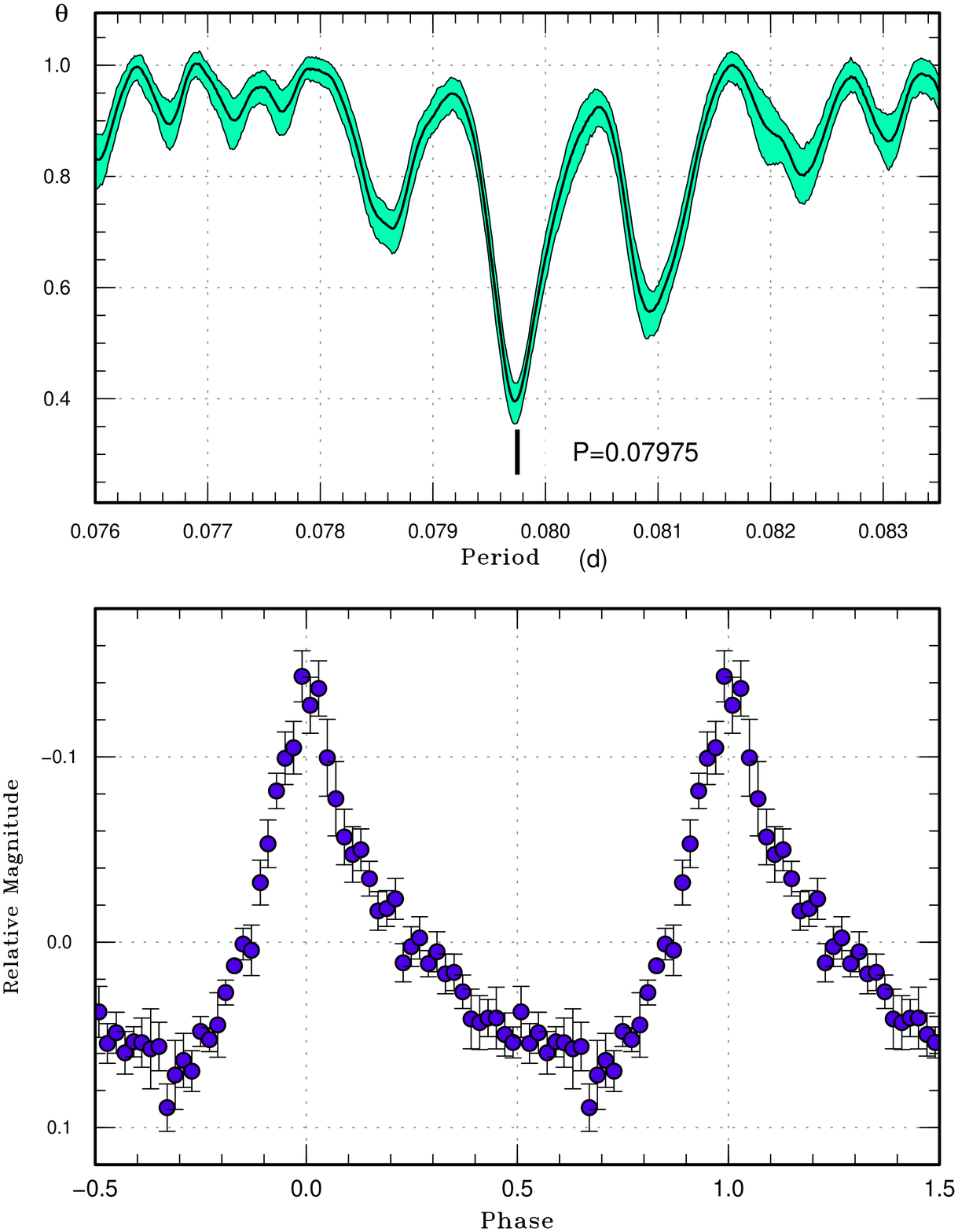}
  \end{center}
  \caption{Superhumps in ASASSN-13cz (2016).
     (Upper): PDM analysis.
     (Lower): Phase-averaged profile.}
  \label{fig:asassn13czshpdm}
\end{figure}

   The 2016 superoutburst was detected by the ASAS-SN
team at $V$=14.47 on July 27.  Subsequent observations
detected superhumps (vsnet-alert 20023, 20042).
The times of superhump maxima are listed in
table \ref{tab:asassn13czoc2016}.  We suspect that
stages B and C were partially observed
(cf. figure \ref{fig:asassn13czcomp}).
We provide a superhump profile (figure \ref{fig:asassn13czshpdm}),
which was determined much better than in 2013.

\begin{table}
\caption{Superhump maxima of ASASSN-13cz (2016)}\label{tab:asassn13czoc2016}
\begin{center}
\begin{tabular}{rp{55pt}p{40pt}r@{.}lr}
\hline
\multicolumn{1}{c}{$E$} & \multicolumn{1}{c}{max\commenta} & \multicolumn{1}{c}{error} & \multicolumn{2}{c}{$O-C$\commentb} & \multicolumn{1}{c}{$N$\commentc} \\
\hline
0 & 57598.4226 & 0.0003 & $-$0&0026 & 59 \\
1 & 57598.5024 & 0.0004 & $-$0&0024 & 73 \\
12 & 57599.3835 & 0.0004 & 0&0020 & 90 \\
13 & 57599.4647 & 0.0040 & 0&0034 & 15 \\
62 & 57603.3653 & 0.0008 & $-$0&0010 & 90 \\
63 & 57603.4508 & 0.0073 & 0&0048 & 28 \\
74 & 57604.3152 & 0.0051 & $-$0&0074 & 20 \\
75 & 57604.4041 & 0.0008 & 0&0018 & 131 \\
76 & 57604.4834 & 0.0014 & 0&0014 & 112 \\
\hline
  \multicolumn{6}{l}{\commenta BJD$-$2400000.} \\
  \multicolumn{6}{l}{\commentb Against max $= 2457598.4252 + 0.079695 E$.} \\
  \multicolumn{6}{l}{\commentc Number of points used to determine the maximum.} \\
\end{tabular}
\end{center}
\end{table}

\subsection{ASASSN-14gg}\label{obj:asassn14gg}

\begin{figure}
  \begin{center}
    \FigureFile(85mm,110mm){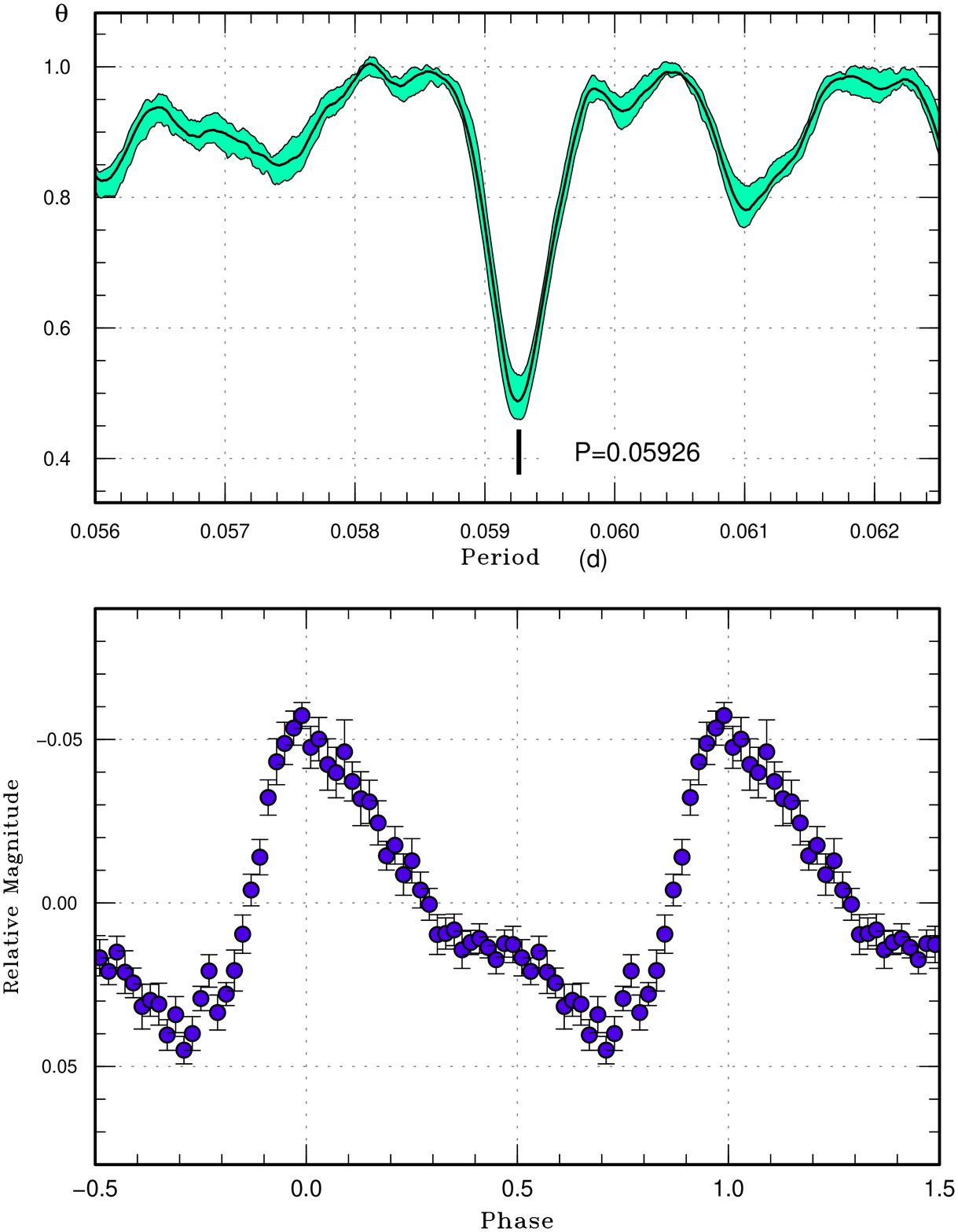}
  \end{center}
  \caption{Superhumps in ASASSN-14gg (2016).
     (Upper): PDM analysis.
     (Lower): Phase-averaged profile.}
  \label{fig:asassn14ggshpdm}
\end{figure}

   This object was detected as a transient
at $V$=14.8 on 2014 August 23 by the ASAS-SN team.
The 2016 outburst was detected by the ASAS-SN team
at $V$=13.95 on August 11.  Subsequent observations
detected superhumps (vsnet-alert 20079;
figure \ref{fig:asassn14ggshpdm}).
The times of superhump maxima are listed in
table \ref{tab:asassn14ggoc2016}.  The positive $P_{\rm dot}$
for stage B superhumps is a common feature
in many short-$P_{\rm SH}$ systems.


\begin{table}
\caption{Superhump maxima of ASASSN-14gg (2016)}\label{tab:asassn14ggoc2016}
\begin{center}
\begin{tabular}{rp{55pt}p{40pt}r@{.}lr}
\hline
\multicolumn{1}{c}{$E$} & \multicolumn{1}{c}{max\commenta} & \multicolumn{1}{c}{error} & \multicolumn{2}{c}{$O-C$\commentb} & \multicolumn{1}{c}{$N$\commentc} \\
\hline
0 & 57614.5584 & 0.0003 & 0&0054 & 59 \\
1 & 57614.6181 & 0.0004 & 0&0057 & 43 \\
31 & 57616.3897 & 0.0003 & $-$0&0020 & 56 \\
32 & 57616.4486 & 0.0003 & $-$0&0024 & 61 \\
33 & 57616.5077 & 0.0004 & $-$0&0026 & 61 \\
34 & 57616.5667 & 0.0005 & $-$0&0030 & 59 \\
35 & 57616.6274 & 0.0010 & $-$0&0015 & 34 \\
54 & 57617.7523 & 0.0006 & $-$0&0036 & 57 \\
55 & 57617.8131 & 0.0006 & $-$0&0021 & 59 \\
56 & 57617.8714 & 0.0010 & $-$0&0031 & 38 \\
66 & 57618.4646 & 0.0007 & $-$0&0030 & 61 \\
67 & 57618.5282 & 0.0011 & 0&0012 & 59 \\
68 & 57618.5824 & 0.0007 & $-$0&0039 & 60 \\
70 & 57618.7119 & 0.0070 & 0&0070 & 31 \\
71 & 57618.7615 & 0.0012 & $-$0&0027 & 58 \\
72 & 57618.8234 & 0.0014 & $-$0&0001 & 59 \\
87 & 57619.7210 & 0.0023 & 0&0078 & 47 \\
88 & 57619.7722 & 0.0013 & $-$0&0003 & 57 \\
89 & 57619.8351 & 0.0022 & 0&0033 & 59 \\
\hline
  \multicolumn{6}{l}{\commenta BJD$-$2400000.} \\
  \multicolumn{6}{l}{\commentb Against max $= 2457614.5531 + 0.059311 E$.} \\
  \multicolumn{6}{l}{\commentc Number of points used to determine the maximum.} \\
\end{tabular}
\end{center}
\end{table}

\subsection{ASASSN-15cr}\label{obj:asassn15cr}

\begin{figure}
  \begin{center}
    \FigureFile(85mm,110mm){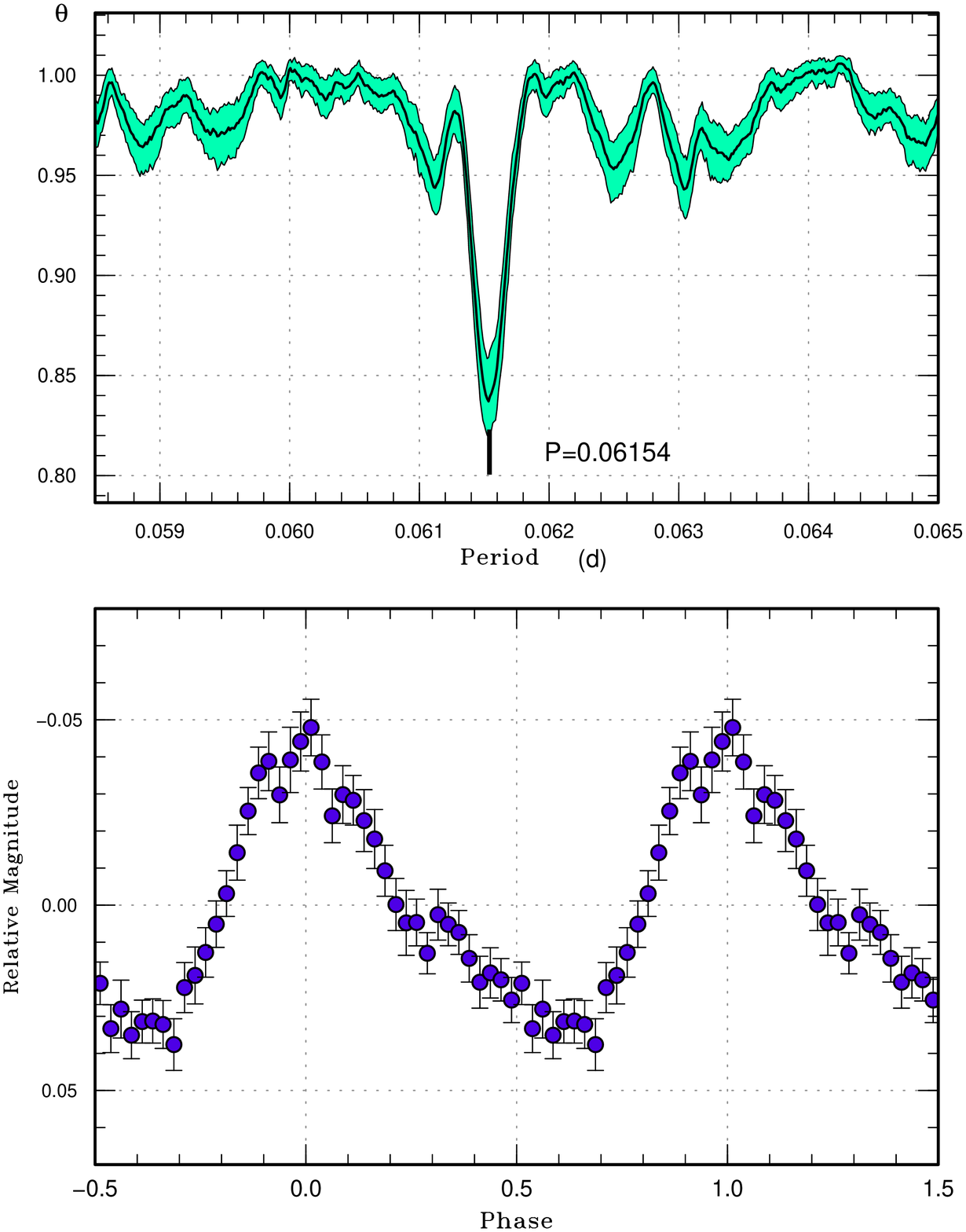}
  \end{center}
  \caption{Superhumps in ASASSN-15cr (2017).
     (Upper): PDM analysis.
     (Lower): Phase-averaged profile.}
  \label{fig:asassn15crshpdm}
\end{figure}

   This object was detected as a transient
at $V$=14.9 on 2015 February 7 by the ASAS-SN team.
The 2017 outburst was detected by the ASAS-SN team
at $V$=14.73 on January 9.
Subsequent observations detected superhumps
(vsnet-alert 20558, 20590; figure \ref{fig:asassn15crshpdm}).
The times of superhump maxima are listed in
table \ref{tab:asassn15croc2017}.
Despite that observational coverage was not
sufficient, all stages of A-C were recorded
thanks to the early detection by the ASAS-SN team.


\begin{table}
\caption{Superhump maxima of ASASSN-15cr (2017)}\label{tab:asassn15croc2017}
\begin{center}
\begin{tabular}{rp{55pt}p{40pt}r@{.}lr}
\hline
\multicolumn{1}{c}{$E$} & \multicolumn{1}{c}{max\commenta} & \multicolumn{1}{c}{error} & \multicolumn{2}{c}{$O-C$\commentb} & \multicolumn{1}{c}{$N$\commentc} \\
\hline
0 & 57764.2386 & 0.0018 & $-$0&0035 & 107 \\
1 & 57764.2945 & 0.0013 & $-$0&0091 & 130 \\
2 & 57764.3613 & 0.0010 & $-$0&0039 & 129 \\
3 & 57764.4223 & 0.0011 & $-$0&0044 & 95 \\
4 & 57764.4852 & 0.0012 & $-$0&0030 & 120 \\
15 & 57765.1730 & 0.0008 & 0&0079 & 35 \\
16 & 57765.2378 & 0.0004 & 0&0113 & 60 \\
46 & 57767.0734 & 0.0006 & 0&0010 & 53 \\
47 & 57767.1359 & 0.0010 & 0&0020 & 36 \\
50 & 57767.3198 & 0.0014 & 0&0013 & 60 \\
51 & 57767.3771 & 0.0068 & $-$0&0029 & 33 \\
65 & 57768.2382 & 0.0005 & $-$0&0032 & 62 \\
66 & 57768.2986 & 0.0020 & $-$0&0043 & 35 \\
80 & 57769.1604 & 0.0009 & $-$0&0040 & 64 \\
81 & 57769.2209 & 0.0010 & $-$0&0049 & 65 \\
82 & 57769.2877 & 0.0022 & 0&0003 & 18 \\
114 & 57771.2557 & 0.0007 & $-$0&0006 & 98 \\
130 & 57772.2419 & 0.0017 & 0&0012 & 64 \\
131 & 57772.3037 & 0.0011 & 0&0014 & 69 \\
132 & 57772.3703 & 0.0009 & 0&0066 & 63 \\
133 & 57772.4353 & 0.0013 & 0&0100 & 49 \\
134 & 57772.4886 & 0.0011 & 0&0018 & 59 \\
135 & 57772.5505 & 0.0013 & 0&0021 & 63 \\
136 & 57772.6142 & 0.0009 & 0&0043 & 63 \\
137 & 57772.6745 & 0.0014 & 0&0031 & 63 \\
146 & 57773.2292 & 0.0006 & 0&0040 & 33 \\
147 & 57773.2901 & 0.0008 & 0&0034 & 31 \\
148 & 57773.3523 & 0.0007 & 0&0041 & 31 \\
149 & 57773.4131 & 0.0006 & 0&0033 & 28 \\
200 & 57776.5375 & 0.0013 & $-$0&0102 & 22 \\
217 & 57777.5786 & 0.0020 & $-$0&0151 & 24 \\
\hline
  \multicolumn{6}{l}{\commenta BJD$-$2400000.} \\
  \multicolumn{6}{l}{\commentb Against max $= 2457764.2421 + 0.061528 E$.} \\
  \multicolumn{6}{l}{\commentc Number of points used to determine the maximum.} \\
\end{tabular}
\end{center}
\end{table}

\subsection{ASASSN-16da}\label{obj:asassn16da}

   This object was detected as a transient
at $V$=16.1 on 2016 March 8 by the ASAS-SN team.
The outburst was confirmed and announced on March 12,
when the object was at $V$=15.5.  The brightness peak
was on March 10 at $V$=15.1.  The object was identified
with an $g$=21.5 mag SDSS object.  The large outburst
amplitude received attention.

\begin{figure}
  \begin{center}
    \FigureFile(85mm,110mm){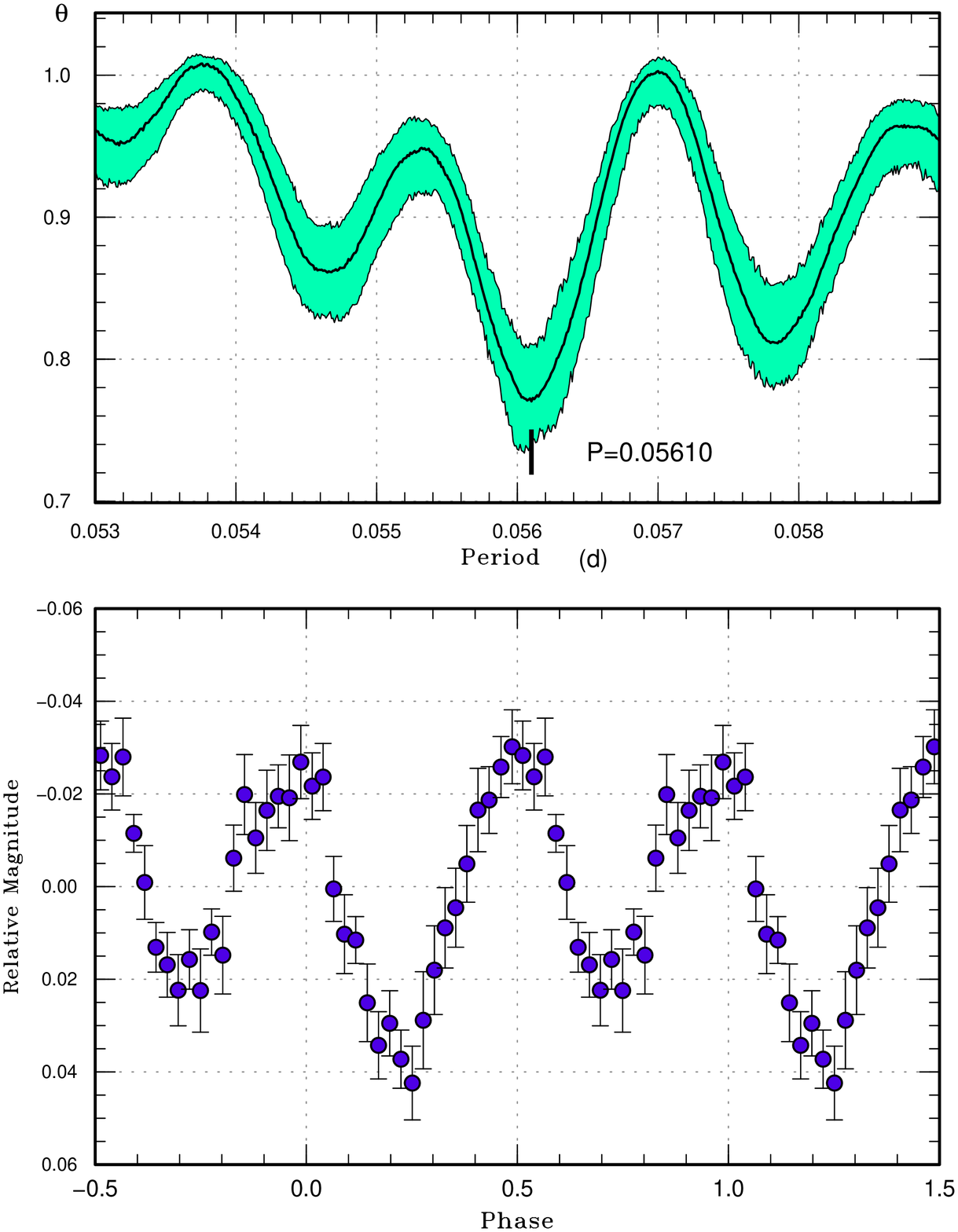}
  \end{center}
  \caption{Early superhumps in ASASSN-16da (2016).
     (Upper): PDM analysis.
     (Lower): Phase-averaged profile.}
  \label{fig:asassn16daeshpdm}
\end{figure}

\begin{figure}
  \begin{center}
    \FigureFile(85mm,110mm){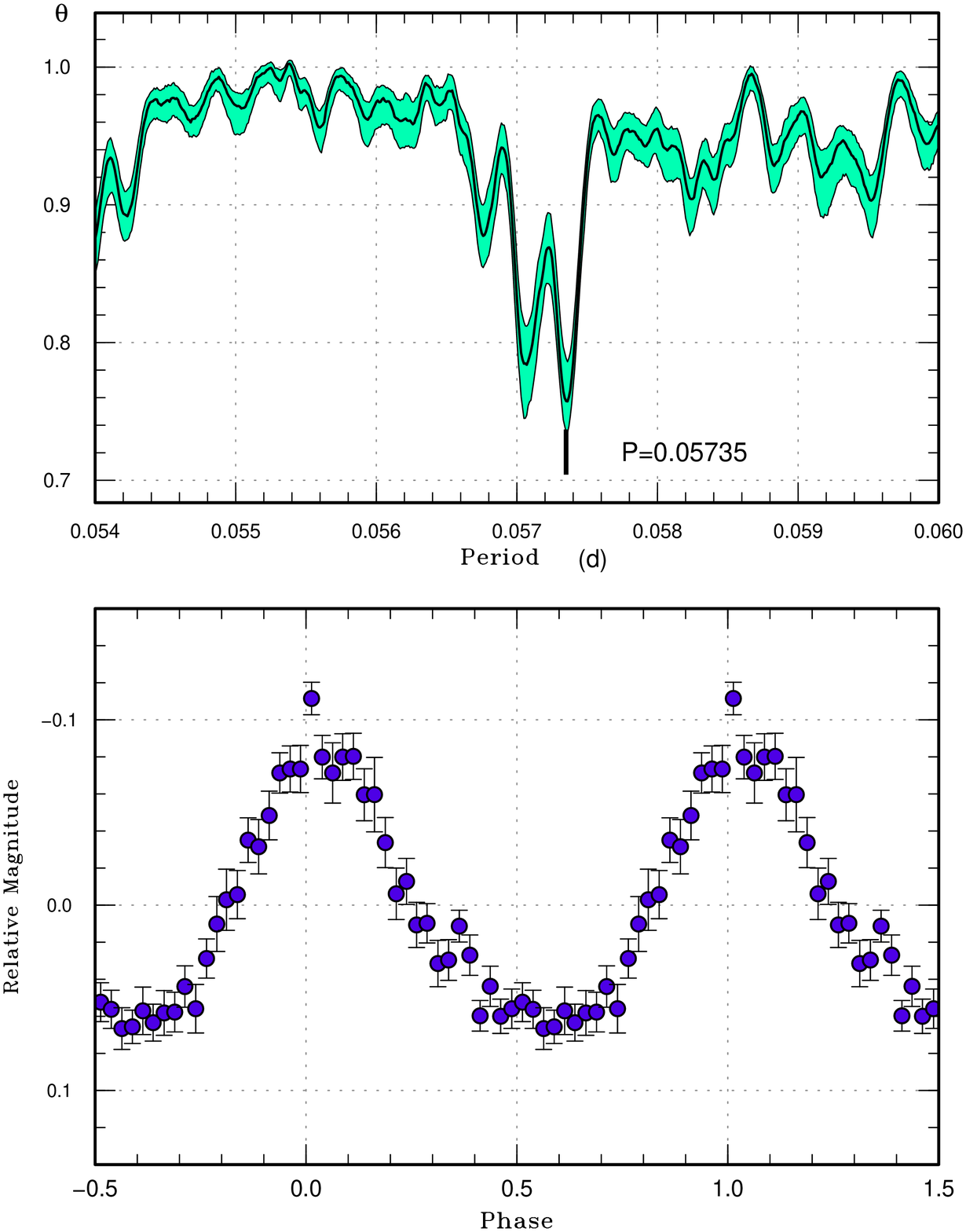}
  \end{center}
  \caption{Ordinary superhumps in ASASSN-16da (2016).
     (Upper): PDM analysis.
     (Lower): Phase-averaged profile.}
  \label{fig:asassn16dashpdm}
\end{figure}

   The object showed double-wave early superhumps
on March 13 and 14 (vsnet-alert 19579, 19592;
figure \ref{fig:asassn16daeshpdm}).
On March 15 (5~d after the brightness peak),
the object started to show ordinary superhumps
(vsnet-alert 19598, 19617, 19653; figure \ref{fig:asassn16dashpdm}).
The times of superhump maxima are listed in
table \ref{tab:asassn16daoc2016}.  The epochs for
$E \le$2 and $E \ge$203 were apparently those of
stage A and C superhumps, respectively.
If we consider that stage A just ended at $E$=10
(which may not be a bad assumption as compared with
$O-C$ diagrams of well-observed objects), the period
of stage A superhumps was 0.05858(10)~d.
The resultant $\epsilon^*$ of 0.042(2) corresponds to
$q$=0.12(1).  This relatively large $q$ for a WZ Sge-type
dwarf nova is consistent with the appearance of
stage C superhumps, short duration of stage A,
relatively large $P_{\rm dot}$ in stage B
[$+7.5(0.9) \times 10^{-5}$], and relatively early
appearance of ordinary superhumps.
This object is probably close to the borderline of
WZ Sge-type dwarf novae and ordinary SU UMa-type
dwarf novae.


\begin{table}
\caption{Superhump maxima of ASASSN-16da (2016)}\label{tab:asassn16daoc2016}
\begin{center}
\begin{tabular}{rp{55pt}p{40pt}r@{.}lr}
\hline
\multicolumn{1}{c}{$E$} & \multicolumn{1}{c}{max\commenta} & \multicolumn{1}{c}{error} & \multicolumn{2}{c}{$O-C$\commentb} & \multicolumn{1}{c}{$N$\commentc} \\
\hline
0 & 57463.7303 & 0.0006 & $-$0&0037 & 56 \\
1 & 57463.7898 & 0.0005 & $-$0&0016 & 56 \\
2 & 57463.8468 & 0.0006 & $-$0&0019 & 56 \\
10 & 57464.3162 & 0.0005 & 0&0087 & 41 \\
11 & 57464.3725 & 0.0004 & 0&0076 & 83 \\
12 & 57464.4302 & 0.0003 & 0&0079 & 88 \\
13 & 57464.4869 & 0.0003 & 0&0072 & 88 \\
14 & 57464.5434 & 0.0005 & 0&0064 & 56 \\
15 & 57464.6012 & 0.0007 & 0&0069 & 39 \\
21 & 57464.9407 & 0.0013 & 0&0023 & 27 \\
28 & 57465.3407 & 0.0005 & 0&0007 & 38 \\
29 & 57465.3973 & 0.0005 & $-$0&0000 & 38 \\
30 & 57465.4511 & 0.0008 & $-$0&0035 & 39 \\
31 & 57465.5124 & 0.0006 & 0&0004 & 34 \\
35 & 57465.7381 & 0.0007 & $-$0&0033 & 27 \\
36 & 57465.7975 & 0.0006 & $-$0&0013 & 25 \\
37 & 57465.8534 & 0.0007 & $-$0&0028 & 27 \\
52 & 57466.7106 & 0.0029 & $-$0&0058 & 27 \\
53 & 57466.7723 & 0.0054 & $-$0&0015 & 26 \\
54 & 57466.8299 & 0.0013 & $-$0&0012 & 22 \\
63 & 57467.3402 & 0.0062 & $-$0&0072 & 18 \\
64 & 57467.3988 & 0.0006 & $-$0&0059 & 39 \\
65 & 57467.4560 & 0.0010 & $-$0&0061 & 43 \\
66 & 57467.5136 & 0.0010 & $-$0&0058 & 41 \\
67 & 57467.5808 & 0.0028 & 0&0041 & 10 \\
70 & 57467.7395 & 0.0033 & $-$0&0093 & 17 \\
117 & 57470.4310 & 0.0093 & $-$0&0135 & 28 \\
118 & 57470.4957 & 0.0020 & $-$0&0062 & 54 \\
119 & 57470.5599 & 0.0029 & 0&0007 & 54 \\
171 & 57473.5506 & 0.0039 & 0&0089 & 31 \\
174 & 57473.7204 & 0.0017 & 0&0067 & 54 \\
175 & 57473.7768 & 0.0029 & 0&0057 & 56 \\
203 & 57475.3872 & 0.0023 & 0&0103 & 31 \\
204 & 57475.4405 & 0.0012 & 0&0062 & 32 \\
205 & 57475.4939 & 0.0010 & 0&0023 & 32 \\
206 & 57475.5518 & 0.0013 & 0&0028 & 32 \\
207 & 57475.6112 & 0.0011 & 0&0048 & 32 \\
221 & 57476.4068 & 0.0028 & $-$0&0025 & 32 \\
222 & 57476.4643 & 0.0022 & $-$0&0024 & 31 \\
223 & 57476.5205 & 0.0056 & $-$0&0036 & 32 \\
238 & 57477.3807 & 0.0033 & $-$0&0036 & 23 \\
239 & 57477.4338 & 0.0020 & $-$0&0080 & 26 \\
\hline
  \multicolumn{6}{l}{\commenta BJD$-$2400000.} \\
  \multicolumn{6}{l}{\commentb Against max $= 2457463.7340 + 0.057355 E$.} \\
  \multicolumn{6}{l}{\commentc Number of points used to determine the maximum.} \\
\end{tabular}
\end{center}
\end{table}

\subsection{ASASSN-16dk}\label{obj:asassn16dk}

   This object was detected as a transient
at $V$=16.4 on 2016 March 21 by the ASAS-SN team.
The outburst was confirmed and announced on March 24,
when the object was at $V$=15.1.
Subsequent observations detected superhumps.
The times of superhump maxima are listed in
table \ref{tab:asassn16dkoc2016}.
Since the observations were apparently obtained
during the late phase of the superoutburst,
the superhump stage was probably C.  The lack of
period variation is consistent with this interpretation.


\begin{figure}
  \begin{center}
    \FigureFile(85mm,110mm){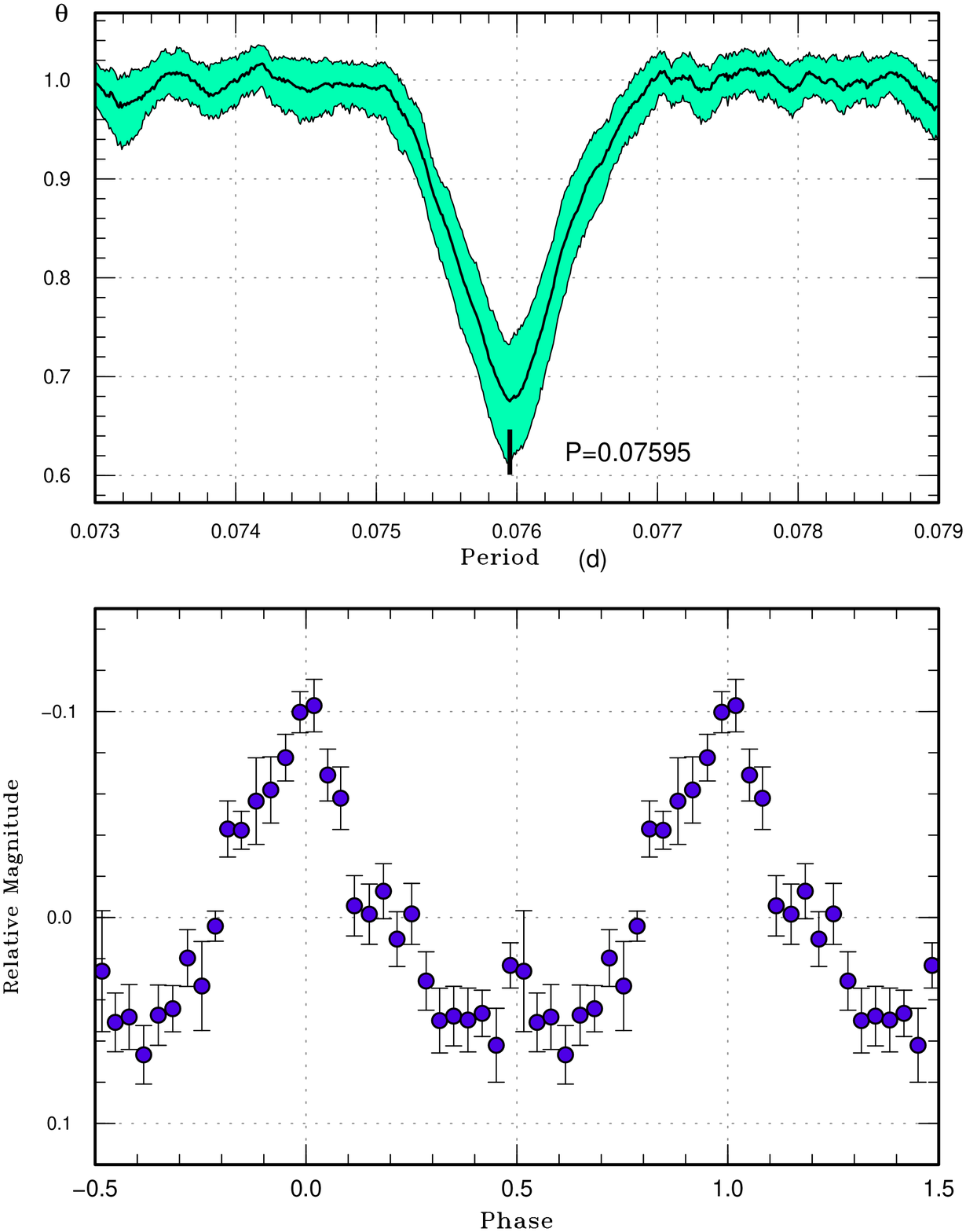}
  \end{center}
  \caption{Superhumps in ASASSN-16dk (2016).
     (Upper): PDM analysis.
     (Lower): Phase-averaged profile.}
  \label{fig:asassn16dkshpdm}
\end{figure}


\begin{table}
\caption{Superhump maxima of ASASSN-16dk (2016)}\label{tab:asassn16dkoc2016}
\begin{center}
\begin{tabular}{rp{55pt}p{40pt}r@{.}lr}
\hline
\multicolumn{1}{c}{$E$} & \multicolumn{1}{c}{max\commenta} & \multicolumn{1}{c}{error} & \multicolumn{2}{c}{$O-C$\commentb} & \multicolumn{1}{c}{$N$\commentc} \\
\hline
0 & 57476.5381 & 0.0012 & $-$0&0047 & 37 \\
1 & 57476.6159 & 0.0008 & $-$0&0029 & 48 \\
2 & 57476.7012 & 0.0023 & 0&0065 & 12 \\
13 & 57477.5314 & 0.0017 & 0&0016 & 30 \\
14 & 57477.6037 & 0.0007 & $-$0&0020 & 49 \\
15 & 57477.6839 & 0.0035 & 0&0022 & 19 \\
26 & 57478.5228 & 0.0157 & 0&0059 & 25 \\
27 & 57478.5874 & 0.0013 & $-$0&0054 & 50 \\
28 & 57478.6703 & 0.0070 & 0&0016 & 28 \\
40 & 57479.5760 & 0.0025 & $-$0&0037 & 51 \\
41 & 57479.6536 & 0.0049 & $-$0&0021 & 38 \\
53 & 57480.5667 & 0.0029 & 0&0000 & 45 \\
54 & 57480.6472 & 0.0025 & 0&0046 & 37 \\
66 & 57481.5497 & 0.0031 & $-$0&0040 & 43 \\
67 & 57481.6321 & 0.0040 & 0&0024 & 40 \\
\hline
  \multicolumn{6}{l}{\commenta BJD$-$2400000.} \\
  \multicolumn{6}{l}{\commentb Against max $= 2457476.5428 + 0.075923 E$.} \\
  \multicolumn{6}{l}{\commentc Number of points used to determine the maximum.} \\
\end{tabular}
\end{center}
\end{table}

\subsection{ASASSN-16ds}\label{obj:asassn16ds}

   This object was detected as a transient
at $V$=14.7 on 2016 April 1 by the ASAS-SN team.
Subsequent observations detected growing superhumps
(vsnet-alert 19680),
which later became fully developed ones.
The times of superhump maxima are listed in
table \ref{tab:asassn16dsoc2016}.
The $O-C$ values indicates typical stages A and B
(vsnet-alert 19704).
Although individual times of superhump maxima
could not be measured after BJD 2457495,
a PDM analysis of the corresponding segment
yielded a signal of 0.06723(5)~d, which is
included in table \ref{tab:perlist}.


\begin{figure}
  \begin{center}
    \FigureFile(85mm,110mm){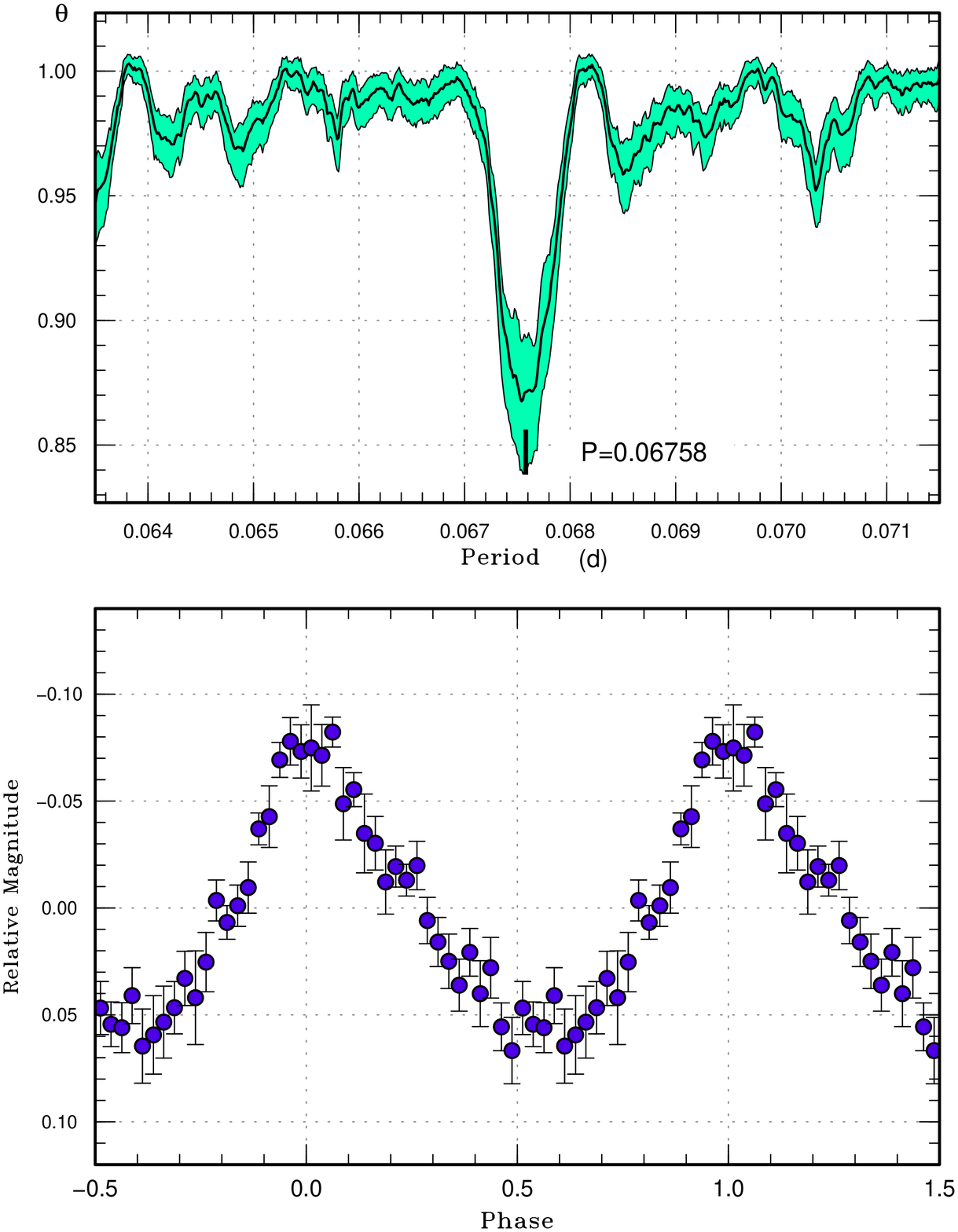}
  \end{center}
  \caption{Superhumps in ASASSN-16ds (2016).
     (Upper): PDM analysis.
     (Lower): Phase-averaged profile.}
  \label{fig:asassn16dsshpdm}
\end{figure}


\begin{table}
\caption{Superhump maxima of ASASSN-16ds (2016)}\label{tab:asassn16dsoc2016}
\begin{center}
\begin{tabular}{rp{55pt}p{40pt}r@{.}lr}
\hline
\multicolumn{1}{c}{$E$} & \multicolumn{1}{c}{max\commenta} & \multicolumn{1}{c}{error} & \multicolumn{2}{c}{$O-C$\commentb} & \multicolumn{1}{c}{$N$\commentc} \\
\hline
0 & 57481.5700 & 0.0006 & $-$0&0031 & 156 \\
1 & 57481.6348 & 0.0017 & $-$0&0060 & 76 \\
3 & 57481.7760 & 0.0007 & $-$0&0003 & 17 \\
4 & 57481.8496 & 0.0009 & 0&0055 & 14 \\
5 & 57481.9135 & 0.0008 & 0&0017 & 11 \\
10 & 57482.2567 & 0.0009 & 0&0062 & 33 \\
13 & 57482.4617 & 0.0003 & 0&0080 & 156 \\
14 & 57482.5290 & 0.0004 & 0&0076 & 132 \\
15 & 57482.5977 & 0.0004 & 0&0085 & 121 \\
33 & 57483.8143 & 0.0006 & 0&0058 & 32 \\
34 & 57483.8810 & 0.0005 & 0&0047 & 40 \\
47 & 57484.7613 & 0.0017 & 0&0044 & 21 \\
48 & 57484.8259 & 0.0004 & 0&0013 & 35 \\
58 & 57485.5005 & 0.0008 & $-$0&0015 & 131 \\
59 & 57485.5678 & 0.0006 & $-$0&0020 & 156 \\
60 & 57485.6361 & 0.0004 & $-$0&0014 & 156 \\
62 & 57485.7657 & 0.0009 & $-$0&0073 & 13 \\
63 & 57485.8377 & 0.0004 & $-$0&0031 & 41 \\
64 & 57485.9058 & 0.0005 & $-$0&0027 & 25 \\
73 & 57486.5121 & 0.0006 & $-$0&0060 & 117 \\
74 & 57486.5817 & 0.0006 & $-$0&0042 & 156 \\
75 & 57486.6507 & 0.0013 & $-$0&0029 & 70 \\
87 & 57487.4620 & 0.0008 & $-$0&0046 & 121 \\
88 & 57487.5275 & 0.0007 & $-$0&0068 & 156 \\
89 & 57487.5961 & 0.0005 & $-$0&0060 & 156 \\
90 & 57487.6611 & 0.0010 & $-$0&0086 & 108 \\
107 & 57488.8155 & 0.0007 & $-$0&0059 & 34 \\
108 & 57488.8843 & 0.0008 & $-$0&0048 & 30 \\
136 & 57490.7870 & 0.0020 & 0&0011 & 20 \\
137 & 57490.8520 & 0.0032 & $-$0&0016 & 12 \\
181 & 57493.8452 & 0.0023 & 0&0110 & 19 \\
195 & 57494.7958 & 0.0021 & 0&0132 & 19 \\
\hline
  \multicolumn{6}{l}{\commenta BJD$-$2400000.} \\
  \multicolumn{6}{l}{\commentb Against max $= 2457481.5731 + 0.067741 E$.} \\
  \multicolumn{6}{l}{\commentc Number of points used to determine the maximum.} \\
\end{tabular}
\end{center}
\end{table}

\subsection{ASASSN-16dz}\label{obj:asassn16dz}

\begin{figure}
  \begin{center}
    \FigureFile(85mm,110mm){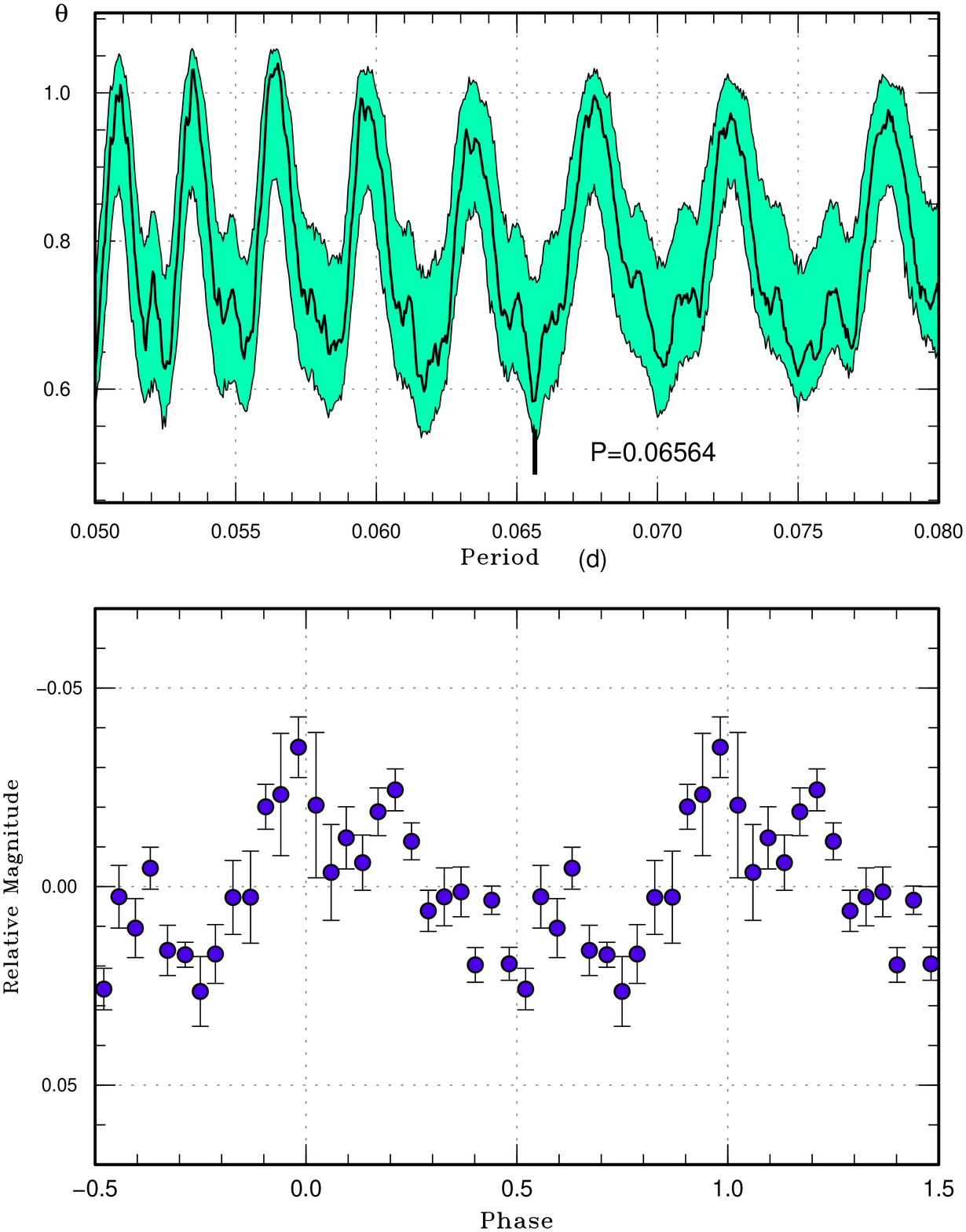}
  \end{center}
  \caption{Superhumps in ASASSN-16dz (2016).
     (Upper): PDM analysis.
     (Lower): Phase-averaged profile.}
  \label{fig:asassn16dzshpdm}
\end{figure}

   This object was detected as a transient
at $V$=15.0 on 2016 April 2 by the ASAS-SN team.
The outburst was announced after the observation
on April 3 at $V$=14.2.
The object had been listed as an emission-line
object IPHAS2 J064225.58$+$082546.7 \citep{wit08IPHAS}.
Although superhumps were detected, the period
was not well determined due to short runs and
the limited coverage only on two nights
(figure \ref{fig:asassn16dzshpdm}).
The times of superhump maxima are listed in
table \ref{tab:asassn16dzoc2016}, which is based
on one of the aliases giving the smallest
$O-C$ scatter.


\begin{table}
\caption{Superhump maxima of ASASSN-16dz (2016)}\label{tab:asassn16dzoc2016}
\begin{center}
\begin{tabular}{rp{55pt}p{40pt}r@{.}lr}
\hline
\multicolumn{1}{c}{$E$} & \multicolumn{1}{c}{max\commenta} & \multicolumn{1}{c}{error} & \multicolumn{2}{c}{$O-C$\commentb} & \multicolumn{1}{c}{$N$\commentc} \\
\hline
0 & 57484.3064 & 0.0052 & 0&0011 & 22 \\
1 & 57484.3703 & 0.0012 & $-$0&0012 & 63 \\
16 & 57485.3655 & 0.0019 & 0&0001 & 47 \\
\hline
  \multicolumn{6}{l}{\commenta BJD$-$2400000.} \\
  \multicolumn{6}{l}{\commentb Against max $= 2457484.3053 + 0.066259 E$.} \\
  \multicolumn{6}{l}{\commentc Number of points used to determine the maximum.} \\
\end{tabular}
\end{center}
\end{table}

\subsection{ASASSN-16ez}\label{obj:asassn16ez}

   This object was detected as a transient
at $V$=14.3 on 2016 May 6 by the ASAS-SN team.
The large outburst amplitude suggested an SU UMa-type
superoutburst (cf. vsnet-alert 19804).

\begin{figure}
  \begin{center}
    \FigureFile(85mm,110mm){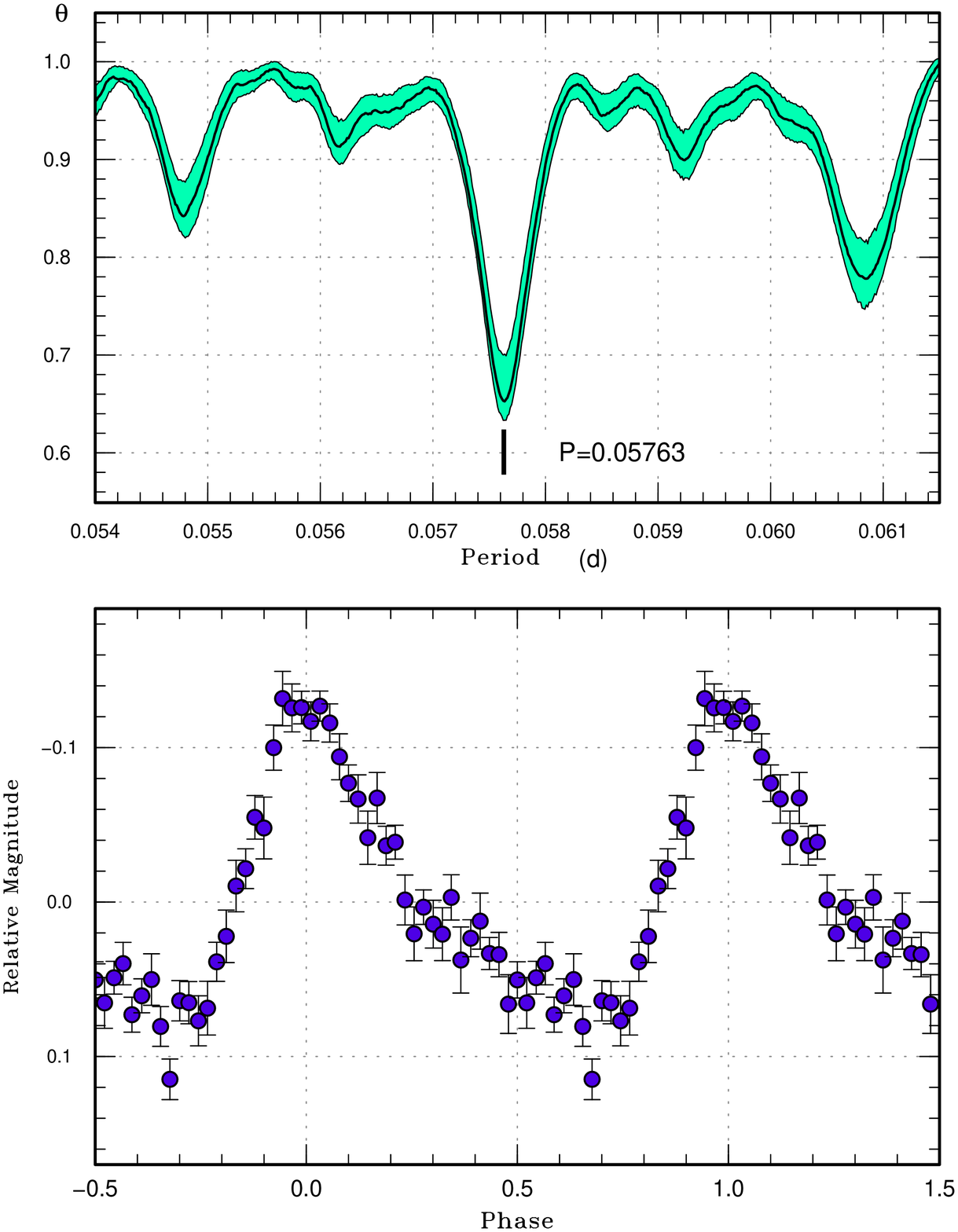}
  \end{center}
  \caption{Superhumps in ASASSN-16ez (2016).
     (Upper): PDM analysis.
     (Lower): Phase-averaged profile.}
  \label{fig:asassn16ezshpdm}
\end{figure}

   On May 12, this object started to show large-amplitude
superhumps (vsnet-alert 19823, 19829).
The times of superhump maxima are listed in
table \ref{tab:asassn16ezoc2016}.  These superhumps
were stage B ones and there was little period variation.
The object was still in outburst on May 23
(17~d after the outburst detection and 11~d after
the first detection of superhumps).
Although there were some variations before May 12,
we could not confidently detect stage A superhumps
(probably due to a 1.5-d gap in the observation).
Although the small $P_{\rm dot}$ might suggest
a small $q$ (cf. \cite{kat15wzsge}), the relatively
large amplitude of superhumps (figure \ref{fig:asassn16ezshpdm})
and the lack of long duration of stage A do not
support this interpretation.  The seemingly small
$P_{\rm dot}$ may be a result of a relatively
short observational coverage of 4~d.


\begin{table}
\caption{Superhump maxima of ASASSN-16ez (2016)}\label{tab:asassn16ezoc2016}
\begin{center}
\begin{tabular}{rp{55pt}p{40pt}r@{.}lr}
\hline
\multicolumn{1}{c}{$E$} & \multicolumn{1}{c}{max\commenta} & \multicolumn{1}{c}{error} & \multicolumn{2}{c}{$O-C$\commentb} & \multicolumn{1}{c}{$N$\commentc} \\
\hline
0 & 57521.1188 & 0.0010 & $-$0&0002 & 117 \\
1 & 57521.1731 & 0.0005 & $-$0&0035 & 170 \\
2 & 57521.2352 & 0.0011 & 0&0009 & 153 \\
7 & 57521.5236 & 0.0024 & 0&0012 & 30 \\
8 & 57521.5837 & 0.0003 & 0&0037 & 61 \\
18 & 57522.1562 & 0.0050 & $-$0&0001 & 14 \\
20 & 57522.2717 & 0.0016 & 0&0002 & 37 \\
22 & 57522.3882 & 0.0014 & 0&0015 & 32 \\
23 & 57522.4453 & 0.0004 & 0&0010 & 58 \\
24 & 57522.4995 & 0.0025 & $-$0&0024 & 19 \\
40 & 57523.4212 & 0.0004 & $-$0&0027 & 94 \\
41 & 57523.4799 & 0.0005 & $-$0&0017 & 85 \\
42 & 57523.5410 & 0.0006 & 0&0018 & 59 \\
59 & 57524.5158 & 0.0005 & $-$0&0028 & 59 \\
60 & 57524.5759 & 0.0005 & $-$0&0004 & 56 \\
61 & 57524.6349 & 0.0016 & 0&0010 & 22 \\
74 & 57525.3826 & 0.0007 & $-$0&0004 & 28 \\
75 & 57525.4416 & 0.0007 & 0&0009 & 30 \\
76 & 57525.4972 & 0.0006 & $-$0&0010 & 33 \\
77 & 57525.5589 & 0.0007 & 0&0030 & 32 \\
\hline
  \multicolumn{6}{l}{\commenta BJD$-$2400000.} \\
  \multicolumn{6}{l}{\commentb Against max $= 2457521.1190 + 0.057621 E$.} \\
  \multicolumn{6}{l}{\commentc Number of points used to determine the maximum.} \\
\end{tabular}
\end{center}
\end{table}

\subsection{ASASSN-16fr}\label{obj:asassn16fr}

\begin{figure}
  \begin{center}
    \FigureFile(85mm,110mm){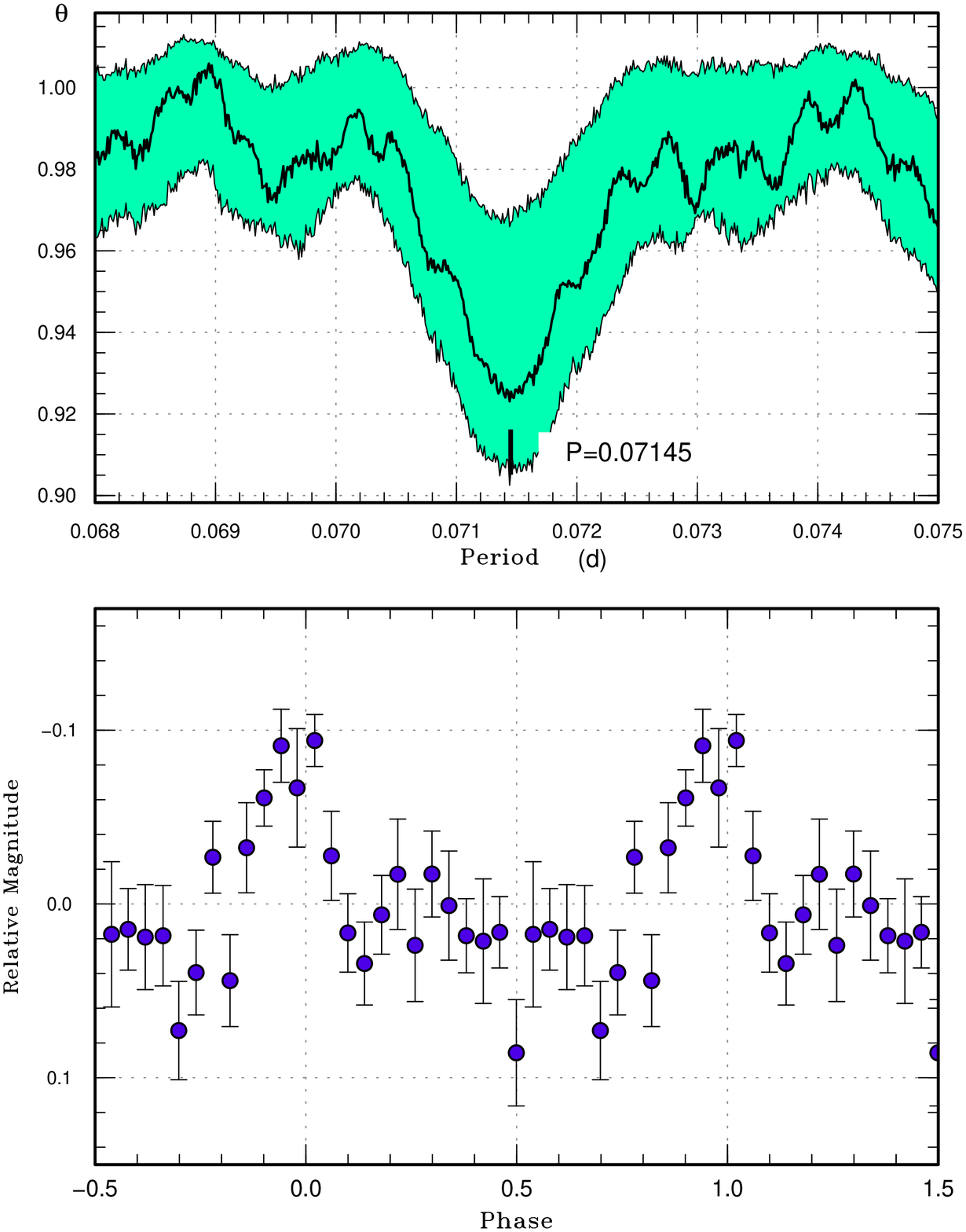}
  \end{center}
  \caption{Superhumps in ASASSN-16fr (2016).
     (Upper): PDM analysis.
     (Lower): Phase-averaged profile.}
  \label{fig:asassn16frshpdm}
\end{figure}

   This object was detected as a transient
at $V$=16.6 on 2016 May 30 by the ASAS-SN team.
Subsequent observations detected superhumps
(vsnet-alert 19863; figure \ref{fig:asassn16frshpdm}).
The times of superhump maxima are listed in
table \ref{tab:asassn16froc2016}.
Due to the faintness of the object (17 mag around
the observations), the statistics was not good.


\begin{table}
\caption{Superhump maxima of ASASSN-16fr (2016)}\label{tab:asassn16froc2016}
\begin{center}
\begin{tabular}{rp{55pt}p{40pt}r@{.}lr}
\hline
\multicolumn{1}{c}{$E$} & \multicolumn{1}{c}{max\commenta} & \multicolumn{1}{c}{error} & \multicolumn{2}{c}{$O-C$\commentb} & \multicolumn{1}{c}{$N$\commentc} \\
\hline
0 & 57541.1447 & 0.0015 & $-$0&0026 & 60 \\
1 & 57541.2179 & 0.0032 & $-$0&0008 & 70 \\
14 & 57542.1502 & 0.0014 & 0&0034 & 141 \\
15 & 57542.2167 & 0.0018 & $-$0&0016 & 129 \\
21 & 57542.6523 & 0.0036 & 0&0057 & 16 \\
35 & 57543.6420 & 0.0028 & $-$0&0041 & 17 \\
\hline
  \multicolumn{6}{l}{\commenta BJD$-$2400000.} \\
  \multicolumn{6}{l}{\commentb Against max $= 2457541.1473 + 0.071394 E$.} \\
  \multicolumn{6}{l}{\commentc Number of points used to determine the maximum.} \\
\end{tabular}
\end{center}
\end{table}

\subsection{ASASSN-16fu}\label{obj:asassn16fu}

\begin{figure}
  \begin{center}
    \FigureFile(85mm,110mm){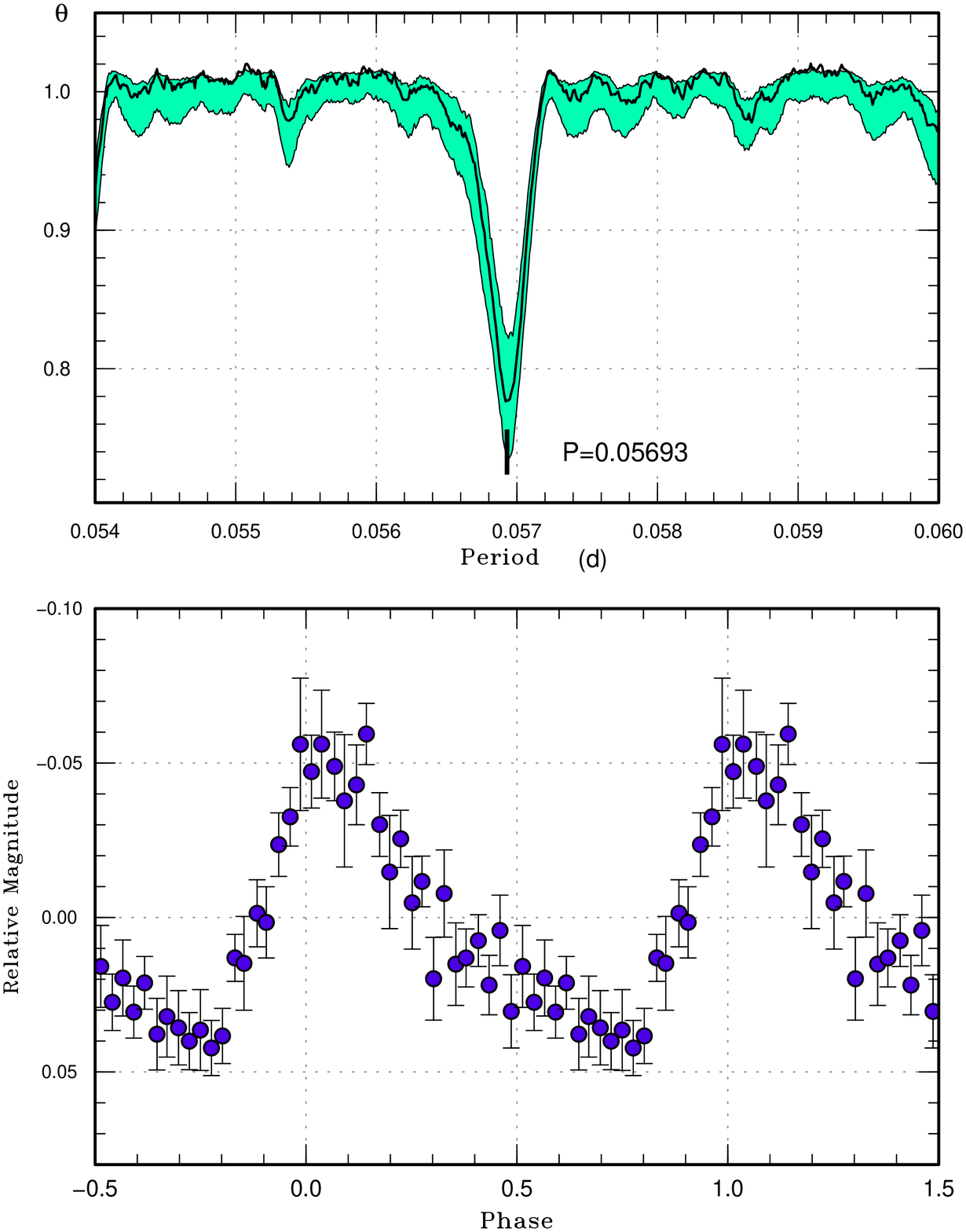}
  \end{center}
  \caption{Ordinary superhumps in ASASSN-16fu during stage B (2016).
     (Upper): PDM analysis.
     (Lower): Phase-averaged profile.}
  \label{fig:asassn16fushpdm}
\end{figure}

\begin{figure}
  \begin{center}
    \FigureFile(85mm,110mm){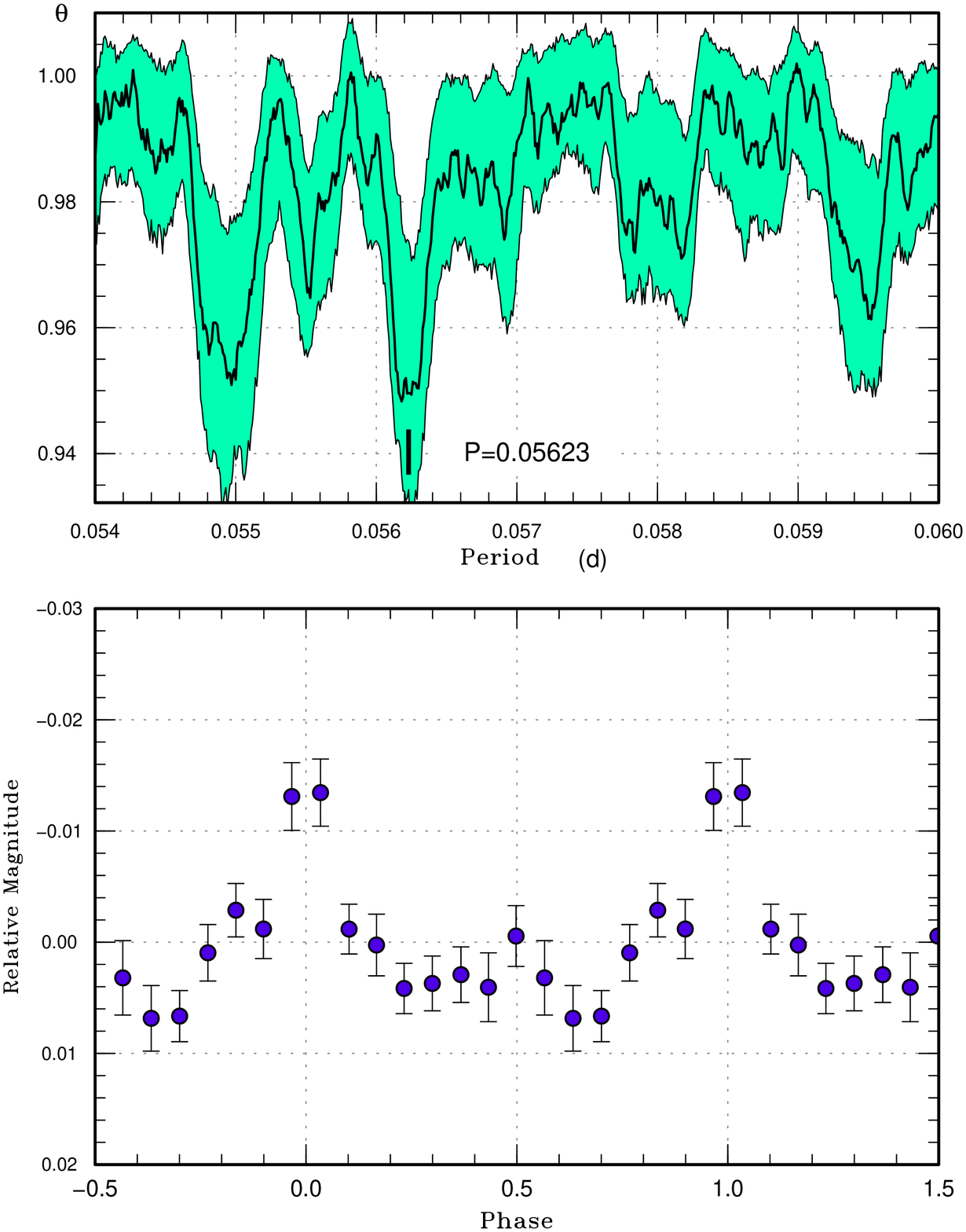}
  \end{center}
  \caption{Possible early superhumps in ASASSN-16fu (2016).
     (Upper): PDM analysis.
     (Lower): Phase-averaged profile.}
  \label{fig:asassn16fueshpdm}
\end{figure}

   This object was detected as a transient
at $V$=13.9 on 2016 June 5 by the ASAS-SN team.
The large outburst amplitude received attention
(cf. vsnet-alert 19864).
On June 14 (9~d after the outburst detection),
this object showed fully developed superhumps
(vsnet-alert 19899, 19901; figure \ref{fig:asassn16fushpdm}).
The times of superhump maxima are listed in
table \ref{tab:asassn16fuoc2016}.
Although the maxima for $E \le$1 were stage A
superhumps, the period of stage A superhumps was
not determined due to the unfortunate 2~d gap
before the full growth of the superhumps.
The relatively small $P_{\rm dot}$ in stage B
[$+4.6(0.6) \times 10^{-5}$] suggests a relatively
small $q$ [$q \sim$0.08(1) according to equation (6) in
\citet{kat15wzsge}].  This expectation is consistent
with the small amplitude of the superhumps
(figure \ref{fig:asassn16fushpdm}).
A retrospective analysis of the data before ordinary
superhumps appeared (BJD before 2457553) detected
possible early superhumps with a period of 0.05623(3)~d
(figure \ref{fig:asassn16fueshpdm}).  We consider that
this object is a WZ Sge-type dwarf nova, which appears
consistent with the small $q$ as inferred from
the small $P_{\rm dot}$.


\begin{table}
\caption{Superhump maxima of ASASSN-16fu (2016)}\label{tab:asassn16fuoc2016}
\begin{center}
\begin{tabular}{rp{55pt}p{40pt}r@{.}lr}
\hline
\multicolumn{1}{c}{$E$} & \multicolumn{1}{c}{max\commenta} & \multicolumn{1}{c}{error} & \multicolumn{2}{c}{$O-C$\commentb} & \multicolumn{1}{c}{$N$\commentc} \\
\hline
0 & 57551.7331 & 0.0018 & $-$0&0139 & 17 \\
1 & 57551.7932 & 0.0017 & $-$0&0108 & 16 \\
35 & 57553.7478 & 0.0006 & 0&0070 & 17 \\
36 & 57553.8032 & 0.0008 & 0&0054 & 16 \\
37 & 57553.8641 & 0.0009 & 0&0093 & 16 \\
38 & 57553.9174 & 0.0009 & 0&0056 & 16 \\
53 & 57554.7719 & 0.0021 & 0&0056 & 16 \\
54 & 57554.8269 & 0.0005 & 0&0036 & 16 \\
55 & 57554.8843 & 0.0009 & 0&0040 & 17 \\
70 & 57555.7347 & 0.0008 & $-$0&0001 & 14 \\
71 & 57555.7901 & 0.0019 & $-$0&0017 & 16 \\
72 & 57555.8531 & 0.0025 & 0&0044 & 16 \\
73 & 57555.9081 & 0.0011 & 0&0024 & 16 \\
89 & 57556.8175 & 0.0021 & 0&0003 & 16 \\
90 & 57556.8733 & 0.0012 & $-$0&0008 & 16 \\
106 & 57557.7815 & 0.0012 & $-$0&0041 & 16 \\
108 & 57557.8972 & 0.0007 & $-$0&0024 & 17 \\
123 & 57558.7476 & 0.0013 & $-$0&0065 & 16 \\
124 & 57558.8044 & 0.0033 & $-$0&0067 & 16 \\
125 & 57558.8685 & 0.0020 & 0&0005 & 17 \\
140 & 57559.7208 & 0.0015 & $-$0&0017 & 13 \\
141 & 57559.7797 & 0.0023 & 0&0002 & 16 \\
143 & 57559.8915 & 0.0010 & $-$0&0020 & 17 \\
159 & 57560.7995 & 0.0052 & $-$0&0054 & 16 \\
160 & 57560.8631 & 0.0017 & 0&0011 & 17 \\
161 & 57560.9148 & 0.0014 & $-$0&0041 & 14 \\
175 & 57561.7171 & 0.0017 & 0&0006 & 13 \\
176 & 57561.7713 & 0.0059 & $-$0&0021 & 16 \\
178 & 57561.8893 & 0.0020 & 0&0019 & 17 \\
193 & 57562.7473 & 0.0031 & 0&0053 & 16 \\
194 & 57562.8003 & 0.0055 & 0&0014 & 16 \\
195 & 57562.8596 & 0.0025 & 0&0037 & 17 \\
\hline
  \multicolumn{6}{l}{\commenta BJD$-$2400000.} \\
  \multicolumn{6}{l}{\commentb Against max $= 2457551.7470 + 0.056969 E$.} \\
  \multicolumn{6}{l}{\commentc Number of points used to determine the maximum.} \\
\end{tabular}
\end{center}
\end{table}

\subsection{ASASSN-16gh}\label{obj:asassn16gh}

\begin{figure}
  \begin{center}
    \FigureFile(85mm,110mm){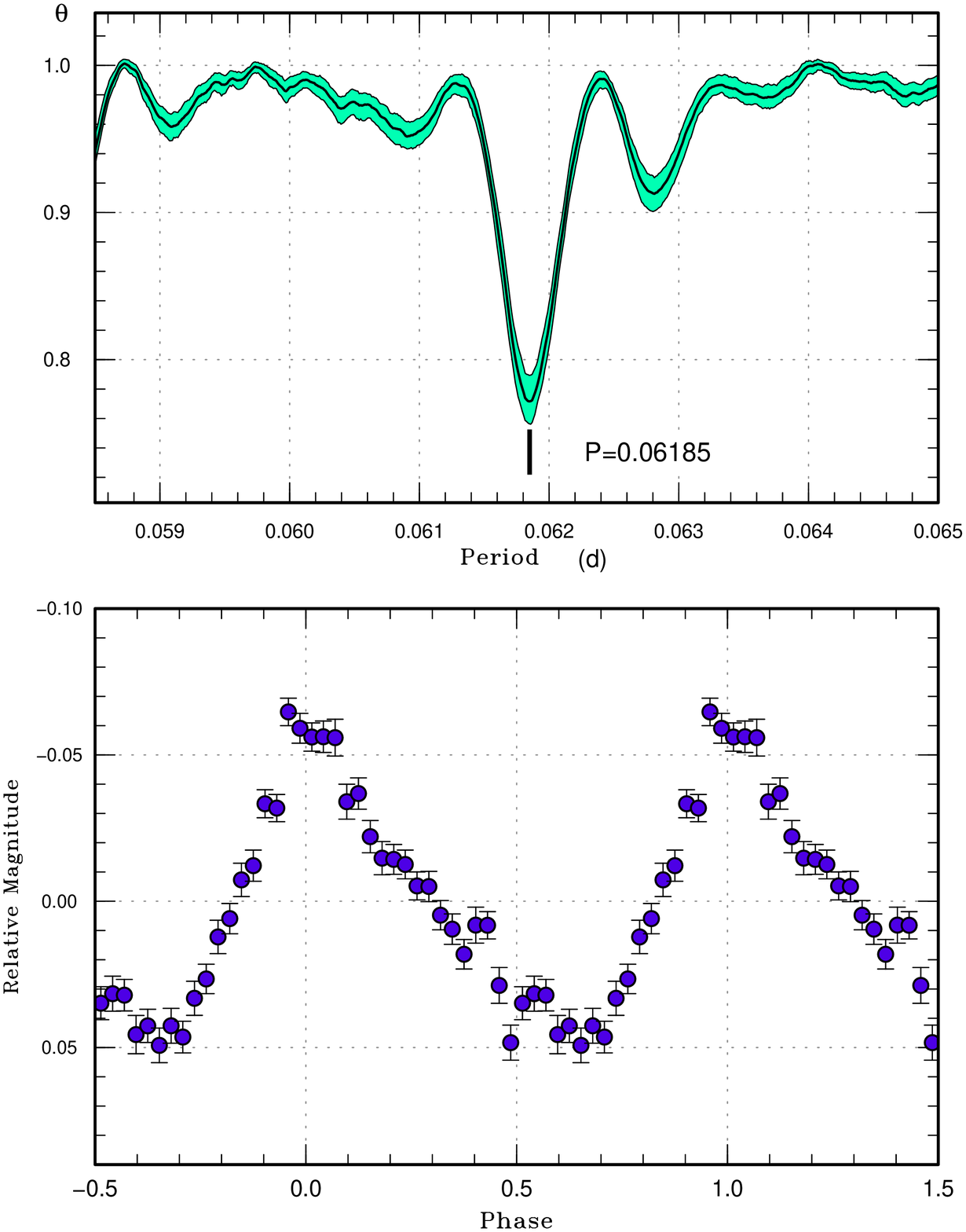}
  \end{center}
  \caption{Ordinary superhumps in ASASSN-16gh during stage B (2016).
     (Upper): PDM analysis.
     (Lower): Phase-averaged profile.}
  \label{fig:asassn16ghshpdm}
\end{figure}

   This object was detected as a transient
at $V$=15.5 on 2016 June 16 by the ASAS-SN team.
The outburst was announced on June 18, when the object
further brightened to $V$=14.3
No strong superhumps were detected up to observations
on June 24.  Growing superhumps were recorded
on June 28 (12~d after the outburst detection;
vsnet-alert 19935).  Further development of superhumps
were observed (vsnet-alert 19943, 19952;
figure \ref{fig:asassn16ghshpdm}).
Although we could not detect early superhumps,
the object may be a WZ Sge-type dwarf nova
since the waiting time (12~d) of the growth of
the superhumps was long.  If this identification
is true, the object appears to be a candidate for
a period bouncer.  The relatively small amplitude
of the superhumps (figure \ref{fig:asassn16ghshpdm})
and the large outburst amplitude (there was no known
quiescent counterpart) would favor this possibility.


\begin{table}
\caption{Superhump maxima of ASASSN-16gh (2016)}\label{tab:asassn16ghoc2016}
\begin{center}
\begin{tabular}{rp{55pt}p{40pt}r@{.}lr}
\hline
\multicolumn{1}{c}{$E$} & \multicolumn{1}{c}{max\commenta} & \multicolumn{1}{c}{error} & \multicolumn{2}{c}{$O-C$\commentb} & \multicolumn{1}{c}{$N$\commentc} \\
\hline
0 & 57568.3036 & 0.0012 & $-$0&0060 & 78 \\
1 & 57568.3688 & 0.0007 & $-$0&0026 & 136 \\
2 & 57568.4292 & 0.0007 & $-$0&0041 & 141 \\
3 & 57568.4950 & 0.0009 & $-$0&0002 & 143 \\
16 & 57569.3045 & 0.0006 & 0&0050 & 109 \\
17 & 57569.3628 & 0.0005 & 0&0015 & 141 \\
18 & 57569.4252 & 0.0006 & 0&0019 & 142 \\
19 & 57569.4851 & 0.0008 & 0&0000 & 136 \\
20 & 57569.5482 & 0.0008 & 0&0012 & 127 \\
21 & 57569.6117 & 0.0007 & 0&0028 & 132 \\
22 & 57569.6678 & 0.0014 & $-$0&0029 & 102 \\
31 & 57570.2316 & 0.0008 & 0&0040 & 137 \\
32 & 57570.2933 & 0.0011 & 0&0039 & 137 \\
33 & 57570.3534 & 0.0014 & 0&0021 & 139 \\
34 & 57570.4134 & 0.0012 & 0&0002 & 141 \\
35 & 57570.4785 & 0.0014 & 0&0034 & 136 \\
36 & 57570.5365 & 0.0014 & $-$0&0005 & 137 \\
65 & 57572.3319 & 0.0009 & 0&0007 & 142 \\
66 & 57572.3929 & 0.0010 & $-$0&0002 & 141 \\
67 & 57572.4564 & 0.0008 & 0&0015 & 143 \\
68 & 57572.5139 & 0.0016 & $-$0&0030 & 141 \\
69 & 57572.5744 & 0.0069 & $-$0&0043 & 49 \\
81 & 57573.3185 & 0.0009 & $-$0&0026 & 142 \\
82 & 57573.3832 & 0.0014 & 0&0002 & 143 \\
83 & 57573.4425 & 0.0016 & $-$0&0024 & 141 \\
84 & 57573.5055 & 0.0015 & $-$0&0013 & 143 \\
85 & 57573.5644 & 0.0020 & $-$0&0043 & 142 \\
86 & 57573.6275 & 0.0018 & $-$0&0030 & 133 \\
87 & 57573.6893 & 0.0026 & $-$0&0031 & 42 \\
97 & 57574.3115 & 0.0029 & 0&0004 & 142 \\
98 & 57574.3753 & 0.0019 & 0&0023 & 143 \\
99 & 57574.4372 & 0.0012 & 0&0023 & 143 \\
100 & 57574.5036 & 0.0033 & 0&0069 & 106 \\
\hline
  \multicolumn{6}{l}{\commenta BJD$-$2400000.} \\
  \multicolumn{6}{l}{\commentb Against max $= 2457568.3096 + 0.061872 E$.} \\
  \multicolumn{6}{l}{\commentc Number of points used to determine the maximum.} \\
\end{tabular}
\end{center}
\end{table}

\subsection{ASASSN-16gj}\label{obj:asassn16gj}

\begin{figure}
  \begin{center}
    \FigureFile(85mm,110mm){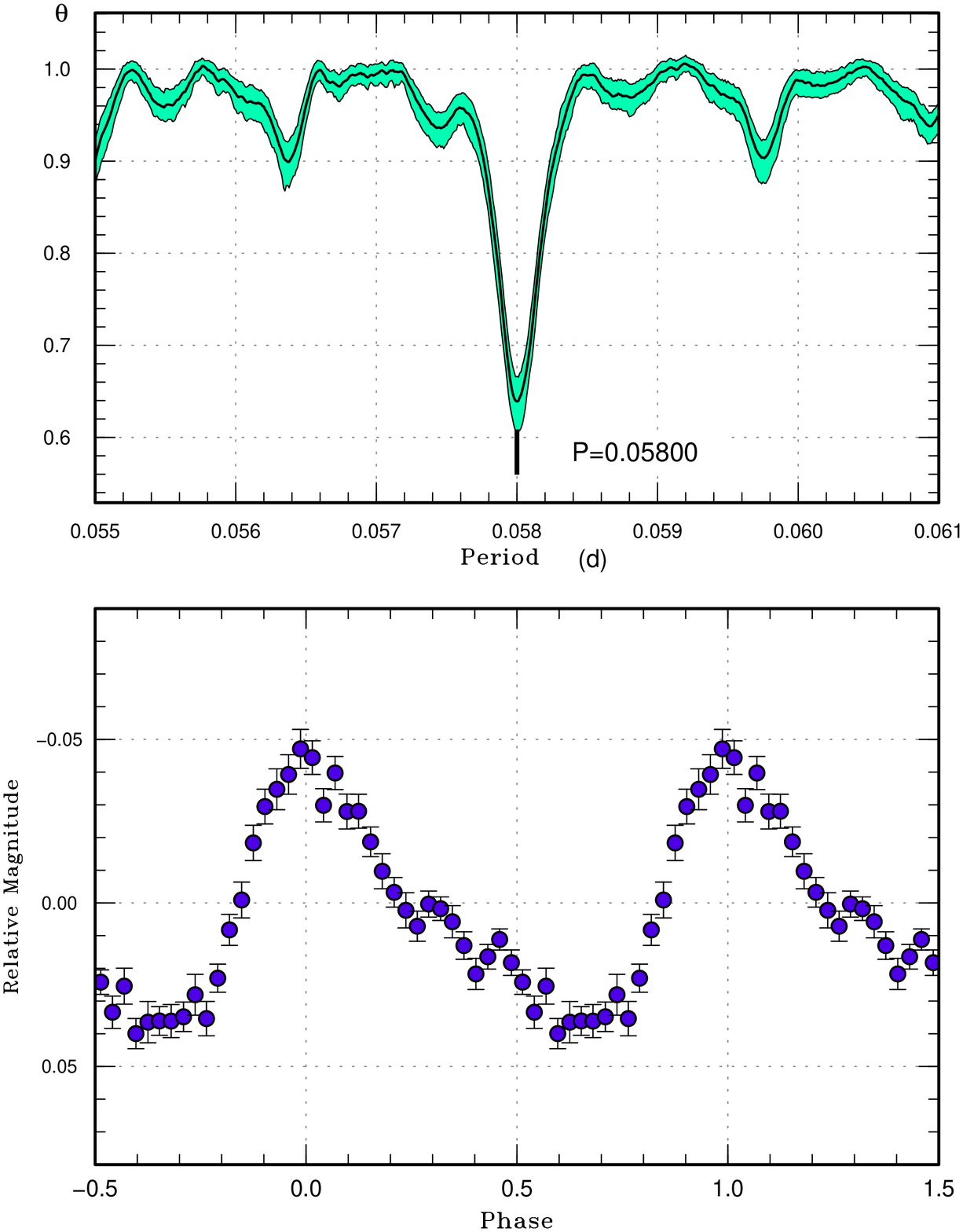}
  \end{center}
  \caption{Ordinary superhumps in ASASSN-16gj during stage B (2016).
     (Upper): PDM analysis.
     (Lower): Phase-averaged profile.}
  \label{fig:asassn16gjshpdm}
\end{figure}

\begin{figure}
  \begin{center}
    \FigureFile(85mm,100mm){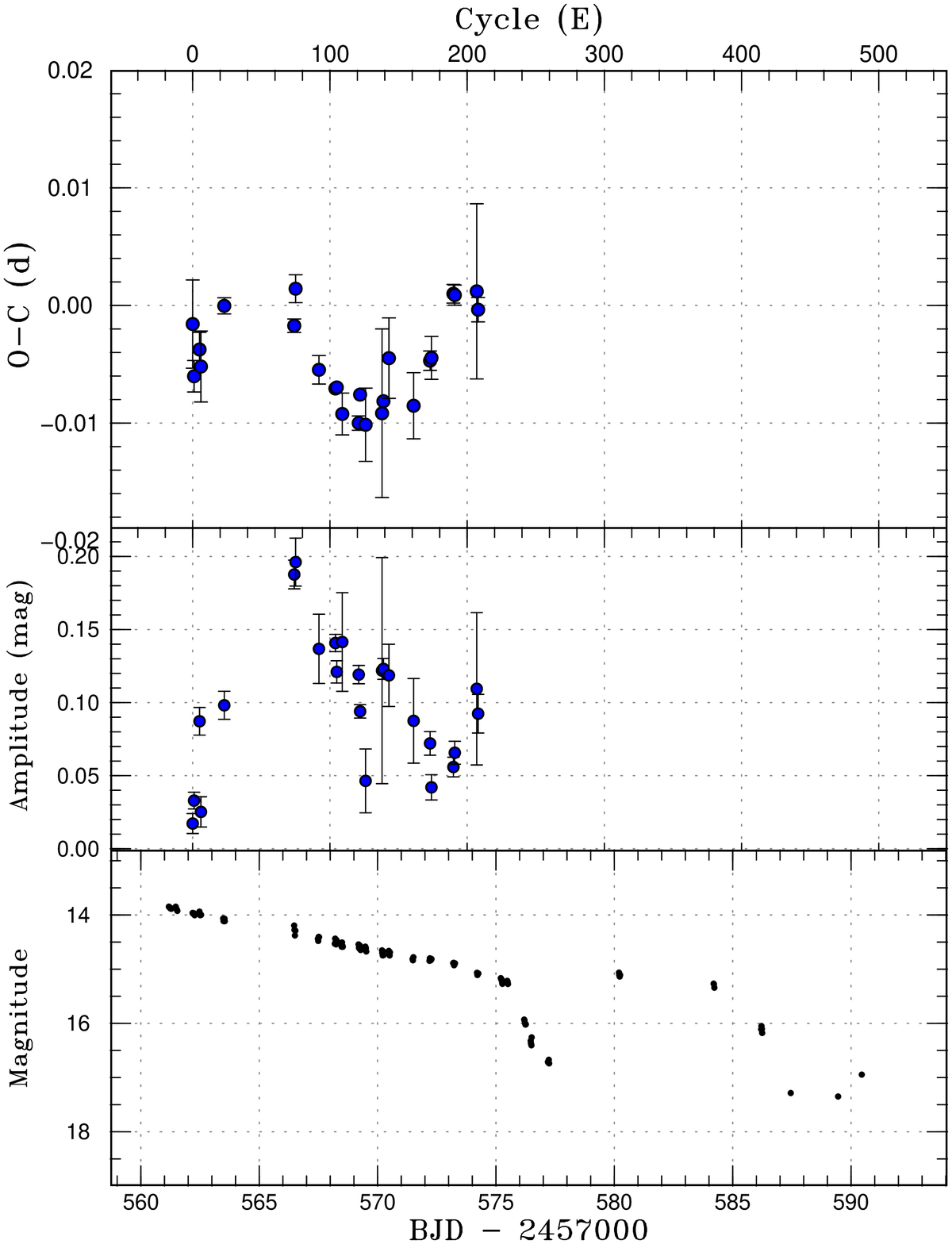}
  \end{center}
  \caption{$O-C$ diagram of superhumps in ASASSN-16gj (2016).
     (Upper:) $O-C$ diagram.
     We used a period of 0.05796~d for calculating the $O-C$ residuals.
     (Middle:) Amplitudes of superhumps.
     (Lower:) Light curve.  The data were binned to 0.019~d.
  }
  \label{fig:asassn16gjhumpall}
\end{figure}

   This object was detected as a transient
at $V$=13.3 on 2016 June 18 by the ASAS-SN team
(cf. vsnet-alert 19905).
The object was also detected on June 17 by the MASTER
network (independent detection, \cite{bal16asassn16gyatel9174}).
The object was undetected on June 12 \citep{bal16asassn16gyatel9174}
and June 11 (ASAS-SN data).
Observations on June 21 already recorded superhumps
(vsnet-alert 19927).  The superhumps grew further
(vsnet-alert 19934, 19936, 19953;
figure \ref{fig:asassn16gjshpdm},
figure \ref{fig:asassn16gjhumpall}).
It took, however, some time to establish
the superhump period since nightly observations were short.
The times of superhump maxima are listed in
table \ref{tab:asassn16gjoc2016}.  The maxima
for $E \le$23 were most likely stage A superhumps
as judged from the growing amplitudes
and $O-C$ values (figure \ref{fig:asassn16gjhumpall}).
The period of stage A superhumps, however,
was not convincingly determined due to the shortness
of nightly observations.

   The object showed a likely plateau-type rebrightening
(figure \ref{fig:asassn16gjhumpall};
the plateau-type rebrightening was favored by the lack
of rising/fading trends in the nightly light curves
in the rebreightening phase).  Although we
could not observe early superhumps, we consider that
this object is a WZ Sge-type dwarf nova as judged
from the long rebrightening (cf. \cite{kat15wzsge}).
The duration of the phase of early superhumps,
if it was present, must have been shorter than 9~d.
This shortness suggests that this object may not be
an extreme WZ Sge-type dwarf nova.  There was,
however, the case of the 2015 superoutburst of AL Com
\citep{kim16alcom}, in which no early superhumps
were observed despite that the same object showed
long phases of early superhumps in previous superoutbursts.
It may be premature to draw any conclusion about
the evolutionary state of ASASSN-16gj
only from the present observation.


\begin{table}
\caption{Superhump maxima of ASASSN-16gj (2016)}\label{tab:asassn16gjoc2016}
\begin{center}
\begin{tabular}{rp{55pt}p{40pt}r@{.}lr}
\hline
\multicolumn{1}{c}{$E$} & \multicolumn{1}{c}{max\commenta} & \multicolumn{1}{c}{error} & \multicolumn{2}{c}{$O-C$\commentb} & \multicolumn{1}{c}{$N$\commentc} \\
\hline
0 & 57562.1891 & 0.0038 & 0&0034 & 88 \\
1 & 57562.2427 & 0.0013 & $-$0&0010 & 126 \\
5 & 57562.4768 & 0.0015 & 0&0013 & 15 \\
6 & 57562.5333 & 0.0030 & $-$0&0002 & 21 \\
23 & 57563.5238 & 0.0007 & 0&0049 & 24 \\
74 & 57566.4780 & 0.0006 & 0&0030 & 16 \\
75 & 57566.5391 & 0.0012 & 0&0062 & 12 \\
92 & 57567.5176 & 0.0012 & $-$0&0008 & 21 \\
104 & 57568.2115 & 0.0003 & $-$0&0024 & 134 \\
105 & 57568.2695 & 0.0004 & $-$0&0024 & 114 \\
109 & 57568.4991 & 0.0018 & $-$0&0046 & 26 \\
121 & 57569.1939 & 0.0006 & $-$0&0054 & 80 \\
122 & 57569.2543 & 0.0004 & $-$0&0030 & 133 \\
126 & 57569.4835 & 0.0031 & $-$0&0056 & 18 \\
138 & 57570.1800 & 0.0072 & $-$0&0047 & 42 \\
139 & 57570.2390 & 0.0004 & $-$0&0037 & 130 \\
143 & 57570.4745 & 0.0034 & $-$0&0000 & 14 \\
161 & 57571.5138 & 0.0028 & $-$0&0041 & 18 \\
173 & 57572.2131 & 0.0008 & $-$0&0003 & 133 \\
174 & 57572.2713 & 0.0018 & $-$0&0001 & 87 \\
190 & 57573.2041 & 0.0008 & 0&0053 & 125 \\
191 & 57573.2620 & 0.0009 & 0&0052 & 109 \\
207 & 57574.1896 & 0.0074 & 0&0054 & 56 \\
208 & 57574.2460 & 0.0010 & 0&0039 & 122 \\
\hline
  \multicolumn{6}{l}{\commenta BJD$-$2400000.} \\
  \multicolumn{6}{l}{\commentb Against max $= 2457562.1857 + 0.057964 E$.} \\
  \multicolumn{6}{l}{\commentc Number of points used to determine the maximum.} \\
\end{tabular}
\end{center}
\end{table}

\subsection{ASASSN-16gl}\label{obj:asassn16gl}

\begin{figure}
  \begin{center}
    \FigureFile(85mm,110mm){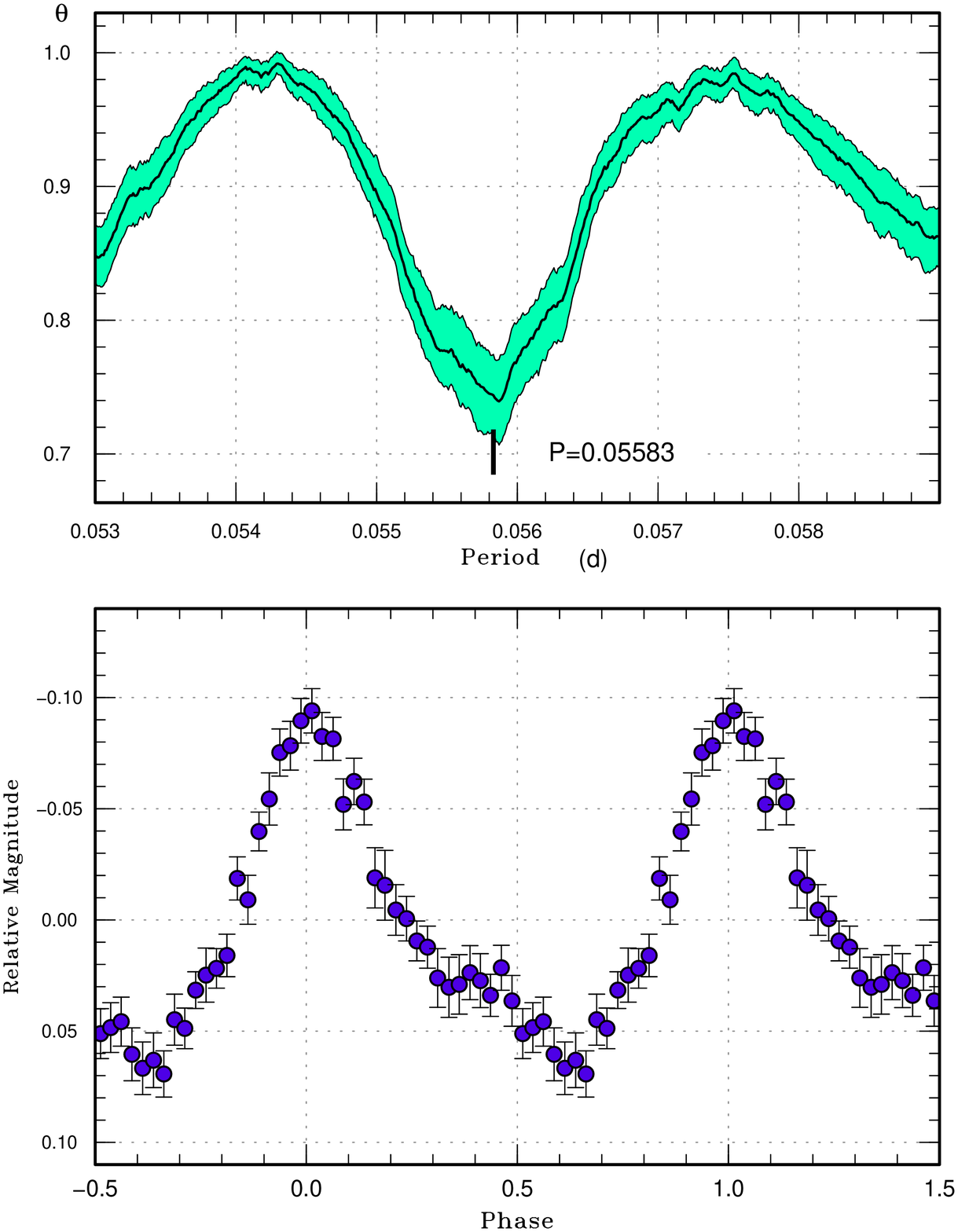}
  \end{center}
  \caption{Ordinary superhumps in ASASSN-16gl (2016).
     (Upper): PDM analysis.
     (Lower): Phase-averaged profile.}
  \label{fig:asassn16glshpdm}
\end{figure}

   This object was detected as a transient
at $V$=14.8 on 2016 June 19 by the ASAS-SN team.
The outburst was announced on June 22, when the object
faded to $V$=15.4.  Subsequent observations detected
superhumps (vsnet-alert 19940, 19942, 19949;
figure \ref{fig:asassn16glshpdm}).
The times of superhump maxima are listed in
table \ref{tab:asassn16gloc2016}.
The object was still in outburst on July 11,
and the entire duration of the superoutburst exceeded
22~d.  Although our initial observation was already 8~d
after the outburst detection, the possibility that
we observed stage C superhumps may be rather
small considering the long
duration of the superoutburst (stage B-C transitions
usually occur very late in such systems,
e.g. SW UMa in figure \ref{fig:stagerev}).


\begin{table}
\caption{Superhump maxima of ASASSN-16gl (2016)}\label{tab:asassn16gloc2016}
\begin{center}
\begin{tabular}{rp{55pt}p{40pt}r@{.}lr}
\hline
\multicolumn{1}{c}{$E$} & \multicolumn{1}{c}{max\commenta} & \multicolumn{1}{c}{error} & \multicolumn{2}{c}{$O-C$\commentb} & \multicolumn{1}{c}{$N$\commentc} \\
\hline
0 & 57569.2036 & 0.0009 & $-$0&0003 & 15 \\
2 & 57569.3180 & 0.0006 & 0&0024 & 128 \\
3 & 57569.3714 & 0.0005 & $-$0&0001 & 127 \\
4 & 57569.4287 & 0.0006 & 0&0013 & 128 \\
5 & 57569.4846 & 0.0008 & 0&0015 & 122 \\
6 & 57569.5388 & 0.0008 & $-$0&0002 & 116 \\
7 & 57569.5933 & 0.0016 & $-$0&0015 & 74 \\
8 & 57569.6507 & 0.0015 & 0&0000 & 88 \\
9 & 57569.7068 & 0.0010 & 0&0003 & 17 \\
10 & 57569.7630 & 0.0011 & 0&0007 & 17 \\
11 & 57569.8171 & 0.0011 & $-$0&0011 & 14 \\
20 & 57570.3185 & 0.0008 & $-$0&0021 & 128 \\
21 & 57570.3732 & 0.0010 & $-$0&0033 & 115 \\
22 & 57570.4315 & 0.0008 & $-$0&0008 & 124 \\
23 & 57570.4890 & 0.0015 & 0&0009 & 86 \\
24 & 57570.5436 & 0.0010 & $-$0&0004 & 98 \\
25 & 57570.6036 & 0.0028 & 0&0038 & 30 \\
27 & 57570.7118 & 0.0011 & 0&0003 & 17 \\
28 & 57570.7663 & 0.0023 & $-$0&0011 & 17 \\
45 & 57571.7162 & 0.0013 & $-$0&0003 & 17 \\
100 & 57574.7851 & 0.0026 & $-$0&0022 & 14 \\
117 & 57575.7377 & 0.0033 & 0&0011 & 13 \\
118 & 57575.7936 & 0.0042 & 0&0013 & 12 \\
\hline
  \multicolumn{6}{l}{\commenta BJD$-$2400000.} \\
  \multicolumn{6}{l}{\commentb Against max $= 2457569.2040 + 0.055834 E$.} \\
  \multicolumn{6}{l}{\commentc Number of points used to determine the maximum.} \\
\end{tabular}
\end{center}
\end{table}

\subsection{ASASSN-16hi}\label{obj:asassn16hi}

\begin{figure}
  \begin{center}
    \FigureFile(85mm,110mm){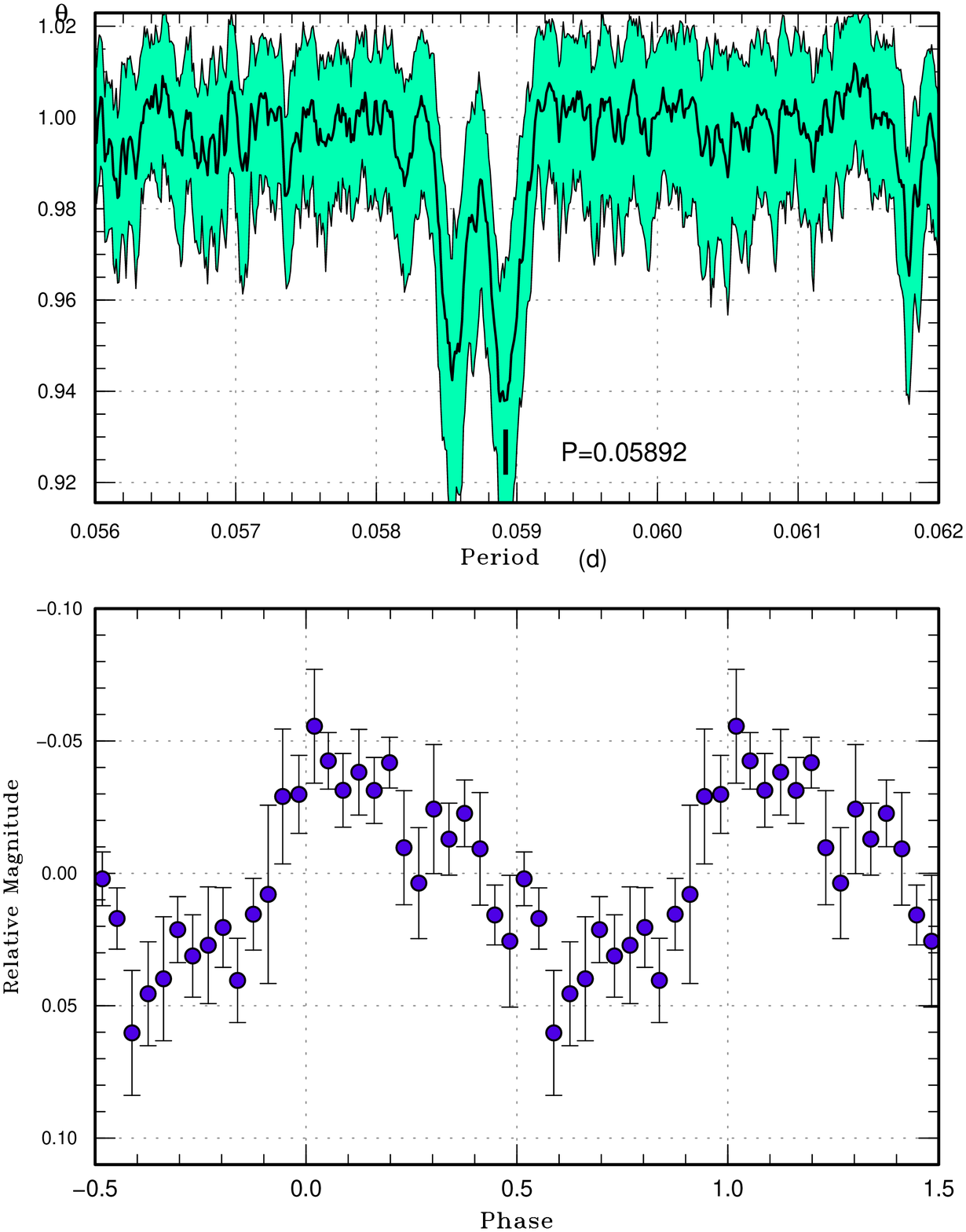}
  \end{center}
  \caption{Superhumps in ASASSN-16hi (2016).
     (Upper): PDM analysis.
     (Lower): Phase-averaged profile.}
  \label{fig:asassn16hishpdm}
\end{figure}

   This object was detected as a transient
at $V$=15.5 on 2016 July 15 by the ASAS-SN team.
Subsequent observations detected superhumps
(vsnet-alert 20002, 20018; figure \ref{fig:asassn16hishpdm}).
The times of superhump maxima are listed in
table \ref{tab:asassn16hioc2016}.  The observed
maxima well illustrate typical stages B and C.
Although the outburst was rather well recorded,
the faintness (around 16 mag) made the quality of
the averaged superhump profile rather poor.


\begin{table}
\caption{Superhump maxima of ASASSN-16hi (2016)}\label{tab:asassn16hioc2016}
\begin{center}
\begin{tabular}{rp{55pt}p{40pt}r@{.}lr}
\hline
\multicolumn{1}{c}{$E$} & \multicolumn{1}{c}{max\commenta} & \multicolumn{1}{c}{error} & \multicolumn{2}{c}{$O-C$\commentb} & \multicolumn{1}{c}{$N$\commentc} \\
\hline
0 & 57589.8022 & 0.0018 & 0&0037 & 15 \\
1 & 57589.8600 & 0.0019 & 0&0025 & 16 \\
2 & 57589.9157 & 0.0011 & $-$0&0007 & 12 \\
17 & 57590.7958 & 0.0033 & $-$0&0049 & 15 \\
18 & 57590.8560 & 0.0018 & $-$0&0037 & 16 \\
19 & 57590.9144 & 0.0034 & $-$0&0043 & 13 \\
34 & 57591.7973 & 0.0015 & $-$0&0056 & 19 \\
51 & 57592.8014 & 0.0024 & $-$0&0036 & 19 \\
52 & 57592.8569 & 0.0027 & $-$0&0071 & 20 \\
53 & 57592.9204 & 0.0048 & $-$0&0025 & 14 \\
68 & 57593.8115 & 0.0043 & 0&0043 & 19 \\
69 & 57593.8664 & 0.0035 & 0&0003 & 19 \\
85 & 57594.8077 & 0.0027 & $-$0&0017 & 16 \\
86 & 57594.8664 & 0.0080 & $-$0&0019 & 16 \\
102 & 57595.8167 & 0.0033 & 0&0052 & 15 \\
103 & 57595.8718 & 0.0041 & 0&0013 & 15 \\
118 & 57596.7673 & 0.0024 & 0&0126 & 20 \\
119 & 57596.8236 & 0.0024 & 0&0099 & 27 \\
120 & 57596.8825 & 0.0019 & 0&0099 & 27 \\
121 & 57596.9419 & 0.0014 & 0&0102 & 11 \\
135 & 57597.7583 & 0.0052 & 0&0014 & 17 \\
136 & 57597.8210 & 0.0016 & 0&0052 & 27 \\
137 & 57597.8786 & 0.0019 & 0&0038 & 26 \\
138 & 57597.9378 & 0.0063 & 0&0041 & 8 \\
152 & 57598.7549 & 0.0028 & $-$0&0042 & 16 \\
153 & 57598.8157 & 0.0035 & $-$0&0024 & 26 \\
154 & 57598.8743 & 0.0031 & $-$0&0027 & 26 \\
169 & 57599.7574 & 0.0047 & $-$0&0038 & 19 \\
170 & 57599.8143 & 0.0054 & $-$0&0059 & 26 \\
171 & 57599.8754 & 0.0018 & $-$0&0038 & 26 \\
187 & 57600.8147 & 0.0044 & $-$0&0076 & 27 \\
188 & 57600.8734 & 0.0080 & $-$0&0079 & 27 \\
\hline
  \multicolumn{6}{l}{\commenta BJD$-$2400000.} \\
  \multicolumn{6}{l}{\commentb Against max $= 2457589.7986 + 0.058951 E$.} \\
  \multicolumn{6}{l}{\commentc Number of points used to determine the maximum.} \\
\end{tabular}
\end{center}
\end{table}

\subsection{ASASSN-16hj}\label{obj:asassn16hj}

\begin{figure}
  \begin{center}
    \FigureFile(85mm,110mm){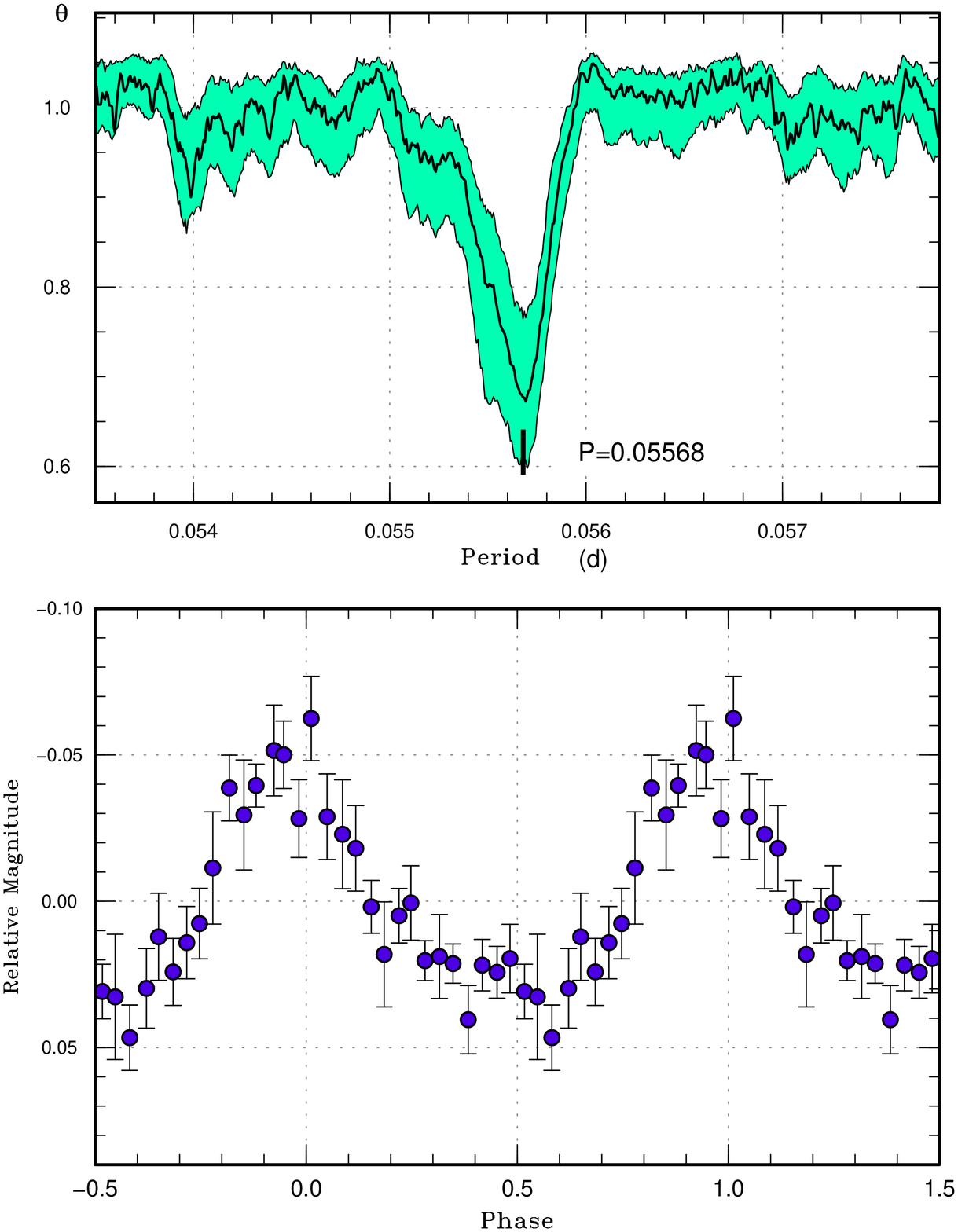}
  \end{center}
  \caption{Ordinary superhumps in ASASSN-16hj before the dip (2016).
     The data segment BJD 2457593--2457604 was used.
     (Upper): PDM analysis.
     (Lower): Phase-averaged profile.}
  \label{fig:asassn16hjshpdm}
\end{figure}

\begin{figure}
  \begin{center}
    \FigureFile(85mm,110mm){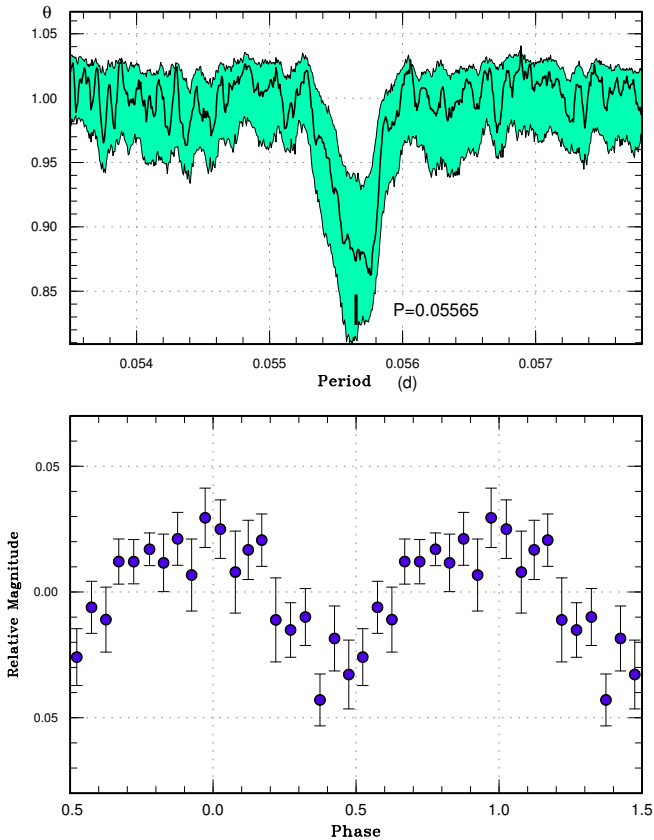}
  \end{center}
  \caption{Ordinary superhumps in ASASSN-16hj during the
     long rebrightening phase (2016).
     The data segment BJD 2457605--2457612 was used.
     (Upper): PDM analysis.
     (Lower): Phase-averaged profile.}
  \label{fig:asassn16hjshrebpdm}
\end{figure}

\begin{figure}
  \begin{center}
    \FigureFile(85mm,100mm){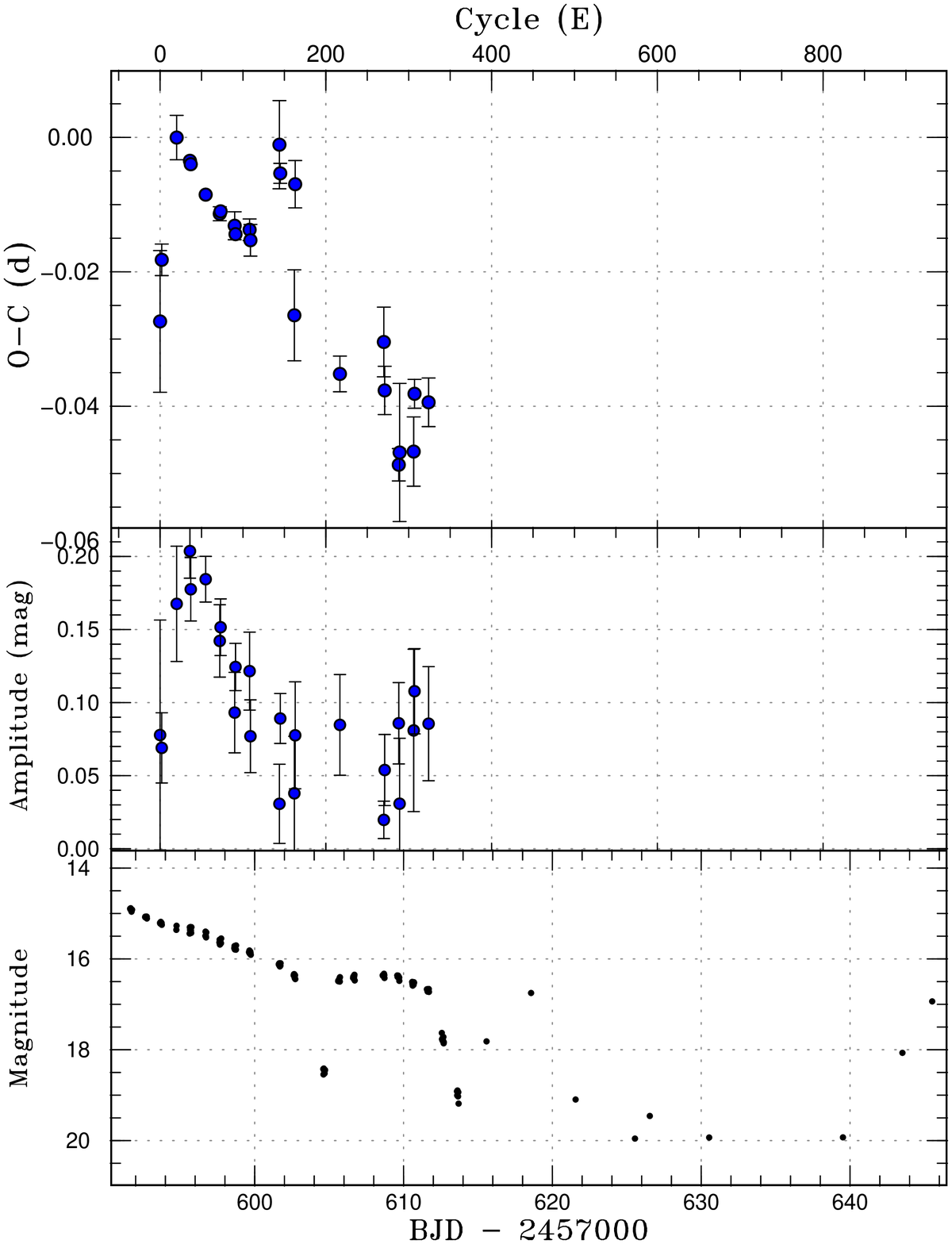}
  \end{center}
  \caption{$O-C$ diagram of superhumps in ASASSN-16hj (2016).
     (Upper:) $O-C$ diagram.
     We used a period of 0.05568~d for calculating the $O-C$ residuals.
     (Middle:) Amplitudes of superhumps.
     (Lower:) Light curve.  The data were binned to 0.018~d.
  }
  \label{fig:asassn16hjhumpall}
\end{figure}

   This object was detected as a transient
at $V$=14.2 on 2016 July 18 by the ASAS-SN team.
Subsequent observations detected likely early superhumps
and ordinary superhumps (vsnet-alert 20006, 20029;
figure \ref{fig:asassn16hjshpdm}).
On August 2--3, the object faded rapidly to
a temporary dip (around 18.5 mag, 
figure \ref{fig:asassn16hjhumpall}).  The object
then entered a plateau-type rebrightening
(vsnet-alert 20049), during which ordinary
superhumps were present (figure \ref{fig:asassn16hjshrebpdm}).
The object showed another separate rebrightening
on August 18 (vsnet-alert 20089).
Although later observations suggested another
rebrightening on September 11--13, the reality of
this rebrightening needs to be confirmed since
it was long after the previous rebrightening and
the observations were at the end of the observing season
(figure \ref{fig:asassn16hjhumpall}).

   The times of superhump maxima are listed in
table \ref{tab:asassn16hjoc2016}.
This table includes superhump maxima after
a short dip.  There were apparent stages A--C,
although observations of stage C were rather poor
(figure \ref{fig:asassn16hjhumpall}).

\begin{figure}
  \begin{center}
    \FigureFile(85mm,110mm){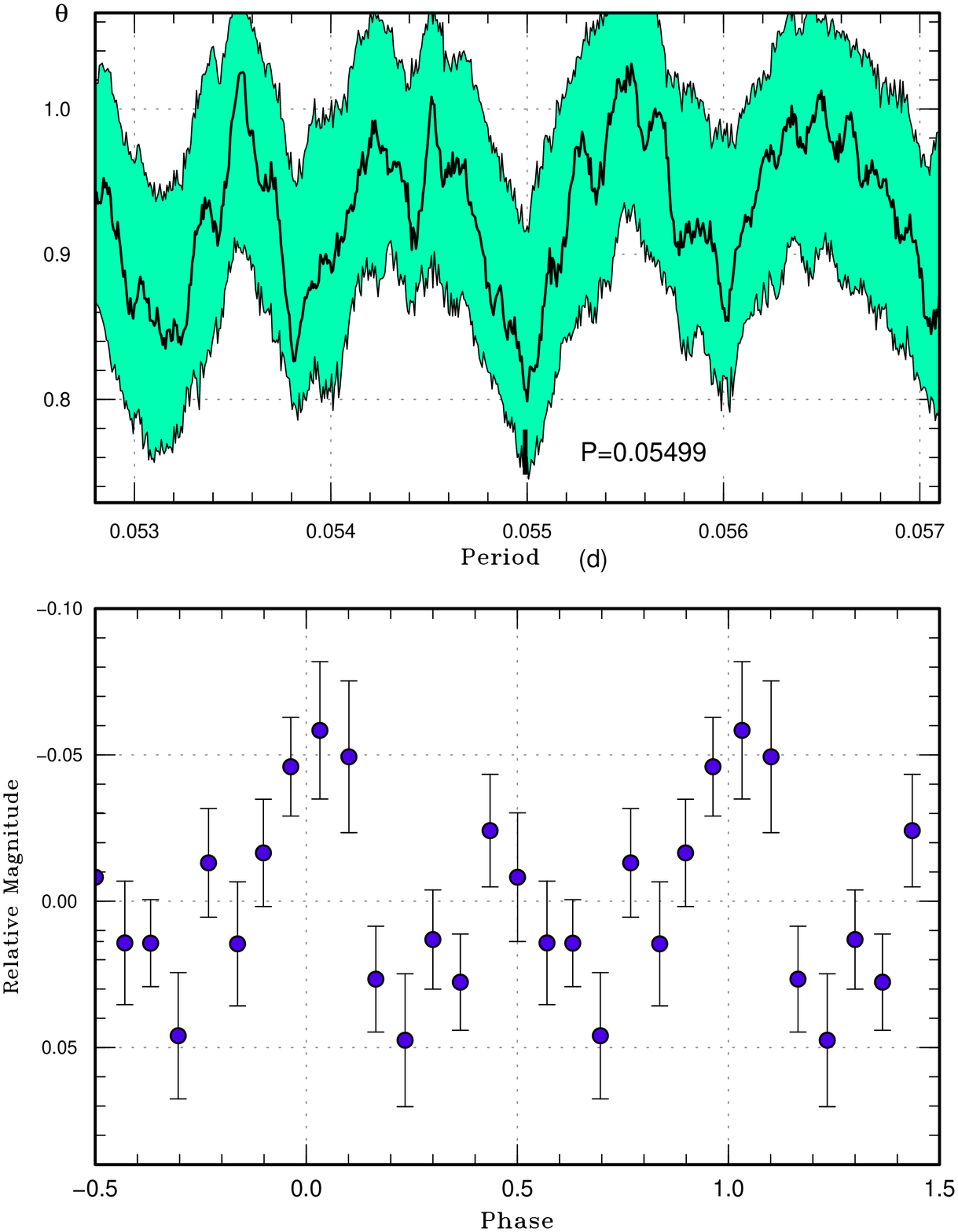}
  \end{center}
  \caption{Possible early superhumps in ASASSN-16hj (2016).
     (Upper): PDM analysis.
     (Lower): Phase-averaged profile.}
  \label{fig:asassn16hjeshpdm}
\end{figure}

   We give the possible signal of early superhumps
in figure \ref{fig:asassn16hjeshpdm}.  Although
the signal was close to the detection limit,
the period appears to be consistent with the
superhump period and the profile is also consistent
with that of early superhumps.
The period with the PDM method was 0.05499(6)~d.
The $\epsilon^*$ for stage A superhumps was 0.034(7),
which corresponds to $q$=0.09(2).


\begin{table}
\caption{Superhump maxima of ASASSN-16hj (2016)}\label{tab:asassn16hjoc2016}
\begin{center}
\begin{tabular}{rp{55pt}p{40pt}r@{.}lr}
\hline
\multicolumn{1}{c}{$E$} & \multicolumn{1}{c}{max\commenta} & \multicolumn{1}{c}{error} & \multicolumn{2}{c}{$O-C$\commentb} & \multicolumn{1}{c}{$N$\commentc} \\
\hline
0 & 57593.6178 & 0.0105 & $-$0&0239 & 6 \\
2 & 57593.7383 & 0.0023 & $-$0&0145 & 16 \\
20 & 57594.7588 & 0.0033 & 0&0058 & 7 \\
36 & 57595.6462 & 0.0005 & 0&0042 & 12 \\
37 & 57595.7014 & 0.0008 & 0&0038 & 13 \\
55 & 57596.6991 & 0.0005 & 0&0014 & 12 \\
72 & 57597.6428 & 0.0010 & 0&0006 & 12 \\
73 & 57597.6988 & 0.0009 & 0&0010 & 13 \\
90 & 57598.6433 & 0.0021 & 0&0009 & 14 \\
91 & 57598.6977 & 0.0009 & $-$0&0002 & 14 \\
108 & 57599.6449 & 0.0016 & 0&0024 & 14 \\
109 & 57599.6990 & 0.0024 & 0&0010 & 14 \\
144 & 57601.6620 & 0.0066 & 0&0193 & 15 \\
145 & 57601.7135 & 0.0015 & 0&0151 & 15 \\
162 & 57602.6389 & 0.0068 & $-$0&0040 & 15 \\
163 & 57602.7141 & 0.0035 & 0&0156 & 16 \\
217 & 57605.6926 & 0.0027 & $-$0&0062 & 22 \\
270 & 57608.6484 & 0.0052 & 0&0047 & 20 \\
271 & 57608.6968 & 0.0036 & $-$0&0024 & 21 \\
288 & 57609.6323 & 0.0024 & $-$0&0114 & 18 \\
289 & 57609.6898 & 0.0103 & $-$0&0095 & 22 \\
306 & 57610.6365 & 0.0052 & $-$0&0073 & 15 \\
307 & 57610.7008 & 0.0021 & 0&0014 & 15 \\
324 & 57611.6461 & 0.0036 & 0&0021 & 15 \\
\hline
  \multicolumn{6}{l}{\commenta BJD$-$2400000.} \\
  \multicolumn{6}{l}{\commentb Against max $= 2457593.6417 + 0.055563 E$.} \\
  \multicolumn{6}{l}{\commentc Number of points used to determine the maximum.} \\
\end{tabular}
\end{center}
\end{table}

\subsection{ASASSN-16ia}\label{obj:asassn16ia}

   This object was detected as a transient
at $V$=14.6 on 2016 August 1 by the ASAS-SN team.
The object was also detected by Gaia (Gaia16azd)
at a magnitude of 16.71 on August 7.\footnote{
  $<$http://gsaweb.ast.cam.ac.uk/alerts/alert/Gaia16azd/$>$.
}  The coordinates of the object were taken from
this Gaia detection.
The object once faded to $V$=17.1 on August 5.
It was observed bright (16.0 mag) again
on August 7 and showed strong early superhumps
(vsnet-alert 20055, 20069, 20076).  The object
was followed until August 15, when early superhumps
were still present.
A transition to ordinary superhumps was not
observed since the object became too faint.

\begin{figure}
  \begin{center}
    \FigureFile(85mm,110mm){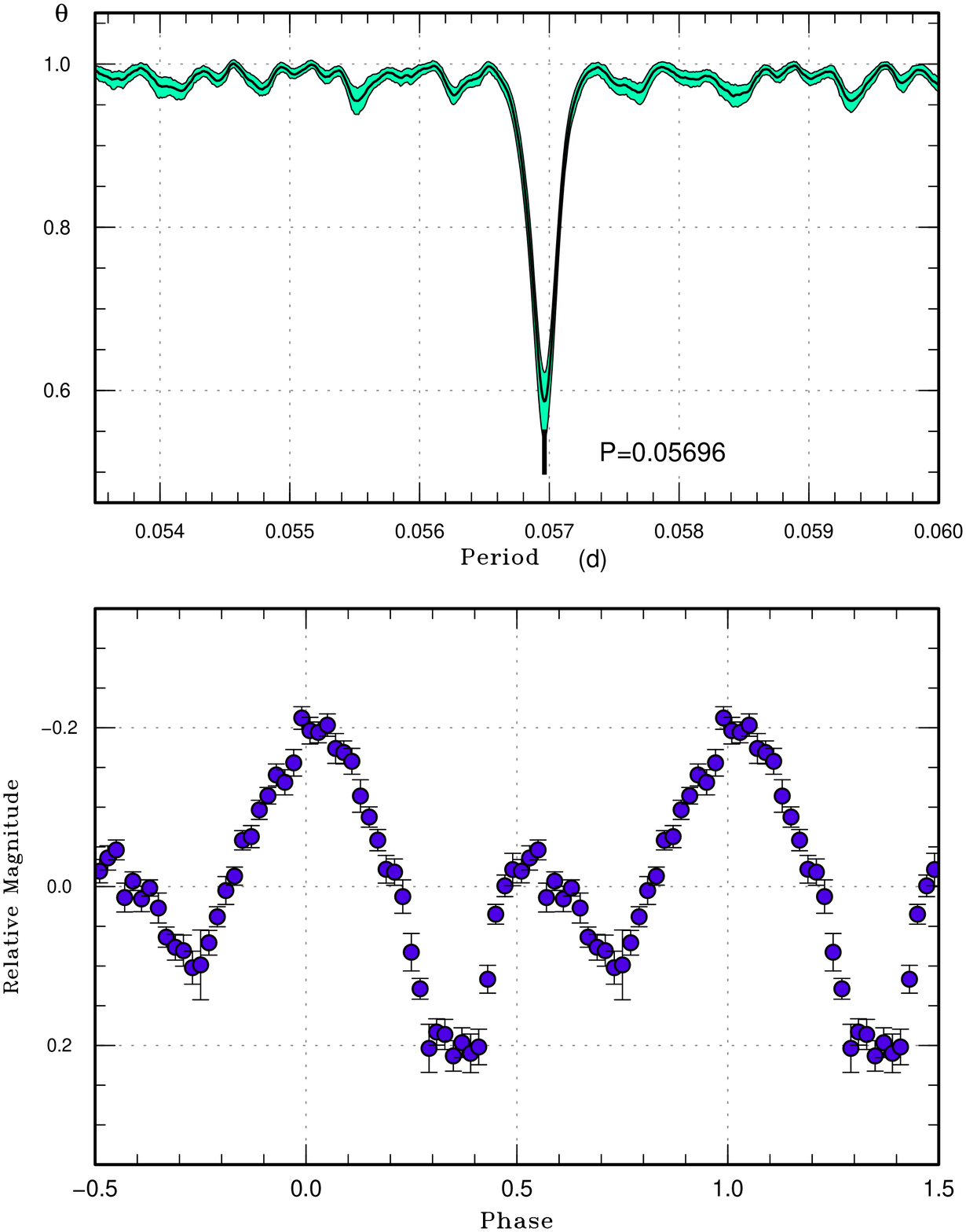}
  \end{center}
  \caption{Early superhumps in ASASSN-16ia (2016).
     (Upper): PDM analysis.
     (Lower): Phase-averaged profile.}
  \label{fig:asassn16iaeshpdm}
\end{figure}

\begin{figure}
  \begin{center}
    \FigureFile(85mm,180mm){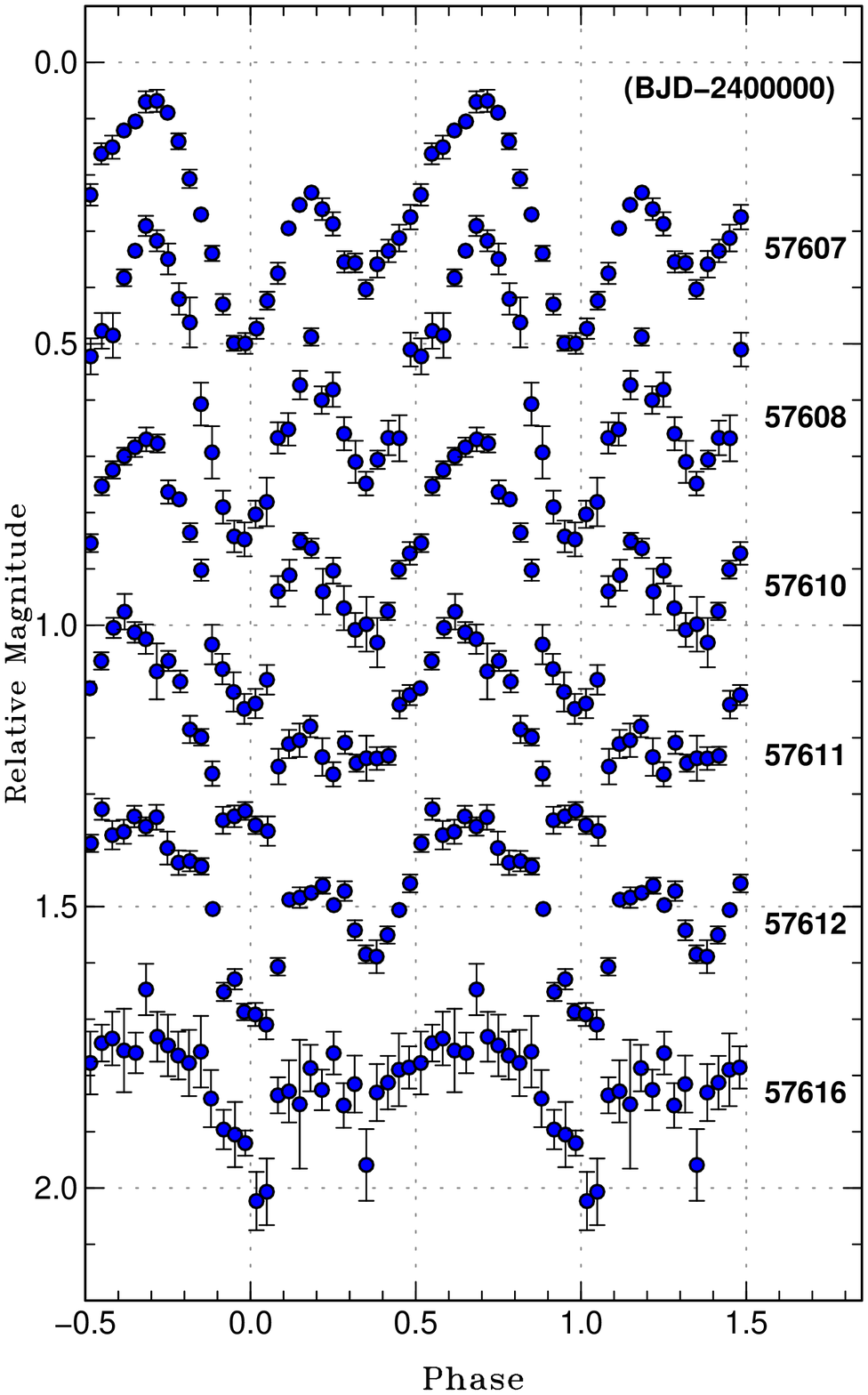}
  \end{center}
  \caption{Evolution of profile of early superhumps in 
     ASASSN-16ia (2016).  A period of 0.056962~d was
     used to draw this figure.  The zero phase
     was defined to be BJD 2457607.428}
  \label{fig:asassn16iaprof}
\end{figure}

   The mean profile of early superhumps is shown
in figure \ref{fig:asassn16iaeshpdm}.  The large
(0.42 mag) full amplitude is exceptional and is
largest among the known WZ Sge-type dwarf novae
(cf. figure 15 in \cite{kat15wzsge}).
The deeper minimum around phase 0.3 in figure
\ref{fig:asassn16iaeshpdm} is somewhat flat-bottomed,
which may be suggestive of an eclipsing component
(see numerical model for MASTER OT J005740.99$+$443101.5
in \cite{Pdot6}).
Nightly variation of early superhumps is shown in
figure \ref{fig:asassn16iaprof}.  It is noteworthy
that the amplitudes of early superhump remained
sufficiently large even 8~d after our initial
observation.  The systematic shift of the phase of
the deeper minimum may reflect the varying degree
of the contribution of the eclipsing component.
Since the object is expected to have a very high
inclination, detailed observations in quiescence
are desired to determine the system parameters.

   It is noteworthy that a precursor outburst
was apparently present before the phase of
the early superhumps.  This is probably the first
case in WZ Sge-type dwarf novae and the reason
why the cooling wave started during the initial
peak needs to be clarified.


\subsection{ASASSN-16ib}\label{obj:asassn16ib}

\begin{figure}
  \begin{center}
    \FigureFile(85mm,110mm){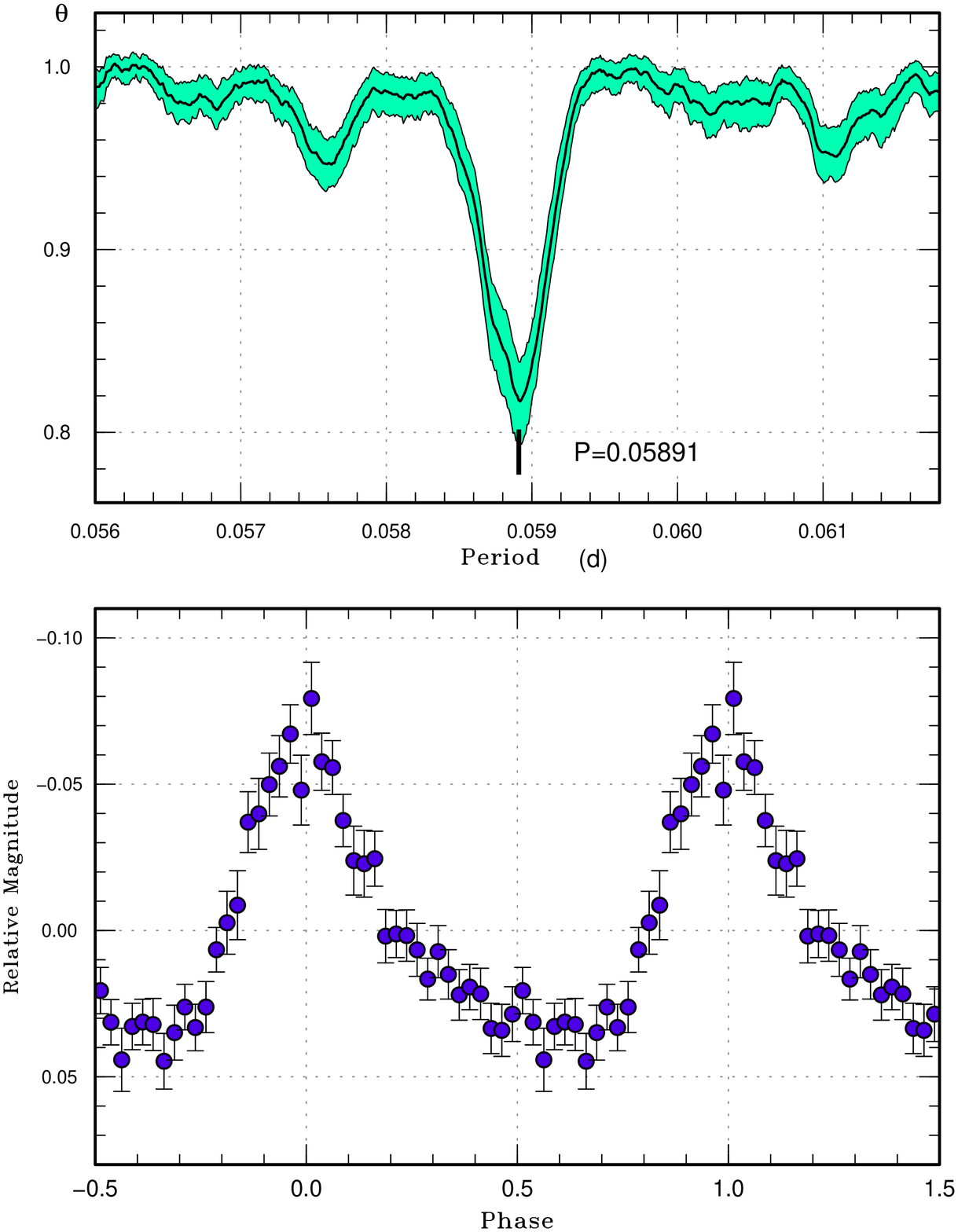}
  \end{center}
  \caption{Superhumps in ASASSN-16ib (2016).
     (Upper): PDM analysis.
     (Lower): Phase-averaged profile.}
  \label{fig:asassn16ibshpdm}
\end{figure}

   This object was detected as a transient
at $V$=14.2 on 2016 August 5 by the ASAS-SN team.
Subsequent observations detected growing superhumps
(vsnet-alert 20066, 20083).
The times of superhump maxima are listed in
table \ref{tab:asassn16ib2016}.  During the epochs for
$E \le$14, the amplitudes of superhumps grew,
and these superhumps can be safely identified
as stage A superhumps.  The distinction of stages
B and C was unclear.  We listed a value for
47 $\le E \le$133 as stage B in table \ref{tab:perlist}.
The mean profile of the superhumps is shown
in figure \ref{fig:asassn16ibshpdm}.


\begin{table}
\caption{Superhump maxima of ASASSN-16ib (2016)}\label{tab:asassn16ib2016}
\begin{center}
\begin{tabular}{rp{55pt}p{40pt}r@{.}lr}
\hline
\multicolumn{1}{c}{$E$} & \multicolumn{1}{c}{max\commenta} & \multicolumn{1}{c}{error} & \multicolumn{2}{c}{$O-C$\commentb} & \multicolumn{1}{c}{$N$\commentc} \\
\hline
0 & 57608.4691 & 0.0119 & $-$0&0160 & 14 \\
13 & 57609.2543 & 0.0007 & 0&0032 & 133 \\
14 & 57609.3136 & 0.0006 & 0&0036 & 134 \\
47 & 57611.2595 & 0.0005 & 0&0051 & 136 \\
48 & 57611.3181 & 0.0006 & 0&0047 & 135 \\
64 & 57612.2579 & 0.0008 & 0&0017 & 136 \\
65 & 57612.3166 & 0.0011 & 0&0015 & 120 \\
68 & 57612.4919 & 0.0017 & 0&0001 & 18 \\
97 & 57614.2021 & 0.0120 & 0&0015 & 60 \\
98 & 57614.2605 & 0.0010 & 0&0010 & 131 \\
99 & 57614.3172 & 0.0013 & $-$0&0012 & 133 \\
119 & 57615.4961 & 0.0042 & $-$0&0009 & 21 \\
133 & 57616.3198 & 0.0054 & $-$0&0021 & 79 \\
187 & 57619.5039 & 0.0020 & 0&0003 & 26 \\
204 & 57620.5057 & 0.0058 & 0&0004 & 26 \\
221 & 57621.5042 & 0.0053 & $-$0&0029 & 39 \\
\hline
  \multicolumn{6}{l}{\commenta BJD$-$2400000.} \\
  \multicolumn{6}{l}{\commentb Against max $= 2457608.4851 + 0.058923 E$.} \\
  \multicolumn{6}{l}{\commentc Number of points used to determine the maximum.} \\
\end{tabular}
\end{center}
\end{table}

\subsection{ASASSN-16ik}\label{obj:asassn16ik}

\begin{figure}
  \begin{center}
    \FigureFile(85mm,110mm){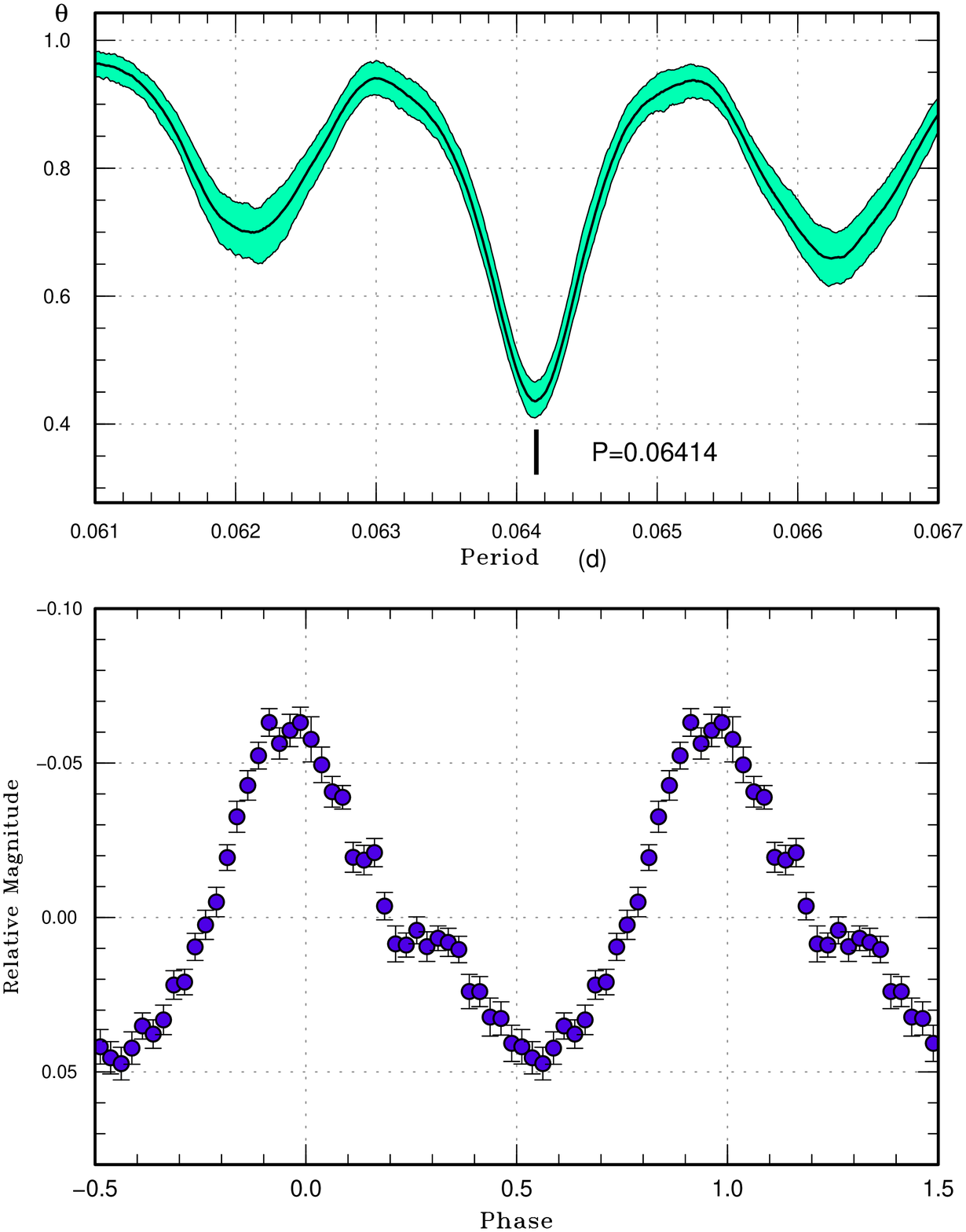}
  \end{center}
  \caption{Superhumps in ASASSN-16ik during stage B (2016).
     (Upper): PDM analysis.
     (Lower): Phase-averaged profile.}
  \label{fig:asassn16ikshpdm}
\end{figure}

   This object was detected as a transient
at $V$=15.26 on 2016 August 6 by the ASAS-SN team.
The object further brightened to $V$=13.9 on
August 8.
The object started to show superhumps on August
11--12 (vsnet-alert 20082; figure \ref{fig:asassn16ikshpdm}).
The times of superhump maxima are listed in
table \ref{tab:asassn16ikoc2016}.  The data very
clearly show stages A (growing superhumps) and B.
The object showed a rebrightening on August 25
(vsnet-alert 20109), which faded rapidly.
During this rebrightening phase, a weak superhump
signal was detected with a period of 0.0649(3)~d.


\begin{table}
\caption{Superhump maxima of ASASSN-16ik (2016)}\label{tab:asassn16ikoc2016}
\begin{center}
\begin{tabular}{rp{55pt}p{40pt}r@{.}lr}
\hline
\multicolumn{1}{c}{$E$} & \multicolumn{1}{c}{max\commenta} & \multicolumn{1}{c}{error} & \multicolumn{2}{c}{$O-C$\commentb} & \multicolumn{1}{c}{$N$\commentc} \\
\hline
0 & 57611.4511 & 0.0031 & $-$0&0274 & 147 \\
1 & 57611.5177 & 0.0020 & $-$0&0253 & 148 \\
2 & 57611.5843 & 0.0017 & $-$0&0231 & 163 \\
14 & 57612.3862 & 0.0011 & 0&0056 & 146 \\
15 & 57612.4513 & 0.0018 & 0&0064 & 88 \\
17 & 57612.5819 & 0.0008 & 0&0081 & 13 \\
18 & 57612.6469 & 0.0010 & 0&0087 & 17 \\
33 & 57613.6169 & 0.0005 & 0&0122 & 16 \\
34 & 57613.6832 & 0.0008 & 0&0141 & 17 \\
45 & 57614.3886 & 0.0004 & 0&0108 & 146 \\
46 & 57614.4517 & 0.0004 & 0&0094 & 147 \\
47 & 57614.5158 & 0.0004 & 0&0092 & 144 \\
48 & 57614.5835 & 0.0006 & 0&0124 & 74 \\
49 & 57614.6439 & 0.0006 & 0&0084 & 23 \\
64 & 57615.6038 & 0.0010 & 0&0018 & 22 \\
65 & 57615.6699 & 0.0008 & 0&0036 & 22 \\
76 & 57616.3758 & 0.0005 & 0&0008 & 148 \\
77 & 57616.4376 & 0.0006 & $-$0&0018 & 147 \\
78 & 57616.5021 & 0.0021 & $-$0&0018 & 60 \\
80 & 57616.6340 & 0.0009 & 0&0013 & 22 \\
81 & 57616.7004 & 0.0041 & 0&0032 & 10 \\
95 & 57617.5932 & 0.0014 & $-$0&0060 & 21 \\
96 & 57617.6612 & 0.0014 & $-$0&0024 & 22 \\
111 & 57618.6254 & 0.0018 & $-$0&0046 & 22 \\
112 & 57618.6849 & 0.0015 & $-$0&0095 & 13 \\
126 & 57619.5827 & 0.0022 & $-$0&0138 & 19 \\
\hline
  \multicolumn{6}{l}{\commenta BJD$-$2400000.} \\
  \multicolumn{6}{l}{\commentb Against max $= 2457611.4786 + 0.064428 E$.} \\
  \multicolumn{6}{l}{\commentc Number of points used to determine the maximum.} \\
\end{tabular}
\end{center}
\end{table}

\subsection{ASASSN-16is}\label{obj:asassn16is}

\begin{figure}
  \begin{center}
    \FigureFile(85mm,110mm){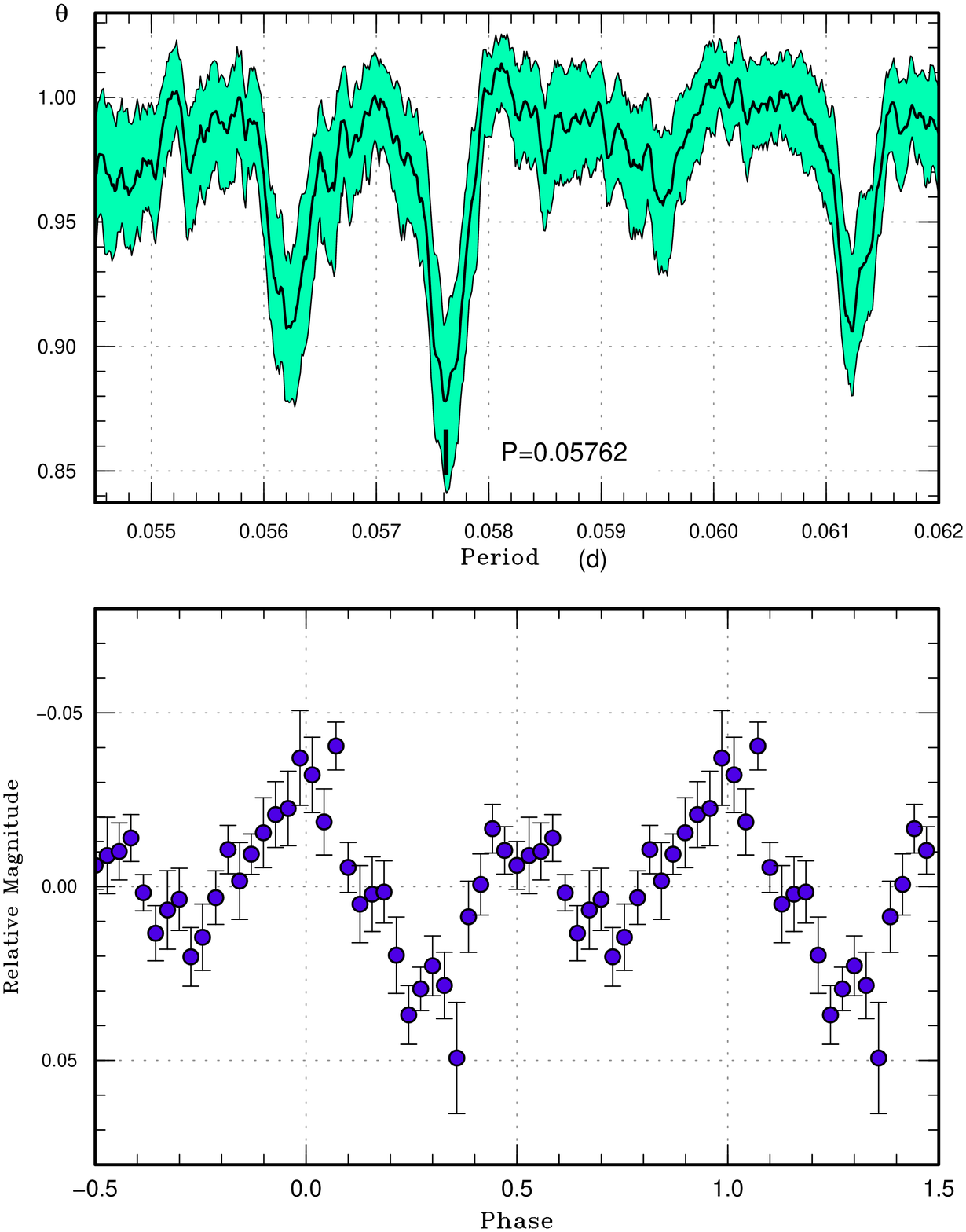}
  \end{center}
  \caption{Early superhumps in ASASSN-16is (2016).
     (Upper): PDM analysis.
     (Lower): Phase-averaged profile.}
  \label{fig:asassn16iseshpdm}
\end{figure}

\begin{figure}
  \begin{center}
    \FigureFile(85mm,110mm){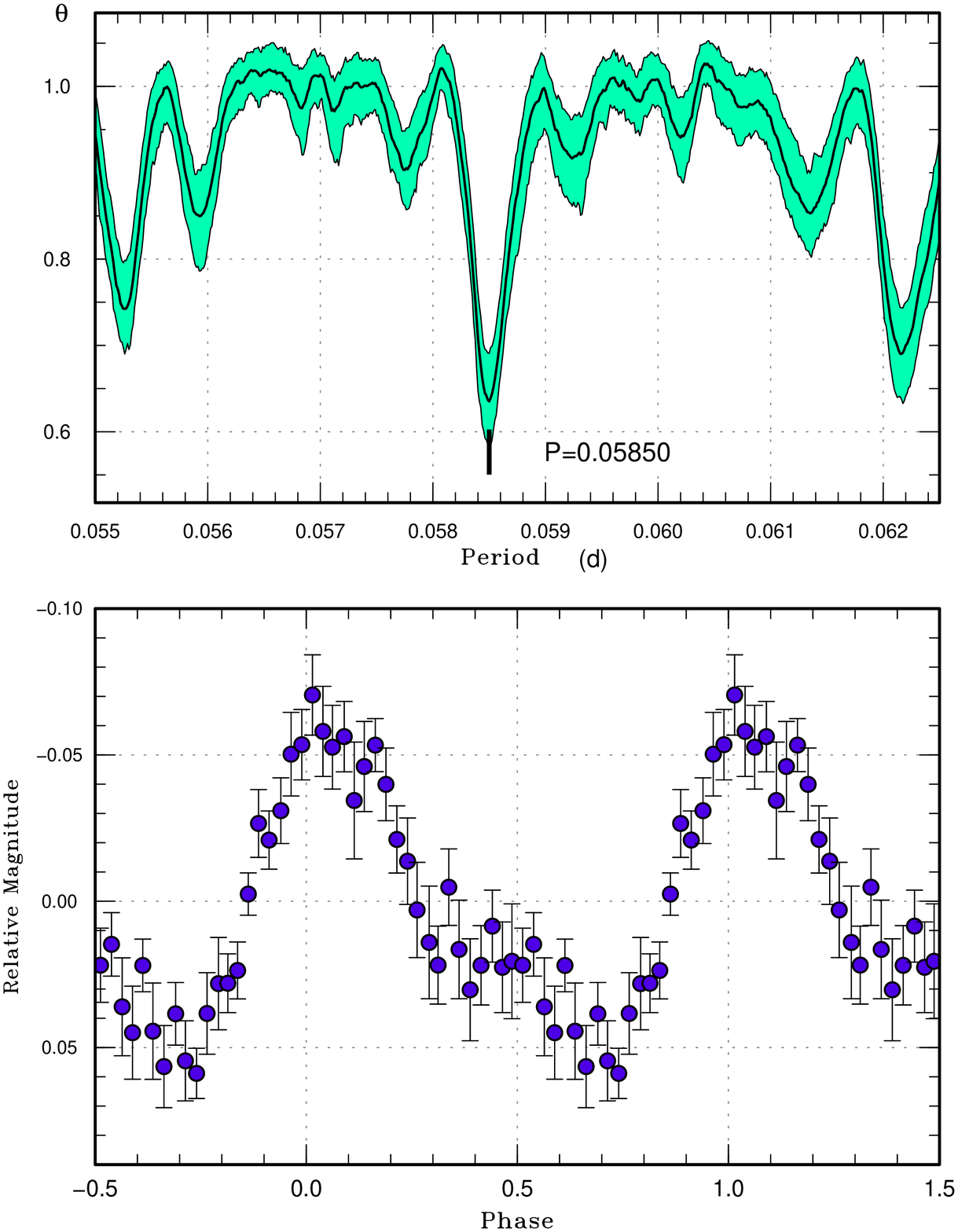}
  \end{center}
  \caption{Ordinary superhumps in ASASSN-16is (2016).
     (Upper): PDM analysis.
     (Lower): Phase-averaged profile.}
  \label{fig:asassn16isshpdm}
\end{figure}

   This object was detected as a transient
at $V$=14.9 on 2016 August 9 by the ASAS-SN team.
Initial observations detected double-wave modulations
attributable to early superhumps
(vsnet-alert 20078, 20084; figure \ref{fig:asassn16iseshpdm}).
The period of early superhumps was 0.05762(2)~d.
The object started to show ordinary superhumps
at least on August 20 (vsnet-alert 20101, 20106;
figure \ref{fig:asassn16isshpdm}).
The times of superhump maxima are listed in
table \ref{tab:asassn16isoc2016}.
The superoutburst plateau was terminated by
rapid fading on August 28.
The object is confirmed to be a WZ Sge-type dwarf nova.


\begin{table}
\caption{Superhump maxima of ASASSN-16is (2016)}\label{tab:asassn16isoc2016}
\begin{center}
\begin{tabular}{rp{55pt}p{40pt}r@{.}lr}
\hline
\multicolumn{1}{c}{$E$} & \multicolumn{1}{c}{max\commenta} & \multicolumn{1}{c}{error} & \multicolumn{2}{c}{$O-C$\commentb} & \multicolumn{1}{c}{$N$\commentc} \\
\hline
0 & 57621.3260 & 0.0007 & 0&0012 & 20 \\
1 & 57621.3835 & 0.0008 & 0&0003 & 26 \\
20 & 57622.4946 & 0.0006 & 0&0002 & 39 \\
21 & 57622.5531 & 0.0005 & 0&0002 & 54 \\
69 & 57625.3568 & 0.0005 & $-$0&0033 & 34 \\
70 & 57625.4167 & 0.0007 & $-$0&0019 & 37 \\
87 & 57626.4111 & 0.0007 & $-$0&0018 & 43 \\
88 & 57626.4727 & 0.0019 & 0&0013 & 22 \\
103 & 57627.3522 & 0.0008 & 0&0036 & 30 \\
104 & 57627.4073 & 0.0009 & 0&0001 & 50 \\
105 & 57627.4658 & 0.0009 & 0&0002 & 39 \\
\hline
  \multicolumn{6}{l}{\commenta BJD$-$2400000.} \\
  \multicolumn{6}{l}{\commentb Against max $= 2457621.3248 + 0.058484 E$.} \\
  \multicolumn{6}{l}{\commentc Number of points used to determine the maximum.} \\
\end{tabular}
\end{center}
\end{table}

\subsection{ASASSN-16iu}\label{obj:asassn16iu}

\begin{figure}
  \begin{center}
    \FigureFile(85mm,110mm){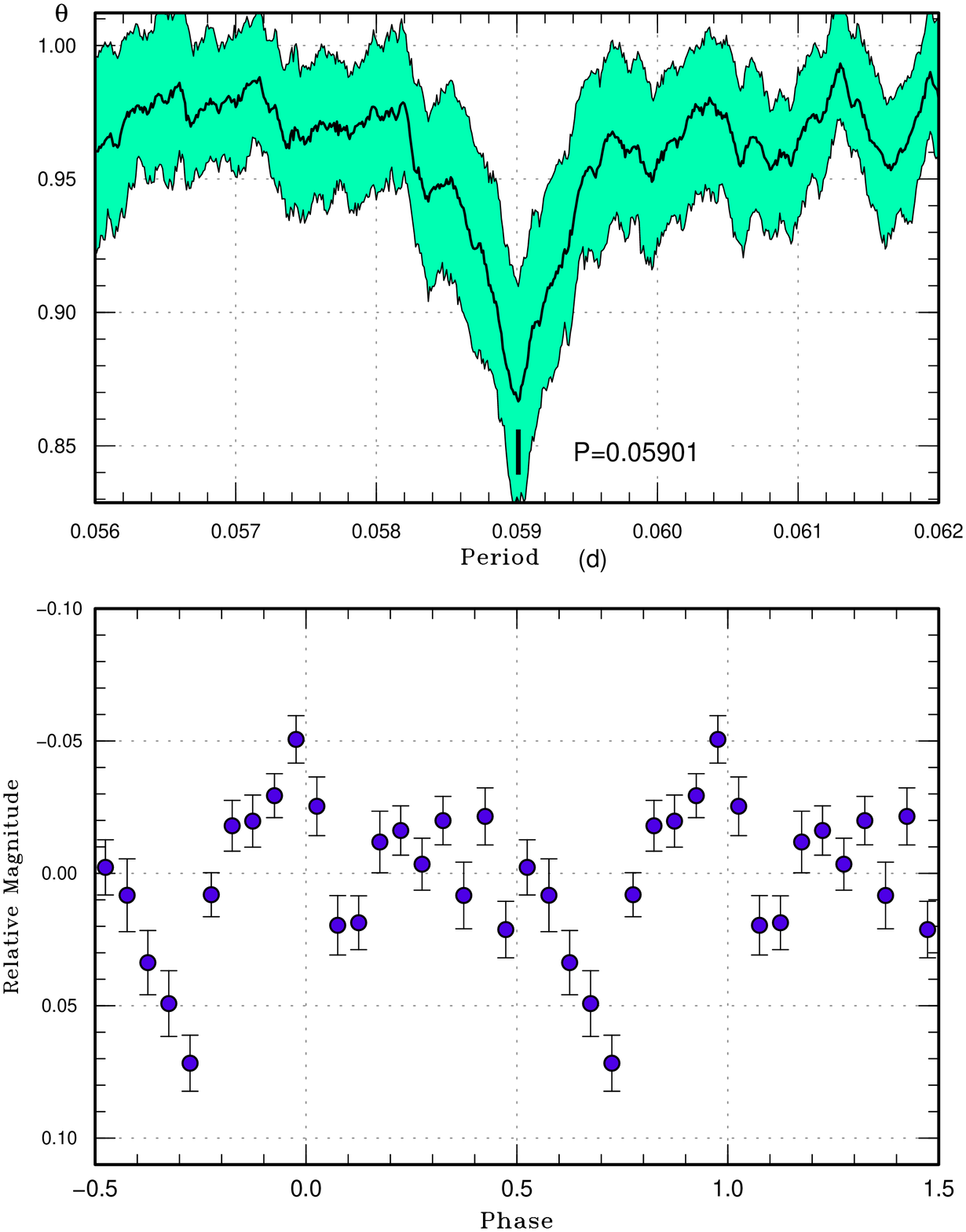}
  \end{center}
  \caption{Superhumps in ASASSN-16iu (2016).
     The data segment BJD 2457615--2457619 was used
     when superhumps were continuously detected.
     (Upper): PDM analysis.
     (Lower): Phase-averaged profile.}
  \label{fig:asassn16iushpdm}
\end{figure}

   This object was detected as a transient
at $V$=15.3 on 2016 August 4 by the ASAS-SN team.
The object once faded to fainter than $V$=17.6
on August 6 and brightened again to $V$=15.2
on August 9.  The detection of the outburst
was announced after this brightening.
Superhumps were soon detected on August 11
(vsnet-alert 20075).  The amplitudes of superhumps
decreased and they became less prominent
on subsequent nights.  They became detectable
again on August 15 (figure \ref{fig:asassn16iushpdm}).
The times of superhump maxima are listed in
table \ref{tab:asassn16iuoc2016}.
Due to the long period of undetectable superhumps,
the $P_{\rm dot}$ for stage B superhumps
is very uncertain.
The period for stage C given in
table \ref{tab:perlist} is very approximate
due to the short baseline.


\begin{table}
\caption{Superhump maxima of ASASSN-16iu (2016)}\label{tab:asassn16iuoc2016}
\begin{center}
\begin{tabular}{rp{55pt}p{40pt}r@{.}lr}
\hline
\multicolumn{1}{c}{$E$} & \multicolumn{1}{c}{max\commenta} & \multicolumn{1}{c}{error} & \multicolumn{2}{c}{$O-C$\commentb} & \multicolumn{1}{c}{$N$\commentc} \\
\hline
0 & 57611.7578 & 0.0015 & 0&0071 & 37 \\
1 & 57611.8157 & 0.0011 & 0&0062 & 36 \\
2 & 57611.8753 & 0.0009 & 0&0072 & 39 \\
68 & 57615.7425 & 0.0014 & $-$0&0048 & 22 \\
69 & 57615.7942 & 0.0031 & $-$0&0118 & 21 \\
70 & 57615.8549 & 0.0012 & $-$0&0099 & 22 \\
71 & 57615.9147 & 0.0027 & $-$0&0089 & 13 \\
82 & 57616.5640 & 0.0024 & $-$0&0061 & 137 \\
83 & 57616.6224 & 0.0012 & $-$0&0064 & 129 \\
85 & 57616.7401 & 0.0029 & $-$0&0063 & 22 \\
86 & 57616.7998 & 0.0028 & $-$0&0054 & 21 \\
87 & 57616.8573 & 0.0025 & $-$0&0067 & 23 \\
102 & 57617.7467 & 0.0101 & 0&0011 & 21 \\
103 & 57617.8182 & 0.0043 & 0&0139 & 21 \\
104 & 57617.8752 & 0.0041 & 0&0121 & 23 \\
119 & 57618.7466 & 0.0024 & 0&0019 & 20 \\
120 & 57618.8082 & 0.0034 & 0&0047 & 20 \\
121 & 57618.8745 & 0.0108 & 0&0122 & 22 \\
\hline
  \multicolumn{6}{l}{\commenta BJD$-$2400000.} \\
  \multicolumn{6}{l}{\commentb Against max $= 2457611.7506 + 0.058774 E$.} \\
  \multicolumn{6}{l}{\commentc Number of points used to determine the maximum.} \\
\end{tabular}
\end{center}
\end{table}

\subsection{ASASSN-16iw}\label{obj:asassn16iw}

   This object was detected as a transient
at $V$=13.9 on 2016 August 10 by the ASAS-SN team.
There was a faint ($g$=21.9) SDSS counterpart (there were
17 measurements in the SDSS data with a range
of 21.8--22.2 in $g$) and the large outburst amplitude
suggested a WZ Sge-type dwarf nova.

\begin{figure}
  \begin{center}
    \FigureFile(85mm,110mm){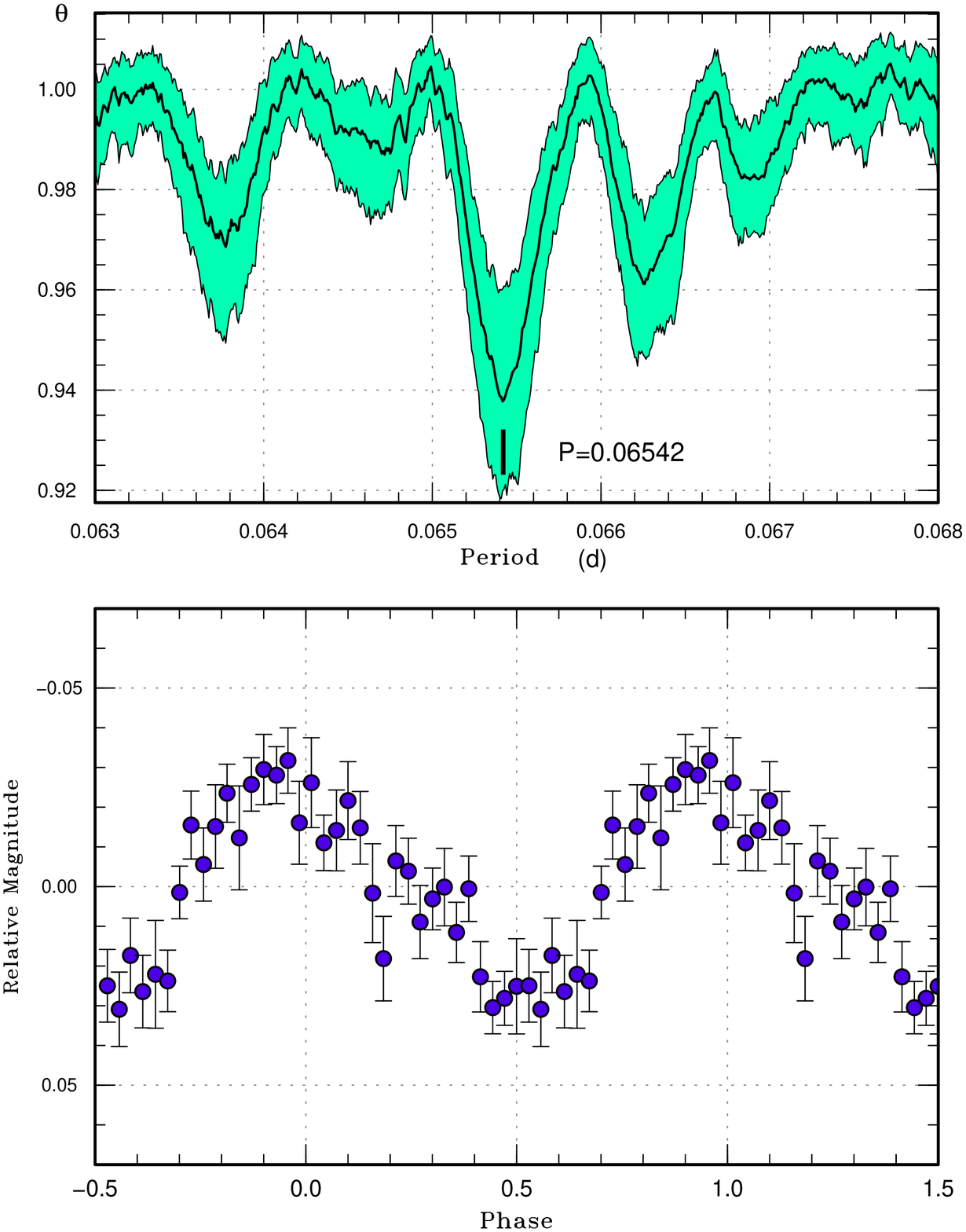}
  \end{center}
  \caption{Ordinary superhumps in ASASSN-16iw (2016).
     The data segment BJD 2457617.5--2457628 was used.
     (Upper): PDM analysis.
     (Lower): Phase-averaged profile.}
  \label{fig:asassn16iwshpdm}
\end{figure}

\begin{figure}
  \begin{center}
    \FigureFile(85mm,100mm){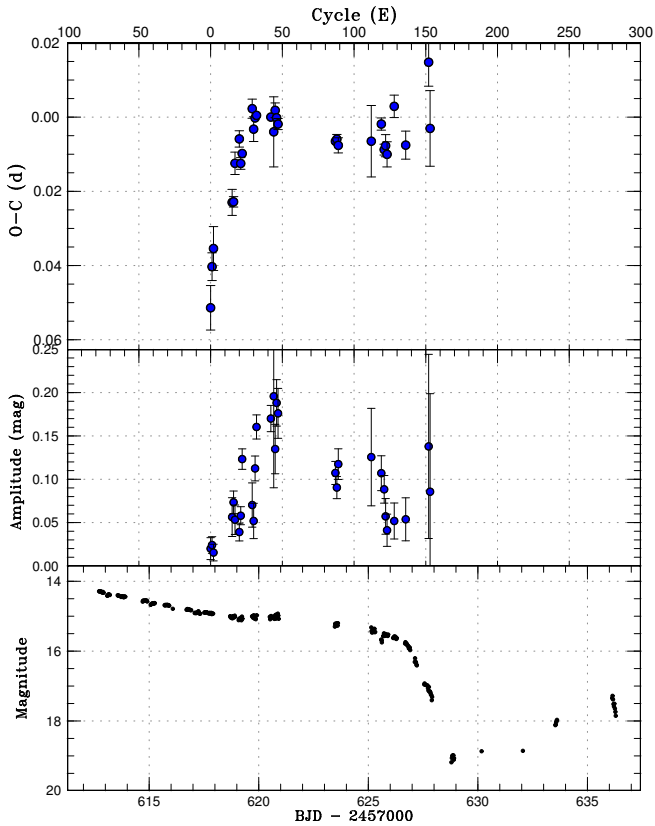}
  \end{center}
  \caption{$O-C$ diagram of superhumps in ASASSN-16iw (2016).
     (Upper:) $O-C$ diagram.
     We used a period of 0.06546~d for calculating the $O-C$ residuals.
     (Middle:) Amplitudes of superhumps.
     (Lower:) Light curve.  The data were binned to 0.022~d.
     The final part of this figure (BJD 2457636) corresponds
     to the fading part from the first rebrightening.
  }
  \label{fig:asassn16iwhumpall}
\end{figure}

   The object started to show superhumps on August 17
(vsnet-alert 20086, 20091, 20100;
figure \ref{fig:asassn16iwshpdm}).
The times of superhump maxima are listed in
table \ref{tab:asassn16iwoc2016}.
The superhumps grew slowly and it took at least
47 cycles to reach the full superhump amplitude.
Based on $O-C$ variations, we have identified
$E \le$32 to be stage A superhumps
(figure \ref{fig:asassn16iwhumpall}).

\begin{figure}
  \begin{center}
    \FigureFile(85mm,70mm){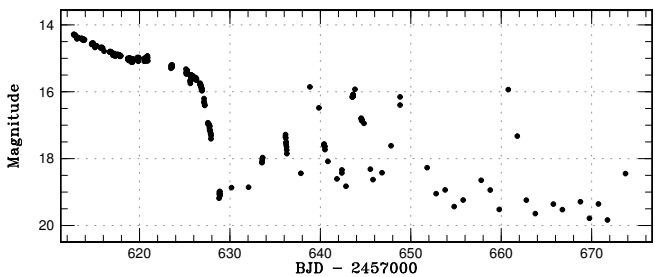}
  \end{center}
  \caption{Overall light curve of ASASSN-16iw (2016).
     The data were binned to 0.022~d.  The data on
     BJD 2457636 represent observations of the fading
     barnch of the first rebrightening.  There may
     have been a rebrightening on BJD 2457674,
     which was not well sampled.
  }
  \label{fig:asassn16iwoutlc}
\end{figure}

   The object showed at least five post-superoutburst
rebrightenings (vsnet-alert 20129, 20147, 20164, 20167:
figure \ref{fig:asassn16iwoutlc}).

\begin{figure}
  \begin{center}
    \FigureFile(85mm,110mm){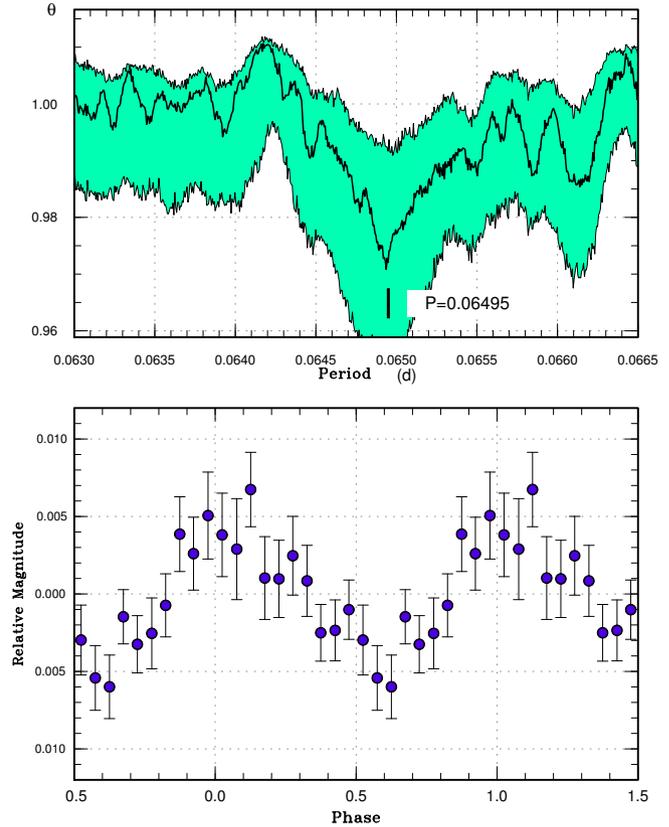}
  \end{center}
  \caption{Possible early superhumps in ASASSN-16iw (2016).
     The data segment before BJD 2457617.5 was used.
     (Upper): PDM analysis.
     (Lower): Phase-averaged profile.}
  \label{fig:asassn16iweshpdm}
\end{figure}

   An analysis of the early part of the light curve
detected a possible signal of early superhumps
(figure \ref{fig:asassn16iweshpdm}).  Although
the signal was weak (the amplitude was smaller than 0.01 mag)
and the profile was not doubly humped as expected
for early superhumps, we suspect that this is
a candidate period of early superhumps since the
period excess was close to what is expected for
a WZ Sge-type dwarf nova.  The period was 0.06495(5)~d.
The $\epsilon^*$ of 0.029(1) for stage A superhumps
corresponds to $q$=0.079(2).  This $q$ value is not
as small as expected for a period bouncer at this
orbital period.  The relatively large $P_{\rm dot}$
for stage B superhump may also be suggestive for
a relatively large $q$.  The object may be similar
to WZ Sge-type dwarf novae with multiple rebrightenings
with relatively large $q$, such as MASTER OT J211258.65+242145.4
and MASTER OT J203749.39+552210.3 \citep{nak13j2112j2037}.


\begin{table}
\caption{Superhump maxima of ASASSN-16iw (2016)}\label{tab:asassn16iwoc2016}
\begin{center}
\begin{tabular}{rp{55pt}p{40pt}r@{.}lr}
\hline
\multicolumn{1}{c}{$E$} & \multicolumn{1}{c}{max\commenta} & \multicolumn{1}{c}{error} & \multicolumn{2}{c}{$O-C$\commentb} & \multicolumn{1}{c}{$N$\commentc} \\
\hline
0 & 57617.7525 & 0.0060 & $-$0&0338 & 24 \\
1 & 57617.8290 & 0.0037 & $-$0&0228 & 24 \\
2 & 57617.8994 & 0.0059 & $-$0&0181 & 26 \\
15 & 57618.7628 & 0.0035 & $-$0&0074 & 22 \\
16 & 57618.8284 & 0.0014 & $-$0&0074 & 23 \\
17 & 57618.9043 & 0.0030 & 0&0028 & 20 \\
20 & 57619.1072 & 0.0022 & 0&0090 & 41 \\
21 & 57619.1661 & 0.0016 & 0&0022 & 41 \\
22 & 57619.2342 & 0.0010 & 0&0048 & 35 \\
29 & 57619.7045 & 0.0026 & 0&0159 & 19 \\
30 & 57619.7644 & 0.0033 & 0&0102 & 22 \\
31 & 57619.8329 & 0.0011 & 0&0131 & 24 \\
32 & 57619.8991 & 0.0007 & 0&0137 & 21 \\
42 & 57620.5532 & 0.0008 & 0&0118 & 57 \\
44 & 57620.6801 & 0.0094 & 0&0075 & 20 \\
45 & 57620.7514 & 0.0020 & 0&0132 & 30 \\
46 & 57620.8149 & 0.0010 & 0&0111 & 44 \\
47 & 57620.8786 & 0.0015 & 0&0092 & 45 \\
87 & 57623.4924 & 0.0009 & $-$0&0010 & 59 \\
88 & 57623.5584 & 0.0013 & $-$0&0005 & 53 \\
89 & 57623.6222 & 0.0020 & $-$0&0023 & 29 \\
112 & 57625.1289 & 0.0096 & $-$0&0044 & 35 \\
119 & 57625.5918 & 0.0016 & $-$0&0008 & 40 \\
121 & 57625.7158 & 0.0015 & $-$0&0079 & 20 \\
122 & 57625.7823 & 0.0030 & $-$0&0070 & 20 \\
123 & 57625.8454 & 0.0034 & $-$0&0095 & 21 \\
128 & 57626.1857 & 0.0030 & 0&0027 & 99 \\
136 & 57626.6989 & 0.0038 & $-$0&0088 & 17 \\
152 & 57627.7686 & 0.0065 & 0&0113 & 20 \\
153 & 57627.8162 & 0.0102 & $-$0&0067 & 22 \\
\hline
  \multicolumn{6}{l}{\commenta BJD$-$2400000.} \\
  \multicolumn{6}{l}{\commentb Against max $= 2457617.7863 + 0.065599 E$.} \\
  \multicolumn{6}{l}{\commentc Number of points used to determine the maximum.} \\
\end{tabular}
\end{center}
\end{table}

\subsection{ASASSN-16jb}\label{obj:asassn16jb}

\begin{figure}
  \begin{center}
    \FigureFile(85mm,110mm){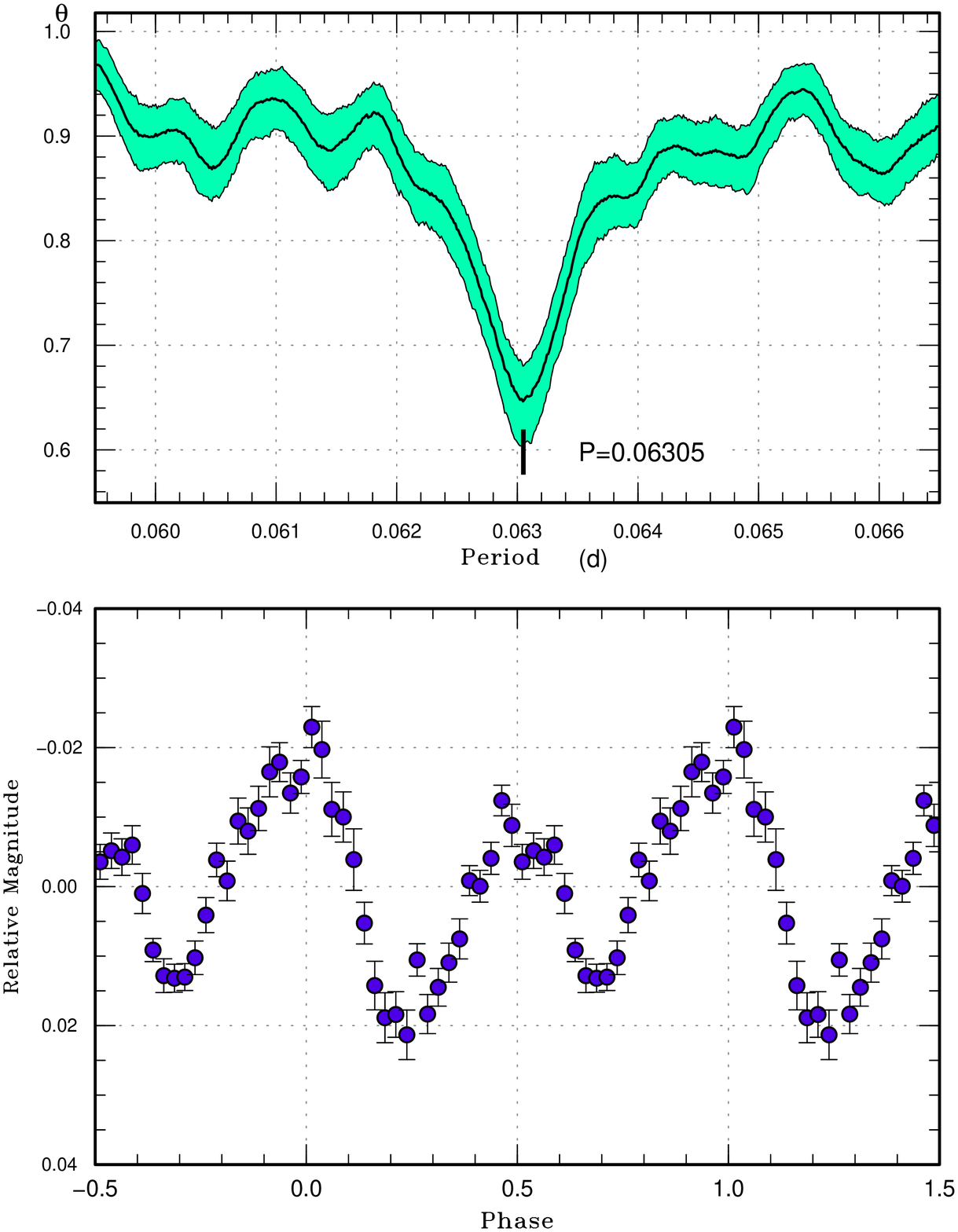}
  \end{center}
  \caption{Early superhumps in ASASSN-16jb (2016).
     (Upper): PDM analysis.
     (Lower): Phase-averaged profile.}
  \label{fig:asassn16jbeshpdm}
\end{figure}

\begin{figure}
  \begin{center}
    \FigureFile(85mm,110mm){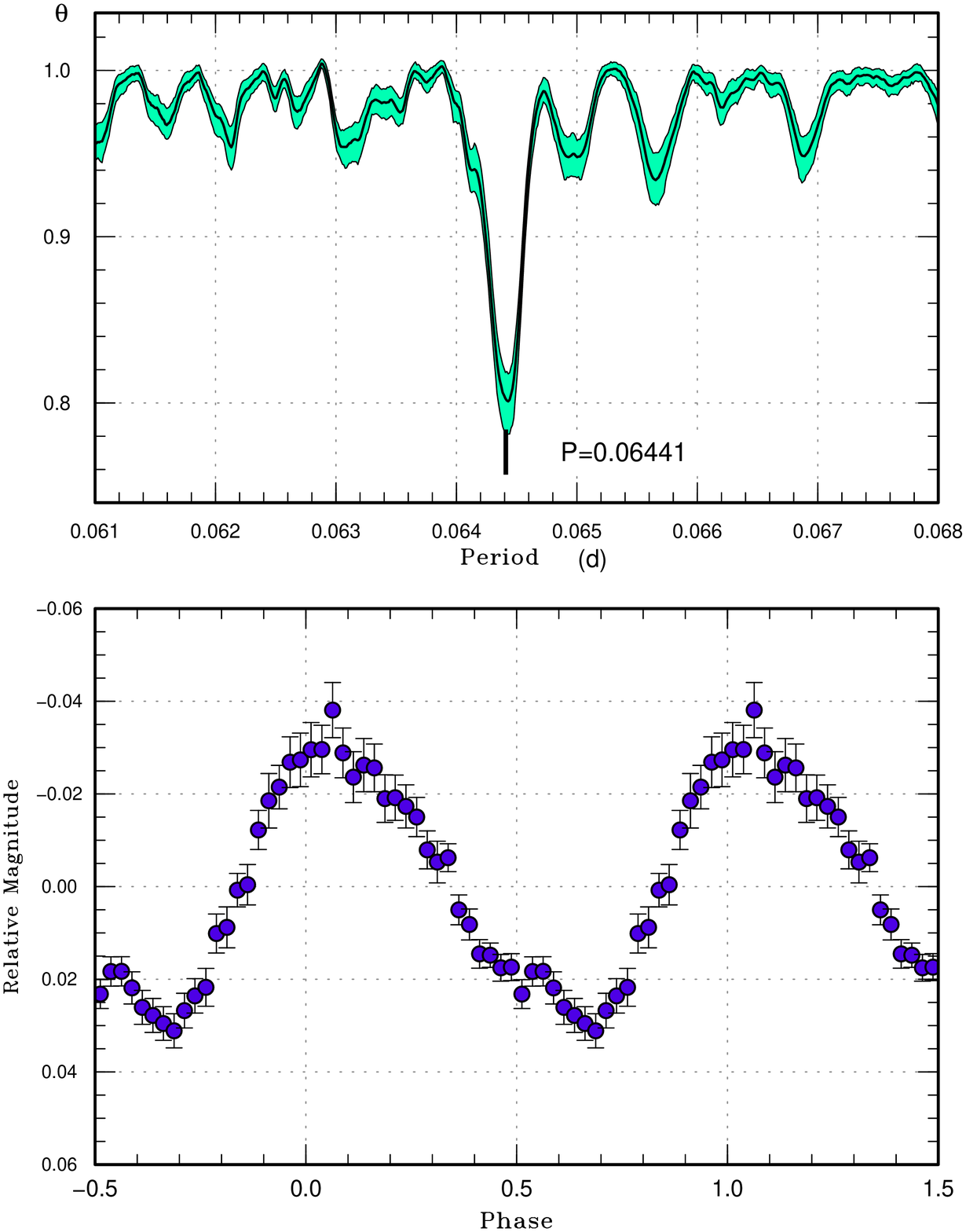}
  \end{center}
  \caption{Ordinary superhumps in ASASSN-16jb (2016).
     (Upper): PDM analysis.
     (Lower): Phase-averaged profile.}
  \label{fig:asassn16jbshpdm}
\end{figure}

   This object was detected as a transient
at $V$=13.3 on 2016 August 18 by the ASAS-SN team.
The object was caught on the rise to the maximum.
The object was initially suspected to be a
Galactic nova (cf. vsnet-alert 20092).  The object
was confirmed to be blue, confirming
the dwarf nova-type nature (vsnet-alert 20094).
Subsequent observations detected early superhumps
(vsnet-alert 20098, 20095; figure \ref{fig:asassn16jbeshpdm}).
The object started to show ordinary superhumps
(figure \ref{fig:asassn16jbshpdm})
on August 25 and showed behavior similar to
a short-period SU UMa-type dwarf nova rather than
an extreme WZ Sge-type dwarf nova
(vsnet-alert 20112, 20125, 20154).

   The times of superhump maxima are listed in
table \ref{tab:asassn16jboc2016}.
All stages (A--C) are clearly seen.
The period of stage A superhumps listed in
table \ref{tab:perlist} was determined by
the PDM method.

   The period of early superhumps was 0.06305(2)~d
(figure \ref{fig:asassn16jbeshpdm}).
The fractional superhump excess of stage A superhumps
$\epsilon^*$ was 0.0321(5), which corresponds to
$q$=0.088(1).  This value is larger than what is
expected for a period bouncer having this orbital
period.  The $O-C$ behavior (positive $P_{\rm dot}$
for stage B and the appearance of stage C) is also
consistent with an object having an intermediately
low $q$.

   The object was also detected in outburst by
ASAS-3 on 2006 March 10 ($V$=13.66, superoutburst),
and possibly on 2009 November 3 ($V$=13.63, single
observation at the end of the observing season).
The presence of earlier outbursts also seems to
exclude the possibility of a period bouncer.

   The identification in AAVSO VSX with
UGPS J175044.95$-$255837.2 appears to be doubtful
considering its red color ($J-K$=2.7).  This
supposed identification likely came from the initial
proposed classification as a classical nova.
We adopted coordinates by the ASAS-SN team.


\begin{table}
\caption{Superhump maxima of ASASSN-16jb (2016)}\label{tab:asassn16jboc2016}
\begin{center}
\begin{tabular}{rp{55pt}p{40pt}r@{.}lr}
\hline
\multicolumn{1}{c}{$E$} & \multicolumn{1}{c}{max\commenta} & \multicolumn{1}{c}{error} & \multicolumn{2}{c}{$O-C$\commentb} & \multicolumn{1}{c}{$N$\commentc} \\
\hline
0 & 57624.6304 & 0.0047 & $-$0&0112 & 35 \\
10 & 57625.2808 & 0.0004 & $-$0&0048 & 148 \\
11 & 57625.3462 & 0.0005 & $-$0&0038 & 149 \\
11 & 57625.3459 & 0.0005 & $-$0&0040 & 148 \\
14 & 57625.5548 & 0.0011 & 0&0116 & 22 \\
15 & 57625.6074 & 0.0013 & $-$0&0002 & 35 \\
16 & 57625.6691 & 0.0010 & $-$0&0029 & 21 \\
21 & 57625.9981 & 0.0007 & 0&0041 & 20 \\
22 & 57626.0629 & 0.0004 & 0&0046 & 33 \\
30 & 57626.5830 & 0.0003 & 0&0094 & 35 \\
31 & 57626.6469 & 0.0003 & 0&0089 & 33 \\
46 & 57627.6099 & 0.0005 & 0&0058 & 35 \\
53 & 57628.0585 & 0.0003 & 0&0036 & 40 \\
58 & 57628.3837 & 0.0029 & 0&0069 & 66 \\
68 & 57629.0183 & 0.0006 & $-$0&0025 & 44 \\
69 & 57629.0824 & 0.0003 & $-$0&0029 & 80 \\
76 & 57629.5392 & 0.0018 & 0&0031 & 20 \\
77 & 57629.5982 & 0.0005 & $-$0&0023 & 32 \\
92 & 57630.5609 & 0.0007 & $-$0&0056 & 35 \\
93 & 57630.6241 & 0.0007 & $-$0&0069 & 35 \\
107 & 57631.5255 & 0.0084 & $-$0&0070 & 18 \\
108 & 57631.5927 & 0.0007 & $-$0&0043 & 35 \\
109 & 57631.6651 & 0.0015 & 0&0038 & 11 \\
118 & 57632.2403 & 0.0089 & $-$0&0006 & 25 \\
119 & 57632.2994 & 0.0006 & $-$0&0060 & 123 \\
120 & 57632.3640 & 0.0005 & $-$0&0058 & 148 \\
121 & 57632.4301 & 0.0013 & $-$0&0041 & 70 \\
123 & 57632.5574 & 0.0012 & $-$0&0056 & 35 \\
124 & 57632.6227 & 0.0013 & $-$0&0046 & 35 \\
138 & 57633.5300 & 0.0026 & 0&0010 & 21 \\
139 & 57633.5928 & 0.0017 & $-$0&0006 & 35 \\
140 & 57633.6577 & 0.0023 & $-$0&0001 & 12 \\
154 & 57634.5602 & 0.0011 & 0&0007 & 36 \\
155 & 57634.6248 & 0.0012 & 0&0010 & 33 \\
169 & 57635.5316 & 0.0011 & 0&0061 & 26 \\
170 & 57635.5929 & 0.0016 & 0&0030 & 35 \\
185 & 57636.5614 & 0.0013 & 0&0055 & 29 \\
186 & 57636.6254 & 0.0010 & 0&0051 & 21 \\
193 & 57637.0777 & 0.0031 & 0&0066 & 7 \\
201 & 57637.5882 & 0.0007 & 0&0019 & 29 \\
216 & 57638.5490 & 0.0017 & $-$0&0033 & 21 \\
217 & 57638.6155 & 0.0014 & $-$0&0013 & 18 \\
232 & 57639.5803 & 0.0029 & $-$0&0025 & 18 \\
\hline
  \multicolumn{6}{l}{\commenta BJD$-$2400000.} \\
  \multicolumn{6}{l}{\commentb Against max $= 2457624.6415 + 0.064402 E$.} \\
  \multicolumn{6}{l}{\commentc Number of points used to determine the maximum.} \\
\end{tabular}
\end{center}
\end{table}

\subsection{ASASSN-16jd}\label{obj:asassn16jd}

\begin{figure}
  \begin{center}
    \FigureFile(85mm,110mm){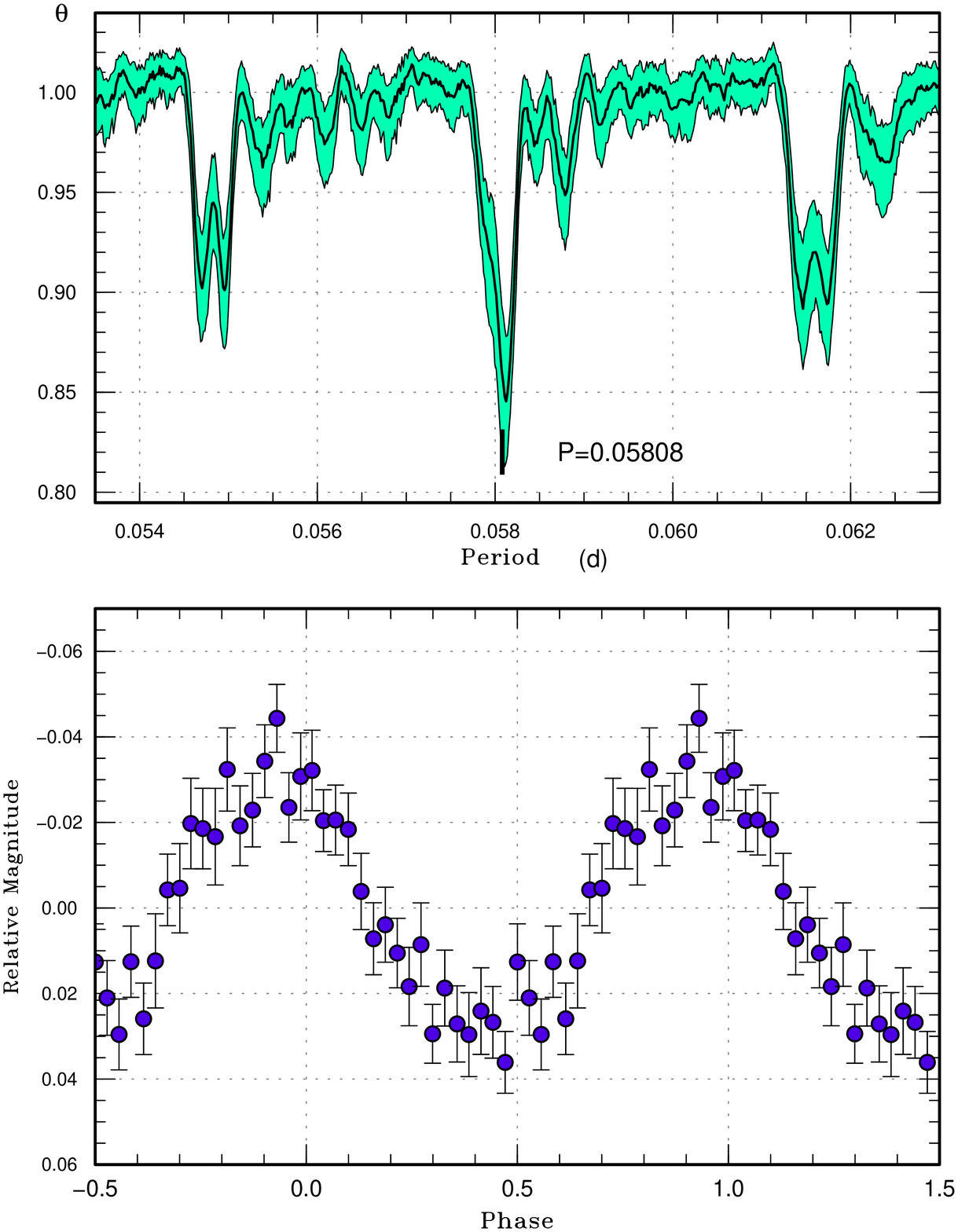}
  \end{center}
  \caption{Ordinary superhumps in ASASSN-16jd (2016).
     (Upper): PDM analysis.
     (Lower): Phase-averaged profile.}
  \label{fig:asassn16jdshpdm}
\end{figure}

   This object was detected as a transient
at $V$=13.6 on 2016 August 20 by the ASAS-SN team.
Superhumps started to appear on August 25
(vsnet-alert 20108, 20113; figure \ref{fig:asassn16jdshpdm}).
The times of superhump maxima are listed in
table \ref{tab:asassn16jdoc2016}.
The period reported in vsnet-alert 20113 was
a one-day alias of the true period.  The period
in this paper has been confirmed by the PDM analysis
(figure \ref{fig:asassn16jdshpdm}) and the much
smoother $O-C$ diagram than obtained using
the former period [the case is the same as presented
in subsection 2.2 in \citet{Pdot7}].


\begin{table}
\caption{Superhump maxima of ASASSN-16jd (2016)}\label{tab:asassn16jdoc2016}
\begin{center}
\begin{tabular}{rp{55pt}p{40pt}r@{.}lr}
\hline
\multicolumn{1}{c}{$E$} & \multicolumn{1}{c}{max\commenta} & \multicolumn{1}{c}{error} & \multicolumn{2}{c}{$O-C$\commentb} & \multicolumn{1}{c}{$N$\commentc} \\
\hline
0 & 57625.5891 & 0.0010 & 0&0003 & 32 \\
1 & 57625.6514 & 0.0017 & 0&0044 & 25 \\
17 & 57626.5792 & 0.0007 & 0&0018 & 32 \\
18 & 57626.6391 & 0.0012 & 0&0035 & 31 \\
34 & 57627.5778 & 0.0003 & 0&0117 & 32 \\
35 & 57627.6351 & 0.0005 & 0&0109 & 31 \\
69 & 57629.6035 & 0.0009 & 0&0020 & 28 \\
86 & 57630.5850 & 0.0007 & $-$0&0052 & 32 \\
87 & 57630.6446 & 0.0008 & $-$0&0037 & 26 \\
103 & 57631.5677 & 0.0010 & $-$0&0110 & 33 \\
104 & 57631.6292 & 0.0006 & $-$0&0077 & 32 \\
120 & 57632.5643 & 0.0012 & $-$0&0031 & 32 \\
121 & 57632.6180 & 0.0017 & $-$0&0075 & 32 \\
137 & 57633.5461 & 0.0014 & $-$0&0100 & 29 \\
138 & 57633.6084 & 0.0014 & $-$0&0058 & 27 \\
154 & 57634.5405 & 0.0046 & $-$0&0042 & 24 \\
155 & 57634.5946 & 0.0012 & $-$0&0082 & 31 \\
206 & 57637.5748 & 0.0015 & 0&0060 & 26 \\
207 & 57637.6316 & 0.0028 & 0&0047 & 21 \\
223 & 57638.5721 & 0.0024 & 0&0147 & 19 \\
240 & 57639.5516 & 0.0022 & 0&0056 & 16 \\
241 & 57639.6047 & 0.0015 & 0&0006 & 16 \\
258 & 57640.5933 & 0.0025 & 0&0004 & 16 \\
\hline
  \multicolumn{6}{l}{\commenta BJD$-$2400000.} \\
  \multicolumn{6}{l}{\commentb Against max $= 2457625.5888 + 0.058155 E$.} \\
  \multicolumn{6}{l}{\commentc Number of points used to determine the maximum.} \\
\end{tabular}
\end{center}
\end{table}

\subsection{ASASSN-16jk}\label{obj:asassn16jk}

   This object was detected as a transient
at $V$=13.9 on 2016 August 27 by the ASAS-SN team.
The object has a blue $g$=20.72 SDSS counterpart.
A neural network analysis of SDSS colors \citep{kat12DNSDSS}
yielded an expected orbital period shorter than
0.06~d.  One long outburst reaching 14.36 mag (unfiltered
CCD) was recorded by the CRTS team on 2007 May 7.
This outburst lasted at least until May 21.
There was another detection at $r$=13.87 by
the Carlsberg Meridian telescope \citep{CMC15}.

\begin{figure}
  \begin{center}
    \FigureFile(85mm,110mm){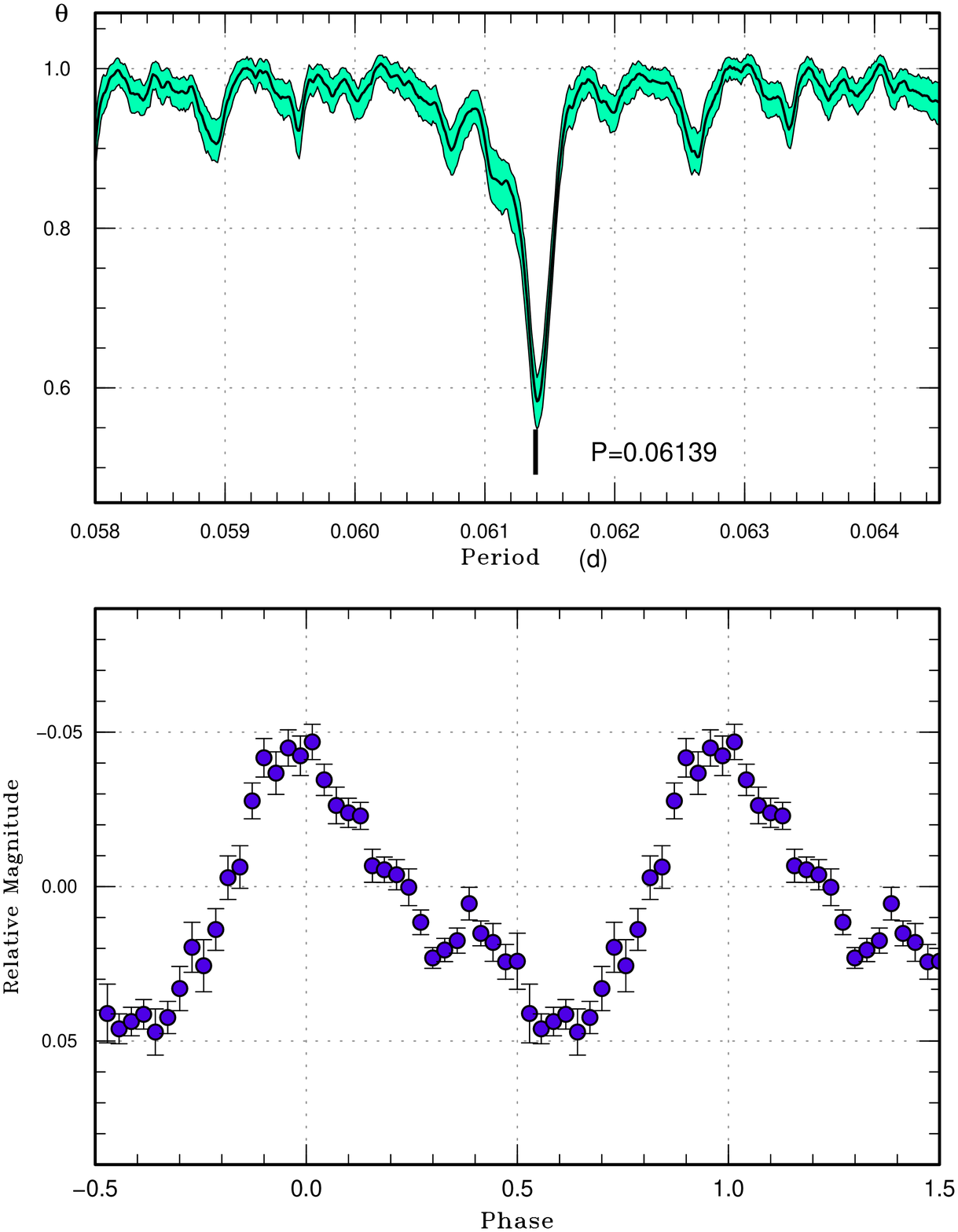}
  \end{center}
  \caption{Superhumps in ASASSN-16jk (2016).
     (Upper): PDM analysis.
     (Lower): Phase-averaged profile.}
  \label{fig:asassn16jkshpdm}
\end{figure}

   Subsequent observations of the 2016 outburst
detected superhumps (vsnet-alert 20119, 20124;
figure \ref{fig:asassn16jkshpdm}).
The times of superhump maxima are listed in
table \ref{tab:asassn16jkoc2016}.  Although the maxima
for $E <$16 are clearly stage A superhumps,
the period was not determined due to
the shortness of the segment.
Although there was apparent stage B-C transition
around $E$=146, the period of stage C superhumps
was not well determined.


\begin{table}
\caption{Superhump maxima of ASASSN-16jk (2016)}\label{tab:asassn16jkoc2016}
\begin{center}
\begin{tabular}{rp{55pt}p{40pt}r@{.}lr}
\hline
\multicolumn{1}{c}{$E$} & \multicolumn{1}{c}{max\commenta} & \multicolumn{1}{c}{error} & \multicolumn{2}{c}{$O-C$\commentb} & \multicolumn{1}{c}{$N$\commentc} \\
\hline
0 & 57631.2947 & 0.0004 & 0&0002 & 47 \\
1 & 57631.3572 & 0.0004 & 0&0012 & 62 \\
16 & 57632.2814 & 0.0002 & 0&0045 & 32 \\
17 & 57632.3438 & 0.0011 & 0&0054 & 25 \\
32 & 57633.2605 & 0.0002 & 0&0011 & 55 \\
33 & 57633.3210 & 0.0005 & 0&0001 & 62 \\
48 & 57634.2378 & 0.0007 & $-$0&0041 & 37 \\
49 & 57634.2984 & 0.0004 & $-$0&0048 & 56 \\
97 & 57637.2444 & 0.0005 & $-$0&0061 & 52 \\
98 & 57637.3071 & 0.0006 & $-$0&0048 & 66 \\
113 & 57638.2352 & 0.0010 & 0&0022 & 41 \\
114 & 57638.2917 & 0.0008 & $-$0&0027 & 66 \\
130 & 57639.2770 & 0.0011 & 0&0002 & 65 \\
146 & 57640.2646 & 0.0008 & 0&0054 & 57 \\
179 & 57642.2834 & 0.0016 & $-$0&0020 & 13 \\
211 & 57644.2496 & 0.0011 & $-$0&0008 & 34 \\
212 & 57644.3168 & 0.0022 & 0&0050 & 23 \\
\hline
  \multicolumn{6}{l}{\commenta BJD$-$2400000.} \\
  \multicolumn{6}{l}{\commentb Against max $= 2457631.2946 + 0.061402 E$.} \\
  \multicolumn{6}{l}{\commentc Number of points used to determine the maximum.} \\
\end{tabular}
\end{center}
\end{table}

\subsection{ASASSN-16js}\label{obj:asassn16js}

\begin{figure}
  \begin{center}
    \FigureFile(85mm,110mm){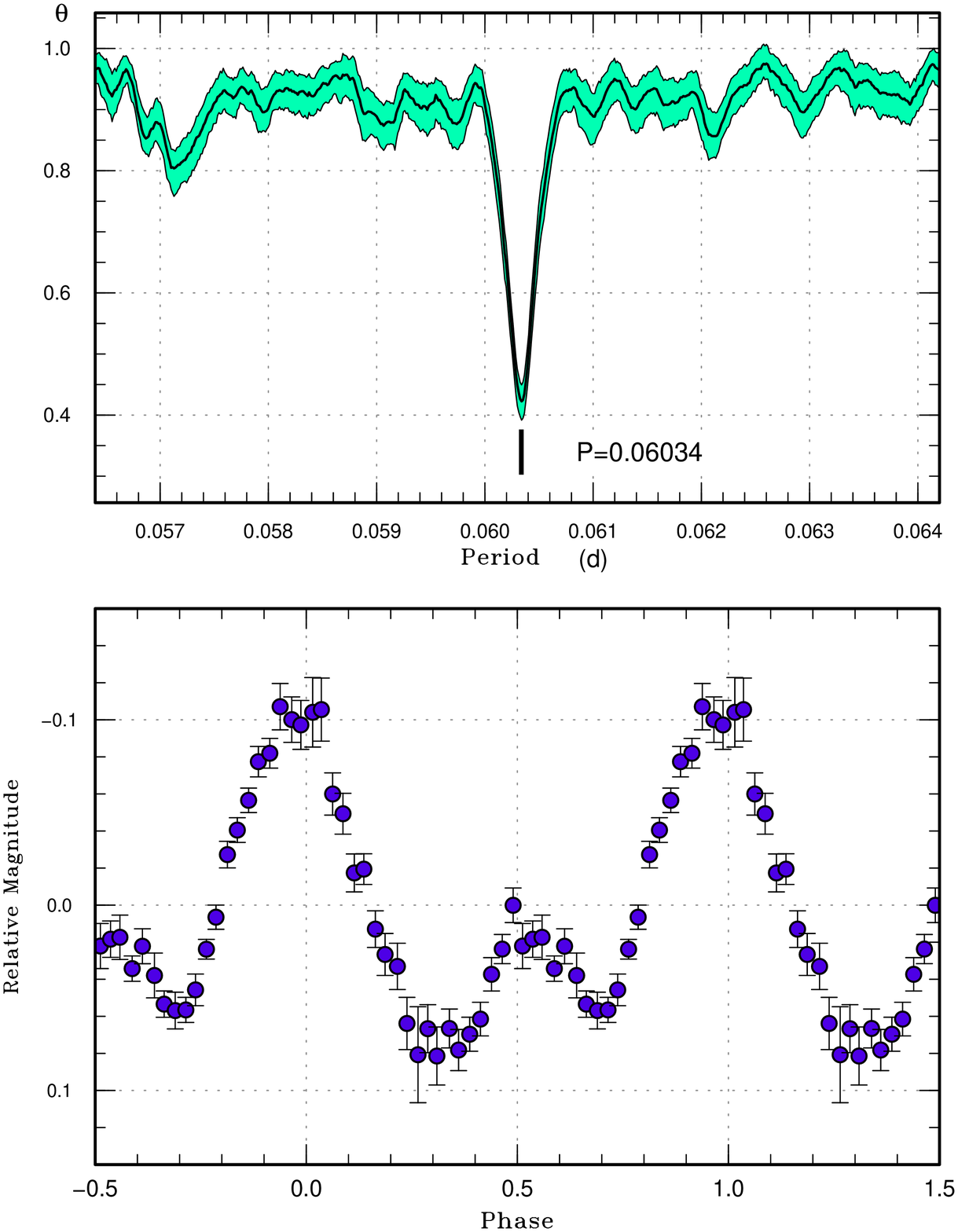}
  \end{center}
  \caption{Early superhumps in ASASSN-16js (2016).
     (Upper): PDM analysis.
     (Lower): Phase-averaged profile.}
  \label{fig:asassn16jseshpdm}
\end{figure}

\begin{figure}
  \begin{center}
    \FigureFile(85mm,110mm){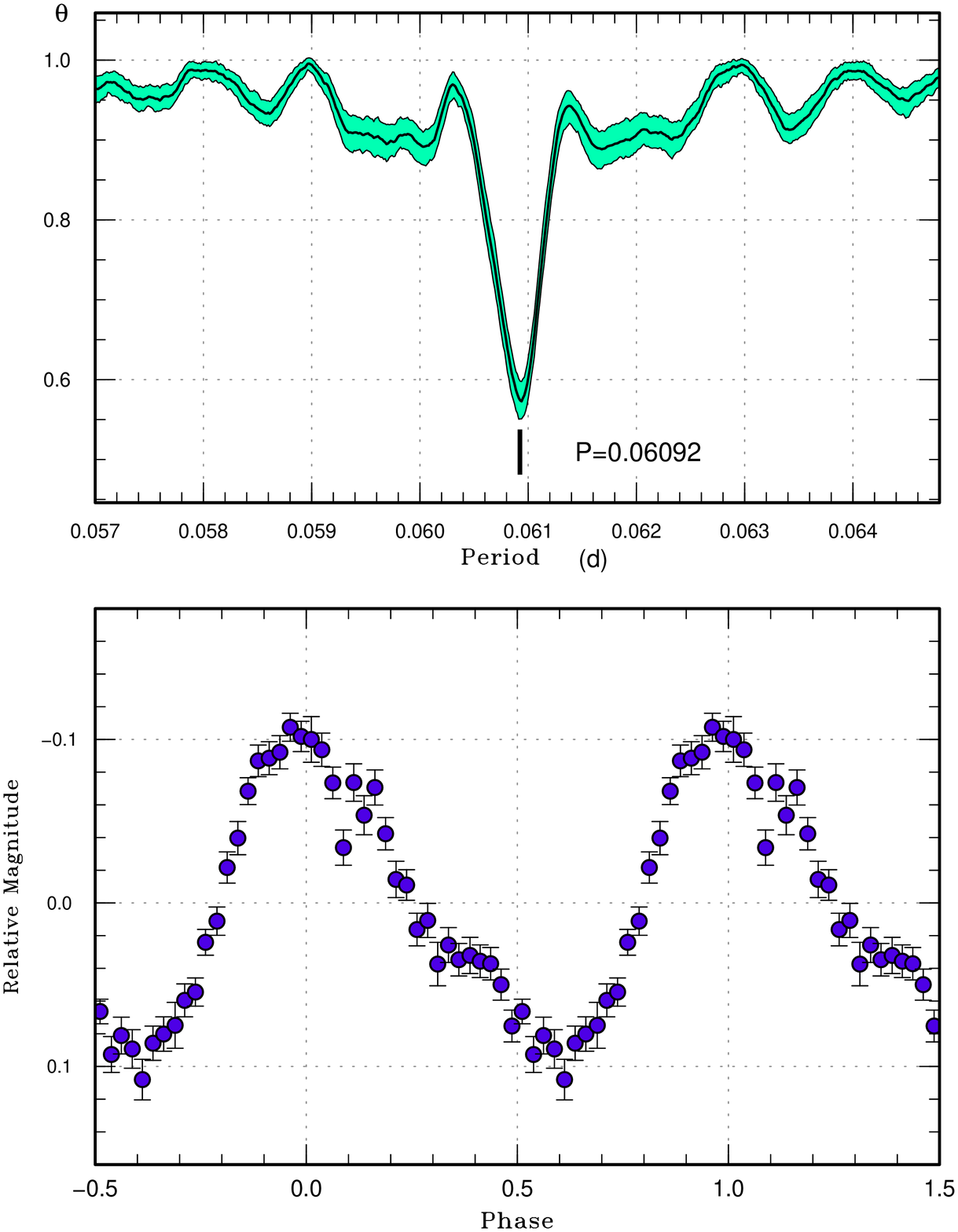}
  \end{center}
  \caption{Ordinary superhumps in ASASSN-16js (2016).
     (Upper): PDM analysis.
     (Lower): Phase-averaged profile.}
  \label{fig:asassn16jsshpdm}
\end{figure}

   This object was detected as a transient
at $V$=13.0 on 2016 August 30 by the ASAS-SN team.
Early superhumps were immediately detected
(vsnet-alert 20122, 20126, 20155;
figure \ref{fig:asassn16jseshpdm}).
Ordinary superhump emerged on September 9
(vsnet-alert 20165, 20187).
The period of early superhumps was
0.060337(5)~d.
The times of superhump maxima are listed in
table \ref{tab:asassn16jsoc2016}.
The $O-C$ somewhat flattened after $E$=32
and there was a rather smooth transition
to stage B, which started at around $E$=48.
The mean profile of ordinary superhumps is
given in figure \ref{fig:asassn16jsshpdm}.

   The fractional superhump excess $\epsilon^*$
for stage A superhumps was 0.0213(16),
which corresponds to $q$=0.056(5).
The small $q$ and an orbital period significantly
longer than the period minimum suggest that
this object is a period bouncer.
Although the mean superhump amplitude is larger
than those of period bouncer candidates,
this may have been due to the high orbital
inclination as suggested by the strong
early superhumps.


\begin{table*}
\caption{Superhump maxima of ASASSN-16js (2016)}\label{tab:asassn16jsoc2016}
\begin{center}
\begin{tabular}{rp{55pt}p{40pt}r@{.}lrrp{55pt}p{40pt}r@{.}lr}
\hline
\multicolumn{1}{c}{$E$} & \multicolumn{1}{c}{max\commenta} & \multicolumn{1}{c}{error} & \multicolumn{2}{c}{$O-C$\commentb} & \multicolumn{1}{c}{$N$\commentc} & \multicolumn{1}{c}{$E$} & \multicolumn{1}{c}{max\commenta} & \multicolumn{1}{c}{error} & \multicolumn{2}{c}{$O-C$\commentb} & \multicolumn{1}{c}{$N$\commentc} \\
\hline
0 & 57640.7131 & 0.0033 & $-$0&0166 & 13 & 67 & 57644.8191 & 0.0019 & 0&0017 & 13 \\
1 & 57640.7772 & 0.0023 & $-$0&0135 & 13 & 68 & 57644.8796 & 0.0014 & 0&0012 & 14 \\
2 & 57640.8339 & 0.0015 & $-$0&0179 & 13 & 81 & 57645.6749 & 0.0016 & 0&0034 & 13 \\
16 & 57641.6950 & 0.0022 & $-$0&0109 & 13 & 82 & 57645.7340 & 0.0017 & 0&0015 & 13 \\
17 & 57641.7580 & 0.0015 & $-$0&0089 & 13 & 83 & 57645.7951 & 0.0032 & 0&0016 & 13 \\
18 & 57641.8198 & 0.0016 & $-$0&0081 & 12 & 84 & 57645.8553 & 0.0013 & 0&0007 & 16 \\
19 & 57641.8865 & 0.0047 & $-$0&0024 & 13 & 92 & 57646.3434 & 0.0007 & 0&0008 & 140 \\
32 & 57642.6874 & 0.0011 & 0&0054 & 14 & 93 & 57646.4027 & 0.0018 & $-$0&0009 & 65 \\
33 & 57642.7478 & 0.0012 & 0&0047 & 13 & 97 & 57646.6522 & 0.0018 & 0&0045 & 14 \\
34 & 57642.8092 & 0.0012 & 0&0051 & 12 & 98 & 57646.7101 & 0.0022 & 0&0014 & 13 \\
35 & 57642.8701 & 0.0010 & 0&0051 & 16 & 99 & 57646.7715 & 0.0024 & 0&0018 & 13 \\
39 & 57643.1152 & 0.0007 & 0&0061 & 38 & 100 & 57646.8290 & 0.0013 & $-$0&0018 & 14 \\
40 & 57643.1765 & 0.0006 & 0&0064 & 38 & 101 & 57646.8942 & 0.0024 & 0&0024 & 10 \\
41 & 57643.2370 & 0.0004 & 0&0059 & 37 & 113 & 57647.6223 & 0.0022 & $-$0&0016 & 14 \\
42 & 57643.2988 & 0.0007 & 0&0067 & 24 & 114 & 57647.6794 & 0.0028 & $-$0&0054 & 18 \\
44 & 57643.4201 & 0.0005 & 0&0060 & 140 & 115 & 57647.7406 & 0.0017 & $-$0&0053 & 17 \\
45 & 57643.4815 & 0.0004 & 0&0063 & 139 & 116 & 57647.8035 & 0.0028 & $-$0&0033 & 14 \\
46 & 57643.5439 & 0.0004 & 0&0078 & 132 & 117 & 57647.8668 & 0.0047 & $-$0&0011 & 16 \\
48 & 57643.6665 & 0.0012 & 0&0083 & 14 & 120 & 57648.0456 & 0.0008 & $-$0&0054 & 38 \\
49 & 57643.7251 & 0.0008 & 0&0059 & 13 & 121 & 57648.1078 & 0.0010 & $-$0&0041 & 38 \\
50 & 57643.7851 & 0.0012 & 0&0048 & 13 & 122 & 57648.1671 & 0.0009 & $-$0&0059 & 38 \\
51 & 57643.8457 & 0.0013 & 0&0045 & 15 & 123 & 57648.2278 & 0.0011 & $-$0&0062 & 30 \\
55 & 57644.0893 & 0.0007 & 0&0040 & 26 & 130 & 57648.6533 & 0.0027 & $-$0&0078 & 19 \\
56 & 57644.1499 & 0.0005 & 0&0036 & 33 & 131 & 57648.7183 & 0.0044 & $-$0&0037 & 17 \\
57 & 57644.2111 & 0.0005 & 0&0038 & 37 & 133 & 57648.8389 & 0.0032 & $-$0&0051 & 15 \\
58 & 57644.2718 & 0.0004 & 0&0035 & 37 & 149 & 57649.8181 & 0.0024 & $-$0&0021 & 13 \\
64 & 57644.6350 & 0.0012 & 0&0007 & 12 & 172 & 57651.2298 & 0.0035 & 0&0063 & 38 \\
65 & 57644.6988 & 0.0017 & 0&0035 & 13 & 173 & 57651.2832 & 0.0054 & $-$0&0012 & 15 \\
66 & 57644.7600 & 0.0014 & 0&0036 & 13 & \multicolumn{1}{c}{--} & \multicolumn{1}{c}{--} & \multicolumn{1}{c}{--} & \multicolumn{2}{c}{--} & \multicolumn{1}{c}{--}\\
\hline
  \multicolumn{12}{l}{\commenta BJD$-$2400000.} \\
  \multicolumn{12}{l}{\commentb Against max $= 2457640.7297 + 0.061010 E$.} \\
  \multicolumn{12}{l}{\commentc Number of points used to determine the maximum.} \\
\end{tabular}
\end{center}
\end{table*}

\subsection{ASASSN-16jz}\label{obj:asassn16jz}

\begin{figure}
  \begin{center}
    \FigureFile(85mm,110mm){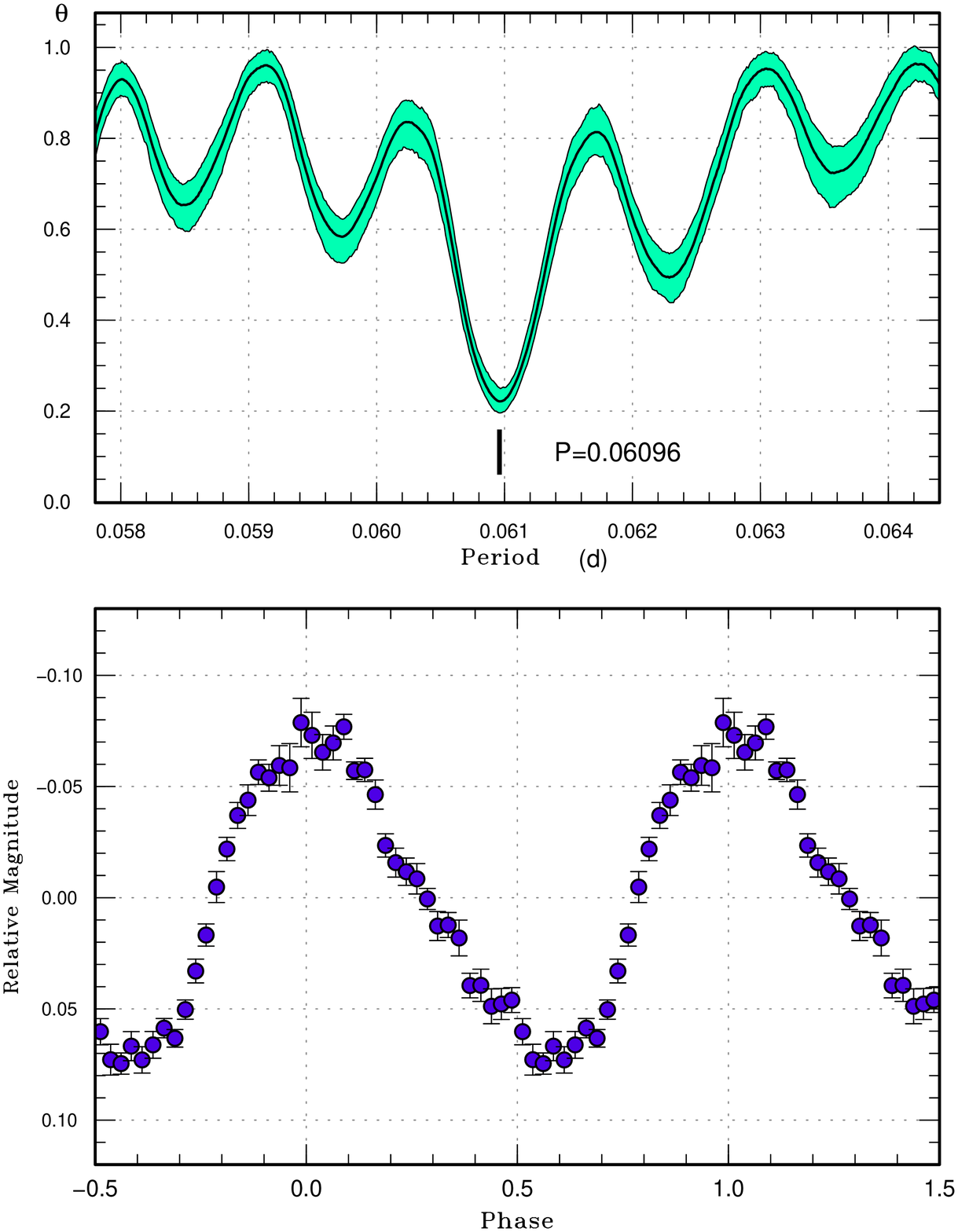}
  \end{center}
  \caption{Ordinary superhumps in ASASSN-16jz (2016).
     (Upper): PDM analysis.
     (Lower): Phase-averaged profile.}
  \label{fig:asassn16jzshpdm}
\end{figure}

   This object was detected as a transient
at $V$=15.7 on 2016 September 5 by the ASAS-SN team.
Observations immediately detected superhumps
(vsnet-alert 20142; figure \ref{fig:asassn16jzshpdm}).
The time of superhump maxima are listed in
table \ref{tab:asassn16jzoc2016}.  There was some
hint of decrease in the $O-C$ values, which may
reflect a stage transition.  The exact identification
of the superhump stage was impossible due to
the short baseline.


\begin{table}
\caption{Superhump maxima of ASASSN-16jz (2016)}\label{tab:asassn16jzoc2016}
\begin{center}
\begin{tabular}{rp{55pt}p{40pt}r@{.}lr}
\hline
\multicolumn{1}{c}{$E$} & \multicolumn{1}{c}{max\commenta} & \multicolumn{1}{c}{error} & \multicolumn{2}{c}{$O-C$\commentb} & \multicolumn{1}{c}{$N$\commentc} \\
\hline
0 & 57638.3632 & 0.0004 & 0&0001 & 59 \\
1 & 57638.4232 & 0.0004 & $-$0&0008 & 61 \\
2 & 57638.4856 & 0.0005 & 0&0006 & 62 \\
3 & 57638.5458 & 0.0004 & $-$0&0001 & 62 \\
4 & 57638.6062 & 0.0004 & $-$0&0007 & 62 \\
16 & 57639.3390 & 0.0004 & 0&0010 & 60 \\
49 & 57641.3501 & 0.0008 & 0&0012 & 62 \\
50 & 57641.4098 & 0.0012 & $-$0&0001 & 61 \\
51 & 57641.4695 & 0.0011 & $-$0&0013 & 42 \\
\hline
  \multicolumn{6}{l}{\commenta BJD$-$2400000.} \\
  \multicolumn{6}{l}{\commentb Against max $= 2457638.3631 + 0.060936 E$.} \\
  \multicolumn{6}{l}{\commentc Number of points used to determine the maximum.} \\
\end{tabular}
\end{center}
\end{table}

\subsection{ASASSN-16kg}\label{obj:asassn16kg}

\begin{figure}
  \begin{center}
    \FigureFile(85mm,110mm){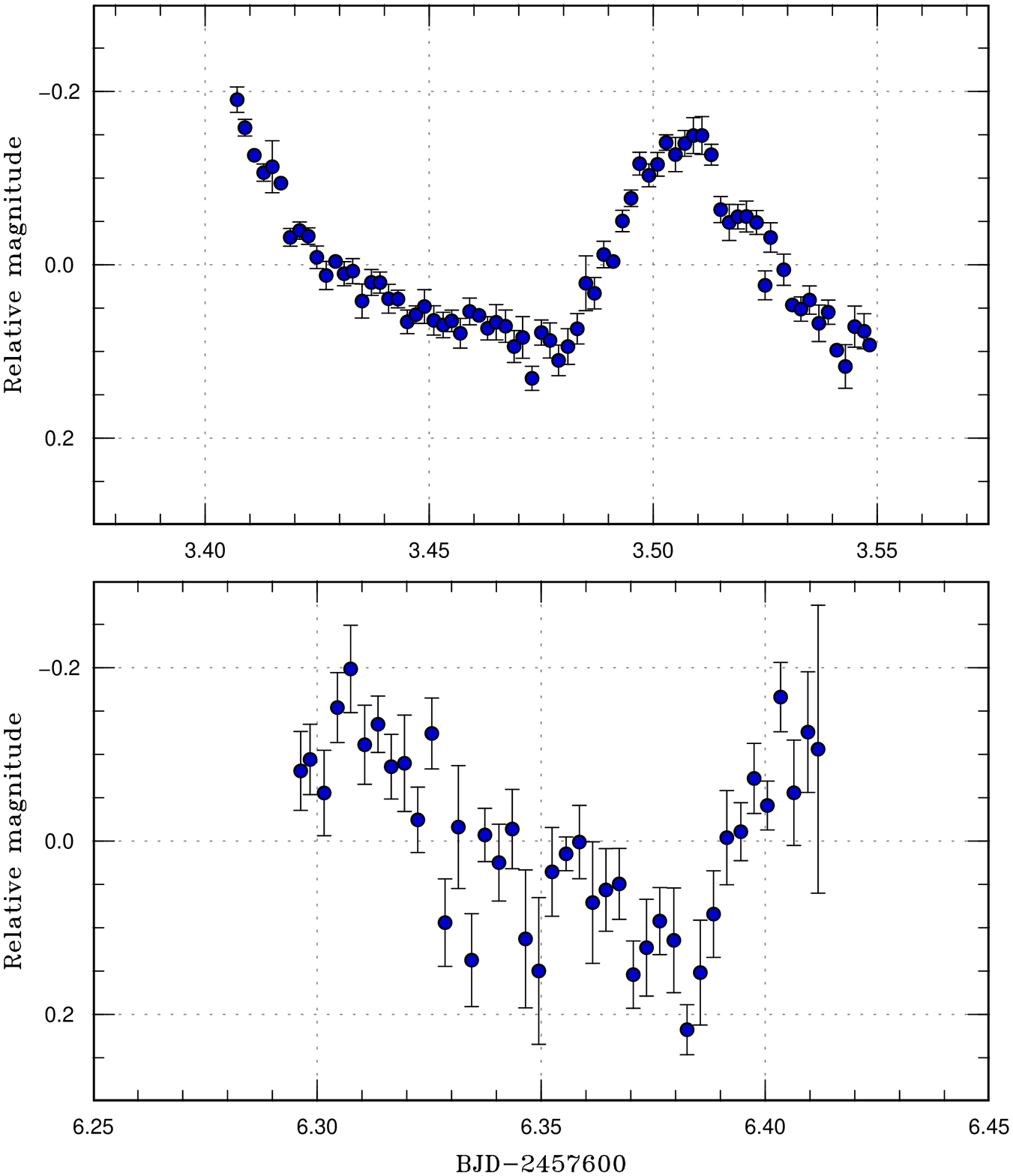}
  \end{center}
  \caption{Superhumps in ASASSN-16kg (2016).
  The data were binned to 0.003~d.
  }
  \label{fig:asassn16kgshlc}
\end{figure}

\begin{figure}
  \begin{center}
    \FigureFile(85mm,110mm){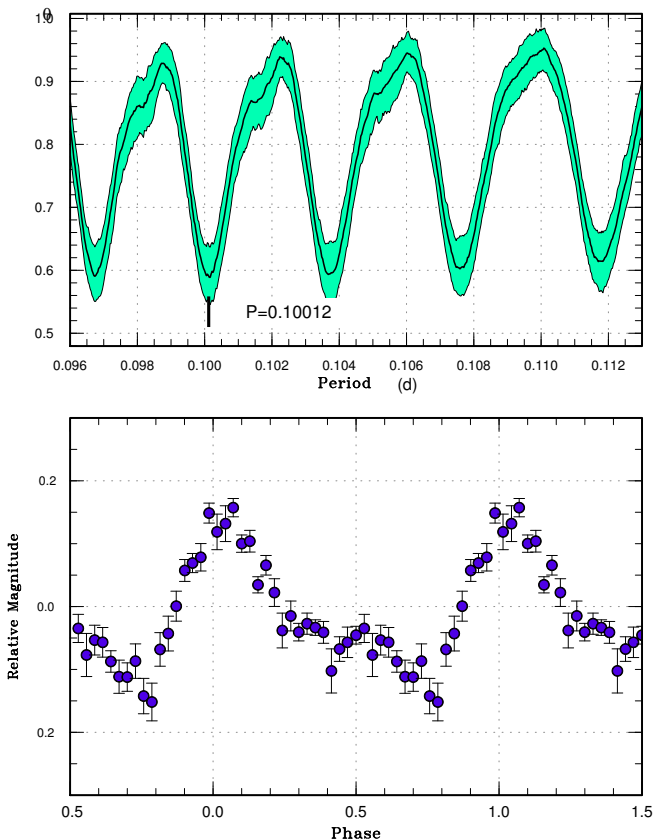}
  \end{center}
  \caption{PDM analysis of superhumps in ASASSN-16kg (2016).
     (Upper): PDM analysis.  The alias selection was
     one of the possibilities.
     (Lower): Phase-averaged profile.}
  \label{fig:asassn16kgshpdm}
\end{figure}

   This object was detected as a transient
at $V$=16.1 on 2016 September 7 by the ASAS-SN team.
The object brightened to $V$=15.2 on September 8
and the outburst was announced.
There was no quiescent counterpart recorded in previous
plates, and the large outburst amplitude received
attention.  Subsequent observations detected
superhumps (vsnet-alert 20182; figure \ref{fig:asassn16kgshlc}).
The superhump period was around 0.10~d, which was
not expected from the large outburst amplitude.
Since there was a 3-d gap in the observation and individual
runs were comparable to one superhump cycle, it was
impossible to select the alias uniquely.
The candidate periods by the PDM methods
(figure \ref{fig:asassn16kgshpdm}) 
are 0.09676(4)~d, 0.10013(4)~d,
0.10373(5)~d, 0.107610(5)~d and 0.11178(5)~d.
Other aliases can be likely rejected because they
give large $O-C$ scatters.  Among them, we have selected
0.10013(4)~d which gives the smallest $\theta$
in the PDM analysis to make cycle counts in
table \ref{tab:asassn16kgoc2016}.  One should note
that there remains cycle count ambiguities due
to the ambiguity in the alias selection.
The object, however, is certainly located
in the period gap.  It might be worth noting that
such a large-amplitude dwarf nova exists
in the period gap.

   The object was detected in outburst at an
unfiltered CCD magnitude of 17.27 on September 26
by the CRTS team (=CSS160926:213630$-$251348).
Since the object had already faded to 18.0 mag
on September 20, this CRTS observation appears to have
detected a post-superoutburst rebrightening.

\begin{table}
\caption{Superhump maxima of ASASSN-16kg (2016)}\label{tab:asassn16kgoc2016}
\begin{center}
\begin{tabular}{rp{55pt}p{40pt}r@{.}lr}
\hline
\multicolumn{1}{c}{$E$} & \multicolumn{1}{c}{max\commenta} & \multicolumn{1}{c}{error} & \multicolumn{2}{c}{$O-C$\commentb} & \multicolumn{1}{c}{$N$\commentc} \\
\hline
0 & 57643.3970 & 0.0026 & $-$0&0050 & 94 \\
1 & 57643.5076 & 0.0004 & 0&0053 & 224 \\
29 & 57646.3094 & 0.0022 & $-$0&0020 & 153 \\
30 & 57646.4135 & 0.0038 & 0&0018 & 108 \\
\hline
  \multicolumn{6}{l}{\commenta BJD$-$2400000.} \\
  \multicolumn{6}{l}{\commentb Against max $= 2457643.4020 + 0.100324 E$.} \\
  \multicolumn{6}{l}{\commentc Number of points used to determine the maximum.} \\
\end{tabular}
\end{center}
\end{table}

\subsection{ASASSN-16kx}\label{obj:asassn16kx}

\begin{figure}
  \begin{center}
    \FigureFile(85mm,110mm){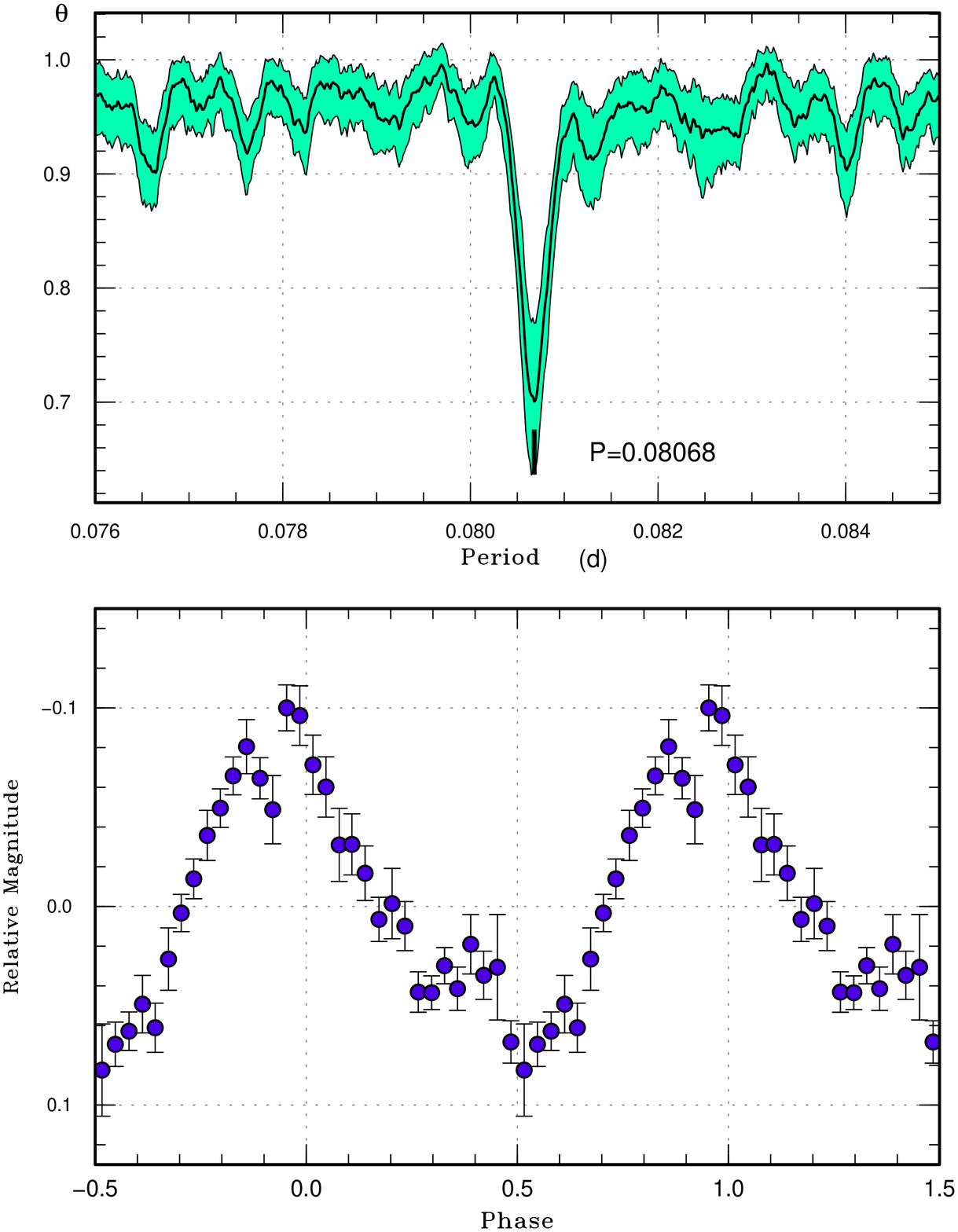}
  \end{center}
  \caption{Superhumps in ASASSN-16kx (2016).
     (Upper): PDM analysis.
     (Lower): Phase-averaged profile.}
  \label{fig:asassn16kxshpdm}
\end{figure}

   This object was detected as a transient
at $V$=14.8 on 2016 September 26 by the ASAS-SN team.
Subsequent observations detected superhumps
(vsnet-alert 20210; figure \ref{fig:asassn16kxshpdm}).
The times of superhump maxima are listed in
table \ref{tab:asassn16kxoc2016}.


\begin{table}
\caption{Superhump maxima of ASASSN-16kx (2016)}\label{tab:asassn16kxoc2016}
\begin{center}
\begin{tabular}{rp{55pt}p{40pt}r@{.}lr}
\hline
\multicolumn{1}{c}{$E$} & \multicolumn{1}{c}{max\commenta} & \multicolumn{1}{c}{error} & \multicolumn{2}{c}{$O-C$\commentb} & \multicolumn{1}{c}{$N$\commentc} \\
\hline
0 & 57662.4690 & 0.0074 & 0&0003 & 64 \\
1 & 57662.5451 & 0.0006 & $-$0&0043 & 185 \\
2 & 57662.6234 & 0.0021 & $-$0&0066 & 104 \\
4 & 57662.7892 & 0.0011 & $-$0&0021 & 20 \\
5 & 57662.8715 & 0.0019 & $-$0&0005 & 17 \\
17 & 57663.8416 & 0.0007 & 0&0020 & 27 \\
29 & 57664.8090 & 0.0009 & 0&0017 & 27 \\
30 & 57664.8904 & 0.0035 & 0&0025 & 11 \\
41 & 57665.7775 & 0.0038 & 0&0025 & 19 \\
42 & 57665.8563 & 0.0025 & 0&0007 & 23 \\
54 & 57666.8270 & 0.0011 & 0&0037 & 27 \\
79 & 57668.8412 & 0.0016 & 0&0020 & 18 \\
91 & 57669.8137 & 0.0020 & 0&0067 & 26 \\
103 & 57670.7746 & 0.0019 & 0&0000 & 16 \\
104 & 57670.8574 & 0.0027 & 0&0022 & 13 \\
116 & 57671.8205 & 0.0023 & $-$0&0024 & 24 \\
128 & 57672.7909 & 0.0029 & 0&0003 & 24 \\
129 & 57672.8663 & 0.0040 & $-$0&0049 & 15 \\
141 & 57673.8394 & 0.0028 & 0&0005 & 23 \\
153 & 57674.8022 & 0.0045 & $-$0&0043 & 24 \\
\hline
  \multicolumn{6}{l}{\commenta BJD$-$2400000.} \\
  \multicolumn{6}{l}{\commentb Against max $= 2457662.4687 + 0.080639 E$.} \\
  \multicolumn{6}{l}{\commentc Number of points used to determine the maximum.} \\
\end{tabular}
\end{center}
\end{table}

\subsection{ASASSN-16le}\label{obj:asassn16le}

\begin{figure}
  \begin{center}
    \FigureFile(85mm,110mm){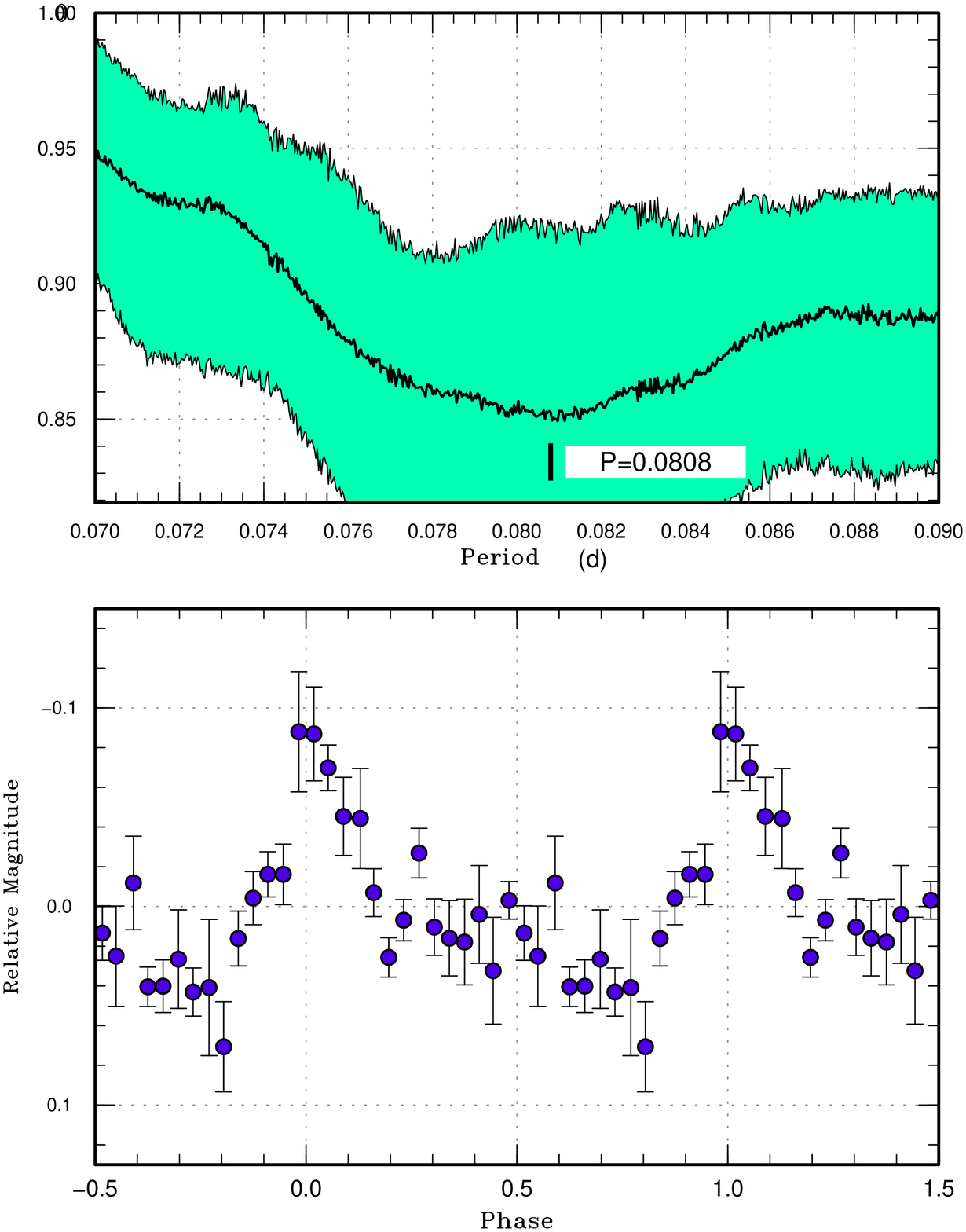}
  \end{center}
  \caption{Superhumps in ASASSN-16le (2016).
     (Upper): PDM analysis.
     (Lower): Phase-averaged profile.}
  \label{fig:asassn16leshpdm}
\end{figure}

   This object was detected as a transient
at $V$=15.5 on 2016 October 2 by the ASAS-SN team.
Subsequent observations detected superhumps
(vsnet-alert 20214).
Since the object was observed only on one night,
only three superhump maxima were recorded:
BJD 2457668.1480(38) ($N$=26), 2457668.2486(13) ($N$=176)
and 2457668.3286(20) ($N$=147).
The best period determined by the PDM method
was 0.0808(13)~d (figure \ref{fig:asassn16leshpdm}).
The object faded to fainter than 17.5 mag on October 14.


\subsection{ASASSN-16lj}\label{obj:asassn16lj}

\begin{figure}
  \begin{center}
    \FigureFile(85mm,110mm){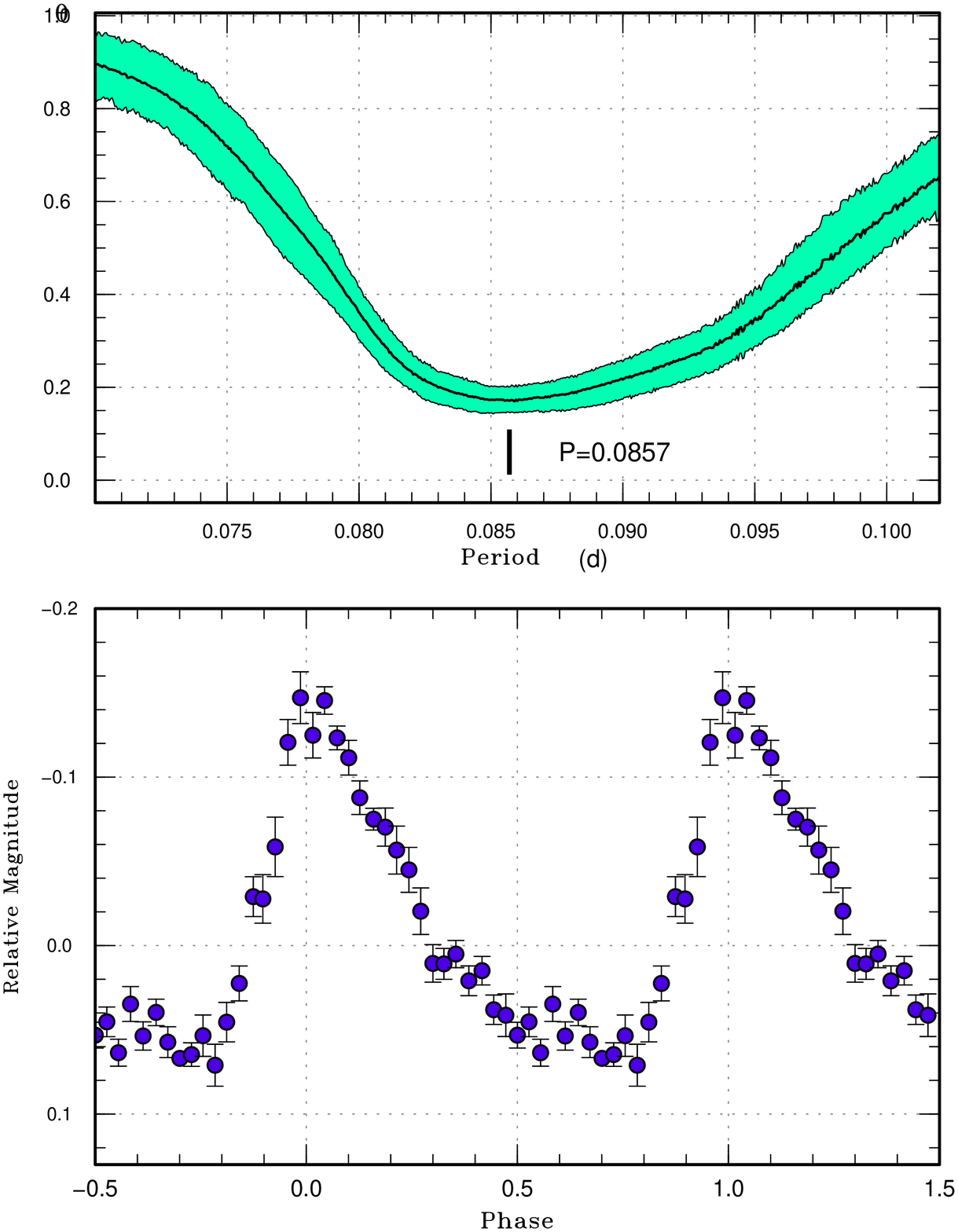}
  \end{center}
  \caption{Superhumps in ASASSN-16lj (2016).
     (Upper): PDM analysis.
     (Lower): Phase-averaged profile.}
  \label{fig:asassn16ljshpdm}
\end{figure}

   This object was detected as a transient
at $V$=15.8 on 2016 October 6 by the ASAS-SN team.
Subsequent observations detected superhumps
(vsnet-alert 20217).
Since the object was observed only on one night,
only three superhump maxima were recorded:
BJD 2457670.3611(5) ($N$=82), 2457670.4447(5) ($N$=87)
and 2457670.5325(22) ($N$=45).
The best period determined by the PDM method
was 0.0857(4)~d (figure \ref{fig:asassn16ljshpdm}).


\subsection{ASASSN-16lo}\label{obj:asassn16lo}

\begin{figure}
  \begin{center}
    \FigureFile(85mm,110mm){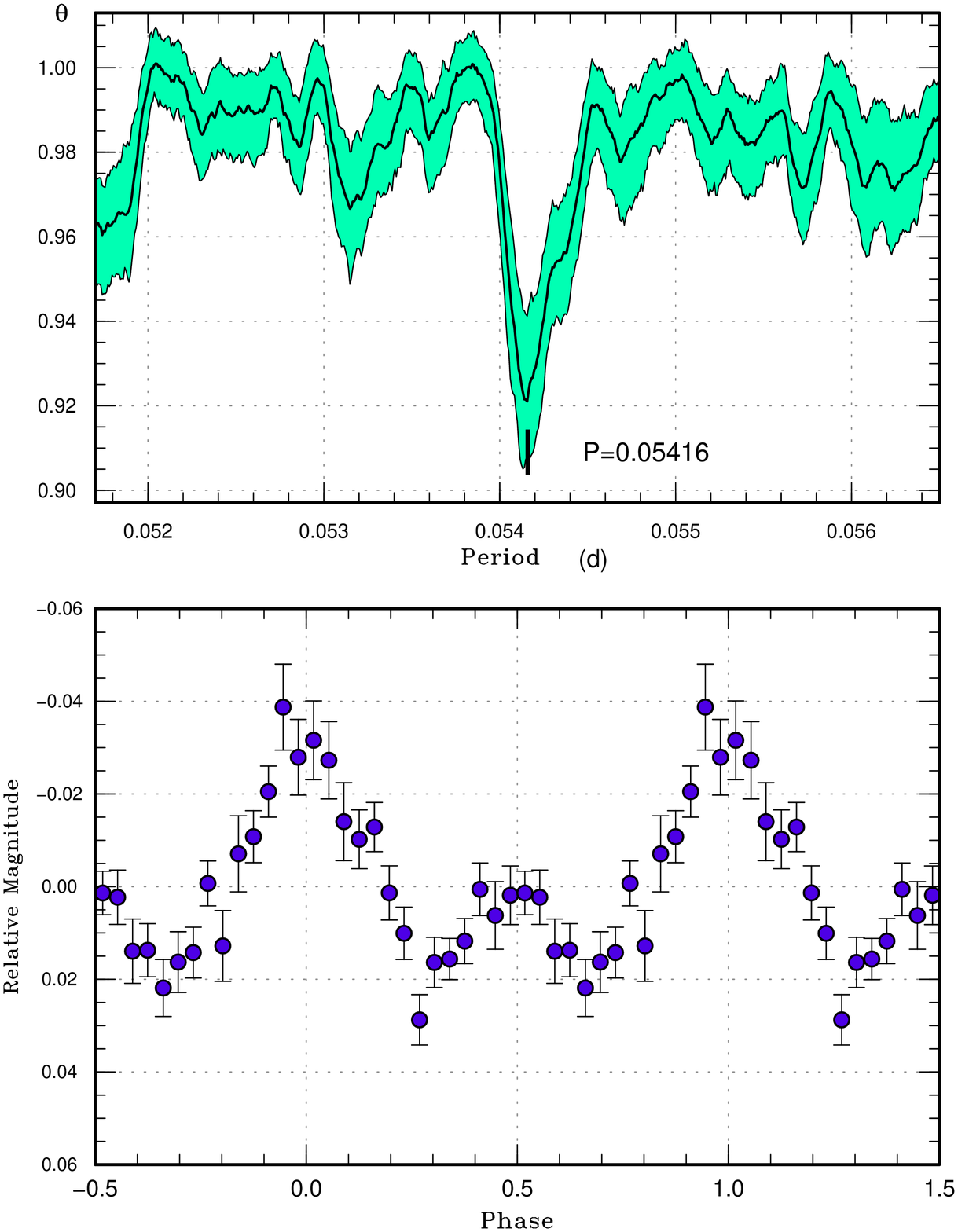}
  \end{center}
  \caption{Early superhumps in ASASSN-16lo (2016).
     (Upper): PDM analysis.  The data before
     BJD 2457678 were used.
     (Lower): Phase-averaged profile.}
  \label{fig:asassn16loeshpdm}
\end{figure}

\begin{figure}
  \begin{center}
    \FigureFile(85mm,110mm){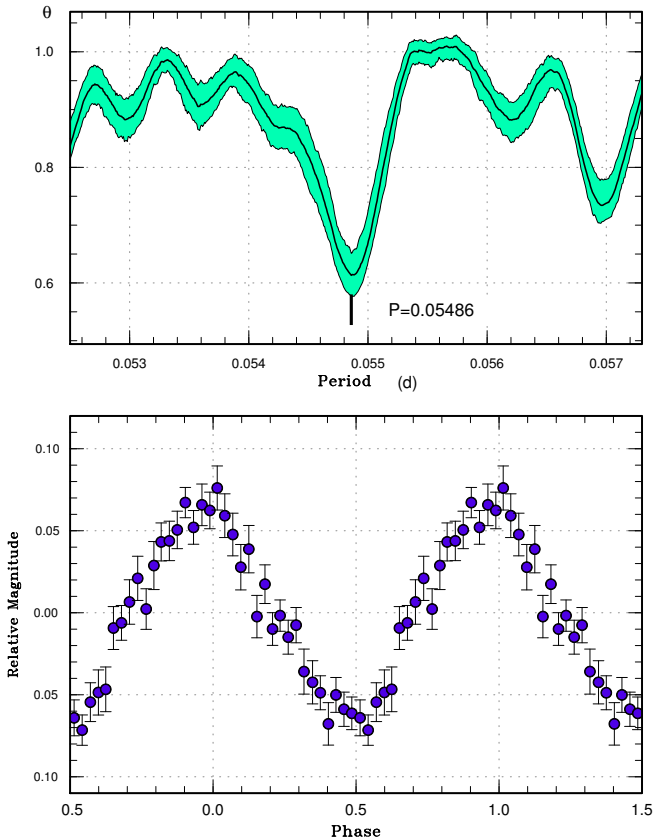}
  \end{center}
  \caption{Ordinary superhumps in ASASSN-16lo (2016).
     (Upper): PDM analysis.  The interval of
     BJD 2457681--2457686 was used.
     (Lower): Phase-averaged profile.}
  \label{fig:asassn16loshpdm}
\end{figure}

   This object was detected as a transient
at $V$=14.3 on 2016 October 8 by the ASAS-SN team.
Subsequent observations detected early superhumps
(vsnet-alert 20219, 20222; figure \ref{fig:asassn16loeshpdm})
confirming that this object is a WZ Sge-type dwarf nova.
The best period of early superhump was
0.05416(1)~d.
Ordinary superhump developed on October 19
(vsnet-alert 20230, 20243; figure \ref{fig:asassn16loshpdm}).
The times of superhump maxima are listed in
table \ref{tab:asassn16looc2016}.
Although epochs $E \le$2 were stage A superhumps,
the period of stage A superhumps could not be determined.
Although observations were present after BJD 2457690,
superhumps were not clearly detected since
the object was very faint (16.5--17.0 mag).
The outburst lasted at least until November 7.


\begin{table}
\caption{Superhump maxima of ASASSN-16lo (2016)}\label{tab:asassn16looc2016}
\begin{center}
\begin{tabular}{rp{55pt}p{40pt}r@{.}lr}
\hline
\multicolumn{1}{c}{$E$} & \multicolumn{1}{c}{max\commenta} & \multicolumn{1}{c}{error} & \multicolumn{2}{c}{$O-C$\commentb} & \multicolumn{1}{c}{$N$\commentc} \\
\hline
0 & 57681.3009 & 0.0007 & $-$0&0064 & 57 \\
1 & 57681.3580 & 0.0009 & $-$0&0041 & 56 \\
2 & 57681.4156 & 0.0021 & $-$0&0014 & 27 \\
38 & 57683.3986 & 0.0005 & 0&0066 & 57 \\
39 & 57683.4529 & 0.0004 & 0&0060 & 54 \\
54 & 57684.2738 & 0.0005 & 0&0040 & 45 \\
55 & 57684.3278 & 0.0005 & 0&0032 & 59 \\
56 & 57684.3832 & 0.0006 & 0&0037 & 58 \\
57 & 57684.4378 & 0.0005 & 0&0034 & 56 \\
84 & 57685.9076 & 0.0041 & $-$0&0080 & 10 \\
85 & 57685.9654 & 0.0007 & $-$0&0051 & 140 \\
86 & 57686.0235 & 0.0020 & $-$0&0019 & 72 \\
\hline
  \multicolumn{6}{l}{\commenta BJD$-$2400000.} \\
  \multicolumn{6}{l}{\commentb Against max $= 2457681.3073 + 0.054861 E$.} \\
  \multicolumn{6}{l}{\commentc Number of points used to determine the maximum.} \\
\end{tabular}
\end{center}
\end{table}

\subsection{ASASSN-16mo}\label{obj:asassn16mo}

\begin{figure}
  \begin{center}
    \FigureFile(85mm,110mm){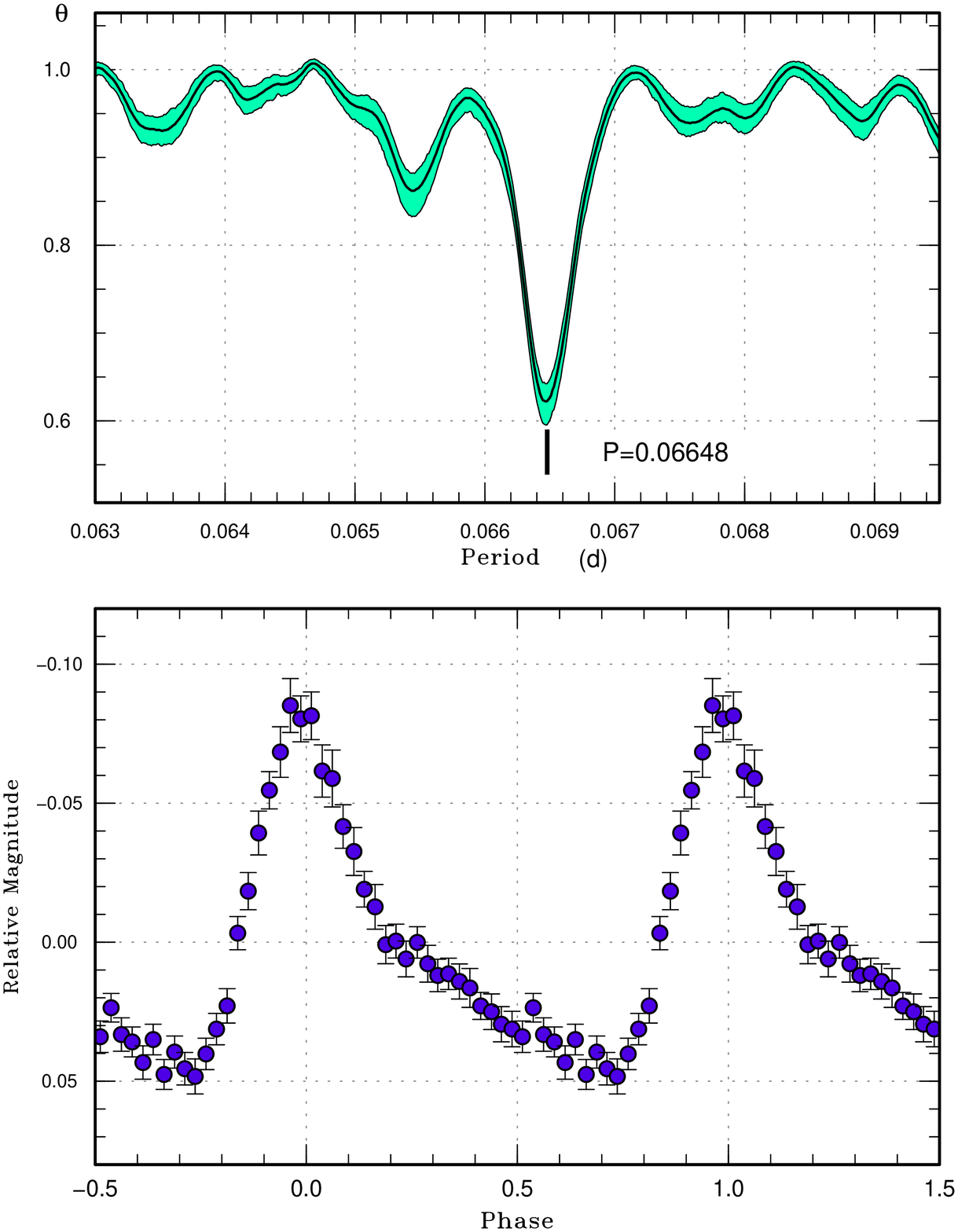}
  \end{center}
  \caption{Superhumps in ASASSN-16mo (2016).
     (Upper): PDM analysis.
     (Lower): Phase-averaged profile.}
  \label{fig:asassn16moshpdm}
\end{figure}

   This object was detected as a transient
at $V$=15.0 on 2016 October 28 by the ASAS-SN team.
Subsequent observations detected superhumps
(vsnet-alert 20280, 20294, 20311;
figure \ref{fig:asassn16moshpdm}).
The times of superhump maxima are listed
in table \ref{tab:asassn16mooc2016}.
Although there was likely stage B-C transition
after $E$=84, later observations could not
detect superhumps very clearly and we did not
determine the period of stage C superhumps.


\begin{table}
\caption{Superhump maxima of ASASSN-16mo (2016)}\label{tab:asassn16mooc2016}
\begin{center}
\begin{tabular}{rp{55pt}p{40pt}r@{.}lr}
\hline
\multicolumn{1}{c}{$E$} & \multicolumn{1}{c}{max\commenta} & \multicolumn{1}{c}{error} & \multicolumn{2}{c}{$O-C$\commentb} & \multicolumn{1}{c}{$N$\commentc} \\
\hline
0 & 57692.5886 & 0.0005 & 0&0042 & 57 \\
13 & 57693.4488 & 0.0004 & 0&0001 & 60 \\
17 & 57693.7134 & 0.0007 & $-$0&0012 & 46 \\
23 & 57694.1115 & 0.0003 & $-$0&0021 & 147 \\
24 & 57694.1804 & 0.0004 & 0&0004 & 146 \\
39 & 57695.1761 & 0.0004 & $-$0&0013 & 113 \\
40 & 57695.2417 & 0.0005 & $-$0&0022 & 120 \\
67 & 57697.0402 & 0.0004 & 0&0012 & 73 \\
68 & 57697.1059 & 0.0004 & 0&0003 & 73 \\
69 & 57697.1740 & 0.0004 & 0&0020 & 74 \\
81 & 57697.9692 & 0.0007 & $-$0&0007 & 74 \\
82 & 57698.0352 & 0.0004 & $-$0&0011 & 74 \\
83 & 57698.1027 & 0.0005 & $-$0&0002 & 74 \\
84 & 57698.1683 & 0.0006 & $-$0&0011 & 50 \\
99 & 57699.1652 & 0.0030 & $-$0&0015 & 50 \\
112 & 57700.0344 & 0.0048 & 0&0033 & 50 \\
\hline
  \multicolumn{6}{l}{\commenta BJD$-$2400000.} \\
  \multicolumn{6}{l}{\commentb Against max $= 2457692.5844 + 0.066488 E$.} \\
  \multicolumn{6}{l}{\commentc Number of points used to determine the maximum.} \\
\end{tabular}
\end{center}
\end{table}

\subsection{ASASSN-16my}\label{obj:asassn16my}

\begin{figure}
  \begin{center}
    \FigureFile(85mm,110mm){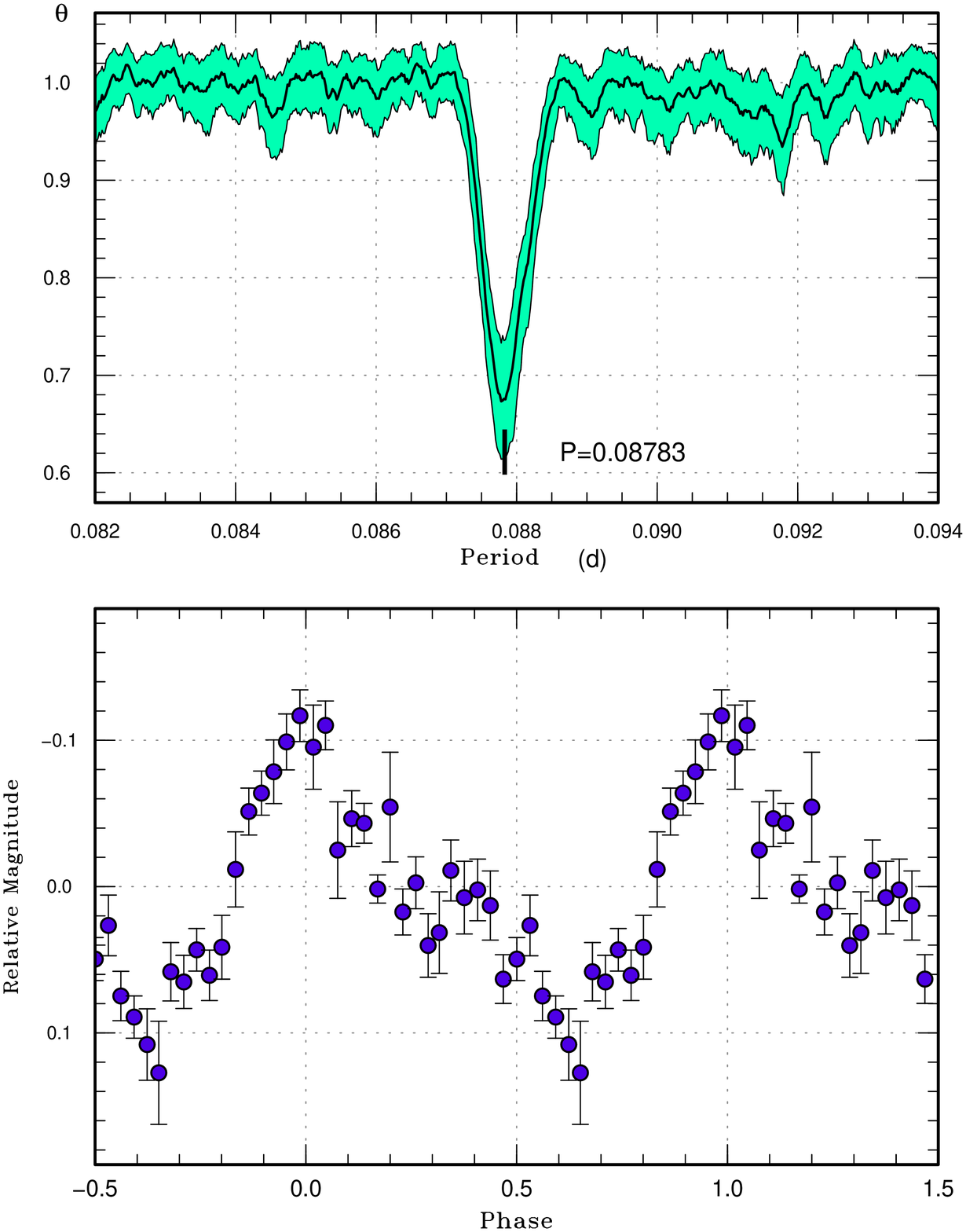}
  \end{center}
  \caption{Superhumps in ASASSN-16my (2016).
     (Upper): PDM analysis.
     (Lower): Phase-averaged profile.}
  \label{fig:asassn16myshpdm}
\end{figure}

   This object was detected as a transient
at $V$=14.4 on 2016 November 6 by the ASAS-SN team.
Subsequent observations detected superhumps
(vsnet-alert 20325, 20357;
figure \ref{fig:asassn16myshpdm}).
The times of superhump maxima are listed in
table \ref{tab:asassn16myoc2016}.
Although the $O-C$ values suggest that $E \le$12
were still stage A superhumps, we did not determine
the period of stage A superhumps since superhump
amplitudes were already large (0.3--0.4 mag)
and it is likely that the period was already affected
by the pressure effect (i.e. transition to stage B).


\begin{table}
\caption{Superhump maxima of ASASSN-16my (2016)}\label{tab:asassn16myoc2016}
\begin{center}
\begin{tabular}{rp{55pt}p{40pt}r@{.}lr}
\hline
\multicolumn{1}{c}{$E$} & \multicolumn{1}{c}{max\commenta} & \multicolumn{1}{c}{error} & \multicolumn{2}{c}{$O-C$\commentb} & \multicolumn{1}{c}{$N$\commentc} \\
\hline
0 & 57700.7424 & 0.0006 & $-$0&0065 & 17 \\
1 & 57700.8318 & 0.0005 & $-$0&0049 & 17 \\
12 & 57701.8010 & 0.0009 & $-$0&0013 & 19 \\
23 & 57702.7675 & 0.0012 & $-$0&0004 & 24 \\
24 & 57702.8628 & 0.0016 & 0&0071 & 8 \\
34 & 57703.7418 & 0.0029 & 0&0083 & 23 \\
35 & 57703.8262 & 0.0013 & 0&0049 & 20 \\
46 & 57704.7908 & 0.0029 & 0&0039 & 19 \\
52 & 57705.3052 & 0.0025 & $-$0&0084 & 44 \\
57 & 57705.7561 & 0.0019 & 0&0035 & 25 \\
58 & 57705.8428 & 0.0029 & 0&0025 & 14 \\
68 & 57706.7180 & 0.0042 & $-$0&0002 & 18 \\
69 & 57706.8033 & 0.0024 & $-$0&0027 & 23 \\
80 & 57707.7724 & 0.0027 & 0&0009 & 27 \\
81 & 57707.8540 & 0.0061 & $-$0&0054 & 12 \\
91 & 57708.7356 & 0.0029 & $-$0&0015 & 28 \\
92 & 57708.8251 & 0.0032 & 0&0002 & 21 \\
\hline
  \multicolumn{6}{l}{\commenta BJD$-$2400000.} \\
  \multicolumn{6}{l}{\commentb Against max $= 2457700.7489 + 0.087783 E$.} \\
  \multicolumn{6}{l}{\commentc Number of points used to determine the maximum.} \\
\end{tabular}
\end{center}
\end{table}

\subsection{ASASSN-16ni}\label{obj:asassn16ni}

\begin{figure}
  \begin{center}
    \FigureFile(85mm,110mm){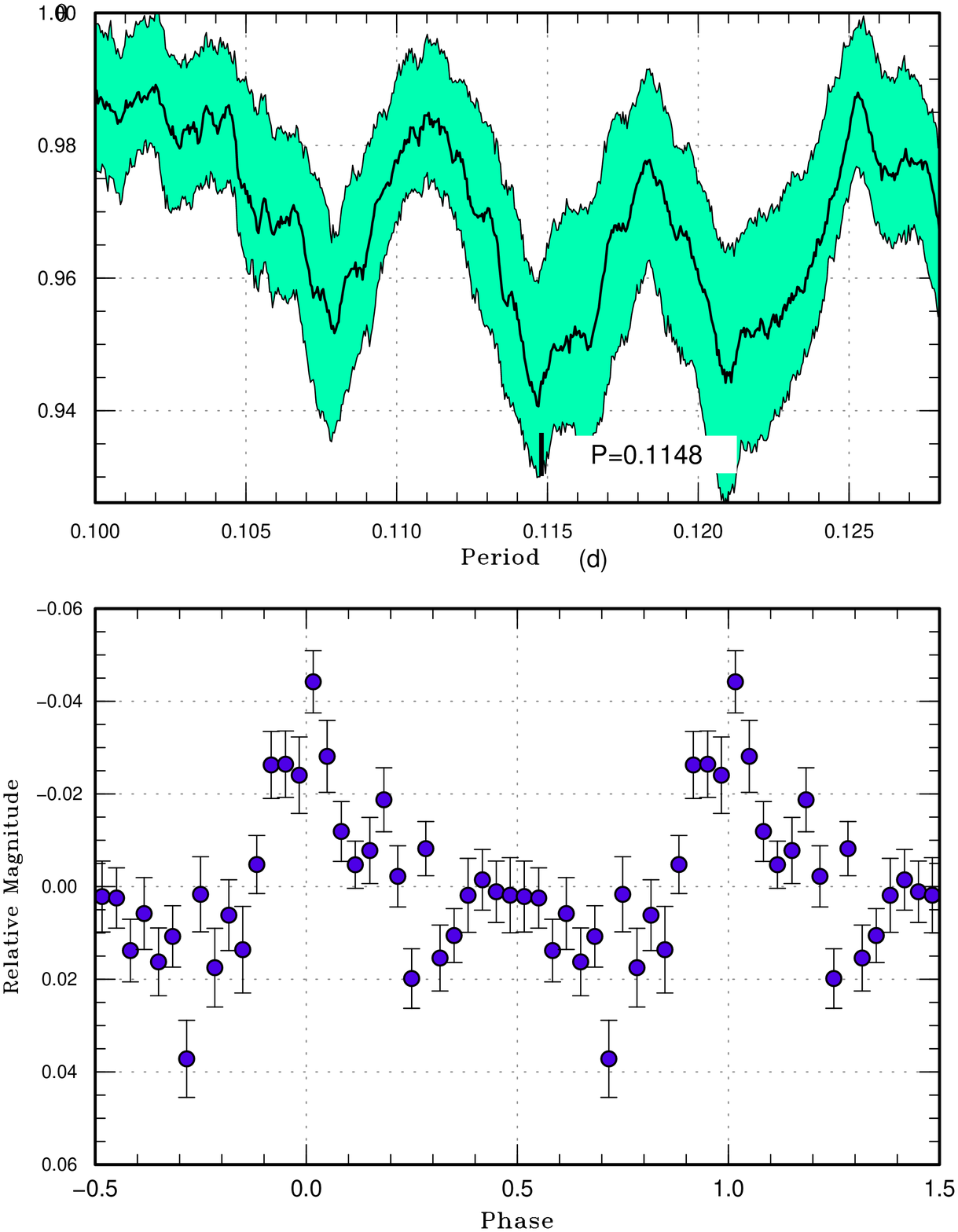}
  \end{center}
  \caption{Superhumps in ASASSN-16ni (2016).
     (Upper): PDM analysis.
     (Lower): Phase-averaged profile.}
  \label{fig:asassn16nishpdm}
\end{figure}

   This object was detected as a transient
at $V$=16.5 on 2016 November 16 by the ASAS-SN team.
The large outburst amplitude received attention.
Subsequent observations detected long-period
superhumps (vsnet-alert 20446; the period reported
in vsnet-alert 20398 was probably in error).
Such a long period [0.1152(4)~d determined in this
paper] was unexpected for
a large-amplitude dwarf nova.
Observations were, however, insufficient to determine
the period uniquely due to the faintness.
We selected the most likely
alias in calculating the $O-C$ values
in table \ref{tab:asassn16nioc2016}.
The superhump amplitudes were growing on the first
two nights, and the period reported here may
refer to that of stage A superhumps.
The small amplitudes (figure \ref{fig:asassn16nishpdm})
may also support this stage identification.

   The quiescent counterpart was originally
proposed to be a very faint ($g$=22.9) object
SDSS J050500.40$+$605455.3.  This object, however,
has a red color ($u-g$=$+$2.6) and it is unlikely
a CV.  The true quiescent counterpart should be
fainter than $g$=23.  We adopted the coordinates
by the ASAS-SN team.


\begin{table}
\caption{Superhump maxima of ASASSN-16ni (2016)}\label{tab:asassn16nioc2016}
\begin{center}
\begin{tabular}{rp{55pt}p{40pt}r@{.}lr}
\hline
\multicolumn{1}{c}{$E$} & \multicolumn{1}{c}{max\commenta} & \multicolumn{1}{c}{error} & \multicolumn{2}{c}{$O-C$\commentb} & \multicolumn{1}{c}{$N$\commentc} \\
\hline
0 & 57717.0368 & 0.0025 & $-$0&0037 & 257 \\
1 & 57717.1598 & 0.0026 & 0&0040 & 256 \\
2 & 57717.2712 & 0.0026 & 0&0001 & 256 \\
11 & 57718.3078 & 0.0040 & $-$0&0004 & 337 \\
\hline
  \multicolumn{6}{l}{\commenta BJD$-$2400000.} \\
  \multicolumn{6}{l}{\commentb Against max $= 2457717.0406 + 0.115242 E$.} \\
  \multicolumn{6}{l}{\commentc Number of points used to determine the maximum.} \\
\end{tabular}
\end{center}
\end{table}

\subsection{ASASSN-16nq}\label{obj:asassn16nq}

\begin{figure}
  \begin{center}
    \FigureFile(85mm,110mm){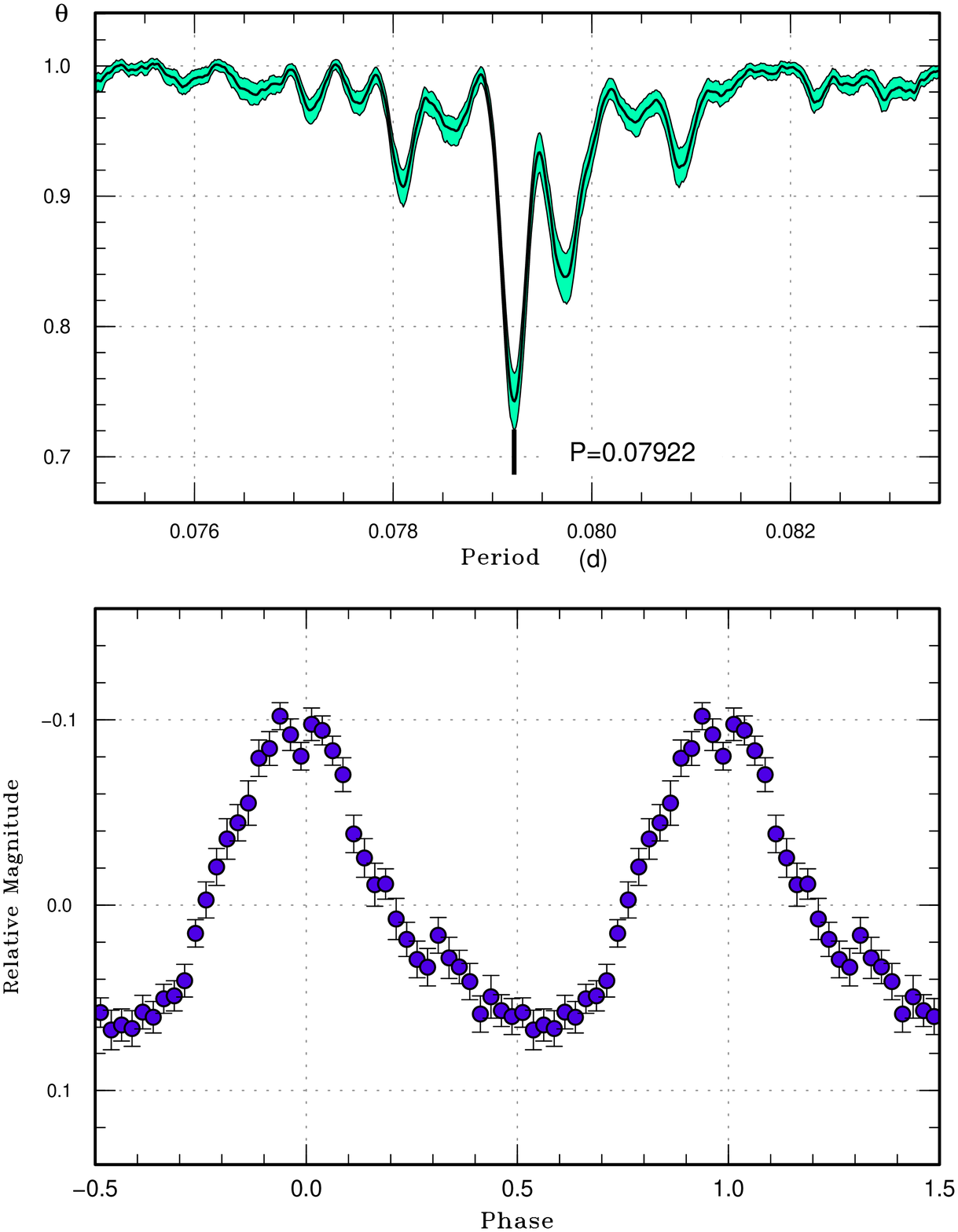}
  \end{center}
  \caption{Superhumps in ASASSN-16nq (2016).
     (Upper): PDM analysis.
     (Lower): Phase-averaged profile.}
  \label{fig:asassn16nqshpdm}
\end{figure}

\begin{figure}
  \begin{center}
    \FigureFile(85mm,100mm){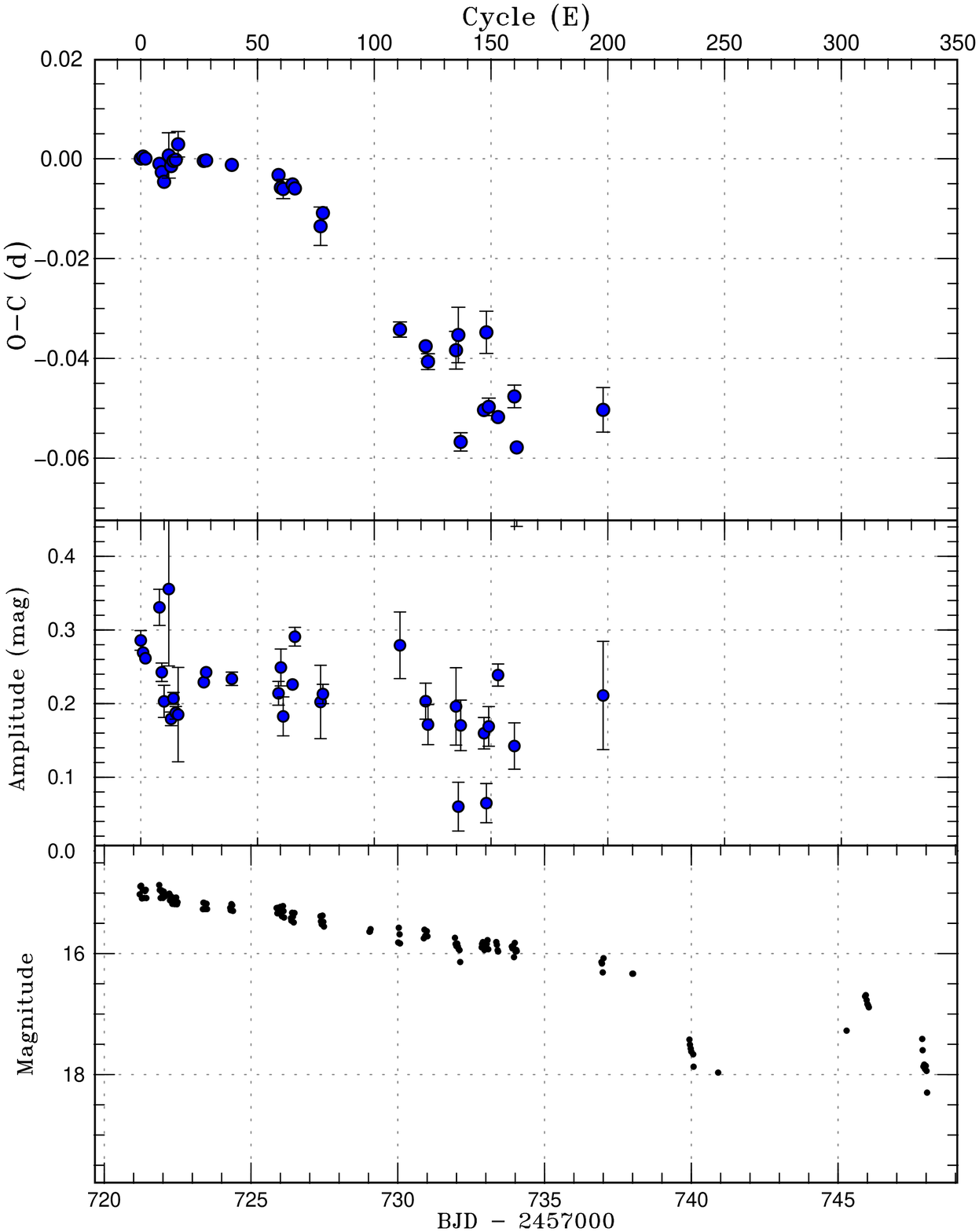}
  \end{center}
  \caption{$O-C$ diagram of superhumps in ASASSN-16nq (2016).
     (Upper:) $O-C$ diagram.
     We used a period of 0.05796~d for calculating the $O-C$ residuals.
     (Middle:) Amplitudes of superhumps.
     (Lower:) Light curve.  The data were binned to 0.026~d.
  }
  \label{fig:asassn16nqhumpall}
\end{figure}

   This object was detected as a transient
at $V$=15.0 on 2016 November 26 by the ASAS-SN team.
Observations on November 28 detected superhumps
(vsnet-alert 20411; figure \ref{fig:asassn16nqshpdm}).
Although the observations were
obtained two nights after the outburst detection,
stage A was already over (vsnet-alert 20431, 20472;
figure \ref{fig:asassn16nqhumpall}).
The times of superhump maxima are listed in
table \ref{tab:asassn16nqoc2016}.
Both stages B and C can be recognized.
There was one post-superoutburst rebrightening
on December 23 (figure \ref{fig:asassn16nqhumpall}),
which is relatively rare for such a long $P_{\rm SH}$ system.


\begin{table}
\caption{Superhump maxima of ASASSN-16nq (2016)}\label{tab:asassn16nqoc2016}
\begin{center}
\begin{tabular}{rp{55pt}p{40pt}r@{.}lr}
\hline
\multicolumn{1}{c}{$E$} & \multicolumn{1}{c}{max\commenta} & \multicolumn{1}{c}{error} & \multicolumn{2}{c}{$O-C$\commentb} & \multicolumn{1}{c}{$N$\commentc} \\
\hline
0 & 57721.2429 & 0.0005 & $-$0&0061 & 74 \\
1 & 57721.3229 & 0.0002 & $-$0&0053 & 128 \\
2 & 57721.4021 & 0.0002 & $-$0&0054 & 147 \\
8 & 57721.8784 & 0.0006 & $-$0&0044 & 47 \\
9 & 57721.9563 & 0.0005 & $-$0&0058 & 81 \\
10 & 57722.0339 & 0.0011 & $-$0&0074 & 82 \\
12 & 57722.1983 & 0.0045 & $-$0&0014 & 27 \\
13 & 57722.2757 & 0.0005 & $-$0&0032 & 164 \\
14 & 57722.3563 & 0.0004 & $-$0&0018 & 252 \\
15 & 57722.4361 & 0.0005 & $-$0&0013 & 154 \\
16 & 57722.5188 & 0.0025 & 0&0022 & 22 \\
27 & 57723.3906 & 0.0002 & 0&0025 & 87 \\
28 & 57723.4702 & 0.0003 & 0&0029 & 88 \\
39 & 57724.3445 & 0.0004 & 0&0057 & 81 \\
59 & 57725.9337 & 0.0007 & 0&0104 & 83 \\
60 & 57726.0107 & 0.0010 & 0&0082 & 82 \\
61 & 57726.0900 & 0.0019 & 0&0083 & 53 \\
65 & 57726.4091 & 0.0003 & 0&0105 & 87 \\
66 & 57726.4879 & 0.0007 & 0&0101 & 45 \\
77 & 57727.3555 & 0.0039 & 0&0062 & 28 \\
78 & 57727.4377 & 0.0006 & 0&0092 & 86 \\
111 & 57730.0398 & 0.0015 & $-$0&0031 & 77 \\
122 & 57730.9117 & 0.0012 & $-$0&0028 & 80 \\
123 & 57730.9881 & 0.0016 & $-$0&0055 & 79 \\
135 & 57731.9451 & 0.0038 & 0&0008 & 78 \\
136 & 57732.0277 & 0.0056 & 0&0042 & 149 \\
137 & 57732.0859 & 0.0018 & $-$0&0169 & 75 \\
147 & 57732.8879 & 0.0013 & $-$0&0072 & 76 \\
148 & 57732.9830 & 0.0042 & 0&0087 & 158 \\
149 & 57733.0476 & 0.0018 & $-$0&0059 & 56 \\
153 & 57733.3638 & 0.0006 & $-$0&0066 & 64 \\
160 & 57733.9249 & 0.0023 & $-$0&0001 & 82 \\
161 & 57733.9942 & 0.0009 & $-$0&0100 & 81 \\
198 & 57736.9455 & 0.0045 & 0&0100 & 52 \\
\hline
  \multicolumn{6}{l}{\commenta BJD$-$2400000.} \\
  \multicolumn{6}{l}{\commentb Against max $= 2457721.2490 + 0.079225 E$.} \\
  \multicolumn{6}{l}{\commentc Number of points used to determine the maximum.} \\
\end{tabular}
\end{center}
\end{table}

\subsection{ASASSN-16nr}\label{obj:asassn16nr}

\begin{figure}
  \begin{center}
    \FigureFile(85mm,110mm){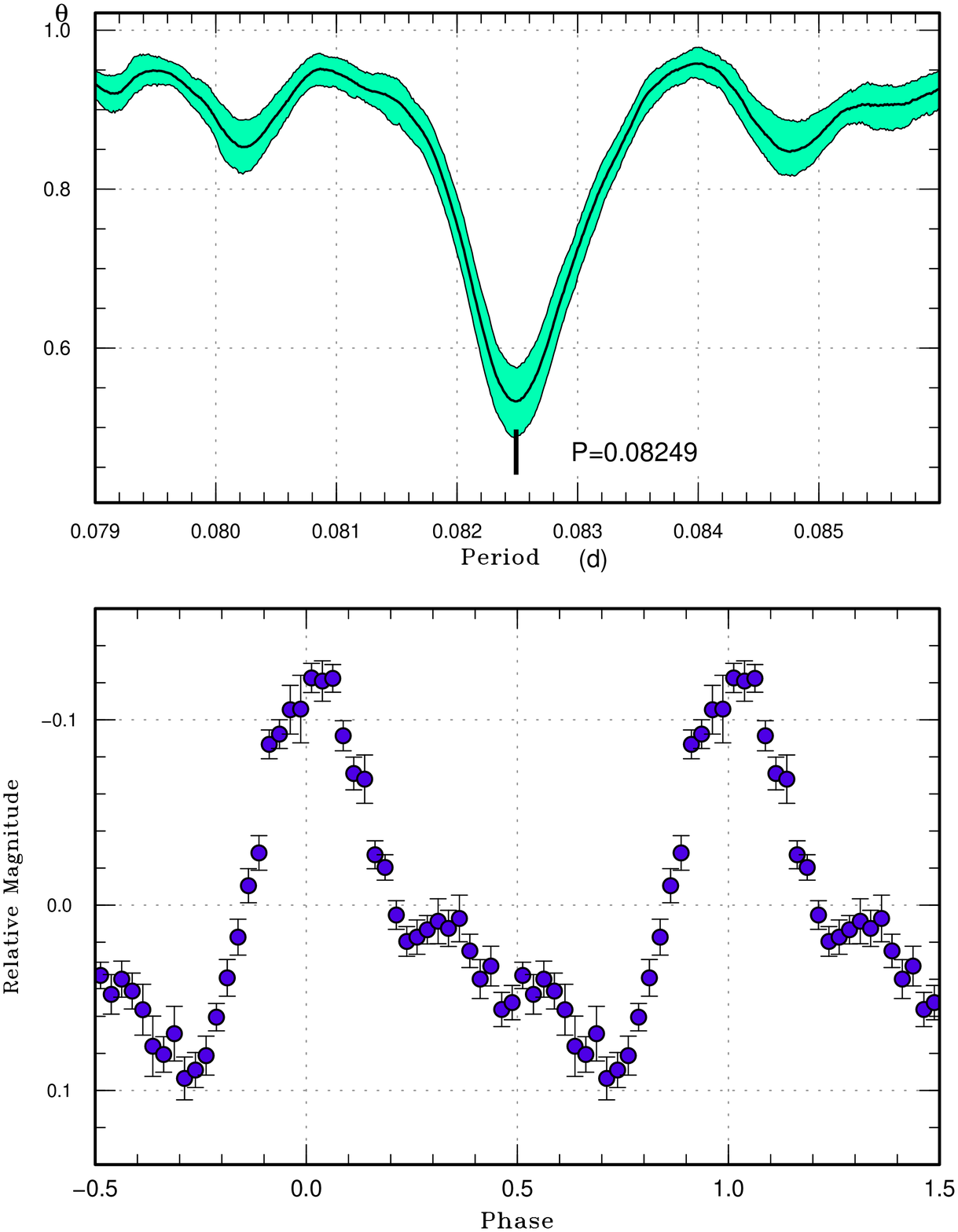}
  \end{center}
  \caption{Superhumps in ASASSN-16nr (2016).
     (Upper): PDM analysis.
     (Lower): Phase-averaged profile.}
  \label{fig:asassn16nrshpdm}
\end{figure}

   This object was detected as a transient
at $V$=15.1 on 2016 November 26 by the ASAS-SN team.
The outburst was announced after the observation
on November 27 at $V$=15.1.
The object showed superhumps (vsnet-alert 20420,
20432; figure \ref{fig:asassn16nrshpdm}).
The times of superhump maxima are listed in
table \ref{tab:asassn16nroc2016}.
Although a large outburst amplitude ($\sim$7 mag)
was suggested, the object was not a WZ Sge-type
dwarf nova since it showed well-developed
ordinary superhumps immediately after the outburst
detection.


\begin{table}
\caption{Superhump maxima of ASASSN-16nr (2016)}\label{tab:asassn16nroc2016}
\begin{center}
\begin{tabular}{rp{55pt}p{40pt}r@{.}lr}
\hline
\multicolumn{1}{c}{$E$} & \multicolumn{1}{c}{max\commenta} & \multicolumn{1}{c}{error} & \multicolumn{2}{c}{$O-C$\commentb} & \multicolumn{1}{c}{$N$\commentc} \\
\hline
0 & 57720.6594 & 0.0047 & $-$0&0044 & 10 \\
1 & 57720.7443 & 0.0010 & $-$0&0022 & 22 \\
2 & 57720.8298 & 0.0026 & 0&0006 & 16 \\
12 & 57721.6464 & 0.0072 & $-$0&0098 & 8 \\
13 & 57721.7419 & 0.0020 & 0&0030 & 21 \\
14 & 57721.8241 & 0.0020 & 0&0024 & 18 \\
21 & 57722.4024 & 0.0005 & 0&0018 & 161 \\
22 & 57722.4874 & 0.0007 & 0&0040 & 165 \\
23 & 57722.5674 & 0.0005 & 0&0013 & 182 \\
32 & 57723.3206 & 0.0078 & 0&0102 & 56 \\
33 & 57723.3915 & 0.0008 & $-$0&0016 & 190 \\
34 & 57723.4754 & 0.0007 & $-$0&0005 & 155 \\
49 & 57724.7288 & 0.0131 & 0&0123 & 29 \\
50 & 57724.7952 & 0.0024 & $-$0&0040 & 30 \\
58 & 57725.4552 & 0.0005 & $-$0&0056 & 163 \\
59 & 57725.5361 & 0.0006 & $-$0&0074 & 189 \\
\hline
  \multicolumn{6}{l}{\commenta BJD$-$2400000.} \\
  \multicolumn{6}{l}{\commentb Against max $= 2457720.6638 + 0.082709 E$.} \\
  \multicolumn{6}{l}{\commentc Number of points used to determine the maximum.} \\
\end{tabular}
\end{center}
\end{table}

\subsection{ASASSN-16nw}\label{obj:asassn16nw}

\begin{figure}
  \begin{center}
    \FigureFile(85mm,110mm){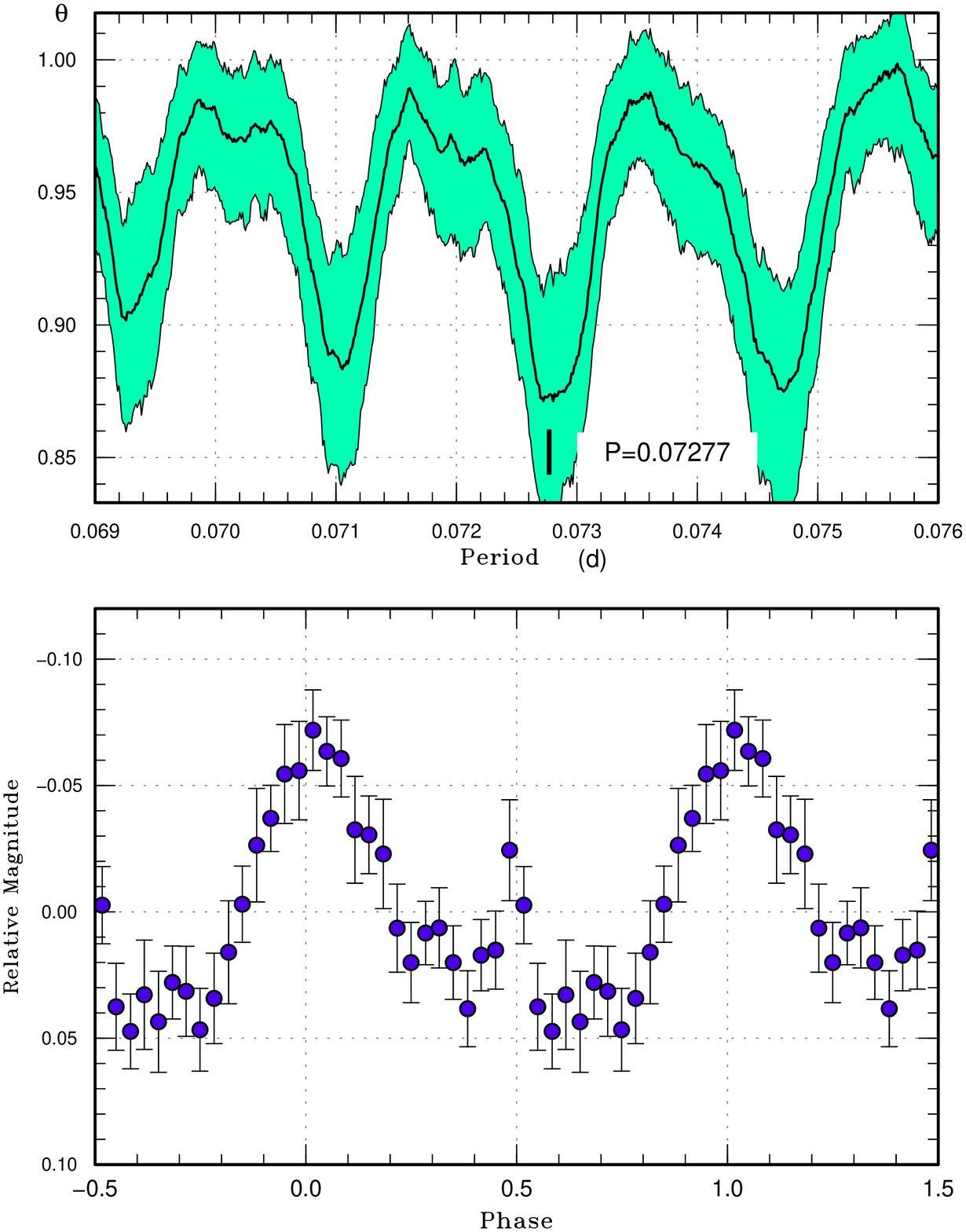}
  \end{center}
  \caption{Superhumps in ASASSN-16nw (2016).
     (Upper): PDM analysis.
     (Lower): Phase-averaged profile.}
  \label{fig:asassn16nwshpdm}
\end{figure}

   This object was detected as a transient
at $V$=15.6 on 2016 November 23 by the ASAS-SN team.
The outburst was announced after the observation
on November 27 at $V$=16.1 and November 29 at $V$=16.3.
Superhumps were detected in observations on
two nights (vsnet-alert 20445; figure \ref{fig:asassn16nwshpdm}).
The alias selection is most likely based on
the single long run on the first night giving
a period of 0.073(1)~d.
The times of superhump maxima are listed in
table \ref{tab:asassn16nwoc2016}.


\begin{table}
\caption{Superhump maxima of ASASSN-16nw (2016)}\label{tab:asassn16nwoc2016}
\begin{center}
\begin{tabular}{rp{55pt}p{40pt}r@{.}lr}
\hline
\multicolumn{1}{c}{$E$} & \multicolumn{1}{c}{max\commenta} & \multicolumn{1}{c}{error} & \multicolumn{2}{c}{$O-C$\commentb} & \multicolumn{1}{c}{$N$\commentc} \\
\hline
0 & 57724.3969 & 0.0011 & $-$0&0028 & 72 \\
1 & 57724.4743 & 0.0015 & 0&0018 & 79 \\
2 & 57724.5471 & 0.0012 & 0&0018 & 80 \\
3 & 57724.6175 & 0.0032 & $-$0&0007 & 35 \\
39 & 57727.2412 & 0.0067 & 0&0018 & 80 \\
40 & 57727.3116 & 0.0023 & $-$0&0006 & 78 \\
41 & 57727.3802 & 0.0060 & $-$0&0048 & 74 \\
42 & 57727.4585 & 0.0029 & 0&0007 & 73 \\
43 & 57727.5334 & 0.0036 & 0&0027 & 36 \\
\hline
  \multicolumn{6}{l}{\commenta BJD$-$2400000.} \\
  \multicolumn{6}{l}{\commentb Against max $= 2457724.3997 + 0.072813 E$.} \\
  \multicolumn{6}{l}{\commentc Number of points used to determine the maximum.} \\
\end{tabular}
\end{center}
\end{table}

\subsection{ASASSN-16ob}\label{obj:asassn16ob}

\begin{figure}
  \begin{center}
    \FigureFile(85mm,110mm){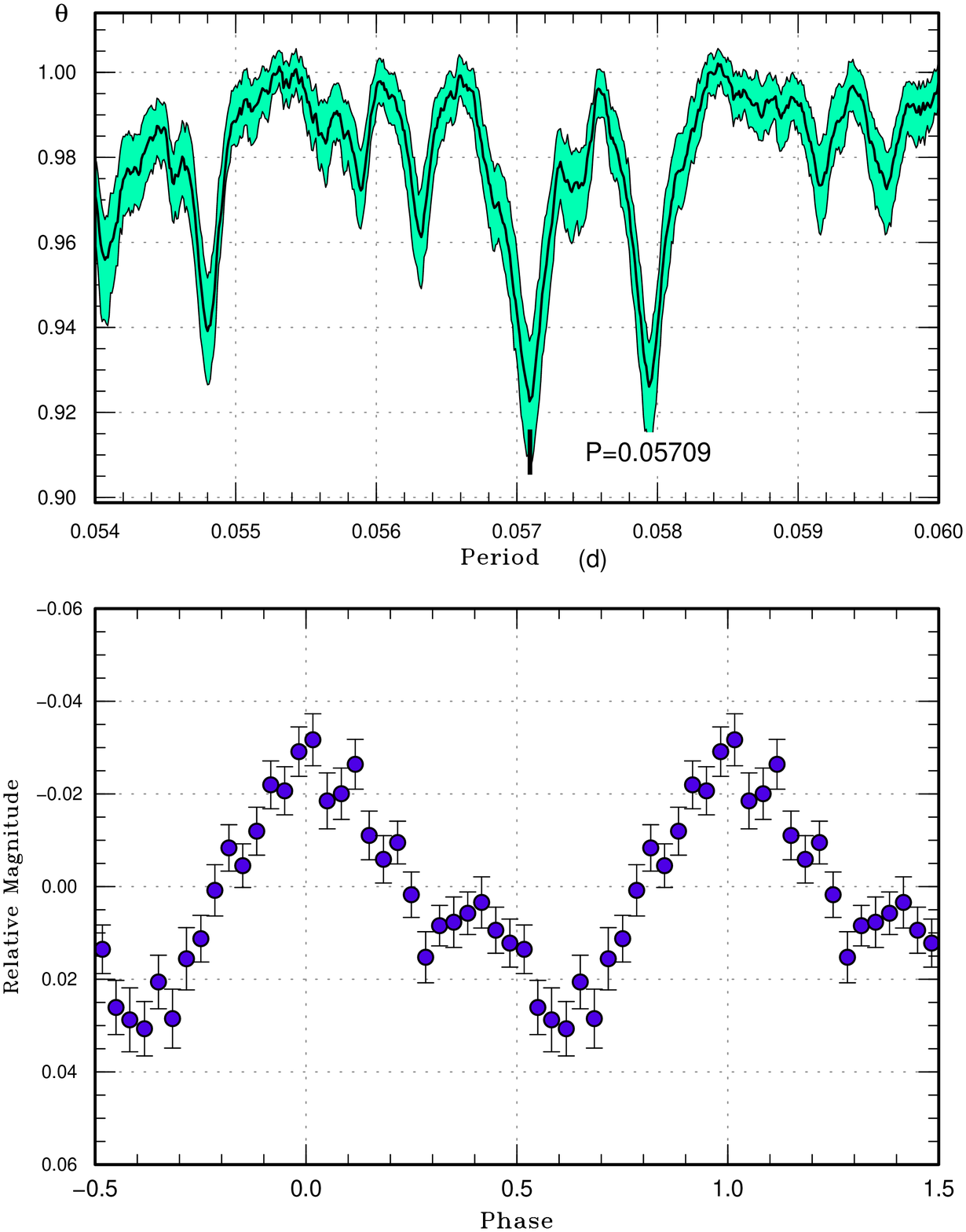}
  \end{center}
  \caption{Ordinary superhumps in ASASSN-16ob (2016).
     The segment of BJD 2457734--2457749 was used.
     (Upper): PDM analysis.
     (Lower): Phase-averaged profile.}
  \label{fig:asassn16obshpdm}
\end{figure}

\begin{figure}
  \begin{center}
    \FigureFile(85mm,110mm){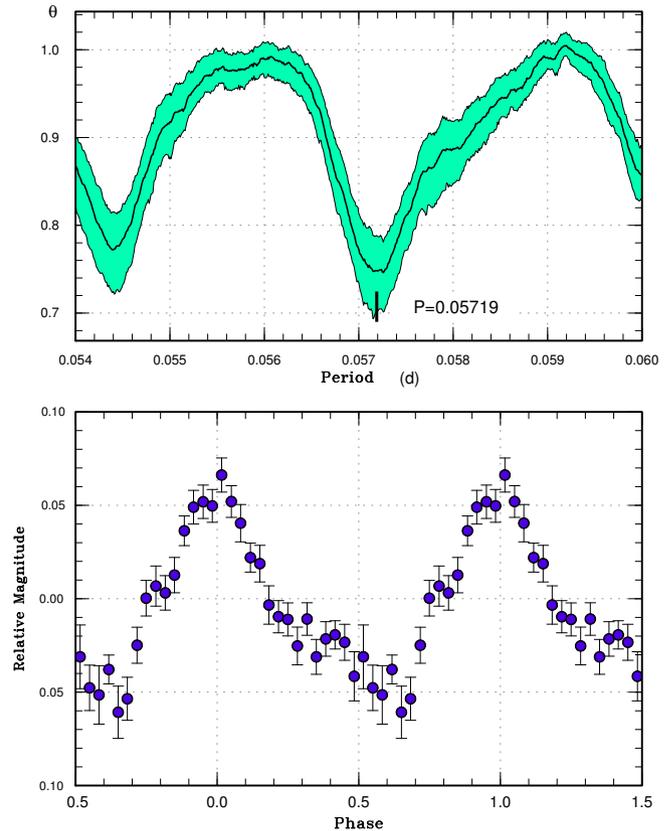}
  \end{center}
  \caption{Ordinary superhumps in ASASSN-16ob (2016).
     The best-observed segment of BJD 2457736--2457740 was used.
     (Upper): PDM analysis.
     (Lower): Phase-averaged profile.}
  \label{fig:asassn16obshpdm2}
\end{figure}

   This object was detected as a transient
at $V$=14.3 on 2016 November 28 by the ASAS-SN team.
The outburst was announced after its further brightening
to $V$=13.8 on November 30.  Although the object was
initially identified with a $B$=18.4 mag star in
USNO catalog, B. Monard obtained outburst astrometry
which indicated that the true counterpart is much
fainter (vsnet-alert 20438).
The object was also detected by Gaia (Gaia16bzl)\footnote{
  $<$http://gsaweb.ast.cam.ac.uk/alerts/alert/Gaia16bzl/$>$.
}
at a magnitude of 13.84 on November 30 and was
announced (with an identification with ASASSN-16ob)
on December 7.  The large outburst amplitude
suggested a WZ Sge-type dwarf nova.
On December 11, low-amplitude superhumps were detected
(vsnet-alert 20464, 20465, 20481), which grew to full
superhumps (vsnet-alert 20484, 20498).
The times of superhump maxima are listed in
table \ref{tab:asassn16oboc2016}.  The maxima for
$E \le$36 correspond to stage A superhumps.
Although a PDM analysis of ordinary superhumps
gave several candidate periods (figure \ref{fig:asassn16obshpdm}),
we consider them false signals due to
low signal-to-noise ratio since an analysis
restricted to better observation quality gave
a single period (figure \ref{fig:asassn16obshpdm2};
one-day aliases can be safely ruled out
by $O-C$ analysis).

   By using the data before BJD 2457734, we could not
detect early superhumps.  The upper limit of
the amplitude of early superhumps was 0.01 mag.
Although we could not detect early superhumps,
we consider that this object belongs to WZ Sge-type
dwarf novae based on its long waiting time
(13~d) for ordinary superhumps to appear and
the large outburst amplitude.
Using the empirical relation between $q$ and
$P_{\rm dot}$ for stage B superhumps
[equation (6) in \cite{kat15wzsge}], the expected
$q$ is 0.069(3) (the error reflects the error
in $P_{\rm dot}$).


\begin{table}
\caption{Superhump maxima of ASASSN-16ob (2016)}\label{tab:asassn16oboc2016}
\begin{center}
\begin{tabular}{rp{55pt}p{40pt}r@{.}lr}
\hline
\multicolumn{1}{c}{$E$} & \multicolumn{1}{c}{max\commenta} & \multicolumn{1}{c}{error} & \multicolumn{2}{c}{$O-C$\commentb} & \multicolumn{1}{c}{$N$\commentc} \\
\hline
0 & 57734.3298 & 0.0094 & $-$0&0111 & 131 \\
1 & 57734.3770 & 0.0028 & $-$0&0210 & 132 \\
2 & 57734.4519 & 0.0037 & $-$0&0033 & 132 \\
3 & 57734.4903 & 0.0031 & $-$0&0221 & 131 \\
34 & 57736.2939 & 0.0009 & 0&0087 & 107 \\
35 & 57736.3500 & 0.0011 & 0&0076 & 131 \\
36 & 57736.4074 & 0.0012 & 0&0078 & 131 \\
52 & 57737.3262 & 0.0009 & 0&0117 & 106 \\
53 & 57737.3825 & 0.0007 & 0&0108 & 131 \\
54 & 57737.4414 & 0.0007 & 0&0125 & 131 \\
104 & 57740.2935 & 0.0009 & 0&0054 & 106 \\
105 & 57740.3464 & 0.0010 & 0&0010 & 130 \\
106 & 57740.4061 & 0.0010 & 0&0036 & 131 \\
107 & 57740.4636 & 0.0014 & 0&0039 & 131 \\
108 & 57740.5184 & 0.0065 & 0&0016 & 32 \\
109 & 57740.5774 & 0.0019 & 0&0034 & 30 \\
110 & 57740.6307 & 0.0015 & $-$0&0006 & 16 \\
111 & 57740.6879 & 0.0036 & $-$0&0005 & 10 \\
112 & 57740.7462 & 0.0023 & 0&0006 & 10 \\
126 & 57741.5406 & 0.0046 & $-$0&0056 & 29 \\
127 & 57741.6035 & 0.0013 & 0&0001 & 25 \\
128 & 57741.6570 & 0.0032 & $-$0&0036 & 11 \\
129 & 57741.7198 & 0.0038 & 0&0020 & 10 \\
130 & 57741.7789 & 0.0075 & 0&0040 & 10 \\
248 & 57748.5146 & 0.0021 & $-$0&0082 & 144 \\
249 & 57748.5714 & 0.0020 & $-$0&0086 & 151 \\
\hline
  \multicolumn{6}{l}{\commenta BJD$-$2400000.} \\
  \multicolumn{6}{l}{\commentb Against max $= 2457734.3409 + 0.057185 E$.} \\
  \multicolumn{6}{l}{\commentc Number of points used to determine the maximum.} \\
\end{tabular}
\end{center}
\end{table}

\subsection{ASASSN-16oi}\label{obj:asassn16oi}

\begin{figure}
  \begin{center}
    \FigureFile(85mm,110mm){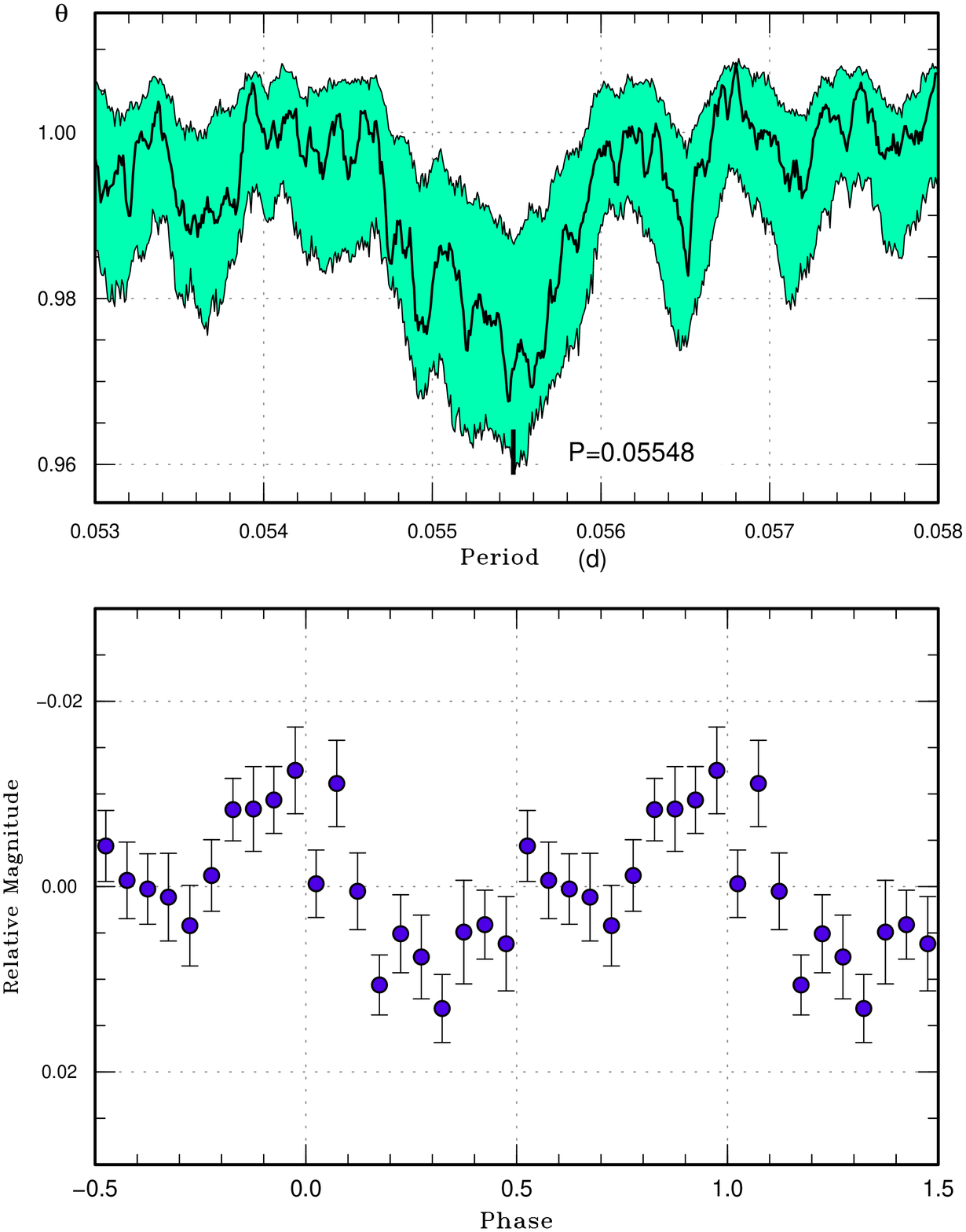}
  \end{center}
  \caption{Early superhumps in ASASSN-16oi (2016).
     (Upper): PDM analysis.
     (Lower): Phase-averaged profile.}
  \label{fig:asassn16oieshpdm}
\end{figure}

\begin{figure}
  \begin{center}
    \FigureFile(85mm,110mm){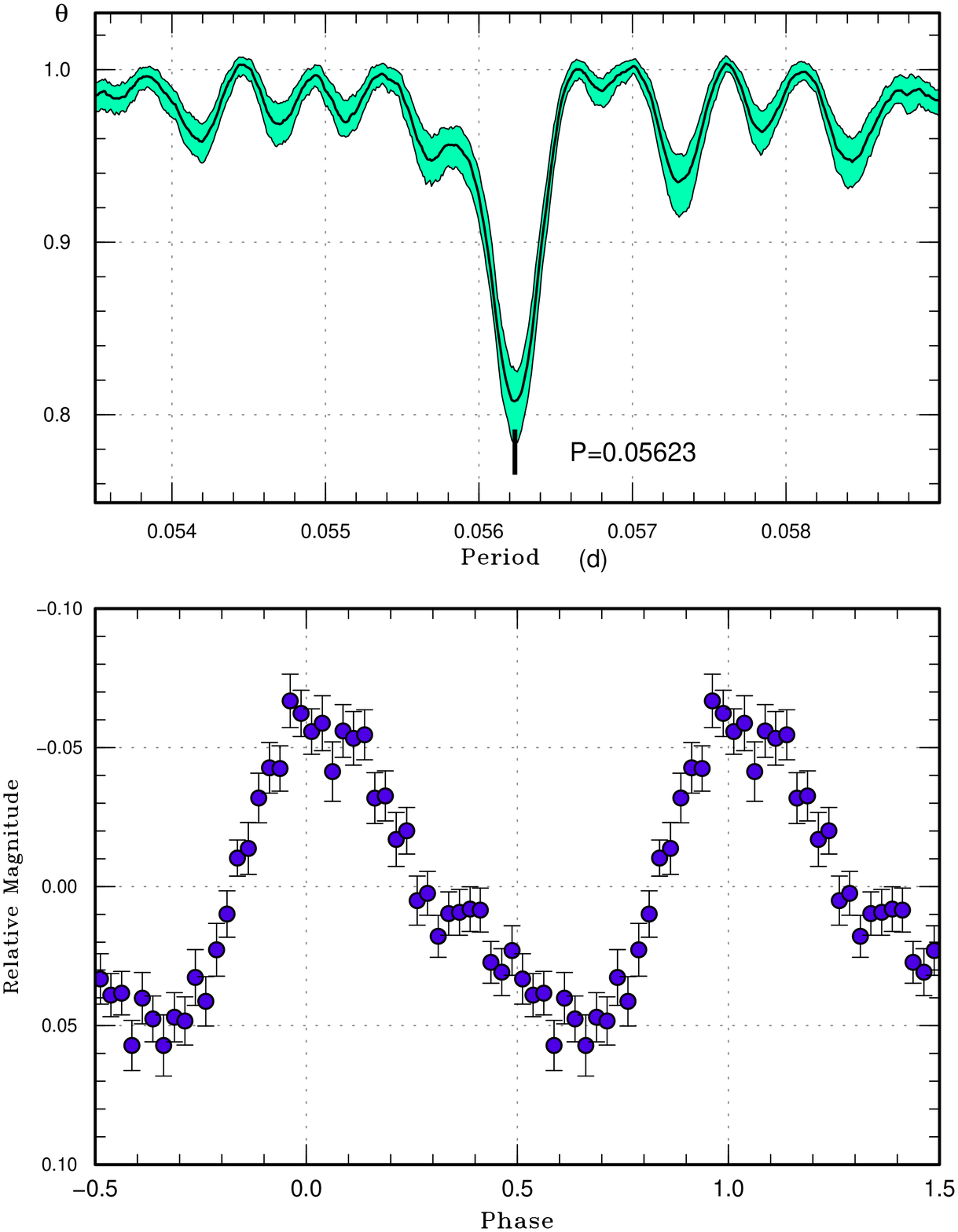}
  \end{center}
  \caption{Ordinary superhumps in ASASSN-16oi (2016).
     The segment of BJD 2457733--2457741 was used.
     (Upper): PDM analysis.
     (Lower): Phase-averaged profile.}
  \label{fig:asassn16oishpdm}
\end{figure}

   This object was detected as a transient
at $V$=13.4 on 2016 December 3 by the ASAS-SN team
(vsnet-alert 20443).
Low-amplitude early superhumps were detected
(vsnet-alert 20466; figure \ref{fig:asassn16oieshpdm}).
The object subsequently showed ordinary superhumps
(vsnet-alert 20466, 20485; figure \ref{fig:asassn16oishpdm}).
The object is confirmed to be a WZ Sge-type dwarf nova.
The times of superhump maxima are listed in
table \ref{tab:asassn16oioc2016}.
The best period of early superhumps with the PDM
method is 0.05548(7)~d.
The fractional superhump excess $\epsilon^*$ for
stage A superhumps is 0.033(2), which gives
$q$=0.091(7).  The relatively large $q$ is consistent
with a relatively large $P_{\rm dot}$ for stage B
superhumps and the relatively large amplitude of superhumps.


\begin{table}
\caption{Superhump maxima of ASASSN-16oi (2016)}\label{tab:asassn16oioc2016}
\begin{center}
\begin{tabular}{rp{55pt}p{40pt}r@{.}lr}
\hline
\multicolumn{1}{c}{$E$} & \multicolumn{1}{c}{max\commenta} & \multicolumn{1}{c}{error} & \multicolumn{2}{c}{$O-C$\commentb} & \multicolumn{1}{c}{$N$\commentc} \\
\hline
0 & 57733.6808 & 0.0038 & $-$0&0088 & 25 \\
2 & 57733.7938 & 0.0012 & $-$0&0082 & 23 \\
12 & 57734.3675 & 0.0033 & 0&0027 & 52 \\
13 & 57734.4273 & 0.0003 & 0&0062 & 129 \\
14 & 57734.4830 & 0.0003 & 0&0056 & 129 \\
24 & 57735.0428 & 0.0004 & 0&0027 & 25 \\
25 & 57735.0991 & 0.0003 & 0&0027 & 34 \\
26 & 57735.1548 & 0.0004 & 0&0021 & 35 \\
27 & 57735.2104 & 0.0003 & 0&0014 & 35 \\
46 & 57736.2774 & 0.0038 & $-$0&0007 & 69 \\
47 & 57736.3351 & 0.0009 & 0&0007 & 129 \\
48 & 57736.3894 & 0.0008 & $-$0&0013 & 130 \\
64 & 57737.2917 & 0.0031 & 0&0006 & 71 \\
65 & 57737.3464 & 0.0010 & $-$0&0009 & 130 \\
66 & 57737.4020 & 0.0010 & $-$0&0017 & 130 \\
67 & 57737.4573 & 0.0013 & $-$0&0026 & 74 \\
118 & 57740.3246 & 0.0015 & $-$0&0054 & 129 \\
119 & 57740.3852 & 0.0012 & $-$0&0010 & 130 \\
120 & 57740.4389 & 0.0021 & $-$0&0036 & 130 \\
121 & 57740.5039 & 0.0017 & 0&0051 & 129 \\
122 & 57740.5595 & 0.0013 & 0&0044 & 129 \\
\hline
  \multicolumn{6}{l}{\commenta BJD$-$2400000.} \\
  \multicolumn{6}{l}{\commentb Against max $= 2457733.6895 + 0.056275 E$.} \\
  \multicolumn{6}{l}{\commentc Number of points used to determine the maximum.} \\
\end{tabular}
\end{center}
\end{table}

\subsection{ASASSN-16os}\label{obj:asassn16os}

\begin{figure}
  \begin{center}
    \FigureFile(85mm,110mm){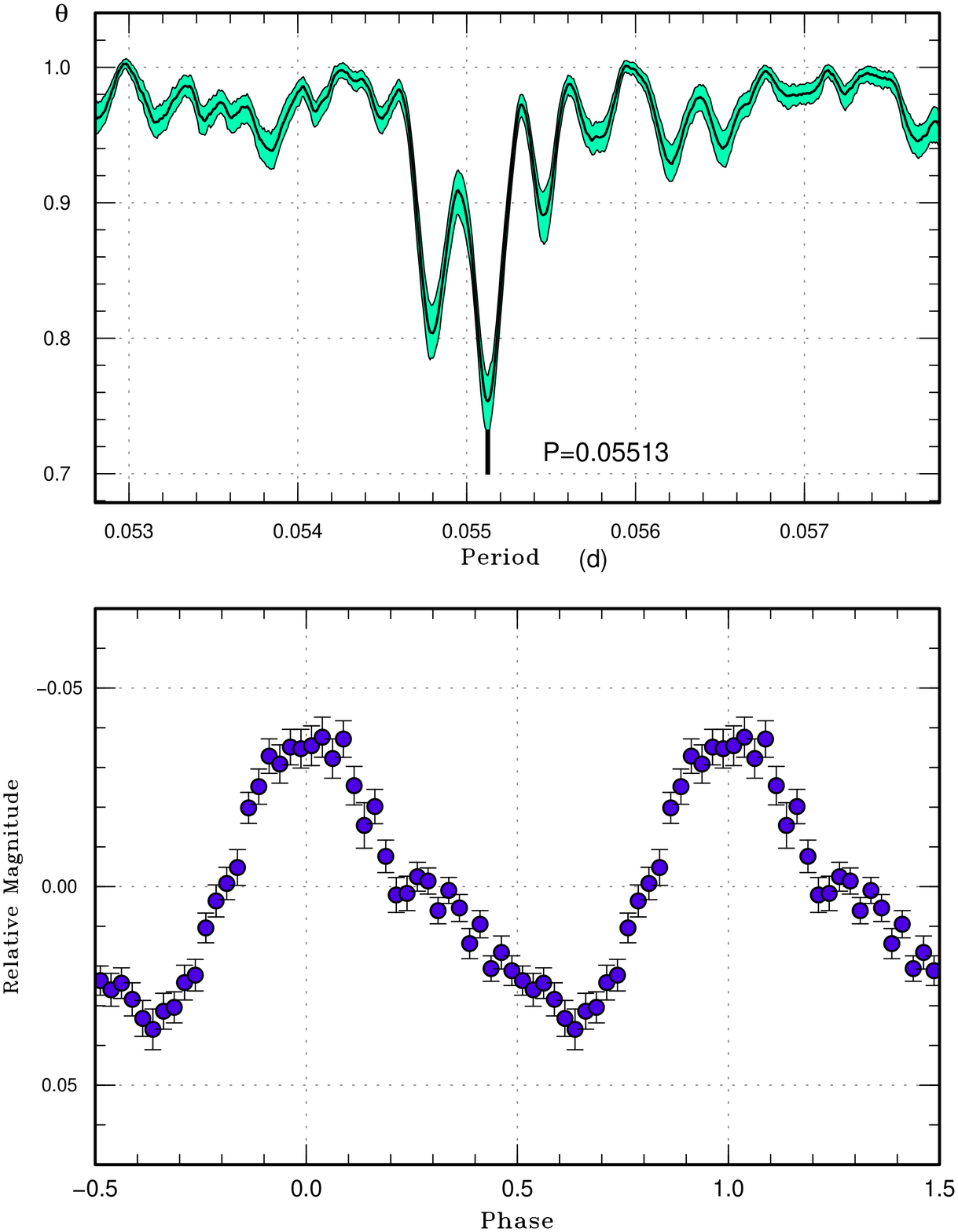}
  \end{center}
  \caption{Ordinary superhumps in ASASSN-16os (2016).
     (Upper): PDM analysis.
     (Lower): Phase-averaged profile.}
  \label{fig:asassn16osshpdm}
\end{figure}

\begin{figure}
  \begin{center}
    \FigureFile(85mm,100mm){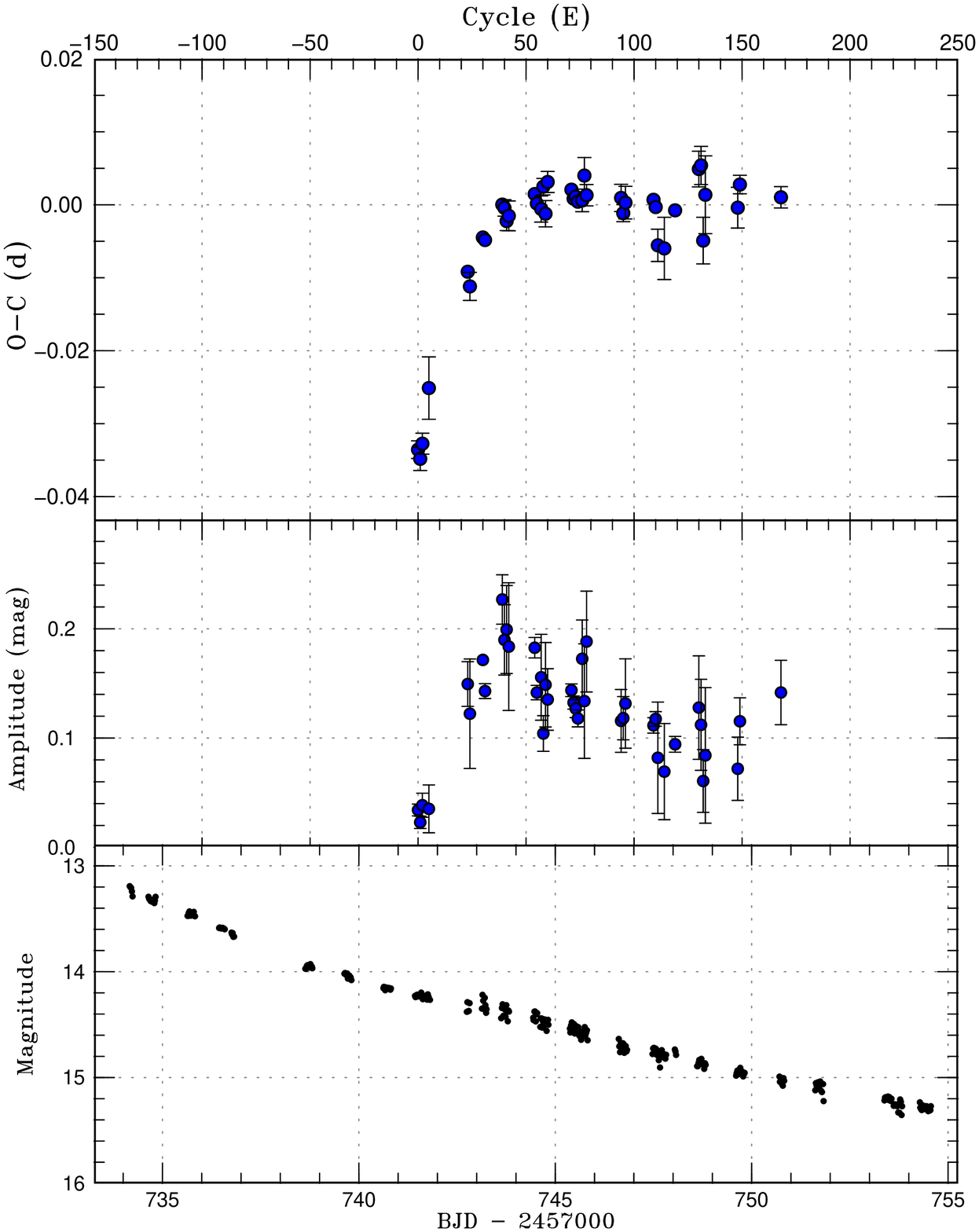}
  \end{center}
  \caption{$O-C$ diagram of superhumps in ASASSN-16os (2016).
     (Upper:) $O-C$ diagram.
     We used a period of 0.05499~d for calculating the $O-C$ residuals.
     (Middle:) Amplitudes of superhumps.
     (Lower:) Light curve.  The data were binned to 0.017~d.
  }
  \label{fig:asassn16oshumpall}
\end{figure}

   This object was detected as a transient
at $V$=13.6 on 2016 December 10 by the ASAS-SN team.
The large outburst amplitude ($\sim$8 mag) attracted
attention.
The object started to show ordinary superhumps
on December 18 (vsnet-alert 20501).
These superhumps grew further (vsnet-alert 20504, 20508;
figure \ref{fig:asassn16osshpdm}).
The times of superhump maxima are listed in
table \ref{tab:asassn16osoc2016}.
Both stages A and B were very clearly recorded
(figure \ref{fig:asassn16oshumpall}).

\begin{figure}
  \begin{center}
    \FigureFile(85mm,110mm){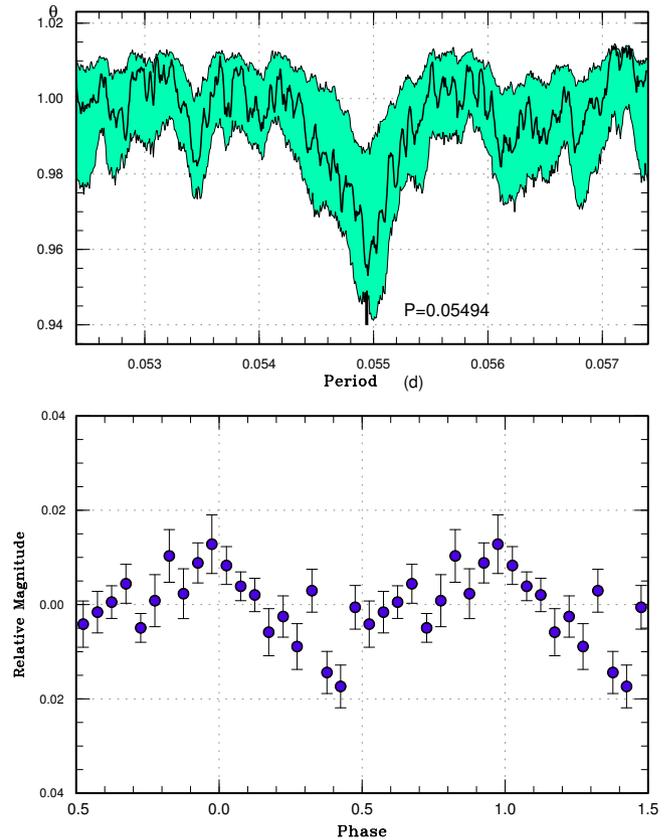}
  \end{center}
  \caption{Possible early superhumps in ASASSN-16os (2016).
     The segment of before BJD 2457741 was used.
     (Upper): PDM analysis.
     (Lower): Phase-averaged profile.}
  \label{fig:asassn16oseshpdm}
\end{figure}

   A PDM analysis of the early part of the data
yielded a signal which may be early superhumps
(figure \ref{fig:asassn16oseshpdm}).
This period was close to that of ordinary superhumps
and we checked a possible contamination of
ordinary superhumps by testing different segments.
Although the test suggested that the period was
not from a contamination of ordinary superhumps,
we were not very confident about the reality of
the signal since the amplitude was small and mostly
only low time-resolution observations were obtained.
If the detected period, 0.05494(6)~d, is that of
early superhumps, the fractional superhump excess
for stage A superhumps is $\epsilon^*$=0.018(1).
Although this value corresponds to $q$=0.047(3),
it needs to be treated with caution due to
the limitation of the quality of observations.
The overall behavior, however, suggests that
this object is a rather extreme WZ Sge-type
dwarf nova.


\begin{table}
\caption{Superhump maxima of ASASSN-16os (2016)}\label{tab:asassn16osoc2016}
\begin{center}
\begin{tabular}{rp{55pt}p{40pt}r@{.}lr}
\hline
\multicolumn{1}{c}{$E$} & \multicolumn{1}{c}{max\commenta} & \multicolumn{1}{c}{error} & \multicolumn{2}{c}{$O-C$\commentb} & \multicolumn{1}{c}{$N$\commentc} \\
\hline
0 & 57741.4738 & 0.0012 & $-$0&0195 & 127 \\
1 & 57741.5276 & 0.0016 & $-$0&0209 & 127 \\
2 & 57741.5846 & 0.0014 & $-$0&0190 & 78 \\
5 & 57741.7572 & 0.0043 & $-$0&0118 & 10 \\
23 & 57742.7630 & 0.0008 & 0&0017 & 7 \\
24 & 57742.8160 & 0.0019 & $-$0&0005 & 5 \\
30 & 57743.1526 & 0.0001 & 0&0054 & 34 \\
31 & 57743.2073 & 0.0004 & 0&0049 & 30 \\
39 & 57743.6520 & 0.0007 & 0&0086 & 10 \\
40 & 57743.7066 & 0.0011 & 0&0080 & 10 \\
41 & 57743.7597 & 0.0013 & 0&0061 & 9 \\
42 & 57743.8155 & 0.0020 & 0&0067 & 7 \\
54 & 57744.4783 & 0.0003 & 0&0080 & 88 \\
55 & 57744.5320 & 0.0003 & 0&0065 & 109 \\
57 & 57744.6412 & 0.0018 & 0&0055 & 9 \\
58 & 57744.6993 & 0.0011 & 0&0084 & 8 \\
59 & 57744.7506 & 0.0018 & 0&0046 & 8 \\
60 & 57744.8099 & 0.0014 & 0&0088 & 8 \\
71 & 57745.4138 & 0.0003 & 0&0062 & 126 \\
72 & 57745.4675 & 0.0003 & 0&0048 & 124 \\
73 & 57745.5228 & 0.0005 & 0&0050 & 126 \\
74 & 57745.5771 & 0.0004 & 0&0041 & 112 \\
76 & 57745.6872 & 0.0015 & 0&0040 & 9 \\
77 & 57745.7456 & 0.0025 & 0&0073 & 8 \\
78 & 57745.7979 & 0.0015 & 0&0045 & 8 \\
94 & 57746.6774 & 0.0019 & 0&0019 & 8 \\
95 & 57746.7303 & 0.0012 & $-$0&0004 & 8 \\
96 & 57746.7867 & 0.0022 & 0&0009 & 8 \\
109 & 57747.5020 & 0.0004 & $-$0&0005 & 127 \\
110 & 57747.5560 & 0.0004 & $-$0&0016 & 127 \\
111 & 57747.6057 & 0.0022 & $-$0&0070 & 24 \\
114 & 57747.7703 & 0.0043 & $-$0&0078 & 9 \\
119 & 57748.0504 & 0.0004 & $-$0&0033 & 40 \\
130 & 57748.6610 & 0.0024 & 0&0008 & 9 \\
131 & 57748.7165 & 0.0026 & 0&0012 & 10 \\
132 & 57748.7612 & 0.0032 & $-$0&0093 & 10 \\
133 & 57748.8224 & 0.0053 & $-$0&0031 & 7 \\
148 & 57749.6455 & 0.0028 & $-$0&0070 & 10 \\
149 & 57749.7037 & 0.0013 & $-$0&0040 & 9 \\
168 & 57750.7467 & 0.0015 & $-$0&0084 & 10 \\
\hline
  \multicolumn{6}{l}{\commenta BJD$-$2400000.} \\
  \multicolumn{6}{l}{\commentb Against max $= 2457741.4933 + 0.055130 E$.} \\
  \multicolumn{6}{l}{\commentc Number of points used to determine the maximum.} \\
\end{tabular}
\end{center}
\end{table}

\subsection{ASASSN-16ow}\label{obj:asassn16ow}

\begin{figure}
  \begin{center}
    \FigureFile(85mm,110mm){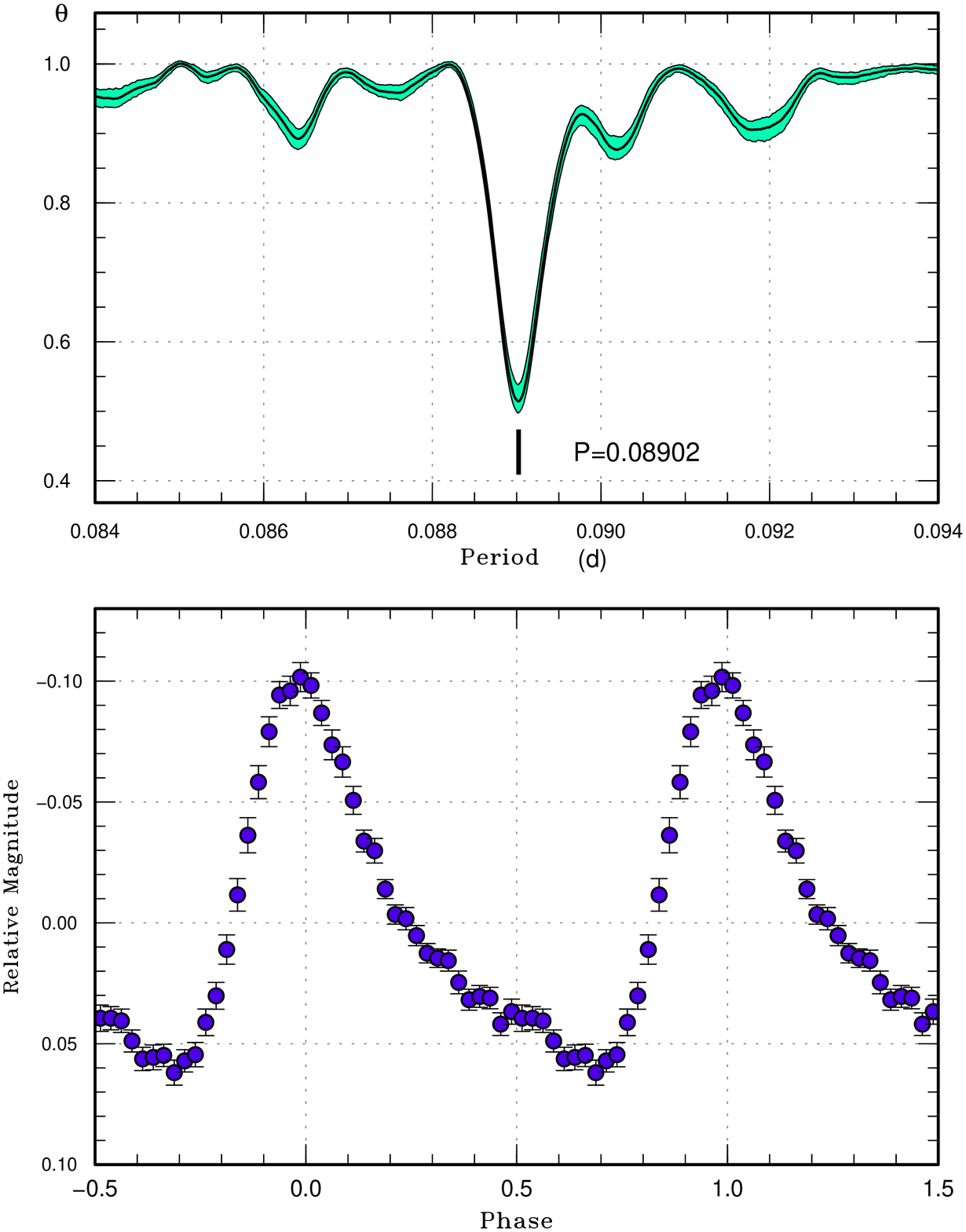}
  \end{center}
  \caption{Superhumps in ASASSN-16ow (2016).
     (Upper): PDM analysis.
     (Lower): Phase-averaged profile.}
  \label{fig:asassn16owshpdm}
\end{figure}

   This object was detected as a transient
at $V$=13.9 on 2016 December 13 by the ASAS-SN team.
Since the object was near the Galactic plane,
it was also suspected to be a nova.  The presence
of a GALEX UV counterpart and an H$\alpha$ emission
in IPHAS catalog (IPHAS2 J063047.05$+$023931.4)
suggested a dwarf nova (vsnet-alert 20474, 20476).
The dwarf nova-type nature was confirmed by spectroscopy
\citep{siv16asassn16owatel9862}.
Subsequent observations detected superhumps
(vsnet-alert 20483, 20497, 20503, 20510;
figure \ref{fig:asassn16owshpdm}).
The times of superhump maxima are listed in
table \ref{tab:asassn16owoc2016}.
The superoutburst lasted at least up to December 25.

   In contrast to many long-$P_{\rm SH}$ SU UMa-type
dwarf novae, this object showed a post-superoutburst
rebrightening on Decmber 29--31 (vsnet-alert 20571).
Although modulations were detected during this
rebrightening, we could not detect a secure
signal of superhumps.


\begin{table}
\caption{Superhump maxima of ASASSN-16ow (2016)}\label{tab:asassn16owoc2016}
\begin{center}
\begin{tabular}{rp{55pt}p{40pt}r@{.}lr}
\hline
\multicolumn{1}{c}{$E$} & \multicolumn{1}{c}{max\commenta} & \multicolumn{1}{c}{error} & \multicolumn{2}{c}{$O-C$\commentb} & \multicolumn{1}{c}{$N$\commentc} \\
\hline
0 & 57737.5297 & 0.0003 & $-$0&0078 & 147 \\
1 & 57737.6182 & 0.0004 & $-$0&0083 & 151 \\
17 & 57739.0492 & 0.0004 & $-$0&0015 & 175 \\
18 & 57739.1439 & 0.0023 & 0&0041 & 61 \\
21 & 57739.4039 & 0.0006 & $-$0&0029 & 90 \\
22 & 57739.4951 & 0.0012 & $-$0&0007 & 52 \\
29 & 57740.1203 & 0.0004 & 0&0014 & 138 \\
30 & 57740.2094 & 0.0005 & 0&0015 & 213 \\
31 & 57740.2984 & 0.0005 & 0&0015 & 275 \\
34 & 57740.5679 & 0.0004 & 0&0040 & 86 \\
35 & 57740.6541 & 0.0004 & 0&0012 & 88 \\
40 & 57741.1031 & 0.0011 & 0&0051 & 92 \\
55 & 57742.4369 & 0.0005 & 0&0037 & 43 \\
56 & 57742.5257 & 0.0005 & 0&0035 & 83 \\
57 & 57742.6176 & 0.0035 & 0&0064 & 21 \\
62 & 57743.0572 & 0.0014 & 0&0010 & 87 \\
66 & 57743.4140 & 0.0004 & 0&0017 & 109 \\
67 & 57743.5031 & 0.0004 & 0&0018 & 117 \\
68 & 57743.5939 & 0.0005 & 0&0036 & 81 \\
79 & 57744.5678 & 0.0021 & $-$0&0017 & 22 \\
88 & 57745.3689 & 0.0005 & $-$0&0017 & 96 \\
89 & 57745.4573 & 0.0004 & $-$0&0023 & 146 \\
90 & 57745.5452 & 0.0004 & $-$0&0034 & 162 \\
96 & 57746.0808 & 0.0019 & $-$0&0018 & 65 \\
97 & 57746.1686 & 0.0010 & $-$0&0031 & 77 \\
100 & 57746.4366 & 0.0005 & $-$0&0021 & 85 \\
101 & 57746.5261 & 0.0007 & $-$0&0016 & 106 \\
102 & 57746.6155 & 0.0016 & $-$0&0013 & 47 \\
\hline
  \multicolumn{6}{l}{\commenta BJD$-$2400000.} \\
  \multicolumn{6}{l}{\commentb Against max $= 2457737.5375 + 0.089011 E$.} \\
  \multicolumn{6}{l}{\commentc Number of points used to determine the maximum.} \\
\end{tabular}
\end{center}
\end{table}

\subsection{ASASSN-17aa}\label{obj:asassn17aa}

\begin{figure}
  \begin{center}
    \FigureFile(85mm,110mm){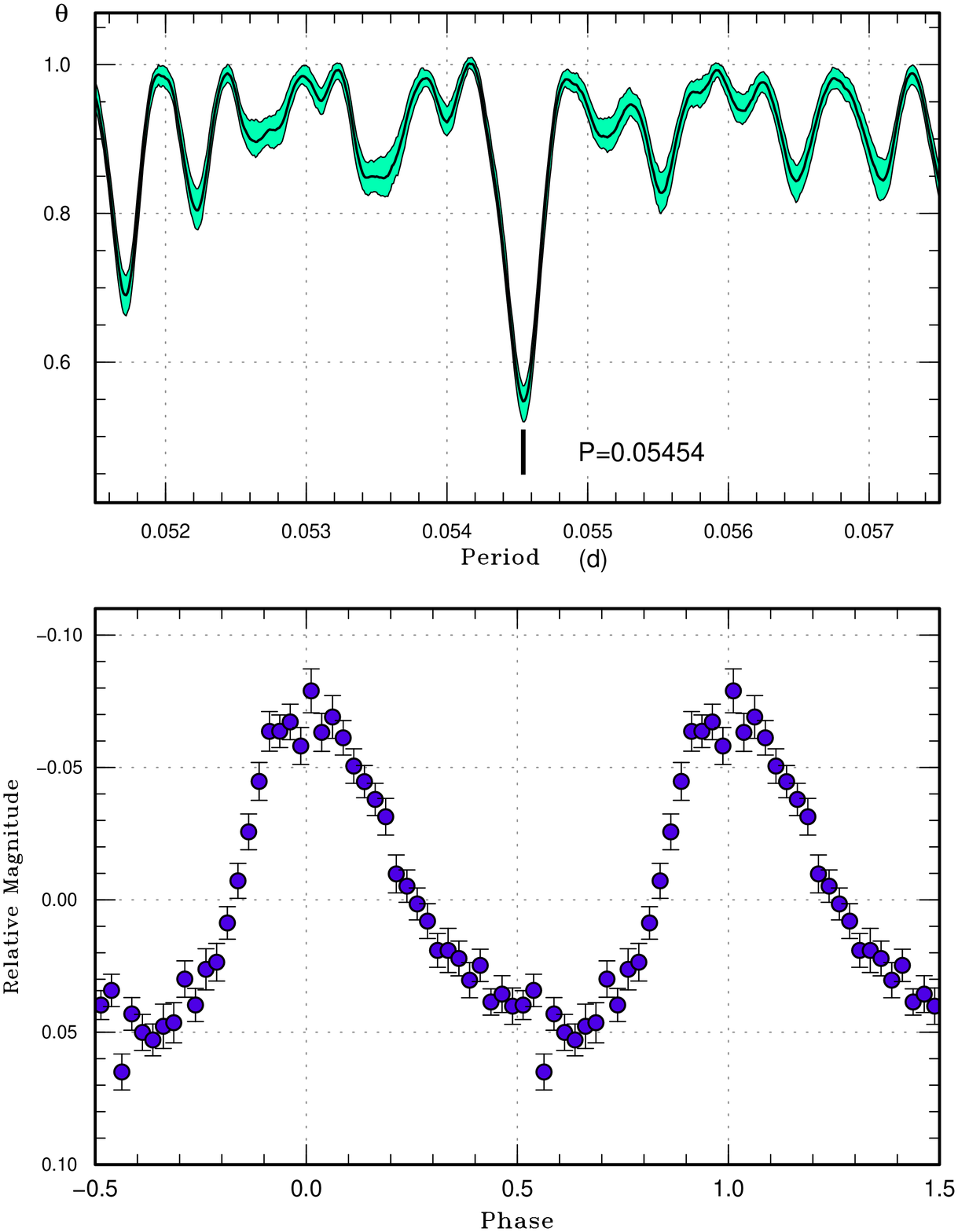}
  \end{center}
  \caption{Ordinary superhumps in ASASSN-17aa (2017).
     (Upper): PDM analysis.
     (Lower): Phase-averaged profile.}
  \label{fig:asassn17aashpdm}
\end{figure}

\begin{figure}
  \begin{center}
    \FigureFile(85mm,110mm){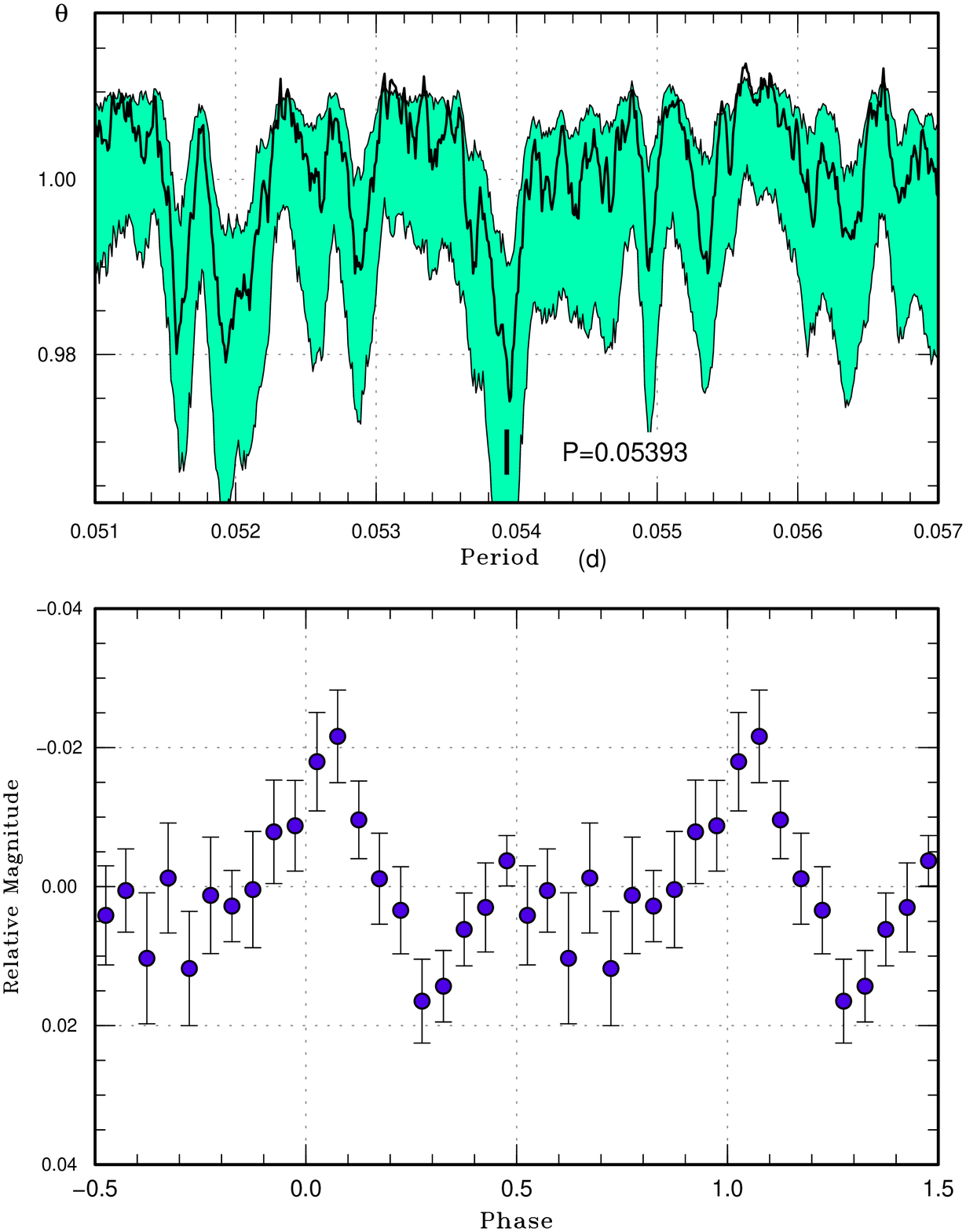}
  \end{center}
  \caption{Possible early superhumps in ASASSN-17aa (2017).
     (Upper): PDM analysis.
     (Lower): Phase-averaged profile.}
  \label{fig:asassn17aaeshpdm}
\end{figure}

   This object was detected as a transient
at $V$=13.9 on 2017 January 2 by the ASAS-SN team
(vsnet-alert 20527).
On January 11, superhumps were finally observed
(vsnet-alert 20562; figure \ref{fig:asassn17aashpdm}).
Although these superhumps were originally suspected
to be stage A ones (vsnet-alert 20572, 20573),
the large superhump amplitudes suggest that
they were already stage B ones.  The cycle numbers in 
table \ref{tab:asassn17aaoc2017} follows this interpretation.
There were possibly low-amplitude early superhumps 
(figure \ref{fig:asassn17aaeshpdm}) with a period
of 0.05393(3)~d.  These properties suggest
the WZ Sge-type classification.


\begin{table}
\caption{Superhump maxima of ASASSN-17aa (2017)}\label{tab:asassn17aaoc2017}
\begin{center}
\begin{tabular}{rp{55pt}p{40pt}r@{.}lr}
\hline
\multicolumn{1}{c}{$E$} & \multicolumn{1}{c}{max\commenta} & \multicolumn{1}{c}{error} & \multicolumn{2}{c}{$O-C$\commentb} & \multicolumn{1}{c}{$N$\commentc} \\
\hline
0 & 57765.2954 & 0.0008 & 0&0036 & 106 \\
1 & 57765.3520 & 0.0005 & 0&0056 & 126 \\
2 & 57765.4052 & 0.0004 & 0&0041 & 104 \\
51 & 57768.0754 & 0.0004 & $-$0&0006 & 49 \\
56 & 57768.3470 & 0.0005 & $-$0&0019 & 125 \\
57 & 57768.3998 & 0.0004 & $-$0&0037 & 112 \\
69 & 57769.0551 & 0.0006 & $-$0&0035 & 49 \\
70 & 57769.1111 & 0.0005 & $-$0&0020 & 49 \\
71 & 57769.1652 & 0.0007 & $-$0&0026 & 49 \\
72 & 57769.2200 & 0.0007 & $-$0&0024 & 48 \\
92 & 57770.3135 & 0.0023 & $-$0&0007 & 78 \\
93 & 57770.3664 & 0.0005 & $-$0&0024 & 125 \\
111 & 57771.3500 & 0.0006 & $-$0&0013 & 101 \\
112 & 57771.4045 & 0.0006 & $-$0&0015 & 126 \\
180 & 57775.1205 & 0.0016 & 0&0024 & 29 \\
181 & 57775.1763 & 0.0009 & 0&0036 & 49 \\
182 & 57775.2306 & 0.0008 & 0&0033 & 49 \\
\hline
  \multicolumn{6}{l}{\commenta BJD$-$2400000.} \\
  \multicolumn{6}{l}{\commentb Against max $= 2457765.2918 + 0.054591 E$.} \\
  \multicolumn{6}{l}{\commentc Number of points used to determine the maximum.} \\
\end{tabular}
\end{center}
\end{table}

\subsection{ASASSN-17ab}\label{obj:asassn17ab}

\begin{figure}
  \begin{center}
    \FigureFile(85mm,110mm){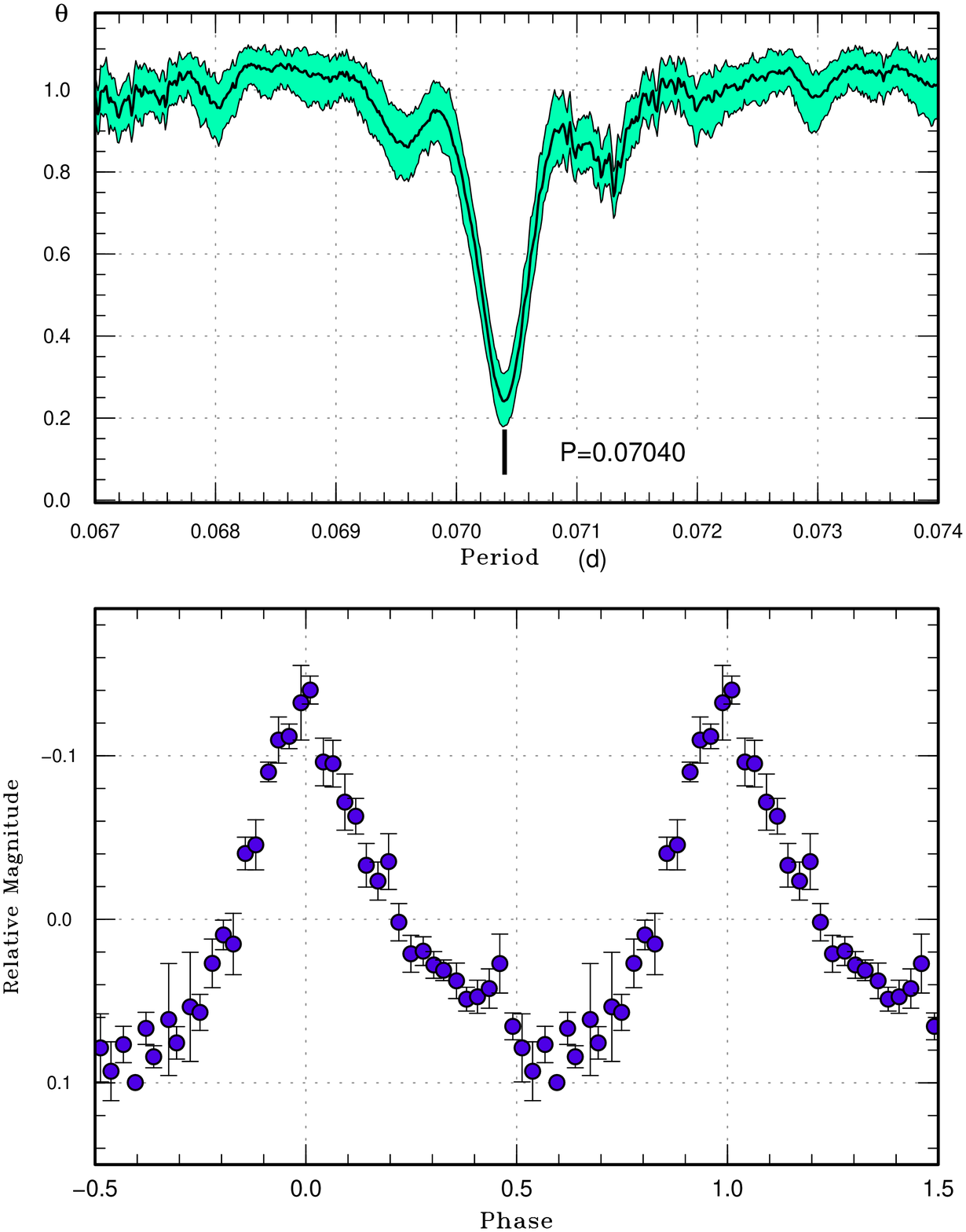}
  \end{center}
  \caption{Superhumps in ASASSN-17ab (2017).
     (Upper): PDM analysis.
     (Lower): Phase-averaged profile.}
  \label{fig:asassn17abshpdm}
\end{figure}

   This object was detected as a transient
at $V$=13.4 on 2017 January 2 by the ASAS-SN team
(vsnet-alert 20527).  The object was already in
outburst at $V$=13.1 on January 1.
Subsequent observations detected superhumps
(vsnet-alert 20531; figure \ref{fig:asassn17abshpdm}).
The times of superhump maxima are listed in
table \ref{tab:asassn17aboc2017}.
Although the maxima for $E \le$2 were stage A
superhumps, we could not determine the period.
   The object was also detected by Gaia
(Gaia17aep)\footnote{
  $<$http://gsaweb.ast.cam.ac.uk/alerts/alert/Gaia17aep/$>$.
}
at a magnitude of 17.33 on January 18.


\begin{table}
\caption{Superhump maxima of ASASSN-17ab (2017)}\label{tab:asassn17aboc2017}
\begin{center}
\begin{tabular}{rp{55pt}p{40pt}r@{.}lr}
\hline
\multicolumn{1}{c}{$E$} & \multicolumn{1}{c}{max\commenta} & \multicolumn{1}{c}{error} & \multicolumn{2}{c}{$O-C$\commentb} & \multicolumn{1}{c}{$N$\commentc} \\
\hline
0 & 57756.6178 & 0.0026 & $-$0&0016 & 9 \\
1 & 57756.6869 & 0.0006 & $-$0&0029 & 13 \\
2 & 57756.7610 & 0.0004 & 0&0008 & 12 \\
15 & 57757.6768 & 0.0009 & 0&0014 & 13 \\
16 & 57757.7470 & 0.0008 & 0&0011 & 13 \\
17 & 57757.8183 & 0.0003 & 0&0020 & 12 \\
29 & 57758.6617 & 0.0006 & 0&0007 & 12 \\
30 & 57758.7314 & 0.0008 & $-$0&0001 & 13 \\
31 & 57758.8006 & 0.0009 & $-$0&0012 & 13 \\
44 & 57759.7163 & 0.0010 & $-$0&0007 & 13 \\
45 & 57759.7882 & 0.0010 & 0&0009 & 13 \\
86 & 57762.6714 & 0.0045 & $-$0&0023 & 12 \\
87 & 57762.7456 & 0.0013 & 0&0014 & 12 \\
88 & 57762.8169 & 0.0013 & 0&0024 & 13 \\
101 & 57763.7288 & 0.0015 & $-$0&0009 & 11 \\
102 & 57763.7990 & 0.0008 & $-$0&0011 & 14 \\
\hline
  \multicolumn{6}{l}{\commenta BJD$-$2400000.} \\
  \multicolumn{6}{l}{\commentb Against max $= 2457756.6194 + 0.070399 E$.} \\
  \multicolumn{6}{l}{\commentc Number of points used to determine the maximum.} \\
\end{tabular}
\end{center}
\end{table}

\subsection{ASASSN-17az}\label{obj:asassn17az}

\begin{figure}
  \begin{center}
    \FigureFile(85mm,110mm){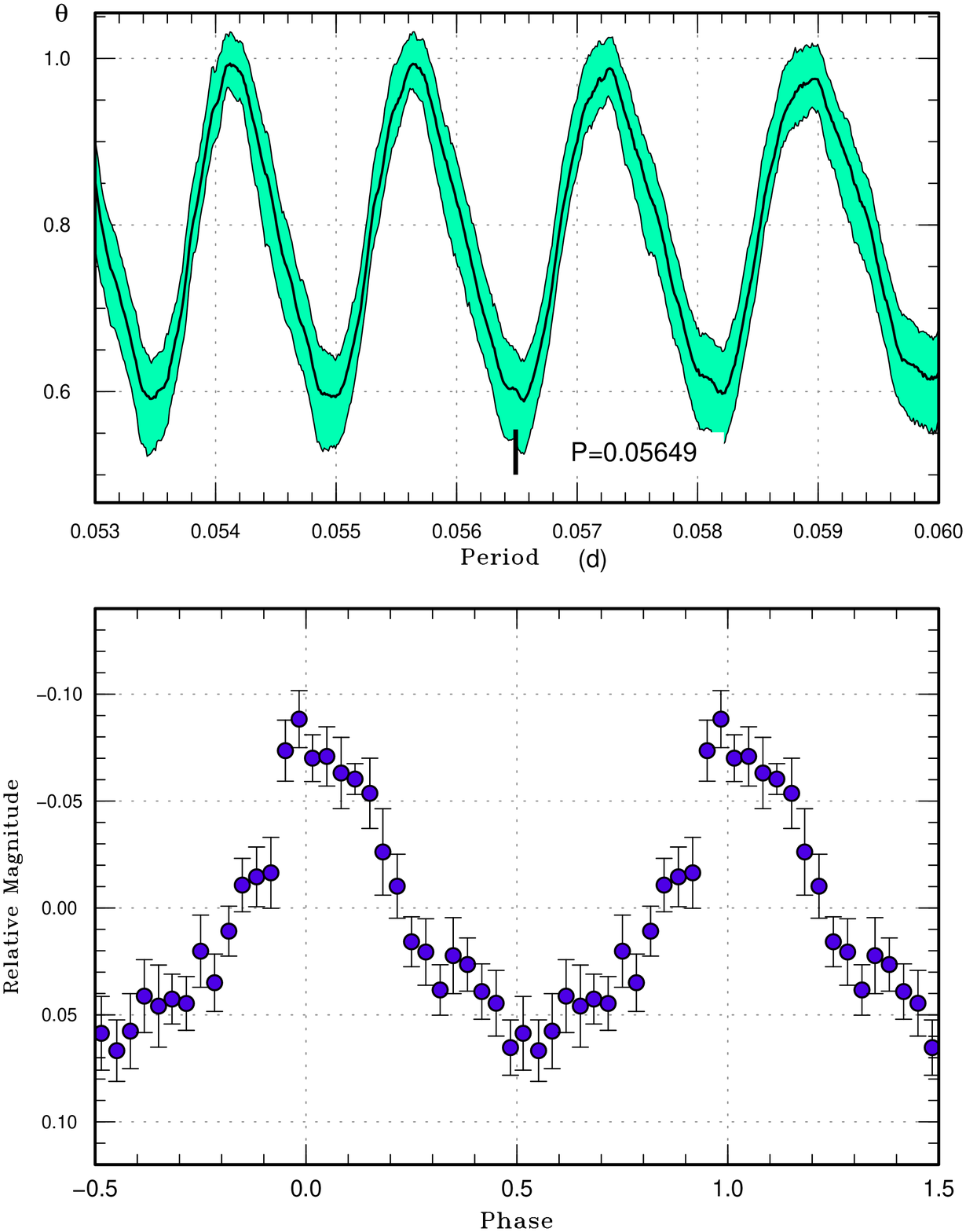}
  \end{center}
  \caption{Superhumps in ASASSN-17az (2017).
     (Upper): PDM analysis.  The most likely alias
     based on $O-C$ analysis was chosen.
     (Lower): Phase-averaged profile.}
  \label{fig:asassn17azshpdm}
\end{figure}

   This object was detected as a transient
at $V$=14.4 on 2017 January 19 by the ASAS-SN team.
Subsequent observations detected superhumps
(vsnet-alert 20626; figure \ref{fig:asassn17azshpdm}).
The times of superhump maxima are listed in
table \ref{tab:asassn17azoc2017}.
Since there was a 2-d gap between observations,
there remained viable aliases.  Among these aliases,
the listed period gave the smallest $O-C$ values
and we selected it as the most likely one.


\begin{table}
\caption{Superhump maxima of ASASSN-17az (2017)}\label{tab:asassn17azoc2017}
\begin{center}
\begin{tabular}{rp{55pt}p{40pt}r@{.}lr}
\hline
\multicolumn{1}{c}{$E$} & \multicolumn{1}{c}{max\commenta} & \multicolumn{1}{c}{error} & \multicolumn{2}{c}{$O-C$\commentb} & \multicolumn{1}{c}{$N$\commentc} \\
\hline
0 & 57783.3017 & 0.0005 & $-$0&0006 & 130 \\
1 & 57783.3594 & 0.0022 & 0&0006 & 76 \\
35 & 57785.2807 & 0.0016 & 0&0012 & 82 \\
36 & 57785.3348 & 0.0011 & $-$0&0012 & 129 \\
\hline
  \multicolumn{6}{l}{\commenta BJD$-$2400000.} \\
  \multicolumn{6}{l}{\commentb Against max $= 2457783.3023 + 0.056492 E$.} \\
  \multicolumn{6}{l}{\commentc Number of points used to determine the maximum.} \\
\end{tabular}
\end{center}
\end{table}

\subsection{ASASSN-17bl}\label{obj:asassn17bl}

\begin{figure}
  \begin{center}
    \FigureFile(85mm,110mm){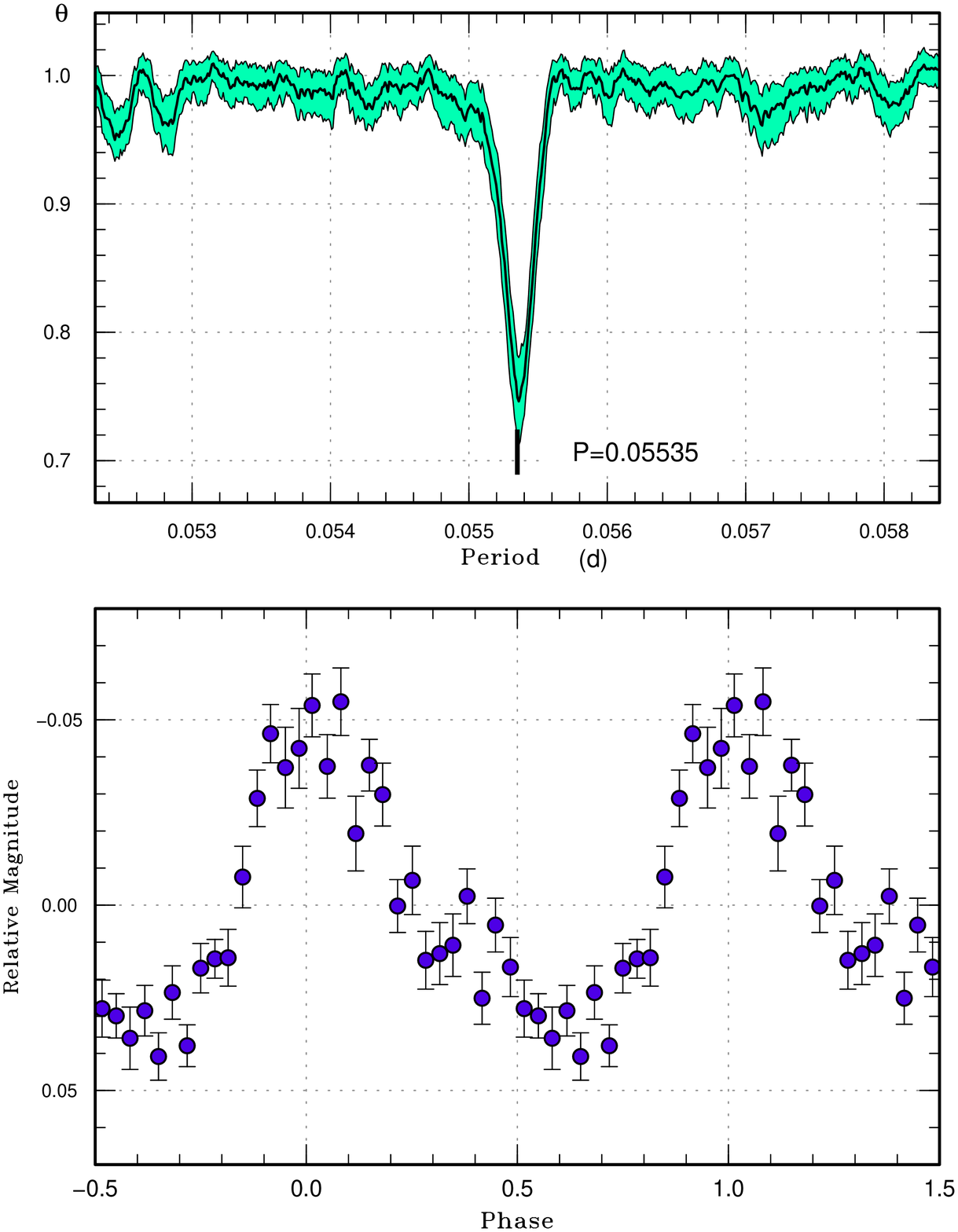}
  \end{center}
  \caption{Ordinary superhumps in ASASSN-17bl (2017).
     (Upper): PDM analysis.
     (Lower): Phase-averaged profile.}
  \label{fig:asassn17blshpdm}
\end{figure}

   This object was detected as a transient
at $V$=13.7 on 2017 January 24 by the ASAS-SN team.
On February 4, the object started to show
superhumps (vsnet-alert 20641).
The long waiting time (11~d) of
superhumps strongly suggested a WZ Sge-type object.
Further development of superhumps was recorded
(vsnet-alert 20646; figure \ref{fig:asassn17blshpdm}).
The times of superhump maxima are listed in
table \ref{tab:asassn17bloc2017}.
Although individual superhumps were not very
well covered (most of them had only 10 observations
or even less; the maxima could be reasonably
determined since the object was bright),
the overall $O-C$ diagram indicates the clear
presence of stage A and stage B with a positive
$P_{\rm dot}$.  The stage A-B transition was
rather uncertain due to the lack of observations
in the initial part.

\begin{figure}
  \begin{center}
    \FigureFile(85mm,110mm){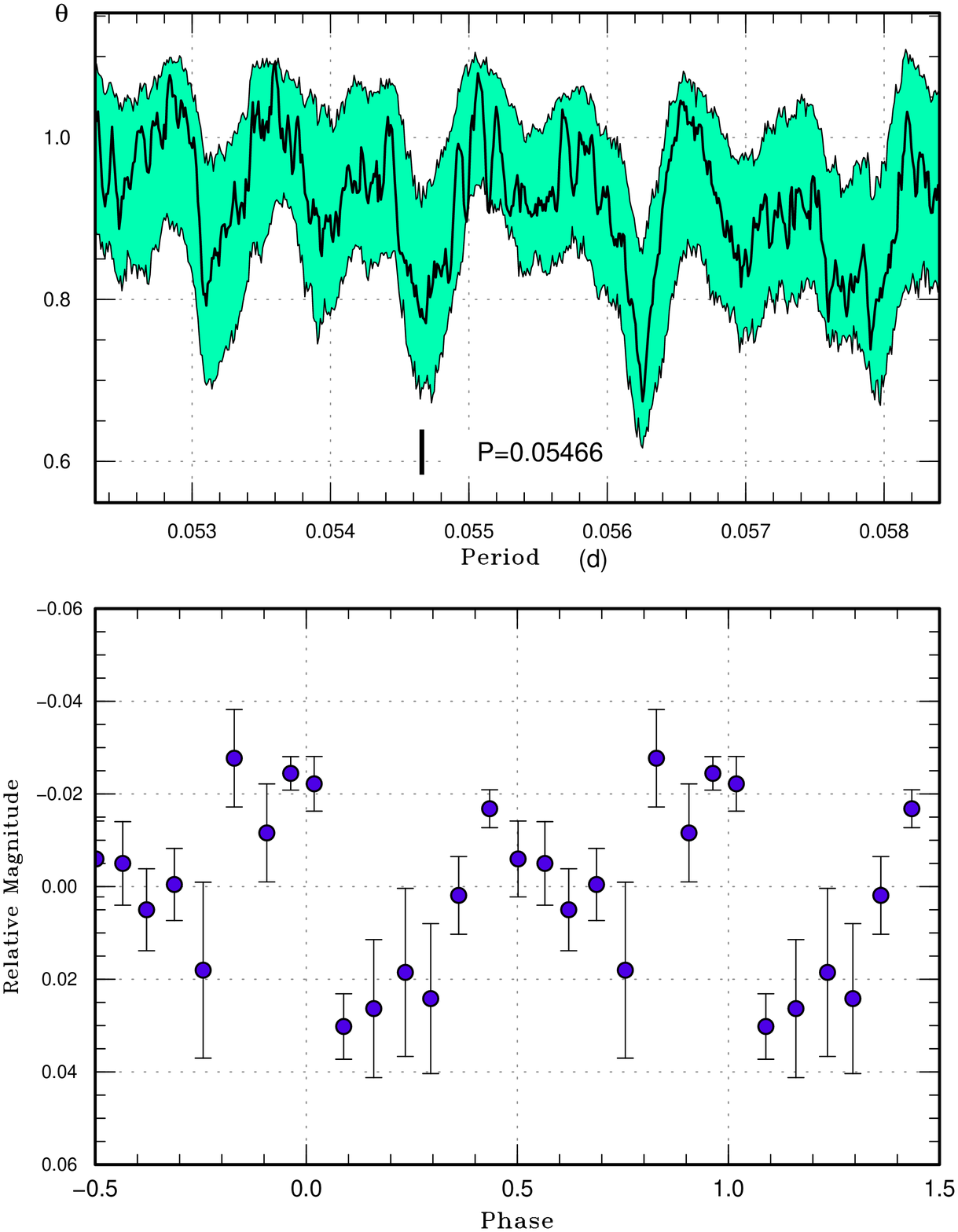}
  \end{center}
  \caption{Possible early superhumps in ASASSN-17bl (2017).
     (Upper): PDM analysis.
     (Lower): Phase-averaged profile.}
  \label{fig:asassn17bleshpdm}
\end{figure}

   An analysis of the early part of the superoutburst
yielded a possible signal of early superhumps
(figure \ref{fig:asassn17bleshpdm}).  Although
there was a stronger signal around 0.0563~d,
we consider it a false alias since it does not
match the period of ordinary superhumps
(the period of early superhumps should be shorter
than that of ordinary superhumps).
The suggested period by the PDM method was
0.05467(5)~d.  The $\epsilon^*$ of stage A superhumps
determined using this period was 0.0235(9),
which corresponds to $q$=0.062(3).
This value, however, could have a larger uncertainty
since both the orbital period and the period
of stage A superhumps were determined from insufficient
observations.  The small $q$ value, however,
appears to be consistent with the WZ Sge-type
behavior, the small amplitude of ordinary
superhumps and the small $P_{\rm dot}$ for
stage B superhumps.


\begin{table}
\caption{Superhump maxima of ASASSN-17bl (2017)}\label{tab:asassn17bloc2017}
\begin{center}
\begin{tabular}{rp{55pt}p{40pt}r@{.}lr}
\hline
\multicolumn{1}{c}{$E$} & \multicolumn{1}{c}{max\commenta} & \multicolumn{1}{c}{error} & \multicolumn{2}{c}{$O-C$\commentb} & \multicolumn{1}{c}{$N$\commentc} \\
\hline
0 & 57788.6964 & 0.0020 & $-$0&0151 & 10 \\
1 & 57788.7494 & 0.0020 & $-$0&0176 & 12 \\
2 & 57788.8101 & 0.0046 & $-$0&0123 & 12 \\
35 & 57790.6538 & 0.0029 & 0&0035 & 9 \\
36 & 57790.7135 & 0.0027 & 0&0078 & 8 \\
37 & 57790.7666 & 0.0011 & 0&0055 & 9 \\
38 & 57790.8238 & 0.0008 & 0&0073 & 9 \\
43 & 57791.0939 & 0.0027 & 0&0004 & 12 \\
53 & 57791.6557 & 0.0029 & 0&0083 & 9 \\
54 & 57791.7093 & 0.0015 & 0&0065 & 8 \\
55 & 57791.7620 & 0.0026 & 0&0039 & 8 \\
56 & 57791.8193 & 0.0024 & 0&0057 & 8 \\
71 & 57792.6505 & 0.0024 & 0&0060 & 9 \\
72 & 57792.7069 & 0.0048 & 0&0071 & 9 \\
73 & 57792.7601 & 0.0023 & 0&0049 & 9 \\
74 & 57792.8141 & 0.0017 & 0&0035 & 10 \\
75 & 57792.8664 & 0.0015 & 0&0004 & 6 \\
79 & 57793.0867 & 0.0050 & $-$0&0009 & 13 \\
89 & 57793.6408 & 0.0022 & $-$0&0007 & 10 \\
90 & 57793.6989 & 0.0046 & 0&0020 & 9 \\
91 & 57793.7539 & 0.0036 & 0&0016 & 9 \\
92 & 57793.8126 & 0.0045 & 0&0049 & 10 \\
93 & 57793.8611 & 0.0016 & $-$0&0021 & 8 \\
97 & 57794.0852 & 0.0004 & 0&0005 & 33 \\
98 & 57794.1388 & 0.0006 & $-$0&0013 & 35 \\
99 & 57794.1944 & 0.0006 & $-$0&0011 & 35 \\
100 & 57794.2504 & 0.0006 & $-$0&0005 & 34 \\
110 & 57794.8038 & 0.0020 & $-$0&0010 & 10 \\
132 & 57796.0206 & 0.0012 & $-$0&0029 & 31 \\
133 & 57796.0729 & 0.0019 & $-$0&0059 & 34 \\
147 & 57796.8495 & 0.0019 & $-$0&0049 & 35 \\
162 & 57797.6856 & 0.0018 & 0&0004 & 8 \\
163 & 57797.7424 & 0.0024 & 0&0017 & 10 \\
164 & 57797.7939 & 0.0027 & $-$0&0022 & 14 \\
180 & 57798.6742 & 0.0036 & $-$0&0082 & 9 \\
182 & 57798.7861 & 0.0016 & $-$0&0071 & 24 \\
183 & 57798.8476 & 0.0022 & $-$0&0009 & 17 \\
189 & 57799.1785 & 0.0011 & $-$0&0024 & 24 \\
191 & 57799.2947 & 0.0045 & 0&0030 & 13 \\
199 & 57799.7322 & 0.0022 & $-$0&0026 & 10 \\
200 & 57799.7852 & 0.0016 & $-$0&0051 & 25 \\
201 & 57799.8441 & 0.0035 & $-$0&0015 & 20 \\
204 & 57800.0146 & 0.0021 & 0&0028 & 9 \\
217 & 57800.7303 & 0.0031 & $-$0&0016 & 9 \\
235 & 57801.7365 & 0.0044 & 0&0075 & 11 \\
236 & 57801.7843 & 0.0029 & $-$0&0001 & 18 \\
237 & 57801.8429 & 0.0040 & 0&0031 & 16 \\
\hline
  \multicolumn{6}{l}{\commenta BJD$-$2400000.} \\
  \multicolumn{6}{l}{\commentb Against max $= 2457788.7115 + 0.055393 E$.} \\
  \multicolumn{6}{l}{\commentc Number of points used to determine the maximum.} \\
\end{tabular}
\end{center}
\end{table}

\subsection{ASASSN-17bm}\label{obj:asassn17bm}

\begin{figure}
  \begin{center}
    \FigureFile(85mm,110mm){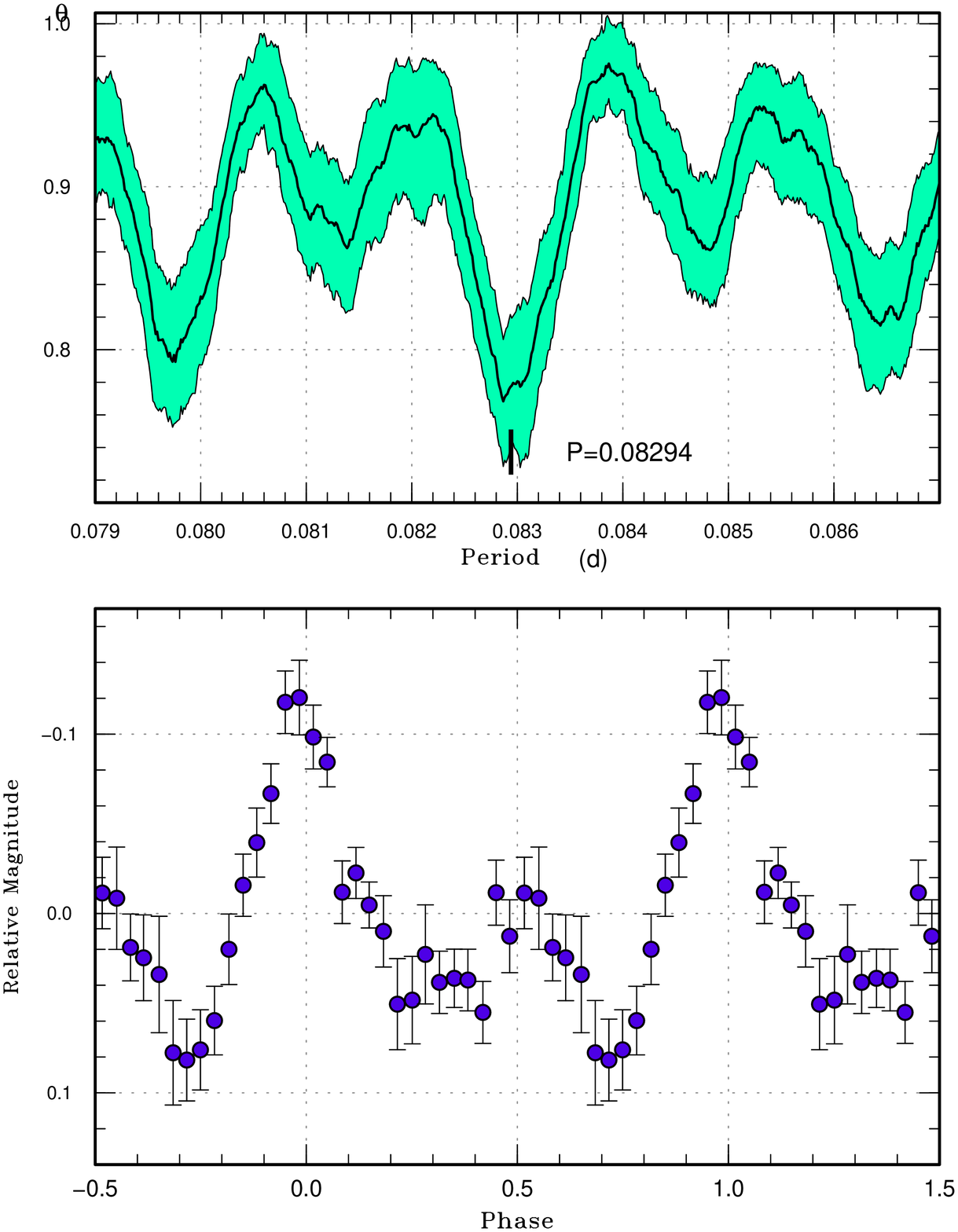}
  \end{center}
  \caption{Superhumps in ASASSN-17bm (2017).
     (Upper): PDM analysis.
     (Lower): Phase-averaged profile.}
  \label{fig:asassn17bmshpdm}
\end{figure}

   This object was detected as a transient
at $V$=15.9 on 2017 January 25 by the ASAS-SN team.
The outburst was announced after confirmation
on January 27.  Subsequent observations
detected superhumps (vsnet-alert 20627;
figure \ref{fig:asassn17bmshpdm}).
The times of superhump maxima are listed in
table \ref{tab:asassn17bmoc2017}.
The period in table \ref{tab:perlist} was determined
by the PDM method since individual maxima were
not very well determined.


\begin{table}
\caption{Superhump maxima of ASASSN-17bm (2017)}\label{tab:asassn17bmoc2017}
\begin{center}
\begin{tabular}{rp{55pt}p{40pt}r@{.}lr}
\hline
\multicolumn{1}{c}{$E$} & \multicolumn{1}{c}{max\commenta} & \multicolumn{1}{c}{error} & \multicolumn{2}{c}{$O-C$\commentb} & \multicolumn{1}{c}{$N$\commentc} \\
\hline
0 & 57783.3970 & 0.0007 & 0&0005 & 152 \\
24 & 57785.3910 & 0.0017 & 0&0018 & 155 \\
25 & 57785.4720 & 0.0011 & $-$0&0002 & 191 \\
28 & 57785.7252 & 0.0039 & 0&0038 & 12 \\
29 & 57785.7979 & 0.0058 & $-$0&0065 & 15 \\
52 & 57787.7031 & 0.0036 & $-$0&0110 & 12 \\
53 & 57787.8088 & 0.0073 & 0&0117 & 18 \\
\hline
  \multicolumn{6}{l}{\commenta BJD$-$2400000.} \\
  \multicolumn{6}{l}{\commentb Against max $= 2457783.3965 + 0.083032 E$.} \\
  \multicolumn{6}{l}{\commentc Number of points used to determine the maximum.} \\
\end{tabular}
\end{center}
\end{table}

\subsection{ASASSN-17bv}\label{obj:asassn17bv}

\begin{figure}
  \begin{center}
    \FigureFile(85mm,110mm){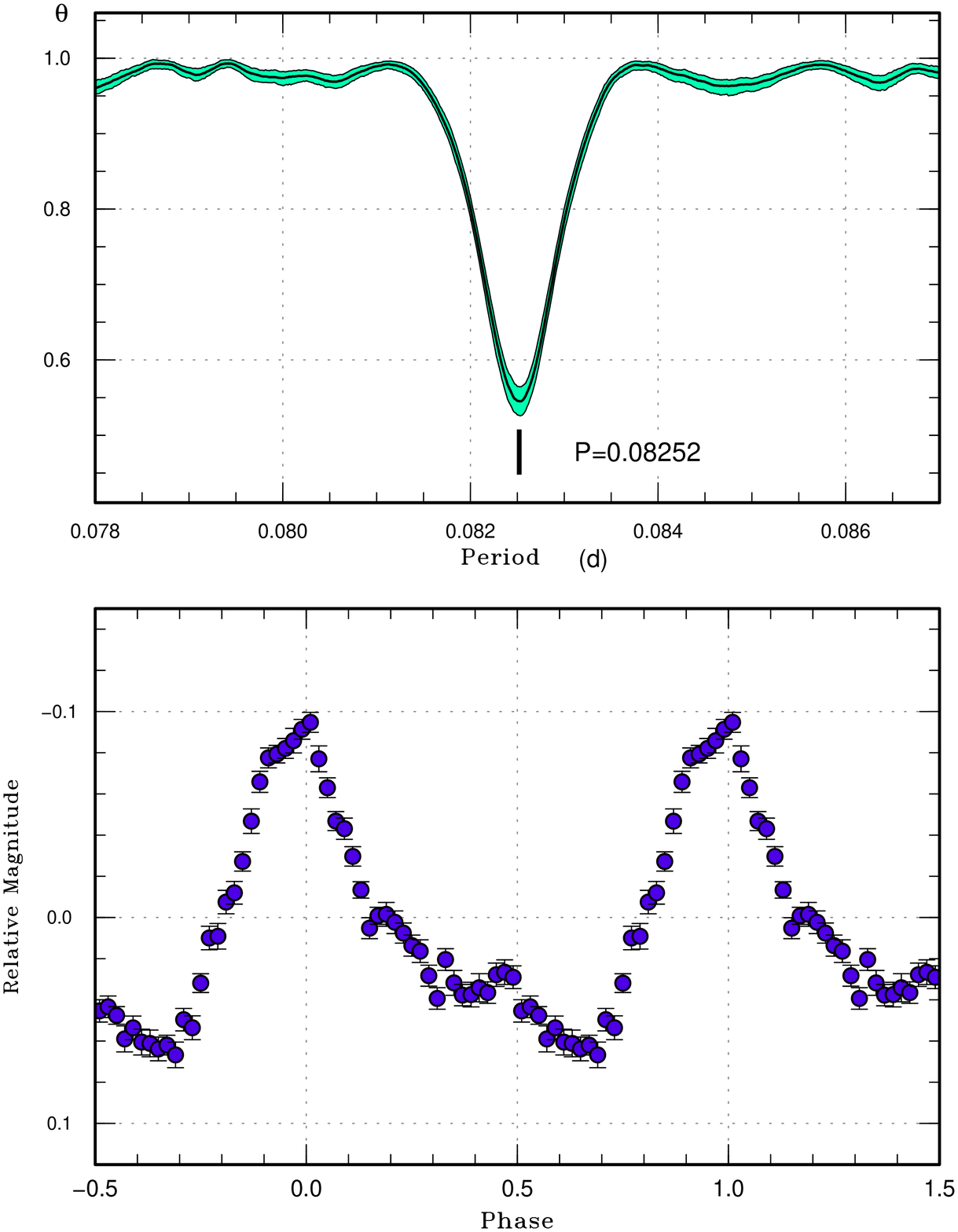}
  \end{center}
  \caption{Superhumps in ASASSN-17bv (2017).
     (Upper): PDM analysis.
     (Lower): Phase-averaged profile.}
  \label{fig:asassn17bvshpdm}
\end{figure}

   This object was detected as a transient
at $V$=15.0 on 2017 January 31 by the ASAS-SN team.
The outburst was announced after confirmation
at $V$=14.9 on February 1.
Subsequent observations starting on February 3
detected superhumps (vsnet-alert 20634;
figure \ref{fig:asassn17bvshpdm}).
The times of superhump maxima are listed in
table \ref{tab:asassn17bvoc2017}.
The $O-C$ values suggest that the maxima for
$E \le$1 were stage A superhumps, although
the amplitudes were already large.
The period of stage C superhumps in table
\ref{tab:perlist} is rather uncertain due to
the large scatter in the final part of
observations.
The object faded to 19 mag on February 13.


\begin{table}
\caption{Superhump maxima of ASASSN-17bv (2017)}\label{tab:asassn17bvoc2017}
\begin{center}
\begin{tabular}{rp{55pt}p{40pt}r@{.}lr}
\hline
\multicolumn{1}{c}{$E$} & \multicolumn{1}{c}{max\commenta} & \multicolumn{1}{c}{error} & \multicolumn{2}{c}{$O-C$\commentb} & \multicolumn{1}{c}{$N$\commentc} \\
\hline
0 & 57787.7449 & 0.0007 & $-$0&0074 & 18 \\
1 & 57787.8269 & 0.0011 & $-$0&0079 & 17 \\
12 & 57788.7425 & 0.0015 & $-$0&0007 & 17 \\
13 & 57788.8231 & 0.0011 & $-$0&0027 & 17 \\
33 & 57790.4778 & 0.0013 & 0&0005 & 100 \\
34 & 57790.5623 & 0.0005 & 0&0024 & 156 \\
35 & 57790.6439 & 0.0015 & 0&0014 & 11 \\
36 & 57790.7266 & 0.0019 & 0&0015 & 12 \\
37 & 57790.8088 & 0.0012 & 0&0012 & 13 \\
40 & 57791.0590 & 0.0003 & 0&0036 & 62 \\
44 & 57791.3876 & 0.0007 & 0&0019 & 135 \\
45 & 57791.4710 & 0.0005 & 0&0028 & 190 \\
46 & 57791.5532 & 0.0005 & 0&0023 & 196 \\
47 & 57791.6370 & 0.0018 & 0&0036 & 9 \\
48 & 57791.7178 & 0.0020 & 0&0018 & 12 \\
49 & 57791.8009 & 0.0017 & 0&0023 & 13 \\
52 & 57792.0484 & 0.0008 & 0&0021 & 42 \\
58 & 57792.5468 & 0.0025 & 0&0050 & 22 \\
60 & 57792.7110 & 0.0023 & 0&0040 & 13 \\
61 & 57792.7923 & 0.0016 & 0&0028 & 14 \\
62 & 57792.8777 & 0.0032 & 0&0056 & 6 \\
64 & 57793.0381 & 0.0003 & 0&0009 & 150 \\
65 & 57793.1201 & 0.0004 & 0&0003 & 136 \\
66 & 57793.2012 & 0.0004 & $-$0&0012 & 148 \\
67 & 57793.2832 & 0.0009 & $-$0&0018 & 167 \\
70 & 57793.5330 & 0.0009 & 0&0003 & 176 \\
71 & 57793.6156 & 0.0012 & 0&0003 & 146 \\
76 & 57794.0267 & 0.0004 & $-$0&0014 & 124 \\
77 & 57794.1092 & 0.0007 & $-$0&0015 & 120 \\
78 & 57794.1930 & 0.0006 & $-$0&0004 & 148 \\
79 & 57794.2704 & 0.0011 & $-$0&0055 & 141 \\
80 & 57794.3575 & 0.0016 & $-$0&0010 & 187 \\
81 & 57794.4385 & 0.0009 & $-$0&0026 & 189 \\
82 & 57794.5217 & 0.0012 & $-$0&0020 & 189 \\
88 & 57795.0138 & 0.0063 & $-$0&0053 & 92 \\
89 & 57795.1017 & 0.0036 & 0&0000 & 94 \\
90 & 57795.1805 & 0.0011 & $-$0&0037 & 148 \\
91 & 57795.2671 & 0.0011 & 0&0003 & 110 \\
96 & 57795.6842 & 0.0029 & 0&0045 & 10 \\
97 & 57795.7538 & 0.0050 & $-$0&0085 & 10 \\
100 & 57796.0153 & 0.0024 & 0&0053 & 32 \\
102 & 57796.1715 & 0.0033 & $-$0&0037 & 44 \\
103 & 57796.2583 & 0.0022 & 0&0005 & 46 \\
\hline
  \multicolumn{6}{l}{\commenta BJD$-$2400000.} \\
  \multicolumn{6}{l}{\commentb Against max $= 2457787.7523 + 0.082578 E$.} \\
  \multicolumn{6}{l}{\commentc Number of points used to determine the maximum.} \\
\end{tabular}
\end{center}
\end{table}

\subsection{ASASSN-17ce}\label{obj:asassn17ce}

\begin{figure}
  \begin{center}
    \FigureFile(85mm,110mm){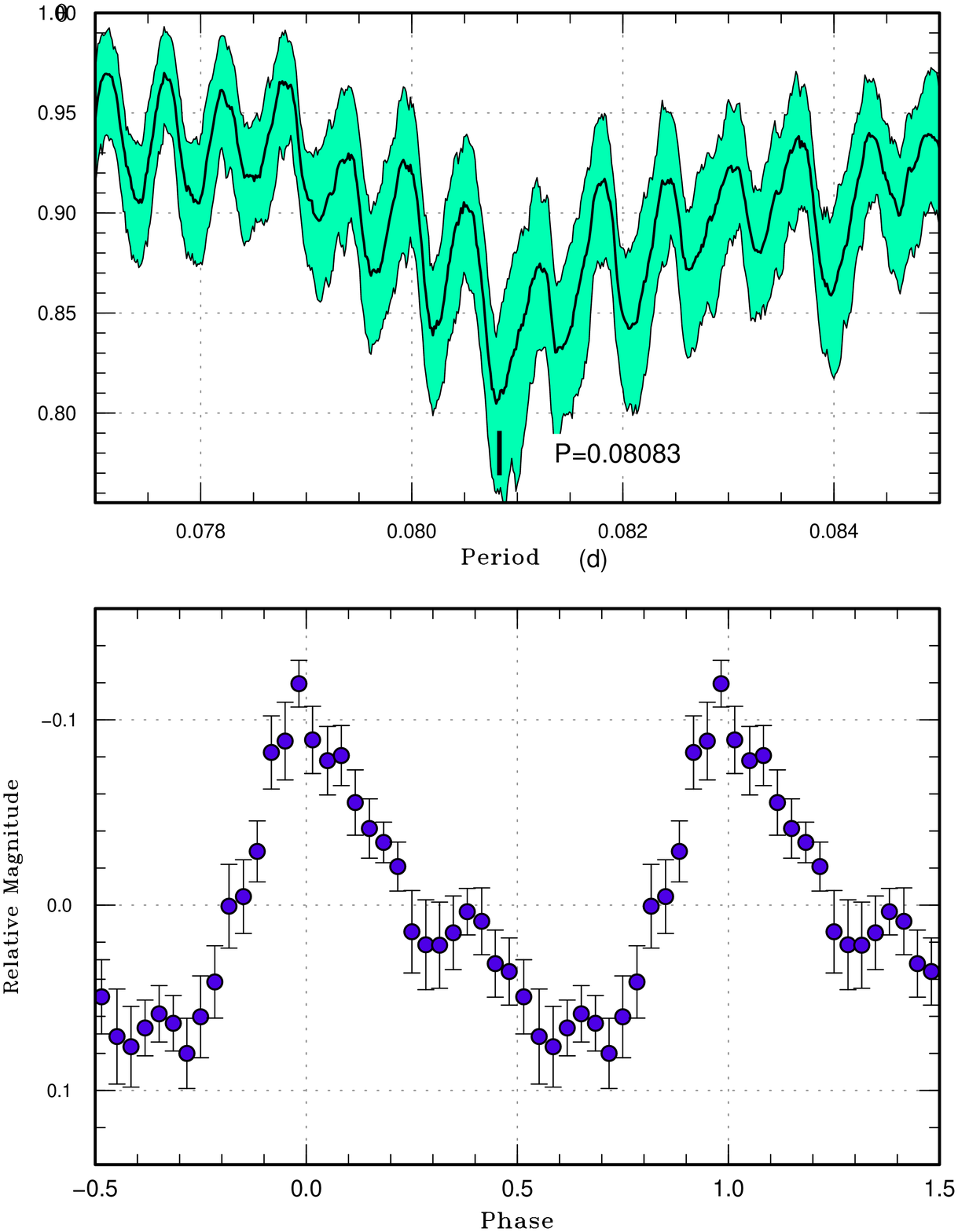}
  \end{center}
  \caption{Superhumps in ASASSN-17ce (2017).
     (Upper): PDM analysis.
     (Lower): Phase-averaged profile.}
  \label{fig:asassn17ceshpdm}
\end{figure}

   This object was detected as a transient
at $V$=14.6 on 2017 February 13 by the ASAS-SN team.
Subsequent observations detected superhumps
(vsnet-alert 20668, 20676; figure \ref{fig:asassn17ceshpdm}).
The times of superhump maxima are listed in
table \ref{tab:asassn17ceoc2017}.
The maxima for $E \le$6 likely correspond to
a short stage B usually seen in long-$P_{\rm orb}$
systems \citep{Pdot}.  The object faded close to
18 mag on February 26.


\begin{table}
\caption{Superhump maxima of ASASSN-17ce (2017)}\label{tab:asassn17ceoc2017}
\begin{center}
\begin{tabular}{rp{55pt}p{40pt}r@{.}lr}
\hline
\multicolumn{1}{c}{$E$} & \multicolumn{1}{c}{max\commenta} & \multicolumn{1}{c}{error} & \multicolumn{2}{c}{$O-C$\commentb} & \multicolumn{1}{c}{$N$\commentc} \\
\hline
0 & 57800.0338 & 0.0003 & $-$0&0028 & 73 \\
1 & 57800.1148 & 0.0002 & $-$0&0026 & 73 \\
4 & 57800.3530 & 0.0030 & $-$0&0069 & 74 \\
5 & 57800.4405 & 0.0004 & $-$0&0003 & 186 \\
6 & 57800.5221 & 0.0007 & 0&0005 & 128 \\
21 & 57801.7390 & 0.0014 & 0&0047 & 17 \\
22 & 57801.8220 & 0.0009 & 0&0069 & 24 \\
26 & 57802.1414 & 0.0008 & 0&0029 & 25 \\
59 & 57804.8064 & 0.0011 & $-$0&0001 & 26 \\
88 & 57807.1428 & 0.0018 & $-$0&0083 & 50 \\
89 & 57807.2411 & 0.0018 & 0&0092 & 45 \\
90 & 57807.3114 & 0.0057 & $-$0&0013 & 13 \\
138 & 57811.1967 & 0.0051 & 0&0033 & 44 \\
139 & 57811.2692 & 0.0065 & $-$0&0051 & 47 \\
\hline
  \multicolumn{6}{l}{\commenta BJD$-$2400000.} \\
  \multicolumn{6}{l}{\commentb Against max $= 2457800.0365 + 0.080847 E$.} \\
  \multicolumn{6}{l}{\commentc Number of points used to determine the maximum.} \\
\end{tabular}
\end{center}
\end{table}

\subsection{ASASSN-17ck}\label{obj:asassn17ck}

\begin{figure}
  \begin{center}
    \FigureFile(85mm,110mm){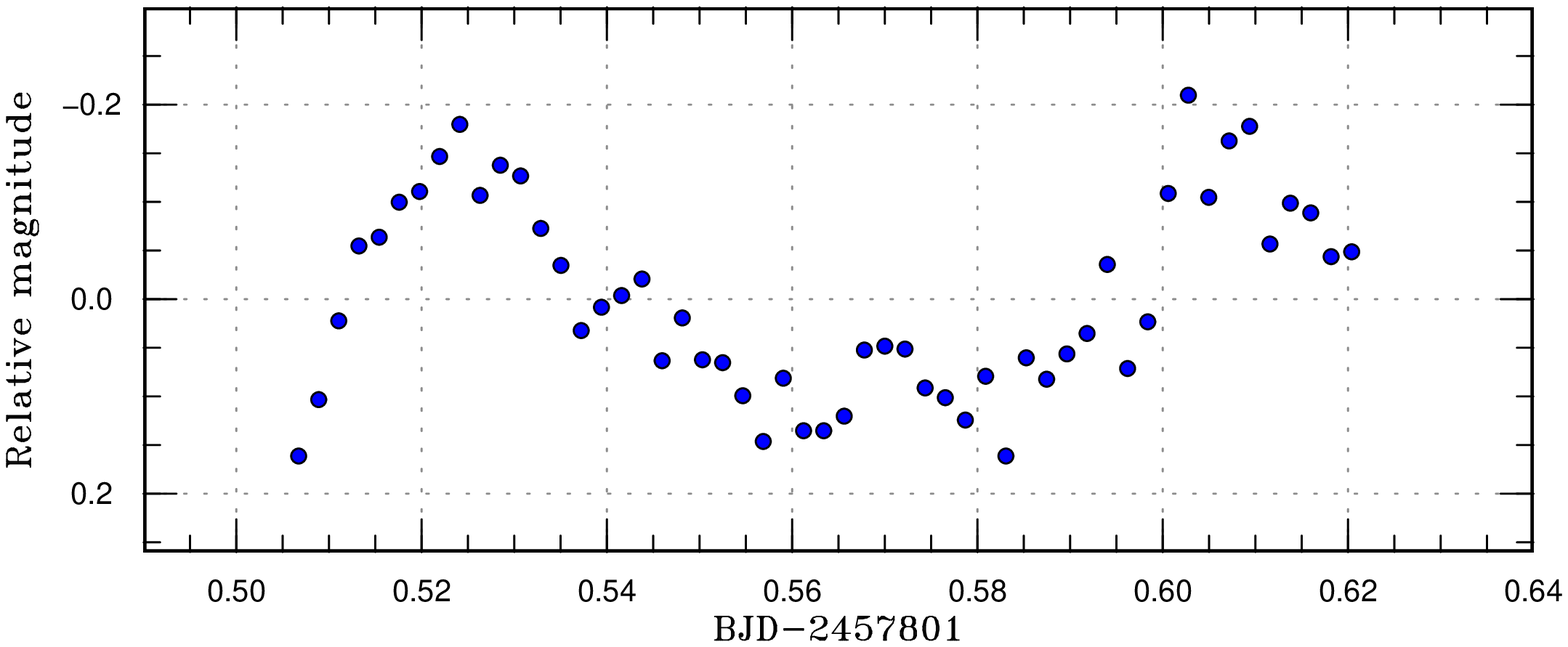}
  \end{center}
  \caption{Superhumps in ASASSN-17ck (2017).
  }
  \label{fig:asassn17ckshlc}
\end{figure}

   This object was detected as a transient
at $V$=16.6 on 2017 February 15 by the ASAS-SN team.
The object was already in outburst at $V$=16.5 on
February 13.  Single-night observations on
February 17 detected superhumps (vsnet-alert 20680,
figure \ref{fig:asassn17ckshlc}).
The times of maxima were BJD 2457801.5269(7) ($N$=25)
and 2457801.6099(17) ($N$=21).
The superhump period by the PDM analysis
was 0.083(1)~d.

\subsection{ASASSN-17cn}\label{obj:asassn17cn}

\begin{figure}
  \begin{center}
    \FigureFile(85mm,110mm){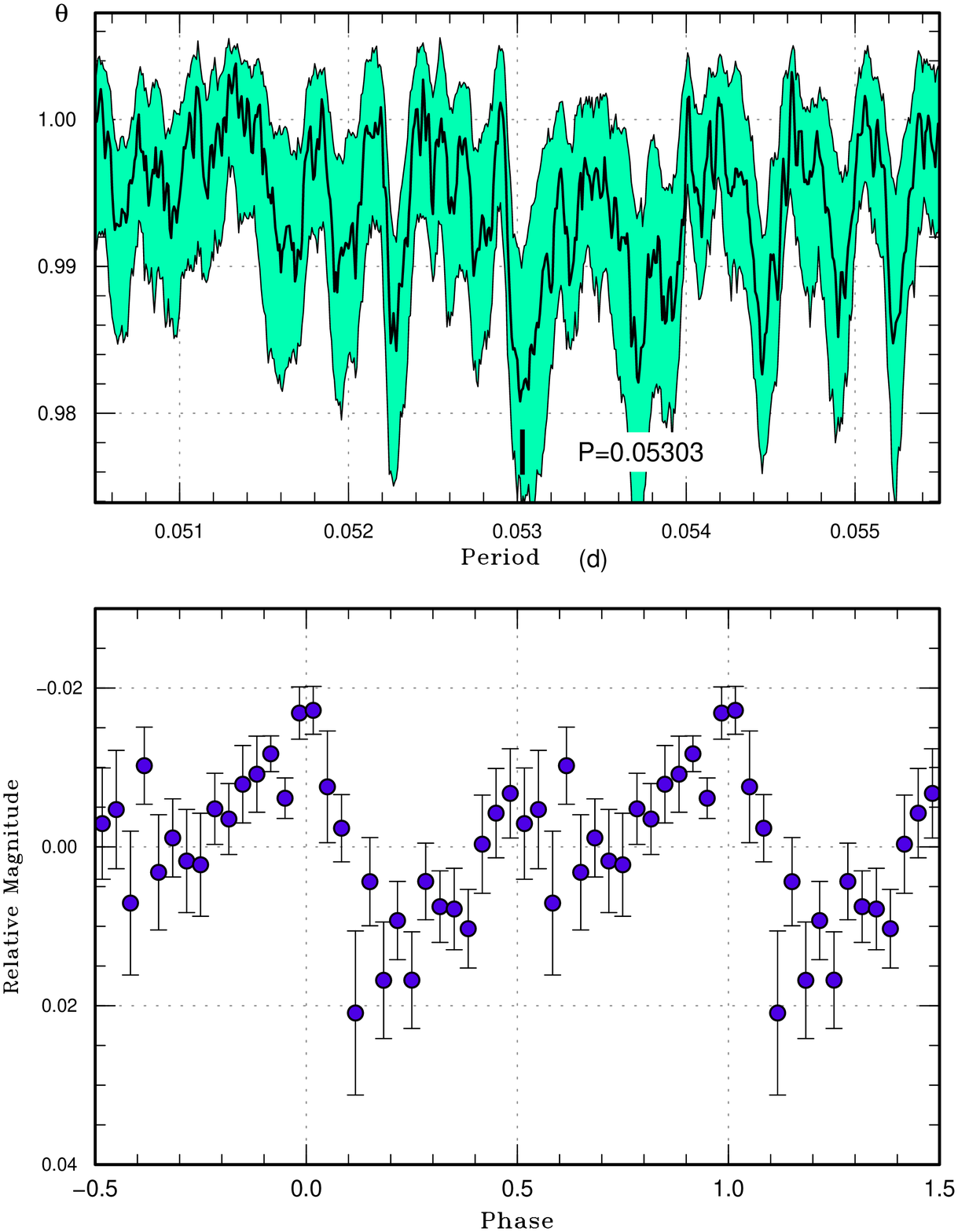}
  \end{center}
  \caption{Early superhumps in ASASSN-17cn (2017).
     (Upper): PDM analysis.
     (Lower): Phase-averaged profile.}
  \label{fig:asassn17cneshpdm}
\end{figure}

\begin{figure}
  \begin{center}
    \FigureFile(85mm,110mm){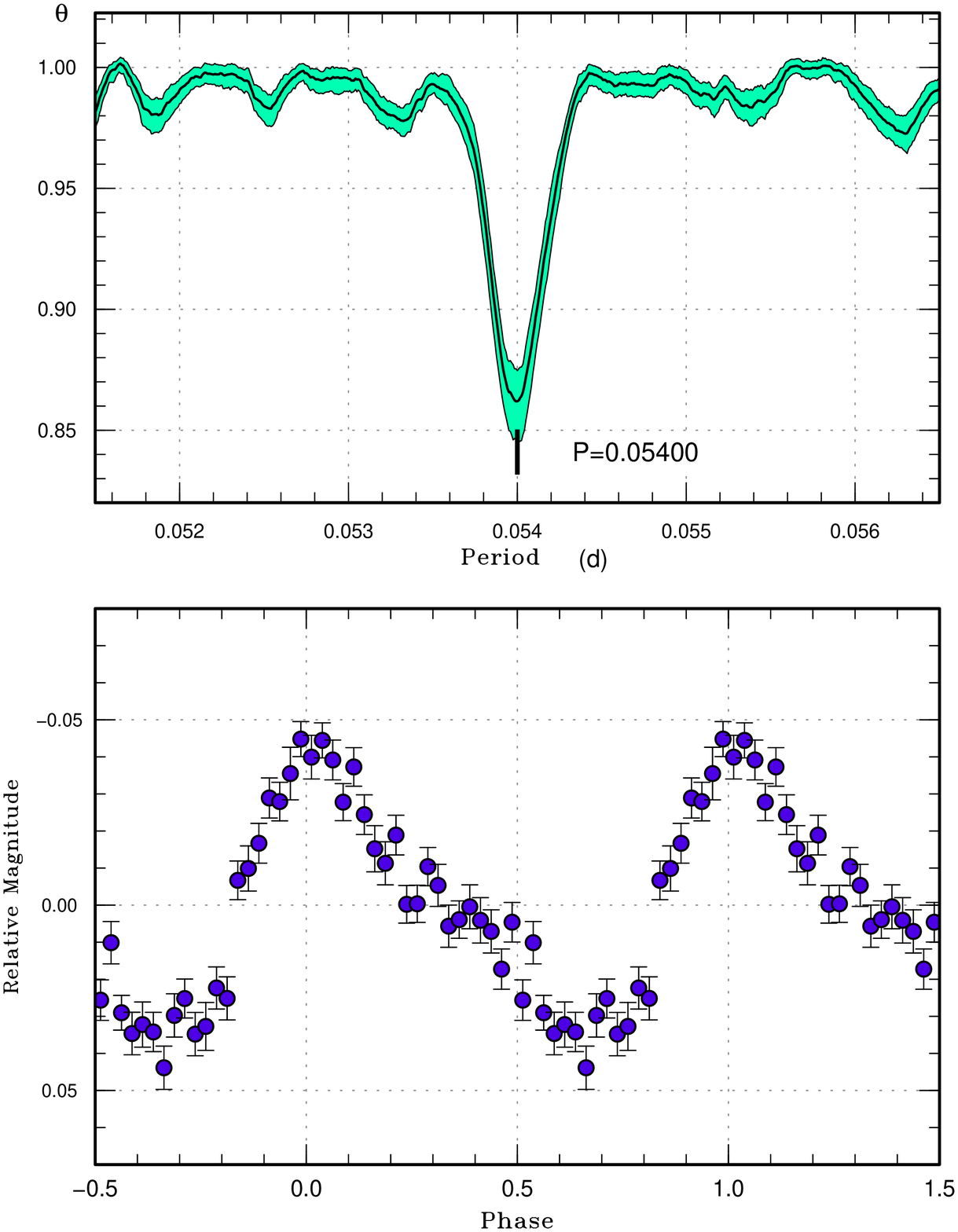}
  \end{center}
  \caption{Ordinary superhumps in ASASSN-17cn (2017).
     (Upper): PDM analysis.
     (Lower): Phase-averaged profile.}
  \label{fig:asassn17cnshpdm}
\end{figure}

   This object was detected as a transient
at $V$=13.7 on 2017 February 13 by the ASAS-SN team.
The outburst announcement was made after an observation
of $V$=13.2 on February 16.
The object showed low-amplitude double-wave
early superhumps (figure \ref{fig:asassn17cneshpdm})
and then ordinary superhumps (vsnet-alert 20750, 20755;
figure \ref{fig:asassn17cnshpdm}).
The behavior was typical for a WZ Sge-type dwarf nova.
There was a 7~d gap in the observations, which
hindered the detection of stage A superhumps.
The times of superhump maxima are listed in
table \ref{tab:asassn17cnoc2017}.
Although the maxima after $E$=137 were stage C
superhumps, we could not determine the period
due to the limited quality of the data.
The period of early superhumps by the PDM method
was 0.05303(2)~d.
The object was also detected by Gaia (Gaia17arq)\footnote{
  $<$http://gsaweb.ast.cam.ac.uk/alerts/alert/Gaia17arq/$>$.
}
at a magnitude of 16.07 on March 14.  This detection
was made during the superoutburst plateau.

   The long waiting time to develop ordinary superhumps
($\geq$9~d counted from the outburst peak, $\geq$12~d
from the outburst detection) and the large outburst
amplitude ($\gtrsim$9 mag) suggest that the object
is a rather extreme WZ Sge-type dwarf nova.


\begin{table}
\caption{Superhump maxima of ASASSN-17cn (2017)}\label{tab:asassn17cnoc2017}
\begin{center}
\begin{tabular}{rp{55pt}p{40pt}r@{.}lr}
\hline
\multicolumn{1}{c}{$E$} & \multicolumn{1}{c}{max\commenta} & \multicolumn{1}{c}{error} & \multicolumn{2}{c}{$O-C$\commentb} & \multicolumn{1}{c}{$N$\commentc} \\
\hline
0 & 57818.2910 & 0.0011 & 0&0081 & 113 \\
1 & 57818.3398 & 0.0008 & 0&0030 & 124 \\
2 & 57818.3934 & 0.0010 & 0&0026 & 123 \\
3 & 57818.4483 & 0.0013 & 0&0033 & 95 \\
19 & 57819.3097 & 0.0010 & 0&0004 & 124 \\
20 & 57819.3626 & 0.0009 & $-$0&0007 & 124 \\
21 & 57819.4217 & 0.0014 & 0&0043 & 123 \\
22 & 57819.4730 & 0.0014 & 0&0016 & 125 \\
23 & 57819.5254 & 0.0008 & $-$0&0000 & 124 \\
37 & 57820.2801 & 0.0009 & $-$0&0016 & 114 \\
38 & 57820.3336 & 0.0009 & $-$0&0022 & 124 \\
39 & 57820.3899 & 0.0011 & 0&0001 & 124 \\
40 & 57820.4428 & 0.0010 & $-$0&0011 & 123 \\
41 & 57820.4971 & 0.0008 & $-$0&0007 & 124 \\
42 & 57820.5506 & 0.0007 & $-$0&0013 & 124 \\
51 & 57821.0406 & 0.0030 & 0&0025 & 39 \\
52 & 57821.0889 & 0.0007 & $-$0&0032 & 52 \\
53 & 57821.1420 & 0.0008 & $-$0&0041 & 49 \\
54 & 57821.1980 & 0.0005 & $-$0&0021 & 49 \\
55 & 57821.2512 & 0.0017 & $-$0&0030 & 36 \\
60 & 57821.5229 & 0.0014 & $-$0&0014 & 26 \\
63 & 57821.6834 & 0.0024 & $-$0&0029 & 21 \\
74 & 57822.2754 & 0.0015 & $-$0&0053 & 120 \\
75 & 57822.3341 & 0.0026 & $-$0&0006 & 124 \\
76 & 57822.3882 & 0.0017 & $-$0&0005 & 124 \\
77 & 57822.4413 & 0.0031 & $-$0&0014 & 124 \\
79 & 57822.5503 & 0.0017 & $-$0&0004 & 125 \\
81 & 57822.6557 & 0.0023 & $-$0&0030 & 12 \\
94 & 57823.3604 & 0.0012 & $-$0&0007 & 124 \\
95 & 57823.4122 & 0.0015 & $-$0&0029 & 124 \\
96 & 57823.4674 & 0.0014 & $-$0&0018 & 124 \\
97 & 57823.5220 & 0.0020 & $-$0&0012 & 125 \\
137 & 57825.6893 & 0.0031 & 0&0052 & 18 \\
152 & 57826.5008 & 0.0018 & 0&0063 & 16 \\
154 & 57826.6050 & 0.0030 & 0&0025 & 17 \\
173 & 57827.6311 & 0.0024 & 0&0021 & 17 \\
\hline
  \multicolumn{6}{l}{\commenta BJD$-$2400000.} \\
  \multicolumn{6}{l}{\commentb Against max $= 2457818.2828 + 0.054024 E$.} \\
  \multicolumn{6}{l}{\commentc Number of points used to determine the maximum.} \\
\end{tabular}
\end{center}
\end{table}

\subsection{ASASSN-17cx}\label{obj:asassn17cx}

   This object was detected as a transient
at $V$=16.4 on 2017 February 21 by the ASAS-SN team.
The object was already in outburst at $V$=16.6
on February 18 and the outburst was announced
after an observation at $V$=16.7 on February 23.
Single-night observations on February 24 detected
superhumps.  The maxima were BJD 2457809.0621(9) ($N$=52),
2457809.1406(6) ($N$=48) and 2457809.2151(14) ($N$=50).
The superhump period by the PDM method was
0.0761(7)~d.


\begin{figure}
  \begin{center}
    \FigureFile(85mm,110mm){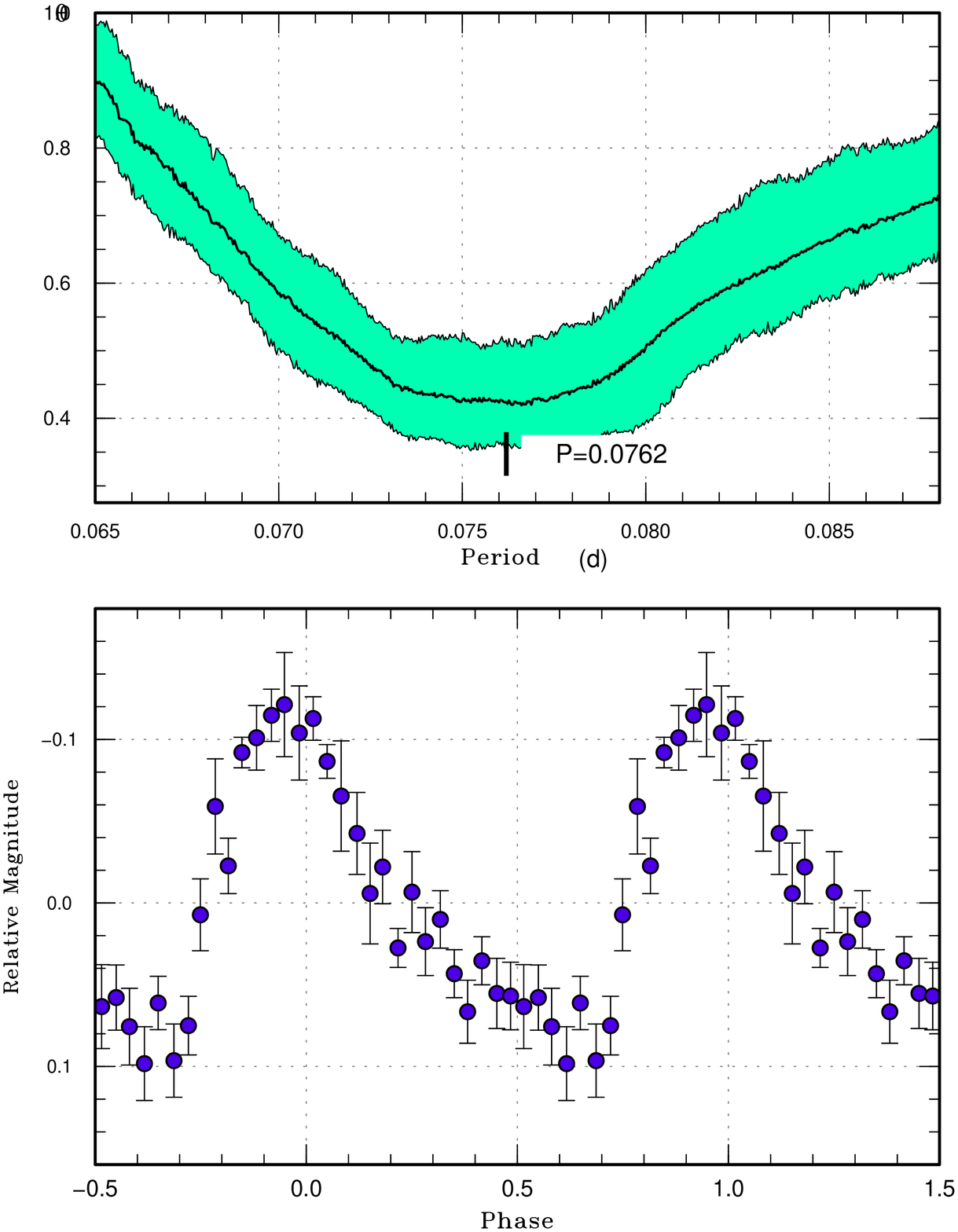}
  \end{center}
  \caption{Superhumps in ASASSN-17cx (2017).
     (Upper): PDM analysis.
     (Lower): Phase-averaged profile.}
  \label{fig:asassn17cxshpdm}
\end{figure}

\subsection{ASASSN-17dg}\label{obj:asassn17dg}

   This object was detected as a transient
at $V$=13.8 on 2017 March 7 by the ASAS-SN team.
The object was already in outburst at $V$=13.8
on March 6.  The last negative observation was
on February 23.  There is an ROSAT X-ray counterpart
1RXS J160232.8$-$603240.  There was also an outburst
with a maximum of $V$=13.03 on 2002 September 25,
which lasted at least for 6~d in the ASAS-3 data.
The actual maximum may have been even brighter
since ASAS-3 did not observe this field for 6~d
before this detection.  The object has
a bright ($J$=13.68) and blue ($J-K$=$+$0.10)
2MASS counterpart, indicating that the object was
in outburst during 2MASS scans.

\begin{figure}
  \begin{center}
    \FigureFile(85mm,110mm){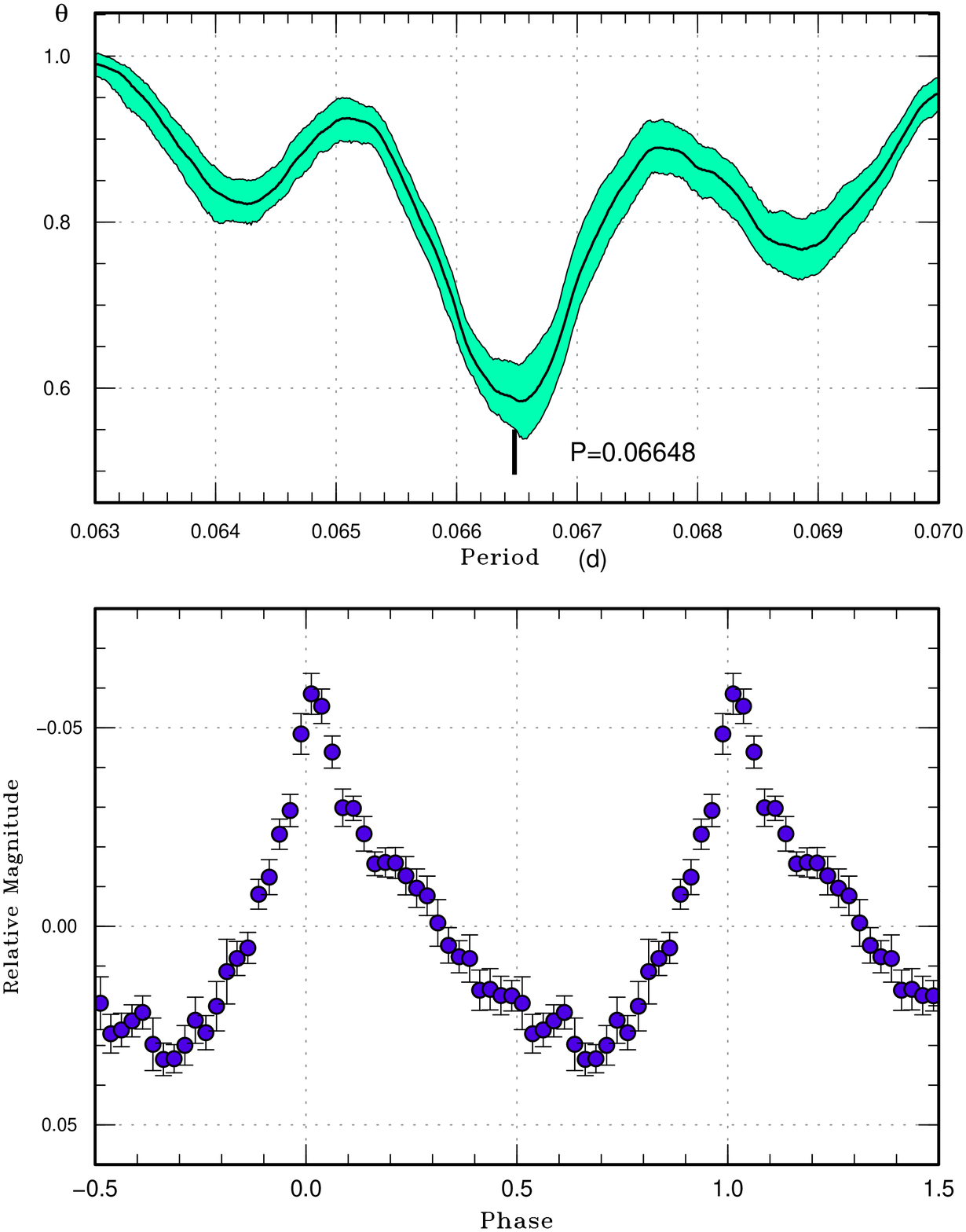}
  \end{center}
  \caption{Superhumps in ASASSN-17dg (2017).
     (Upper): PDM analysis.
     (Lower): Phase-averaged profile.}
  \label{fig:asassn17dgshpdm}
\end{figure}

   Observations started on March 9 and superhumps
were immediately detected (vsnet-alert 20760;
figure \ref{fig:asassn17dgshpdm}).
The object started fading rapidly already on
March 11.  It was most likely the true maximum
was missed by ASAS-SN observations for more than $\sim$5~d.
The times of superhump maxima are listed in
table \ref{tab:asassn17dgoc2017}.
These superhumps were most likely stage C ones
since observations were performed in the final
phase of the superoutburst.  The low amplitudes
of superhumps (vsnet-alert 20760) suggested
that superhumps were already decaying.

   There was one post-superoutburst rebrightening
on March 19 ($V$=14.3, vsnet-alert 20809).
A PDM analysis of the post-superoutburst data
(BJD 2457825.7--2457842.9) yielded a period of
0.06655(5)~d, which is likely a continuation
of stage C superhumps.


\begin{table}
\caption{Superhump maxima of ASASSN-17dg (2017)}\label{tab:asassn17dgoc2017}
\begin{center}
\begin{tabular}{rp{55pt}p{40pt}r@{.}lr}
\hline
\multicolumn{1}{c}{$E$} & \multicolumn{1}{c}{max\commenta} & \multicolumn{1}{c}{error} & \multicolumn{2}{c}{$O-C$\commentb} & \multicolumn{1}{c}{$N$\commentc} \\
\hline
0 & 57821.8464 & 0.0010 & 0&0006 & 26 \\
3 & 57822.0427 & 0.0008 & $-$0&0026 & 60 \\
4 & 57822.1111 & 0.0006 & $-$0&0006 & 60 \\
5 & 57822.1761 & 0.0007 & $-$0&0021 & 60 \\
6 & 57822.2435 & 0.0007 & $-$0&0012 & 60 \\
8 & 57822.3801 & 0.0007 & 0&0025 & 154 \\
9 & 57822.4440 & 0.0006 & $-$0&0001 & 153 \\
10 & 57822.5110 & 0.0005 & 0&0004 & 154 \\
11 & 57822.5782 & 0.0005 & 0&0012 & 153 \\
15 & 57822.8451 & 0.0014 & 0&0021 & 27 \\
21 & 57823.2440 & 0.0010 & 0&0021 & 31 \\
33 & 57824.0393 & 0.0010 & $-$0&0003 & 60 \\
34 & 57824.1098 & 0.0019 & 0&0037 & 60 \\
35 & 57824.1679 & 0.0014 & $-$0&0047 & 60 \\
36 & 57824.2382 & 0.0011 & $-$0&0009 & 60 \\
\hline
  \multicolumn{6}{l}{\commenta BJD$-$2400000.} \\
  \multicolumn{6}{l}{\commentb Against max $= 2457821.8458 + 0.066482 E$.} \\
  \multicolumn{6}{l}{\commentc Number of points used to determine the maximum.} \\
\end{tabular}
\end{center}
\end{table}

\subsection{ASASSN-17dq}\label{obj:asassn17dq}

\begin{figure}
  \begin{center}
    \FigureFile(85mm,110mm){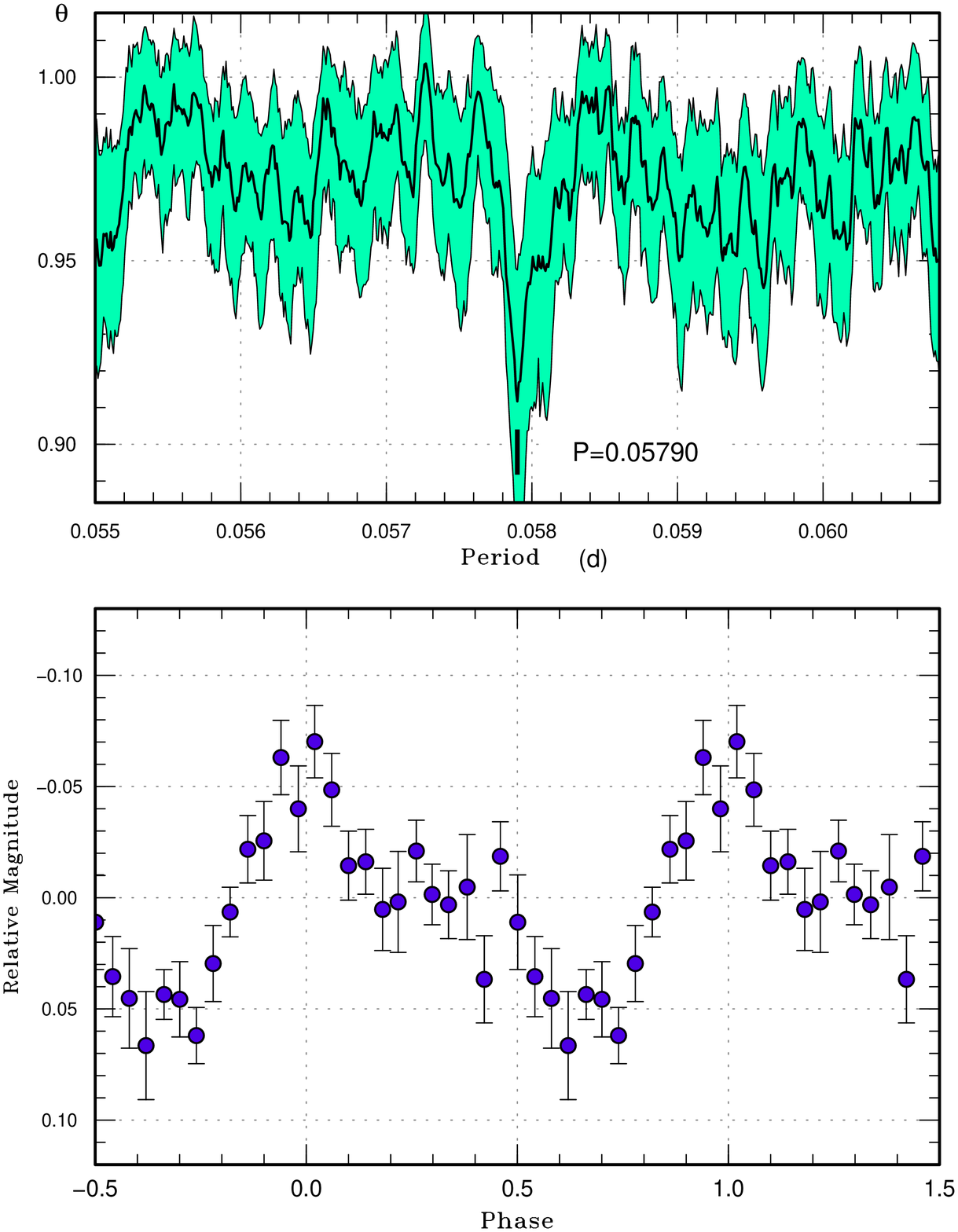}
  \end{center}
  \caption{Superhumps in ASASSN-17dq (2017).
     (Upper): PDM analysis.
     (Lower): Phase-averaged profile.}
  \label{fig:asassn17dqshpdm}
\end{figure}

   This object was detected as a transient
at $V$=15.4 on 2017 March 11 by the ASAS-SN team.
The outburst was announced after the observation
at $V$=15.2 on March 14.  Observations starting
on March 15 recorded superhumps (vsnet-alert 20791,
20810; figure \ref{fig:asassn17dqshpdm}).
The times of superhump maxima are listed in
table \ref{tab:asassn17dqoc2017}.
The object started fading rapidly on March 24.


\begin{table}
\caption{Superhump maxima of ASASSN-17dq (2017)}\label{tab:asassn17dqoc2017}
\begin{center}
\begin{tabular}{rp{55pt}p{40pt}r@{.}lr}
\hline
\multicolumn{1}{c}{$E$} & \multicolumn{1}{c}{max\commenta} & \multicolumn{1}{c}{error} & \multicolumn{2}{c}{$O-C$\commentb} & \multicolumn{1}{c}{$N$\commentc} \\
\hline
0 & 57828.3203 & 0.0006 & 0&0010 & 113 \\
1 & 57828.3769 & 0.0010 & $-$0&0004 & 72 \\
2 & 57828.4339 & 0.0011 & $-$0&0014 & 103 \\
38 & 57830.5179 & 0.0009 & $-$0&0050 & 38 \\
40 & 57830.6359 & 0.0015 & $-$0&0028 & 21 \\
55 & 57831.5074 & 0.0017 & $-$0&0011 & 39 \\
57 & 57831.6253 & 0.0015 & 0&0008 & 21 \\
74 & 57832.6074 & 0.0030 & $-$0&0029 & 21 \\
90 & 57833.5426 & 0.0016 & 0&0045 & 14 \\
91 & 57833.6081 & 0.0019 & 0&0120 & 15 \\
92 & 57833.6552 & 0.0027 & 0&0012 & 15 \\
93 & 57833.7177 & 0.0033 & 0&0056 & 7 \\
108 & 57834.5796 & 0.0022 & $-$0&0022 & 17 \\
109 & 57834.6387 & 0.0023 & $-$0&0011 & 18 \\
110 & 57834.7000 & 0.0017 & 0&0022 & 16 \\
142 & 57836.5428 & 0.0034 & $-$0&0105 & 23 \\
\hline
  \multicolumn{6}{l}{\commenta BJD$-$2400000.} \\
  \multicolumn{6}{l}{\commentb Against max $= 2457828.3194 + 0.057986 E$.} \\
  \multicolumn{6}{l}{\commentc Number of points used to determine the maximum.} \\
\end{tabular}
\end{center}
\end{table}

\subsection{CRTS J000130.5$+$050624}\label{obj:j0001}

   This object (=CSS101127:000130$+$050624, hereafter
CRTS J000130) was detected by the CRTS team
at an unfiltered CCD magnitude of 15.68 on
2010 November 27.

   The 2016 outburst was detected by the ASAS-SN
team at $V$=16.73 on September 3 while the object
was still rising.  The detection announcement was made
when it reached $V$=15.47 on September 9.
The past outbursts in the ASAS-SN data suggested
an SU UMa-type dwarf nova.
Subsequent observations detected superhumps
(vsnet-alert 20152, 20176).
The times of superhump maxima are listed in
table \ref{tab:j0001oc2016}.
The data were insufficient to give a solid value
of $P_{\rm dot}$.  The superhump period of
0.09477(1)~d (by the PDM method) places the object
in the period gap.


\begin{figure}
  \begin{center}
    \FigureFile(85mm,110mm){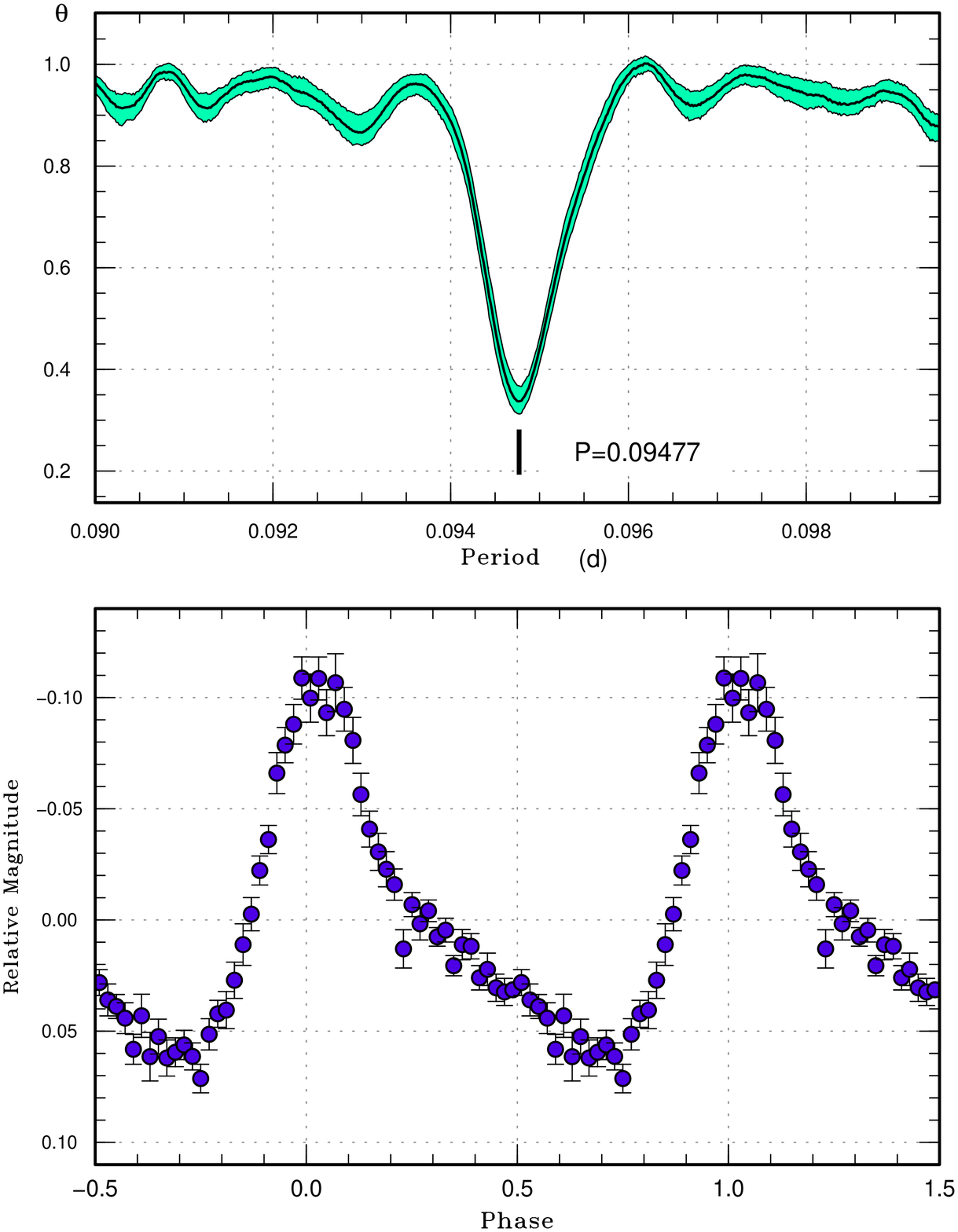}
  \end{center}
  \caption{Superhumps in CRTS J000130 (2016).
     (Upper): PDM analysis.
     (Lower): Phase-averaged profile.}
  \label{fig:j0001shpdm}
\end{figure}


\begin{table}
\caption{Superhump maxima of CRTS J000130 (2016)}\label{tab:j0001oc2016}
\begin{center}
\begin{tabular}{rp{55pt}p{40pt}r@{.}lr}
\hline
\multicolumn{1}{c}{$E$} & \multicolumn{1}{c}{max\commenta} & \multicolumn{1}{c}{error} & \multicolumn{2}{c}{$O-C$\commentb} & \multicolumn{1}{c}{$N$\commentc} \\
\hline
0 & 57639.5745 & 0.0003 & 0&0020 & 96 \\
1 & 57639.6745 & 0.0009 & 0&0073 & 33 \\
10 & 57640.5097 & 0.0031 & $-$0&0103 & 34 \\
11 & 57640.6146 & 0.0004 & $-$0&0001 & 95 \\
21 & 57641.5612 & 0.0005 & $-$0&0010 & 96 \\
22 & 57641.6603 & 0.0010 & 0&0033 & 54 \\
30 & 57642.4117 & 0.0010 & $-$0&0033 & 31 \\
31 & 57642.5077 & 0.0011 & $-$0&0021 & 26 \\
41 & 57643.4551 & 0.0009 & $-$0&0021 & 37 \\
51 & 57644.4083 & 0.0009 & 0&0036 & 177 \\
52 & 57644.4989 & 0.0014 & $-$0&0005 & 102 \\
53 & 57644.6001 & 0.0011 & 0&0060 & 97 \\
63 & 57645.5388 & 0.0020 & $-$0&0029 & 25 \\
\hline
  \multicolumn{6}{l}{\commenta BJD$-$2400000.} \\
  \multicolumn{6}{l}{\commentb Against max $= 2457639.5725 + 0.094749 E$.} \\
  \multicolumn{6}{l}{\commentc Number of points used to determine the maximum.} \\
\end{tabular}
\end{center}
\end{table}

\subsection{CRTS J015321.5$+$340857}\label{obj:j0153}

   This object (=CSS081026:015321$+$340857,
hereafter CRTS J015321) was discovered by
the CRTS team on 2008 October 26.  The SU UMa-type
nature was established during the 2012 superoutburst
(see \cite{Pdot5} for more history).
The 2016 superoutburst was detected by the ASAS-SN
team at $V$=15.95 on November 16.
One superhump maximum at BJD 2457710.3850(10) ($N$=72)
was observed.

\subsection{CRTS J023638.0$+$111157}\label{obj:j0236}

   This object (=CSS091106:023638$+$111157, hereafter
CRTS J023638) was detected by the CRTS team
at an unfiltered CCD magnitude of 16.23 on
2009 November 6.  Seven outbursts were detected
in the CRTS data and there was a brighter ($V$=15.17)
outburst in 2013 (vsnet-alert 16477).

\begin{figure}
  \begin{center}
    \FigureFile(85mm,110mm){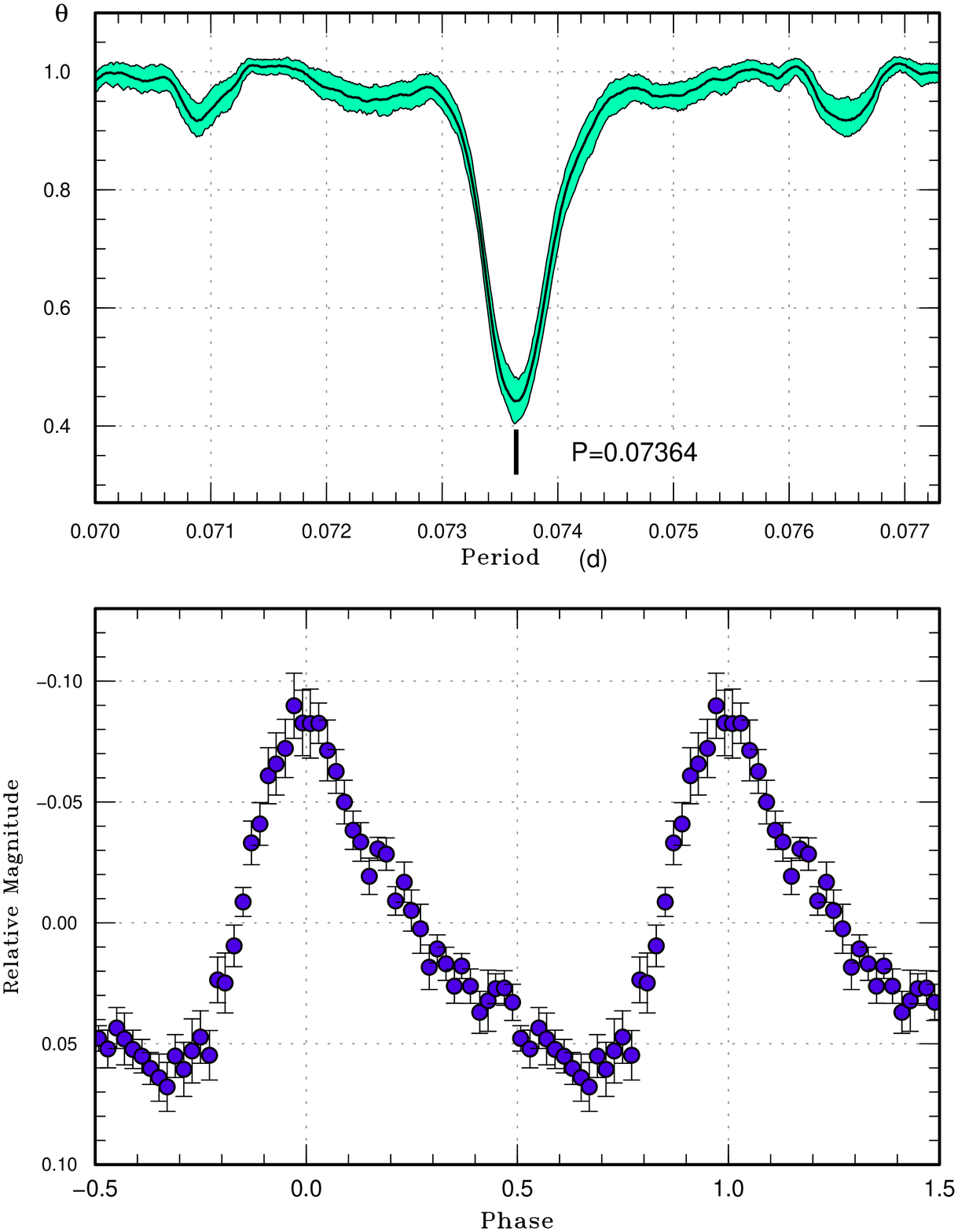}
  \end{center}
  \caption{Superhumps in CRTS J023638 (2016).
     (Upper): PDM analysis.
     The data during the superoutburst plateau
     (before BJD 2457641) were used.
     (Lower): Phase-averaged profile.}
  \label{fig:j0236shpdm}
\end{figure}

   The 2016 outburst was detected by the ASAS-SN team
at $V$=14.90 on August 29.  Subsequent observations
detected superhumps (vsnet-alert 20118, 20128, 20153;
figure \ref{fig:j0236shpdm}).
The times of superhump maxima are listed in
table \ref{tab:j0236oc2016}.  The maxima for
$E \ge$137 were post-superoutburst ones.  There was
most likely a phase jump between $E$=80 and $E$=137
and the humps for $E \ge$137 were likely traditional
late superhumps.  The transition between stages B and C
was rather smooth as in other relatively long
$P_{\rm SH}$ systems.


\begin{table}
\caption{Superhump maxima of CRTS J023638 (2016)}\label{tab:j0236oc2016}
\begin{center}
\begin{tabular}{rp{55pt}p{40pt}r@{.}lr}
\hline
\multicolumn{1}{c}{$E$} & \multicolumn{1}{c}{max\commenta} & \multicolumn{1}{c}{error} & \multicolumn{2}{c}{$O-C$\commentb} & \multicolumn{1}{c}{$N$\commentc} \\
\hline
0 & 57632.5227 & 0.0034 & $-$0&0111 & 41 \\
1 & 57632.5933 & 0.0003 & $-$0&0138 & 73 \\
14 & 57633.5485 & 0.0004 & $-$0&0105 & 82 \\
15 & 57633.6227 & 0.0006 & $-$0&0095 & 52 \\
28 & 57634.5818 & 0.0005 & $-$0&0023 & 68 \\
29 & 57634.6560 & 0.0004 & $-$0&0014 & 49 \\
40 & 57635.4657 & 0.0017 & 0&0029 & 21 \\
41 & 57635.5409 & 0.0007 & 0&0049 & 29 \\
42 & 57635.6194 & 0.0033 & 0&0101 & 11 \\
55 & 57636.5723 & 0.0007 & 0&0111 & 69 \\
56 & 57636.6441 & 0.0005 & 0&0097 & 48 \\
68 & 57637.5268 & 0.0021 & 0&0137 & 27 \\
68 & 57637.5267 & 0.0020 & 0&0136 & 27 \\
80 & 57638.4083 & 0.0016 & 0&0166 & 27 \\
137 & 57642.5488 & 0.0009 & $-$0&0167 & 31 \\
138 & 57642.6214 & 0.0013 & $-$0&0173 & 25 \\
\hline
  \multicolumn{6}{l}{\commenta BJD$-$2400000.} \\
  \multicolumn{6}{l}{\commentb Against max $= 2457632.5338 + 0.073224 E$.} \\
  \multicolumn{6}{l}{\commentc Number of points used to determine the maximum.} \\
\end{tabular}
\end{center}
\end{table}

\subsection{CRTS J033349.8$-$282244}\label{obj:j0333}

\begin{figure}
  \begin{center}
    \FigureFile(85mm,110mm){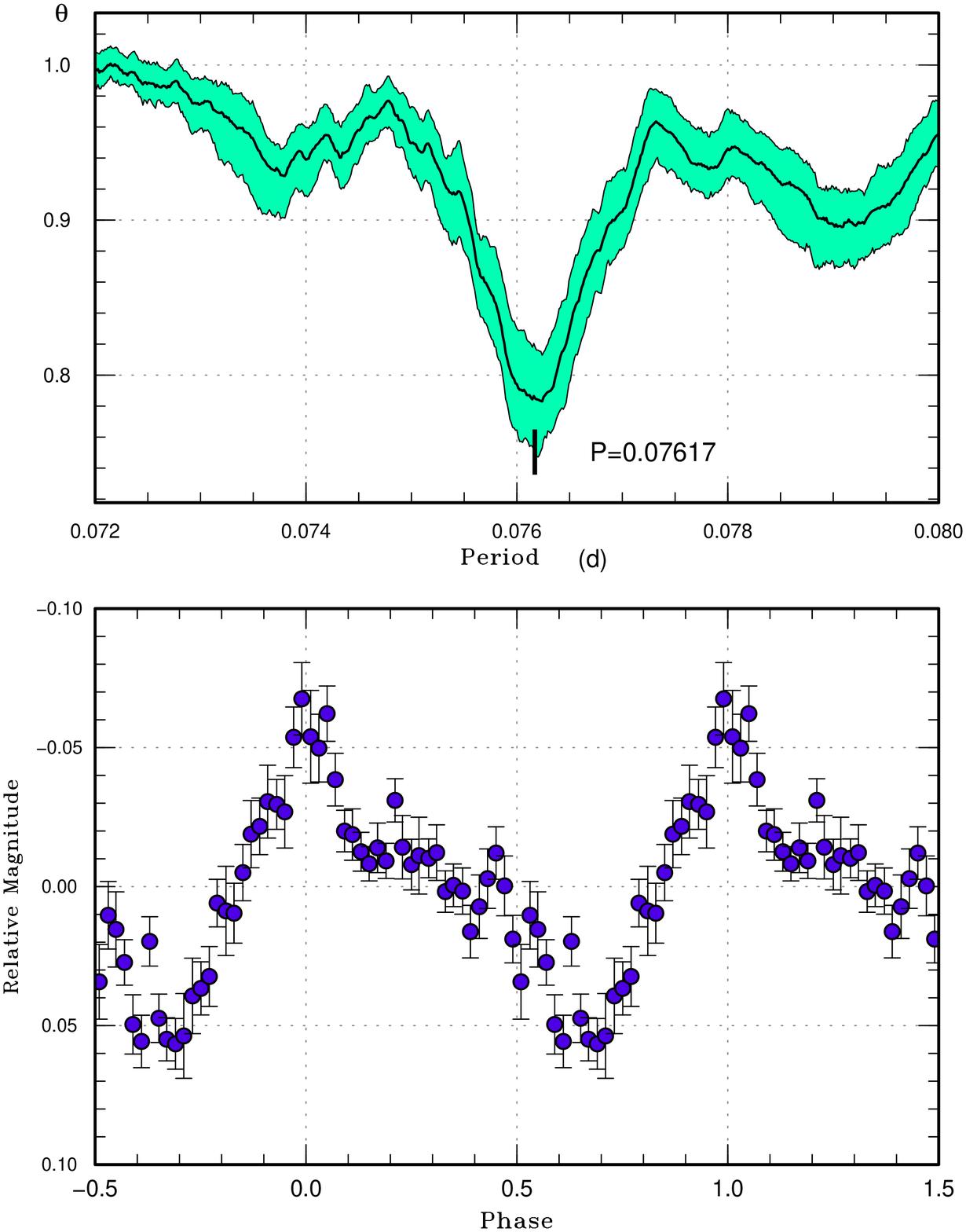}
  \end{center}
  \caption{Superhumps in CRTS J033349 (2016).
     (Upper): PDM analysis.
     (Lower): Phase-averaged profile.}
  \label{fig:j0333shpdm}
\end{figure}

   This object (=SSS110224:033350$-$282244, hereafter
CRTS J033349) was discovered by the CRTS team at
an unfiltered CCD magnitude of 15.06 on 2011 February 24.
The bright outburst in 2016 November was detected
by the ASAS-SN team at $V$=14.46 on November 19.
Subsequent observations detected superhumps
(vsnet-alert 20400, 20403; figure \ref{fig:j0333shpdm}).
The times of superhump maxima are listed in
table \ref{tab:j0333oc2016}.  The observations were
performed during the later course of the superoutburst
and the period probably refers to stage C one.

   We listed well-defined superoutbursts in the ASAS-SN
data since 2014 in table \ref{tab:j0333out}.
These superoutbursts can be well expressed by
a supercycle of 108(1)~d with maximum $|O-C|$ values
of 10~d.  There have been typically two normal outbursts
between superoutbursts.  These features closely
resemble those of V503 Cyg \citep{har95v503cyg}
(vsnet-alert 20386).
V503 Cyg, however, sometimes showed frequent
normal outbursts (e.g. \cite{kat02v503cyg})
and these alternations between phases of different
number of normal outbursts have been considered
to be a result of a disk tilt, which is considered
to suppress normal outbursts (see the subsection
of 1RXS J161659, subsection \ref{obj:j1616}).
Detection of negative superhumps is expected
in CRTS J033349.


\begin{table}
\caption{Superhump maxima of CRTS J033349 (2016)}\label{tab:j0333oc2016}
\begin{center}
\begin{tabular}{rp{55pt}p{40pt}r@{.}lr}
\hline
\multicolumn{1}{c}{$E$} & \multicolumn{1}{c}{max\commenta} & \multicolumn{1}{c}{error} & \multicolumn{2}{c}{$O-C$\commentb} & \multicolumn{1}{c}{$N$\commentc} \\
\hline
0 & 57718.0808 & 0.0038 & 0&0002 & 76 \\
1 & 57718.1532 & 0.0011 & $-$0&0036 & 166 \\
5 & 57718.4653 & 0.0013 & 0&0039 & 154 \\
6 & 57718.5405 & 0.0012 & 0&0030 & 127 \\
8 & 57718.6937 & 0.0033 & 0&0039 & 19 \\
9 & 57718.7615 & 0.0027 & $-$0&0046 & 20 \\
10 & 57718.8430 & 0.0058 & 0&0008 & 12 \\
21 & 57719.6798 & 0.0024 & $-$0&0002 & 24 \\
22 & 57719.7597 & 0.0043 & 0&0036 & 25 \\
23 & 57719.8323 & 0.0024 & 0&0001 & 19 \\
31 & 57720.4414 & 0.0014 & $-$0&0002 & 174 \\
34 & 57720.6668 & 0.0020 & $-$0&0032 & 27 \\
35 & 57720.7436 & 0.0017 & $-$0&0026 & 20 \\
36 & 57720.8164 & 0.0039 & $-$0&0060 & 20 \\
47 & 57721.6592 & 0.0027 & $-$0&0009 & 28 \\
48 & 57721.7310 & 0.0032 & $-$0&0052 & 20 \\
49 & 57721.8172 & 0.0049 & 0&0047 & 20 \\
60 & 57722.6564 & 0.0040 & 0&0062 & 27 \\
\hline
  \multicolumn{6}{l}{\commenta BJD$-$2400000.} \\
  \multicolumn{6}{l}{\commentb Against max $= 2457718.0806 + 0.076159 E$.} \\
  \multicolumn{6}{l}{\commentc Number of points used to determine the maximum.} \\
\end{tabular}
\end{center}
\end{table}

\begin{table}
\caption{List of superoutbursts of CRTS J033349}\label{tab:j0333out}
\begin{center}
\begin{tabular}{ccccc}
\hline
Year & Month & Day & max\commenta & $V$-mag \\
\hline
2014 & 10 & 28 & 56959 & 14.26 \\
2015 &  2 & 16 & 57070 & 14.49 \\
2015 &  9 & 16 & 57282 & 14.50 \\
2015 & 12 & 29 & 57386 & 14.38 \\
2016 &  8 & 17 & 57618 & 14.41 \\
2016 & 11 & 18 & 57711 & 14.46 \\
\hline
  \multicolumn{5}{l}{\commenta JD$-$2400000.} \\
\end{tabular}
\end{center}
\end{table}

\subsection{CRTS J044636.9$+$083033}\label{obj:j0446}

   This object (=CSS130201:044637$+$083033, hereafter
CRTS J044637) was detected by the CRTS team at
an unfiltered CCD magnitude of 17.70 on 2013 February 1.
There was a bright outburst at $V$=15.12 on 2017
January 12 detected by the ASAS-SN team.
Subsequent observations detected two superhumps
on a single night (vsnet-alert 20580).  The maxima were
BJD 2457768.9773(11) ($N$=97) and 2457769.0716(14) ($N$=98).
A PDM analysis yielded a period of 0.093(1)~d.
Although there were single-night observations 4~d later,
the observing condition was not sufficient to
detect superhumps.

\begin{figure}
  \begin{center}
    \FigureFile(85mm,110mm){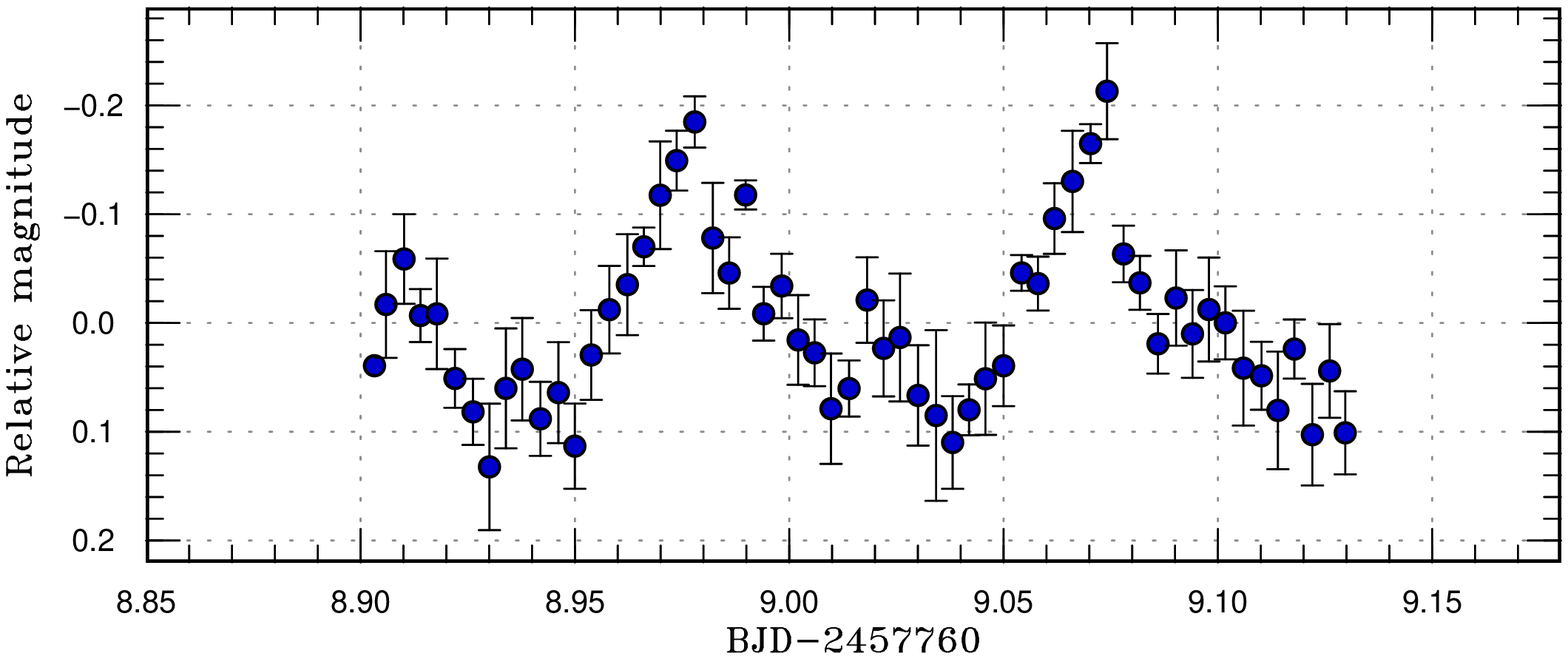}
  \end{center}
  \caption{Superhump in CRTS J044637 (2017).
  The data were binned to 0.004~d.
  }
  \label{fig:j0446shlc}
\end{figure}

\subsection{CRTS J082603.7$+$113821}\label{obj:j0826}

   This object (=CSS110124:082604$+$113821, hereafter
CRTS J082603) was detected by the CRTS team at
an unfiltered CCD magnitude of 15.95 on 2011 January 24
\citep{dra14CRTSCVs}.

   The 2017 outburst was detected by the ASAS-SN team
at $V$=14.9 on January 3.  The observation on
January 5 detected superhumps (vsnet-alert 20542).
The best period with the PDM method was
0.0719(4)~d.  Two superhump maxima were measured:
BJD 2457759.4542(5) ($N$=72) and 2457759.5245(5)
($N$=68).


\begin{figure}
  \begin{center}
    \FigureFile(85mm,110mm){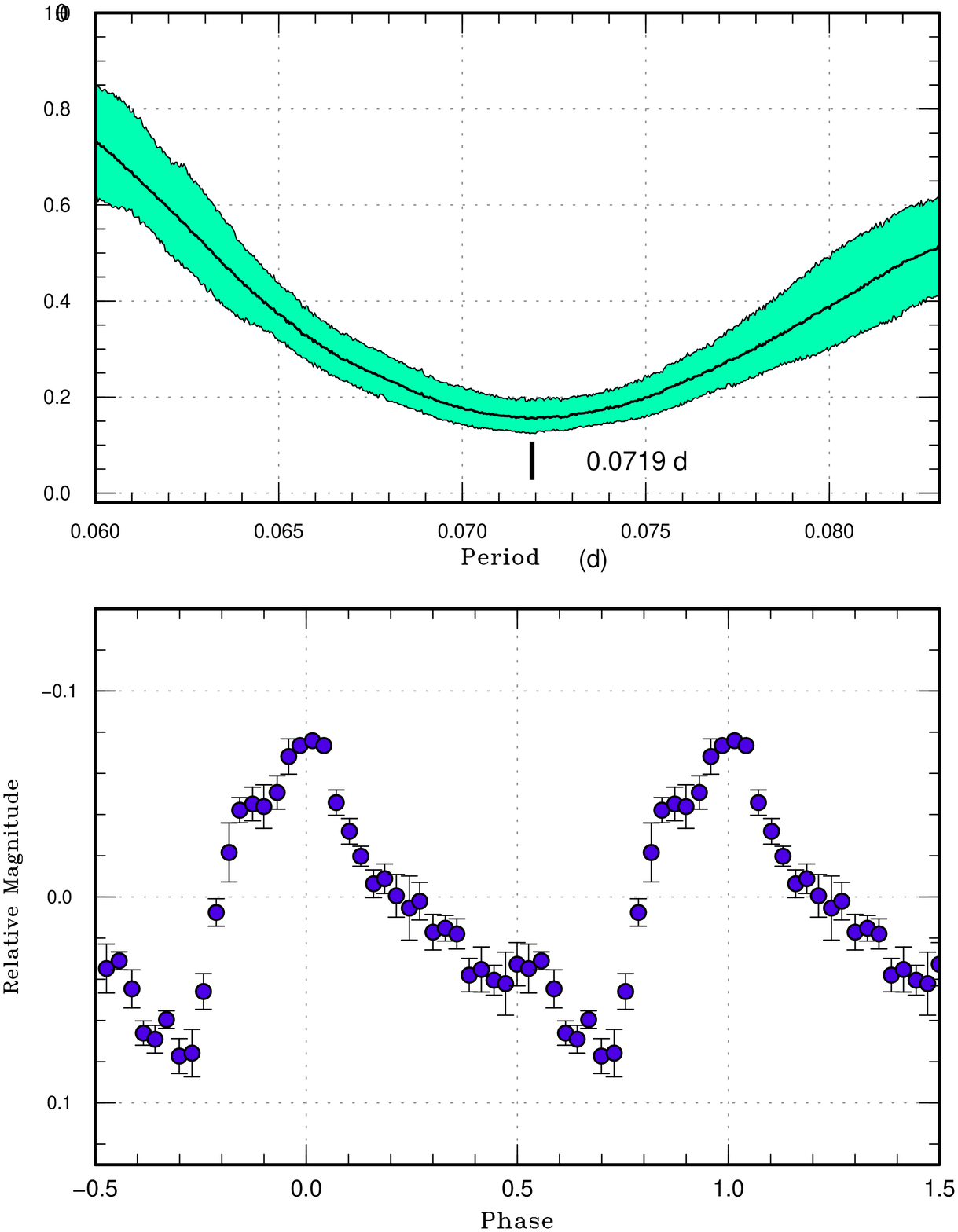}
  \end{center}
  \caption{Superhumps in CRTS J082603 (2017).
     (Upper): PDM analysis.
     (Lower): Phase-averaged profile.}
  \label{fig:j0826shpdm}
\end{figure}

\subsection{CRTS J085113.4$+$344449}\label{obj:j0851}

\begin{figure}
  \begin{center}
    \FigureFile(85mm,110mm){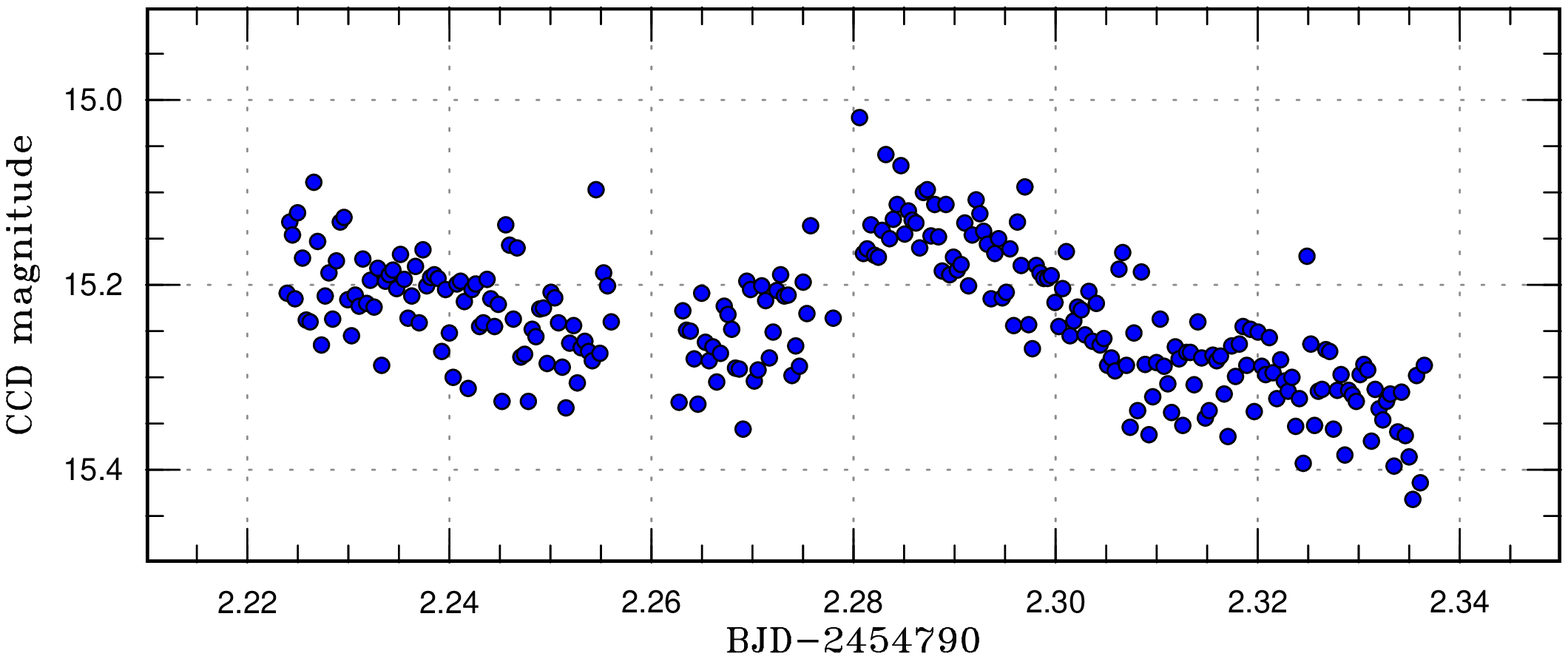}
  \end{center}
  \caption{Superhump in CRTS J085113 (2008).
  }
  \label{fig:j0851shlc2008}
\end{figure}

   This object (=CSS080401:085113$+$344449, hereafter
CRTS J085113) was detected by the CRTS team at
an unfiltered CCD magnitude of 16.4 on 2008 April 1
\citep{dra08atel1479}.  The past data suggested
that there was a bright ($I$=14) outburst in the past
(vsnet-alert 10009).
There was a bright outburst at unfiltered CCD
magnitudes of 14.14--14.73 on 2008 November 20,
detected by the CRTS team (cf .vsnet-alert 10717).
The outburst was suspected to be a superoutburst.
Subsequent observations detected a superhump with
a period of $\sim$0.08~d (vsnet-alert 10723;
figure \ref{fig:j0851shlc2008}).
There have been eight outbursts (up to 2016)
in the CRTS database.
The SDSS colors in quiescence suggested an orbital
period of 0.08--0.12~d \citep{kat12DNSDSS}.

\begin{figure}
  \begin{center}
    \FigureFile(85mm,110mm){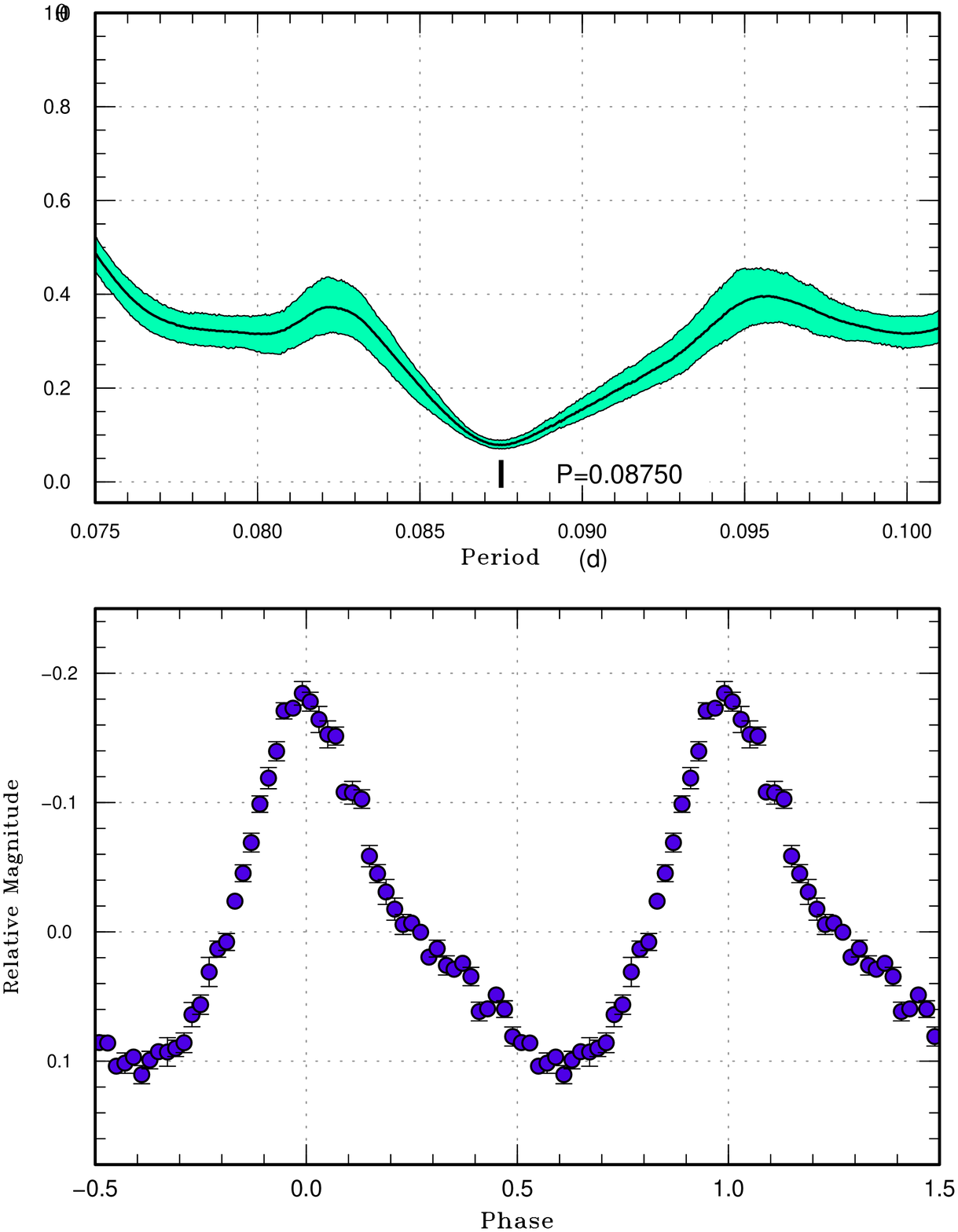}
  \end{center}
  \caption{Superhumps in CRTS J085113 (2016).
     (Upper): PDM analysis.
     (Lower): Phase-averaged profile.}
  \label{fig:j0851shpdm}
\end{figure}

   The 2016 superoutburst was detected by the ASAS-SN
team at $V$=13.85 on November 1.
Subsequent observations detected superhumps
(vsnet-alert 20315; figure \ref{fig:j0851shpdm}).
The times of superhump maxima were BJD 2457697.2246(2) ($N$=181)
and 2457697.3114(3) ($N$=176).
The best superhump period by the PDM method was
0.08750(9)~d.

\subsection{CRTS J085603.8+322109}\label{obj:j0856}

   This object (=CSS100508:085604$+$322109, hereafter
CRTS J085603) was detected by the CRTS team at
an unfiltered CCD magnitude of 16.20 on 2010 May 8.
There is a $g$=19.6-mag SDSS counterpart and
its colors yielded an expected orbital period
of 0.067(1)~d \citep{kat12DNSDSS}.

\begin{figure}
  \begin{center}
    \FigureFile(85mm,110mm){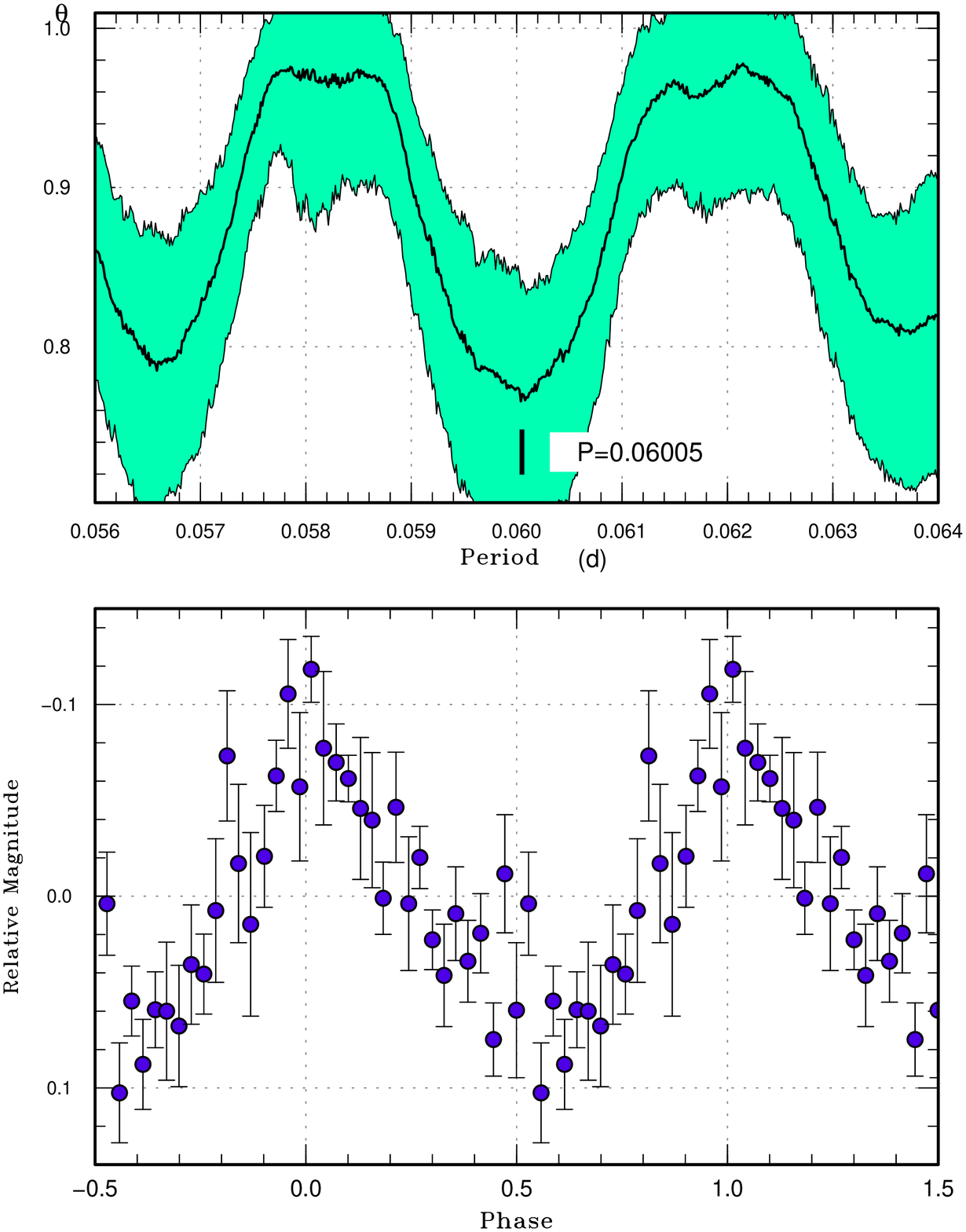}
  \end{center}
  \caption{Superhumps in CRTS J085603 (2016).
     (Upper): PDM analysis.
     (Lower): Phase-averaged profile.}
  \label{fig:j0856shpdm}
\end{figure}

   The 2016 outburst was detected by the ASAS-SN team
at $V$=16.62 on November 26.
Subsequent observations detected superhumps
(vsnet-alert 20428; figure \ref{fig:j0856shpdm}).
The times of superhump maxima are listed in
table \ref{tab:j0856oc2016}.

   The ASAS-SN data indicate that past outbursts
occurred rather regularly.  We listed outburst maxima
(they are likely superoutburst as judged from
the brightness) in table \ref{tab:j0856out}.
These maxima can be expressed by a supercycle
of 232(10)~d, with the maximum $|O-C|$ of 33~d.
It was likely that the peak of the 2016 superoutburst
was not covered by ASAS-SN observations.


\begin{table}
\caption{Superhump maxima of CRTS J085603 (2016)}\label{tab:j0856oc2016}
\begin{center}
\begin{tabular}{rp{55pt}p{40pt}r@{.}lr}
\hline
\multicolumn{1}{c}{$E$} & \multicolumn{1}{c}{max\commenta} & \multicolumn{1}{c}{error} & \multicolumn{2}{c}{$O-C$\commentb} & \multicolumn{1}{c}{$N$\commentc} \\
\hline
0 & 57721.4212 & 0.0034 & $-$0&0039 & 29 \\
1 & 57721.4868 & 0.0005 & 0&0017 & 61 \\
2 & 57721.5477 & 0.0004 & 0&0025 & 61 \\
17 & 57722.4482 & 0.0041 & 0&0024 & 56 \\
18 & 57722.5033 & 0.0017 & $-$0&0026 & 57 \\
\hline
  \multicolumn{6}{l}{\commenta BJD$-$2400000.} \\
  \multicolumn{6}{l}{\commentb Against max $= 2457721.4251 + 0.060043 E$.} \\
  \multicolumn{6}{l}{\commentc Number of points used to determine the maximum.} \\
\end{tabular}
\end{center}
\end{table}

\begin{table}
\caption{List of likely superoutbursts of CRTS J085603}\label{tab:j0856out}
\begin{center}
\begin{tabular}{ccccc}
\hline
Year & Month & Day & max\commenta & $V$-mag \\
\hline
2014 &  5 & 10 & 56788 & 15.67 \\
2015 &  2 & 17 & 57071 & 15.67 \\
2015 & 10 & 22 & 57317 & 15.61 \\
2016 &  5 & 17 & 57526 & 15.75 \\
2016 & 11 & 26 & 57718 & 16.62 \\
\hline
  \multicolumn{5}{l}{\commenta JD$-$2400000.} \\
\end{tabular}
\end{center}
\end{table}

\subsection{CRTS J164950.4$+$035835}\label{obj:j1649}

   This object (=CSS100707:164950$+$035835, hereafter
CRTS J164950) was detected by the CRTS team at
an unfiltered CCD magnitude of 14.1 on 2010 July 7.
There were seven outbursts recorded in the CRTS data.

   The 2015 superoutburst was detected by the ASAS-SN
team at $V$=13.80 on April 13.  Subsequent observations
detected superhumps (vsnet-alert 18545).
The times of superhump maxima are listed in
table \ref{tab:j1649oc2015}.

   The 2016 superoutburst was detected by the ASAS-SN
team at $V$=13.41 on August 31.  Subsequent observations
detected superhumps (vsnet-alert 20132).
The times of superhump maxima are listed in
table \ref{tab:j1649oc2016}.

\begin{figure}
  \begin{center}
    \FigureFile(88mm,70mm){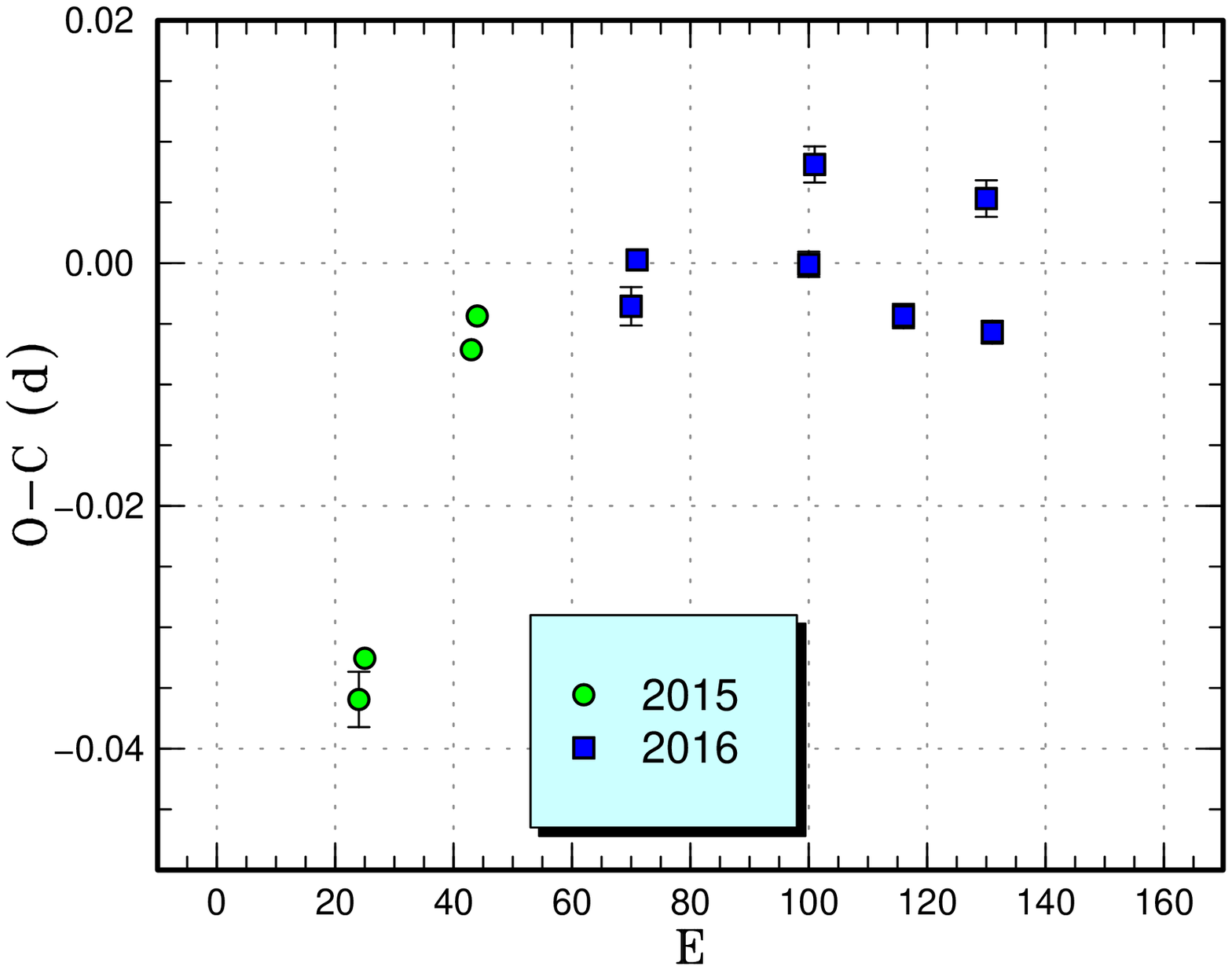}
  \end{center}
  \caption{Comparison of $O-C$ diagrams of CRTS J164950
  between different superoutbursts.
  A period of 0.06490~d was used to draw this figure.
  Approximate cycle counts ($E$) after the start of the superoutburst
  were used.
  }
  \label{fig:j1649comp}
\end{figure}

\begin{figure}
  \begin{center}
    \FigureFile(85mm,110mm){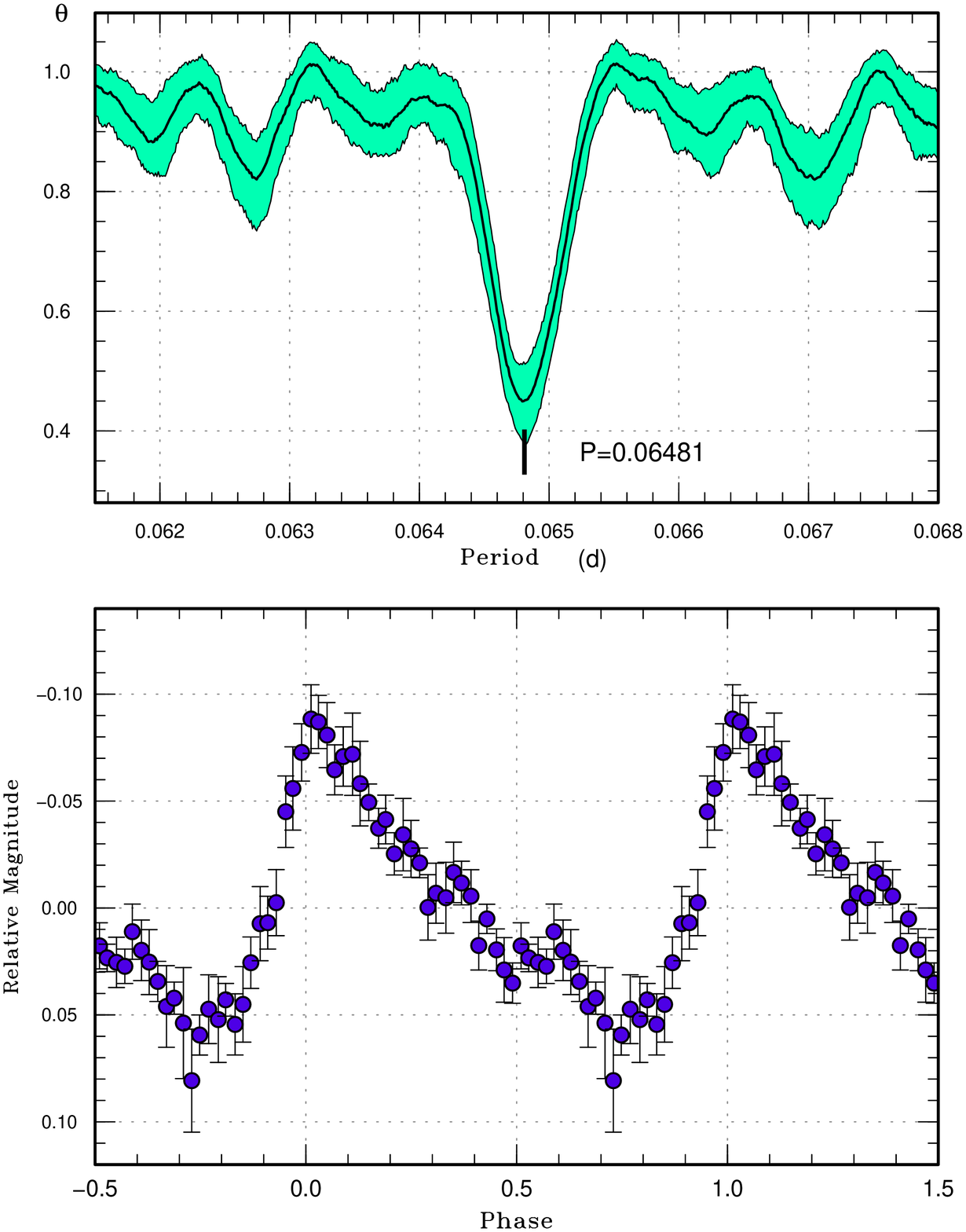}
  \end{center}
  \caption{Superhumps in CRTS J164950 (2016).
     (Upper): PDM analysis.
     (Lower): Phase-averaged profile.}
  \label{fig:j1649shpdm}
\end{figure}

   The period for the 2015 observations was much longer
than in 2016.  The 2015 observations were carried out
soon after the outburst detection and they may have
recorded stage A superhumps.
This interpretation is illustrated in figure \ref{fig:j1649comp}.
A mean superhump profile is given for the better
observed 2016 superoutburst (figure \ref{fig:j1649shpdm}).

\begin{table}
\caption{Superhump maxima of CRTS J164950 (2015)}\label{tab:j1649oc2015}
\begin{center}
\begin{tabular}{rp{55pt}p{40pt}r@{.}lr}
\hline
\multicolumn{1}{c}{$E$} & \multicolumn{1}{c}{max\commenta} & \multicolumn{1}{c}{error} & \multicolumn{2}{c}{$O-C$\commentb} & \multicolumn{1}{c}{$N$\commentc} \\
\hline
0 & 57127.5058 & 0.0023 & $-$0&0009 & 24 \\
1 & 57127.5740 & 0.0004 & 0&0010 & 64 \\
19 & 57128.7677 & 0.0002 & $-$0&0007 & 93 \\
20 & 57128.8353 & 0.0002 & 0&0006 & 111 \\
\hline
  \multicolumn{6}{l}{\commenta BJD$-$2400000.} \\
  \multicolumn{6}{l}{\commentb Against max $= 2457127.5067 + 0.066405 E$.} \\
  \multicolumn{6}{l}{\commentc Number of points used to determine the maximum.} \\
\end{tabular}
\end{center}
\end{table}


\begin{table}
\caption{Superhump maxima of CRTS J164950 (2016)}\label{tab:j1649oc2016}
\begin{center}
\begin{tabular}{rp{55pt}p{40pt}r@{.}lr}
\hline
\multicolumn{1}{c}{$E$} & \multicolumn{1}{c}{max\commenta} & \multicolumn{1}{c}{error} & \multicolumn{2}{c}{$O-C$\commentb} & \multicolumn{1}{c}{$N$\commentc} \\
\hline
0 & 57636.3464 & 0.0016 & $-$0&0034 & 15 \\
1 & 57636.4151 & 0.0005 & 0&0004 & 61 \\
30 & 57638.2969 & 0.0010 & $-$0&0001 & 39 \\
31 & 57638.3700 & 0.0015 & 0&0081 & 29 \\
46 & 57639.3310 & 0.0010 & $-$0&0044 & 66 \\
60 & 57640.2493 & 0.0015 & 0&0052 & 38 \\
61 & 57640.3032 & 0.0009 & $-$0&0058 & 66 \\
\hline
  \multicolumn{6}{l}{\commenta BJD$-$2400000.} \\
  \multicolumn{6}{l}{\commentb Against max $= 2457636.3498 + 0.064905 E$.} \\
  \multicolumn{6}{l}{\commentc Number of points used to determine the maximum.} \\
\end{tabular}
\end{center}
\end{table}

\subsection{CSS J062450.1$+$503114}\label{obj:j0624}

\begin{figure}
  \begin{center}
    \FigureFile(85mm,110mm){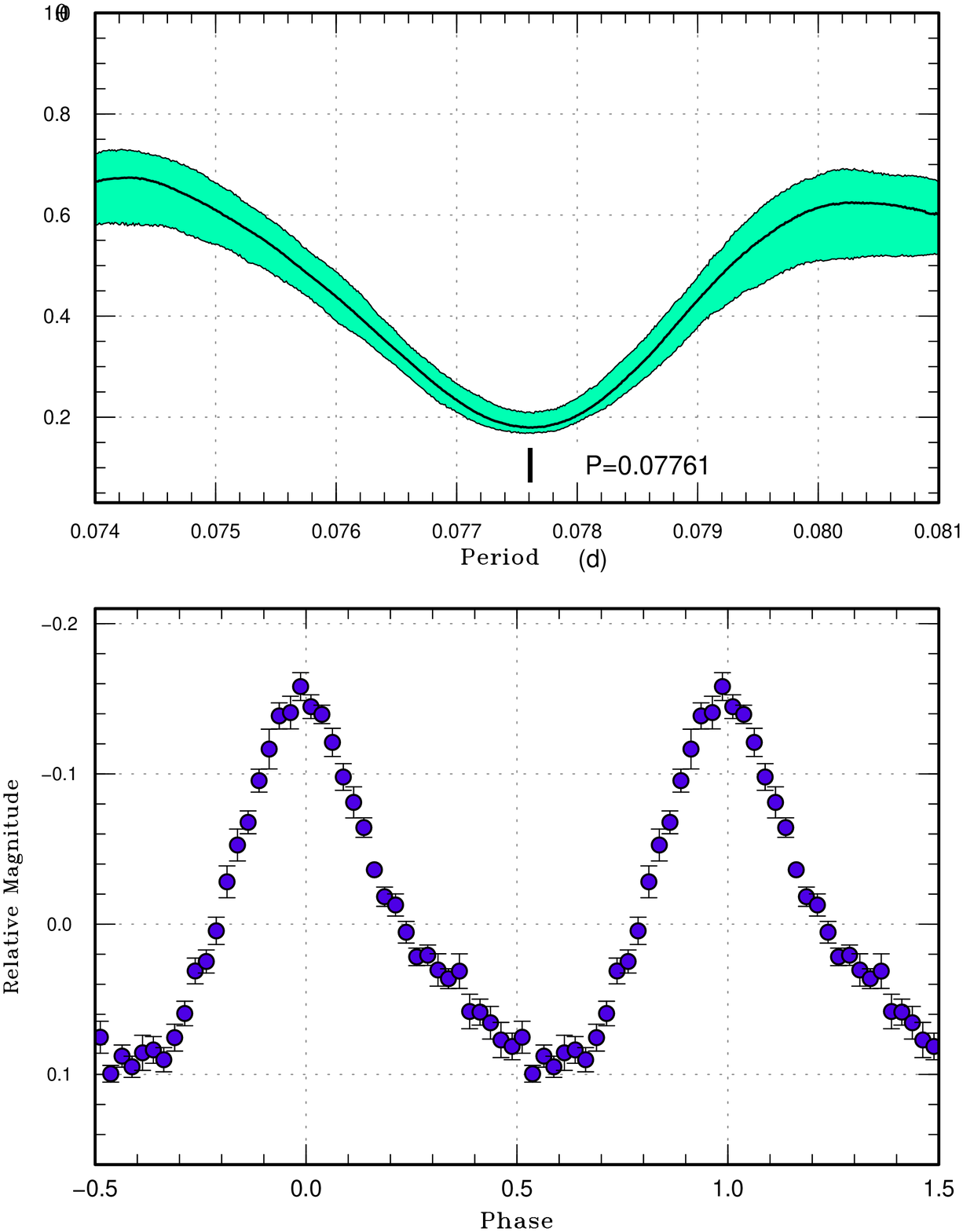}
  \end{center}
  \caption{Superhumps in CSS J062450 (2016).
     (Upper): PDM analysis.
     (Lower): Phase-averaged profile.}
  \label{fig:j0624shpdm}
\end{figure}

   This object (=CSS131223:062450$+$503111, hereafter
CSS J062450) was detected by the CRTS team at
an unfiltered CCD magnitude of 14.76 on 2013 December 23.
The 2017 outburst was detected by the ASAS-SN team
at $V$=14.59 on March 11.  Subsequent observations
detected superhumps (vsnet-alert 20766, 20770;
figure \ref{fig:j0624shpdm}).
The times of superhump maxima are listed in
table \ref{tab:j0624oc2017}.

   According to the ASAS-SN data, this object showed
relatively regular superoutbursts (table \ref{tab:j0624out}).
These superoutburst can be expressed by a supercycle
of 128(2)~d with maximum $|O-C|$ of 20~d.
The interval between the 2016 September and 2017 March
superoutbursts was 166~d, which was rather unusually
long for this object.


\begin{table}
\caption{Superhump maxima of CSS J062450 (2017)}\label{tab:j0624oc2017}
\begin{center}
\begin{tabular}{rp{55pt}p{40pt}r@{.}lr}
\hline
\multicolumn{1}{c}{$E$} & \multicolumn{1}{c}{max\commenta} & \multicolumn{1}{c}{error} & \multicolumn{2}{c}{$O-C$\commentb} & \multicolumn{1}{c}{$N$\commentc} \\
\hline
0 & 57825.3383 & 0.0002 & $-$0&0010 & 109 \\
1 & 57825.4167 & 0.0003 & $-$0&0002 & 109 \\
2 & 57825.4944 & 0.0003 & $-$0&0000 & 171 \\
3 & 57825.5735 & 0.0009 & 0&0015 & 39 \\
14 & 57826.4251 & 0.0006 & $-$0&0003 & 88 \\
\hline
  \multicolumn{6}{l}{\commenta BJD$-$2400000.} \\
  \multicolumn{6}{l}{\commentb Against max $= 2457825.3393 + 0.077577 E$.} \\
  \multicolumn{6}{l}{\commentc Number of points used to determine the maximum.} \\
\end{tabular}
\end{center}
\end{table}

\begin{table}
\caption{List of likely superoutbursts of CSS J062450 in the ASAS-SN data}\label{tab:j0624out}
\begin{center}
\begin{tabular}{ccccc}
\hline
Year & Month & Day & max\commenta & $V$-mag \\
\hline
2013 & 12 & 24 & 56651 & 14.31 \\
2014 &  9 &  7 & 56908 & 14.61 \\
2015 &  1 & 12 & 57035 & 14.70 \\
2015 &  9 & 17 & 57283 & 14.79 \\
2016 &  1 & 29 & 57416 & 14.74 \\
2016 &  9 & 25 & 57657 & 14.94 \\
2017 &  3 & 11 & 57823 & 14.59 \\
\hline
  \multicolumn{5}{l}{\commenta JD$-$2400000.} \\
\end{tabular}
\end{center}
\end{table}

\subsection{DDE 26}\label{obj:dde26}

\begin{figure}
  \begin{center}
    \FigureFile(85mm,70mm){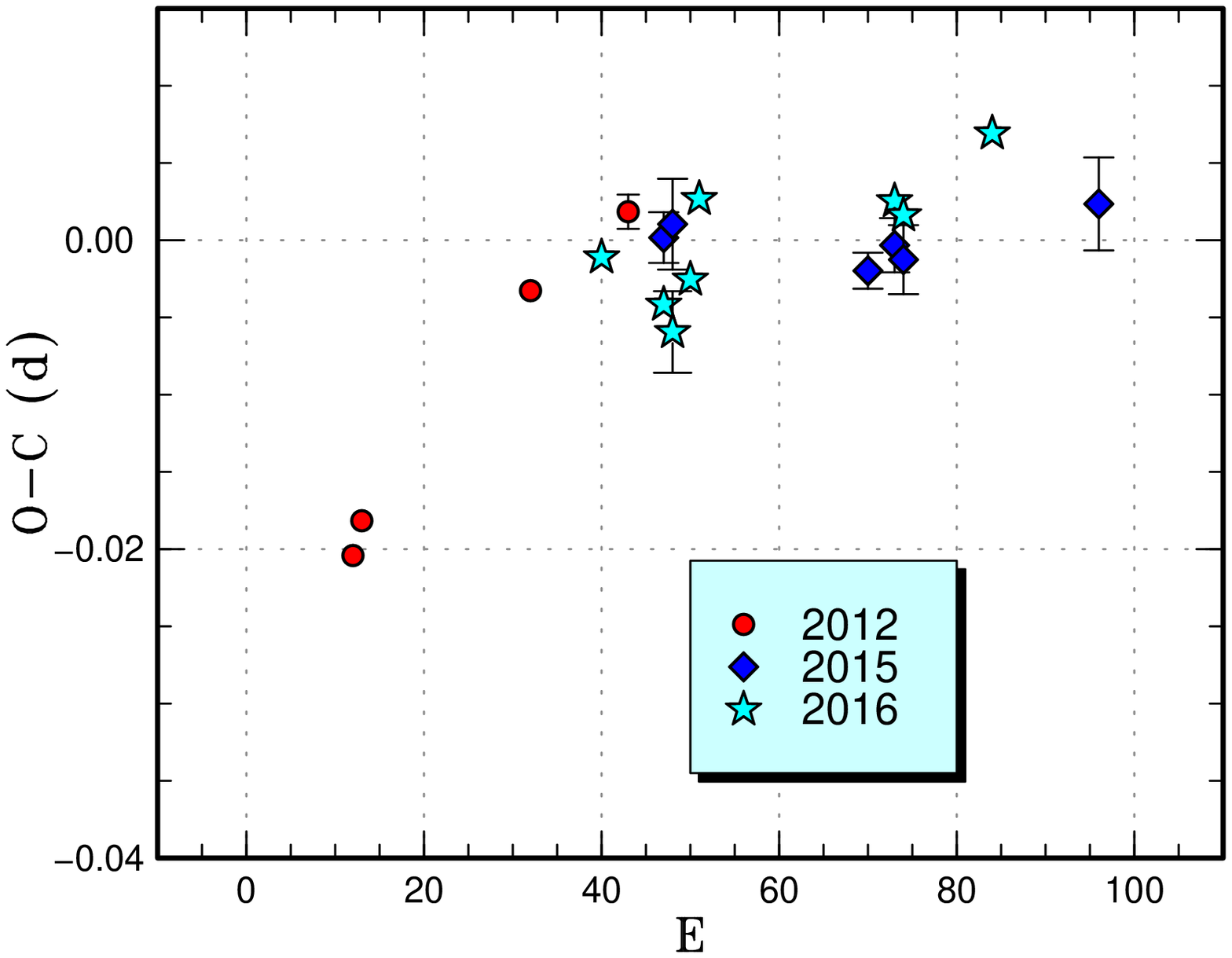}
  \end{center}
  \caption{Comparison of $O-C$ diagrams of DDE 26 between different
  superoutbursts.  A period of 0.08860~d was used to draw this figure.
  Approximate cycle counts ($E$) after the start of the superoutburst
  were used.
  }
  \label{fig:dde26comp2}
\end{figure}

   DDE 26 is a dwarf nova discovered by \citet{den12USNOCVs}.
See \citet{Pdot5} for more information.
The 2016 superoutburst was detected by the ASAS-SN
team at $V$=15.95 on August 1.  Superhumps were
recorded (vsnet-alert 20067).
The times of superhump maxima are listed in
table \ref{tab:dde26oc2016}.  The observation probably
covered stage B (figure \ref{fig:dde26comp2}).

\begin{table}
\caption{Superhump maxima of DDE 26 (2016)}\label{tab:dde26oc2016}
\begin{center}
\begin{tabular}{rp{55pt}p{40pt}r@{.}lr}
\hline
\multicolumn{1}{c}{$E$} & \multicolumn{1}{c}{max\commenta} & \multicolumn{1}{c}{error} & \multicolumn{2}{c}{$O-C$\commentb} & \multicolumn{1}{c}{$N$\commentc} \\
\hline
0 & 57605.5307 & 0.0011 & 0&0026 & 28 \\
7 & 57606.1479 & 0.0011 & $-$0&0018 & 96 \\
8 & 57606.2347 & 0.0026 & $-$0&0038 & 87 \\
10 & 57606.4153 & 0.0011 & $-$0&0008 & 107 \\
11 & 57606.5091 & 0.0006 & 0&0042 & 128 \\
33 & 57608.4582 & 0.0004 & $-$0&0005 & 104 \\
34 & 57608.5459 & 0.0004 & $-$0&0016 & 104 \\
44 & 57609.4372 & 0.0009 & 0&0017 & 186 \\
\hline
  \multicolumn{6}{l}{\commenta BJD$-$2400000.} \\
  \multicolumn{6}{l}{\commentb Against max $= 2457605.5281 + 0.088804 E$.} \\
  \multicolumn{6}{l}{\commentc Number of points used to determine the maximum.} \\
\end{tabular}
\end{center}
\end{table}

\subsection{DDE 48}\label{obj:dde48}

\begin{figure}
  \begin{center}
    \FigureFile(85mm,110mm){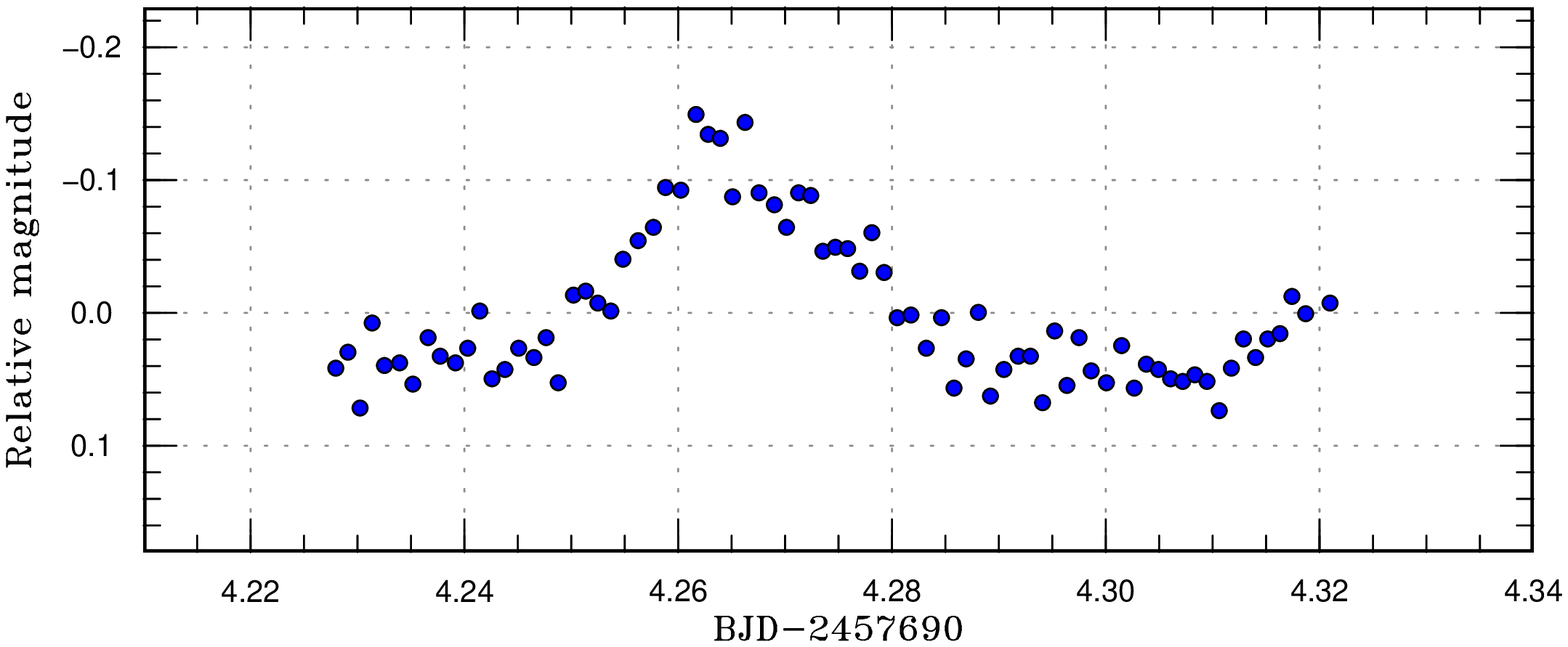}
  \end{center}
  \caption{Superhump in DDE 48 (2016).
  }
  \label{fig:dde48shlc}
\end{figure}

   DDE 48 is a dwarf nova discovered by D. Denisenko
(vsnet-alert 20146) in the vicinity
of the dwarf nova MASTER OT J204627.96$+$242218.0
\citep{shu16j2046atel9470}.
N. Mishevskiy monitored this object in 2016
and detected a bright outburst at $V$=15.5
on November 1 (vsnet-alert 20290).
Subsequent observations detected a superhump
(vsnet-alert 20291).  Although this superhump was
recorded only on a single night, the profile
suggests a genuine superhump (figure \ref{fig:dde48shlc}).
The superhump maximum was at BJD 2457694.2662(6) ($N$=43).
The superhump became undetectable on two nights
during the same superoutburst.

\begin{figure}
  \begin{center}
    \FigureFile(85mm,110mm){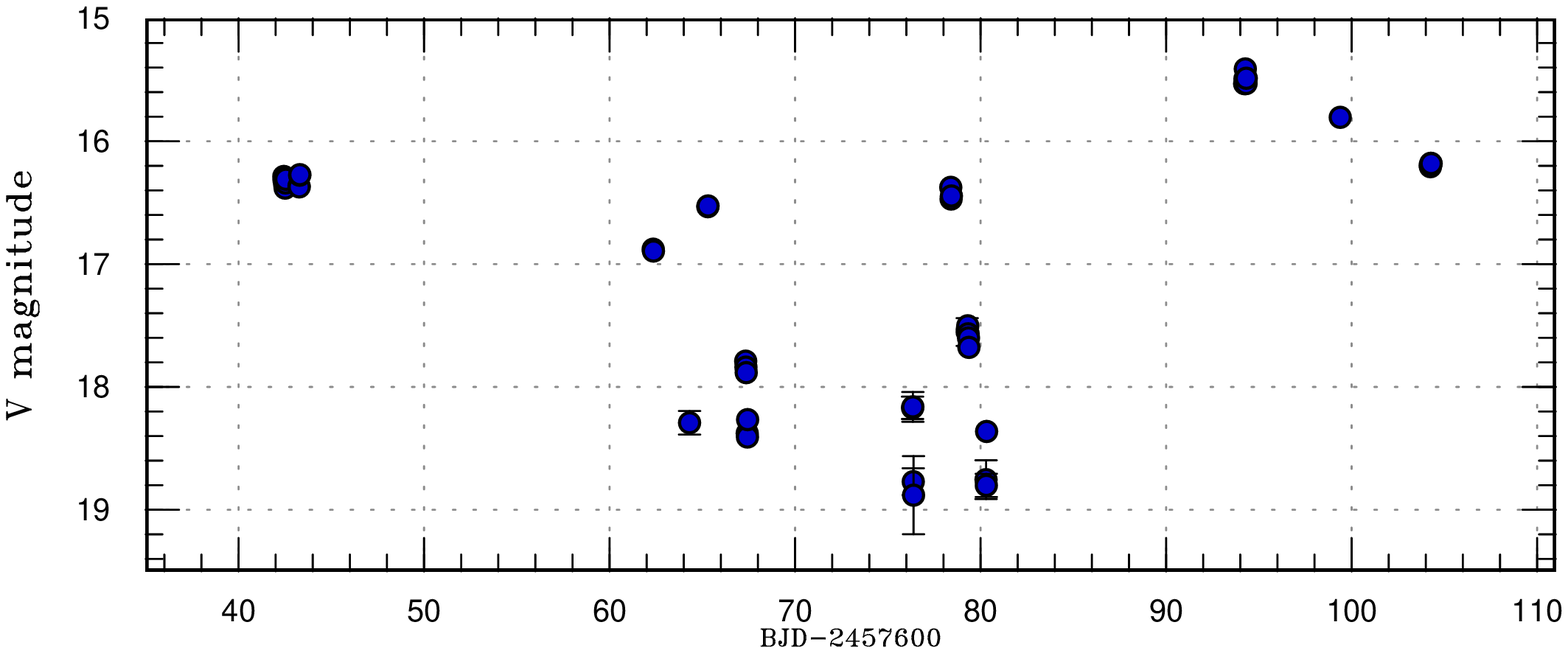}
  \end{center}
  \caption{Light curve of DDE 48 (2016).
  The data were binned to 0.002~d.
  }
  \label{fig:dde48lc}
\end{figure}

   This object shows frequent outbursts
(cf. vsnet-alert 20291; figure \ref{fig:dde48lc}).
The shortest interval of outbursts was 3~d.
The initial long outburst in figure \ref{fig:dde48lc}
was also likely a superoutburst recorded in its
the final phase.  If it is indeed the case,
the supercycle is around 62~d.  The object
may belong to ER UMa-type dwarf novae
(\cite{kat95eruma}; \cite{rob95eruma}).
(see also vsnet-alert 20291).
Future continuous observations to determine
the outburst characteristics, duty cycle and
superhump period are desired.

\subsection{MASTER OT J021315.37$+$533822.7}\label{obj:j0213}

\begin{figure}
  \begin{center}
    \FigureFile(85mm,110mm){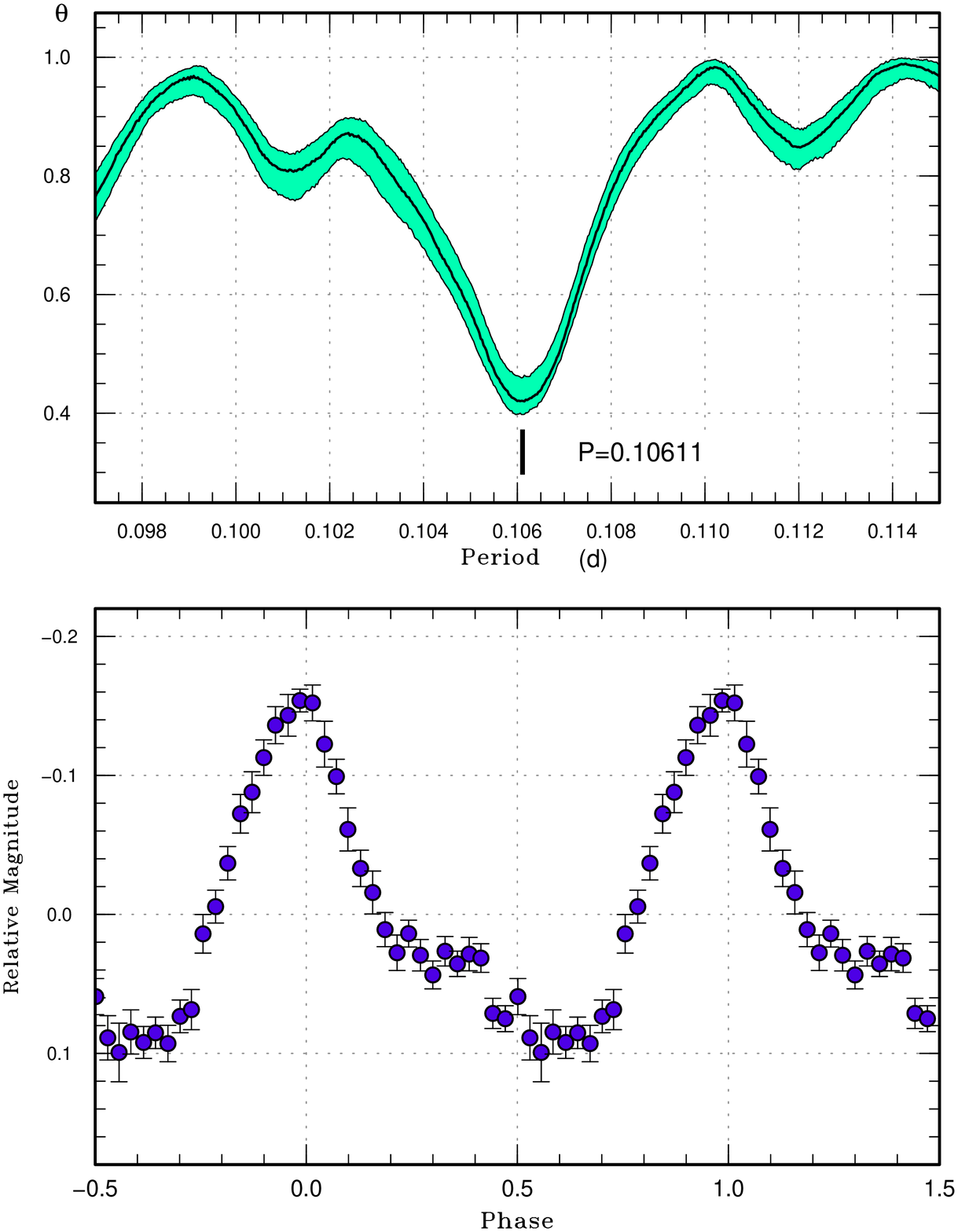}
  \end{center}
  \caption{Superhumps in MASTER J021315 (2016).
     (Upper): PDM analysis.
     (Lower): Phase-averaged profile.}
  \label{fig:j0213shpdm}
\end{figure}

   This object (hereafter MASTER J021315) was discovered
as a transient at an unfiltered CCD magnitude of
16.8 mag on 2013 November 1 by the MASTER network
\citep{yec13j0213atel5536}.
The 2016 outburst was detected by the ASAS-SN team
at $V$=16.39 on October 2.  The ASAS-SN also detected
the 2013 outburst and its duration was long (at least 8~d).
During the 2016 outburst, long-period superhumps
were detected (vsnet-alert 20218; figure \ref{fig:j0213shpdm}).
The period indicates that the object is in the period gap.
The times of superhump maxima are listed in
table \ref{tab:j0213oc2016}.
The period markedly decreased with
a global $P_{\rm dot}$ of $-205(35) \times 10^{-5}$.  
As recently recognized in many long-$P_{\rm orb}$
objects, such a large period decrease is most likely
a result of stage A-B transition
(cf. V1006 Cyg and MN Dra: \cite{kat16v1006cyg}; 
CRTS J214738.4$+$244554 and OT J064833.4$+$065624:
\cite{Pdot7};
KK Tel, possibly V452 Cas and ASASSN-15cl: \cite{Pdot8}).
The case is also likely since the initial observation
of MASTER J021315 started only 1~d after the outburst
detection.  We gave values in table \ref{tab:perlist}
following this interpretation.
The ASAS-SN data suggest that outbursts in this system
were relatively rare (only two were known with
a separation of $\sim$3~yr).  The object should have
a low mass-transfer rate.


\begin{table}
\caption{Superhump maxima of MASTER J021315}\label{tab:j0213oc2016}
\begin{center}
\begin{tabular}{rp{55pt}p{40pt}r@{.}lr}
\hline
\multicolumn{1}{c}{$E$} & \multicolumn{1}{c}{max\commenta} & \multicolumn{1}{c}{error} & \multicolumn{2}{c}{$O-C$\commentb} & \multicolumn{1}{c}{$N$\commentc} \\
\hline
0 & 57665.4175 & 0.0008 & $-$0&0077 & 84 \\
1 & 57665.5308 & 0.0043 & $-$0&0006 & 40 \\
10 & 57666.4920 & 0.0012 & 0&0055 & 62 \\
11 & 57666.5966 & 0.0006 & 0&0039 & 105 \\
12 & 57666.7072 & 0.0046 & 0&0084 & 26 \\
18 & 57667.3382 & 0.0009 & 0&0026 & 103 \\
19 & 57667.4387 & 0.0009 & $-$0&0031 & 98 \\
20 & 57667.5458 & 0.0008 & $-$0&0021 & 98 \\
21 & 57667.6470 & 0.0010 & $-$0&0069 & 100 \\
\hline
  \multicolumn{6}{l}{\commenta BJD$-$2400000.} \\
  \multicolumn{6}{l}{\commentb Against max $= 2457665.4252 + 0.106129 E$.} \\
  \multicolumn{6}{l}{\commentc Number of points used to determine the maximum.} \\
\end{tabular}
\end{center}
\end{table}

\subsection{MASTER OT J030205.67$+$254834.3}\label{obj:j0302}

\begin{figure}
  \begin{center}
    \FigureFile(85mm,110mm){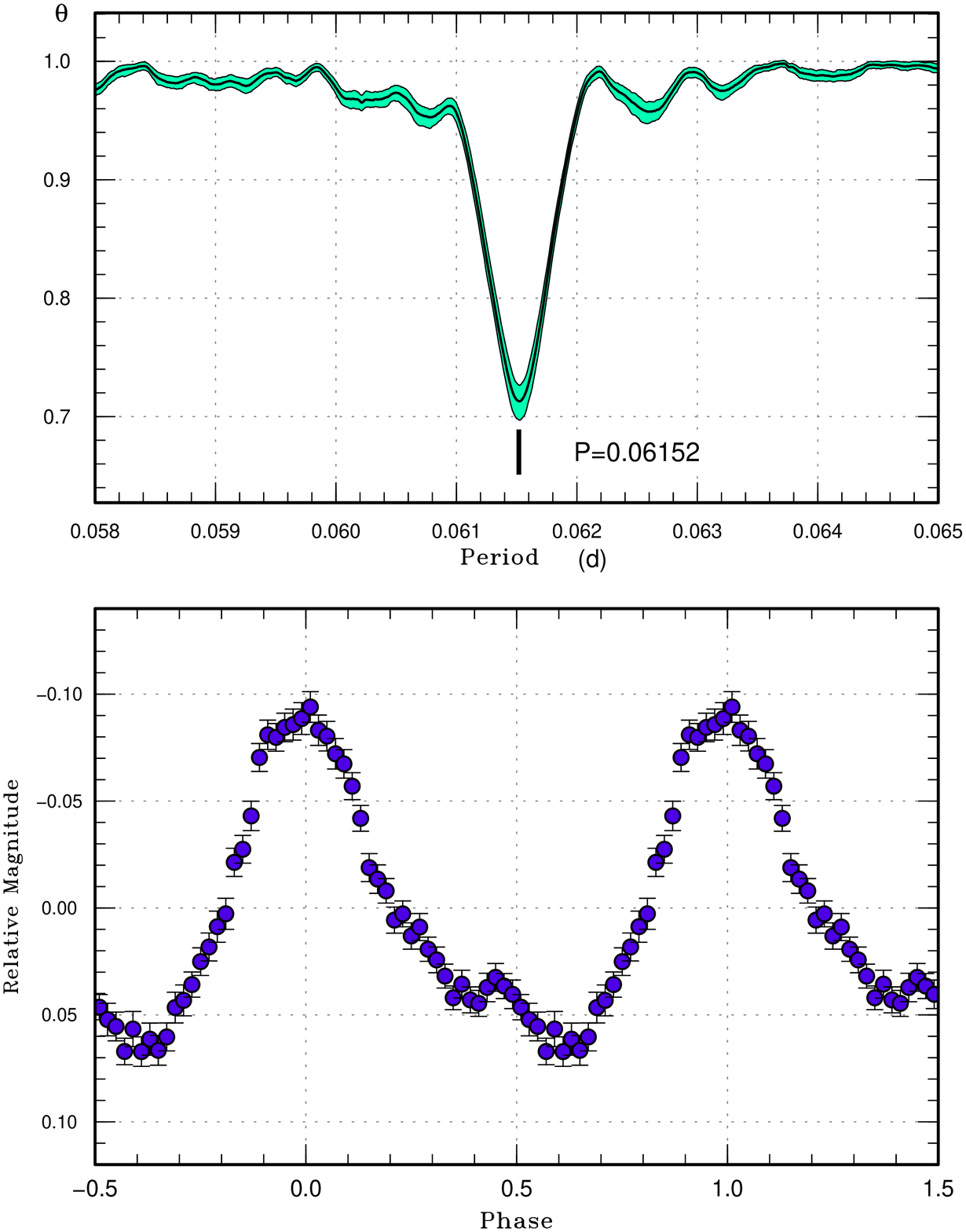}
  \end{center}
  \caption{Superhumps in MASTER J030205 (2016).
     (Upper): PDM analysis.
     (Lower): Phase-averaged profile.}
  \label{fig:j0302shpdm}
\end{figure}

   This object (hereafter MASTER J030205) was discovered
as a transient at an unfiltered CCD magnitude of
13.7 mag on 2016 December 4 by the MASTER network
\citep{bal16j0302atel9824}.
Although initial observations suggested the presence
of early superhumps of the WZ Sge-type dwarf nova
(vsnet-alert 20447), they were later identified
as developing superhumps (stage A) with double maxima
(vsnet-alert 20449, 20451).  
Further development of superhumps were reported
(vsnet-alert 20456, 20471; figure \ref{fig:j0302shpdm}).
The times of superhump maxima are listed in
table \ref{tab:j0302oc2016}.  Stages A and B
can be recognized and the $P_{\rm dot}$ of
stage B superhumps is positive, which is expected
for this $P_{\rm SH}$.  The period of stage A
superhump in table \ref{tab:perlist} was
determined by the PDM method for the data
before BJD 2457728.7.

\begin{figure}
  \begin{center}
    \FigureFile(85mm,110mm){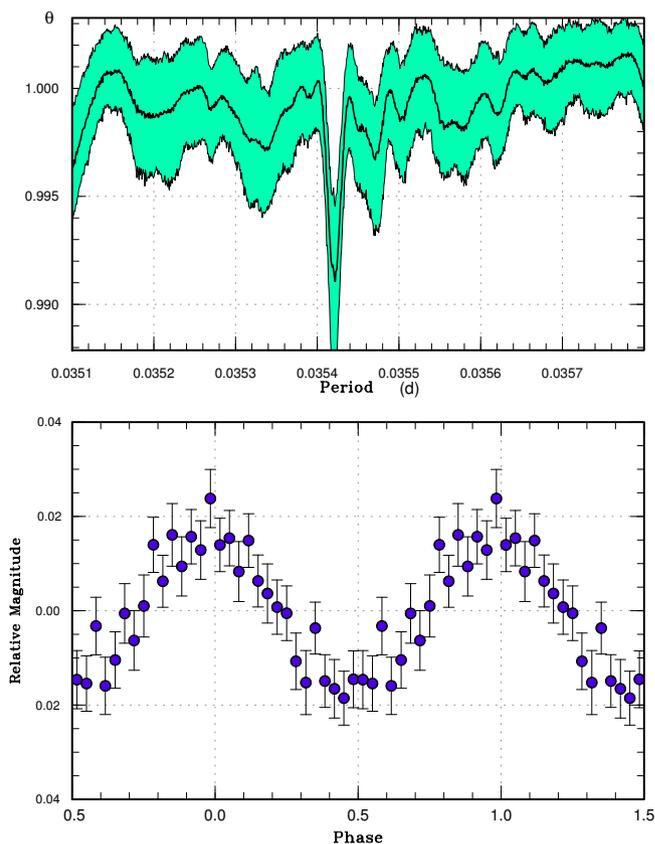}
  \end{center}
  \caption{Possible intermediate-polar-type signal
     in MASTER J030205 (2016).
     (Upper): PDM analysis.
     (Lower): Phase-averaged profile.}
  \label{fig:j0302spinpdm}
\end{figure}

   Short-term periodic oscillations were reported
(vsnet-alert 20452, 20457).  An analysis of the entire
data confirmed the presence of a coherent signal
with a period 0.0035420(2)~d [306.03(2)~s]
as originally reported (vsnet-alert 20457)
(figure \ref{fig:j0302spinpdm}).
Given the sharpness (high coherence) of the signal,
it may be an intermediate-polar (IP) signal
rather than quasi-periodic oscillations
(vsnet-alert 20458).  Since IPs are relatively rare
in SU UMa-type dwarf novae [see table 1 in
\citet{ham17DNIP}; the only well-established
SU UMa-type dwarf nova (not including WZ Sge-type
one) is CC Scl \citep{kat15ccscl}],
further confirmation of the signal in this system
is desired.


\begin{table}
\caption{Superhump maxima of MASTER J030205 (2016)}\label{tab:j0302oc2016}
\begin{center}
\begin{tabular}{rp{55pt}p{40pt}r@{.}lr}
\hline
\multicolumn{1}{c}{$E$} & \multicolumn{1}{c}{max\commenta} & \multicolumn{1}{c}{error} & \multicolumn{2}{c}{$O-C$\commentb} & \multicolumn{1}{c}{$N$\commentc} \\
\hline
0 & 57728.2331 & 0.0010 & $-$0&0010 & 94 \\
1 & 57728.2942 & 0.0010 & $-$0&0013 & 100 \\
2 & 57728.3565 & 0.0008 & $-$0&0006 & 92 \\
3 & 57728.4166 & 0.0020 & $-$0&0021 & 47 \\
4 & 57728.4758 & 0.0055 & $-$0&0044 & 21 \\
6 & 57728.6009 & 0.0010 & $-$0&0025 & 53 \\
12 & 57728.9791 & 0.0005 & 0&0064 & 131 \\
13 & 57729.0356 & 0.0004 & 0&0013 & 131 \\
14 & 57729.1002 & 0.0003 & 0&0044 & 246 \\
15 & 57729.1643 & 0.0014 & 0&0069 & 83 \\
16 & 57729.2225 & 0.0009 & 0&0035 & 54 \\
17 & 57729.2878 & 0.0008 & 0&0073 & 75 \\
18 & 57729.3452 & 0.0003 & 0&0031 & 144 \\
19 & 57729.4060 & 0.0003 & 0&0024 & 145 \\
20 & 57729.4671 & 0.0003 & 0&0020 & 145 \\
21 & 57729.5283 & 0.0006 & 0&0016 & 61 \\
22 & 57729.5901 & 0.0006 & 0&0018 & 62 \\
24 & 57729.7127 & 0.0005 & 0&0014 & 122 \\
25 & 57729.7714 & 0.0006 & $-$0&0015 & 122 \\
26 & 57729.8386 & 0.0010 & 0&0041 & 88 \\
34 & 57730.3253 & 0.0006 & $-$0&0016 & 87 \\
35 & 57730.3870 & 0.0004 & $-$0&0014 & 119 \\
36 & 57730.4466 & 0.0005 & $-$0&0034 & 131 \\
37 & 57730.5139 & 0.0009 & 0&0024 & 49 \\
44 & 57730.9401 & 0.0004 & $-$0&0024 & 131 \\
45 & 57730.9987 & 0.0004 & $-$0&0052 & 131 \\
46 & 57731.0600 & 0.0003 & $-$0&0055 & 129 \\
50 & 57731.3076 & 0.0016 & $-$0&0042 & 72 \\
51 & 57731.3677 & 0.0012 & $-$0&0056 & 72 \\
52 & 57731.4258 & 0.0008 & $-$0&0090 & 68 \\
53 & 57731.4949 & 0.0011 & $-$0&0015 & 72 \\
54 & 57731.5572 & 0.0011 & $-$0&0008 & 72 \\
61 & 57731.9860 & 0.0011 & $-$0&0029 & 135 \\
62 & 57732.0482 & 0.0013 & $-$0&0021 & 188 \\
63 & 57732.1110 & 0.0006 & $-$0&0009 & 175 \\
64 & 57732.1765 & 0.0009 & 0&0030 & 99 \\
65 & 57732.2313 & 0.0006 & $-$0&0037 & 134 \\
66 & 57732.3015 & 0.0012 & 0&0049 & 148 \\
67 & 57732.3500 & 0.0010 & $-$0&0081 & 129 \\
68 & 57732.4179 & 0.0007 & $-$0&0018 & 77 \\
69 & 57732.4736 & 0.0008 & $-$0&0077 & 132 \\
70 & 57732.5429 & 0.0006 & 0&0001 & 127 \\
71 & 57732.5997 & 0.0009 & $-$0&0047 & 42 \\
77 & 57732.9768 & 0.0011 & 0&0031 & 195 \\
86 & 57733.5226 & 0.0006 & $-$0&0051 & 77 \\
87 & 57733.5912 & 0.0014 & 0&0019 & 40 \\
92 & 57733.9053 & 0.0017 & 0&0083 & 79 \\
93 & 57733.9625 & 0.0011 & 0&0039 & 131 \\
94 & 57734.0291 & 0.0006 & 0&0090 & 131 \\
95 & 57734.0879 & 0.0012 & 0&0062 & 130 \\
96 & 57734.1450 & 0.0012 & 0&0018 & 94 \\
\hline
  \multicolumn{6}{l}{\commenta BJD$-$2400000.} \\
  \multicolumn{6}{l}{\commentb Against max $= 2457728.2340 + 0.061554 E$.} \\
  \multicolumn{6}{l}{\commentc Number of points used to determine the maximum.} \\
\end{tabular}
\end{center}
\end{table}

\subsection{MASTER OT J042609.34$+$354144.8}\label{obj:j0426}

   This object (hereafter MASTER J042609) was discovered
as a transient at an unfiltered CCD magnitude of 12.9
on 2012 September 30 by the MASTER network
\citep{den12j0426atel4441}.  The SU UMa-type nature
was confirmed during this superoutburst \citep{Pdot5}.
This object is also a grazing eclipser.
Fore more information and history, see \citet{Pdot5}.

\begin{figure}
  \begin{center}
    \FigureFile(88mm,70mm){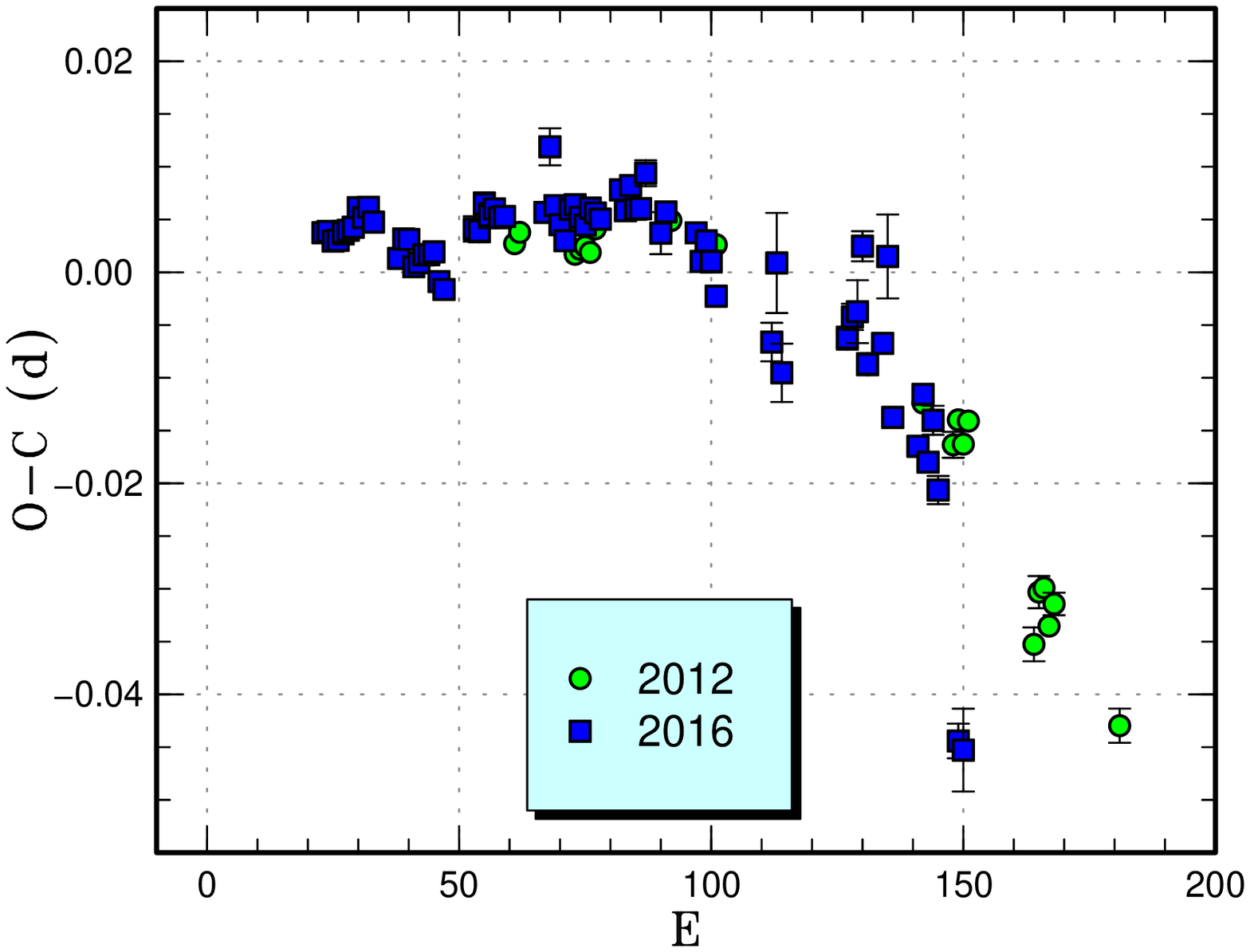}
  \end{center}
  \caption{Comparison of $O-C$ diagrams of MASTER J042609
  between different superoutbursts.
  A period of 0.06756~d was used to draw this figure.
  Approximate cycle counts ($E$) after the outburst detection
  were used.  The 2012 superoutburst was shifted by
  20 cycles to match the 2016 one.
  }
  \label{fig:j0426comp}
\end{figure}

\begin{figure}
  \begin{center}
    \FigureFile(85mm,110mm){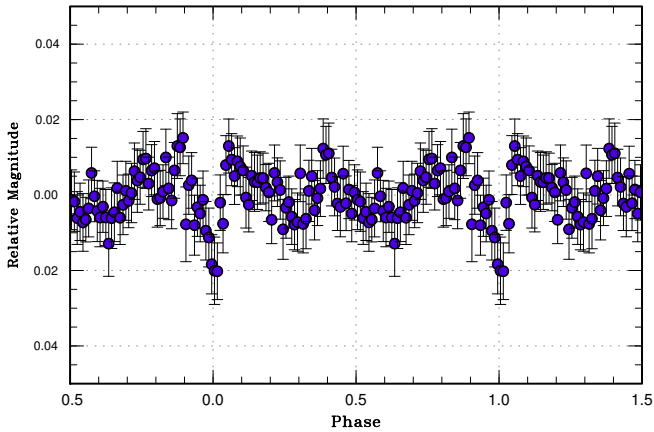}
  \end{center}
  \caption{Eclipse profile in MASTER J042609 (2016).
     The superhumps were mostly removed by using LOWESS.
     The phase-averaged profile was drawn against
     the ephemeris BJD 2456276.6430$+$0.06550168$E$.
     }
  \label{fig:j0426eclph}
\end{figure}

   The 2016 superoutburst was detected by E. Muyllaert
at a visual magnitude of 13.8 on December 26.
The last observation before this detection was on
$V$=15.35--15.51 on December 23 (ASAS-SN).
It was not clear when the outburst started.
Superhumps were observed with a period change
(vsnet-alert 20532, 20533).
The times of superhump maxima are listed in
table \ref{tab:j0426oc2016}.
The period change observed during the 2016 superoutburst
probably reflected stage B-C transition
(figure \ref{fig:j0426comp}).  Although the last two points
may have been traditional late superhumps,
no clear superhumps were observed after them
despite relatively good observational coverage.
Although no clear eclipses were visible during this
superoutburst, a phase-averaged light curve
with a period of 0.06550168~d and an epoch of
BJD 2456276.6430 \citep{Pdot5} yielded a shallow
eclipse (0.03 mag) at the expected phase
(figure \ref{fig:j0426eclph}).
We consider 0.06550168(1)~d to be the refined orbital
period.

\begin{table*}
\caption{Superhump maxima of MASTER J042609}\label{tab:j0426oc2016}
\begin{center}
\begin{tabular}{rp{50pt}p{30pt}r@{.}lcrrp{50pt}p{30pt}r@{.}lcr}
\hline
\multicolumn{1}{c}{$E$} & \multicolumn{1}{c}{max\commenta} & \multicolumn{1}{c}{error} & \multicolumn{2}{c}{$O-C$\commentb} & \multicolumn{1}{c}{phase\commentc} & \multicolumn{1}{c}{$N$\commentd} & \multicolumn{1}{c}{$E$} & \multicolumn{1}{c}{max\commenta} & \multicolumn{1}{c}{error} & \multicolumn{2}{c}{$O-C$\commentb} & \multicolumn{1}{c}{phase\commentc} & \multicolumn{1}{c}{$N$\commentd} \\
\hline
1 & 57751.0076 & 0.0003 & $-$0&0054 & 0.81 & 110 & 52 & 57754.4538 & 0.0003 & 0&0039 & 0.42 & 142 \\
2 & 57751.0742 & 0.0004 & $-$0&0062 & 0.82 & 123 & 53 & 57754.5229 & 0.0006 & 0&0056 & 0.47 & 142 \\
3 & 57751.1419 & 0.0004 & $-$0&0059 & 0.86 & 124 & 54 & 57754.5900 & 0.0005 & 0&0053 & 0.50 & 142 \\
4 & 57751.2102 & 0.0011 & $-$0&0050 & 0.90 & 87 & 55 & 57754.6570 & 0.0008 & 0&0049 & 0.52 & 130 \\
5 & 57751.2779 & 0.0004 & $-$0&0046 & 0.93 & 75 & 59 & 57754.9300 & 0.0005 & 0&0084 & 0.69 & 125 \\
6 & 57751.3458 & 0.0004 & $-$0&0042 & 0.97 & 159 & 60 & 57754.9956 & 0.0006 & 0&0066 & 0.69 & 170 \\
7 & 57751.4152 & 0.0004 & $-$0&0021 & 0.03 & 155 & 61 & 57755.0655 & 0.0007 & 0&0091 & 0.76 & 183 \\
8 & 57751.4818 & 0.0005 & $-$0&0029 & 0.05 & 141 & 62 & 57755.1309 & 0.0006 & 0&0071 & 0.75 & 198 \\
9 & 57751.5504 & 0.0003 & $-$0&0018 & 0.09 & 141 & 63 & 57755.1985 & 0.0007 & 0&0073 & 0.79 & 220 \\
10 & 57751.6165 & 0.0004 & $-$0&0030 & 0.10 & 139 & 64 & 57755.2694 & 0.0012 & 0&0108 & 0.87 & 78 \\
15 & 57751.9509 & 0.0006 & $-$0&0056 & 0.21 & 143 & 67 & 57755.4664 & 0.0020 & 0&0056 & 0.88 & 89 \\
16 & 57752.0203 & 0.0005 & $-$0&0036 & 0.27 & 192 & 68 & 57755.5360 & 0.0008 & 0&0078 & 0.94 & 133 \\
17 & 57752.0878 & 0.0003 & $-$0&0035 & 0.30 & 205 & 74 & 57755.9393 & 0.0006 & 0&0069 & 0.10 & 141 \\
18 & 57752.1528 & 0.0003 & $-$0&0059 & 0.29 & 208 & 75 & 57756.0042 & 0.0009 & 0&0043 & 0.09 & 180 \\
19 & 57752.2207 & 0.0004 & $-$0&0053 & 0.33 & 83 & 76 & 57756.0737 & 0.0008 & 0&0064 & 0.15 & 207 \\
20 & 57752.2890 & 0.0003 & $-$0&0044 & 0.37 & 353 & 77 & 57756.1393 & 0.0010 & 0&0047 & 0.15 & 206 \\
21 & 57752.3566 & 0.0003 & $-$0&0042 & 0.40 & 415 & 78 & 57756.2036 & 0.0008 & 0&0015 & 0.13 & 143 \\
22 & 57752.4244 & 0.0004 & $-$0&0038 & 0.44 & 360 & 89 & 57756.9424 & 0.0018 & $-$0&0009 & 0.41 & 189 \\
23 & 57752.4892 & 0.0005 & $-$0&0064 & 0.42 & 412 & 90 & 57757.0174 & 0.0047 & 0&0067 & 0.56 & 194 \\
24 & 57752.5560 & 0.0005 & $-$0&0070 & 0.44 & 120 & 91 & 57757.0746 & 0.0028 & $-$0&0035 & 0.43 & 204 \\
30 & 57752.9671 & 0.0012 & $-$0&0002 & 0.72 & 196 & 104 & 57757.9562 & 0.0011 & 0&0020 & 0.89 & 83 \\
31 & 57753.0344 & 0.0011 & $-$0&0003 & 0.75 & 188 & 105 & 57758.0257 & 0.0012 & 0&0042 & 0.95 & 131 \\
32 & 57753.1046 & 0.0011 & 0&0025 & 0.82 & 208 & 106 & 57758.0938 & 0.0030 & 0&0048 & 0.99 & 203 \\
33 & 57753.1710 & 0.0012 & 0&0015 & 0.83 & 83 & 107 & 57758.1676 & 0.0014 & 0&0112 & 0.12 & 122 \\
34 & 57753.2392 & 0.0007 & 0&0023 & 0.88 & 260 & 108 & 57758.2239 & 0.0011 & 0&0002 & 0.98 & 80 \\
35 & 57753.3059 & 0.0005 & 0&0017 & 0.89 & 259 & 111 & 57758.4286 & 0.0008 & 0&0027 & 0.10 & 45 \\
36 & 57753.3736 & 0.0007 & 0&0019 & 0.93 & 291 & 112 & 57758.5044 & 0.0040 & 0&0111 & 0.26 & 38 \\
44 & 57753.9145 & 0.0008 & 0&0037 & 0.18 & 124 & 113 & 57758.5567 & 0.0007 & $-$0&0040 & 0.06 & 54 \\
45 & 57753.9883 & 0.0018 & 0&0101 & 0.31 & 101 & 118 & 57758.8918 & 0.0006 & $-$0&0059 & 0.17 & 96 \\
46 & 57754.0502 & 0.0008 & 0&0047 & 0.26 & 179 & 119 & 57758.9642 & 0.0010 & $-$0&0008 & 0.28 & 125 \\
47 & 57754.1160 & 0.0008 & 0&0030 & 0.26 & 203 & 120 & 57759.0254 & 0.0009 & $-$0&0070 & 0.21 & 109 \\
48 & 57754.1820 & 0.0007 & 0&0017 & 0.27 & 228 & 121 & 57759.0969 & 0.0014 & $-$0&0029 & 0.30 & 126 \\
49 & 57754.2526 & 0.0003 & 0&0048 & 0.35 & 273 & 122 & 57759.1578 & 0.0013 & $-$0&0093 & 0.23 & 123 \\
50 & 57754.3206 & 0.0004 & 0&0054 & 0.38 & 207 & 126 & 57759.4043 & 0.0016 & $-$0&0324 & 1.00 & 59 \\
51 & 57754.3869 & 0.0004 & 0&0044 & 0.40 & 133 & 127 & 57759.4710 & 0.0039 & $-$0&0331 & 0.01 & 74 \\
\hline
  \multicolumn{14}{l}{\commenta BJD$-$2400000.} \\
  \multicolumn{14}{l}{\commentb Against max $= 2457750.9456 + 0.067390 E$.} \\
  \multicolumn{14}{l}{\commentc Orbital phase.} \\
  \multicolumn{14}{l}{\commentd Number of points used to determine the maximum.} \\
\end{tabular}
\end{center}
\end{table*}

\subsection{MASTER OT J043220.15$+$784913.8}\label{obj:j0432}

\begin{figure}
  \begin{center}
    \FigureFile(85mm,110mm){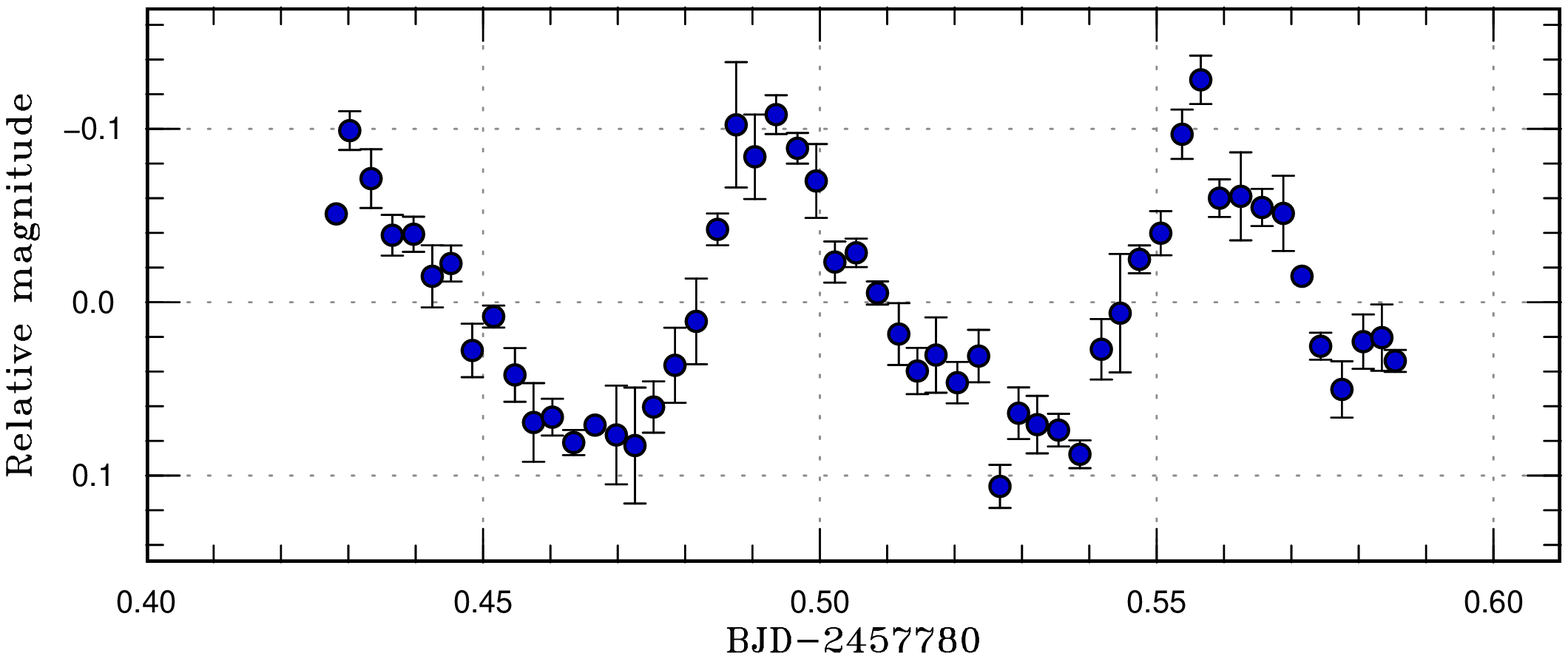}
  \end{center}
  \caption{Superhump in MASTER J043220 (2017).
  The data were binned to 0.003~d.
  }
  \label{fig:j0432shlc}
\end{figure}

   This object (hereafter MASTER J043220) was discovered
as a transient at an unfiltered CCD magnitude of 16.8
on 2013 December 12 by the MASTER network
\citep{shu13j0432atel5657}.
The 2017 outburst was detected by the ASAS-SN team
at $V$=16.29 on January 25.  The object was already
seen in outburst at $V$=16.52 in the ASAS-SN data.
Subsequent observations detected superhumps
(vsnet-alert 20614; figure \ref{fig:j0432shlc}).
The times of superhump maxima were BJD 2457780.4941(5) ($N$=64)
and 2457780.5584(5) ($N$=64).  A PDM analysis
yielded a period of 0.0640(6)~d.

\subsection{MASTER OT J043915.60$+$424232.3}\label{obj:j0439}

   This object (hereafter MASTER J043915) was discovered
as a transient at an unfiltered CCD magnitude of 15.7
on 2014 January 21 by the MASTER network
\citep{bal14j0439atel5787}.  The SU UMa-type nature
was confirmed during this superoutburst \citep{Pdot7}.
Fore more history, see \citet{Pdot7}.

\begin{figure}
  \begin{center}
    \FigureFile(88mm,70mm){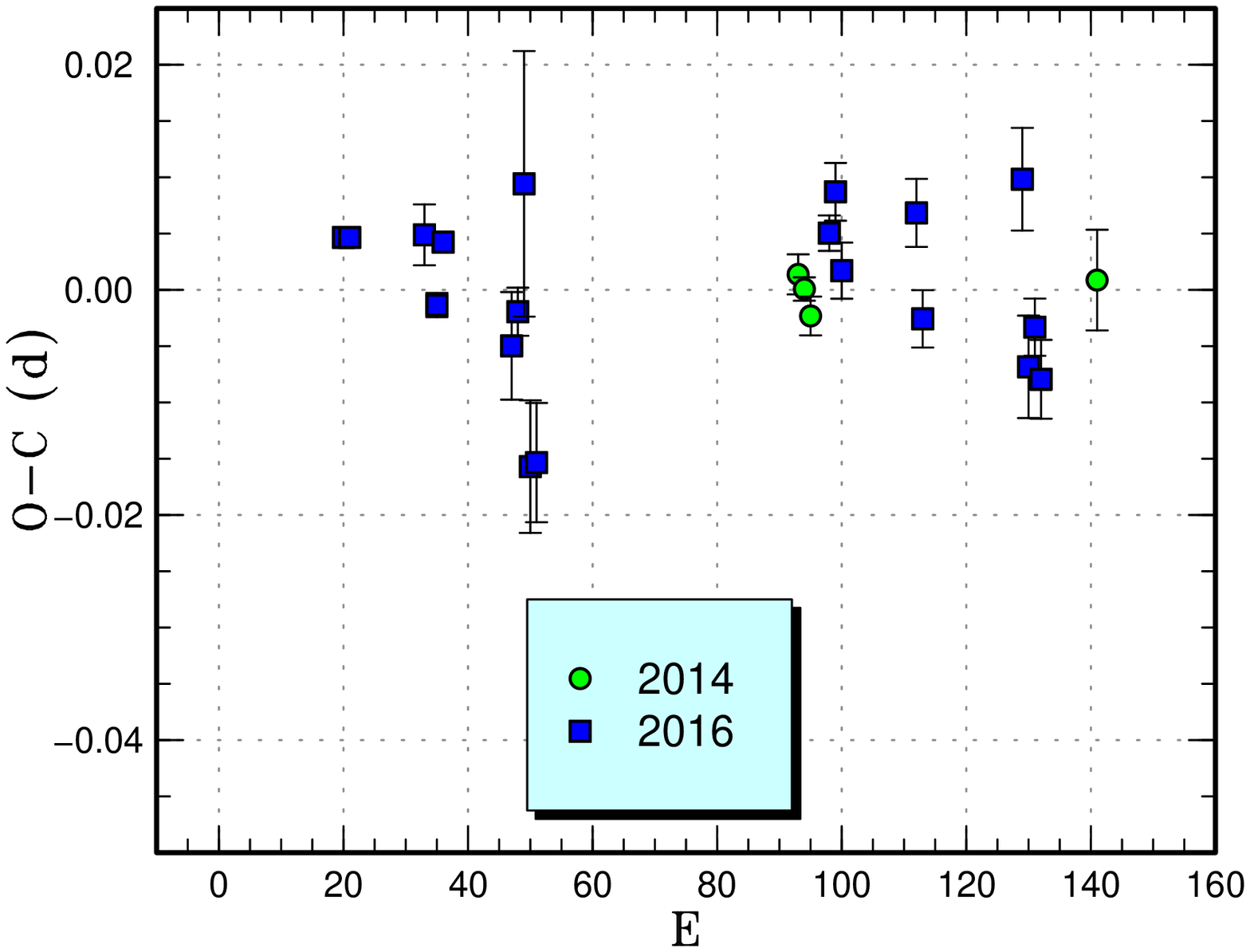}
  \end{center}
  \caption{Comparison of $O-C$ diagrams of MASTER J043915
  between different superoutbursts.
  A period of 0.06243~d was used to draw this figure.
  Approximate cycle counts ($E$) after the start of the superoutburst
  were used.
  }
  \label{fig:j0439comp}
\end{figure}

   The 2016 superoutburst was detected by the ASAS-SN team
at $V$=16.38 on December 22.  Superhumps were subsequently
recorded (vsnet-alert 20509).  The times of superhump
maxima are listed in table \ref{tab:j0439oc2016}.
Although there were not disagreement between the
2014 and 2016 observations, we could not detect
clear stages B and C, which are expected for this short
$P_{\rm SH}$ (figure \ref{fig:j0439comp}).

\begin{table}
\caption{Superhump maxima of MASTER J043915 (2016)}\label{tab:j0439oc2016}
\begin{center}
\begin{tabular}{rp{55pt}p{40pt}r@{.}lr}
\hline
\multicolumn{1}{c}{$E$} & \multicolumn{1}{c}{max\commenta} & \multicolumn{1}{c}{error} & \multicolumn{2}{c}{$O-C$\commentb} & \multicolumn{1}{c}{$N$\commentc} \\
\hline
0 & 57746.2392 & 0.0003 & 0&0045 & 150 \\
1 & 57746.3016 & 0.0003 & 0&0045 & 149 \\
13 & 57747.0510 & 0.0027 & 0&0048 & 47 \\
15 & 57747.1696 & 0.0011 & $-$0&0015 & 83 \\
16 & 57747.2376 & 0.0004 & 0&0041 & 137 \\
27 & 57747.9151 & 0.0048 & $-$0&0050 & 42 \\
28 & 57747.9806 & 0.0021 & $-$0&0020 & 43 \\
29 & 57748.0544 & 0.0118 & 0&0094 & 43 \\
30 & 57748.0917 & 0.0059 & $-$0&0158 & 42 \\
31 & 57748.1545 & 0.0053 & $-$0&0154 & 32 \\
78 & 57751.1091 & 0.0016 & 0&0051 & 42 \\
79 & 57751.1752 & 0.0026 & 0&0088 & 43 \\
80 & 57751.2306 & 0.0025 & 0&0018 & 15 \\
92 & 57751.9849 & 0.0030 & 0&0069 & 43 \\
93 & 57752.0379 & 0.0026 & $-$0&0025 & 43 \\
109 & 57753.0492 & 0.0046 & 0&0100 & 24 \\
110 & 57753.0950 & 0.0046 & $-$0&0067 & 43 \\
111 & 57753.1609 & 0.0025 & $-$0&0032 & 42 \\
112 & 57753.2187 & 0.0035 & $-$0&0078 & 33 \\
\hline
  \multicolumn{6}{l}{\commenta BJD$-$2400000.} \\
  \multicolumn{6}{l}{\commentb Against max $= 2457746.2346 + 0.062428 E$.} \\
  \multicolumn{6}{l}{\commentc Number of points used to determine the maximum.} \\
\end{tabular}
\end{center}
\end{table}

\subsection{MASTER OT J054746.81$+$762018.9}\label{obj:j0547}

\begin{figure}
  \begin{center}
    \FigureFile(85mm,110mm){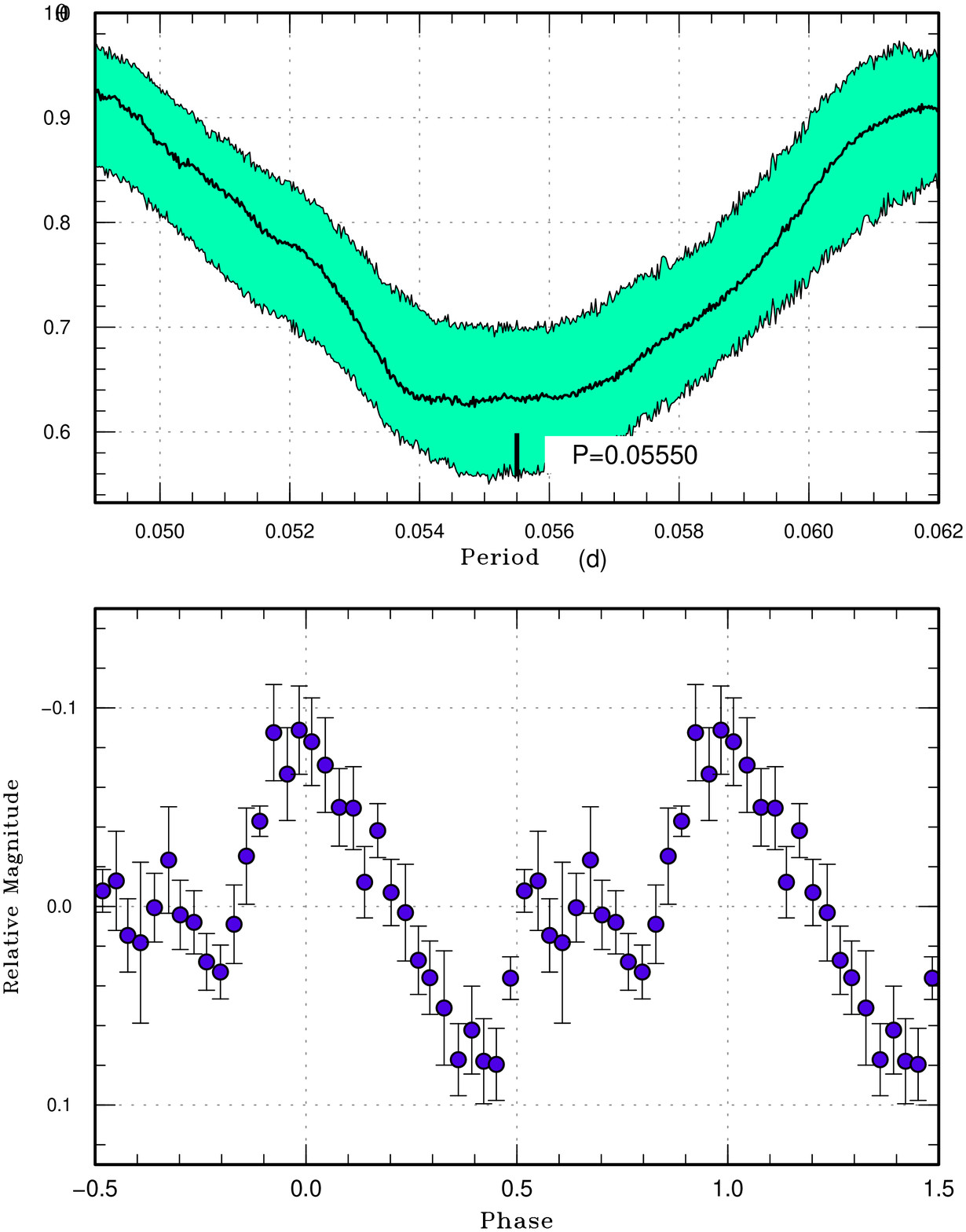}
  \end{center}
  \caption{Superhumps in MASTER J054746 (2016).
     (Upper): PDM analysis.
     (Lower): Phase-averaged profile.}
  \label{fig:j0547shpdm}
\end{figure}

   This object (hereafter MASTER J054746) was discovered
as a transient at an unfiltered CCD magnitude of
16.7 mag on 2016 October 12 by the MASTER network
\citep{shu16j0547atel9616}.  Subsequent observations
detected superhumps (vsnet-alert 20227;
figure \ref{fig:j0547shpdm}).
The times of superhump maxima are listed in
table \ref{tab:j0547oc2016}.  The best superhump period
with the PDM method is listed in table \ref{tab:perlist}.


\begin{table}
\caption{Superhump maxima of MASTER J054746 (2016)}\label{tab:j0547oc2016}
\begin{center}
\begin{tabular}{rp{55pt}p{40pt}r@{.}lr}
\hline
\multicolumn{1}{c}{$E$} & \multicolumn{1}{c}{max\commenta} & \multicolumn{1}{c}{error} & \multicolumn{2}{c}{$O-C$\commentb} & \multicolumn{1}{c}{$N$\commentc} \\
\hline
0 & 57674.4439 & 0.0012 & 0&0008 & 52 \\
1 & 57674.4982 & 0.0016 & $-$0&0008 & 44 \\
2 & 57674.5539 & 0.0014 & $-$0&0009 & 56 \\
3 & 57674.6117 & 0.0015 & 0&0009 & 56 \\
\hline
  \multicolumn{6}{l}{\commenta BJD$-$2400000.} \\
  \multicolumn{6}{l}{\commentb Against max $= 2457674.4430 + 0.055927 E$.} \\
  \multicolumn{6}{l}{\commentc Number of points used to determine the maximum.} \\
\end{tabular}
\end{center}
\end{table}

\subsection{MASTER OT J055348.98$+$482209.0}\label{obj:j0553}

\begin{figure}
  \begin{center}
    \FigureFile(85mm,110mm){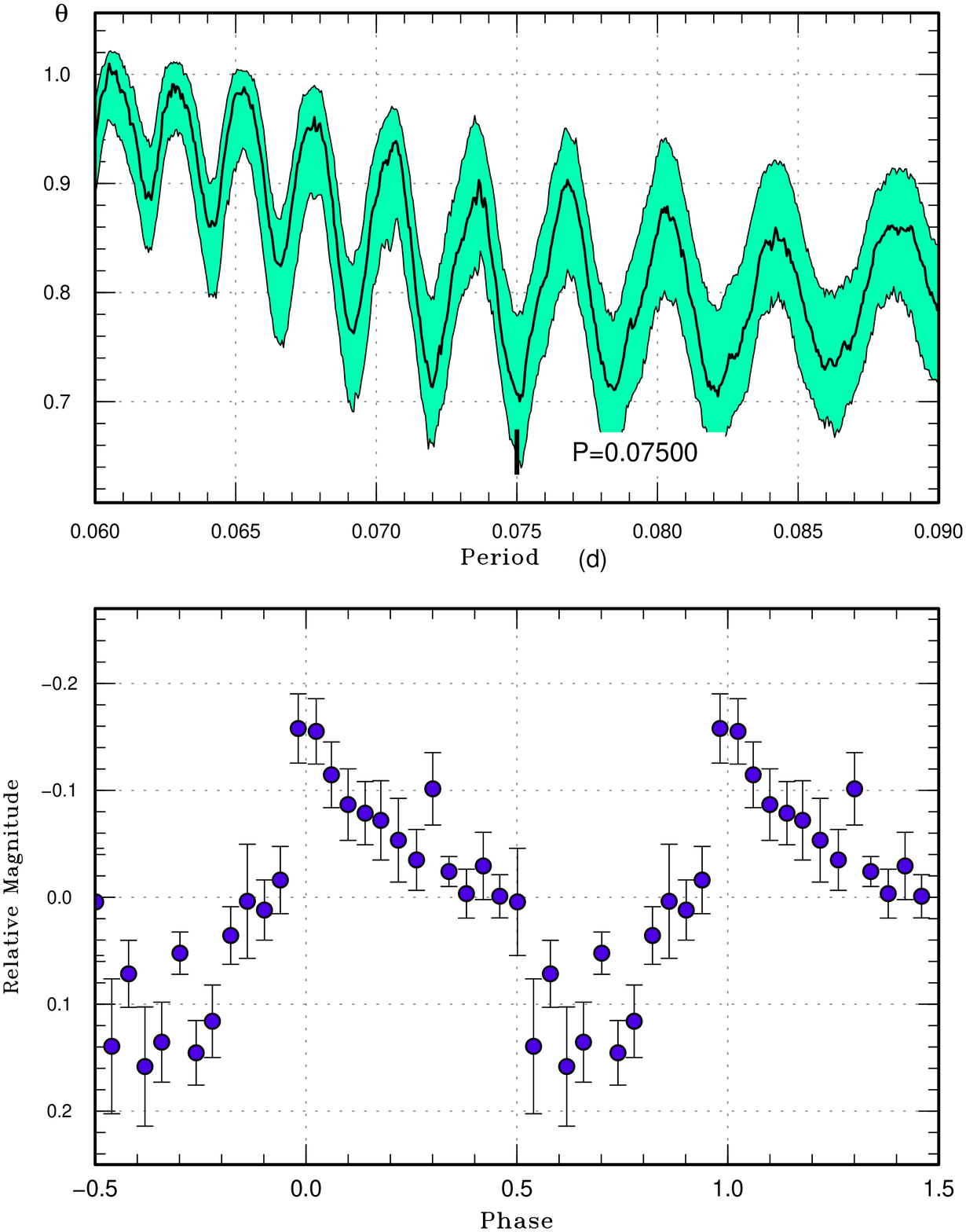}
  \end{center}
  \caption{Superhumps in MASTER J055348 (2017).
     (Upper): PDM analysis.  The selected period was
     one of the possibilities.
     (Lower): Phase-averaged profile.}
  \label{fig:j0553shpdm}
\end{figure}

   This object (hereafter MASTER J055348) was discovered
as a transient at an unfiltered CCD magnitude of
16.5 mag on 2014 March 13 by the MASTER network
\citep{vla14j0553atel5983}.
The 2017 outburst was detected by the ASAS-SN team
at $V$=16.52 on February 16.
Subsequent observations detected superhumps
(vsnet-alert 20681).  Although observations on two
nights were reported, neither data were of sufficient
quality to determine the superhump period
(the object already faded below 17 mag).
The period used to calculated epochs in
table \ref{tab:j0553oc2017} was one of the possibilities
giving smallest $O-C$ residuals.  Other candidate
aliases were 0.0784(1)~d and 0.0720(1)~d
(figure \ref{fig:j0553shpdm}).
The period of 0.0666~d reported in vsnet-alert 20681
could not express the second-night observation.


\clearpage 


\begin{table}
\caption{Superhump maxima of MASTER J055348 (2017)}\label{tab:j0553oc2017}
\begin{center}
\begin{tabular}{rp{55pt}p{40pt}r@{.}lr}
\hline
\multicolumn{1}{c}{$E$} & \multicolumn{1}{c}{max\commenta} & \multicolumn{1}{c}{error} & \multicolumn{2}{c}{$O-C$\commentb} & \multicolumn{1}{c}{$N$\commentc} \\
\hline
0 & 57803.3236 & 0.0010 & 0&0035 & 47 \\
1 & 57803.3917 & 0.0010 & $-$0&0035 & 53 \\
23 & 57805.0467 & 0.0022 & $-$0&0014 & 65 \\
24 & 57805.1248 & 0.0028 & 0&0015 & 77 \\
\hline
  \multicolumn{6}{l}{\commenta BJD$-$2400000.} \\
  \multicolumn{6}{l}{\commentb Against max $= 2457803.3201 + 0.075130 E$.} \\
  \multicolumn{6}{l}{\commentc Number of points used to determine the maximum.} \\
\end{tabular}
\end{center}
\end{table}

\subsection{MASTER OT J055845.55$+$391533.4}\label{obj:j0558}

   This optical transient (hereafter MASTER J055845)
was detected on 2014 February 19 at a magnitude of 14.4
\citep{yec14j0558atel5905}.  During the 2014 superoutburst,
single-night observations detected superhumps (likely
stage C ones) with a period of 0.0563(4)~d \citep{Pdot7}.

\begin{figure}
  \begin{center}
    \FigureFile(85mm,110mm){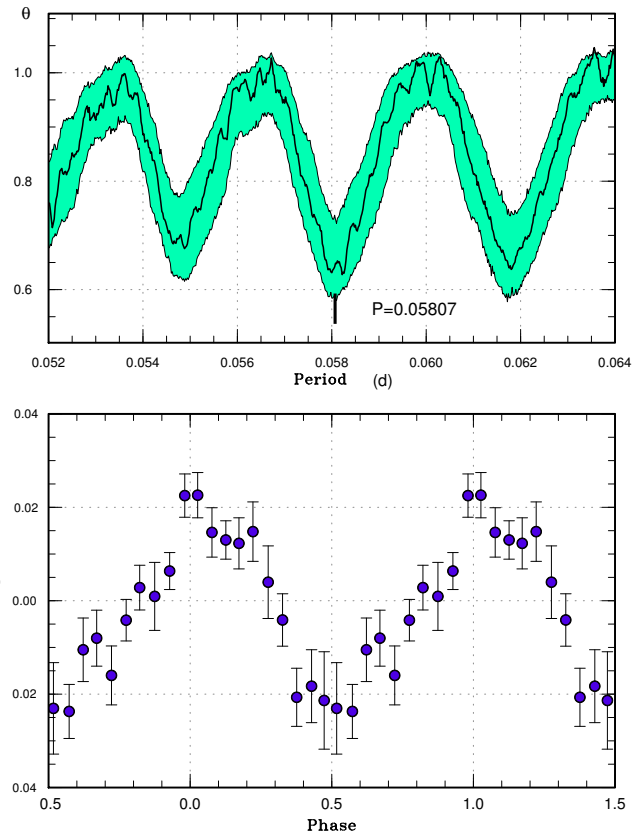}
  \end{center}
  \caption{Superhumps in MASTER J055845 (2016).
     (Upper): PDM analysis.
     (Lower): Phase-averaged profile.}
  \label{fig:j0558shpdm}
\end{figure}

   The 2016 outburst was detected by the ASAS-SN team
at $V$=15.24 on September 14.  The object was on the rise
at $V$=16.51 on September 7.  Rather queerly, the object
was also detected at $V$=14.16 on August 30.  There were
no observations between August 30 and September 7.
Three-night observations starting on September 16 detected
superhumps (vsnet-alert 20186).  During these observations,
the object brightened from 15.4 mag (September 16) to
15.2 mag (September 18).  Superhumps were recorded on
the first two nights (table \ref{tab:j0558oc2017}).
The period in table \ref{tab:perlist} was determined
by the PDM method (figure \ref{fig:j0558shpdm}).
Since the recorded outburst behavior
was rather strange, we could not determine the superhump
stage.  The reason of the large difference of superhump
periods between 2014 and 2016 is unclear.  The 2014
observations were single-night ones and there were no
possibility of an alias and the 2014 period could not
satisfy the 2016 data.  More observations are apparently
needed.


\begin{table}
\caption{Superhump maxima of MASTER J055845 (2016)}\label{tab:j0558oc2017}
\begin{center}
\begin{tabular}{rp{55pt}p{40pt}r@{.}lr}
\hline
\multicolumn{1}{c}{$E$} & \multicolumn{1}{c}{max\commenta} & \multicolumn{1}{c}{error} & \multicolumn{2}{c}{$O-C$\commentb} & \multicolumn{1}{c}{$N$\commentc} \\
\hline
0 & 57647.5355 & 0.0018 & 0&0015 & 13 \\
1 & 57647.5904 & 0.0016 & $-$0&0017 & 35 \\
17 & 57648.5226 & 0.0017 & 0&0009 & 42 \\
18 & 57648.5800 & 0.0016 & 0&0001 & 30 \\
19 & 57648.6370 & 0.0023 & $-$0&0009 & 26 \\
\hline
  \multicolumn{6}{l}{\commenta BJD$-$2400000.} \\
  \multicolumn{6}{l}{\commentb Against max $= 2457647.5340 + 0.058102 E$.} \\
  \multicolumn{6}{l}{\commentc Number of points used to determine the maximum.} \\
\end{tabular}
\end{center}
\end{table}

\subsection{MASTER OT J064725.70$+$491543.9}\label{obj:j0647}

   This object (hereafter MASTER J064725) was discovered
as a transient at an unfiltered CCD magnitude of
13.2 mag on 2013 March 7 by the MASTER network
\citep{tiu13j0647atel4871}.  Subsequent observations
detected superhumps \citep{Pdot5}.

\begin{figure}
  \begin{center}
    \FigureFile(88mm,70mm){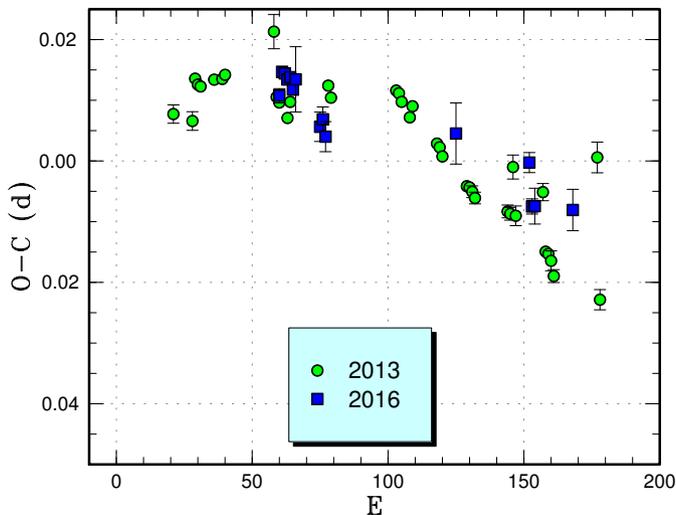}
  \end{center}
  \caption{Comparison of $O-C$ diagrams of MASTER J064725
  between different superoutbursts.
  A period of 0.06777~d was used to draw this figure.
  Approximate cycle counts ($E$) after the start of the superoutburst
  were used.
  }
  \label{fig:j0647comp}
\end{figure}

   The 2016 superoutburst was detected by the ASAS-SN
team at $V$=13.99 on December 13.  Time-resolved
photometry started on December 17 and stage A superhumps
were not recorded.  The times of superhump maxima
are listed in table \ref{tab:j0647oc2016}.
The 2016 observations, which were obtained in poorer
conditions than in the 2013 observations, likely
resulted a mixture of stages B and C
(figure \ref{fig:j0647comp}).  Due to the limited
number of superhumps maxima, we could not determine
the periods for these stages individually.

\begin{table}
\caption{Superhump maxima of MASTER J064725 (2016)}\label{tab:j0647oc2016}
\begin{center}
\begin{tabular}{rp{55pt}p{40pt}r@{.}lr}
\hline
\multicolumn{1}{c}{$E$} & \multicolumn{1}{c}{max\commenta} & \multicolumn{1}{c}{error} & \multicolumn{2}{c}{$O-C$\commentb} & \multicolumn{1}{c}{$N$\commentc} \\
\hline
0 & 57739.9678 & 0.0012 & $-$0&0017 & 39 \\
1 & 57740.0395 & 0.0007 & 0&0024 & 74 \\
2 & 57740.1070 & 0.0007 & 0&0023 & 76 \\
3 & 57740.1738 & 0.0008 & 0&0015 & 76 \\
4 & 57740.2419 & 0.0008 & 0&0021 & 75 \\
5 & 57740.3076 & 0.0008 & 0&0002 & 76 \\
6 & 57740.3771 & 0.0054 & 0&0021 & 25 \\
15 & 57740.9792 & 0.0024 & $-$0&0040 & 58 \\
16 & 57741.0482 & 0.0020 & $-$0&0026 & 65 \\
17 & 57741.1131 & 0.0025 & $-$0&0053 & 67 \\
65 & 57744.3666 & 0.0050 & 0&0042 & 49 \\
92 & 57746.1916 & 0.0017 & 0&0044 & 21 \\
93 & 57746.2522 & 0.0013 & $-$0&0026 & 25 \\
94 & 57746.3200 & 0.0030 & $-$0&0024 & 20 \\
108 & 57747.2681 & 0.0034 & $-$0&0004 & 13 \\
\hline
  \multicolumn{6}{l}{\commenta BJD$-$2400000.} \\
  \multicolumn{6}{l}{\commentb Against max $= 2457739.9695 + 0.067584 E$.} \\
  \multicolumn{6}{l}{\commentc Number of points used to determine the maximum.} \\
\end{tabular}
\end{center}
\end{table}

\subsection{MASTER OT J065330.46$+$251150.9}\label{obj:j0653}

\begin{figure}
  \begin{center}
    \FigureFile(85mm,110mm){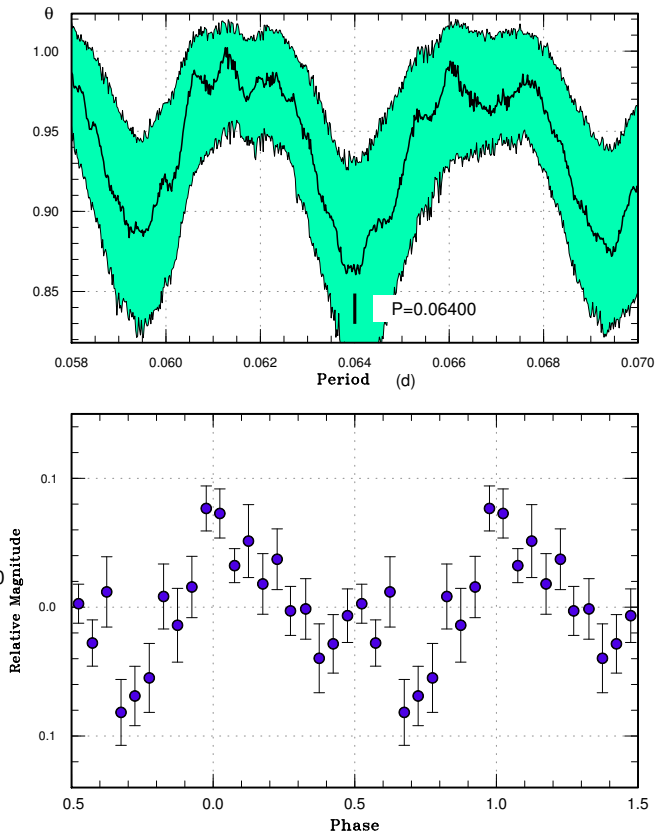}
  \end{center}
  \caption{Superhumps in MASTER J065330 (2017).
     (Upper): PDM analysis.
     (Lower): Phase-averaged profile.}
  \label{fig:j0653shpdm}
\end{figure}

   This object (hereafter MASTER J065330) was discovered
as a transient at an unfiltered CCD magnitude of
15.9 mag on 2014 February 16 by the MASTER network
\citep{ech14j0653atel5898}.
The 2017 outburst was detected by the ASAS-SN team
at $V$=15.89 on January 23.
Subsequent observations detected superhumps
(vsnet-alert 20612; figure \ref{fig:j0653shpdm}).
The times of superhump maxima are listed in
table \ref{tab:j0653oc2017}.
According to the ASAS-SN data, there was another
long outburst (superoutburst) on 2015 September 17.


\begin{table}
\caption{Superhump maxima of MASTER J065330 (2017)}\label{tab:j0653oc2017}
\begin{center}
\begin{tabular}{rp{55pt}p{40pt}r@{.}lr}
\hline
\multicolumn{1}{c}{$E$} & \multicolumn{1}{c}{max\commenta} & \multicolumn{1}{c}{error} & \multicolumn{2}{c}{$O-C$\commentb} & \multicolumn{1}{c}{$N$\commentc} \\
\hline
0 & 57780.2944 & 0.0008 & 0&0004 & 64 \\
1 & 57780.3575 & 0.0010 & $-$0&0006 & 54 \\
12 & 57781.0643 & 0.0128 & 0&0021 & 40 \\
13 & 57781.1244 & 0.0027 & $-$0&0018 & 67 \\
\hline
  \multicolumn{6}{l}{\commenta BJD$-$2400000.} \\
  \multicolumn{6}{l}{\commentb Against max $= 2457780.2941 + 0.064012 E$.} \\
  \multicolumn{6}{l}{\commentc Number of points used to determine the maximum.} \\
\end{tabular}
\end{center}
\end{table}

\subsection{MASTER OT J075450.18$+$091020.2}\label{obj:j0754}

\begin{figure}
  \begin{center}
    \FigureFile(85mm,110mm){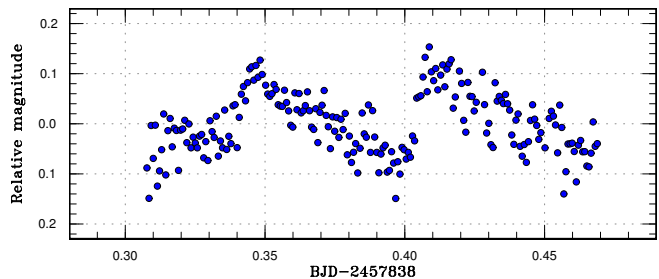}
  \end{center}
  \caption{Superhump in MASTER J075450 (2017).
  }
  \label{fig:j0754shlc}
\end{figure}

   This object (hereafter MASTER J075450) was discovered
as a transient at an unfiltered CCD magnitude of
16.0 mag on 2013 November 7 by the MASTER network
\citep{vla13j0754atel5585}.
The 2017 outburst was detected by the ASAS-SN team
at $V$=16.36 on March 24.  The object was then found
to be already in outburst at $V$=16.39 on March 22.
Observations on March 25--26 detected superhumps
(vsnet-alert 20821; figure \ref{fig:j0754shlc}).
The times of superhump maxima were BJD 2457838.3515(8) ($N$=71)
and 2457838.4172(8) ($N$=68).
The superhump period determined by the PDM method
was 0.0664(5)~d.
There was also a most likely superoutburst in the ASAS-SN
data on 2015 January 20 with a maximum of $V$=16.08.

\subsection{MASTER OT J150518.03$-$143933.6}\label{obj:j1505}

   This object (hereafter MASTER J150518) was discovered
as a transient at an unfiltered CCD magnitude of
15.5 mag on 2017 February 8 by the MASTER network
\citep{gre17j1505atel10061}.  This transient was
also detected by the ASAS-SN team (ASASSN-17cb)
at $V$=15.4 on the same night, but the announcement
was made after confirmation at $V$=15.1 on February 11.
Although only the late course of the outburst was
observed, superhumps were recorded (vsnet-alert 20660).
Since observations only recorded one superhump maximum
on each night, the one-day alias could not be resolved.
We selected one of them to minimize the $\theta$ of
the PDM analysis to make cycle counts in table
\ref{tab:j1505oc2017}.  Although the large negative
global $P_{\rm dot}$ may have reflected stage B-C
transition, the quality of the data were insufficient
to confirm it.


\begin{figure}
  \begin{center}
    \FigureFile(85mm,110mm){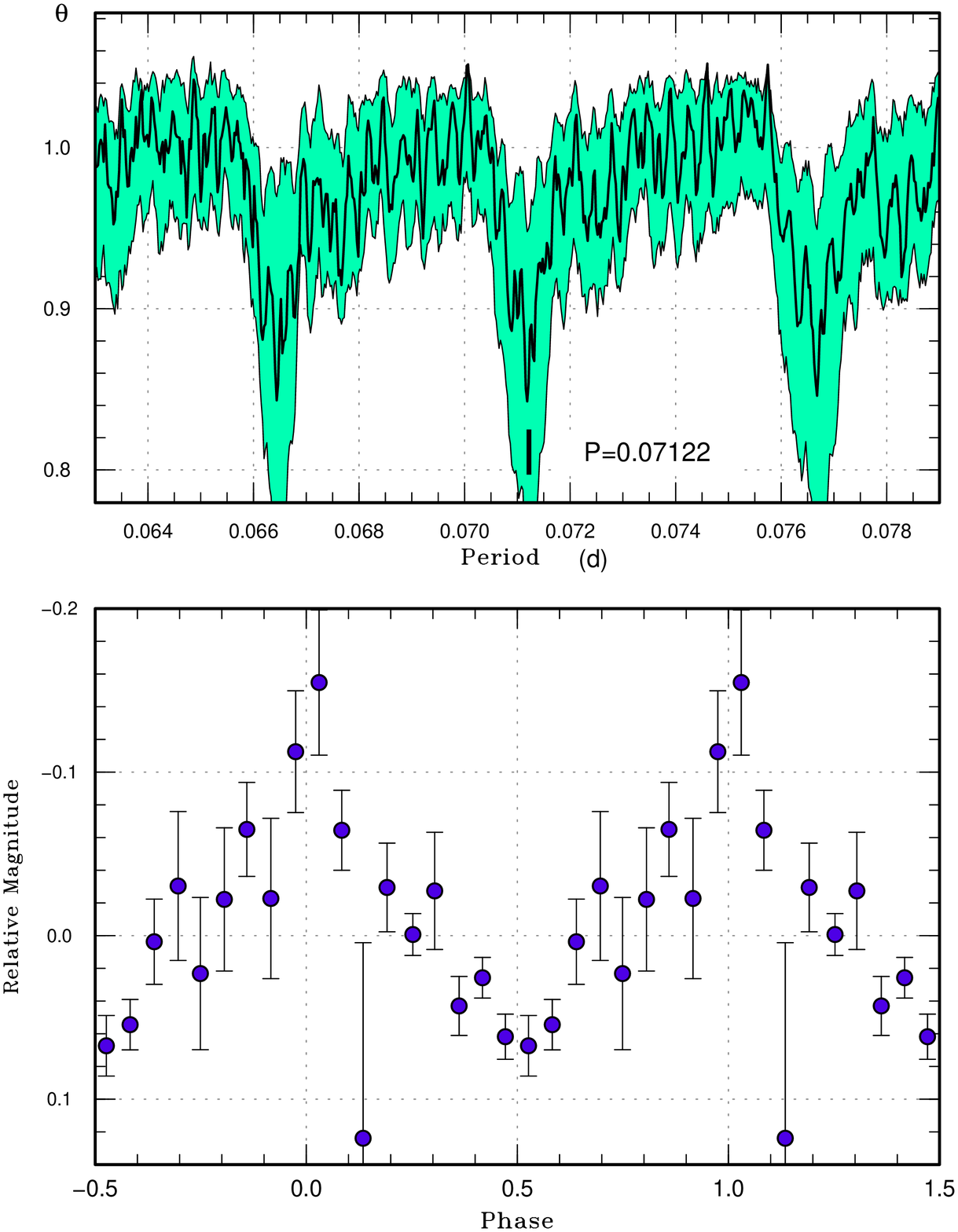}
  \end{center}
  \caption{Superhumps in MASTER J150518 (2017).
     (Upper): PDM analysis.
     (Lower): Phase-averaged profile.}
  \label{fig:j1505shpdm}
\end{figure}


\begin{table}
\caption{Superhump maxima of MASTER J150518 (2017)}\label{tab:j1505oc2017}
\begin{center}
\begin{tabular}{rp{55pt}p{40pt}r@{.}lr}
\hline
\multicolumn{1}{c}{$E$} & \multicolumn{1}{c}{max\commenta} & \multicolumn{1}{c}{error} & \multicolumn{2}{c}{$O-C$\commentb} & \multicolumn{1}{c}{$N$\commentc} \\
\hline
0 & 57797.8081 & 0.0013 & $-$0&0040 & 18 \\
14 & 57798.8103 & 0.0011 & 0&0022 & 31 \\
28 & 57799.8090 & 0.0026 & 0&0048 & 30 \\
56 & 57801.7933 & 0.0038 & $-$0&0030 & 23 \\
\hline
  \multicolumn{6}{l}{\commenta BJD$-$2400000.} \\
  \multicolumn{6}{l}{\commentb Against max $= 2457797.8121 + 0.071145 E$.} \\
  \multicolumn{6}{l}{\commentc Number of points used to determine the maximum.} \\
\end{tabular}
\end{center}
\end{table}

\subsection{MASTER OT J151126.74$-$400751.9}\label{obj:j1511}

\begin{figure}
  \begin{center}
    \FigureFile(85mm,110mm){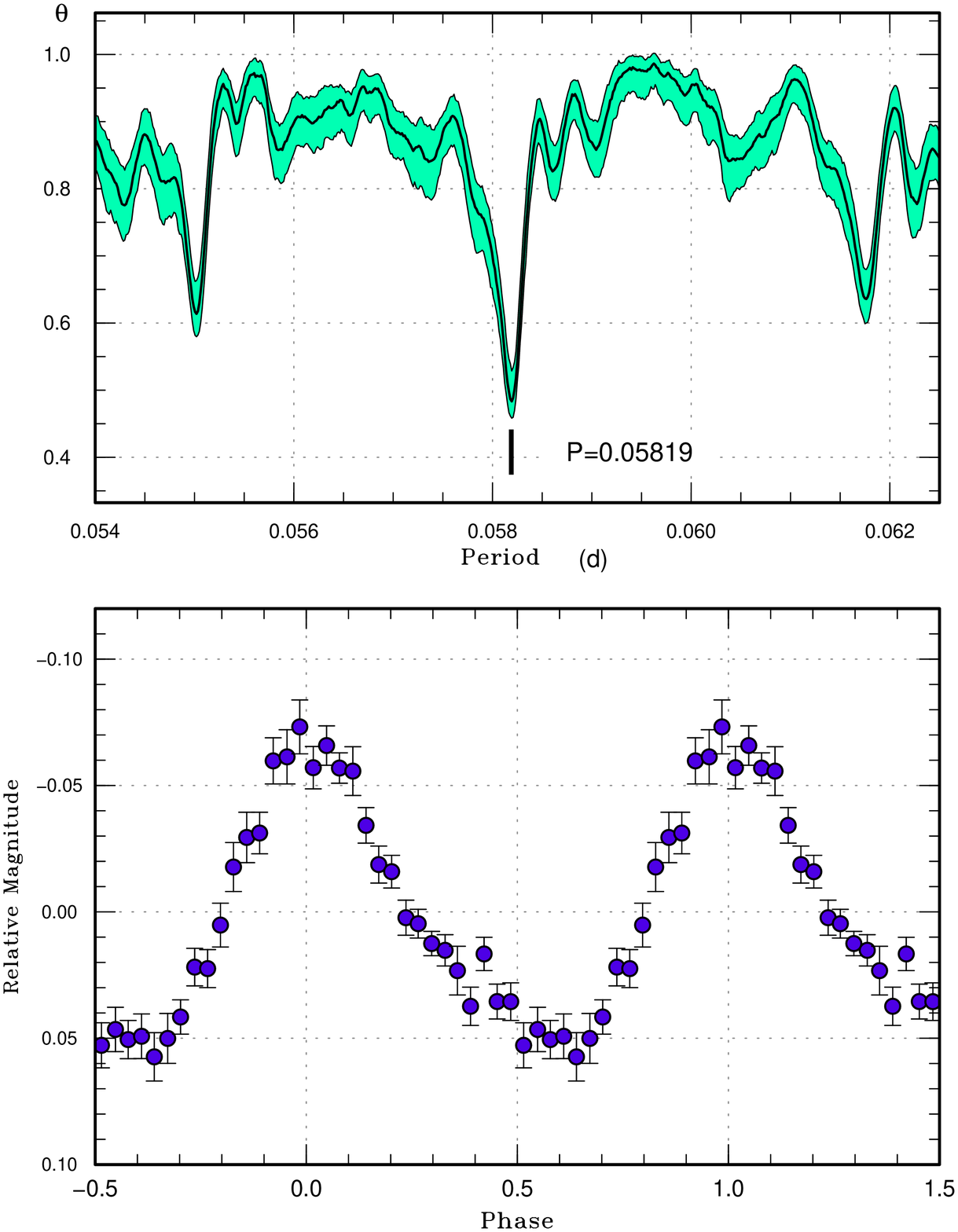}
  \end{center}
  \caption{Superhumps in MASTER J151126 (2016).
     (Upper): PDM analysis.
     The data during the superoutburst plateau
     (before BJD 2457480) were used.
     (Lower): Phase-averaged profile.}
  \label{fig:j1511shpdm}
\end{figure}

   This object (hereafter MASTER J151126) was discovered
as a transient at an unfiltered CCD magnitude of
14.0 mag on 2016 March 18 by the MASTER network
\citep{pop16j1511atel8843}.
Subsequent observations detected superhumps
(vsnet-alert 19614, 19630; figure \ref{fig:j1511shpdm}).
The times of superhump maxima are listed in
table \ref{tab:j1511oc2016}.
We interpreted that most of our observations
recorded stage B as judged from a positive $P_{\rm dot}$
expected for this $P_{\rm SH}$.
The outburst faded on April 3.  The duration of
the outburst was at least 16~d.


\begin{table}
\caption{Superhump maxima of MASTER J151126 (2016)}\label{tab:j1511oc2016}
\begin{center}
\begin{tabular}{rp{55pt}p{40pt}r@{.}lr}
\hline
\multicolumn{1}{c}{$E$} & \multicolumn{1}{c}{max\commenta} & \multicolumn{1}{c}{error} & \multicolumn{2}{c}{$O-C$\commentb} & \multicolumn{1}{c}{$N$\commentc} \\
\hline
0 & 57468.8045 & 0.0006 & $-$0&0032 & 14 \\
16 & 57469.7426 & 0.0006 & 0&0039 & 28 \\
17 & 57469.8009 & 0.0007 & 0&0039 & 19 \\
30 & 57470.5560 & 0.0011 & 0&0027 & 72 \\
31 & 57470.6139 & 0.0003 & 0&0023 & 134 \\
32 & 57470.6697 & 0.0016 & $-$0&0001 & 51 \\
51 & 57471.7718 & 0.0009 & $-$0&0036 & 25 \\
84 & 57473.6907 & 0.0011 & $-$0&0049 & 30 \\
85 & 57473.7514 & 0.0011 & $-$0&0023 & 30 \\
102 & 57474.7389 & 0.0009 & $-$0&0040 & 18 \\
153 & 57477.7113 & 0.0014 & 0&0008 & 32 \\
154 & 57477.7670 & 0.0056 & $-$0&0018 & 18 \\
170 & 57478.7027 & 0.0011 & 0&0029 & 32 \\
171 & 57478.7614 & 0.0016 & 0&0035 & 19 \\
\hline
  \multicolumn{6}{l}{\commenta BJD$-$2400000.} \\
  \multicolumn{6}{l}{\commentb Against max $= 2457468.8077 + 0.058188 E$.} \\
  \multicolumn{6}{l}{\commentc Number of points used to determine the maximum.} \\
\end{tabular}
\end{center}
\end{table}

\subsection{MASTER OT J162323.48$+$782603.3}\label{obj:j1623}

   This object (hereafter MASTER J162323) was detected
as a transient at an unfiltered CCD magnitude of
13.2 mag on 2013 December 9 by the MASTER network
\citep{den13j1623atel5643}.
The 2013 superoutburst was well observed \citep{Pdot6}.

\begin{figure}
  \begin{center}
    \FigureFile(85mm,110mm){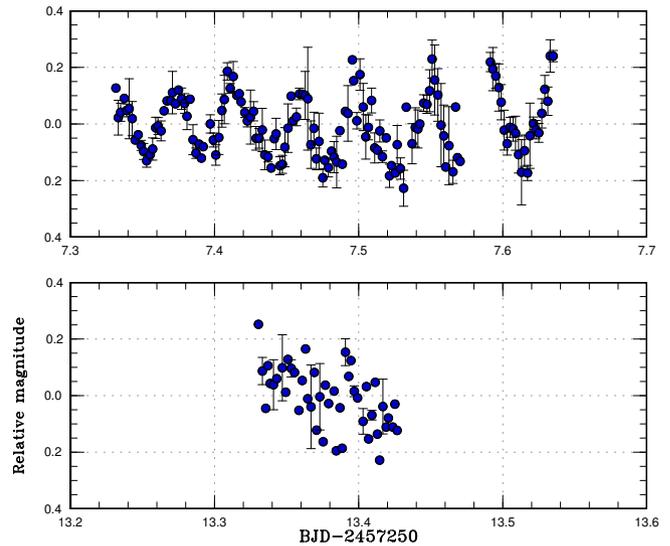}
  \end{center}
  \caption{Superhump-like double wave modulations in
  MASTER J162323 (2015).  The data were binned to 0.002~d.
  The variations were only recorded on BJD 2457255
  (August 22) and they disappeared 6~d later.
  }
  \label{fig:j1623shlc2015}
\end{figure}

   The 2015 superoutburst was detected at $V$=13.49
on August 10 by the ASAS-SN team.  Double-wave modulations
were recorded on August 22 (vsnet-alert 19004).
These variations disappeared 6~d later
(see figure \ref{fig:j1623shlc2015}).
Since these observations covered only the last
(and likely post-superoutburst) phase of
the superoutburst and the nature of humps is
unclear, we did not use these data for comparison
with other superoutbursts.

\begin{figure}
  \begin{center}
    \FigureFile(88mm,70mm){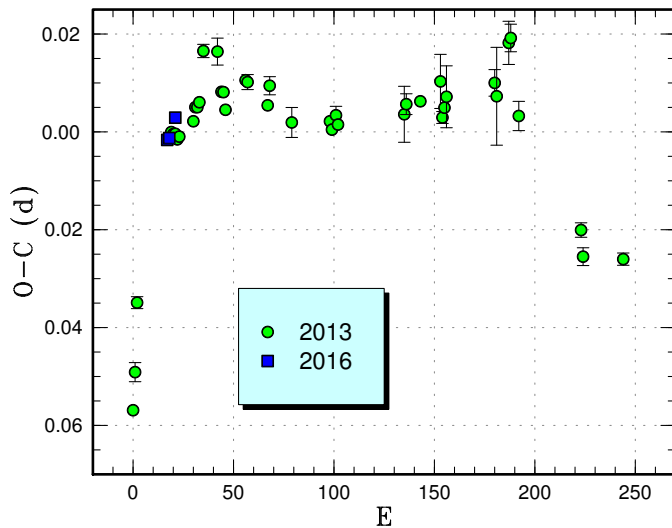}
  \end{center}
  \caption{Comparison of $O-C$ diagrams of MASTER J162323
  between different superoutbursts.
  A period of 0.08866~d was used to draw this figure.
  Approximate cycle counts ($E$) after the start of the superoutburst
  were used.  We shifted the 2016 $O-C$ diagram
  by 25 cycles to match the well-observed 2015 one.
  }
  \label{fig:j1623comp}
\end{figure}

   There was an outburst at $V$=13.46 on 2016 April 22
(ASAS-SN detection).  Subsequent observations did not
detect superhumps (observers: Shugarov team and Akazawa).
There was another outburst at $V$=13.34 on 2016 September 26
(ASAS-SN detection).  Three superhumps were recorded
during this superoutburst (table \ref{tab:j1623oc2016}).
These superhumps were likely obtained around transition
from stage A to B (figure \ref{fig:j1623comp})
and the period [0.09013(7)~d, PDM method] is not
listed in table \ref{tab:perlist}.

\begin{table}
\caption{Superhump maxima of MASTER J162323 (2016)}\label{tab:j1623oc2016}
\begin{center}
\begin{tabular}{rp{55pt}p{40pt}r@{.}lr}
\hline
\multicolumn{1}{c}{$E$} & \multicolumn{1}{c}{max\commenta} & \multicolumn{1}{c}{error} & \multicolumn{2}{c}{$O-C$\commentb} & \multicolumn{1}{c}{$N$\commentc} \\
\hline
0 & 57659.3470 & 0.0001 & 0&0003 & 146 \\
1 & 57659.4361 & 0.0003 & $-$0&0004 & 77 \\
4 & 57659.7062 & 0.0003 & 0&0001 & 174 \\
\hline
  \multicolumn{6}{l}{\commenta BJD$-$2400000.} \\
  \multicolumn{6}{l}{\commentb Against max $= 2457659.3467 + 0.089866 E$.} \\
  \multicolumn{6}{l}{\commentc Number of points used to determine the maximum.} \\
\end{tabular}
\end{center}
\end{table}

\subsection{MASTER OT J165153.86$+$702525.7}\label{obj:j1651}

   This object (hereafter MASTER J165153) was detected
as a transient at an unfiltered CCD magnitude of 15.9
on 2013 May 23 by the MASTER network \citep{shu13asassn13akatel5083}.

\begin{figure}
  \begin{center}
    \FigureFile(85mm,110mm){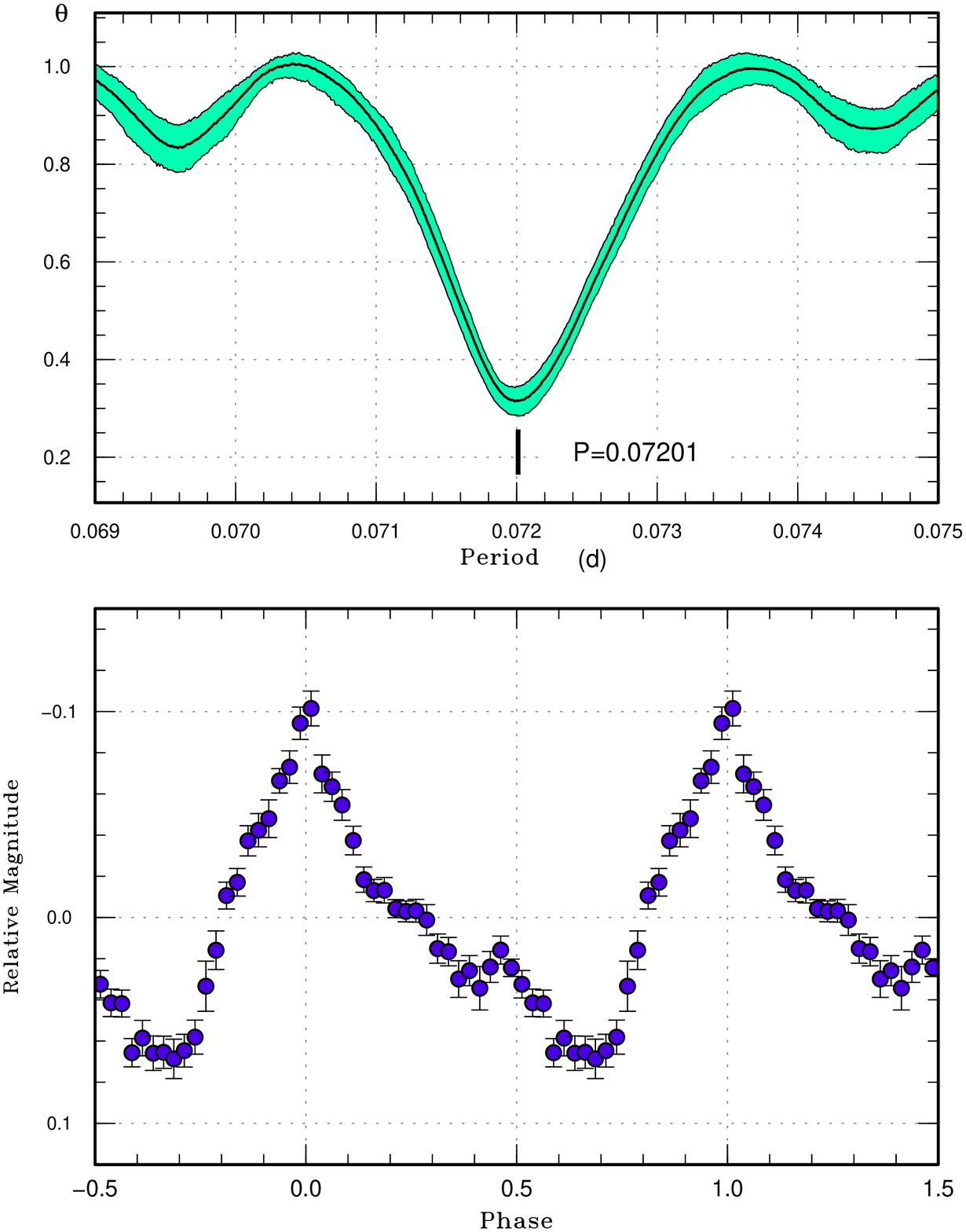}
  \end{center}
  \caption{Superhumps in MASTER J165153 (2017).
     (Upper): PDM analysis.
     (Lower): Phase-averaged profile.}
  \label{fig:j1651shpdm}
\end{figure}

   The 2017 outburst was detected by the ASAS-SN team
at $V$=14.64 on February 4.  Subsequent observations
detected superhumps (vsnet-alert 20659;
figure \ref{fig:j1651shpdm}).
The times of superhump maxima are listed in
table \ref{tab:j1651oc2017}.


\begin{table}
\caption{Superhump maxima of MASTER J165153 (2017)}\label{tab:j1651oc2017}
\begin{center}
\begin{tabular}{rp{55pt}p{40pt}r@{.}lr}
\hline
\multicolumn{1}{c}{$E$} & \multicolumn{1}{c}{max\commenta} & \multicolumn{1}{c}{error} & \multicolumn{2}{c}{$O-C$\commentb} & \multicolumn{1}{c}{$N$\commentc} \\
\hline
0 & 57797.4370 & 0.0007 & $-$0&0015 & 41 \\
1 & 57797.5098 & 0.0004 & $-$0&0007 & 65 \\
2 & 57797.5817 & 0.0004 & $-$0&0007 & 62 \\
16 & 57798.5899 & 0.0009 & 0&0002 & 68 \\
17 & 57798.6654 & 0.0007 & 0&0037 & 66 \\
18 & 57798.7371 & 0.0022 & 0&0034 & 34 \\
29 & 57799.5199 & 0.0044 & $-$0&0052 & 24 \\
30 & 57799.5975 & 0.0005 & 0&0005 & 71 \\
31 & 57799.6694 & 0.0006 & 0&0004 & 71 \\
\hline
  \multicolumn{6}{l}{\commenta BJD$-$2400000.} \\
  \multicolumn{6}{l}{\commentb Against max $= 2457797.4385 + 0.071951 E$.} \\
  \multicolumn{6}{l}{\commentc Number of points used to determine the maximum.} \\
\end{tabular}
\end{center}
\end{table}

\subsection{MASTER OT J174816.22$+$501723.3}\label{obj:j1748}

   This object (hereafter MASTER J174816) was discovered
as a transient at an unfiltered CCD magnitude of
15.6 mag on 2013 June 28 by the MASTER network
\citep{den13j1748atel5182}.  The object has a blue
SDSS counterpart ($g$=17.59).  At least nine outbursts
were recorded in the CRTS data.  The object has
a bright ($J$=15.55) 2MASS counterpart, suggesting
that the object was in outburst during 2MASS scans.

\begin{figure}
  \begin{center}
    \FigureFile(85mm,110mm){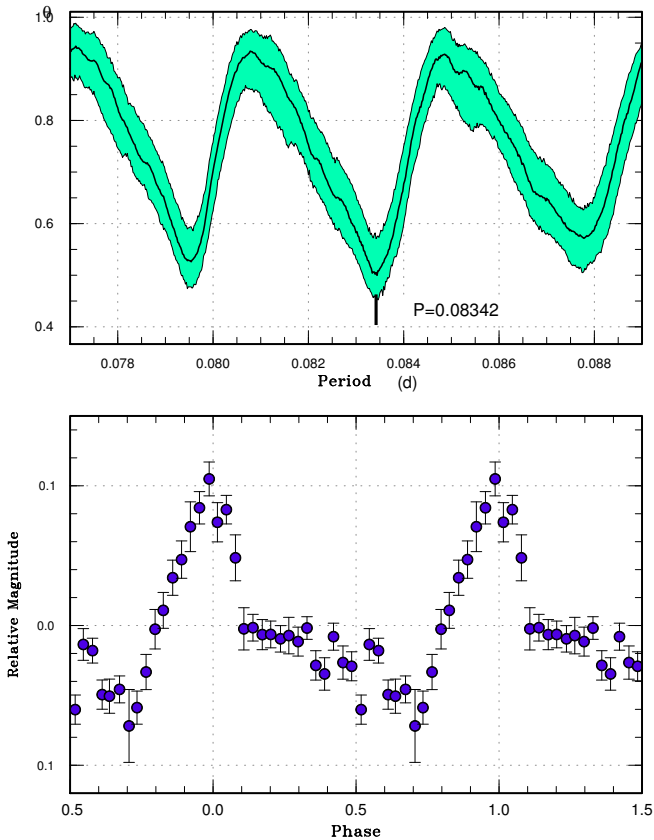}
  \end{center}
  \caption{Superhumps in MASTER J174816 (2016).
     (Upper): PDM analysis.
     (Lower): Phase-averaged profile.}
  \label{fig:j1748shpdm}
\end{figure}

   The 2016 outburst was detected by the ASAS-SN team
at $V$=15.50 on March 25.  Subsequent observations
detected superhumps (vsnet-alert 19642;
figure \ref{fig:j1748shpdm}).
The times of superhump maxima are listed in
table \ref{tab:j1748oc2016}.
Although we adopted a period of 0.08342(4)~d
(PDM method), an alias of 0.07950(4)~d
could not be excluded.


\begin{table}
\caption{Superhump maxima of MASTER J174816 (2016)}\label{tab:j1748oc2016}
\begin{center}
\begin{tabular}{rp{55pt}p{40pt}r@{.}lr}
\hline
\multicolumn{1}{c}{$E$} & \multicolumn{1}{c}{max\commenta} & \multicolumn{1}{c}{error} & \multicolumn{2}{c}{$O-C$\commentb} & \multicolumn{1}{c}{$N$\commentc} \\
\hline
0 & 57473.5497 & 0.0012 & 0&0012 & 83 \\
1 & 57473.6305 & 0.0006 & $-$0&0014 & 84 \\
20 & 57475.2172 & 0.0016 & 0&0021 & 57 \\
21 & 57475.2965 & 0.0017 & $-$0&0019 & 73 \\
\hline
  \multicolumn{6}{l}{\commenta BJD$-$2400000.} \\
  \multicolumn{6}{l}{\commentb Against max $= 2457473.5485 + 0.083328 E$.} \\
  \multicolumn{6}{l}{\commentc Number of points used to determine the maximum.} \\
\end{tabular}
\end{center}
\end{table}

\subsection{MASTER OT J211322.92$+$260647.4}\label{obj:j2113}

\begin{figure}
  \begin{center}
    \FigureFile(85mm,110mm){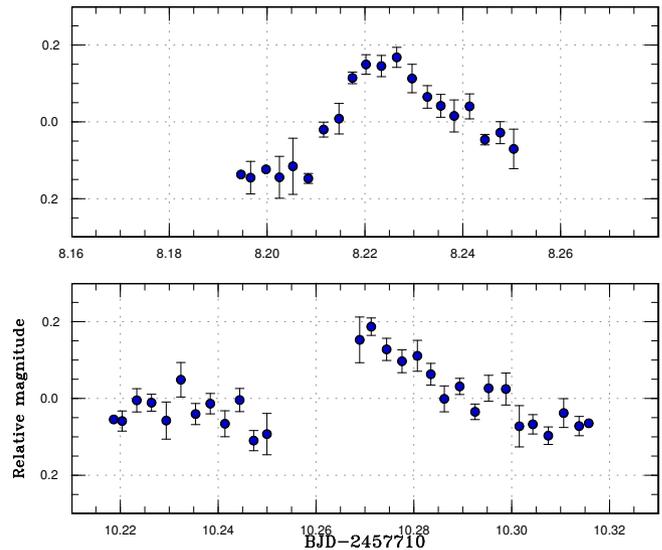}
  \end{center}
  \caption{Superhump in MASTER J211322 (2016).
  }
  \label{fig:j2113shlc}
\end{figure}

   This object (hereafter MASTER J211322) was discovered
by the MASTER network at an unfiltered CCD magnitude
of 15.2 on 2012 December 21 \citep{shu12j2113atel4675}.
The 2016 outburst was detected by the ASAS-SN team
at $V$=15.53 on November 24.
There was another bright outburst reaching $V$=14.91
on 2015 May 29 according to the ASAS-SN data.
Subsequent observations detected superhumps
(vsnet-alert 20453; figure \ref{fig:j2113shlc}).
Although two superhump maxima
were measured to be BJD 2457718.2276(8) ($N$=73) and
2457720.2718(14) ($N$=66, the maximum was missed
and was estimated by template fitting),
only one superhump maximum was recorded on each night
and the period is 2.04(1)$/n$, where $n$ is an integer.

\subsection{MASTER OT J220559.40$-$341434.9}\label{obj:j2205}

\begin{figure}
  \begin{center}
    \FigureFile(85mm,110mm){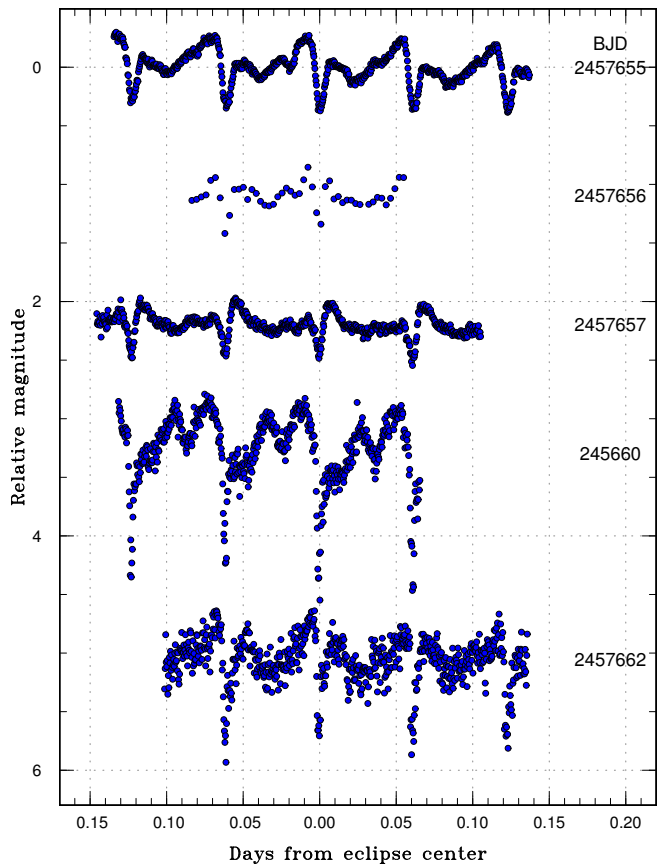}
  \end{center}
  \caption{Eclipses and superhumps in MASTER J220559.
  }
  \label{fig:j2205lc}
\end{figure}

\begin{figure}
  \begin{center}
    \FigureFile(85mm,110mm){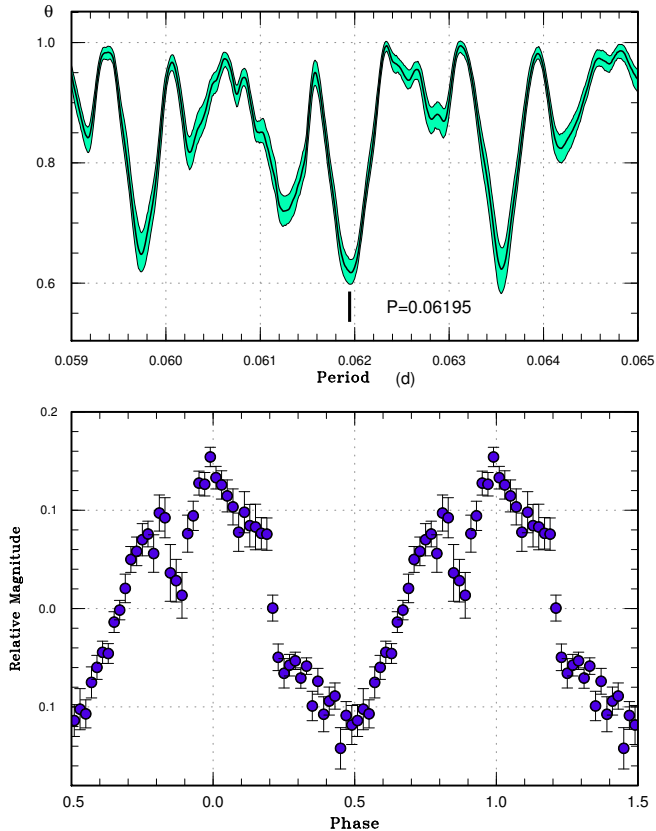}
  \end{center}
  \caption{Superhumps in MASTER J220559 (2016).
     (Upper): PDM analysis.
     The signal at 0.06129~d is the orbital period.
     Other signals are aliases and false ones produced
     in combination with orbital variations.
     (Lower): Phase-averaged profile.}
  \label{fig:j2205shpdm}
\end{figure}

   This object (hereafter MASTER J220559) was discovered
by the MASTER network at an unfiltered CCD magnitude
of 14.5 on 2016 September 19 (\cite{pog16j2205atel9509};
correction in \cite{pog16j2205atel9510}).
The object was also detected by the ASAS-SN team
(ASASSN-16kr) at $V$=14.3 on September 11.  The ASAS-SN
detection was announced after the object brightened
to $V$=13.9 on September 22, 2~d after the MASTER
announcement.
Although the object was initially considered to be
an SS Cyg-type object from the low outburst amplitude
(cf. vsnet-alert 20189), time-resolved photometry
detected superhumps and eclipses
(vsnet-alert 20190, 20196, 20206;
figure \ref{fig:j2205lc},
figure \ref{fig:j2205shpdm}).

   We obtained the eclipse ephemeris
using the MCMC analysis \citep{Pdot4}:
\begin{equation}
{\rm Min(BJD)} = 2457658.72016(3) + 0.0612858(3) E .
\label{equ:j2205ecl}
\end{equation}
This ephemeris is not intended for long-term prediction
of eclipses.  The epoch refers to the center of
the observation.
The times of superhump maxima outside the eclipses
are listed in table \ref{tab:j2205oc2016}.
Although the $O-C$ diagram suggests stage B-C
transition, the periods and $P_{\rm dot}$ may have
not been well determined since the actual start
of the outburst was much earlier than the detection
announcement and determination of superhump maxima
should have been affected by overlapping eclipses
and orbital humps in the late epochs.
A large positive $P_{\rm dot}$, however, is
usual for such a short-$P_{\rm orb}$ SU UMa-type
dwarf nova.

   The small outburst amplitude ($\sim$4.5 mag)
was probably a result of the high orbital inclination.
Since the object has deep eclipses and apparently
shows a large positive $P_{\rm dot}$, it surely
deserves further detailed observations to
clarify the origin of increasing $P_{\rm SH}$
during stage B.  Observations of the early phase
of a superoutburst are also desired to determine $q$
by the stage A superhump method.

\begin{table}
\caption{Superhump maxima of MASTER J220559 (2016)}\label{tab:j2205oc2016}
\begin{center}
\begin{tabular}{rp{50pt}p{30pt}r@{.}lcr}
\hline
$E$ & max\commenta & error & \multicolumn{2}{c}{$O-C$\commentb} & phase\commentc & $N$\commentd \\
\hline
0 & 57655.2787 & 0.0007 & 0&0037 & 0.85 & 48 \\
1 & 57655.3385 & 0.0004 & 0&0016 & 0.82 & 111 \\
2 & 57655.4032 & 0.0006 & 0&0043 & 0.88 & 111 \\
3 & 57655.4630 & 0.0004 & 0&0022 & 0.85 & 109 \\
4 & 57655.5247 & 0.0003 & 0&0020 & 0.86 & 105 \\
23 & 57656.6919 & 0.0026 & $-$0&0077 & 0.90 & 13 \\
32 & 57657.2503 & 0.0006 & $-$0&0067 & 0.02 & 108 \\
33 & 57657.3144 & 0.0006 & $-$0&0046 & 0.06 & 111 \\
34 & 57657.3758 & 0.0005 & $-$0&0051 & 0.06 & 110 \\
35 & 57657.4381 & 0.0005 & $-$0&0048 & 0.08 & 111 \\
71 & 57659.6648 & 0.0034 & $-$0&0079 & 0.41 & 17 \\
72 & 57659.7253 & 0.0032 & $-$0&0093 & 0.40 & 16 \\
81 & 57660.3019 & 0.0022 & 0&0099 & 0.81 & 111 \\
82 & 57660.3663 & 0.0023 & 0&0123 & 0.86 & 111 \\
83 & 57660.4333 & 0.0013 & 0&0174 & 0.95 & 88 \\
87 & 57660.6702 & 0.0019 & 0&0065 & 0.82 & 13 \\
88 & 57660.7305 & 0.0028 & 0&0049 & 0.80 & 14 \\
113 & 57662.2699 & 0.0018 & $-$0&0041 & 0.92 & 111 \\
114 & 57662.3323 & 0.0007 & $-$0&0037 & 0.94 & 110 \\
115 & 57662.3961 & 0.0008 & $-$0&0018 & 0.98 & 110 \\
116 & 57662.4507 & 0.0018 & $-$0&0092 & 0.87 & 100 \\
\hline
  \multicolumn{7}{l}{\commenta BJD$-$2400000.} \\
  \multicolumn{7}{l}{\commentb Against max $= 2457655.2750 + 0.061939 E$.} \\
  \multicolumn{7}{l}{\commentc Orbital phase.} \\
  \multicolumn{7}{l}{\commentd Number of points used to determine the maximum.} \\
\end{tabular}
\end{center}
\end{table}

\subsection{SBS 1108$+$574}\label{obj:sbs1108}

   This object (hereafter SBS 1108) was originally selected as an
ultraviolet-excess object during the course of 
the Second Byurakan Survey (SBS, \cite{mar83SBS1}).
An outburst detected by CRTS on 2012 April 22
(=CSS120422:111127$+$571239) led to an identification
as an SU UMa-type dwarf nova having a period
below the period minimum (cf. \cite{Pdot4}).
\citet{lit13sbs1108} studied this object by spectroscopy
and found He~I emission of comparable strength to 
the Balmer lines, indicating a hydrogen abundance less than
0.1 of ordinary hydrogen-rich CVs but still at least
10 times higher than that in AM CVn stars.
The object received special attention since it is
considered to be a candidate progenitor of an AM CVn system
(also known as EI Psc-type objects) \citep{lit13sbs1108}.

   The 2016 outburst was detected by the ASAS-SN team
at $V$=15.44 on March 17.  Although subsequent observations
detected superhumps (vsnet-alert 19615, 19674),
the 2016 outburst was not as well observed as in
2012 and superhumps were detected only on
two nights (table \ref{tab:sbs1108oc2016}).
We could not make a comparison of $O-C$ diagrams
between the 2012 and 2016 observations due to
the insufficiency of observations in 2016.


\begin{table}
\caption{Superhump maxima of SBS 1108 (2016)}\label{tab:sbs1108oc2016}
\begin{center}
\begin{tabular}{rp{55pt}p{40pt}r@{.}lr}
\hline
\multicolumn{1}{c}{$E$} & \multicolumn{1}{c}{max\commenta} & \multicolumn{1}{c}{error} & \multicolumn{2}{c}{$O-C$\commentb} & \multicolumn{1}{c}{$N$\commentc} \\
\hline
0 & 57466.3045 & 0.0016 & $-$0&0007 & 43 \\
1 & 57466.3458 & 0.0015 & 0&0015 & 43 \\
2 & 57466.3832 & 0.0010 & $-$0&0001 & 43 \\
3 & 57466.4223 & 0.0010 & $-$0&0000 & 44 \\
4 & 57466.4608 & 0.0008 & $-$0&0007 & 37 \\
69 & 57468.9995 & 0.0020 & $-$0&0002 & 34 \\
70 & 57469.0385 & 0.0014 & $-$0&0003 & 125 \\
71 & 57469.0774 & 0.0020 & $-$0&0004 & 85 \\
72 & 57469.1178 & 0.0031 & 0&0009 & 57 \\
\hline
  \multicolumn{6}{l}{\commenta BJD$-$2400000.} \\
  \multicolumn{6}{l}{\commentb Against max $= 2457466.3052 + 0.039051 E$.} \\
  \multicolumn{6}{l}{\commentc Number of points used to determine the maximum.} \\
\end{tabular}
\end{center}
\end{table}

\subsection{SDSS J032015.29$+$441059.3}\label{obj:j0320}

   This object (hereafter SDSS J032015) was originally
selected as a CV by \citet{wil10newCVs} based on
SDSS variability.  The SDSS colors suggested an object
below the period gap \citep{kat12DNSDSS}.

\begin{figure}
  \begin{center}
    \FigureFile(85mm,110mm){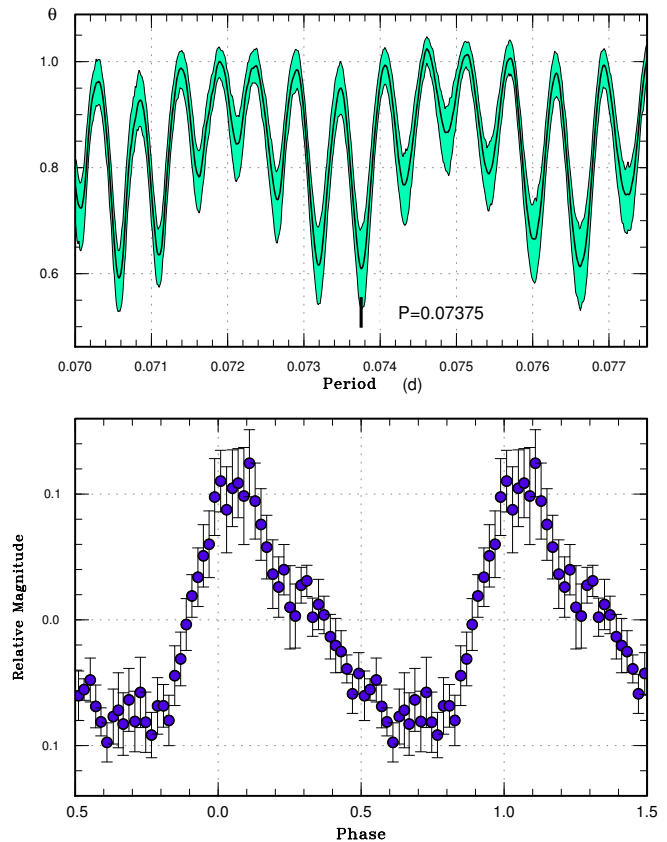}
  \end{center}
  \caption{Superhumps in SDSS J032015 (2016).
     (Upper): PDM analysis.  The alias selection
     was based on $O-C$ analysis and the single-night
     determination by I. Miller (vsnet-alert 20209).
     (Lower): Phase-averaged profile.}
  \label{fig:j0320shpdm}
\end{figure}

   The 2016 outburst was detected by the ASAS-SN team
at $V$=14.84 on September 19.
Subsequent observations detected
superhumps (vsnet-alert 20209; figure \ref{fig:j0320shpdm}).
The times of superhump maxima are listed in
table \ref{tab:j0320oc2016}.
Since the observations were obtained during the final
part of the superoutburst, these superhumps probably
consisted of both stage B and C ones.

   Although there were single-night observations
by C. Littlefield on 2014 October 3, the nature
of this outburst and detected variations were
unknown (cf. vsnet-alert 17801, 17818).


\begin{table}
\caption{Superhump maxima of SDSS J032015 (2016)}\label{tab:j0320oc2016}
\begin{center}
\begin{tabular}{rp{55pt}p{40pt}r@{.}lr}
\hline
\multicolumn{1}{c}{$E$} & \multicolumn{1}{c}{max\commenta} & \multicolumn{1}{c}{error} & \multicolumn{2}{c}{$O-C$\commentb} & \multicolumn{1}{c}{$N$\commentc} \\
\hline
0 & 57653.5127 & 0.0031 & $-$0&0045 & 24 \\
1 & 57653.5964 & 0.0004 & 0&0054 & 70 \\
2 & 57653.6674 & 0.0005 & 0&0027 & 51 \\
26 & 57655.4311 & 0.0026 & $-$0&0038 & 32 \\
27 & 57655.5077 & 0.0005 & $-$0&0009 & 73 \\
134 & 57663.4055 & 0.0008 & 0&0049 & 55 \\
136 & 57663.5504 & 0.0012 & 0&0023 & 73 \\
137 & 57663.6156 & 0.0023 & $-$0&0062 & 59 \\
\hline
  \multicolumn{6}{l}{\commenta BJD$-$2400000.} \\
  \multicolumn{6}{l}{\commentb Against max $= 2457653.5172 + 0.073757 E$.} \\
  \multicolumn{6}{l}{\commentc Number of points used to determine the maximum.} \\
\end{tabular}
\end{center}
\end{table}

\subsection{SDSS J091001.63$+$164820.0}\label{obj:j0910}

   This object (hereafter SDSS J091001) was originally
selected as a CV by the SDSS \citep{szk09SDSSCV7}.
The SDSS colors suggested an object below the period
gap \citep{kat12DNSDSS}.
There was an outburst in 2016 February--March
(cf. vsnet-alert 19539), but CCD observations by
T. Vanmunster and Y. Maeda showed that the outburst
was a rapidly fading normal one.

   The 2017 outburst was detected by the ASAS-SN team
at $V$=14.39 on March 25.  The ASAS-SN data indicated
that the outburst started at $V$=16.28 on March 21
and peaked at $V$=14.04 on March 23.
Subsequent observations detected superhumps
(vsnet-alert 20830).
Three superhump maxima were recorded:
BJD 2457839.3524(3) ($N$=50), 2457839.4252(3) ($N$=76)
and 2457839.4980(3) ($N$=73).
The superhump period determined by the PDM method
was 0.0734(2)~d.

\begin{figure}
  \begin{center}
    \FigureFile(85mm,110mm){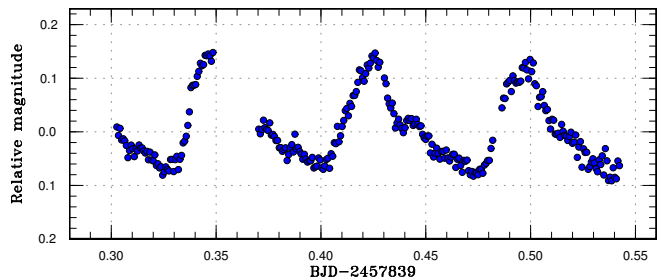}
  \end{center}
  \caption{Superhump in SDSS J091001 (2017).
  }
  \label{fig:j0910shlc}
\end{figure}

\subsection{SDSS J113551.09$+$532246.2}\label{obj:j1135}

   This object (hereafter SDSS J113551) was reported
as an outbursting object discovered by ROTSE-IIIb telescope
at an unfiltered CCD magnitude of 15.1 on 2006 March 24
\citep{qui06j1202atel787}.  \citet{kat12DNSDSS} expected
an orbital period of 0.112(6)~d based on SDSS colors.

   The 2017 outburst was detected by the ASAS-SN team
at $V$=15.34 on February 16.
Although observations on two nights were reported, neither
data were of sufficient quality to determine the superhump period
(due to cloud gaps).
The period used to calculated epochs in
table \ref{tab:j1135oc2017} was one of the possibilities
giving smallest $O-C$ residuals.  Other candidate
aliases were 0.1023(1)~d and 0.0914(1)~d
(table \ref{tab:j1135oc2017}).
In any case, SDSS J113551 is in or close to
the period gap and should be studied further.
According to the ASAS-SN data, there were past
(most likely) superoutbursts on 2012 March 10
($V$=15.53) and 2016 May 15 ($V$=15.68).
There were additional possible ones which were
not well recorded.  The frequency of superoutbursts
was not probably especially low.


\begin{figure}
  \begin{center}
    \FigureFile(85mm,110mm){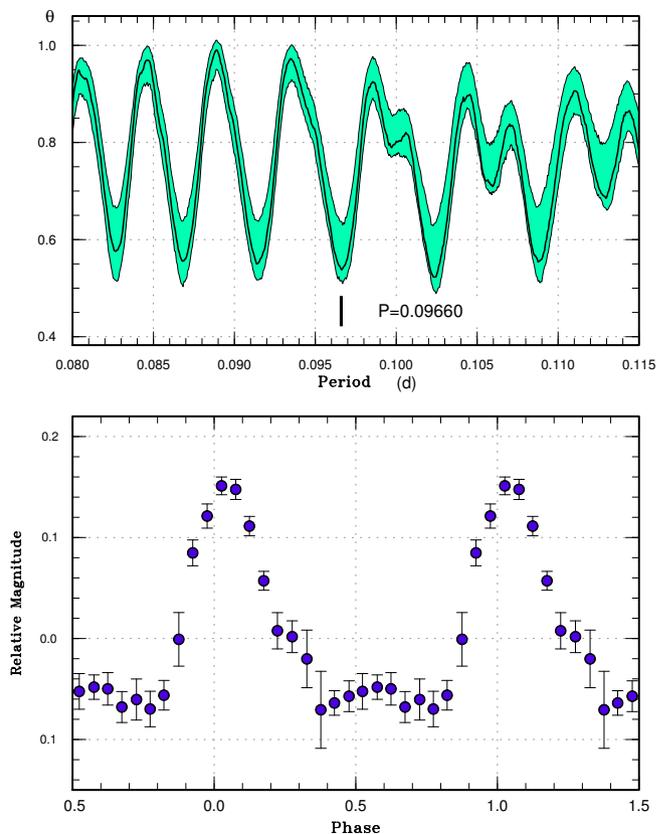}
  \end{center}
  \caption{Superhumps in SDSS J113551 (2017).
     (Upper): PDM analysis.  The selected period was
     one of the possibilities.
     (Lower): Phase-averaged profile.}
  \label{fig:j1135shpdm}
\end{figure}


\begin{table}
\caption{Superhump maxima of SDSS J113551 (2017)}\label{tab:j1135oc2017}
\begin{center}
\begin{tabular}{rp{55pt}p{40pt}r@{.}lr}
\hline
\multicolumn{1}{c}{$E$} & \multicolumn{1}{c}{max\commenta} & \multicolumn{1}{c}{error} & \multicolumn{2}{c}{$O-C$\commentb} & \multicolumn{1}{c}{$N$\commentc} \\
\hline
0 & 57803.4956 & 0.0004 & 0&0021 & 75 \\
1 & 57803.5880 & 0.0007 & $-$0&0023 & 51 \\
18 & 57805.2362 & 0.0007 & 0&0001 & 181 \\
\hline
  \multicolumn{6}{l}{\commenta BJD$-$2400000.} \\
  \multicolumn{6}{l}{\commentb Against max $= 2457803.4935 + 0.096810 E$.} \\
  \multicolumn{6}{l}{\commentc Number of points used to determine the maximum.} \\
\end{tabular}
\end{center}
\end{table}

\subsection{SDSS J115207.00$+$404947.8}\label{obj:j1152}

   This object (hereafter SDSS J115207) was originally
selected as a CV by the SDSS \citep{szk07SDSSCV6}.
Although \citet{szk07SDSSCV6} suspected an eclipsing
system, its nature was established by \citet{sou10SDSSeclCV},
who determined the orbital period of 0.06770(28)~d
and an mass ratio of 0.14(3).  \citet{sav11CVeclmass}
obtained further observations and refined the values
to be 0.067721356(3)~d and 0.155(6), respectively.

   The object was confirmed to be an SU UMa-type
dwarf nova by the detection of superhumps during
the 2009 superoutburst \citep{Pdot2}.  Due to the poor
coverage of the 2009 superoutburst and the limited
knowledge of the orbital period at that time,
we could not determine superhump and orbital periods
precisely in \citet{Pdot2}.

   The 2017 superoutburst was detected by the ASAS-SN
team at $V$=15.51 on February 14.  Superhumps were
subsequently detected (vsnet-alert 20664, 20671, 20688).

\begin{figure}
  \begin{center}
    \FigureFile(85mm,110mm){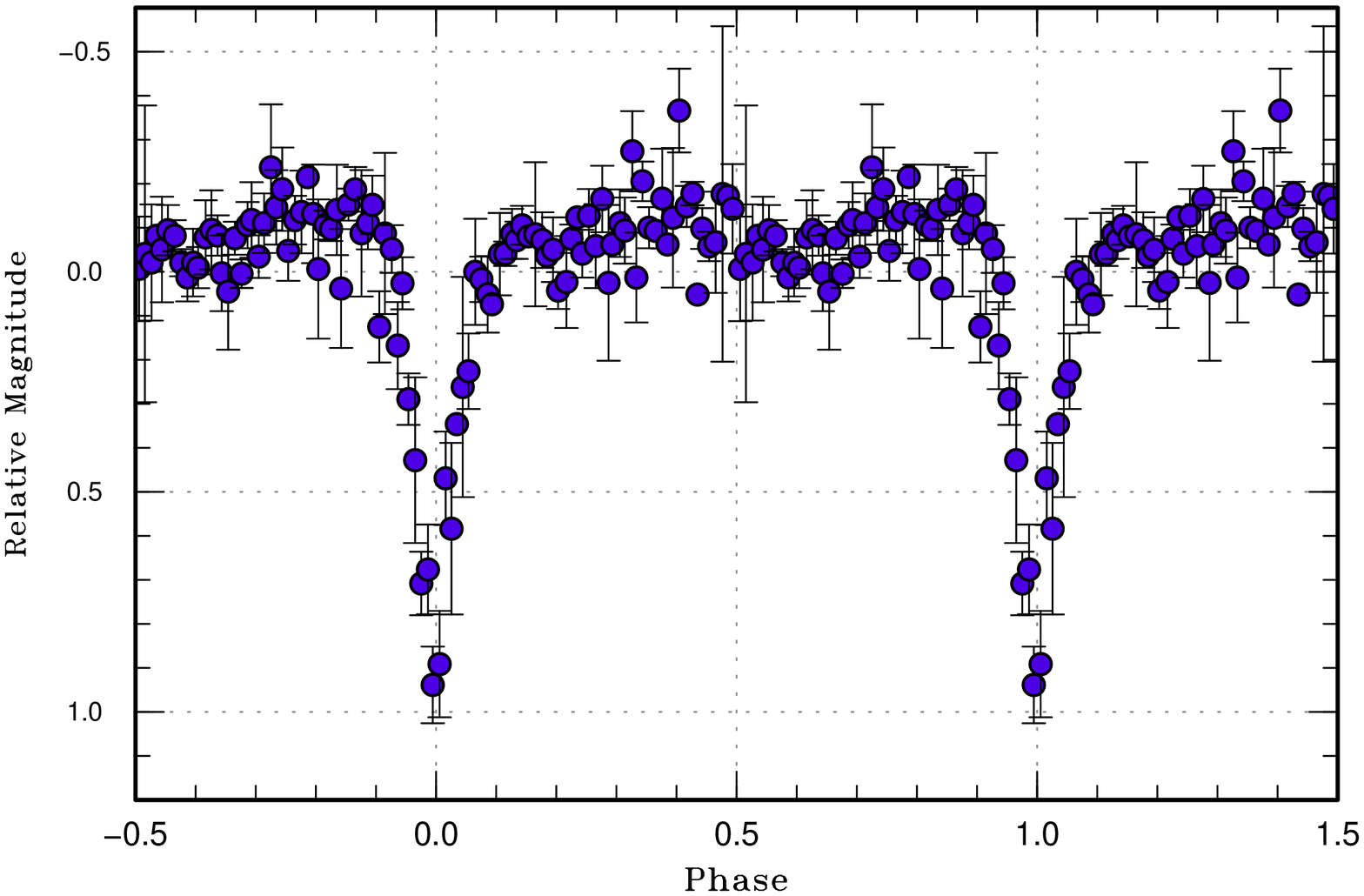}
  \end{center}
  \caption{Eclipse profile in SDSS J115207 (2009).
     The superhumps were mostly removed by using LOWESS.
     The phase-averaged profile was drawn against
     the ephemeris equation (\ref{equ:j1152ecl}).
     }
  \label{fig:j1152eclph2009}
\end{figure}

\begin{figure}
  \begin{center}
    \FigureFile(85mm,110mm){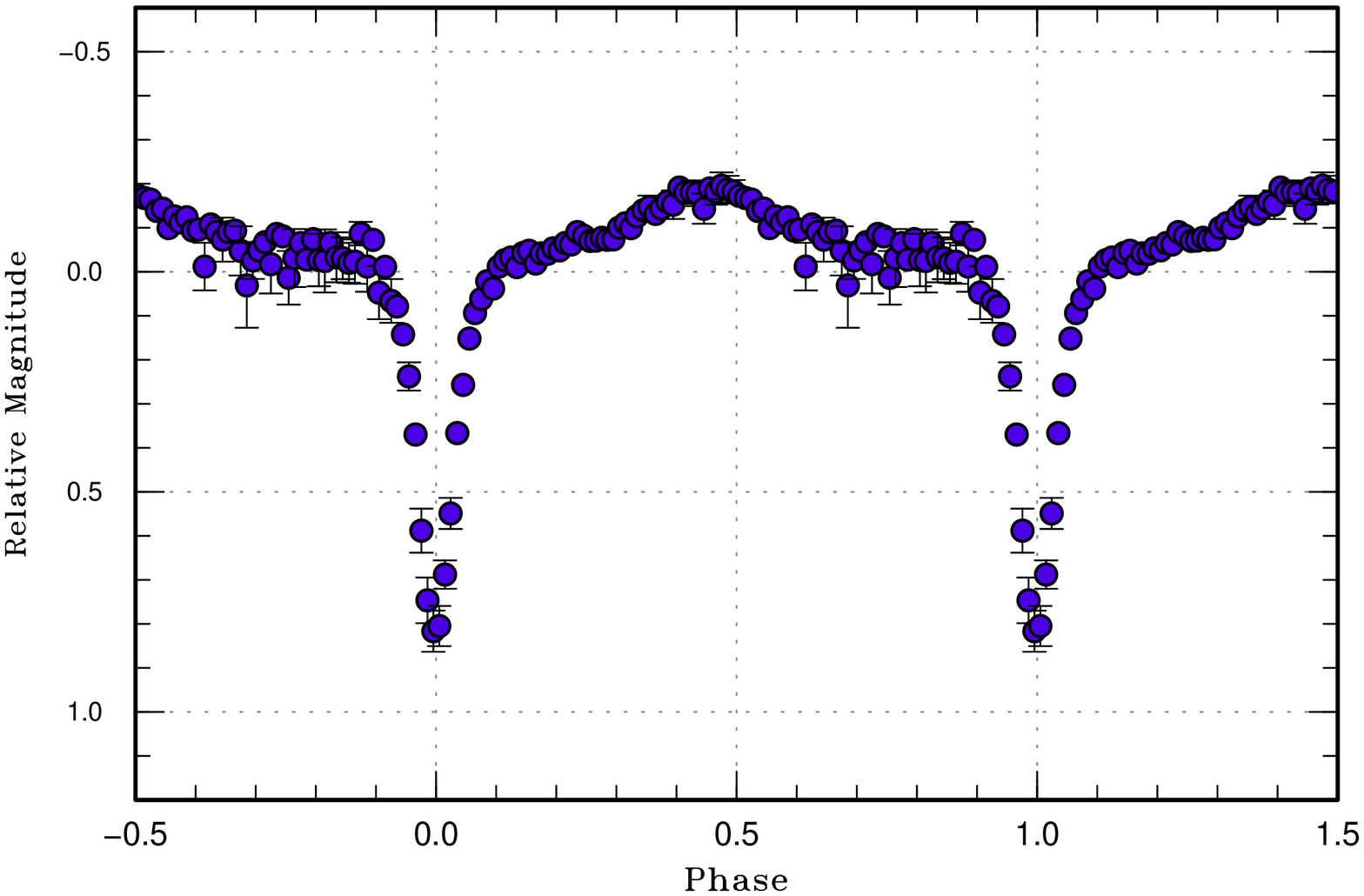}
  \end{center}
  \caption{Eclipse profile in SDSS J115207 (2017).
     The superhumps were mostly removed by using LOWESS.
     The phase-averaged profile was drawn against
     the ephemeris equation (\ref{equ:j1152ecl}).
     }
  \label{fig:j1152eclph2017}
\end{figure}

   We noticed that the ephemeris by \citet{sav11CVeclmass}
could not express our eclipse observations and found
that the period 0.0677497~d satisfy all the data
(\cite{sou10SDSSeclCV}; \cite{sav11CVeclmass};
\cite{Pdot2} and the present observations).
By using our data in 2009 and 2017, 
we have updated the eclipse ephemeris
using the MCMC analysis \citep{Pdot4}:
\begin{equation}
{\rm Min(BJD)} = 2457578.07695(6) + 0.0677497014(14) E .
\label{equ:j1152ecl}
\end{equation}
The epoch corresponds to the center of the entire
combined observation of 2009 and 2017.
This period corresponds to 4797 cycles between
\citet{sou10SDSSeclCV} and \citet{sav11CVeclmass},
which was assumed to be 4799 cycles in \citet{sav11CVeclmass}.
The $O-C$ values against this ephemeris are listed
in table \ref{tab:j1152ecl}.  The times of eclipse centers
by our observations in table \ref{tab:j1152ecl}
were determined by the same MCMC method against
the data segments (2007 and 2017) by fixing
the orbital period.
The eclipse profiles used to determine these minima are shown in
figure \ref{fig:j1152eclph2009} and
figure \ref{fig:j1152eclph2017}.

\begin{figure}
  \begin{center}
    \FigureFile(88mm,70mm){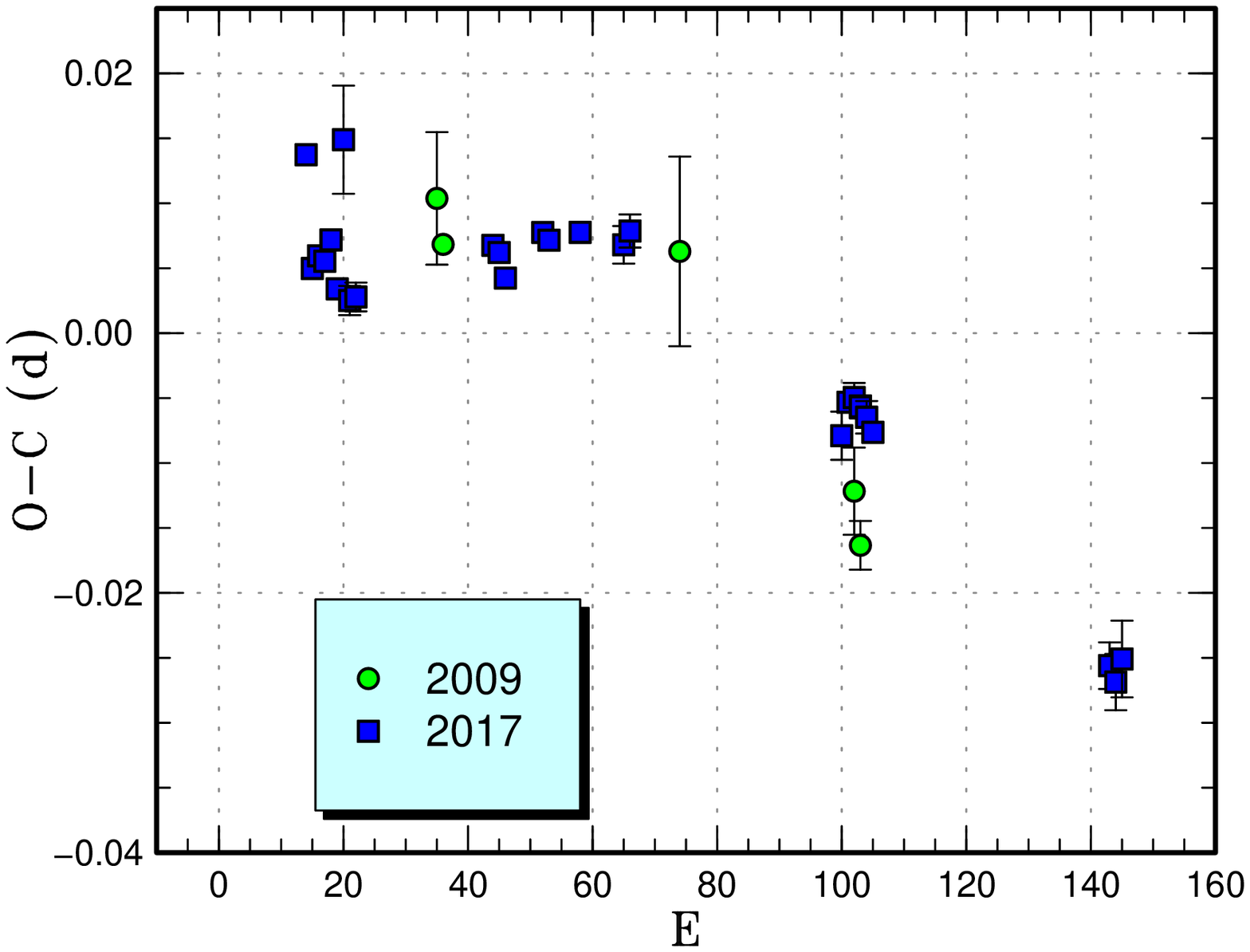}
  \end{center}
  \caption{Comparison of $O-C$ diagrams of SDSS J115207 between different
  superoutbursts.  A period of 0.07036~d was used to draw this figure.
  Approximate cycle counts ($E$) after the start of the superoutburst
  were used.
  }
  \label{fig:j1152comp}
\end{figure}

   The times of superhump maxima are listed in
table \ref{tab:j1152oc2017}.  
Although superhumps were initially
suspected to be stage A ones (vsnet-alert 20671),
they were more likely already stage B ones
(figure \ref{fig:j1152comp}).
Stage B-C transition occurred around $E$=52.
We also provide an updated table of superhump
maxima of the 2009 superoutburst in table \ref{tab:j1152oc2009}.
This table is based on the identification of
the true superhump period and based on the updated
orbital ephemeris.  The 2009 observations likely
recorded a combination of stages B and C.

\begin{table}
\caption{List of eclipse minima in SDSS J115207}\label{tab:j1152ecl}
\begin{center}
\begin{tabular}{cccc}
\hline
$E$ & BJD-2400000 & $O-C$ & Source\commenta \\
\hline
$-$39831 & 54879.5387(2) & 0.0001 & 1 \\
$-$39830 & 54879.6065(2) & 0.0002 & 1 \\
$-$38125 & 54995.11953(6) & $-$0.00005 & 2 \\
$-$35033 & 55204.601324(6) & $-$0.00034 & 3 \\
3321 & 57803.07370(2) & $-$0.00001 & 2 \\
\hline
  \multicolumn{4}{l}{\parbox{190pt}{\commenta 1: \citet{sou10SDSSeclCV},
  2: this work, 3: \citet{sav11CVeclmass}}} \\
\end{tabular}
\end{center}
\end{table}

\begin{table}
\caption{Superhump maxima of SDSS J115207 (2017)}\label{tab:j1152oc2017}
\begin{center}
\begin{tabular}{rp{50pt}p{30pt}r@{.}lcr}
\hline
$E$ & max\commenta & error & \multicolumn{2}{c}{$O-C$\commentb} & phase\commentc & $N$\commentd \\
\hline
0 & 57799.9836 & 0.0007 & 0&0026 & 0.39 & 85 \\
1 & 57800.0452 & 0.0008 & $-$0&0059 & 0.30 & 124 \\
2 & 57800.1165 & 0.0007 & $-$0&0047 & 0.35 & 129 \\
3 & 57800.1865 & 0.0003 & $-$0&0049 & 0.38 & 167 \\
4 & 57800.2585 & 0.0006 & $-$0&0030 & 0.45 & 201 \\
5 & 57800.3251 & 0.0007 & $-$0&0066 & 0.43 & 95 \\
6 & 57800.4069 & 0.0042 & 0&0051 & 0.64 & 40 \\
7 & 57800.4649 & 0.0011 & $-$0&0070 & 0.49 & 37 \\
8 & 57800.5355 & 0.0011 & $-$0&0065 & 0.54 & 35 \\
30 & 57802.0874 & 0.0003 & 0&0024 & 0.44 & 135 \\
31 & 57802.1572 & 0.0003 & 0&0021 & 0.47 & 212 \\
32 & 57802.2256 & 0.0005 & 0&0004 & 0.48 & 100 \\
38 & 57802.6513 & 0.0003 & 0&0053 & 0.77 & 137 \\
39 & 57802.7211 & 0.0003 & 0&0049 & 0.79 & 180 \\
44 & 57803.0735 & 0.0005 & 0&0066 & 1.00 & 85 \\
51 & 57803.5650 & 0.0014 & 0&0073 & 0.25 & 29 \\
52 & 57803.6365 & 0.0013 & 0&0085 & 0.31 & 30 \\
86 & 57806.0129 & 0.0019 & 0&0005 & 0.38 & 21 \\
87 & 57806.0859 & 0.0007 & 0&0033 & 0.46 & 36 \\
88 & 57806.1566 & 0.0011 & 0&0039 & 0.50 & 37 \\
89 & 57806.2262 & 0.0009 & 0&0034 & 0.53 & 91 \\
90 & 57806.2958 & 0.0013 & 0&0028 & 0.56 & 87 \\
91 & 57806.3650 & 0.0008 & 0&0019 & 0.58 & 43 \\
129 & 57809.0207 & 0.0018 & $-$0&0074 & 0.78 & 62 \\
130 & 57809.0898 & 0.0022 & $-$0&0085 & 0.80 & 61 \\
131 & 57809.1619 & 0.0030 & $-$0&0065 & 0.86 & 56 \\
\hline
  \multicolumn{7}{l}{\commenta BJD$-$2400000.} \\
  \multicolumn{7}{l}{\commentb Against max $= 2457799.9810 + 0.070133 E$.} \\
  \multicolumn{7}{l}{\commentc Orbital phase.} \\
  \multicolumn{7}{l}{\commentd Number of points used to determine the maximum.} \\
\end{tabular}
\end{center}
\end{table}


\begin{table}
\caption{Superhump maxima of SDSS J115207 (2009)}\label{tab:j1152oc2009}
\begin{center}
\begin{tabular}{rp{50pt}p{30pt}r@{.}lcr}
\hline
$E$ & max\commenta & error & \multicolumn{2}{c}{$O-C$\commentb} & phase\commentc & $N$\commentd \\
\hline
0 & 54993.9978 & 0.0051 & $-$0&0003 & 0.44 & 72 \\
1 & 54994.0646 & 0.0007 & $-$0&0035 & 0.43 & 136 \\
39 & 54996.7378 & 0.0073 & 0&0086 & 0.88 & 9 \\
67 & 54998.6894 & 0.0034 & $-$0&0005 & 0.69 & 17 \\
68 & 54998.7556 & 0.0019 & $-$0&0044 & 0.67 & 31 \\
\hline
  \multicolumn{7}{l}{\commenta BJD$-$2400000.} \\
  \multicolumn{7}{l}{\commentb Against max $= 2454993.9981 + 0.070028 E$.} \\
  \multicolumn{7}{l}{\commentc Orbital phase.} \\
  \multicolumn{7}{l}{\commentd Number of points used to determine the maximum.} \\
\end{tabular}
\end{center}
\end{table}

\subsection{SDSS J131432.10$+$444138.7}\label{obj:j1314}

\begin{figure}
  \begin{center}
    \FigureFile(85mm,110mm){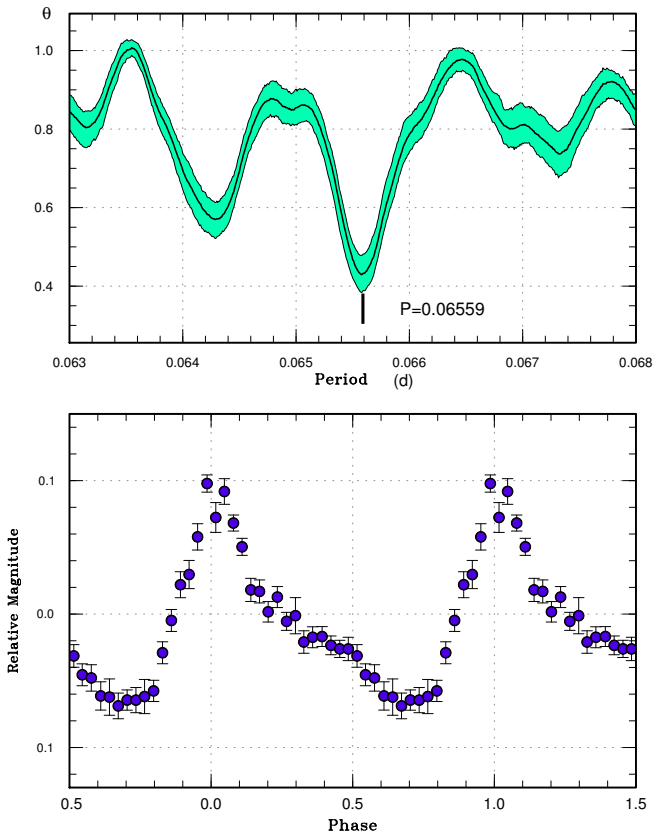}
  \end{center}
  \caption{Superhumps in SDSS J131432 (2017).
     (Upper): PDM analysis.
     (Lower): Phase-averaged profile.}
  \label{fig:j1314shpdm}
\end{figure}

   This object (hereafter SDSS J131432) was originally
selected as a CV by \citet{wil10newCVs} based on SDSS
colors and variability.  The 2017 outburst was detected
by the ASAS-SN team at $V$=15.63 on March 28.
Subsequent observations detected superhumps
(vsnet-alert 20841; figure \ref{fig:j1314shpdm}).
The times of superhump maxima are listed in
table \ref{tab:j1314oc2017}.


\begin{table}
\caption{Superhump maxima of SDSS J131432 (2017)}\label{tab:j1314oc2017}
\begin{center}
\begin{tabular}{rp{55pt}p{40pt}r@{.}lr}
\hline
\multicolumn{1}{c}{$E$} & \multicolumn{1}{c}{max\commenta} & \multicolumn{1}{c}{error} & \multicolumn{2}{c}{$O-C$\commentb} & \multicolumn{1}{c}{$N$\commentc} \\
\hline
0 & 57842.5432 & 0.0004 & 0&0014 & 60 \\
2 & 57842.6754 & 0.0005 & 0&0024 & 43 \\
12 & 57843.3287 & 0.0004 & $-$0&0005 & 54 \\
13 & 57843.3928 & 0.0004 & $-$0&0020 & 62 \\
14 & 57843.4579 & 0.0004 & $-$0&0026 & 56 \\
53 & 57846.0226 & 0.0033 & 0&0030 & 86 \\
54 & 57846.0845 & 0.0008 & $-$0&0008 & 136 \\
55 & 57846.1499 & 0.0009 & $-$0&0009 & 102 \\
\hline
  \multicolumn{6}{l}{\commenta BJD$-$2400000.} \\
  \multicolumn{6}{l}{\commentb Against max $= 2457842.5418 + 0.065620 E$.} \\
  \multicolumn{6}{l}{\commentc Number of points used to determine the maximum.} \\
\end{tabular}
\end{center}
\end{table}

\subsection{SDSS J153015.04$+$094946.3}\label{obj:j1530}

\begin{figure}
  \begin{center}
    \FigureFile(85mm,110mm){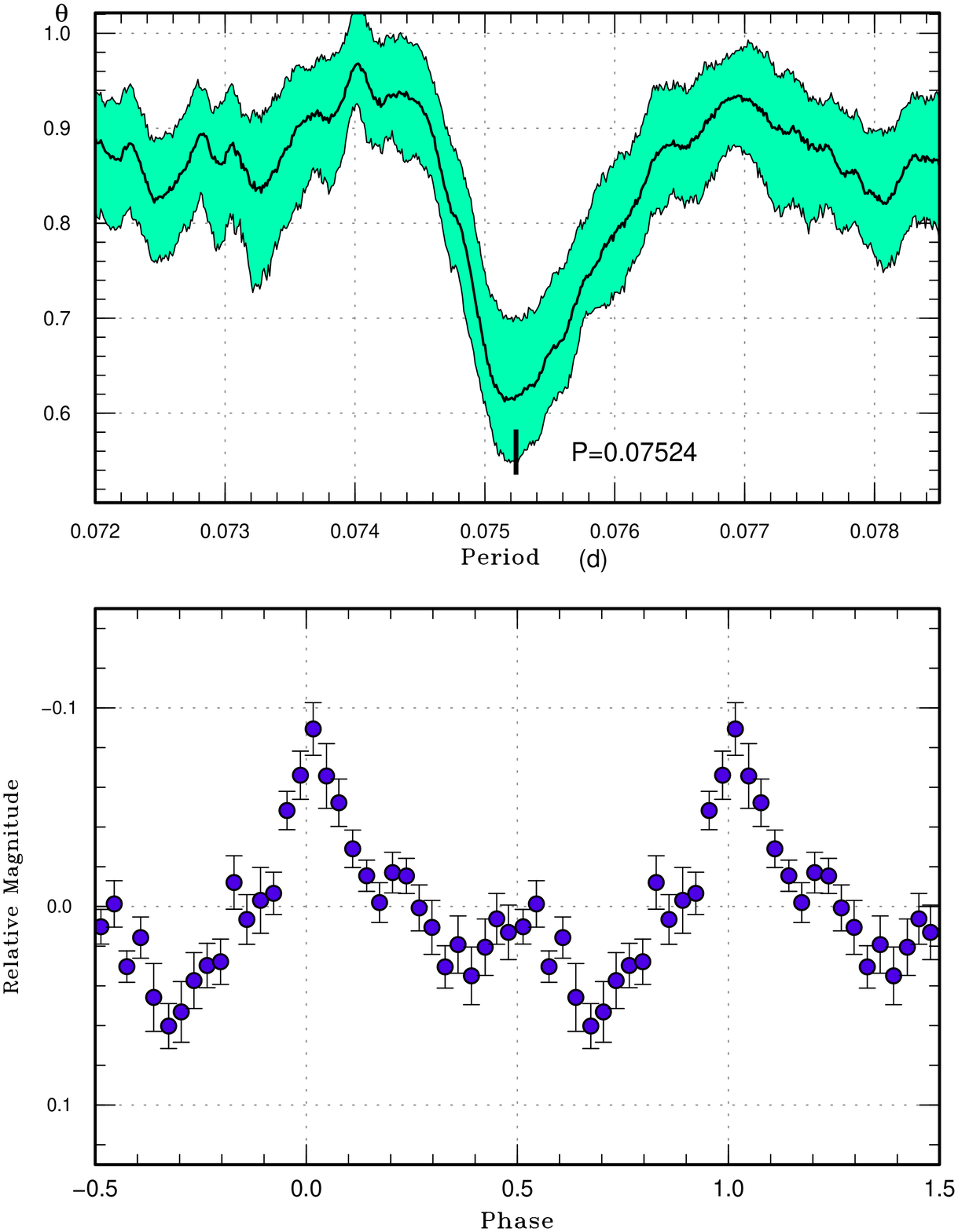}
  \end{center}
  \caption{Superhumps in SDSS J153015 (2017).
     (Upper): PDM analysis.
     (Lower): Phase-averaged profile.}
  \label{fig:j1530shpdm}
\end{figure}

   This object (hereafter SDSS J153015) was originally
selected as a CV by the SDSS \citep{szk09SDSSCV7}.
The dwarf nova-type variation was confirmed by
CRTS observations \citep{dra14CRTSCVs}.
The 2017 outburst was detected by the ASAS-SN team
at $V$=15.84 on March 7.  Based on ASAS-SN observations,
this outburst was likely a precursor one.
The peak was observed on March 10.
Subsequent observations detected superhumps (vsnet-alert 20769;
figure \ref{fig:j1530shpdm}).  The times of superhump maxima 
are listed in table \ref{tab:j1530oc2017}.
The period in table \ref{tab:perlist} was obtained
by the PDM analysis.  Since these superhump observations
were made during the early phase, the period may
refer to that of stage A superhumps.

   ASAS-SN observations indicate that this object
shows superoutburst relatively regularly.
The times of recent (likely) superoutburst are
listed in table \ref{tab:j1530out}.
Assuming that two superoutbursts were not
recorded between 2016 July and 2017 March,
all the superoutbursts were well expressed
by a supercycle of 84.7(1.2)~d with
the maximum $|O-C|$ of 18~d.  The long-term
light curve does not look like that of an ER UMa-type
dwarf nova but resembles that of V503 Cyg
(cf. subsections \ref{obj:nyher}, \ref{obj:j1616}
and \ref{obj:j0333}).  Further observations
to search for negative superhumps are recommended.


\begin{table}
\caption{Superhump maxima of SDSS J153015 (2017)}\label{tab:j1530oc2017}
\begin{center}
\begin{tabular}{rp{55pt}p{40pt}r@{.}lr}
\hline
\multicolumn{1}{c}{$E$} & \multicolumn{1}{c}{max\commenta} & \multicolumn{1}{c}{error} & \multicolumn{2}{c}{$O-C$\commentb} & \multicolumn{1}{c}{$N$\commentc} \\
\hline
0 & 57822.5636 & 0.0025 & 0&0023 & 36 \\
13 & 57823.5380 & 0.0025 & $-$0&0022 & 58 \\
14 & 57823.6147 & 0.0011 & $-$0&0007 & 74 \\
15 & 57823.6903 & 0.0011 & $-$0&0005 & 62 \\
41 & 57825.6497 & 0.0018 & 0&0011 & 44 \\
\hline
  \multicolumn{6}{l}{\commenta BJD$-$2400000.} \\
  \multicolumn{6}{l}{\commentb Against max $= 2457822.5613 + 0.075301 E$.} \\
  \multicolumn{6}{l}{\commentc Number of points used to determine the maximum.} \\
\end{tabular}
\end{center}
\end{table}

\begin{table}
\caption{List of likely superoutbursts of SDSS J153015 since 2015}\label{tab:j1530out}
\begin{center}
\begin{tabular}{ccccc}
\hline
Year & Month & Day & max\commenta & $V$-mag \\
\hline
2015 &  2 & 17 & 57070 & 15.93 \\
2015 &  5 & 12 & 57155 & 15.79 \\
2015 &  7 & 31 & 57235 & 15.57 \\
2016 &  1 & 22 & 57409 & 15.97 \\
2016 &  4 & 15 & 57493 & 15.91 \\
2016 &  7 & 28 & 57597 & 15.77 \\
2017 &  3 & 10 & 57823 & 15.76 \\
\hline
  \multicolumn{5}{l}{\commenta JD$-$2400000.} \\
\end{tabular}
\end{center}
\end{table}

\subsection{SDSS J155720.75$+$180720.2}\label{obj:j1557}

   This object (hereafter SDSS J155720) was originally
selected as a CV by the SDSS \citep{szk09SDSSCV7}.
The spectrum was that of a dwarf nova in quiescence
and \citet{szk09SDSSCV7} suggested a period of $\sim$2.1~hr.
There is an ROSAT X-ray counterpart of 1RXS J155720.3$+$180715.
The object was detected in outburst on 2007 June 12
(14.8 mag) and 2008 September 5 (16.1 mag) by the CRTS
team\footnote{
  $<$http://nesssi.cacr.caltech.edu/catalina/20160317/1603171180824123493.html$>$.
}

\begin{figure}
  \begin{center}
    \FigureFile(85mm,110mm){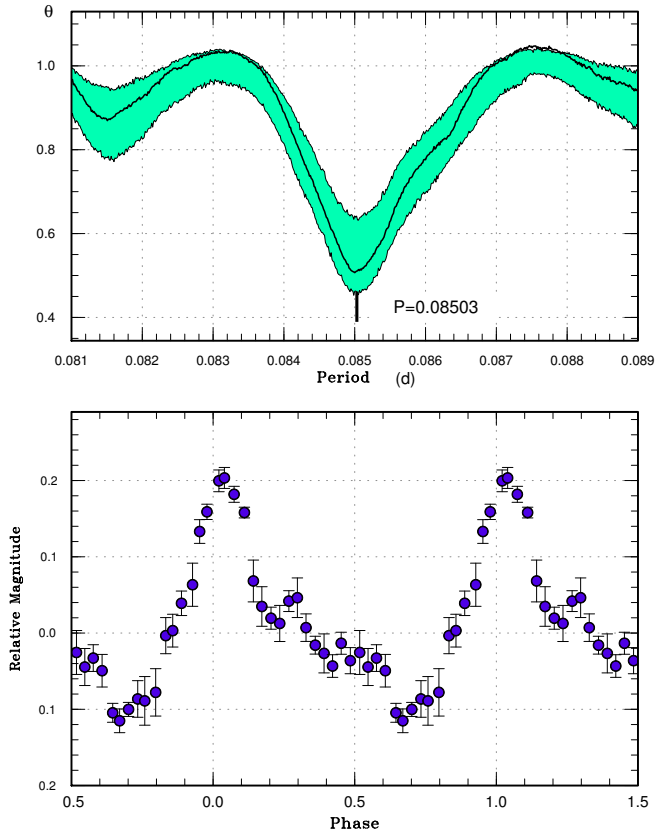}
  \end{center}
  \caption{Superhumps in SDSS J155720 (2016).
     (Upper): PDM analysis.
     The data segment BJD 2457467--2457471 was used.
     (Lower): Phase-averaged profile.}
  \label{fig:j1557shpdm}
\end{figure}

   The 2016 outburst was detected by the CRTS team
at an unfiltered CCD magnitude of 15.37 on March 17
(=CSS160317:155721$+$180720) and by the ASAS-SN team
at $V$=15.12 on the same night.
Subsequent observations detected superhumps
(vsnet-alert 19613; figure \ref{fig:j1557shpdm}).
The times of superhump maxima are listed in
table \ref{tab:j1557oc2016}.


\begin{table}
\caption{Superhump maxima of SDSS J155720 (2016)}\label{tab:j1557oc2016}
\begin{center}
\begin{tabular}{rp{55pt}p{40pt}r@{.}lr}
\hline
\multicolumn{1}{c}{$E$} & \multicolumn{1}{c}{max\commenta} & \multicolumn{1}{c}{error} & \multicolumn{2}{c}{$O-C$\commentb} & \multicolumn{1}{c}{$N$\commentc} \\
\hline
0 & 57467.7552 & 0.0050 & 0&0009 & 17 \\
1 & 57467.8406 & 0.0004 & 0&0009 & 41 \\
2 & 57467.9249 & 0.0025 & $-$0&0005 & 12 \\
12 & 57468.7812 & 0.0009 & 0&0002 & 17 \\
13 & 57468.8640 & 0.0014 & $-$0&0025 & 12 \\
23 & 57469.7278 & 0.0030 & 0&0056 & 25 \\
24 & 57469.8008 & 0.0009 & $-$0&0070 & 30 \\
29 & 57470.2381 & 0.0047 & 0&0025 & 93 \\
\hline
  \multicolumn{6}{l}{\commenta BJD$-$2400000.} \\
  \multicolumn{6}{l}{\commentb Against max $= 2457467.7542 + 0.085565 E$.} \\
  \multicolumn{6}{l}{\commentc Number of points used to determine the maximum.} \\
\end{tabular}
\end{center}
\end{table}

\subsection{SSS J134850.1$-$310835}\label{obj:j1348}

   This object (hereafter SSS J134850) was discovered 
by Stan Howerton during the course of the CRTS SNHunt
(supernova hunt).\footnote{
  $<$https://www.aavso.org/sssj1348501-310835-bright-dwarf-nova-centaurus$>$.
}
The object showed a number of outbursts in the ASAS-3
data in the past.

\begin{figure}
  \begin{center}
    \FigureFile(85mm,110mm){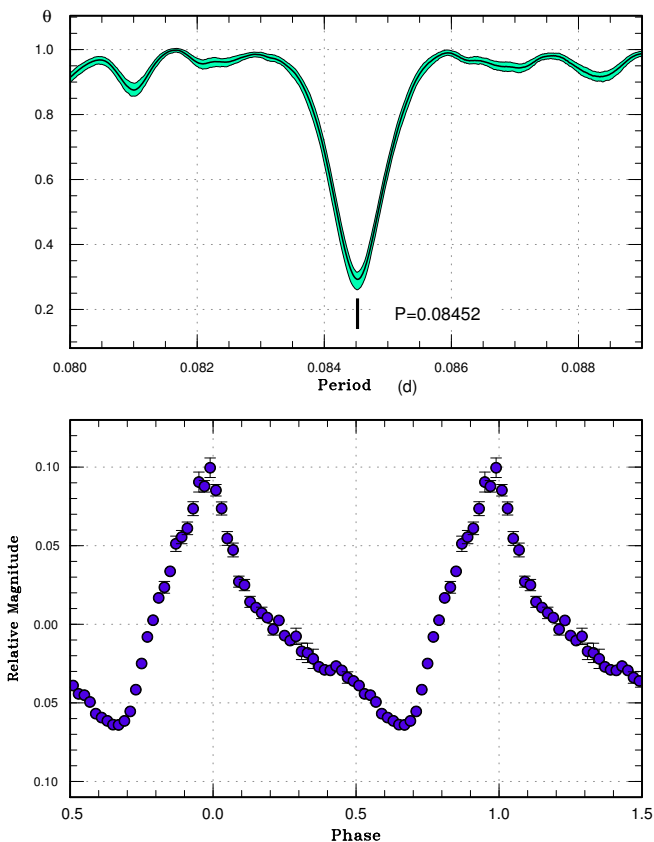}
  \end{center}
  \caption{Superhumps in SSS J134850 during the superoutburst
     plateau (2016).
     (Upper): PDM analysis.
     (Lower): Phase-averaged profile.}
  \label{fig:j1348shpdm}
\end{figure}

   The 2016 outburst was detected by R. Stubbings
at a visual magnitude of 11.8 on April 17.
Subsequent observations detected superhumps
(vsnet-alert 19751; figure \ref{fig:j1348shpdm}).
The times of superhump maxima are listed in
table \ref{tab:j1348oc2016}.
The period variation was rather smooth and we gave
a global $P_{\rm dot}$ rather than giving stages.
An analysis of the post-superoutburst data
did not yield a superhump period.


\begin{table}
\caption{Superhump maxima of SSS J134850 (2016)}\label{tab:j1348oc2016}
\begin{center}
\begin{tabular}{rp{55pt}p{40pt}r@{.}lr}
\hline
\multicolumn{1}{c}{$E$} & \multicolumn{1}{c}{max\commenta} & \multicolumn{1}{c}{error} & \multicolumn{2}{c}{$O-C$\commentb} & \multicolumn{1}{c}{$N$\commentc} \\
\hline
0 & 57499.6580 & 0.0009 & $-$0&0030 & 28 \\
1 & 57499.7437 & 0.0009 & $-$0&0018 & 20 \\
12 & 57500.6755 & 0.0014 & 0&0002 & 29 \\
13 & 57500.7634 & 0.0019 & 0&0035 & 10 \\
24 & 57501.6888 & 0.0008 & $-$0&0009 & 27 \\
32 & 57502.3677 & 0.0003 & 0&0017 & 195 \\
33 & 57502.4507 & 0.0004 & 0&0002 & 195 \\
44 & 57503.3822 & 0.0003 & 0&0018 & 195 \\
45 & 57503.4669 & 0.0003 & 0&0020 & 195 \\
46 & 57503.5504 & 0.0003 & 0&0009 & 193 \\
55 & 57504.3106 & 0.0003 & 0&0003 & 155 \\
56 & 57504.3938 & 0.0003 & $-$0&0010 & 179 \\
67 & 57505.3233 & 0.0003 & $-$0&0014 & 194 \\
68 & 57505.4080 & 0.0003 & $-$0&0013 & 195 \\
69 & 57505.4916 & 0.0004 & $-$0&0022 & 132 \\
79 & 57506.3402 & 0.0003 & 0&0011 & 193 \\
80 & 57506.4236 & 0.0005 & $-$0&0001 & 163 \\
\hline
  \multicolumn{6}{l}{\commenta BJD$-$2400000.} \\
  \multicolumn{6}{l}{\commentb Against max $= 2457499.6609 + 0.084534 E$.} \\
  \multicolumn{6}{l}{\commentc Number of points used to determine the maximum.} \\
\end{tabular}
\end{center}
\end{table}

\subsection{TCP J01375892$+$4951055}\label{obj:j0137}

\begin{figure}
  \begin{center}
    \FigureFile(85mm,110mm){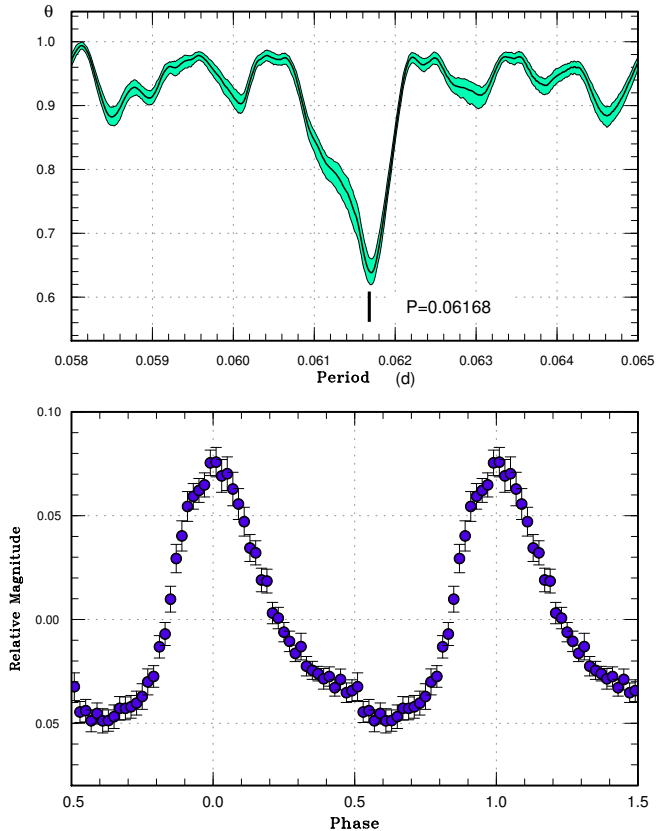}
  \end{center}
  \caption{Superhumps in TCP J013758 (2016).
     (Upper): PDM analysis.
     (Lower): Phase-averaged profile.}
  \label{fig:j0137shpdm}
\end{figure}

\begin{figure}
  \begin{center}
    \FigureFile(85mm,100mm){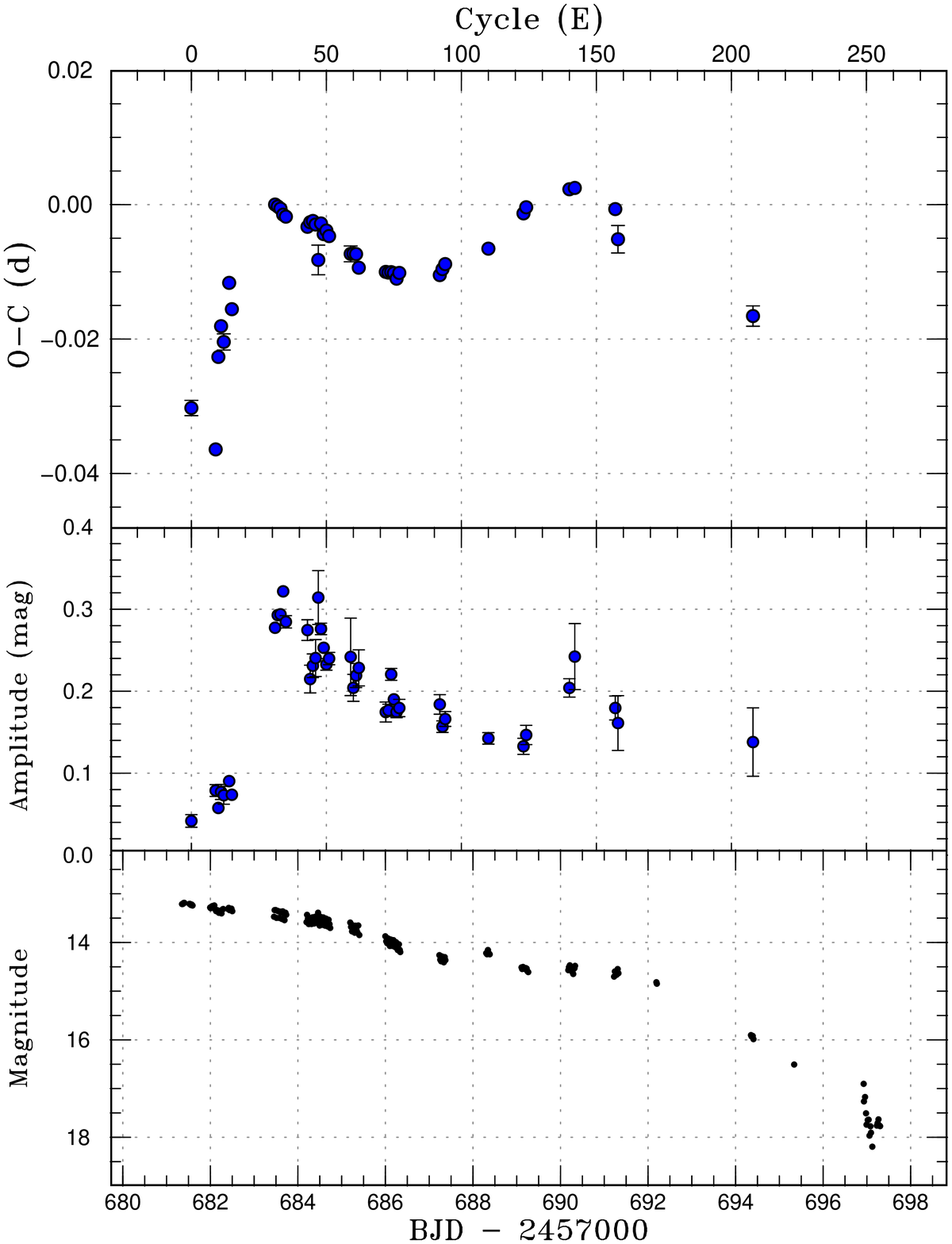}
  \end{center}
  \caption{$O-C$ diagram of superhumps in TCP J013758 (2016).
     (Upper:) $O-C$ diagram.
     We used a period of 0.06169~d for calculating the $O-C$ residuals.
     (Middle:) Amplitudes of superhumps.
     (Lower:) Light curve.  The data were binned to 0.021~d.
  }
  \label{fig:j0137humpall}
\end{figure}

   This object (hereafter TCP J013758) was discovered 
by K. Itagaki at an unfiltered CCD magnitude of 13.2
on 2016 October 19.\footnote{
  $<$http://www.cbat.eps.harvard.edu/unconf/followups/J01375892+4951055.html$>$.
}
There is a blue SDSS counterpart ($g$=19.85 and $u-g$=0.15)
and also a GALEX UV counterpart.  The object was
suspected to be a dwarf nova.
Subsequent observations detected growing superhumps
(vsnet-alert 20238).  Superhumps with a stable period
were observed 2~d after the discovery (vsnet-alert 20242,
20268; figure \ref{fig:j0137shpdm}).
The times of superhump maxima are listed in
table \ref{tab:j0137oc2016}.  The data show clear
stages A--C, with a positive $P_{\rm dot}$ for stage B
characteristic to this short $P_{\rm SH}$
(figure \ref{fig:j0137humpall}).
The period of stage A superhumps was not very well
determined due to a gap around the stage A-B transition.
The behavior suggests a typical SU UMa-type dwarf nova.


\begin{table}
\caption{Superhump maxima of TCP J013758 (2016)}\label{tab:j0137oc2016}
\begin{center}
\begin{tabular}{rp{55pt}p{40pt}r@{.}lr}
\hline
\multicolumn{1}{c}{$E$} & \multicolumn{1}{c}{max\commenta} & \multicolumn{1}{c}{error} & \multicolumn{2}{c}{$O-C$\commentb} & \multicolumn{1}{c}{$N$\commentc} \\
\hline
0 & 57681.5361 & 0.0011 & $-$0&0179 & 52 \\
9 & 57682.0851 & 0.0006 & $-$0&0246 & 92 \\
10 & 57682.1605 & 0.0008 & $-$0&0109 & 115 \\
11 & 57682.2268 & 0.0009 & $-$0&0064 & 114 \\
12 & 57682.2862 & 0.0012 & $-$0&0088 & 93 \\
14 & 57682.4183 & 0.0004 & $-$0&0002 & 54 \\
15 & 57682.4761 & 0.0006 & $-$0&0042 & 68 \\
31 & 57683.4787 & 0.0002 & 0&0104 & 50 \\
32 & 57683.5401 & 0.0001 & 0&0100 & 144 \\
33 & 57683.6015 & 0.0002 & 0&0097 & 96 \\
34 & 57683.6623 & 0.0001 & 0&0087 & 96 \\
35 & 57683.7237 & 0.0002 & 0&0083 & 74 \\
43 & 57684.2157 & 0.0003 & 0&0063 & 77 \\
44 & 57684.2781 & 0.0006 & 0&0070 & 40 \\
45 & 57684.3399 & 0.0005 & 0&0071 & 26 \\
46 & 57684.4011 & 0.0007 & 0&0065 & 24 \\
47 & 57684.4575 & 0.0022 & 0&0011 & 13 \\
48 & 57684.5246 & 0.0002 & 0&0065 & 99 \\
49 & 57684.5847 & 0.0002 & 0&0049 & 95 \\
50 & 57684.6469 & 0.0002 & 0&0053 & 96 \\
51 & 57684.7078 & 0.0002 & 0&0044 & 96 \\
59 & 57685.1987 & 0.0012 & 0&0013 & 15 \\
60 & 57685.2604 & 0.0006 & 0&0013 & 22 \\
61 & 57685.3220 & 0.0005 & 0&0011 & 24 \\
62 & 57685.3817 & 0.0010 & $-$0&0010 & 16 \\
72 & 57685.9980 & 0.0005 & $-$0&0022 & 82 \\
73 & 57686.0596 & 0.0003 & $-$0&0024 & 159 \\
74 & 57686.1213 & 0.0003 & $-$0&0024 & 184 \\
75 & 57686.1829 & 0.0002 & $-$0&0026 & 184 \\
76 & 57686.2437 & 0.0003 & $-$0&0035 & 185 \\
77 & 57686.3063 & 0.0007 & $-$0&0027 & 67 \\
92 & 57687.2313 & 0.0006 & $-$0&0040 & 43 \\
93 & 57687.2939 & 0.0003 & $-$0&0031 & 67 \\
94 & 57687.3563 & 0.0004 & $-$0&0024 & 58 \\
110 & 57688.3457 & 0.0004 & $-$0&0012 & 68 \\
123 & 57689.1529 & 0.0006 & 0&0033 & 113 \\
124 & 57689.2155 & 0.0006 & 0&0041 & 113 \\
140 & 57690.2052 & 0.0004 & 0&0058 & 60 \\
142 & 57690.3288 & 0.0009 & 0&0058 & 32 \\
157 & 57691.2510 & 0.0006 & 0&0017 & 34 \\
158 & 57691.3082 & 0.0020 & $-$0&0028 & 17 \\
208 & 57694.3812 & 0.0015 & $-$0&0174 & 47 \\
\hline
  \multicolumn{6}{l}{\commenta BJD$-$2400000.} \\
  \multicolumn{6}{l}{\commentb Against max $= 2457681.5540 + 0.061753 E$.} \\
  \multicolumn{6}{l}{\commentc Number of points used to determine the maximum.} \\
\end{tabular}
\end{center}
\end{table}

\subsection{TCP J18001854$-$3533149}\label{obj:j1800}

\begin{figure}
  \begin{center}
    \FigureFile(85mm,110mm){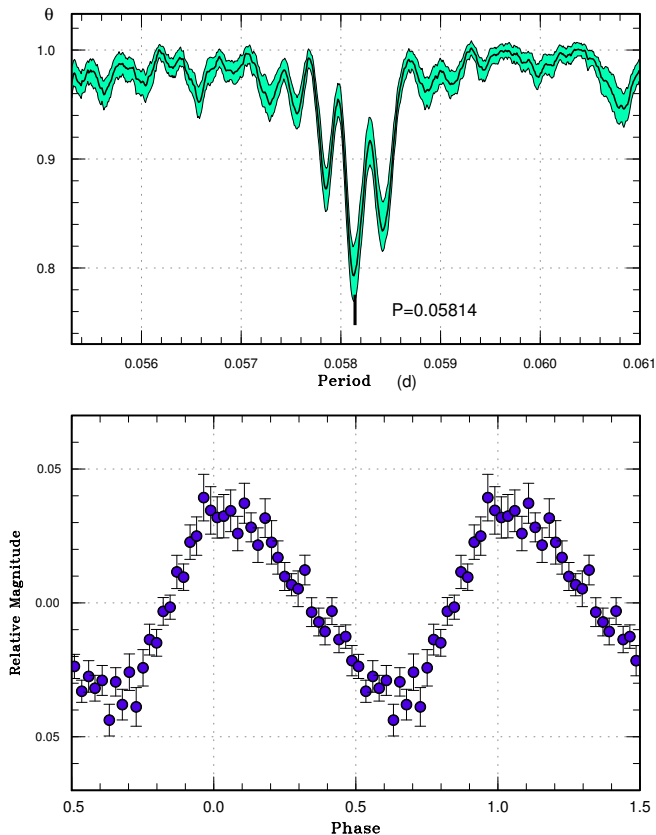}
  \end{center}
  \caption{Superhumps in TCP J180018 (2016).
     (Upper): PDM analysis.
     (Lower): Phase-averaged profile.}
  \label{fig:j1800shpdm}
\end{figure}

\begin{figure}
  \begin{center}
    \FigureFile(85mm,100mm){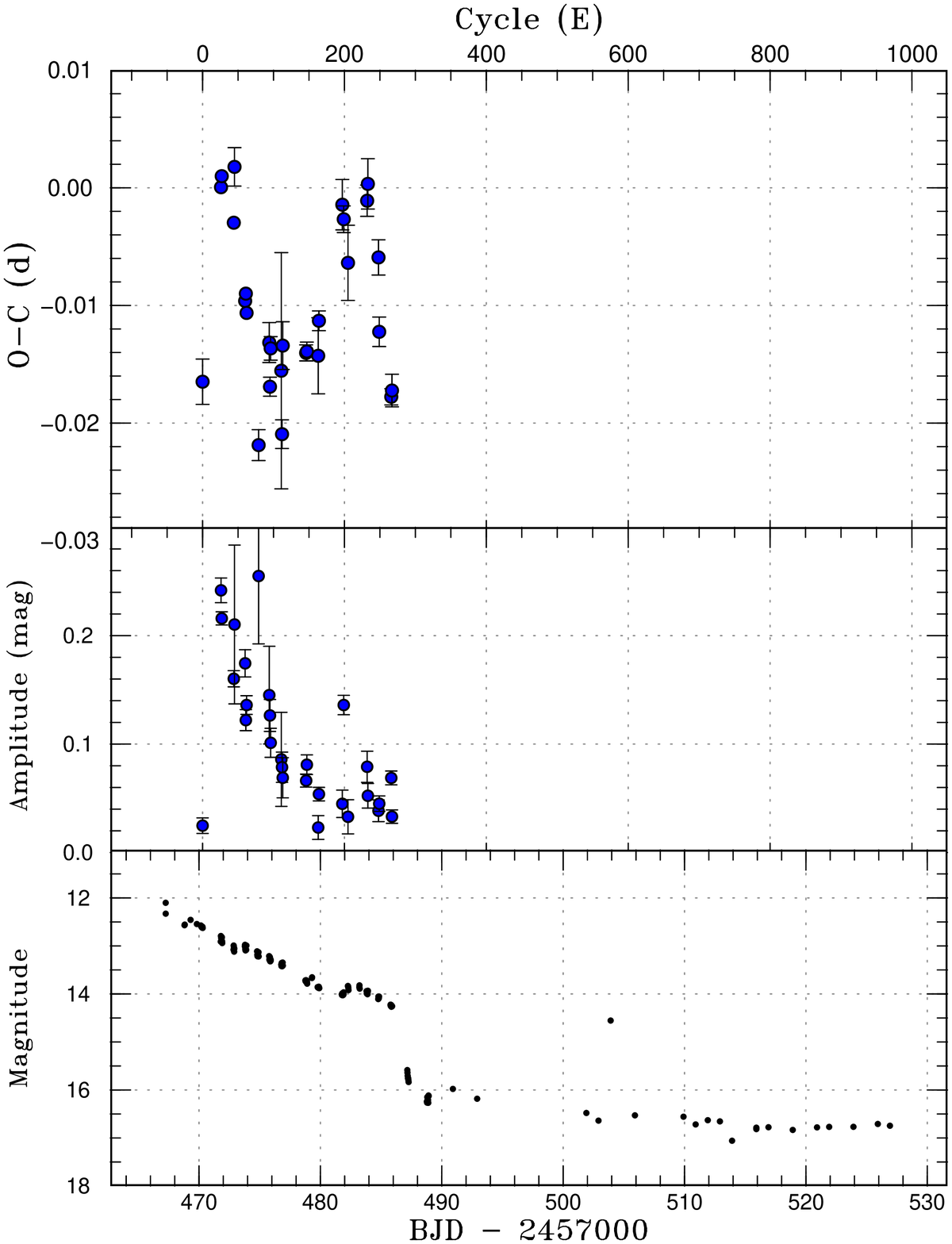}
  \end{center}
  \caption{$O-C$ diagram of superhumps in TCP J180018 (2016).
     (Upper:) $O-C$ diagram.
     We used a period of 0.05844~d for calculating the $O-C$ residuals.
     (Middle:) Amplitudes of superhumps.
     (Lower:) Light curve.  The data were binned to 0.019~d.
  }
  \label{fig:j1800humpall}
\end{figure}

   This object (hereafter TCP J180018) was discovered 
as a transient by K. Nishiyama and F. Kabashima at
an unfiltered CCD magnitude of 12.2 on March 16.\footnote{
  $<$http://www.cbat.eps.harvard.edu/unconf/followups/J18001854-3533149.html$>$.
}
Multicolor photometry by S. Kiyota showed a blue color,
suggesting a dwarf nova-type outburst.
A spectroscopic study by K. Ayani on March 20 showed
Balmer absorption lines (H$\beta$ to H$\delta$).
The H$\alpha$ line was not clear probably because the absorption
is filled with the emission. 
The spectrum indicated a dwarf nova in outburst.
Subsequent observations detected superhumps
(vsnet-alert 19635, 19646, 19662, 19684, 19703;
figure \ref{fig:j1800shpdm}).
The times of superhump maxima are listed in
table \ref{tab:j1800oc2016}.
There were clear stages A--C, with a positive
$P_{\rm dot}$ for stage B, characteristic to
a short-$P_{\rm SH}$ SU UMa-type dwarf nova
(figure \ref{fig:j1800humpall}).

   The outburst faded on April 8 and the duration of
the total outburst was at least 23~d.  The object
showed a single post-superoutburst rebrightening
on April 25 at 14.5 mag (figure \ref{fig:j1800humpall}).


\begin{table}
\caption{Superhump maxima of TCP J180018 (2016)}\label{tab:j1800oc2016}
\begin{center}
\begin{tabular}{rp{55pt}p{40pt}r@{.}lr}
\hline
\multicolumn{1}{c}{$E$} & \multicolumn{1}{c}{max\commenta} & \multicolumn{1}{c}{error} & \multicolumn{2}{c}{$O-C$\commentb} & \multicolumn{1}{c}{$N$\commentc} \\
\hline
0 & 57470.2775 & 0.0019 & $-$0&0068 & 71 \\
26 & 57471.8134 & 0.0003 & 0&0098 & 50 \\
27 & 57471.8728 & 0.0002 & 0&0107 & 68 \\
44 & 57472.8624 & 0.0003 & 0&0068 & 280 \\
45 & 57472.9255 & 0.0016 & 0&0116 & 78 \\
60 & 57473.7907 & 0.0005 & 0&0002 & 50 \\
61 & 57473.8498 & 0.0005 & 0&0008 & 62 \\
62 & 57473.9066 & 0.0005 & $-$0&0008 & 44 \\
79 & 57474.8888 & 0.0013 & $-$0&0120 & 16 \\
94 & 57475.7742 & 0.0017 & $-$0&0033 & 26 \\
95 & 57475.8288 & 0.0008 & $-$0&0070 & 50 \\
96 & 57475.8905 & 0.0010 & $-$0&0038 & 48 \\
111 & 57476.7652 & 0.0100 & $-$0&0056 & 30 \\
112 & 57476.8183 & 0.0012 & $-$0&0110 & 50 \\
113 & 57476.8843 & 0.0020 & $-$0&0035 & 50 \\
146 & 57478.8122 & 0.0007 & $-$0&0040 & 52 \\
147 & 57478.8707 & 0.0008 & $-$0&0039 & 50 \\
163 & 57479.8054 & 0.0032 & $-$0&0042 & 52 \\
164 & 57479.8668 & 0.0008 & $-$0&0013 & 50 \\
197 & 57481.8052 & 0.0021 & 0&0087 & 34 \\
199 & 57481.9208 & 0.0011 & 0&0074 & 12 \\
205 & 57482.2678 & 0.0032 & 0&0037 & 31 \\
232 & 57483.8509 & 0.0013 & 0&0091 & 70 \\
233 & 57483.9108 & 0.0021 & 0&0105 & 42 \\
248 & 57484.7812 & 0.0015 & 0&0043 & 60 \\
249 & 57484.8333 & 0.0013 & $-$0&0020 & 52 \\
266 & 57485.8212 & 0.0007 & $-$0&0075 & 72 \\
267 & 57485.8802 & 0.0014 & $-$0&0070 & 70 \\
\hline
  \multicolumn{6}{l}{\commenta BJD$-$2400000.} \\
  \multicolumn{6}{l}{\commentb Against max $= 2457470.2843 + 0.058438 E$.} \\
  \multicolumn{6}{l}{\commentc Number of points used to determine the maximum.} \\
\end{tabular}
\end{center}
\end{table}

\section{Discussion}\label{sec:discuss}

\subsection{Statistics of objects}\label{sec:stat}

\begin{figure}
  \begin{center}
    \FigureFile(80mm,70mm){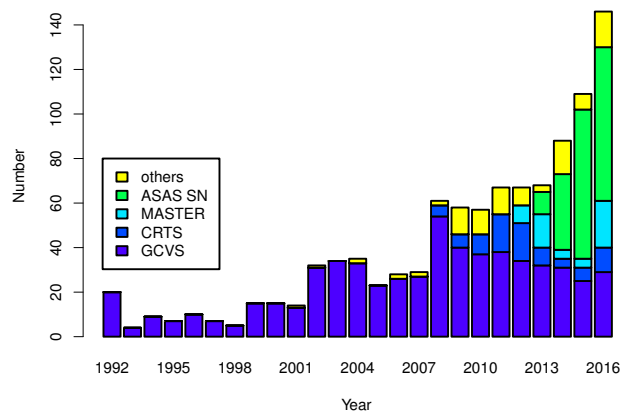}
  \end{center}
  \caption{Object categories in our survey.  Superoutbursts
  with measured superhump periods are included.
  The year represents the year of outburst.
  The year 1992 represents outbursts up to 1992 and the year
  2016 includes the outbursts in 2017, respectively.
  The category GCVS includes the objects named in the General
  Catalog of Variable Stars \citet{GCVS} in the latest version
  and objects named in New Catalog of Suspected Variable Stars
  (NSV: \cite{NSV}).  The categories CRTS, MASTER, ASAS-SN represent
  objects which were discovered in respective surveys.
  A fraction of objects discovered by these surveys
  are already named in GCVS and are included in the category GCVS.
  }
  \label{fig:objsource}
\end{figure}

   Following \citet{Pdot7} and \citet{Pdot8}, we present
statistics of sources of the objects studied in our surveys
(figure \ref{fig:objsource}).  Although ASAS-SN CVs
remained the majority of the objects we studied,
there have also been an increase in MASTER CVs
and CRTS CVs.  The noteworthy recent tendency is
the increase of objects in the Galactic plane
discovered by ASAS-SN.  This region had usually
been avoided by the majority of surveys (the best
examples being SDSS and CRTS) and we can expect a great increase
of dwarf novae if the Galactic plane is thoroughly surveyed
by ASAS-SN.  This increase of CV candidates in
the Galactic plane, however, has made it difficult
to distinguish dwarf novae and classical novae.
Indeed, there have been four Galactic novae
discovered by the ASAS-SN team: ASASSN-16ig = V5853 Sgr
(\cite{sta16v5853sgratel9343}; \cite{wil16v5853sgratel9375}),
ASASSN-16kb (\cite{pri16asassn16kbasassn16kdatel9479}),
ASASSN-16kd (\cite{sta16asassn16kdatel9469};
\cite{boh16asassn16kdatel9477}),
and ASASSN-16ma = V5856 Sgr (\cite{sta16asassn16maatel9669};
\cite{kuc16asassn16maatel9678}) in 2016.
Several dwarf novae studied in this paper were
also flagged as ``could also be a nova'' on
the ASAS-SN Transients page.
Some objects in this paper (ASASSN-16jb and ASASSN-16ow)
were initially suspected to be Galactic novae.
Although they have not been
a serious problem in studying dwarf novae, the supposed
nova classification might cause a delay in time-resolved
photometry to detect superhumps in the earliest
phase, and observers should keep in mind
the dwarf nova-type possibilities of nova candidate
in the Galactic plane.

\subsection{Period distribution}\label{sec:perdist}

\begin{figure}
  \begin{center}
    \FigureFile(80mm,135mm){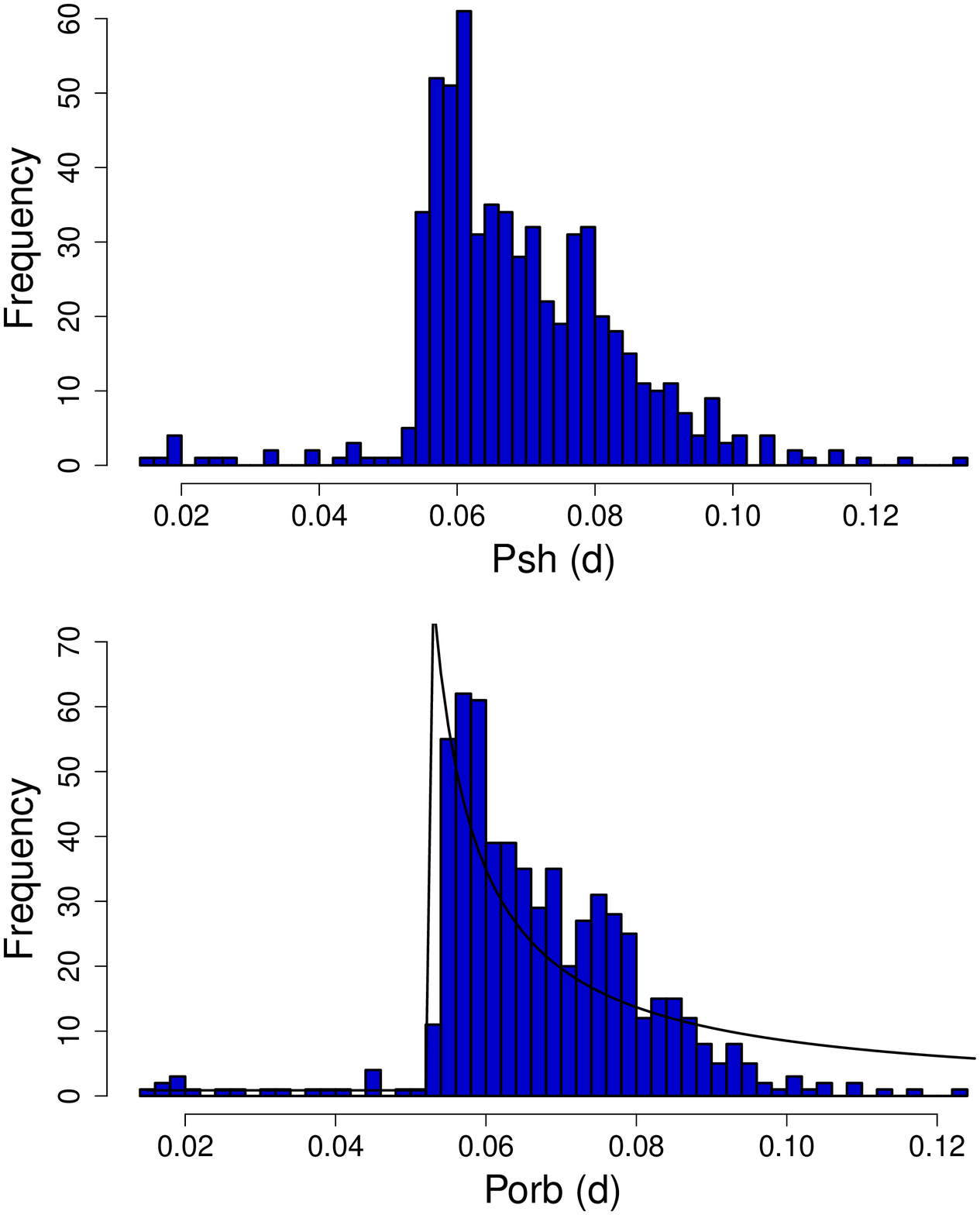}
  \end{center}
  \caption{Distribution of superhump periods in this survey.
  The data are from \citet{Pdot}, \citet{Pdot2}, \citet{Pdot3},
  \citet{Pdot4}, \citet{Pdot5}, \citet{Pdot6}, \citet{Pdot7},
  \citet{Pdot8} and this paper.
  The mean values are used when multiple superoutbursts
  were observed.
  (Upper) distribution of superhump periods.
  (Lower) distribution of orbital periods.  For objects with
  superhump periods shorter than 0.053~d, the orbital periods
  were assumed to be 1\% shorter than superhump periods.
  For objects with superhump periods longer than 0.053~d,
  we used the calibration in \citet{Pdot3} to estimate
  orbital periods.
  The line is the model distribution to determine
  the period minimum (equation \ref{equ:pdistfunc}, see text
  for the details).
  }
  \label{fig:phist}
\end{figure}

   In figure \ref{fig:phist}, we give distributions of
superhump and estimated orbital periods
(see the caption for details) since \citet{Pdot}.
For readers' convenience, we also listed ephemerides of
eclipsing systems newly determined or used in this study
in table \ref{tab:eclipsing}.
When there are multiple observations of superoutbursts
of the same object, we adopted an average of
the measurements.  This figure can be considered
to be a good representation of the distribution of
orbital periods for non-magnetic CVs below the period gap,
since the majority of CVs below the period gap
are SU UMa-type dwarf novae.  The following features reported
in \citet{Pdot8} are apparent:
(1) the sharp cut-off at a period of 0.053~d and
(2) accumulation of objects (``period spike'') just above
the cutoff.

   We determined the location of the sharp cut-off
(period minimum) by using the Bayesian approch.
We assumed the following period distribution
$D(P_{\rm orb})$:
\begin{equation}
  D(P_{\rm orb}) \propto \left\{
    \begin{array}{ll}
      c_1, & \mbox{($P_{\rm orb} \leq P_{\rm min}$)} \\
      1/(P_{\rm orb}-c_2), & \mbox{($P_{\rm orb} > P_{\rm min}$)}. \\
    \end{array}
    \right .
\label{equ:pdistfunc}
\end{equation}
$P_{\rm min}$ is the cut-off and $c_1$, $c_2$
are parameters to be determined.
We defined the likelihood to obtain the entire sample
of our SU UMa-type dwarf novae by using this distribution
(the distribution is normalized for a range of
0.01--0.13~d). 
We obtained the parameters by the MCMC method as follows:
$c_1$ = 1.93(25), $c_2$ = 0.0471(7) and
$P_{\rm min}$ = 0.052897(16).
The value of $P_{\rm min}$ is insensitive to the functional
form above $P_{\rm min}$.  The resulting distribution
is drawn as a line in the lower panel of
figure \ref{fig:phist}.

   Although the model does not properly reproduce
the location of the period spike, the numbers of
dwarf novae are lower than the best fit curve
above $P_{\rm orb} \sim$ 0.09~d.  This appears to
correspond to the period gap, contrary to our finding
in \citet{Pdot8}.

\begin{table}
\caption{Ephemerides of eclipsing systems.}\label{tab:eclipsing}
\begin{center}
\begin{tabular}{cccccc}
\hline
Object & Epoch (BJD) & Period (d) \\
\hline
OY Car & 2457120.49413(2) & 0.0631209131(5) \\
GP CVn & 2455395.37115(4) & 0.0629503676(9) \\
V893 Sco & 2454173.3030(3) & 0.0759614600(16) \\
MASTER J220559 & 2457658.72016(3) & 0.0612858(3) \\
SDSS J115207 & 2457578.07695(6) & 0.0677497014(14) \\
\hline
\end{tabular}
\end{center}
\end{table}

\subsection{Period derivatives during stage B}\label{sec:stagebpdot}

\begin{figure*}
  \begin{center}
    \FigureFile(110mm,88mm){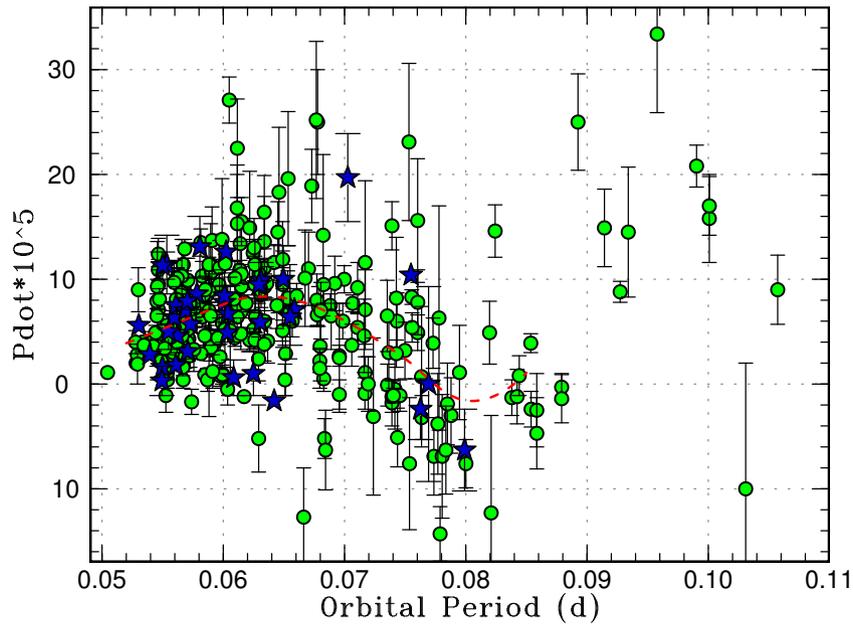}
  \end{center}
  \caption{$P_{\rm dot}$ for stage B versus $P_{\rm orb}$.
  Filled circles and filled stars represent samples in 
  \citet{Pdot}--\citet{Pdot8} and this paper, respectively.
  The curve represents the spline-smoothed global trend.
  }
  \label{fig:pdotporb9}
\end{figure*}

   Figure \ref{fig:pdotporb9} represents updated relation
between $P_{\rm dot}$ for stage B versus $P_{\rm orb}$.
We have omitted poor quality observation (quality C) 
since \citet{Pdot8} and simplified the symbols.
The majority of new objects studied in this paper follow
the trend presented in earlier papers.

\subsection{Mass ratios from stage A superhumps}\label{sec:stagea}

   We list new estimates for $q$ from stage A
superhumps \citep{kat13qfromstageA} in
table \ref{tab:newqstageA}.  This table includes measurements
of the objects in separate papers, which are listed in
table \ref{tab:outobs}.
In table \ref{tab:pera}, we list all stage A superhumps
recorded in the present study.

\begin{figure*}
  \begin{center}
    \FigureFile(110mm,88mm){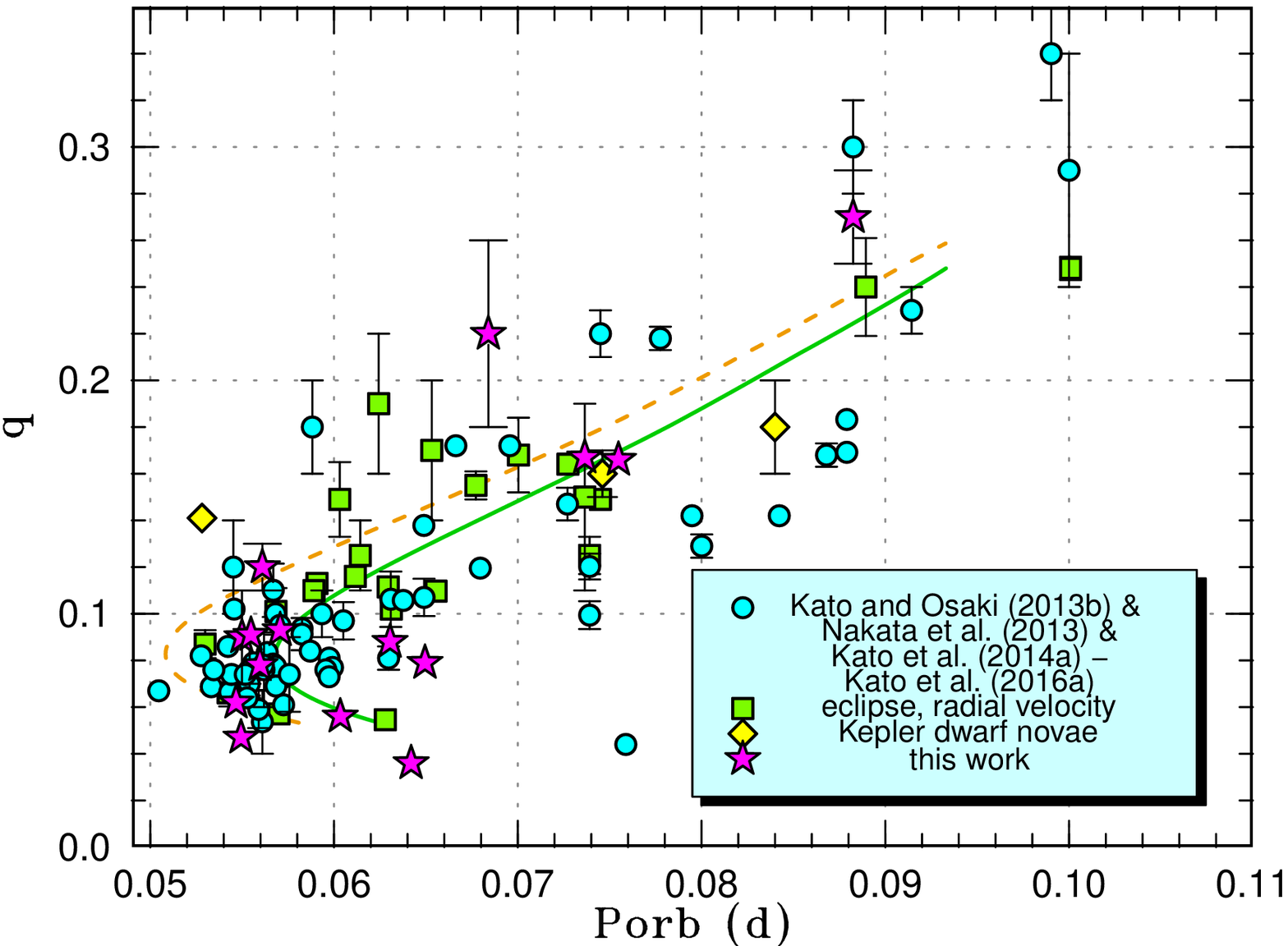}
  \end{center}
  \caption{Mass ratio versus orbital period.
  The dashed and solid curves represent the standard and optimal
  evolutionary tracks in \citet{kni11CVdonor}, respectively.
  The filled circles, filled squares, filled stars, filled diamonds
  represent $q$ values from a combination of the estimates
  from stage A superhumps published in four preceding
  sources (\cite{kat13qfromstageA}; \cite{nak13j2112j2037};
  \cite{Pdot5}; \cite{Pdot6}; \cite{Pdot7}; \cite{Pdot8}
  and references therein),
  known $q$ values from quiescent eclipses or 
  radial-velocity study, $q$ estimated in this work and dwarf novae
  in the Kepler data (see text for the reference),
  respectively.  The objects in ``this work'' includes
  objects studied in separate papers but listed in table \ref{tab:outobs}.}
  \label{fig:qall6}
\end{figure*}

   An updated distribution of mass ratios is shown in
figure \ref{fig:qall6} [for the list of objects, see
\citet{kat13qfromstageA}, \citet{Pdot7} and \citet{Pdot8}].
We have newly added SDSS J105754.25$+$275947.5
(hereafter SDSS J105754) with $P_{\rm orb}$=0.062792~d
and $q$=0.0546(20) (\cite{mca17j1057}, eclipse observation) and
ASASSN-14ag with $P_{\rm orb}$=0.060311~d and
$q$=0.149(16) (\cite{mca17asassn14ag}, eclipse observation).
It would be worth mentioning that \citet{mca17j1057}
classified SDSS J105754 as a period
bouncer and we have two objects (ASASSN-16dt and
ASASSN-16js) near the location of SDSS J105754.
Both objects are likely identified as period bouncers
and these detections demonstrate the efficiency
of the stage A superhump method.
The present study has strengthened the concentration
of WZ Sge-type dwarf novae around $q=0.07$
just above the period minimum, as reported
in \citet{Pdot7} and \citet{Pdot8}.

\begin{table}
\caption{New estimates for the binary mass ratio from stage A superhumps}\label{tab:newqstageA}
\begin{center}
\begin{tabular}{ccc}
\hline
Object         & $\epsilon^*$ (stage A) & $q$ from stage A \\
\hline
HT Cas         & 0.0566(5) & 0.171(2) \\
GS Cet         & 0.0288(8) & 0.078(2) \\
GZ Cnc         & 0.081(3)  & 0.27(2) \\
IR Gem         & 0.068(11) & 0.22(4) \\
HV Vir         & 0.034(1)  & 0.093(3) \\
ASASSN-16da    & 0.042(2)  & 0.12(1) \\
ASASSN-16dt    & 0.0135(7) & 0.036(2) \\
ASASSN-16eg    & 0.0552(6) & 0.166(2) \\
ASASSN-16hj    & 0.034(7)  & 0.09(2) \\
ASASSN-16iw    & 0.029(1)  & 0.079(2) \\
ASASSN-16jb    & 0.0321(5) & 0.088(1) \\
ASASSN-16js    & 0.0213(16) & 0.056(5) \\
ASASSN-16oi    & 0.033(2)  & 0.091(7) \\
ASASSN-16os    & 0.018(1)  & 0.047(3) \\
ASASSN-17bl    & 0.0235(9) & 0.062(3) \\
\hline
\end{tabular}
\end{center}
\end{table}

\begin{table}
\caption{Superhump Periods during Stage A}\label{tab:pera}
\begin{center}
\begin{tabular}{cccc}
\hline
Object & Year & period (d) & err \\
\hline
BB Ari & 2016 & 0.07514 & 0.00007 \\
HT Cas & 2016 & 0.07807 & 0.00004 \\
GS Cet & 2016 & 0.05763 & 0.00027 \\
GZ Cnc & 2017 & 0.09599 & -- \\
V1113 Cyg & 2016 & 0.08030 & 0.00018 \\
IR Gem & 2017 & 0.07315 & 0.00000 \\
HV Vir & 2016 & 0.05824 & 0.00001 \\
ASASSN-13ak & 2016 & 0.08884 & 0.00047 \\
ASASSN-15cr & 2017 & 0.06258 & 0.00014 \\
ASASSN-16da & 2016 & 0.05858 & 0.00010 \\
ASASSN-16ds & 2016 & 0.06856 & 0.00015 \\
ASASSN-16dt & 2016 & 0.06512 & 0.00001 \\
ASASSN-16eg & 2016 & 0.07989 & 0.00004 \\
ASASSN-16hj & 2016 & 0.05691 & 0.00031 \\
ASASSN-16ib & 2016 & 0.06036 & 0.00007 \\
ASASSN-16ik & 2016 & 0.06656 & 0.00010 \\
ASASSN-16iw & 2016 & 0.06691 & 0.00012 \\
ASASSN-16jb & 2016 & 0.06514 & 0.00003 \\
ASASSN-16jd & 2016 & 0.05840 & 0.00012 \\
ASASSN-16js & 2016 & 0.06165 & 0.00010 \\
ASASSN-16ob & 2016 & 0.05785 & 0.00016 \\
ASASSN-16oi & 2016 & 0.05738 & 0.00009 \\
ASASSN-16os & 2016 & 0.05596 & 0.00006 \\
ASASSN-17bl & 2017 & 0.05599 & 0.00004 \\
CRTS J164950 & 2015 & 0.06641 & 0.00006 \\
MASTER J021315 & 2016 & 0.10712 & 0.00025 \\
MASTER J030205 & 2016 & 0.06275 & 0.00015 \\
OT J002656 & 2016 & 0.13320 & 0.00003 \\
TCP J013758 & 2016 & 0.06290 & 0.00056 \\
\hline
\end{tabular}
\end{center}
\end{table}

\subsection{WZ Sge-type objects}\label{sec:wzsgetype}

   WZ Sge-type dwarf novae are a subclass of SU UMa-type
dwarf novae characterized by the presence of early superhumps
(\cite{kat96alcom}; \cite{kat02wzsgeESH};
\cite{ish02wzsgeletter}; see a recent review
\cite{kat15wzsge}).
They are seen during the earliest stages of
a superoutburst, and have period almost equal
to the orbital periods.

   These early superhumps are considered to be a result
of the 2:1 resonance \citep{osa02wzsgehump}.
These objects usually show very rare outbursts
(once in several years to decades) with large
outburst amplitudes (6--9 mag or even more,
\cite{kat15wzsge}) and often have
complex light curves \citep{kat15wzsge}.
The WZ Sge-type dwarf novae are of special astrophysical
interest for several reasons.  We list two of them:
(1) From the point of view of outburst physics,
the origin of the complex light curves, including repetitive
rebrightenings, is not well understood.
They are also considered to be analogous to
black-hole X-ray transients which often show rebrightenings
(cf. \cite{kuu96TOAD}) and there may be common
underlying physics between WZ Sge-type dwarf novae
and black-hole X-ray transients.
(2) From the point of view of CV evolution,
they are considered to represent the terminal stage of
CV evolution and they may have brown-dwarf secondaries.
Studies of WZ Sge-type dwarf novae are indispensable
when discussing the terminal stage of CV evolution,
such as the period minimum and period bouncers
(e.g. \cite{kni06CVsecondary}; \cite{kni11CVdonor};
\cite{pat11CVdistance}; \cite{kat15wzsge}).
We used the period of early superhumps as the approximate
orbital period (\cite{Pdot6}; \cite{kat15wzsge};
labeled as `E' in table \ref{tab:perlist}).
In table \ref{tab:wztab}, we list the parameters of
WZ Sge-type dwarf novae (including likely ones).

\begin{table*}
\caption{Parameters of WZ Sge-type superoutbursts.}\label{tab:wztab}
\begin{center}
\begin{tabular}{cccccccccccc}
\hline
Object & Year & $P_{\rm SH}$\commenta & $P_{\rm orb}$\commentb & $P_{\rm dot}$\commentc & err\commentc & $\epsilon$ & Type\commentd & $N_{\rm reb}$\commente & delay\commentf & Max & Min \\
\hline
GS Cet & 2016 & 0.056645 & 0.05597 & 6.3 & 0.6 & 0.012 & -- & -- & 8 & 13.0 & 20.4 \\
ASASSN-16da & 2016 & 0.057344 & 0.05610 & 7.5 & 0.9 & 0.022 & -- & -- & 5 & 15.1 & 21.5 \\
ASASSN-16dt & 2016 & 0.064507 & 0.064197 & $-$1.6 & 0.5 & 0.005 & E+C & 2 & 23 & 13.4 & 21.1: \\
ASASSN-16eg & 2016 & 0.077880 & 0.075478 & 10.4 & 0.8 & 0.032 & C  & 1  & 6 & 12.7 & 19.4 \\
ASASSN-16fu & 2016 & 0.056936 & 0.05623 & 4.6 & 0.6 & 0.013 & -- & -- & 6 & 13.9 & 21.6 \\
ASASSN-16gh & 2016 & 0.061844 & -- & 6.7 & 2.7 & -- & -- & -- & 12 & 14.3 & [22 \\
ASASSN-16gj & 2016 & 0.057997 & -- & 7.0 & 1.0 & -- & A: & 1 & $\leq$9 & ]13.3 & 21.3 \\
ASASSN-16hg & 2016 & 0.062371 & -- & 0.6 & 1.7 & -- & A: & 1 & $\geq$6 & ]14.1 & 21.6: \\
ASASSN-16hj & 2016 & 0.055644 & 0.05499 & 11.3 & 1.3 & 0.012 & A+B? & 3 & 9 & 14.2 & 21.1: \\
ASASSN-16ia & 2016 & -- & 0.05696 & -- & -- & -- & -- & -- & -- & 14.6 & [22 \\
ASASSN-16is & 2016 & 0.058484 & 0.05762 & 4.2 & 1.7 & 0.015 & -- & -- & 11 & 14.9 & 20.4 \\
ASASSN-16iw & 2016 & 0.065462 & 0.06495 & 10.0 & 3.2 & 0.008 & B & 5 & 7 & 13.9 & 21.9 \\
ASASSN-16jb & 2016 & 0.064397 & 0.06305 & 5.9 & 0.7 & 0.021 & -- & -- & 7 & 13.3 & [21 \\
ASASSN-16js & 2016 & 0.060934 & 0.06034 & 4.9 & 1.0 & 0.010 & -- & -- & 10 & 13.0 & 20.1 \\
ASASSN-16lo & 2016 & 0.054608 & 0.05416 & -- & -- & 0.008 & -- & -- & 11 & 14.3 & 20.7: \\
ASASSN-16ob & 2016 & 0.057087 & -- & 1.8 & 0.5 & -- & -- & -- & 13 & 13.8 & [22 \\
ASASSN-16oi & 2016 & 0.056241 & 0.05548 & 5.0 & 1.7 & 0.014 & -- & -- & 8 & 13.4 & 22.0: \\
ASASSN-16os & 2016 & 0.054992 & 0.05494 & 0.3 & 1.4 & 0.001 & -- & -- & 8 & 13.6 & 21.4: \\
ASASSN-17aa & 2017 & 0.054591 & 0.05393 & 2.8 & 0.3 & 0.012 & -- & -- & 9 & 13.9 & [22 \\
ASASSN-17bl & 2017 & 0.055367 & 0.05467 & 3.6 & 0.6 & 0.013 & -- & -- & 11 & 13.7 & [22 \\
ASASSN-17cn & 2017 & 0.053991 & 0.05303 & 5.6 & 0.8 & 0.018 & -- & -- & $\geq$9 & 13.2 & [22 \\
\hline
  \multicolumn{12}{l}{\commenta Representative value ($P_1$).} \\
  \multicolumn{12}{l}{\commentb Period of early superhumps.} \\
  \multicolumn{12}{l}{\commentc Unit $10^{-5}$.} \\
  \multicolumn{12}{l}{\commentd A: long-lasting rebrightening; B: multiple rebegitehnings; C: single rebrightening; D: no rebrightening.} \\
  \multicolumn{12}{l}{\commente Number of rebrightenings.} \\
  \multicolumn{12}{l}{\commentf Days before ordinary superhumps appeared.} \\
\end{tabular}
\end{center}
\end{table*}

\begin{figure*}
  \begin{center}
    \FigureFile(110mm,88mm){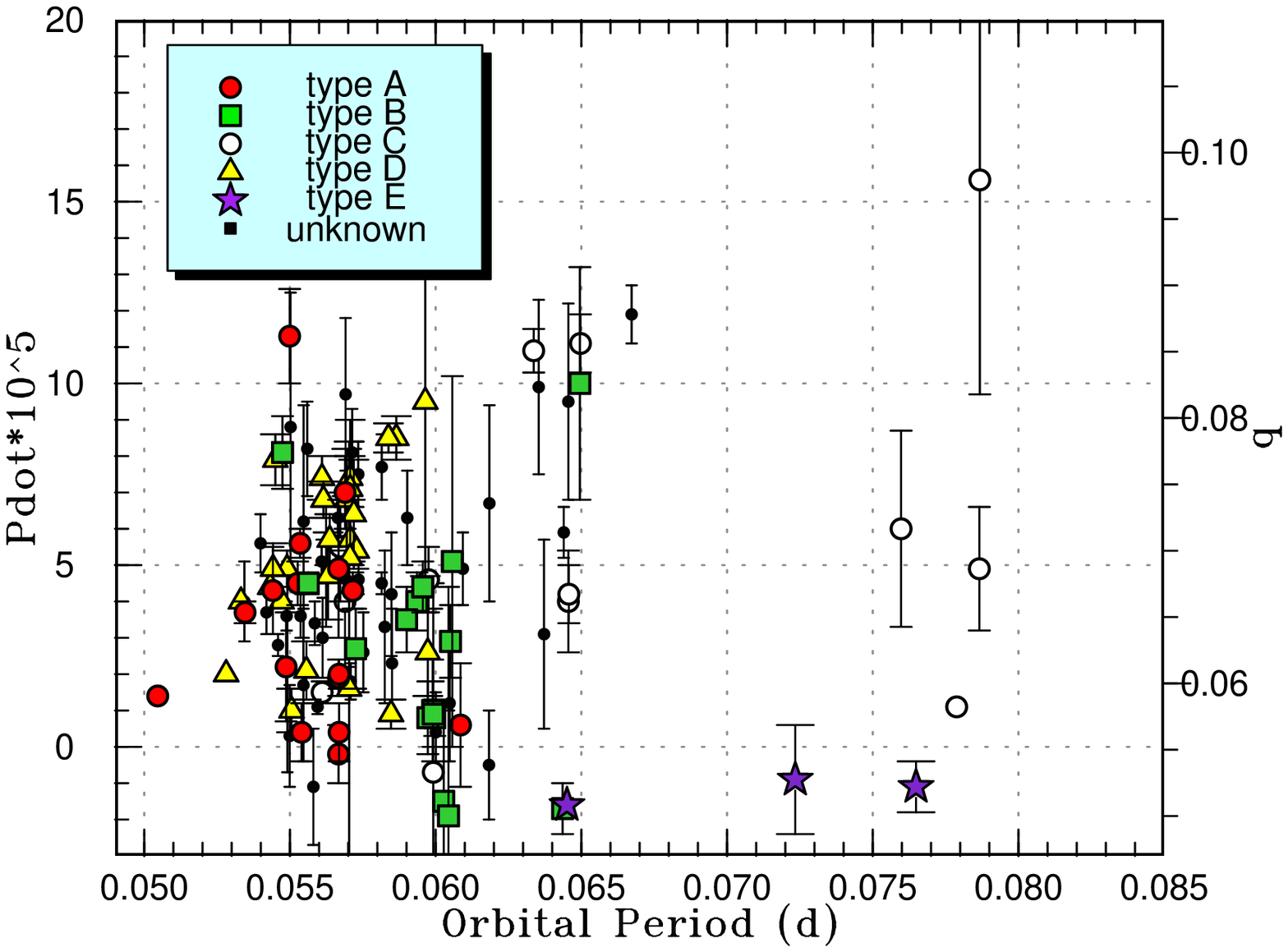}
  \end{center}
  \caption{$P_{\rm dot}$ versus $P_{\rm orb}$ for WZ Sge-type
  dwarf novae.  Symbols represent the type of outburst:
  type-A (filled circles), type-B (filled squares),
  type-C (filled triangles), type-D (open circles)
  and type-E (filled stars) (see text for details).
  On the right side, we show mass ratios estimated
  using equation (6) in \citet{kat15wzsge}.
  We can regard this figure as to represent
  an evolutionary diagram.
  }
  \label{fig:wzpdottype9}
\end{figure*}

   It has been known that $P_{\rm dot}$
and $P_{\rm orb}$ are correlated with the rebrightening type
[starting with figure 36 in \cite{Pdot} and refined in
\citet{Pdot}--\citet{Pdot7} and \citet{kat15wzsge},
and updated in \citet{Pdot8}].
The five types of outbursts based on rebrightenings are:
type-A outbursts [long-duration rebrightening; we include
type-A/B introduced in \citet{kat15wzsge}],
type-B outbursts (multiple rebrightenings),
type-C outbursts (single rebrightening),
type-D outbursts (no rebrightening) and
type-E outbursts (double superoutburst, with ordinary superhumps
only during the second one).
In figure \ref{fig:wzpdottype9}, we show the updated
result up to this paper.
In this figure, we also added objects without known
rebrightening types.  These objects have been confirmed
to follow the same trend, which we consider
to represent the evolutionary track [see subsection 7.6
in \citet{kat15wzsge}].  The outlier around
$P_{\rm orb}$=0.050~d is ASASSN-15po, the object
below the period minimum \citep{nam17asassn15po}.
The two points around $P_{\rm orb}$=0.079~d is RZ Leo.
As shown in subsection 3.17 in \citet{Pdot8}, the two
superoutbursts in 2000 and 2016 were not sufficiently
covered and $P_{\rm dot}$ values had large errors
and we consider that these points are not very reliable.

\subsection{Objects with Very Short Supercycles}\label{sec:v503cyglike}

   In this study, there were a group of four object
with very short supercycles:
NY Her [supercycle 63.5(2)~d, subsection \ref{obj:nyher}],
1RXS J161659 [89(1)~d, subsection \ref{obj:j1616}],
CRTS J033349 [108(1)~d, subsection \ref{obj:j0333}] and
SDSS J153015 [84.7(1.2)~d, subsection \ref{obj:j1530}].
We are not aware whether such a large number of detections
were by chance or as a result of the recent change in
detection policies of transients such as ASAS-SN.
The most notable common features of these objects are
the small number of normal outbursts.  Since the short
supercycle reflects the high mass-transfer rate
(cf. \cite{osa96review}), the small number of normal
outbursts is unexpected.

   A likely solution to this apparent inconsistency is 
the disk tilt, which would prevent the accreted matter
accumulating in the outer edge of the disk and
it would suppress normal outbursts
(\cite{ohs12eruma}; \cite{osa13v1504cygKepler};
\cite{osa13v344lyrv1504cyg}).  It has been also
demonstrated that the prototypical example V503 Cyg
(supercycle 89~d) showed negative superhumps
\citep{har95v503cyg}, which are considered to be
a consequence of a disk tilt (e.g. \cite{woo07negSH};
\cite{mon10negSH}).  The temporary emergence of frequent
normal outbursts in V503 Cyg \citep{kat02v503cyg}
also suggests that normal outbursts were somehow
suppressed, most likely by a disk tilt.
More recent examples in Kepler dwarf novae
V1504 Cyg and V344 Lyr in relation to transiently
appearing negative superhumps were discussed
in \citet{osa13v1504cygKepler}, \citet{osa13v344lyrv1504cyg}
and it has become more evident that the state with
negative superhumps (i.e. the disk is likely tilted)
is associated with the reduced number of normal outbursts.

   It has been proposed that a high mass-transfer
rate is prone to produce a disk tilt in a hydrodynamical
model \citep{mon10disktilt}.  If it is indeed the case,
the large number of SU UMa-type dwarf novae with
few normal outbursts but with short supercycles
may be a result of easy occurrence of a disk tilt
in high-mass transfer systems and may not be surprising.
A search for negative superhumps in the four systems 
reported in this paper is recommended.
Long-term monitoring is also encouraged to see whether
these objects switch the outburst mode as in V503 Cyg.

   We should make a comment on another SU UMa-type
dwarf nova V4140 Sgr with a short supercycle
(80--90~d, \cite{bor05v4140sgr}).
\citet{bor05v4140sgr} and \citet{bap16v4140sgr}
used the eclipse mapping method and derived
a conclusion that the distribution of the disk temperature
in quiescence is incompatible with the disk-instability
model and they interpreted that the outbursts in
this object are caused by mass-transfer bursts
from the secondary.  We noticed that, despite
its short supercycle, this object rarely shows
normal outbursts (for example, there were no
outburst between 2017 March 12 and April 16,
observations by J. Hambsch).  The object appears
to be a V503 Cyg-like one and we can expect
a disk tilt.  The apparent deviation of
the distribution of the disk temperature may have
been caused by an accretion stream hitting
the inner parts of the disk when the disk is
tilted and may not be contradiction with
the picture of the disk-instability model.
Since no other V503 Cyg-like objects are eclipsing,
we have had no observation about the structure of 
the disk in V503 Cyg-like objects.
We propose to study V4140 Sgr more closely
for detecting negative superhumps, and searching
for a switch in the outburst mode to test
the possibility of the V503 Cyg-like nature.

\section{Summary}\label{sec:summary}

   We provided updated statistics of the period 
distribution.  We obtained the period minimum of 0.05290(2)~d
and confirmed the presence of the period gap above
$P_{\rm orb} \sim$ 0.09~d.
We refined the $P_{\rm orb}$--$P_{\rm dot}$
relation of SU UMa-type dwarf novae, the updated evolutionary
track using stage A superhumps and refined relationship between
$P_{\rm orb}$--$P_{\rm dot}$ versus the rebrightening
type in WZ Sge-type dwarf novae.  We also provide
basic observational data of superoutbursts we studied for
SU UMa-type dwarf novae.

   The objects of special interest in this paper can be
summarized as follows:

\begin{itemize}

\item Four objects (NY Her, 1RXS J161659, CRTS J033349
and SDSS J153015) have supercycles shorter than 100~d.
These objects do not resemble ER UMa-type dwarf novae
but show infrequent normal outbursts as in V503 Cyg.
We consider that these properties may be caused by
a tilted disk and we expect to detect negative superhumps
in these systems.

\item DDE 48 is likely an ER UMa-type dwarf nova.
NSV 2026 also has a short supercycle but with frequent
normal outbursts.

\item ASASSN-16ia and ASASSN-16js are WZ Sge-type dwarf novae 
with large-amplitude early superhumps.

\item ASASSN-16ia showed a precursor outburst prior
to the WZ Sge-type superoutburst.  This is the first
certain case of such a precursor outburst in a WZ Sge-type
superoutburst.

\item ASASSN-16js has a low mass ratio and is most likely
a period bouncer.  ASASSN-16gh is also a candidate for
a period bouncer.

\item ASASSN-16iw is a WZ Sge-type dwarf nova
with five post-superoutburst rebrightenings.

\item MASTER J021315 is located in the period gap.
This object likely showed long-lasting phase of stage A.
Outbursts in this system were relatively rare and
it should have a low mass-transfer rate.

\item ASASSN-16kg, ASASSN-16ni, CRTS J000130 and SDSS J113551
are also SU UMa-type dwarf novae in the period gap.
ASASSN-16ni is possibly an SU UMa-type dwarf nova
in or above the period gap.

\item ASASSN-16kg and ASASSN-16ni have large outburst
amplitudes.  ASASSN-16kg showed a rebrighening.
These properties are uncommon among dwarf novae
in the period gap.

\item MASTER J030205 showed a likely spin period
and it is likely a rare intermediate polar
among SU UMa-type dwarf novae.

\item Five objects OY Car, GP CVn, V893 Sco,
MASTER J220559 and SDSS J115207 are eclipsing and
we present refined eclipse ephemerides for some of them.

\end{itemize}

\section*{Acknowledgements}

This work was supported by the Grant-in-Aid
``Initiative for High-Dimensional Data-Driven Science through Deepening
of Sparse Modeling'' (25120007) 
from the Ministry of Education, Culture, Sports, 
Science and Technology (MEXT) of Japan.
This work was also partially supported by
Grant VEGA 2/0008/17 and APVV-15-0458 
(by Shugarov, Chochol, Seker\'a\v{s}), 
NSh-9670.2016.2 (Voloshina, Katysheva)
RFBR 17-52-175300 (Voloshina),
RSF-14-12-00146 (Golysheva for processing observations data 
from Slovak Observatory) and
APVV-15-0458 (Dubovsky, Kudzej).
ASAS-SN is supported by the Gordon and Betty Moore
Foundation through grant GBMF5490 to the Ohio State
University and NSF grant AST-1515927.
The authors are grateful to observers of VSNET Collaboration and
VSOLJ observers who supplied vital data.
We acknowledge with thanks the variable star
observations from the AAVSO International Database contributed by
observers worldwide and used in this research.
We are also grateful to the VSOLJ database.
This work is helped by outburst detections and announcement
by a number of variable star observers worldwide,
including participants of CVNET and BAA VSS alert.
The CCD operation of the Bronberg Observatory is partly sponsored by
the Center for Backyard Astrophysics.
We are grateful to the Catalina Real-time Transient Survey
team for making their real-time detection of transient objects
and the past photometric database available to the public.
We are also grateful to the ASAS-3 team for making
the past photometric database available to the public.
This research has made use of the SIMBAD database,
operated at CDS, Strasbourg, France.
This research has made use of the International Variable Star Index 
(VSX) database, operated at AAVSO, Cambridge, Massachusetts, USA.

\section*{Supporting information}

(In the PASJ verision):
Additional supporting information can be found in the online version
of this article: Tables.
Figures.\\
Supplementary data is available at PASJ Journal online.


\begin{thebibliography}{}

\bibitem[{Ahn} et~al.(2012)]{SDSS9}
  {Ahn}, C.~P., {et~al.}\ 2012, ApJS, 203, 21

\bibitem[Antipin(1999)]{ant99v368pegftcamv367pegv2209cyg}
  Antipin, S.~V.\ 1999, IBVS, 4673, 1

\bibitem[Arenas and Mennickent(1998)]{are98akcnc}
  Arenas, J., \& Mennickent, R.~E.\ 1998, A\&A, 337, 472

\bibitem[{Balanutsa} et~al.(2013)]{bal13j1616}
  {Balanutsa}, P., {Denisenko}, D., {Gorbovskoy}, E., \& {Lipunov}, V.\ 2013,
  Perem.\ Zvezdy, submitted (arXiv/1307.7396)

\bibitem[{Balanutsa} et~al.(2014)]{bal14j0439atel5787}
  {Balanutsa}, P., {et~al.}\ 2014, Astron.\ Telegram, 5787

\bibitem[{Balanutsa} et~al.(2016a)]{bal16asassn16gyatel9174}
  {Balanutsa}, P., {et~al.}\ 2016a, Astron.\ Telegram, 9174

\bibitem[{Balanutsa} et~al.(2016b)]{bal16j0302atel9824}
  {Balanutsa}, P., {et~al.}\ 2016b, Astron.\ Telegram, 9824

\bibitem[{Baptista} et~al.(2016)]{bap16v4140sgr}
  {Baptista}, R., {Borges}, B.~W., \& {Oliveira}, A.~S.\ 2016, MNRAS, 463, 3799

\bibitem[Barwig et~al.(1992)]{bar92hvvir}
  Barwig, H., Mantel, K.~H., \& Ritter, H.\ 1992, A\&A, 266, L5

\bibitem[{Belyavskii}(1936)]{bel36nsv14681}
  {Belyavskii}, S.~I.\ 1936, Perem.\ Zvezdy, 5, 36

\bibitem[{Bernhard} et~al.(2005)]{ber05j0532}
  {Bernhard}, K., {Lloyd}, C., {Berthold}, T., {Kriebel}, W., \& {Renz}, W.\
  2005, IBVS, 5620, 1

\bibitem[{Bohlsen}(2016)]{boh16asassn16kdatel9477}
  {Bohlsen}, T.\ 2016, Astron.\ Telegram, 9477

\bibitem[{Bond}(1978)]{bon78bluevar2}
  {Bond}, H.~E.\ 1978, PASP, 90, 526

\bibitem[{Borges} and {Baptista}(2005)]{bor05v4140sgr}
  {Borges}, B.~W., \& {Baptista}, R.\ 2005, A\&A, 437, 235

\bibitem[{Boyce}(1942)]{boy42v699oph}
  {Boyce}, E.~H.\ 1942, Annals\ of\ the\ Astron.\ Obs.\ of\ Harvard\ Coll.\,
  109, 11

\bibitem[{Boyd} et~al.(2007)]{boy07v337cyg}
  {Boyd}, D., {Krajci}, T., {Shears}, J., \& {Poyner}, G.\ 2007, J.\ Br.\
  Astron.\ Assoc., 117, 198

\bibitem[Bruch et~al.(2000)]{bru00v893sco}
  Bruch, A., Steiner, J.~E., \& Gneiding, C.~D.\ 2000, PASP, 112, 237

\bibitem[{Burenkov} and {Voikhanskaia}(1979)]{bur79DNspec}
  {Burenkov}, A.~N., \& {Voikhanskaia}, N.~F.\ 1979, Soviet\ Astronomy\
  Letters, 5, 452

\bibitem[{Cleveland}(1979)]{LOWESS}
  {Cleveland}, W.~S.\ 1979, J. Amer. Statist. Assoc., 74, 829

\bibitem[{Coppejans} et~al.(2016)]{cop16DNCRTS}
  {Coppejans}, D.~L., {K{\"o}rding}, E.~G., {Knigge}, C., {Pretorius}, M.~L.,
  {Woudt}, P.~A., {Groot}, P.~J., {Van Eck}, C.~L., \& {Drake}, A.~J.\ 2016,
  MNRAS, 456, 4441

\bibitem[{Cutri} et~al.(2003)]{2MASS}
  {Cutri}, R.~M., {et~al.}\ 2003, {2MASS} {All Sky Catalog} of point sources
  (NASA/IPAC Infrared Science Archive)

\bibitem[{Davis} et~al.(2015)]{dav15ASASSNCVAAS}
  {Davis}, A.~B., {Shappee}, B.~J., {Archer Shappee}, B., \& {ASAS-SN}\ 2015,
  American\ Astron.\ Soc.\ Meeting\ Abstracts, 225, \#344.02

\bibitem[{Denisenko} et~al.(2013a)]{den13j1623atel5643}
  {Denisenko}, D., {et~al.}\ 2013a, Astron.\ Telegram, 5643

\bibitem[{Denisenko} et~al.(2012)]{den12j0426atel4441}
  {Denisenko}, D., {et~al.}\ 2012, Astron.\ Telegram, 4441

\bibitem[{Denisenko} et~al.(2013b)]{den13j1748atel5182}
  {Denisenko}, D., {et~al.}\ 2013b, Astron.\ Telegram, 5182

\bibitem[{Denisenko}(2012)]{den12USNOCVs}
  {Denisenko}, D.~V.\ 2012, Astron.\ Lett., 38, 249

\bibitem[{Drake} et~al.(2009)]{CRTS}
  {Drake}, A.~J., {et~al.}\ 2009, ApJ, 696, 870

\bibitem[{Drake} et~al.(2014)]{dra14CRTSCVs}
  {Drake}, A.~J., {et~al.}\ 2014, MNRAS, 441, 1186

\bibitem[{Drake} et~al.(2008)]{dra08atel1479}
  {Drake}, A.~J., {Mahabal}, A., {Djorgovski}, S.~G., {Graham}, M.~J.,
  {Williams}, R., {Beshore}, E.~C., {Larson}, S.~M., \& {Christensen}, E.\
  2008, Astron.\ Telegram, 1479

\bibitem[Duerbeck(1984)]{due84hvvir}
  Duerbeck, H.~W.\ 1984, IBVS, 2502

\bibitem[Duerbeck(1987)]{due87novaatlas}
  Duerbeck, H.~W.\ 1987, Space\ Sci.\ Rev., 45, 1

\bibitem[{Echeistov} et~al.(2014)]{ech14j0653atel5898}
  {Echeistov}, V., {et~al.}\ 2014, Astron.\ Telegram, 5898

\bibitem[Feinswog et~al.(1988)]{fei88irgem}
  Feinswog, L., Szkody, P., \& Garnavich, P.\ 1988, AJ, 96, 1702

\bibitem[Fernie(1989)]{fer89error}
  Fernie, J.~D.\ 1989, PASP, 101, 225

\bibitem[{Gaia Collaboration}(2016)]{GaiaDR1}
  {Gaia Collaboration}\ 2016, VizieR\ Online\ Data\ Catalog, 1337

\bibitem[{Gorbovskoy} et~al.(2013)]{MASTER}
  {Gorbovskoy}, E.~S., {et~al.}\ 2013, Astron.\ Rep., 57, 233

\bibitem[{Gress} et~al.(2017)]{gre17j1505atel10061}
  {Gress}, O., {et~al.}\ 2017, Astron.\ Telegram, 61

\bibitem[{Hameury} and {Lasota}(2017)]{ham17DNIP}
  {Hameury}, J.-M., \& {Lasota}, J.-P.\ 2017, A\&A, in press (arXiv/1703.03563)

\bibitem[{Han} et~al.(2015)]{han15oycar}
  {Han}, Z.-T., {Qian}, S.-B., {Fern{\'a}ndez Laj{\'u}s}, E., {Liao}, W.-P., \&
  {Zhang}, J.\ 2015, New\ Astron., 34, 1

\bibitem[Harvey et~al.(1995)]{har95v503cyg}
  Harvey, D., Skillman, D.~R., Patterson, J., \& Ringwald, F.~A.\ 1995, PASP,
  107, 551

\bibitem[{Hirose} and {Osaki}(1990)]{hir90SHexcess}
  {Hirose}, M., \& {Osaki}, Y.\ 1990, PASJ, 42, 135

\bibitem[{Hirose} and {Osaki}(1993)]{hir93SHperiod}
  {Hirose}, M., \& {Osaki}, Y.\ 1993, PASJ, 45, 595

\bibitem[{Hoffleit}(1935)]{hof35newvar}
  {Hoffleit}, D.\ 1935, Harvard\ Coll.\ Obs.\ Bull., 901, 20

\bibitem[{Hoffmeister}(1949)]{hof49newvar}
  {Hoffmeister}, C.\ 1949, Erg.\ Astron.\ Nachr., 12, 12

\bibitem[{Hoffmeister}(1964)]{hof64an28849}
  {Hoffmeister}, C.\ 1964, Astron.\ Nachr., 288, 49

\bibitem[{Hoffmeister}(1966)]{hof66an289139}
  {Hoffmeister}, C.\ 1966, Astron.\ Nachr., 289, 139

\bibitem[Howell et~al.(1990)]{how90faintCV3}
  Howell, S.~B., Szkody, P., Kreidl, T.~J., Mason, K.~O., \& Puchnarewicz,
  E.~M.\ 1990, PASP, 102, 758

\bibitem[{Imada} et~al.(2009)]{ima09j0532}
  {Imada}, A., {et~al.}\ 2009, PASJ, 61, L17

\bibitem[{Imada} et~al.(2006)]{ima06j0137}
  {Imada}, A., {et~al.}\ 2006, PASJ, 58, 143

\bibitem[Ishioka et~al.(2001)]{ish01ixdra}
  Ishioka, R., Kato, T., Uemura, M., Iwamatsu, H., Matsumoto, K., Martin,
  B.~E., Billings, G.~W., \& Novak, R.\ 2001, PASJ, 53, L51

\bibitem[Ishioka et~al.(2003)]{ish03hvvir}
  Ishioka, R., {et~al.}\ 2003, PASJ, 55, 683

\bibitem[{Ishioka} et~al.(2007)]{ish07CVIR}
  {Ishioka}, R., {Sekiguchi}, K., \& {Maehara}, H.\ 2007, PASJ, 59, 929

\bibitem[Ishioka et~al.(2002)]{ish02wzsgeletter}
  Ishioka, R., {et~al.}\ 2002, A\&A, 381, L41

\bibitem[{Kapusta} and {Thorstensen}(2006)]{kap06j0532}
  {Kapusta}, A.~B., \& {Thorstensen}, J.~R.\ 2006, PASP, 118, 1119

\bibitem[Kato(1994)]{kat94akcnc}
  Kato, T.\ 1994, IBVS, 4136

\bibitem[Kato(2001)]{kat01irgem}
  Kato, T.\ 2001, IBVS, 5122

\bibitem[Kato(2002)]{kat02wzsgeESH}
  Kato, T.\ 2002, PASJ, 54, L11

\bibitem[{Kato}(2015)]{kat15wzsge}
  {Kato}, T.\ 2015, PASJ, 67, 108

\bibitem[{Kato} et~al.(2014a)]{Pdot6}
  {Kato}, T., {et~al.}\ 2014a, PASJ, 66, 90

\bibitem[{Kato} et~al.(2015a)]{Pdot7}
  {Kato}, T., {et~al.}\ 2015a, PASJ, 67, 105

\bibitem[Kato et~al.(2002a)]{kat02gzcncnsv10934}
  Kato, T., {et~al.}\ 2002a, A\&A, 396, 929

\bibitem[{Kato} et~al.(2013)]{Pdot4}
  {Kato}, T., {et~al.}\ 2013, PASJ, 65, 23

\bibitem[{Kato} et~al.(2014b)]{Pdot5}
  {Kato}, T., {et~al.}\ 2014b, PASJ, 66, 30

\bibitem[{Kato} et~al.(2016a)]{Pdot8}
  {Kato}, T., {et~al.}\ 2016a, PASJ, 68, 65

\bibitem[{Kato} et~al.(2015b)]{kat15ccscl}
  {Kato}, T., {Hambsch}, F.-J., {Oksanen}, A., {Starr}, P., \& {Henden}, A.\
  2015b, PASJ, 67, 3

\bibitem[Kato et~al.(2000)]{kat00ssumi}
  Kato, T., Hanson, G., Poyner, G., Muyllaert, E., Reszelski, M., \& Dubovsky,
  P.~A.\ 2000, IBVS, 4932

\bibitem[Kato et~al.(1998)]{kat98v893sco}
  Kato, T., Haseda, K., Takamizawa, K., Kazarovets, E.~V., \& Samus, N.~N.\
  1998, IBVS, 4585

\bibitem[{Kato} et~al.(2009)]{Pdot}
  {Kato}, T., {et~al.}\ 2009, PASJ, 61, S395

\bibitem[Kato et~al.(2002b)]{kat02v503cyg}
  Kato, T., Ishioka, R., \& Uemura, M.\ 2002b, PASJ, 54, 1029

\bibitem[{Kato} and {Kunjaya}(1995)]{kat95eruma}
  {Kato}, T., \& {Kunjaya}, C.\ 1995, PASJ, 47, 163

\bibitem[{Kato} et~al.(2012a)]{Pdot3}
  {Kato}, T., {et~al.}\ 2012a, PASJ, 64, 21

\bibitem[{Kato} et~al.(2012b)]{kat12DNSDSS}
  {Kato}, T., {Maehara}, H., \& {Uemura}, M.\ 2012b, PASJ, 64, 62

\bibitem[{Kato} et~al.(2010)]{Pdot2}
  {Kato}, T., {et~al.}\ 2010, PASJ, 62, 1525

\bibitem[Kato et~al.(1996a)]{kat96alcom}
  Kato, T., Nogami, D., Baba, H., Matsumoto, K., Arimoto, J., Tanabe, K., \&
  Ishikawa, K.\ 1996a, PASJ, 48, L21

\bibitem[Kato et~al.(1996b)]{kat96v1113cyg}
  Kato, T., Nogami, D., Masuda, S., \& Hirata, R.\ 1996b, PASJ, 48, 45

\bibitem[{Kato} and {Osaki}(2013)]{kat13qfromstageA}
  {Kato}, T., \& {Osaki}, Y.\ 2013, PASJ, 65, 115

\bibitem[{Kato} et~al.(2016b)]{kat16v1006cyg}
  {Kato}, T., {et~al.}\ 2016b, PASJ, 68, L4

\bibitem[{Kato} et~al.(2001)]{kat01hvvir}
  {Kato}, T., {Sekine}, Y., \& {Hirata}, R.\ 2001, PASJ, 53, 1191

\bibitem[Kato et~al.(2001a)]{kat01bfara}
  Kato, T., Stubbings, R., Pearce, A., Nelson, P., \& Monard, B.\ 2001a, IBVS,
  5119, 1

\bibitem[{Kato} et~al.(2017)]{kat17j0026}
  {Kato}, T., {et~al.}\ 2017, PASJ, in press (arXiv/1703.00650)

\bibitem[Kato et~al.(2001b)]{kat01gzcnc}
  Kato, T., Uemura, M., Buczynski, D., \& Schmeer, P.\ 2001b, IBVS, 5123

\bibitem[Kato et~al.(2004)]{VSNET}
  Kato, T., Uemura, M., Ishioka, R., Nogami, D., Kunjaya, C., Baba, H., \&
  Yamaoka, H.\ 2004, PASJ, 56, S1

\bibitem[Kholopov et~al.(1985)]{GCVS}
  Kholopov, P.~N., {et~al.}\ 1985, General Catalogue of Variable Stars, fourth
  edition (Moscow: Nauka Publishing House)

\bibitem[{Kimura} et~al.(2017)]{kim17asassn16dt16hg}
  {Kimura}, M., {et~al.}\ 2017, PASJ, submitted

\bibitem[{Kimura} et~al.(2016)]{kim16alcom}
  {Kimura}, M., {et~al.}\ 2016, PASJ, 68, L2

\bibitem[{Knigge}(2006)]{kni06CVsecondary}
  {Knigge}, C.\ 2006, MNRAS, 373, 484

\bibitem[{Knigge} et~al.(2011)]{kni11CVdonor}
  {Knigge}, C., {Baraffe}, I., \& {Patterson}, J.\ 2011, ApJS, 194, 28

\bibitem[{Kukarkin} et~al.(1982)]{NSV}
  {Kukarkin}, B.~V., {et~al.}\ 1982, {New Catalogue of Suspected Variable
  Stars} (Moscow: Nauka Publishing House)

\bibitem[Kuulkers et~al.(1996)]{kuu96TOAD}
  Kuulkers, E., Howell, S.~B., \& van Paradijs, J.\ 1996, ApJ, 462, L87

\bibitem[{Lasker} et~al.(2007)]{GSC232}
  {Lasker}, B., {Lattanzi}, M.~G., {McLean}, B.~J., \& {et al.}\ 2007, VizieR\
  Online\ Data\ Catalog, 1305

\bibitem[{Lazaro} et~al.(1991)]{laz91irgem}
  {Lazaro}, C., {Martinez-Pais}, I.~G., {Arevalo}, M.~J., \& {Solheim}, J.~E.\
  1991, AJ, 101, 196

\bibitem[{L{\'a}zaro} et~al.(1990)]{laz90irgem}
  {L{\'a}zaro}, C., {Martinez-Pais}, I.~G., {Solheim}, J.~E., \& {Ar{\'e}valo},
  M.~J.\ 1990, Ap\&SS, 169, 257

\bibitem[{Leibowitz} et~al.(1994)]{lei94hvvir}
  {Leibowitz}, E.~M., {Mendelson}, H., {Bruch}, A., {Duerbeck}, H.~W.,
  {Seitter}, W.~C., \& {Richter}, G.~A.\ 1994, ApJ, 421, 771

\bibitem[{Littlefair} et~al.(2008)]{lit08eclCV}
  {Littlefair}, S.~P., {Dhillon}, V.~S., {Marsh}, T.~R., {G{\"a}nsicke}, B.~T.,
  {Southworth}, J., {Baraffe}, I., {Watson}, C.~A., \& {Copperwheat}, C.\ 2008,
  MNRAS, 388, 1582

\bibitem[{Littlefield} et~al.(2013)]{lit13sbs1108}
  {Littlefield}, C., {et~al.}\ 2013, AJ, 145, 145

\bibitem[Liu et~al.(1999)]{liu99CVspec2}
  Liu, Wu., Hu, J.~Y., Li, Z.~Y., \& Cao, L.\ 1999, ApJS, 122, 257

\bibitem[{Lubow}(1991)]{lub91SHa}
  {Lubow}, S.~H.\ 1991, ApJ, 381, 259

\bibitem[{Lubow}(1992)]{lub92SH}
  {Lubow}, S.~H.\ 1992, ApJ, 401, 317

\bibitem[{Luckas}(2016)]{kuc16asassn16maatel9678}
  {Luckas}, P.\ 2016, Astron.\ Telegram, 9678

\bibitem[{Markarian} and {Stepanian}(1983)]{mar83SBS1}
  {Markarian}, B.~E., \& {Stepanian}, D.~A.\ 1983, Astrofizika, 19, 639

\bibitem[Mason and Howell(2003)]{mas03faintCV}
  Mason, E., \& Howell, S.\ 2003, A\&A, 403, 699

\bibitem[Matsumoto et~al.(2000)]{mat00v893sco}
  Matsumoto, K., Mennickent, R.~E., \& Kato, T.\ 2000, A\&A, 363, 1029

\bibitem[{Mayall}(1968)]{may68UG}
  {Mayall}, M.~W.\ 1968, JRASC, 62, 141

\bibitem[{Maza} et~al.(1990)]{maz90v344paviauc}
  {Maza}, J., {Hamuy}, M., {Wischnjewsky}, M., {Wells}, L., {Phillips}, M., \&
  {Barros}, S.\ 1990, IAU\ Circ., 5073

\bibitem[{McAllister} et~al.(2017a)]{mca17asassn14ag}
  {McAllister}, M.~J., {et~al.}\ 2017a, MNRAS, 464, 1353

\bibitem[{McAllister} et~al.(2017b)]{mca17j1057}
  {McAllister}, M.~J., {et~al.}\ 2017b, MNRAS, 467, 1024

\bibitem[{Meinunger}(1976)]{mei76irgem}
  {Meinunger}, L.\ 1976, Mitteil.\ Ver{\"{a}}nderl.\ Sterne, 7

\bibitem[Mennickent et~al.(1996)]{men96akcnc}
  Mennickent, R.~E., Nogami, D., Kato, T., \& Worraker, W.\ 1996, A\&A, 315,
  493

\bibitem[{Miller}(1971)]{mil71cygvar}
  {Miller}, W.~J.\ 1971, Ricerche\ Astronomiche, 8, 167

\bibitem[Montgomery(2001)]{mon01SH}
  Montgomery, M.~M.\ 2001, MNRAS, 325, 761

\bibitem[{Montgomery} and {Bisikalo}(2010)]{mon10negSH}
  {Montgomery}, M.~M., \& {Bisikalo}, D.~V.\ 2010, MNRAS, 405, 1397

\bibitem[{Montgomery} and {Martin}(2010)]{mon10disktilt}
  {Montgomery}, M.~M., \& {Martin}, E.~L.\ 2010, ApJ, 722, 989

\bibitem[{Morgenroth}(1933)]{mor33newVS}
  {Morgenroth}, O.\ 1933, Astron.\ Nachr., 250, 75

\bibitem[{Murray}(1998)]{mur98SH}
  {Murray}, J.~R.\ 1998, MNRAS, 297, 323

\bibitem[{Nakata} et~al.(2013)]{nak13j2112j2037}
  {Nakata}, C., {et~al.}\ 2013, PASJ, 65, 117

\bibitem[{Namekata} et~al.(2017)]{nam17asassn15po}
  {Namekata}, K., {et~al.}\ 2017, PASJ, 69, 2

\bibitem[{Niels Bohr Institute} et~al.(2014)]{CMC15}
  {Niels Bohr Institute}, U.~o.~C., {Institute of Astronomy}, UK, C., \& {Real
  Instituto y Observatorio de La Armada en San Fernando}\ 2014, VizieR\ Online\
  Data\ Catalog, 1327

\bibitem[{Ohshima} et~al.(2012)]{ohs12eruma}
  {Ohshima}, T., {et~al.}\ 2012, PASJ, 64, L3

\bibitem[{Olech} et~al.(2006)]{ole06ssumi}
  {Olech}, A., {Mularczyk}, K., {K{\c e}dzierski}, P., {Z{\l}oczewski}, K.,
  {Wi{\'s}niewski}, M., \& {Szaruga}, K.\ 2006, A\&A, 452, 933

\bibitem[{Olech} et~al.(2004)]{ole04ixdra}
  {Olech}, A., {Zloczewski}, K., {Mularczyk}, K., {Kedzierski}, P.,
  {Wisniewski}, M., \& {Stachowski}, G.\ 2004, Acta\ Astron., 54, 57

\bibitem[{Osaki}(1989)]{osa89suuma}
  {Osaki}, Y.\ 1989, PASJ, 41, 1005

\bibitem[{Osaki}(1996)]{osa96review}
  {Osaki}, Y.\ 1996, PASP, 108, 39

\bibitem[{Osaki} and {Kato}(2013a)]{osa13v1504cygKepler}
  {Osaki}, Y., \& {Kato}, T.\ 2013a, PASJ, 65, 50

\bibitem[{Osaki} and {Kato}(2013b)]{osa13v344lyrv1504cyg}
  {Osaki}, Y., \& {Kato}, T.\ 2013b, PASJ, 65, 95

\bibitem[{Osaki} and {Meyer}(2002)]{osa02wzsgehump}
  {Osaki}, Y., \& {Meyer}, F.\ 2002, A\&A, 383, 574

\bibitem[{Otulakowska-Hypka} et~al.(2013)]{otu13ixdra}
  {Otulakowska-Hypka}, M., {Olech}, A., {de Miguel}, E., {Rutkowski}, A.,
  {Koff}, R., \& {B\k{a}kakowska}, K.\ 2013, MNRAS, 429, 868

\bibitem[{Pastukhova}(1988)]{pas88nyher}
  {Pastukhova}, E.~N.\ 1988, Astron.\ Tsirk., 1534, 17

\bibitem[{Patterson}(2011)]{pat11CVdistance}
  {Patterson}, J.\ 2011, MNRAS, 411, 2695

\bibitem[Patterson et~al.(1993)]{pat93vyaqr}
  Patterson, J., Bond, H.~E., Grauer, A.~D., Shafter, A.~W., \& Mattei, J.~A.\
  1993, PASP, 105, 69

\bibitem[Patterson et~al.(1981)]{pat81wzsge}
  Patterson, J., McGraw, J.~T., Coleman, L., \& Africano, J.~L.\ 1981, ApJ,
  248, 1067

\bibitem[Patterson et~al.(2003)]{pat03suumas}
  Patterson, J., {et~al.}\ 2003, PASP, 115, 1308

\bibitem[{Pearson}(2006)]{pea06SH}
  {Pearson}, K.~J.\ 2006, MNRAS, 371, 235

\bibitem[{Pogrosheva} et~al.(2016a)]{pog16j2205atel9509}
  {Pogrosheva}, T., {et~al.}\ 2016a, Astron.\ Telegram, 9509

\bibitem[{Pogrosheva} et~al.(2016b)]{pog16j2205atel9510}
  {Pogrosheva}, T., {et~al.}\ 2016b, Astron.\ Telegram, 9510

\bibitem[{Pojma\'nski}(2002)]{ASAS3}
  {Pojma\'nski}, G.\ 2002, Acta\ Astron., 52, 397

\bibitem[{Popova}(1960)]{pop60irgem}
  {Popova}, A.\ 1960, Mitteil.\ Ver{\"{a}}nderl.\ Sterne, 464

\bibitem[{Popova} et~al.(2016)]{pop16j1511atel8843}
  {Popova}, E., {et~al.}\ 2016, Astron.\ Telegram, 8843

\bibitem[{Popowa}(1961)]{pop61kraur}
  {Popowa}, M.\ 1961, Astron.\ Nachr., 286, 81

\bibitem[{Pretorius} et~al.(2004)]{pre04j0137}
  {Pretorius}, M.~L., {Woudt}, P.~A., {Warner}, B., {Bolt}, G., {Patterson},
  J., \& {Armstrong}, E.\ 2004, MNRAS, 352, 1056

\bibitem[{Prieto} et~al.(2016)]{pri16asassn16kbasassn16kdatel9479}
  {Prieto}, J.~L., {Chomiuk}, L., {Strader}, J., {Morrell}, N., {Stanek},
  K.~Z., \& {Shappee}, B.~J.\ 2016, Astron.\ Telegram, 9479

\bibitem[{Prieto} et~al.(2013)]{pri13asassn13clatel5102}
  {Prieto}, J.~L., {et~al.}\ 2013, Astron.\ Telegram, 5102

\bibitem[{Quimby} and {Mondol}(2006)]{qui06j1202atel787}
  {Quimby}, R., \& {Mondol}, P.\ 2006, Astron.\ Telegram, 787

\bibitem[Robertson et~al.(1995)]{rob95eruma}
  Robertson, J.~W., Honeycutt, R.~K., \& Turner, G.~W.\ 1995, PASP, 107, 443

\bibitem[{Ross}(1927)]{ros27VS5}
  {Ross}, F.~E.\ 1927, AJ, 37, 155

\bibitem[{Satyvoldiev}(1972)]{sat72v893sco}
  {Satyvoldiev}, V.\ 1972, Astron.\ Tsirk., 711, 7

\bibitem[{Savoury} et~al.(2011)]{sav11CVeclmass}
  {Savoury}, C.~D.~J., {et~al.}\ 2011, MNRAS, 415, 2025

\bibitem[Schmeer et~al.(1992)]{sch92hvviriauc}
  Schmeer, P., Hurst, G.~M., Kilmartin, P.~M., \& Gilmore, A.~C.\ 1992, IAU\
  Circ., 5502

\bibitem[Schneller(1931)]{sch31hvvir}
  Schneller, H.\ 1931, Astron.\ Nachr., 243, 335

\bibitem[Shafter et~al.(1984)]{sha84irgem}
  Shafter, A.~W., Cowley, A.~P., \& Szkody, P.\ 1984, ApJ, 282, 236

\bibitem[{Shappee} et~al.(2014)]{ASASSN}
  {Shappee}, B.~J., {et~al.}\ 2014, ApJ, 788, 48

\bibitem[{Shears} et~al.(2008)]{she08j1227}
  {Shears}, J., {Brady}, S., {Foote}, J., {Starkey}, D., \& {Vanmunster}, T.\
  2008, J.\ Br.\ Astron.\ Assoc., 118, 288

\bibitem[{Shumkov} et~al.(2016a)]{shu16j2046atel9470}
  {Shumkov}, V., {et~al.}\ 2016a, Astron.\ Telegram, 9470

\bibitem[{Shumkov} et~al.(2016b)]{shu16j0547atel9616}
  {Shumkov}, V., {et~al.}\ 2016b, Astron.\ Telegram, 9616

\bibitem[{Shurpakov} et~al.(2012)]{shu12j2113atel4675}
  {Shurpakov}, S., {et~al.}\ 2012, Astron.\ Telegram, 4675

\bibitem[{Shurpakov} et~al.(2013a)]{shu13j0432atel5657}
  {Shurpakov}, S., {et~al.}\ 2013a, Astron.\ Telegram, 5657

\bibitem[{Shurpakov} et~al.(2013b)]{shu13asassn13akatel5083}
  {Shurpakov}, S., {et~al.}\ 2013b, Astron.\ Telegram, 5083

\bibitem[{Simonsen}(2011)]{sim11zcamcamp1}
  {Simonsen}, M.\ 2011, J.\ American\ Assoc.\ Variable\ Star\ Obs., 39, 66

\bibitem[{Siviero} and {Munari}(2016)]{siv16asassn16owatel9862}
  {Siviero}, A., \& {Munari}, U.\ 2016, Astron.\ Telegram, 9862

\bibitem[{Smart}(2013)]{IGSL}
  {Smart}, R.~L.\ 2013, VizieR\ Online\ Data\ Catalog, 1324

\bibitem[{Southworth} et~al.(2010)]{sou10SDSSeclCV}
  {Southworth}, J., {Copperwheat}, C.~M., {G{\"a}nsicke}, B.~T., \& {Pyrzas},
  S.\ 2010, A\&A, 510, A100

\bibitem[{Southworth} et~al.(2007)]{sou07SDSSCV2}
  {Southworth}, J., {Marsh}, T.~R., {G{\"a}nsicke}, B.~T., {Aungwerojwit}, A.,
  {Hakala}, P., {de Martino}, D., \& {Lehto}, H.\ 2007, MNRAS, 382, 1145

\bibitem[{Stanek} et~al.(2016a)]{sta16v5853sgratel9343}
  {Stanek}, K.~Z., {et~al.}\ 2016a, Astron.\ Telegram, 9343

\bibitem[{Stanek} et~al.(2016b)]{sta16asassn16maatel9669}
  {Stanek}, K.~Z., {et~al.}\ 2016b, Astron.\ Telegram, 9669

\bibitem[{Stanek} et~al.(2016c)]{sta16asassn16kdatel9469}
  {Stanek}, K.~Z., {et~al.}\ 2016c, Astron.\ Telegram, 9469

\bibitem[{Stanek} et~al.(2013)]{sta13asassn13akatel5082}
  {Stanek}, K.~Z., {et~al.}\ 2013, Astron.\ Telegram, 5082

\bibitem[Stellingwerf(1978)]{PDM}
  Stellingwerf, R.~F.\ 1978, ApJ, 224, 953

\bibitem[Szkody et~al.(2002)]{szk02SDSSCVs}
  Szkody, P., {et~al.}\ 2002, AJ, 123, 430

\bibitem[{Szkody} et~al.(2009)]{szk09SDSSCV7}
  {Szkody}, P., {et~al.}\ 2009, AJ, 137, 4011

\bibitem[{Szkody} et~al.(2003)]{szk03SDSSCV2}
  {Szkody}, P., {et~al.}\ 2003, AJ, 126, 1499

\bibitem[{Szkody} et~al.(2006)]{szk06SDSSCV5}
  {Szkody}, P., {et~al.}\ 2006, AJ, 131, 973

\bibitem[{Szkody} et~al.(2005)]{szk05SDSSCV4}
  {Szkody}, P., {et~al.}\ 2005, AJ, 129, 2386

\bibitem[{Szkody} et~al.(2007)]{szk07SDSSCV6}
  {Szkody}, P., {et~al.}\ 2007, AJ, 134, 185

\bibitem[Szkody and Howell(1992)]{szk92CVspec}
  Szkody, P., \& Howell, S.~B.\ 1992, ApJS, 78, 537

\bibitem[Szkody et~al.(1992)]{szk92hvviriauc}
  Szkody, P., Ingram, D., Schmeer, P., Midtskogen, O., Dahle, H., \& Bortle,
  J.~E.\ 1992, IAU\ Circ., 5516

\bibitem[Tappert and Bianchini(2003)]{tap03gzcnc}
  Tappert, C., \& Bianchini, A.\ 2003, A\&A, 401, 1101

\bibitem[Thorstensen et~al.(2002)]{tho02j2329}
  Thorstensen, J.~R., Fenton, W.~H., Patterson, J.~O., Kemp, J., Krajci, T., \&
  Baraffe, I.\ 2002, ApJ, 567, L49

\bibitem[{Thorstensen} et~al.(2015)]{tho15SDSSCVs}
  {Thorstensen}, J.~R., {Taylor}, C.~J., {Peters}, C.~S., {Skinner}, J.~N.,
  {Southworth}, J., \& {G{\"a}nsicke}, B.~T.\ 2015, AJ, 149, 128

\bibitem[{Tiurina} et~al.(2013)]{tiu13j0647atel4871}
  {Tiurina}, N., {et~al.}\ 2013, Astron.\ Telegram, 4871

\bibitem[Tsesevich(1967)]{GCVS2sup2}
  Tsesevich, V.~P.\ 1967, Second Supplement to {General Catalogue of Variable
  Stars}, second edition (Moscow: Astronomical Council of the Academy of
  Sciences in the USSR)

\bibitem[Uemura et~al.(2002)]{uem02j2329letter}
  Uemura, M., {et~al.}\ 2002, PASJ, 54, L15

\bibitem[{Uemura} et~al.(2004)]{uem04v344pav}
  {Uemura}, M., {Mennickent}, R., \& {Stubbings}, R.\ 2004, IBVS, 5569

\bibitem[{Vladimirov} et~al.(2013)]{vla13j0754atel5585}
  {Vladimirov}, V., {et~al.}\ 2013, Astron.\ Telegram, 5585

\bibitem[{Vladimirov} et~al.(2014)]{vla14j0553atel5983}
  {Vladimirov}, V., {et~al.}\ 2014, Astron.\ Telegram, 5983

\bibitem[{Vogt}(1983)]{vog83lateSH}
  {Vogt}, N.\ 1983, A\&A, 118, 95

\bibitem[Vogt and Bateson(1982)]{vog82atlas}
  Vogt, N., \& Bateson, F.~M.\ 1982, A\&AS, 48, 383

\bibitem[{Wakamatsu} et~al.(2017)]{wak17asassn16eg}
  {Wakamatsu}, Y., {et~al.}\ 2017, PASJ, submitted

\bibitem[Walker and Olmsted(1958)]{wal58CVchart}
  Walker, A.~D., \& Olmsted, M.\ 1958, PASP, 70, 495

\bibitem[Warner(1985)]{war85suuma}
  Warner, B.\ 1985, in Interacting Binaries, ed. P.~P. Eggleton, \& J.~E.
  Pringle (Dordrecht: D.\ Reidel Publishing Company), p.~367

\bibitem[Warner(1995)]{war95book}
  Warner, B.\ 1995, Cataclysmic Variable Stars (Cambridge: Cambridge University
  Press)

\bibitem[Wenzel(1993a)]{wen93akcnccycle}
  Wenzel, W.\ 1993a, Mitteil.\ Ver{\"{a}}nderl.\ Sterne, 12, 153

\bibitem[Wenzel(1993b)]{wen93akcnc}
  Wenzel, W.\ 1993b, IBVS, 3921

\bibitem[Whitehurst(1988)]{whi88tidal}
  Whitehurst, R.\ 1988, MNRAS, 232, 35

\bibitem[Williams(1983)]{wil83CVspec1}
  Williams, G.\ 1983, ApJS, 53, 523

\bibitem[{Williams} and {Darnley}(2016)]{wil16v5853sgratel9375}
  {Williams}, S.~C., \& {Darnley}, M.~J.\ 2016, Astron.\ Telegram, 9375

\bibitem[{Wils} et~al.(2010)]{wil10newCVs}
  {Wils}, P., {G{\"a}nsicke}, B.~T., {Drake}, A.~J., \& {Southworth}, J.\ 2010,
  MNRAS, 402, 436

\bibitem[{Witham} et~al.(2008)]{wit08IPHAS}
  {Witham}, A.~R., {Knigge}, C., {Drew}, J.~E., {Greimel}, R., {Steeghs}, D.,
  {G{\"a}nsicke}, B.~T., {Groot}, P.~J., \& {Mampaso}, A.\ 2008, MNRAS, 384,
  1277

\bibitem[{Wolf} and {Wolf}(1906)]{wol06awsge}
  {Wolf}, M., \& {Wolf}, G.\ 1906, Astron.\ Nachr., 170, 361

\bibitem[{Wood} and {Burke}(2007)]{woo07negSH}
  {Wood}, M.~A., \& {Burke}, C.~J.\ 2007, ApJ, 661, 1042

\bibitem[{Wood} et~al.(2011)]{woo11v344lyr}
  {Wood}, M.~A., {Still}, M.~D., {Howell}, S.~B., {Cannizzo}, J.~K., \&
  {Smale}, A.~P.\ 2011, ApJ, 741, 105

\bibitem[{Yamaoka} et~al.(2008)]{yam08j0406cbet1463}
  {Yamaoka}, H., {Itagaki}, K., {Kaneda}, H., {Jacques}, C., {Pimentel}, E.,
  {Maehara}, H., \& {Bolt}, G.\ 2008, Cent.\ Bur.\ Electron.\ Telegrams, 1463

\bibitem[{Yecheistov} et~al.(2013)]{yec13j0213atel5536}
  {Yecheistov}, V., {et~al.}\ 2013, Astron.\ Telegram, 5536

\bibitem[{Yecheistov} et~al.(2014)]{yec14j0558atel5905}
  {Yecheistov}, V., {et~al.}\ 2014, Astron.\ Telegram, 5905

\bibitem[{Zengin {\c C}amurdan} et~al.(2010)]{zen10v849eclCVs}
  {Zengin {\c C}amurdan}, D., {Ibano{\v g}lu}, C., \& M., {{\c C}amurdan}~C.\
  2010, New\ Astron., 15, 476

\bibitem[{Zwitter} and {Munari}(1996)]{zwi96CVspec}
  {Zwitter}, T., \& {Munari}, U.\ 1996, A\&AS, 117, 449

\end{thebibliography}
\end{document}